\documentclass[11pt,headsepline,cleardoubleempty,twoside,openright]{scrbook}
\pdfoutput=1
\usepackage{LSST_Observing_Strategy_White_Paper}

\begin{document}

\begin{titlepage}
\begin{center}

\vspace*{\stretch{2}}

{\Huge\bfseries\scshape
 Science-Driven Optimization \\ \vspace{\baselineskip}
 of the LSST Observing Strategy}

\vspace*{\stretch{2}}

\vspace*{\stretch{2.5}}

{\Large  Prepared by the LSST Science Collaborations,}\\
\vspace*{\stretch{0.15}}
{\Large with support from the LSST Project. }\\
\vspace*{\stretch{1}}

\begin{center}{
    {\large\bf Version 1.0}\\
    \vspace*{\stretch{0.1}}
    Most recent commit: \href{https://github.com/LSSTScienceCollaborations/ObservingStrategy/commit/fe3d2ad2eb87f6d94b88426d0a990ee5531ae6ad?diff=split}{\texttt{fe3d2ad}}\\
    \vspace*{\stretch{0.1}}
    (Mon, 14 Aug 2017 02:08:33 -0700)}\\
\end{center}

\end{center}
\end{titlepage}

% --------------------------------------------------------------------

\clearemptydoublepage
% Author list
\setcounter{chapter}{0}
\chapter*{Contributing Authors}
\def\chpname{authors}\label{chp:\chpname}
\addcontentsline{toc}{section}{Contributing Authors}
\markboth{}{}

% List of institutions:
\def\adler{Adler Planetarium, Chicago, IL, USA}
\def\aims{African Institute for Mathematical Sciences, 6 Melrose Road, Muizenberg 7945, South Africa}
\def\arizona{University of Arizona, Tucson, AZ, USA}
\def\berkeley{Physics Division, Lawrence Berkeley National Laboratory, 1 Cyclotron Road, Berkeley, CA, 94720, USA}
\def\caltech{California Institute of Technology, Pasadena, CA, USA}
\def\caltechgps{Division of Geological and Planetary Sciences, California Institute of Technology, Pasadena, CA 91125, USA}
\def\centrallancashire{University of Central Lancashire, Preston PR1 2HE, UK}
\def\cfa{Harvard-Smithsonian Center for Astrophysics, Harvard University, Cambridge, MA, USA}
\def\chicago{Department of Astronomy and Astrophysics, University of Chicago, 5640 South Ellis Avenue, Chicago, IL 60637, USA}
\def\cmu{Carnegie Mellon University, Pittsburgh, PA, USA}
\def\columbia{Columbia University, New York, NY, USA}
\def\cook{Cook Astronomical Consulting, USA}
\def\cuny{The City University of New York, New York, NY, USA}
\def\ctio{Cerro Tololo Inter-American Observatory, Casilla 603, La Serena, Chile}
\def\dearborn{University of Michigan–Dearborn, 4901 Evergreen Road, Dearborn, MI 48128, USA}
\def\delaware{University of Delaware, Department of Physics and Astronomy, 104 The Green, Newark, DE 19716, USA}
\def\drexel{Drexel University, Philadelphia, PA, USA}
\def\ennu{Department of Physics \& Astronomy/CIERA, Northwestern University, 2145 Sheridan Road, Evanston, IL, 60208, USA}
\def\fermilab{Fermilab, PO Box 500, Batavia, IL, 60510, USA}
\def\floridagulf{Florida Gulf Coast University, Fort Meyers, FL, USA}
\def\goddard{NASA Goddard Space Flight Center, 8800 Greenbelt Road, Greenbelt, MD 20771, USA}
\def\harvard{Department of Physics \& Department of Astronomy, 17 Oxford Street, Harvard University, Cambridge, MA, 02138, USA}
\def\ifa{Institute for Astronomy, University of Hawaii at Manoa, 2680 Woodlawn Drive, Honolulu, HI 96822, USA}
\def\ipac{IPAC, 770 South Wilson Ave., Pasadena, CA 91125, USA}
\def\irvine{University of California, Irvine, CA, USA}
\def\jpl{Jet Propulsion Laboratory, California Institute of Technology, 4800 Oak Grove Drive, Pasadena, CA 91109, USA}
\def\kicp{Kavli Institute for Cosmological Physics, University of Chicago, Chicago, IL 60637, USA}
\def\lcogt{LCOGT, University of California, Santa Barbara, CA, USA}
\def\lpc{Laboratoire de Physique de Clermont, N2P3/CNRS, 63178 Aubière Cedex, France}
\def\lpnhe{LPNHE, Barre 12-22, 1er s\'{e}tage, 4 Place Jussieu, 75252 Paris Cedex 05, France}
\def\lsst{LSST, 933 N. Cherry Ave., Tucson, AZ 85721, USA}
\def\mssl{Mullard Space Science Laboratory (MSSL), University College London (UCL), Surrey RH5 6NT, UK}
\def\msu{Department of Physics and Astronomy, Michigan State University, 5678 Wilson Road, Lansing, MI 48824, USA}
\def\nau{Dept. of Physics \& Astronomy, Northern Arizona University, NAU Box 6010, Flagstaff, AZ, 86011, USA}
\def\nso{National Solar Observatory, 3004 Telescope Loop, Sunspot, NM 88349, USA}
\def\noao{NOAO, 950 N. Cherry Ave., Tucson, AZ 85719}
\def\nyu{Center for Cosmology and Particle Physics, Department of Physics, New York University, 726 Broadway, 9th Floor, New York, NY 10003, USA}
\def\okc{The Oskar Klein Centre for Cosmoparticle Physics, Stockholm University, Stockholm, Sweden}
\def\nyuc{Center for Cosmology and Particle Physics, Department of Physics, New York University, 726 Broadway, 9th Floor, New York,  NY 10003, USA}
\def\osu{The Ohio State University, Columbus, OH, USA}
\def\oswego{State University of New York at Oswego, 7060 New York 104, Oswego, NY 13126, USA}
\def\oxford{Department of Physics, University of Oxford, Keble Road, Oxford, UK}
\def\penn{Department of Astronomy and Astrophysics, University of Pennsylvania, Philadelphia, PA, USA}
\def\pennstate{Pennsylvania State University, 514A Davey Lab
University Park, PA 16802, USA}
\def\pitt{Pittsburgh Particle Physics, Astrophysics, and Cosmology Center (PITT PACC), Physics and Astronomy Department, University of Pittsburgh, Pittsburgh, PA 15260, USA}
\def\princeton{Department of Astrophysical Sciences, Princeton University, Princeton, NJ 08544, USA}
\def\rice{Department of Physics and Astronomy, Rice University, Houston TX 77005-1892, USA}
\def\rutgers{Department of Physics and Astronomy, Rutgers the State University of New Jersey, 136 Frelinghuysen Road, Piscataway, NJ 08854 USA}
\def\uc{Department of Astronomy and Astrophysics, University of California, Santa Cruz, CA 95064, USA}
\def\scsu{Department of Physics, Southern Connecticut State University, 501 Crescent Street, New Haven, CT 06515, USA}
\def\seti{SETI Institute, 189 N. Bernardo Ave., Mountain View, CA, 94043, USA}
\def\ska{SKA South Africa, 3rd Floor, The Park, Park Road, Pinelands 7405, South Africa}
\def\sofia{SOFIA Science Center, NASA Ames Research Center, MS211-1, Moffett Field, CA 94035, USA}
\def\slac{SLAC National Accelerator Laboratory, 2575 Sand Hill Road, MS29, Menlo Park, CA 94025, USA}
\def\somewhere{Some Institute, Somewhere, \ldots}
\def\soton{School of Physics and Astronomy, University of Southampton, Southampton, SO17 1BJ, UK}
\def\stanford{Physics Department, Stanford University, Stanford, CA, 94305, USA}
\def\stsci{Space Telescope Science Institute, Baltimore, MD, USA}
\def\texastech{Department of Physics, Texas Tech University, Box 41051 Lubbock, TX 79409-1051, USA}
\def\toronto{Dunlap Institute \& Department of Astronomy and Astrophysics, University of Toronto, 50 St George Street, Toronto, ON M5S 3H4, Canada}
\def\ucd{University of California, Davis, CA, USA}
\def\ucl{Department of Physics and Astronomy, University College London, Gower Street, London WC1E 6BT, UK}
\def\unab{Universidad Andr\'{e}s Bello, 13 Nte. 798, Vi\~{n}a del Mar, Región de Valparaíso, Chile}
\def\unt{University of North Texas, 1155 Union Cir, Denton, TX 76203, USA}
\def\usno{US Naval Observatory, 10391 West Naval Observatory Road, Flagstaff, AZ 86001, USA}
\def\utaustin{University of Texas at Austin,  Austin, TX, 78712, USA}
\def\uw{University of Washington, Department of Astronomy, University of Washington, 3910 15th Avenue NE, Seattle, WA, 98195, USA}
\def\uwe{The eScience Institute, University of Washington, Seattle, WA, 98195, USA}
\def\westernwash{Western Washington University, 516 High Street, Bellingham, WA 98225, USA}
\def\vanderbilt{Vanderbilt University, 2201 West End Ave, Nashville, TN 37235, USA}
\def\yale{Department of Astronomy, Yale University, P.O. Box 208101, New Haven, CT 06520-8101, USA}

% List of authors
% ===============
% 1. Chapter editors:
\author{Phil~Marshall}{drphilmarshall}{\slac}
\author{Timo~Anguita}{tanguita}{\unab}
\author{Federica~B.~Bianco}{fedhere}{\nyu}
\author{Eric~C.~Bellm}{ebellm}{\caltech}
\author{Niel~Brandt}{nielbrandt}{\pennstate}
\author{Will~Clarkson}{willclarkson}{\dearborn}
\author{Andy~Connolly}{connolly}{\uw}
\author{Eric~Gawiser}{egawiser}{\rutgers}
\author{\v{Z}eljko~Ivezi\'{c}}{ivezic}{\uw}
\author{Lynne~Jones}{rhiannonlynne}{\uw}
% ...
% {\it your name here},
% {and the LSST Project and Science Collaborations}

\noindent A graphical representation of the contributions made to this white paper can be found on this paper's \href{https://github.com/LSSTScienceCollaborations/ObservingStrategy/graphs/contributors}{GitHub repository}.

\vspace{3\baselineskip}
\hrule
\vspace{-3\baselineskip}
\theendnotes

% --------------------------------------------------------------------

\tableofcontents
\label{toc}

% --------------------------------------------------------------------

\clearemptydoublepage
\setcounter{chapter}{0}
\chapter*{Preface}
\def\chpname{preface}\label{chp:\chpname}
\addcontentsline{toc}{section}{Preface}
\markboth{}{}

\noindent The Large Synoptic Survey Telescope (LSST) is a dedicated
ground-based astronomical facility whose goal is to provide a database
of high fidelity images and object catalogs that enable a wide range of
science investigations. With its 9.6~square degree field of view and
effective aperture of 6.7~meters, it will be able to survey the Southern half of the sky every few nights (on average), building up a 10-year, 900-frame movie of the ever-changing cosmos. Its community of scientists will be able to make major contributions in the fields of extragalactic astronomy and cosmology, the study of our Milky Way, its local environment and its stellar populations, solar system science, and time domain astronomy.

\noindent As its name suggests, LSST is designed to
carry out a large synoptic survey: it has a baseline observing strategy, simulations of which demonstrate that the data required for
the promised science can be delivered.
However, this baseline strategy may well not be the {\it
best} way to schedule the telescope. Smaller, specialized surveys are
likely to provide high scientific value, as is optimizing the pattern of
repeated sky coverage.  The baseline strategy is not set in
stone, and can and will be optimized: even small changes could result in
significant improvements to the overall science yield. How can we design
an observing strategy that maximizes the scientific output of the LSST
system?

\noindent The LSST Observing Strategy community formed in July 2015 to
tackle this problem. Drawn primarily (but not exclusively) from the team of people engaged in the LSST construction Project, and the set of LSST ``Science Collaborations'' who are engaged in preparing to exploit the LSST data, we are working together
to use software tools provided by the LSST Project
to evaluate simulations of the LSST survey (also provided by the
Project) specifically for the science that we each care most about. In
this way, we aim to give sustainable, quantitative feedback about how any
proposed observing strategy would impact the performance of our science
cases, and so enable good decisions to be made when the telescope
schedule is eventually set up.

\noindent This white paper is a compendium of ideas and results
generated by the community, assembled so that everyone can follow along
with the analysis. It is a living document, whose purpose is to bind
together the group of people who are thinking about the LSST observing
strategy problem, and facilitate their collective discussion and
understanding of that problem (a process we might think of as  ``cadence
diplomacy''). Its audience is the LSST science community, and most notable its Science
Advisory Committee and Project Scientist who together will in the end decide what the LSST observing strategy will be. This white paper is {\it
the} vehicle for the community to communicate to the LSST Project, while
the baseline observing strategy continues to be improved.

\noindent The white paper's modular design allows pieces of it to be
split off and published in a series of snapshot journal papers, as the
various metric analyses reach maturity. The white paper itself will be
continuously published on
\href{https://github.com/LSSTScienceCollaborations/ObservingStrategy}{\GitHub}
and advertized periodically on \href{http://arxiv.org}{astro-ph}. This
white paper is large, but we hope that its hyperlinked structure helps
our community quickly find the science cases that they are most interested in,
starting from the \hyperref[toc]{table of contents}.

\noindent The LSST observing strategy evaluation and optimization
process will be as open and inclusive as possible. New community members
are welcome at any time; we explain how to get involved in \autoref{chp:intro}.  We invite all stakeholders to participate.

\vspace{2\baselineskip}

{\raggedleft \credit{drphilmarshall}, \credit{ivezic} and \credit{bethwillman} \\
 \medskip \hspace{0.8\linewidth} \it August 12, 2017.}

\clearpage

% --------------------------------------------------------------------

\clearemptydoublepage
\setcounter{chapter}{0}
\chapter*{Summary}
\def\chpname{summary}\label{chp:\chpname}
\addcontentsline{toc}{section}{Summary}
\markboth{}{}

% Short (c. 250 word) summary for arxiv posting:

\noindent
The Large Synoptic Survey Telescope is designed to provide an 
unprecedented optical imaging dataset that will support investigations 
of our Solar System, Galaxy and Universe, across half the sky and over ten
years of repeated observation.
However, exactly how the LSST observations will be taken (the observing strategy or ``cadence'') is not yet finalized.
In this dynamically-evolving community white paper, we explore how the
detailed performance of the anticipated science investigations is
expected to depend on small changes to the LSST observing strategy.
Using realistic simulations of the LSST schedule and observation
properties, we design and compute diagnostic metrics and Figures of
Merit that provide quantitative evaluations of different observing
strategies, analyzing their impact on a wide range of proposed science
projects.
This is work in progress: we are using this white paper to communicate
to each other the relative merits of the observing strategy choices that
could be made, in an effort to maximize the scientific value of the
survey.
The investigation of some science cases leads to suggestions for new
strategies that could be simulated and potentially adopted.
Notably, we find motivation for exploring departures from a spatially
uniform annual tiling of the sky: focusing instead on different parts of
the survey area in different years in a ``rolling cadence'' is likely to
have significant benefits for a number of time domain and moving object
astronomy projects.
The communal assembly of a suite of quantified and homogeneously coded
metrics is the vital first step towards an automated, systematic,
science-based assessment of any given cadence simulation, that will
enable the scheduling of the LSST to be as well-informed as possible.

\clearpage

% --------------------------------------------------------------------

\chapter[Introduction]{Introduction}
\def\chpname{intro}\label{chp:\chpname}

Chapter editors:
\credit{connolly},
\credit{ivezic},
\credit{drphilmarshall},
\credit{michaelstrauss}.

The Large Synoptic Survey Telescope (LSST) is a dedicated optical
telescope with an effective aperture of 6.7 meters, currently under
construction on Cerro Pach\'on in the Chilean Andes.  The telescope
and camera will have a huge field of view, 9.6 deg$^2$, and the
\'etendue, i.e., the product of collecting area and field of view will
be significantly larger than any other optical telescope.  Thus this telescope
is designed for wide-field deep imaging of the sky; its mantra is
``Wide-Fast-Deeo'', i.e., the ability to cover large swaths of sky
(``Wide'') to faint magnitudes (``Deep'') in a short amount of time
(``Fast''), allowing it to scan the sky repeatedly.  LSST will image
in six broad filters, $ugrizy$, spanning the optical band from the
atmospheric cutoff in the ultraviolet to the limit of CCD sensitivity
in the near-infrared.

The science case for the LSST is based broadly on four science themes:
\begin{itemize}
\item Probing the distribution of dark matter and measuring the effects
  of dark energy (via measurements of gravitational
  lensing, satellite galaxies and streams, large-scale structure,
  clusters of galaxies, and supernovae);
\item Exploring the transient and variable universe;
\item Studying the structure of the Milky Way galaxy and its neighbors
  via resolved stellar populations;
\item Making an inventory of the Solar System, including Near Earth
  Asteroids and Potential Hazardous Objects, Main Belt Asteroids, and
  Kuiper Belt Objects.
\end{itemize}

These themes, together with {\em many} other science applications, are
described in detail in the
\href{http://lsst.org/scientists/scibook}{LSST Science Book}, produced
by the LSST Project Team and Science collaborations in 2009.  The
present white paper represents an important next step in science
planning beyond the Science Book.  In particular, we now need to
quantify how well the LSST (for a given realization of its observing
strategy, or ``{\em cadence}'') will be able to carry out its science
goals; we will then use this quantification to refine and optimize
the cadence itself.  To zeroth order, the large \'etendue of LSST
allows it to meet all its science goals with a single dataset with a
``universal'' cadence.  However, small perturbations to such a universal strategy are expected to yield significant improvements to certain science investigations. This document describes the design of the current ``baseline'' LSST
cadence, and various ways in which it could be further refined to
optimize the overall science output of the survey.  As we describe in detail
below, we quantify the effectiveness of a given cadence realization to
meet science goals by defining a series of quantitative {\em metrics}.
Any given realization will be more favorable for some science areas,
and less so for others; the metrics allow us to quantify this, and
optimize the overall cadence for the broadest range of LSST science
areas.

Since the Science Book was written, some of the
science themes described there have evolved or become obsolete,
while new science opportunities and ideas have arisen.  Moreover, our
understanding of the capabilities (such as system response and
therefore depth, telescope optics, and so on) have matured
considerably.  The present document endeavors to explore the principal
science themes as described in the Science Book, but is not slaved to
them. Where appropriate, we point out relevant updates to the
Science Book.

% --------------------------------------------------------------------

\section{Synoptic Sky Surveying}
\def\secname{intro:baseline}\label{sec:\secname}

\credit{ivezic},
\credit{drphilmarshall},
\credit{michaelstrauss}

The LSST defined a so-called ``baseline cadence'', described in the
\href{http://adsabs.harvard.edu/abs/2008arXiv0805.2366I}{LSST overview
paper} \citep{IvezicEtal2008} and Chapter 3 of the Science Book \citep{2009arXiv0912.0201L}.  This was used
to demonstrate that LSST could meet its basic science goals, and indeed
the formal
\href{https://docushare.lsstcorp.org/docushare/dsweb/Get/LPM-17}{science
requirements}.\footnote{\url{https://docushare.lsstcorp.org/docushare/dsweb/Get/LPM-17}}    As described in these references, the default LSST
exposure is 15 seconds, and all exposures are taken in pairs called  ``{\em visits},'' before the telescope is slewed to a neighboring field.
Any given field is observed twice on a given night to allow
preliminary trajectories of asteroids to be determined.

The baseline cadence optimizes the amount of sky covered in any given
night (subject to the constraint of observing at airmass less than 1.4
throughout), and allows the entire sky visible at any time of the year
to be covered in about three nights.  The cadence is designed to give
uniform coverage at any given time, and reaches the survey goals for
measuring stellar parallax and proper motion over the ten-year survey.
The survey requirements on depth lead to roughly 825 visits (summing
over the six filters) in the 10-year LSST survey to any given point on
the sky.  (The exact return position depends on the dithering strategy assumed; while \OpSim does not include dither patterns explicitly, they can be applied to the field centers in post-processing and before image depth, visit spacing etc.\ are calculated as a function of position on the sky.) The resulting Wide-Fast-Deep (WFD) component of the survey covers
roughly 18,000 deg$^2$ of high Galactic latitude sky, and requires about
85\% of the available observing time in its current baseline realization.

There are obvious science cases that the WFD survey does
not address, and thus the remaining 15\% of the telescope time in the
baseline cadence is devoted to a series of specialized surveys.  They
are as follows:
\begin{itemize}
\item Imaging at low Galactic latitudes.  This is currently defined as
  a wedge which is broader closer to the Galactic Center,
  corresponding roughly to a locus of constant stellar density.  In
  this region, the number of repeat observations is reduced, given the
  confusion limit in the stacked LSST data.
\item Imaging in the South Celestial Cap.  The airmass limit of 1.4
  restricts observations to declination $> -75^\circ$, thus missing
  large fraction of both the Magellanic Clouds.
  Observations are done in the Cap to cover this region of sky, again
  to shallower depth.
\item Imaging in a series of four or more {\em Deep Drilling Fields},   single
  pointings in which
  we will obtain roughly 5 times more exposures in all
  filters in order to go about a magnitude fainter in the stacked
  data, as well as to get better sampled light curves of variable
  objects. See \href{https://www.lsst.org/sites/default/files/docs/sciencebook/SB_2.pdf}{Chapter 2 of the Science Book} \citep{2009arXiv0912.0201L} for more details; the four field positions are listed on the \href{https://www.lsst.org/News/enews/deep-drilling-201202.html}{LSST website}.\footnote{\url{https://www.lsst.org/News/enews/deep-drilling-201202.html}}
\item Imaging in the Northern portions of the Ecliptic Plane.  The
  airmass limit of 1.4 restricts us to declination $< +15^\circ$, which
  means that a significant fraction of the Ecliptic Plane
  is uncovered.  By observing with a reduced cadence close to the
  Ecliptic Plane north of this limit, we will be able to significantly
  increase the fraction of Near-Earth Asteroids and Main Belt
  Asteroids for which LSST obtains orbits.
\end{itemize}

We note that the LSST Science Requirements Document ``...assumes a nominal
10-year duration with about 90\% of the observing time allocated for the main
LSST survey.'', and thus 10\% of observing time is left for all other programs.
However, if the system will perform better than expected, or if science priorities
will change over time, it is conceivable that 90\% could be modified and become
as low as perhaps 80\%, with the observing time for other programs thus doubled.

The LSST Project has developed an ``Operations Simulator'' (\OpSim),
% described in detail in Chapter~\ref{chp:cadexp} of this paper,
which
includes a realistic model of observatory operations, including time
required for camera readout, slew time, filter exchange, as well as time
loss due to clouds.  Given a set of so-called ``proposals''
that set the priorities of which fields to
observe at any given time, \OpSim has developed a series of realizations
of the series of pointings that make up the ten-year LSST survey.
The baseline cadence is a specific realization of the
\OpSim output, which meets the LSST survey requirements, following the
rules briefly outlined above. \OpSim is described in more detail in \autoref{sec:cadexp:opsim} below.

Again, while the baseline cadence demonstrates that the LSST is capable
of meeting its stated science goals, it is not optimized for all
science, and \autoref{chp:cadexp} of this document describes a
series of experiments varying the assumptions in \OpSim.  In
\autoref{chp:specialsurveys}, we explore additional ideas for future
experiments to be done in \OpSim -- many of them, we hope, inspired by
requests from the community.
\OpSim itself may be viewed as a prototype for the LSST Scheduler
software (a part of the Observatory Control System).  \OpSim and Scheduler teams
will develop this vital piece of observatory infrastructure code based on the \OpSim experience; these
teams' role in the continuing exploration of the LSST observing strategy
is laid out in more detail in \autoref{sec:intro:timeline} below.

\navigationbar

% --------------------------------------------------------------------

\section{Evaluating and Optimizing the LSST Observing Strategy}
\def\secname{intro:evaluation}\label{sec:\secname}

\credit{drphilmarshall}

Given a realization of the LSST observing strategy (i.e., an output of
\OpSim), our first task is to quantify how well it supports the (many)
science investigations that LSST will enable.  As the algorithms controlling
\OpSim are varied, some investigations will benefit, while others may suffer.
By quantifying this for each investigation, we can determine which cadence
maximizes the science potential overall of the investigation.

Therefore, we need a {\it science-based evaluation of the baseline LSST
observing strategy and its variants}. After simulating a sample
observing schedule consistent with this strategy (see
\autoref{chp:cadexp}), we then need to quantify its value to each
science investigation team.  This is what the LSST Simulations team's ``Metric
Analysis Framework'' (\MAF\footnote{\url{https://sims-maf.lsst.io/}}) was designed to enable: science case investigators
can now design quantitative evaluations of the outputs of \OpSim, to
answer the question, ``how good would that observing strategy be, for my
science?'' These ``metrics'' can be coded against the \MAF python API, and
shared among the LSST science community at the
\href{https://sims-maf.lsst.io/metricList.html#contributed-mafcontrib-metrics}{\simsMafContrib}
online repository.
All of the community-developed \MAF metrics described in this paper can be found there; a complete listing of all available \MAF metrics is provided in the appendix. The more basic metrics provided by the \MAF package itself allow a wide range of diagnostic analysis; some basic dither patterns may also be applied and investigated within the \MAF.

Once the fiducial strategy has
been evaluated in this way, then any other strategy can be evaluated
in the same terms, using the same code.  We will then be able to iterate towards a
science-optimized strategy.

With this program in mind, it makes sense to define {\it one ``Figure
of Merit'' (FoM) per science investigation}, that captures the value of  the
observing strategy under consideration to that science team. This FoM
will probably be a function of several ``diagnostic metrics'' that quantify
lower-level features of the observing sequence.  For FoMs
to be {\it directly} comparable between disparate science projects,  they would
need to be dimensional and have the same units. Even without such a universal scheme, useful comparisons between science investigations can still be made using the percentage improvement or degradation in each project's FoM.
% \footnote{One natural choice of units for Figures of Merit could be the
% {\it information  gained} by the science team, in bits. This is a
% well-defined statistical quantity, albeit not yet one in common use.}
% A given observing schedule's value would then depend on both this
% information gain, but also {\it how much that information is worth to
% the whole community}. It is at this point that the debate could become
% heated: probably the best we could do in cadence diplomacy would be to
% to quantify all the information gains implied by each proposed change to
% the baseline  observing strategy, combine them to see whether it makes
% everyone happy, and iterate. In this way we might hope to minimize the
% debates about the less quantifiable worth of each piece of information.}

It may not always be straightforward to define a Figure of
Merit for some science cases, but the diagnostic metrics that they will likely depend
upon will be easier to derive. Writing this white paper is an
opportunity to think through the FoM for each science
project that we as a community want to carry out, and how that measure
of success is likely (or even known) to depend on metrics that
summarize the observing sequence presented to us. We return to the
question of FoM design in \autoref{sec:intro:guidelines:metrics} below.

Thinking about the problem in terms of science projects, each with a
Figure of Merit, encourages us to design modular document sections
for this white paper, with
one science project and one FoM per section. These modular
sections ought then to be easily extracted, combined and edited into
publishable articles. They will also naturally lead to the definition of
a suite of \MAF metrics that can be evaluated on any future \OpSim output
database.  Tabulating the values of the diagnostic metrics and the FoM,
for different cadences, for each science case, will be very helpful for
this purpose.

\navigationbar

% --------------------------------------------------------------------

\section{Influencing the LSST Observing Schedule}
\def\secname{intro:schedule}\label{sec:\secname}

What we find regarding the impact of various observing strategies on
science performance will be of great interest to the LSST project, as it
works towards defining the observatory's observing schedule. How will
the findings presented in this paper be taken forward?

In this section we describe the mechanisms by which community input to
the developing observing schedule will be absorbed, and explain how
we will distil the vital information that the project needs from our
\OpSim / \MAF analyses. We then provide a target timeline for the provision of community input.

% - - - - - - - - - - - - - - - - - - - - - - - - - - - - - - - - - - -

\subsection{How will the results of our analyses be used?}
\label{sec:\secname:useage}

\credit{bethwillman}, \credit{connolly}, \credit{ivezic}

Through the end of construction and commissioning, this community
Observing Strategy White Paper will remain a living document that is
{\it the} vehicle for the community to communicate to the LSST Project
regarding the Wide-Fast-Deep and special survey observing strategies.
The Project Scientist will {\it synthesize the results presented in this paper
and develop an appropriate response strategy with the Scheduler and \OpSim
teams,} with support from the Science Advisory Committee and Survey Strategy Committee (see below).

As described in the LSST Operations Plan,\footnote{The LSST Operations Plan is available to LSST project and science community members as document LPM-181, \url{https://docushare.lsstcorp.org/docushare/dsweb/Services/LPM-181}. It contains descriptions for the various individual roles and groups referred to here.} the observing
strategy will continue to be refined and optimized during operations:
the Survey Scientist will chair a Survey Evaluation Working Group that
will evaluate quarterly the current and expected performance of the
survey (using software which we expect to be evolved from \OpSim and \MAF).
This group may include representation
from the Survey Support Scientist, the Pipelines and Data Products
group, the Data Processing group, the Camera team, and science
community.  The science community representation may be implemented as a
sub-group of the Science Advisory Committee.

% - - - - - - - - - - - - - - - - - - - - - - - - - - - - - - - - - - -

\subsection{Communicating via Science Case Conclusions}
\def\secname{intro:evaluation:caseConclusions}\label{sec:\secname}

\credit{ivezic}

In order to consolidate the various constraints on the observing
strategy by different science cases, and provide high signal to noise
data for the project to take forward, each science case in this white paper will conclude by answering the following ten
questions probing different aspects of the observing strategy:

% AUTHORS! COPY AND PASTE THE FOLLOWING INTO YOUR SCIENCE SECTION!
%
% \subsection{Conclusions}
%
% Here we answer the ten questions posed in
% \autoref{sec:intro:evaluation:caseConclusions}:
%
\begin{description}

\item[Q1:] {\it Does the science case place any constraints on the
tradeoff between the sky coverage and coadded depth? For example, should
the sky coverage be maximized (to $\sim$30,000 deg$^2$, as e.g., in
Pan-STARRS) or the number of detected galaxies (the current baseline of 18,000 deg$^2$)?}

% \item[A1:] ...

\item[Q2:] {\it Does the science case place any constraints on the
tradeoff between uniformity of sampling and frequency of  sampling? For
example, a ``rolling cadence'' can provide enhanced sample rates over a part
of the survey or the entire survey for a designated time at the cost of
reduced sample rate the rest of the time (while maintaining the nominal
total visit counts).}

% \item[A2:] ...

\item[Q3:] {\it Does the science case place any constraints on the
tradeoff between the single-visit depth and the number of visits
(especially in the $u$-band where longer exposures would minimize the
impact of the readout noise)?}

% \item[A3:] ...

\item[Q4:] {\it Does the science case place any constraints on the
Galactic plane coverage (spatial coverage, temporal sampling, visits per
band)?}

% \item[A4:] ...

\item[Q5:] {\it Does the science case place any constraints on the
fraction of observing time allocated to each band?}

% \item[A5:] ...

\item[Q6:] {\it Does the science case place any constraints on the
cadence for deep drilling fields?}

% \item[A6:] ...

\item[Q7:] {\it Assuming two visits per night, would the science case
benefit if they are obtained in the same band or not?}

% \item[A7:] ...

\item[Q8:] {\it Will the case science benefit from a special cadence
prescription during commissioning or early in the survey, such as:
acquiring a full 10-year count of visits for a small area (either in all
the bands or in a  selected set); a greatly enhanced cadence for a small
area?}

% \item[A8:] ...

\item[Q9:] {\it Does the science case place any constraints on the
sampling of observing conditions (e.g., seeing, dark sky, airmass),
possibly as a function of band, etc.?}

% \item[A9:] ...

\item[Q10:] {\it Does the case have science drivers that would require
real-time exposure time optimization to obtain nearly constant
single-visit limiting depth?}

% \item[A10:] ...

\end{description}

These questions were designed by the LSST Project Scientist (this section's author) so as to provide the specific details needed to prepare the next generation of \OpSim simulations, as part of an on-going investigation described in the next section.

% - - - - - - - - - - - - - - - - - - - - - - - - - - - - - - - - - - -

\subsection{Timeline}
\def\secname{intro:timeline}\label{sec:\secname}

\credit{connolly},
\credit{ivezic},
\credit{drphilmarshall}

The intersection between the community observing strategy investigation and the scheduler software development,
and the expected support that can be provided by the Project, is
outlined in \autoref{fig:timeline}. From the point of view of the community, this timeline contains a number of interesting features:

\begin{description}

\item{\textbf{Update of the Baseline Cadence and Exploration of Rolling
Cadences.}} During the development of version 1.0 of this white paper, the LSST
Project has been developing an enhanced operations simulator code,
\OpSim~4. This will be used to generate, by the end of 2017, a new set of
observing strategies, including some that have a ``rolling cadence''
component (see \autoref{sec:rolling}). This is in response to the results presented in the science
chapters of this paper. Analysis of these simulations would form the
backbone of an updated, version~2.0 of this white paper, with existing
science cases being updated to include quantitative assessment of the
new \OpSim~4 simulations, and new science cases being identified and
investigated. This updated white paper will be developed throughout 2018.

\item{\textbf{The definition of the Deep Drilling Fields (DDFs) and
associated cadences.}} The 2017 simulations will all continue
to use the baseline DDF cadence while exploring the properties of the main survey. However, by December 2017, the LSST
will issue a call for proposals to define the cadence and properties of
the currently selected DDFs, and to propose a new set of DDFs. To enable
this, the project will publish the known boundary conditions for
additional DDFs (e.g.\ the definition of a DDF, the current division of
survey time, constraints on the number of filter exchanges that can be
accomplished within a night, the expected range of integration times).
This call will include a request to describe the science objectives of
new DDFs, the position on the sky of these DDFs, the depth required as a
function of filter, the required cadence of observations, and the
metrics that will demonstrate that the DDF observations meet their
science requirements (these metrics do not need to be written within the
framework of \MAF). Delivery of these white papers by the community will
be expected by the end of April 1, 2018.  The LSST Observing Strategy GitHub
repository can support the development and aggregation of these DDF
white papers. The LSST Science Advisory Committee (SAC) will be asked to make a recommendation to the
project by the end of May 2018 on which DDFs and cadences should be
considered, and the project will respond to these recommendations by
the end of 2018. The Project's \OpSim team will support
this effort by evaluating the proposed cadences and DDFs. This may be in
the form of simulations (for new cadence proposals) or through an
evaluation of the visibility and properties of the fields relative to
the nominal performance of the LSST system.

\item \textbf{The definition of Figures of Merit (FoMs) for the LSST
survey strategy}. By March 2019 the project will issue a request to
to the community to update this Observing Strategy white paper with
\MAF-coded Figures of Merit, to evaluate both the Wide-Fast-Deep survey and various extensions
(\eg the Galactic plane, Northern Ecliptic Spur, South Celestial
Cap, the DDFs, and a set of community-proposed ``special'' or ``mini-'' surveys) for their impacts on specific science cases. The timeline for mini-survey proposals is given in \autoref{fig:timeline}; some preliminary ideas for special surveys can be found in \autoref{chp:specialsurveys}. These FoMs will be
required for the Project to evaluate the efficacy of different survey
strategies on a range of LSST science (e.g.\ the trade-off between a
rolling cadence for supernova classification vs transient detection or
long period variability will need to be explored quantitatively). The
requested delivery date for these \MAF FoMs into the Observing Strategy
White Paper will be Aug 1, 2019. This will leave time for a Survey
Strategy Committee (SSC; see below) to undertake trade studies that
incorporate the community-provided FoMs. Details of the design of the
FoMs (including units, thresholds, speed) will be described at a later
date (prior to December 2019). If Project resources can be allocated to
the process, then the \OpSim team will support the writing
of the FoMs with advice and tutorials on the use of \OpSim~4, but
the Observing Strategy white paper community will be expected to deliver
their metrics as \MAF code.  By the summer of  2020, just prior to the
start of commissioning phase, MAF and Opsim tools will be finalized and a
series of simulated surveys and supporting documentation delivered
to the SAC and the SSC, which will be asked to recommend the initial
observing strategy. By early 2021, an initial survey strategy will be
announced and a baseline simulation that reproduces that strategy will
be published.

\item \textbf{Establishment of a Survey Strategy Committee (SSC)}. Given
the delivery of the FoMs, the project will establish a committee by March
2020 to evaluate competing survey strategy proposals and to propose a
survey strategy for commissioning (early 2021) and operation (late 2022) of the full LSST camera.
This committee will be comprised of project and non-project personnel, and
will include the LSST Project Scientist. The SAC will be asked to
make recommendations for committee membership. The SSC will report to
the LSST Director until the end of LSST construction and commissioning.
In January 2021, based on the recommendation made by the SSC, the
project will announce an initial survey strategy and publish a baseline
simulation that reproduces that strategy. If Project resources can be
allocated to the process, then the \OpSim team might support
the committee by helping to generate the proposed survey strategies.

\end{description}

%%%%%%%%%%%%%%%%%%%%%%%%%%%%%%%%%
\begin{figure}[t!]
\centering\includegraphics[angle=0,width=0.95\linewidth,clip]{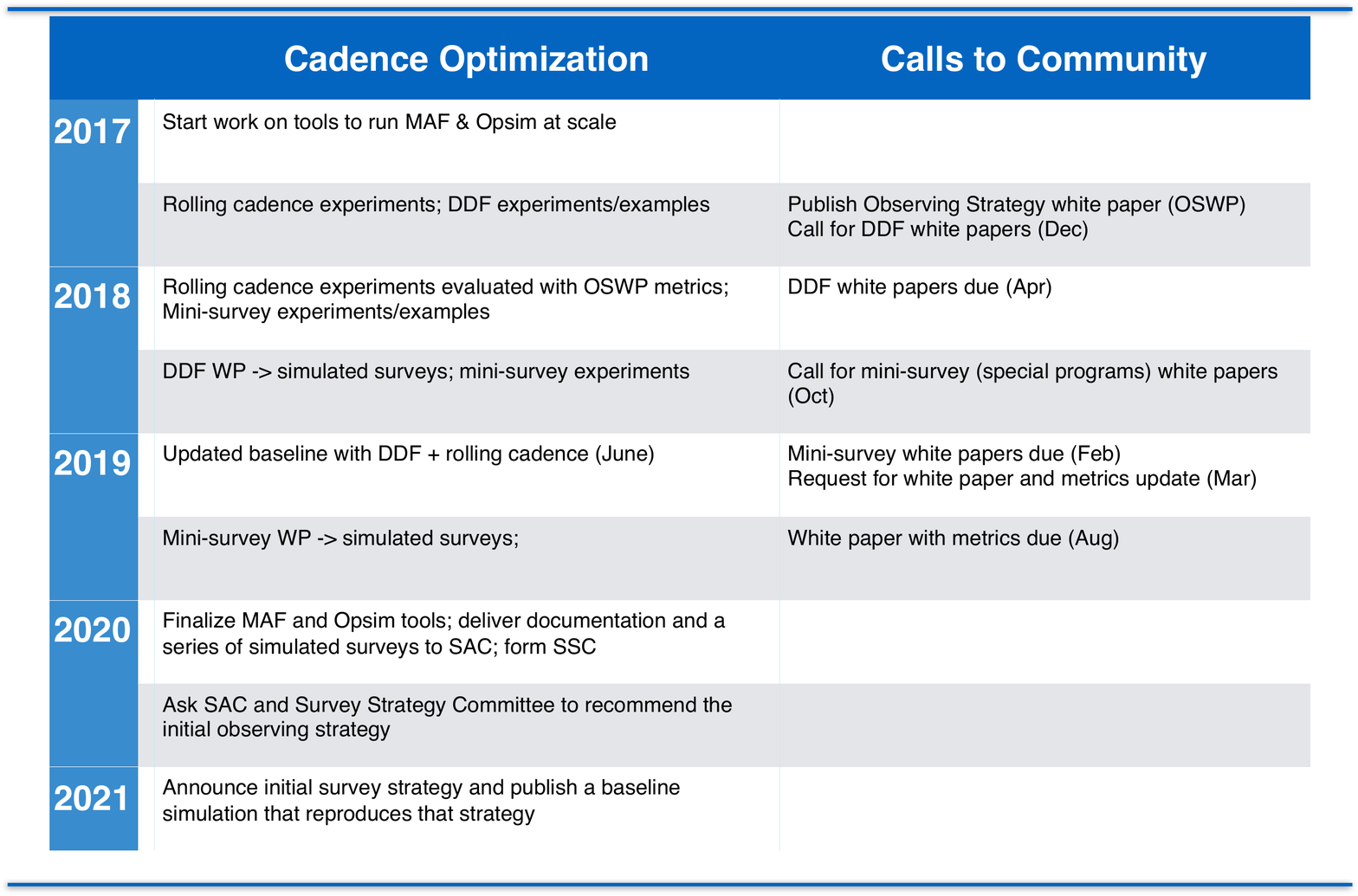}
\vskip -0.4in
\caption{Target timeline for the iterated optimization of the LSST observing strategy through 2021.}
\label{fig:timeline}
\end{figure}
%%%%%%%%%%%%%%%%%%%%%%%%%%%%%%%%%

It's important to note that the dates for this timeline are {\it
targets}. Since the deliverables are dependent on the availability of
project resources, these milestones should be considered as those we
could achieve given our best effort. Likewise, given the limited
availability of resources in the SOCS and Scheduler engineering team,
support of community members who wish to use \OpSim~v4 will be on a best
effort basis. \OpSim v4 will be delivered as a Docker\footnote{\url{https://www.docker.com/what-docker}} container and its
use and operation will be documented, but there will be no guarantee of
support for, or timeliness in response to requests for support from,
community users. The solution to this problem is to work together: the LSST Observing Strategy community, represented here by this white paper, is already developing the skills to perform and analyze LSST operations simulations: by learning from each other, we can produce high quality quantitative conclusions for the Project to act upon.

\navigationbar

% --------------------------------------------------------------------

\section{Guidelines for Contributors}
\def\secname{intro:guidelines}\label{sec:\secname}

\credit{drphilmarshall}

Contributions to this community effort are welcome from everyone. In
this section we give brief guidelines for how to make a contribution,
and how you should structure that contribution.

\subsection{How to Get Involved}

The first thing you should do is browse the current version of the white
paper, which you should be able to \href{http://ls.st/iw2}{view on
\GitHub}. You can download the  continuously-compiled
\href{http://www.slac.stanford.edu/~digel/ObservingStrategy/whitepaper/LSST_Observing_Strategy_White_Paper.pdf}{latest
version of the PDF document}, which is hyper-linked for easy navigation.
You will then be able to provide good feedback, which you should do via
the
\href{https://github.com/LSSTScienceCollaborations/ObservingStrategy/issues}{\GitHub
issues}. Note that the white paper chapters and sections each have a
list of editors and contributors, hyperlinked to the contributing author
list which shows their GitHub usernames: when joining (or starting) a
conversation on the issues, please do {\@}mention people by
their usernames to draw their attention to your comment. The chapter
editors will be especially effective at helping you find answers to
questions and guidance about your science case. Please search the
existing issues first: there might be a
conversation already taking place that you could join. New issues are
most welcome: we'd like to make this white paper as comprehensive as
possible.

To edit the white paper, you'll need to
\href{https://help.github.com/articles/fork-a-repo/}{``fork'' its
repository}. You will then  be able to edit the paper in your own
fork, and when you are ready,  submit a
\href{https://help.github.com/articles/using-pull-requests/}{``pull
request''} explaining what you are doing and the new version that  you
would like to be accepted. It's a good idea to submit this pull
request sooner rather than later, because associated with it will be a
discussion thread that the writing community can use to discuss your
ideas with you. For help getting started with \git and \GitHub, please
see this
\href{https://github.com/drphilmarshall/GettingStarted#top}{handy
guide}.

\subsection{Writing Science Cases}

For a high-level justification of the following design, please see
\autoref{sec:intro:evaluation} above. In short, we're aiming for modular
science sections (that are easy to write in parallel, and then
re-arrange into other publications later) that are each focused on {\it
one science project each}, and quantified by one (or maybe two) Figures
of Merit (which will likely depend on other, lower-level diagnostic
metrics).

At the beginning of each science chapter there will be a brief
\textit{introduction} that outlines the commonality of the key science
projects contained in that chapter. The individual science sections
following this introduction will then need to describe the particular
discoveries and measurements that are being targeted in each
\textit{science case}.

It will be helpful to think of these science
cases as investigations that the section leads {\it actually plan to
do}. Thinking this way means that each individual section can follow the
tried and tested format of an {\it observing proposal}: a brief
description of the investigation, with references, followed by subsections
describing the analysis
of its technical feasibility. The latter is where the \MAF analysis
should go. Like an observing proposal, each section will seek to
demonstrate the science performance achievable given various assumptions
about the time that could be awarded, or in our case, the survey that
could be delivered.

For an example of how all this could look, please see the
\hyperref[sec:lenstimedelays]{lens time delays section}. While the \MAF
analysis in this science case is still in progress, the suggested
structure of the science case can be seen. Template latex files for the
chapters and sections can be found in the
\href{https://github.com/LSSTScienceCollaborations/ObservingStrategy}{\GitHub
repository}.

\subsection{Metric Quantification}
\label{sec:\secname:metrics}

The feasibility of each science case will need to be quantified using
the \MAF framework, via a set of metrics (a Figure of Merit, and some
diagnostic metrics)  that need to be computed for any given observing
strategy in order to quantify the impact of that cadence on the
described science.

In many (or perhaps all) cases, a Figure of Merit will be a
\textit{precision} (\ie a percentage statistical uncertainty) on a
astrophysical model parameter, assuming negligible bias in its
inference. Precision is usually what we need to forecast in order to
convince observatory time allocation committees to give us telescope
time, and so it makes sense to focus on it here too.

Early on in a metric analysis, it may not be possible to compute a
science case's Figure of Merit, most likely because to do so would
require a large simulation program to capture the response of the
parameter measurement to the observing strategy. At this early stage, it
makes sense to look for simple {\it proxies} that scale the same way as
model parameter precision. For example, we might expect the precision on
a set of luminosity function  parameters to scale with the square root
of the number of objects in the sample, and so $\sqrt{N}$ could be a
sensible proxy for the Figure of Merit. Provided we get the scaling
right, we can then compare different observing strategies by looking  at
{\it the percentage change in the Figure of Merit,} and arguing that
this will correspond approximately to the same percentage change in the
ideal case.

Each science section needs to conclude with a discussion of any risks
that have been identified, and how these could be mitigated. What does
this mean? Each science project will have a threshold acceptable
Figure of Merit value, as well as a target (or ``design'') value.  If
an observing strategy gives an FoM value below the threshold, it
is very important that we know about it.  Optimizing all science cases
in such a complex and diverse set is not really the best way of thinking
about LSST's scheduling task: rather, {\it what we are really
trying to do is minimize global unhappiness with the LSST observing
strategy.} The comparisons between different simulated strategies will
help make the case for any changes to the baseline strategy, and in the
short term provide motivation for
\href{https://github.com/LSSTScienceCollaborations/ObservingStrategy/blob/master/opsim/README.md}{proposed
new \OpSim simulation runs.}

For some science sections we will have only a metric design, without an
implementation. As this white paper  evolves, many of these designs will
be realized and put into action. At first, though, the discussion of
risks to these science cases will necessarily be minimal, containing
only predictions for how the Figure of Merit is likely to vary among
observing strategies. These ``ideas'' sections will be presented as
sub-sections of a ``Future Work'' section, on the grounds that the
quantitative analysis is still to come. As its \MAF-based evaluation and
investigation proceeds, a science case will graduate into the main
part of the chapter and become a ``results'' section. The results
sections are clearly visible from the Table of
Contents. To find a Future Work subsection on any particular topic,
you can use the search facility on your PDF viewer.

When does an  ``ideas'' section become a ``results'' section? {\it As
soon as science  performance is quantified using any of the outputs from
\OpSim.} There is  a learning curve associated with the \MAF, but
metrics should be able to be designed before the \MAF documentation is
even opened. And since \MAF metrics always work with \OpSim outputs, any
quantitative analysis that is focused towards those outputs will have
the potential to grow into one of the \MAF analyses we need. The decision to upgrade a further work sub-section into its own science section should be made collectively, with the chapter editors. The editor in chief has the final say: for version 1.0 this role was filled by \credit{drphilmarshall}.

\subsection{Proposing New Simulations}

Before we can optimize the LSST observing strategy we must first
evaluate its current version for all the science cases we care about.
The logical point at which to propose a new \OpSim simulation, capturing
a novel aspect of the observing strategy, is {\it after} evaluating the
baseline  cadence (and others). The discussion section of your science
case is a good place to suggest new \OpSim simulations for further
testing by the community; there is also an
\href{https://github.com/LSSTScienceCollaborations/ObservingStrategy/blob/master/opsim/README.md}{online
suggestion board} for ideas for new  simulations to be registered and
shared.

\navigationbar

% --------------------------------------------------------------------

\section{Outline of This Paper}
\def\secname{intro:outline}\label{sec:\secname}

The rest of this white paper is structured as follows. In
\autoref{chp:cadexp} we describe a number of \OpSim simulated observing
schedules (``cadences'') explored by the LSST Sims team in summer 2015
in preparation for this paper: they include a ``baseline cadence'', and
then some small but interesting perturbations to it. Then, we present
the science cases considered so far, organised into the following
chapters:

\begin{itemize}
    \item \autoref{chp:solarsystem}: \nameref{chp:solarsystem}
    \item \autoref{chp:galaxy}: \nameref{chp:galaxy}
    \item \autoref{chp:variables}: \nameref{chp:variables}
    \item \autoref{chp:transients}: \nameref{chp:transients}
    \item \autoref{chp:MCs}: \nameref{chp:MCs}
    \item \autoref{chp:agn}: \nameref{chp:agn}
    \item \autoref{chp:cosmo}: \nameref{chp:cosmo}
    \item \autoref{chp:specialsurveys}: \nameref{chp:specialsurveys}
    \item \autoref{chp:wfirst}: \nameref{chp:wfirst}
\end{itemize}

Finally, in \autoref{chp:tradeoffs} we bring the results of all the
science metric analyses  together and discuss the tensions between
them, and the trade-offs that we can anticipate having to make.
This final chapter will serve as this work's set of running conclusions.

\navigationbar

% The following chapter shows a template introduction and
% science case section for you to work from. The latter is checked into
% the repository as
% \href{https://github.com/LSSTScienceCollaborations/ObservingStrategy/blob/master/whitepaper/section-template.tex}{\texttt{section-template.tex}}.
% For an example of a section being developed according to the above guidelines,
% please take a look at \autoref{sec:lenstimedelays}.

% % ====================================================================
%
% \setcounter{chapter}{-1}
% \chapter{Template Science Chapter}
% \def\chpname{example}\label{chp:\chpname}
%
% \noindent {\it
% Editor Name(s)
% }
%
% % --------------------------------------------------------------------
%
% \section{Introduction}
% \label{sec:\chpname:intro}
%
% General introduction to the chapter's science projects.
%
% Overview of observing strategy needed by those projects, bringing
% out common themes or points of tension.
%
% % --------------------------------------------------------------------
%
% \input{section-template}
%
% % --------------------------------------------------------------------

% --------------------------------------------------------------------

% --------------------------------------------------------------------

\chapter[Some Example Observing Strategies]{The Operations Simulator and its Outputs}
\def\chpname{cadexp}\label{chp:\chpname}

Chapter editors:
\credit{ivezic},
\credit{yoachim},
\credit{rhiannonlynne}.

Contributing authors:
\credit{kem0cook},
\credit{StephenRidgway},
\credit{drphilmarshall}

% --------------------------------------------------------------------

\section*{Summary}
\addcontentsline{toc}{section}{~~~~~~~~~Summary}

In this chapter
%(which was originally prepared as \href{http://www.astro.washington.edu/users/ivezic/lsst/cadexp2.pdf}{an
%LSST DM Sims standalone report})
we analyze and compare the performance of a number of simulated LSST
observing strategies (``cadences'')
which were developed in support of the LSST 2015 Observing
Strategy Workshop.  The Baseline Cadence,
\opsimdbref{db:baseCadence}, was found to be adequate, and replaces the previous
version (\texttt{opsim3.61}).
Simulations that only implemented the Wide, Fast, Deep Cadence proposal imply a
``best-case scenario'' margin for the number of visits of about 40\% relative to {\it the design
specifications} for the main survey sky coverage (18,000 sq.deg.) and the number of
visits per field (825, summed over all bands) from the Science Requirements Document (SRD),
and assuming {\it perfect} dithering\footnote{With a fill factor of 0.9 for the 9.6 sq.deg. large
field of view, it takes 1.72 million visits to meet the SRD specifications when a perfect
redistribution of the field overlap coverage is assumed.}.
This margin can be used to increase the sky coverage of the main survey, the total
number of visits per field, or to enhance special programs, such as
Deep Drilling fields and Galactic plane coverage. Several  simulations
analyzed here quantitatively explore these strategic options.
Additional simulations show that the effects of variations of the
visit exposure time in the  range 20-60 seconds on survey
efficiency can be predicted using simple efficiency estimates. Various
modifications of baseline cadence (e.g. Pan-STARRS-like cadence,  no
visit pairs, sequences with 3 and 4 visits) indicate a large parameter
space for further optimization, especially for time-domain
investigations and detailed coverage of special sky regions.

% --------------------------------------------------------------------

\section{Introduction}
\def\secname{cadexp:intro}\label{sec:\secname}

With the release of version 3.3.5 of the Operations Simulator (\OpSim, see \autoref{sec:intro:baseline}) code for simulating LSST
deployment, and the active development of the Metrics Analysis Framework
(\MAF,  currently version 0.2) for analyzing \OpSim outputs, we were able to undertake systematic and massive investigations of
various LSST deployment strategies.

The optimization of the ultimate LSST observing strategy will be done
with significant input from  the community. To facilitate this
process, the first of a series of meetings, the ``LSST \& NOAO
Observing Cadences Workshop'', was held during the
\href{https://project.lsst.org/meetings/ocw}{LSST 2014 meeting} in
Phoenix, AZ, August 11-15, 2014. A subsequent workshop, the ``LSST
Observing Strategy Workshop'',  was held
\href{http://lsstsciencecollaborations.github.io/ObservingStrategy/}{after
the LSST 2015 meeting} in Bremerton, WA, August 20-22, 2015.

In part as a preparation for the second workshop, the LSST
Simulations Team and the Project Science Team designed, executed
and analyzed a number of simulated surveys.  The cadence strategies
for these surveys were designed to
study the impact of various strategy variations on the scientific
potential of LSST\@.
Analysis of these
\href{http://opsim.lsst.org:8080}{simulated surveys} is presented here,
based on
\href{https://confluence.lsstcorp.org/display/SIM/MAF+documentation}{\MAF} reports.

\listofopsimdbs

\navigationbar

% --------------------------------------------------------------------

\section{The LSST Operations Simulator, \OpSim}
\def\secname{cadexp:opsim}\label{sec:\secname}

\OpSim is a software tool that runs a survey simulation with given science driven desirables;
a software model of the telescope and its control system; and models of weather and other environmental variables. The
output of such a simulation is an ``observation history,'' which is a record of times, pointings and associated environmental  data  and  telescope  activities  throughout  the  simulated  survey.  This history can be examined to assess
whether the simulated survey would be useful for any particular purpose or interest.

In most of the simulations discussed in this document, the \OpSim scheduler balances several different observing proposals:
\begin{itemize}
\item {\bf Wide, Fast, Deep (WFD):} The WFD is the primary LSST survey, taking 85-95\% of the observing time and covering 18,000 square degrees of sky, in the current implementation spanning the declination range from about $-65$ to about $+5$~degrees (the total
sky area between these limits is about 20,500 square degrees, but a region aligned with the Galactic Plane is not included in WFD).
% @ivezic: Our SAC reviewer suggested including this dec range, and also that we say what the galactic latitude cut is. Can you check the former and add the latter, please? Thanks!
This observing proposal is usually configured to attempt observing pairs spaced $\sim40$\ minutes apart.  This temporal spacing is designed to optimize the detection of moving solar system objects. This proposal typically balances the six $ugrizy$ filters, observing each field every $\sim3$\ days.
\item {\bf North Ecliptic Spur (NES):} The NES is an extension to reach the Ecliptic at higher airmass than the WFD survey typically covers.  The NES typically does not include the $uy$ filters.
\item {\bf Galactic Plane:} This proposal covers the region where LSST is expected to be highly confused by the density of stellar sources.  Typically takes fewer total exposures per field than the WFD survey and does not collect in pairs. This region is defined by the galactic
latitude limit $|b| <  (1-l/90^\circ) \, 10^\circ$ for $0^\circ < l < 90^\circ$ and analogously (mirror image) for $270^\circ < l < 360^\circ$.
\item {\bf South Celestial Pole (SCP):} The SCP is an extension to higher airmass than the WFD to cover the region south of declination
$-65$ degrees.  This proposal includes $ugrizy$, but takes fewer exposures per field than the WFD and does not collect in pairs.
\item {\bf Deep Drilling Fields (DD):} The Deep Drilling Fields are single pointings that are observed in extended sequences. The DD proposals often include certain filter combinations to ensure that near-simultaneous color information is available for variable and transient objects. Four of the LSST Deep Drilling fields have been selected and announced. It is expected that there will be more DD fields selected for the final survey. Most of the simulations here include five DD fields.
\end{itemize}

One of the more unique constraints on the \OpSim scheduler is that it highly penalizes, and thus avoids, filter changes.  With it's large field of view, LSST filter changes take about two minutes to complete. The filters are also large and heavy enough that we want to minimize wear on the filter changing mechanism.

\navigationbar

\section{The Baseline Observing Strategy}
\def\secname{cadexp:baseline}\label{sec:\secname}

The official (managed by the LSST Change Control Board)
Baseline Cadence, \opsimdbref{db:baseCadence},
was produced by the 3.3.5 version of
\OpSim. We first introduce this Baseline Cadence,
and then proceed with the analysis of other simulations that modify the baseline
observing strategy in various informative ways. Suggestions for
further tool development, and a summary of the main cadence questions
addressed here are given in \autoref{sec:cadexp:summary} below.

% - - - - - - - - - - - - - - - - - - - - - - - - - - - - - - - - - -

%%%%%%%%%%%%%%%%%%%%%%%%%%%%%%%%%
\opsimdb[db:baseCadence]{minion\_1016}{The Baseline Cadence.}
%%%%%%%%%%%%%%%%%%%%%%%%%%%%%%%%%
%   RunName minion_1016
%   MinDist2Moon 30
%   MinAlt 20.0
%   MaxAlt 86.5
%   TimeFilterChange 120.0
%   TimeReadout 2.0
%%%%%%%%%%%%%%%%%%%%%%%%%%%%%%%%%

%%%%%%%%%%%%%%%%%%%%%%%%%%%%%%%%%
\begin{figure}[tbh!]
%\vskip -1.3in
\begin{subfigure}[b]{0.49\textwidth}
\includegraphics[angle=0,width=0.99\hsize,clip]{figs/cadence/minion_1016_Median_airmass_r_band_all_props_OPSI_SkyMap.pdf}
\end{subfigure}
\hfill
\begin{subfigure}[b]{0.49\textwidth}
\includegraphics[angle=0,width=0.99\hsize,clip]{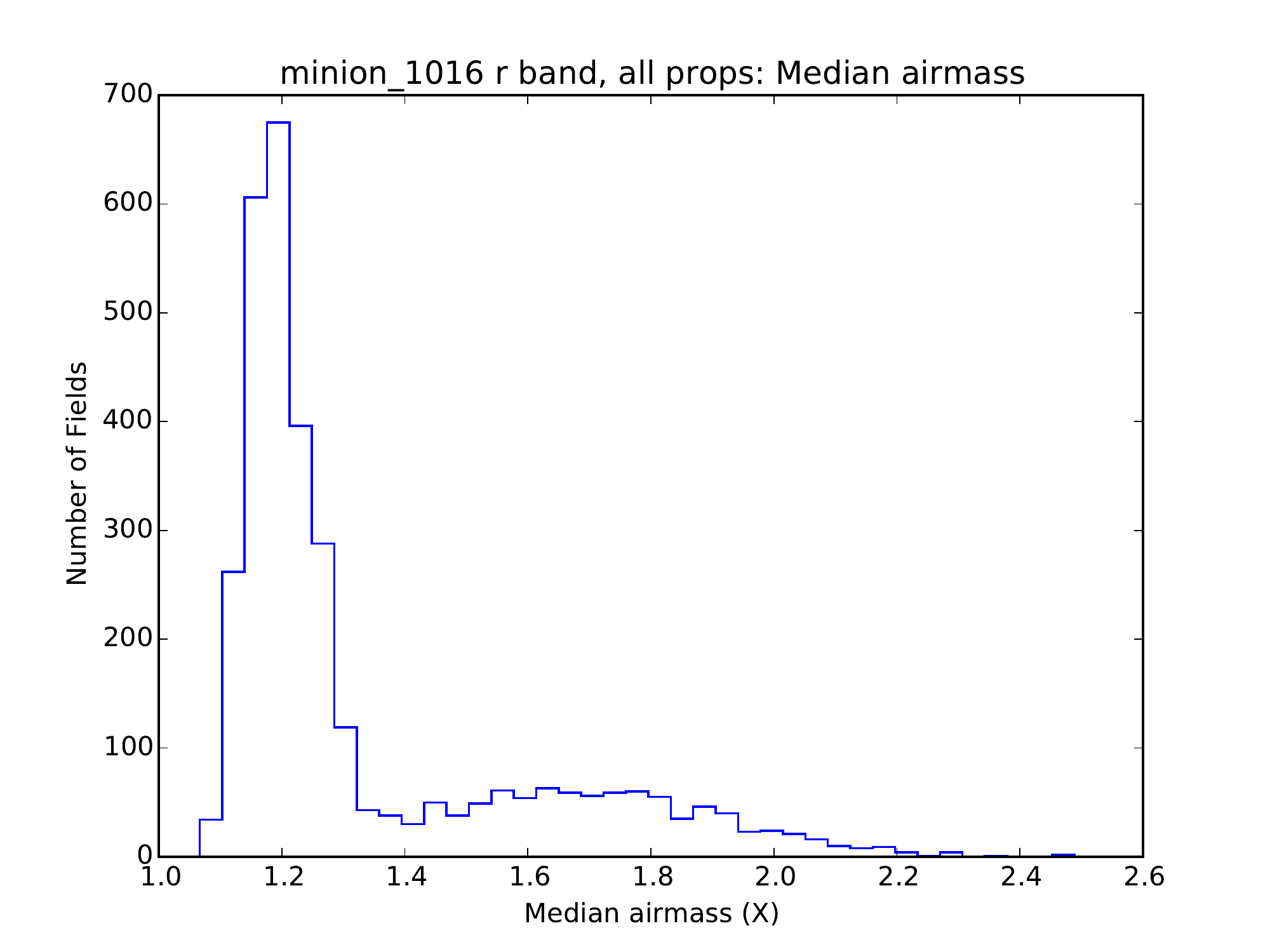}
\end{subfigure}
%\vskip -1.3in
\caption{The median airmass in the $r$ band across the sky for simulated cadence
\opsimdbref{db:baseCadence} is shown in Aitoff projection of equatorial coordinates
in the left panel. The red line shows the Ecliptic and the blue line shows the Galactic
equator. The blue curve splits to enclose the so-called ``Galactic confusion zone''. The corresponding
airmass histogram is shown in the right panel. For the main survey area, the maximum
allowed airmass was set to 1.5. }
\label{fig:airmassenigma}
\end{figure}
%%%%%%%%%%%%%%%%%%%%%%%%%%%%%%%%%

%%%%%%%%%%%%%%%%%%%%%%%%%%%%%%%%%
\begin{figure}[t!]
\vskip -0.1in
\includegraphics[angle=0,width=0.99\hsize,clip]{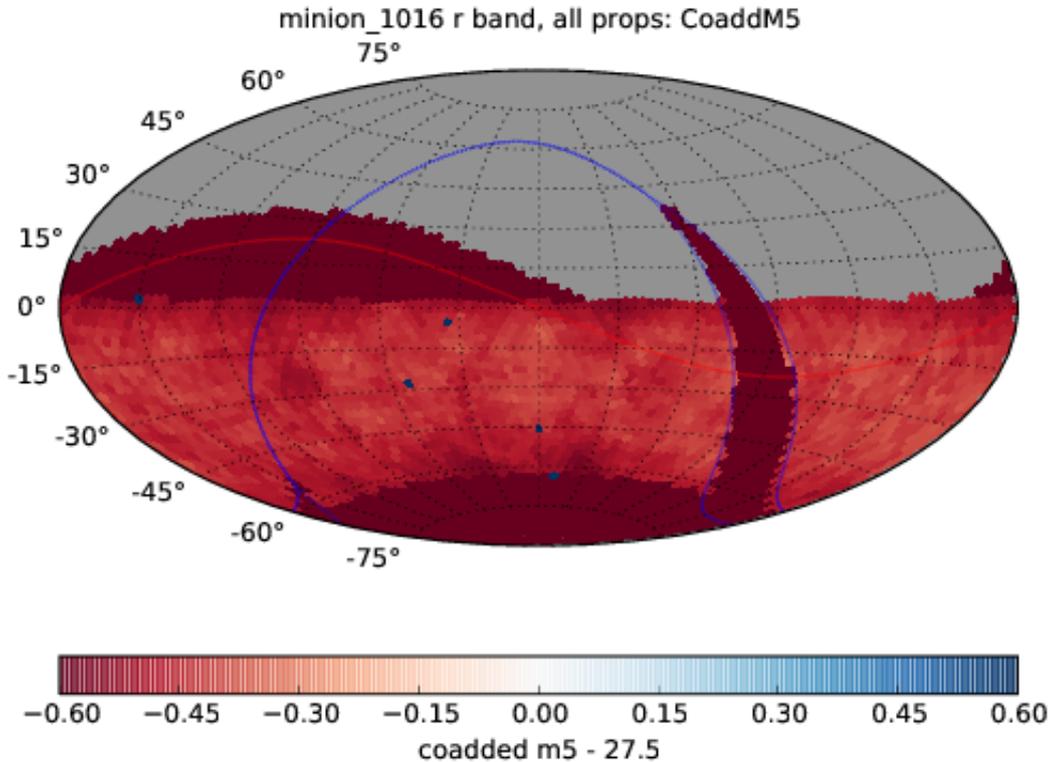}
\vskip -0.5in
\caption{The coadded $5\sigma$ depth for point sources in the $r$ band
across the sky at the end of 10 years for simulated cadence \opsimdbref{db:baseCadence} is shown
in Aitoff projection of equatorial coordinates. The red line shows the Ecliptic and
the blue line shows the Galactic equator (it bifurcates around the so-called
``Galactic confusion zone''). The median value across the WFD Cadence area
is 27.1, with RMS scatter of only 0.04 mag. The small dark dots are Deep Drilling
fields, with a median $5\sigma$ depth of 28.6.}
\label{fig:coaddm5enigma}
\end{figure}
%%%%%%%%%%%%%%%%%%%%%%%%%%%%%%%%%

%%%%%%%%%%%%%%%%%%%%%%%%%%%%%%%%%
\begin{figure}[t!]
\vskip -0.0in
\includegraphics[angle=0,width=0.49\hsize,clip]{figs/cadence/minion_1016_Parallax_Normed_All_Visits_non-dithered_HEAL_SkyMap.pdf}
\includegraphics[angle=0,width=0.49\hsize,clip]{figs/cadence/minion_1016_Proper_Motion_Normed_All_Visits_non-dithered_HEAL_SkyMap.pdf}
\includegraphics[angle=0,width=0.49\hsize,clip]{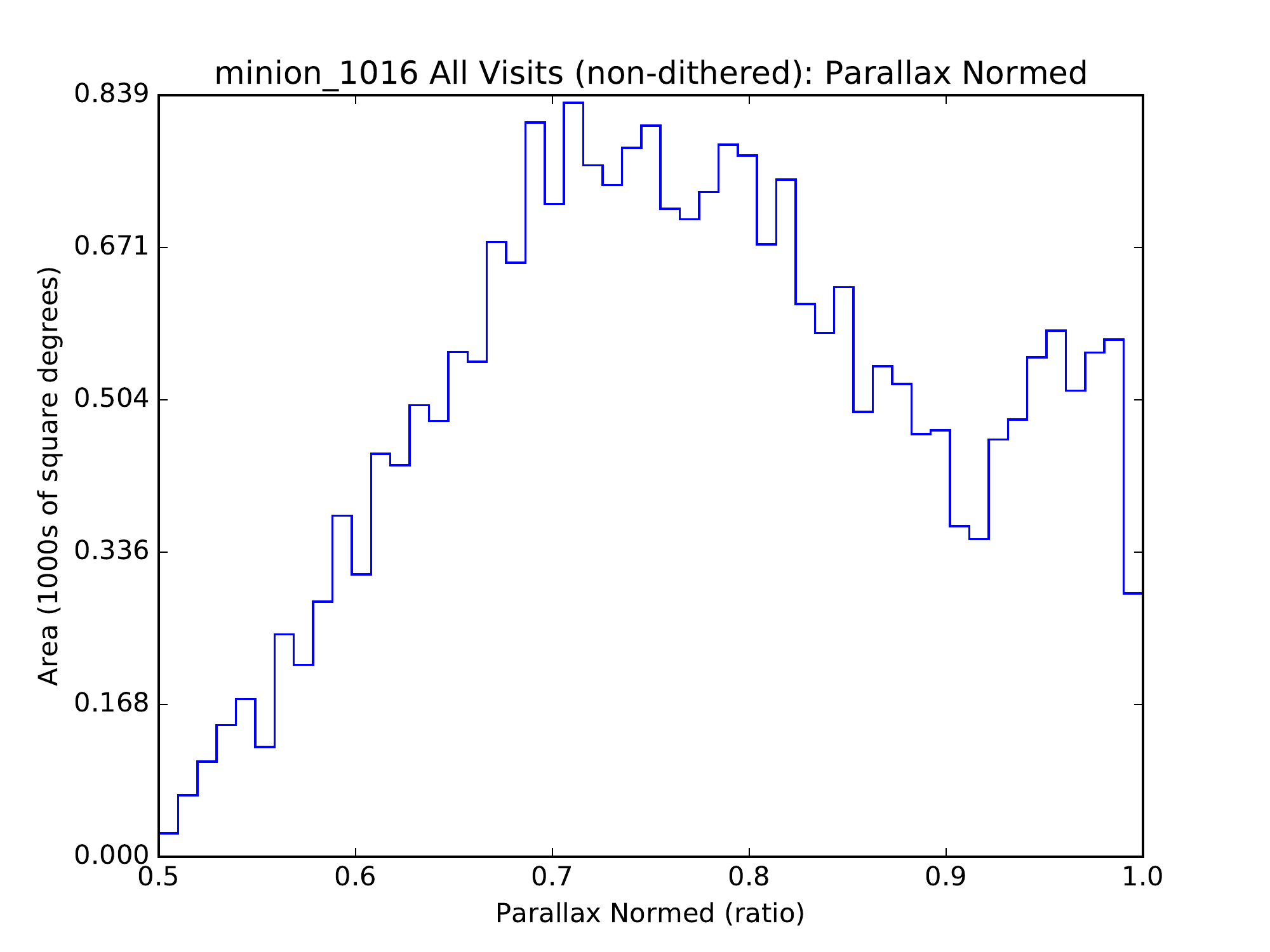}
\includegraphics[angle=0,width=0.49\hsize,clip]{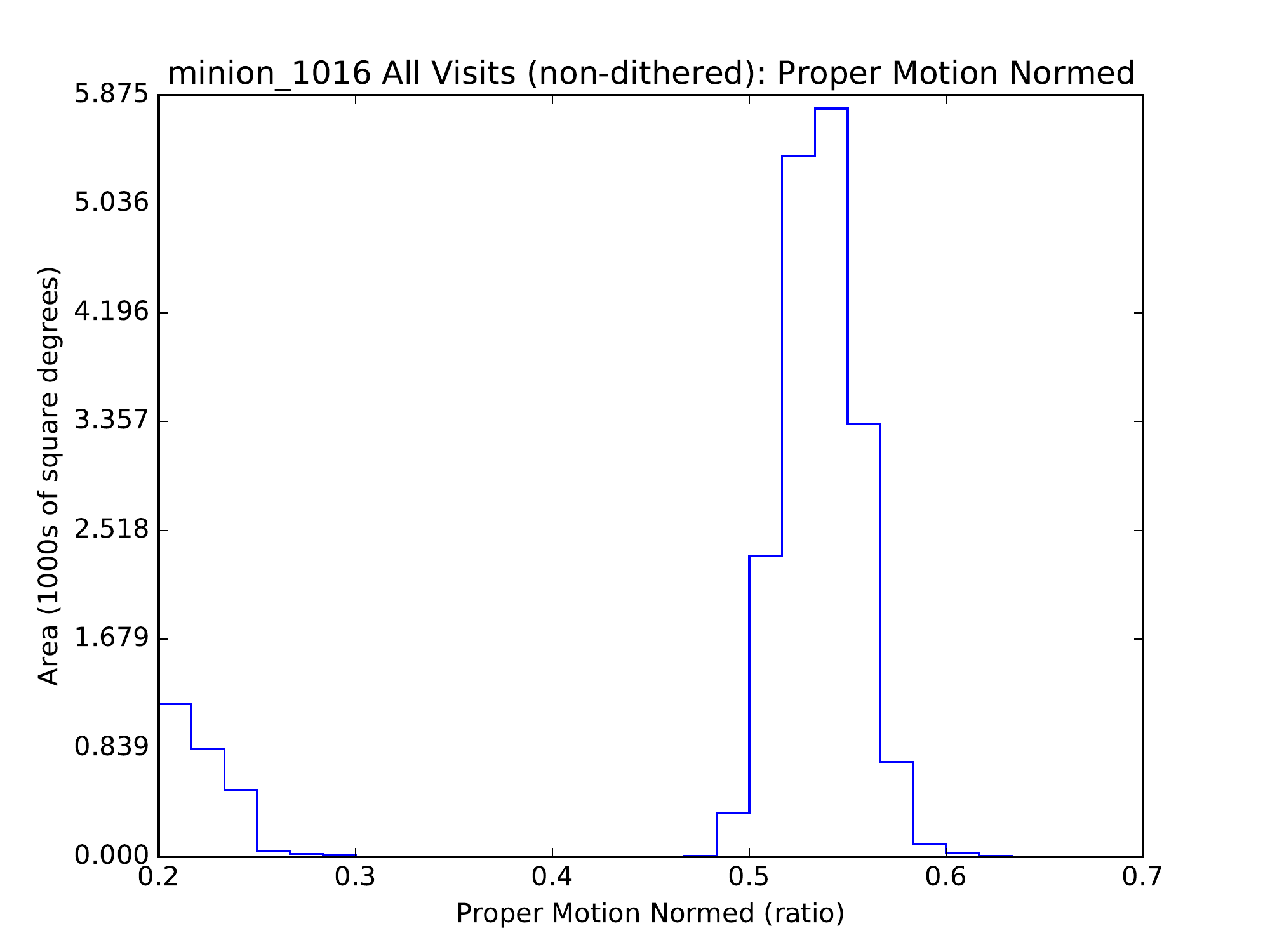}
\vskip -0.1in
\caption{The trigonometric parallax errors (left) and proper motion errors (right), normalized
by the values for idealized perfectly optimized cadences (parallax: all the observations are taken
at maximum parallax factor, resulting in a peak at the South Ecliptic pole; proper motion:
a half of all visits are obtained on the first day and the rest on the last day of the survey),
obtained for simulated cadence \opsimdbref{db:baseCadence} are shown in Aitoff projection of equatorial
coordinates.}
\label{fig:parapmenigma}
\end{figure}
%%%%%%%%%%%%%%%%%%%%%%%%%%%%%%%%%

The Baseline Cadence,
\opsimdbref{db:baseCadence}, has the following basic
properties\footnote{For MAF output, see \url{http://ls.st/tny} and
\url{http://ls.st/67x}}:
\begin{enumerate}
\item The total number of visits is 2,447,931, with 85.1\% spent on
the Universal proposal (the main Wide, Fast, Deep (WFD) survey and henceforth
known as the WFD proposal), 6.5\% on the
North Ecliptic Spur proposal, 1.7\% on the Galactic Plane proposal, 2.2\%
on the South Celestial Pole proposal, and 4.5\% on the Deep Drilling
Cosmology proposal (5 fields).\footnote{The community-contributed white papers leading to the
Deep Drilling fields defined in the Baseline Cadence can be found via
\url{https://community.lsst.org/t/deep-drilling-whitepapers/732}.}
\item The median number of visits per night is 816, the range is
88 to 1,104, with 3,026 observing nights. The mean slew time is 6.8
seconds (median: 4.8 sec) and the total exposure time (after 10 yeras) is 73.4 Msec.
The surveying efficiency, or the median total open shutter time (per night)
as a fraction of the observing time (the ratio of the open shutter time to
the sum of the open shutter time, readout time and slew time) is 73\%.
\item
The 25\%-75\% quartiles for the number of filter changes per night are 2
and 6, with the mean of 4.3 The total number of filter changes through the survey is 14,194.
\item In the $r$ band, the median effective seeing for all proposals is 0.93 arcsec (for the more
traditional geometric FWHM, the median is 0.81 arcsec).  We define ``geometric FWHM'' as the actual full-width-at-half-maximum.  The ``Effective FWHM'' is the FWHM of a single Gaussian describing the PSF and is typically $\sim15$\% larger than the geometric FWHM. The median
airmass for all filters and all proposals is 1.23. The median single-visit $5\sigma$ depth for point sources in $r$ band in the WFD area is 24.16 (using the best
current estimate of the fiducial depth at airmass of one, $m_5(r)=24.39$,
defined by the SRD Table 5). The variation of the median airmass for the $r$
band observations with the position on the sky is shown in
\autoref{fig:airmassenigma}.
\item The median single-visit depths for WFD fields are (23.14, 24.47, 24.16,
23.40, 22.23, 21.57) in the $ugrizy$ bands\footnote{Note that these values
depend on externally supplied values for fiducial zenith dark time single-epoch
$5\sigma$ depths; the following values were used in analysis described
here: (23.62, 24.85, 24.39, 23.94, 23.36, 22.45) in the $ugrizy$
bands, respectively. These values are similar, but not identical, to the values
listed in Table 2 from the latest version (v3.1) of the LSST overview
paper: (23.68, 24.89, 24.43, 24.00, 23.45, 22.60). This discrepancy
is due to continuing improvements in the system performance estimates.}.
These values are shallower than
the zenith dark time values for three main reasons: the sky is expected to be
brighter for non-dark time and away from zenith, the sky brightness model
currently implemented in \OpSim has some shortcomings (a new model has been implemented for version 4), and the moon avoidance is not as aggressive
as it could be (many observations are taken very close to the moon avoidance limit of 30 degrees, rather than farther away where the sky is darker). As a result, the median limiting depths above are brighter than typical zenith dark-time images by close
to 1 mag in the $z$ and $y$ bands, and a few tenths of a magnitude in the
$u$, $g$ and $i$ bands.
\item For the 2,293 (overlapping) fields from the WFD area,
the median number of visits in the $ugrizy$ bands is (62, 88, 199, 201, 180,
180), respectively. Not only do these medians exceed the requested
number of visits (design specification from the SRD\footnote{The LSST
Science Requirements Document (SRD) is available as
\url{http://ls.st/lpm-17}}) of (56, 80, 184, 184, 160, 160) in the $ugrizy$
bands, but the minimum number of visits per field over this area does
so, too. This result is quite encouraging given that
only 85\% of observing time was spent on the WFD Cadence proposal.
\item The median coadded $5\sigma$ depth
for point sources in the $ugrizy$ bands is (25.4, 27.0, 27.1, 26.4,
25.2, 24.4), respectively, for the WFD area. The distribution
of coadded depth across the sky is fairly uniform, as illustrated in \autoref{fig:coaddm5enigma}.
\item For the 2,293 fields from the WFD area, the median
geometric FWHM for seeing is 0.78 arcsec in the $r$ band and 0.77 arcsec
in the $i$ band. The median airmass in the $urz$ bands is 1.25, 1.20 and 1.26
(the maximum allowed airmass for the WFD area was set to
1.5).  The median sky brightness in the $ury$ bands is 22.0 mag/arcsec$^2$,
21.1 mag/arcsec$^2$, and 17.3 mag/arcsec$^2$, respectively (for comparison, the
assumed dark sky brightness at zenith in the $ury$ bands is 23.0, 21.2 and 18.6
mag/arcsec$^2$).  The current model sky brightness in the $y$ band is biased
very high because most $y$ band (and many $z$ band) observations are taken in twilight where \OpSim currently uses a very simple (and bright) sky model.
\item Restricted to the WFD fields, a unique area of
18,000 square degrees received at least 888 visits per field (summed over bands;
the SRD design value is 825).
\item The median trigonometric parallax and proper motion errors are
0.62 mas and 0.17 mas/yr, respectively, for bright sources (limited by
assumed systematic errors in relative astrometry of 10 mas), and 7.9
mas and 2.3 mas/yr for points sources with $r=24$ (assuming flat
spectral energy distribution), over the WFD fields. The
variation of parallax and proper motion errors across the sky is
visualized in \autoref{fig:parapmenigma}.
\end{enumerate}

For comparison, the old Baseline Cadence, \texttt{opsim3.61}
(obtained with an older version of the \OpSim code), delivered 2,651,588 visits, or 8.3\%
more than \opsimdbref{db:baseCadence}  (this is due to known effects and
changes in the code,  such as more pre-scheduled down time in the new
version). Perhaps the most important (and undesired!) difference
between the two simulations is that the Baseline Cadence
spent 6.5\% of the observing time on the North Ecliptic Spur proposal (vs.
4\% spent on the corresponding Universal North proposal in
\texttt{opsim3.61}), and less than 90\% of time on the WFD proposal.

Analysis of the hour angle distribution, shown in
\autoref{fig:HAenigma} and \autoref{fig:AltAzenigma}, reveals a strong
bias towards observations west from the meridian for the main survey.
This pattern is being investigated: it may be
caused by specific features of the cost function implemented in the
\OpSim code. Removing the bias has the potential to increase the survey depth by around $\sim10\%$ (the survey would reach it's current limiting depth in 9 years rather than 10).

%%%%%%%%%%%%%%%%%%%%%%%%%%%%%%%%%
\begin{figure}[t!]
\vskip -0.0in
\includegraphics[angle=0,width=0.49\hsize,clip]{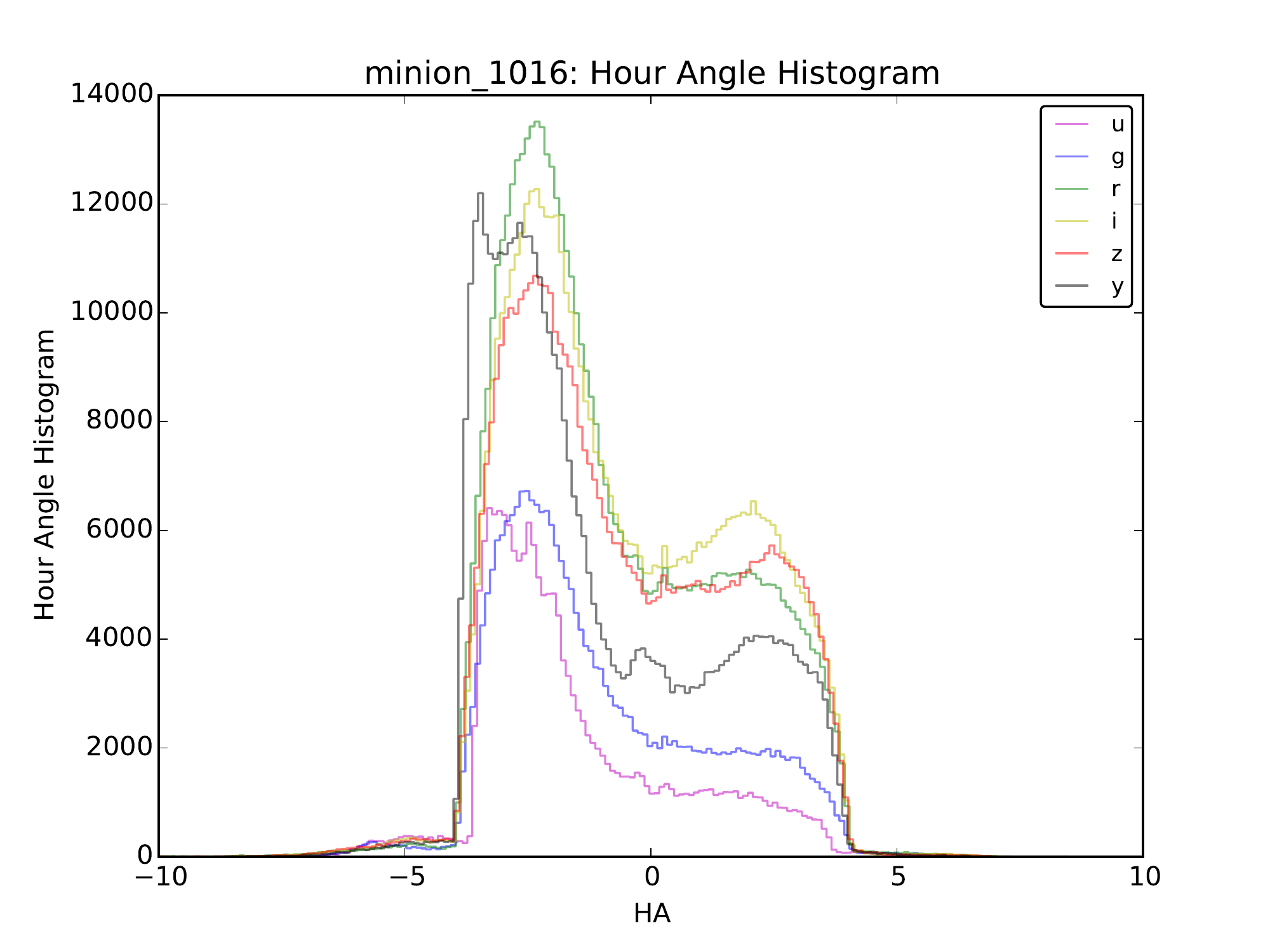}
\includegraphics[angle=0,width=0.49\hsize,clip]{figs/cadence/minion_1016_Mean_HA_r_band_all_props_OPSI_SkyMap.pdf}
\vskip -0.1in
\caption{Histograms in the left panel show the distribution of hour angles (HA) in
6 bands for all proposals from simulated cadence \opsimdbref{db:baseCadence} (the distributions are
similar for WFD fields considered alone). Note the bias towards observations west from
the meridian. The right panel shows the distribution across the sky of the mean HA for
all observations in the $r$ band. }
\label{fig:HAenigma}
\end{figure}
%%%%%%%%%%%%%%%%%%%%%%%%%%%%%%%%%

%%%%%%%%%%%%%%%%%%%%%%%%%%%%%%%%%
\begin{figure}[t!]
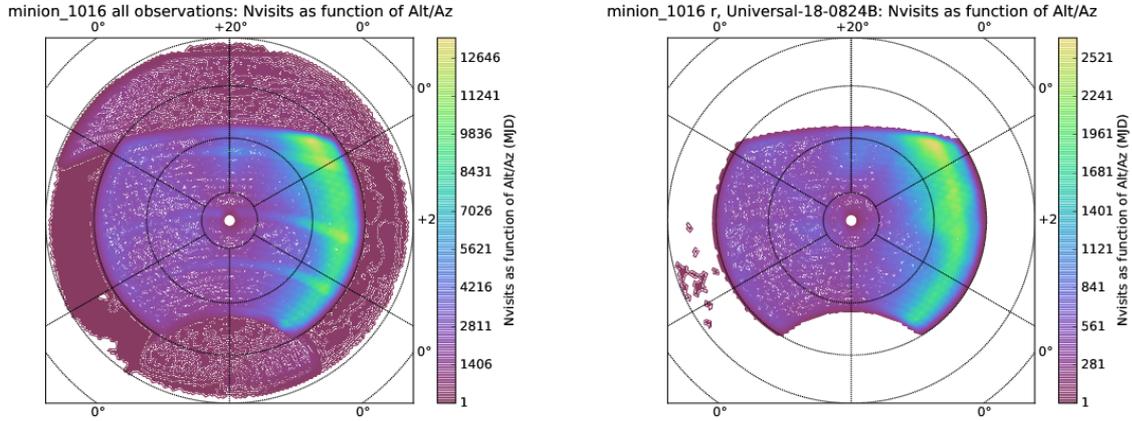

\vskip -0.0in
\includegraphics[angle=0,width=0.49\hsize,clip]{figs/cadence/minion_1016_Nvisits_as_function_of_Alt_Az_all_observations_HEAL_SkyMap.pdf}
\includegraphics[angle=0,width=0.49\hsize,clip]{figs/cadence/minion_1016_Nvisits_as_function_of_Alt_Az_r_Universal-18-0824B_HEAL_SkyMap.pdf}
\vskip -0.1in
\caption{The color-coded map in the left panel shows the visit count from the
Baseline Cadence simulation \opsimdbref{db:baseCadence} in the equal-area Lambert projection of the
horizontal coordinate system (altitude-azimuth), with north on top and west towards the
right, for all six bands and proposals (Wide, Fast, Deep, Galactic Plane, Deep Drilling
fields, North Ecliptic Spur, and South Celestial Pole region). The WFD Cadence was
limited to airmass below 1.5, while other proposals sampled higher airmass, too (see the
histogram in \autoref{fig:airmassenigma}).  Note the strong propensity of fields
for westward observations (the median airmass is about 1.2). The right panel is analogous,
but only shows the $r$ band visits for WFD fields.}
\label{fig:AltAzenigma}
\end{figure}
%%%%%%%%%%%%%%%%%%%%%%%%%%%%%%%%%

%%%%%%%%%%%%%%%%%%%%%%%%%%%%%%%%%
\begin{figure}[th!]
\vskip -0.0in
\includegraphics[angle=0,width=0.49\hsize,clip]{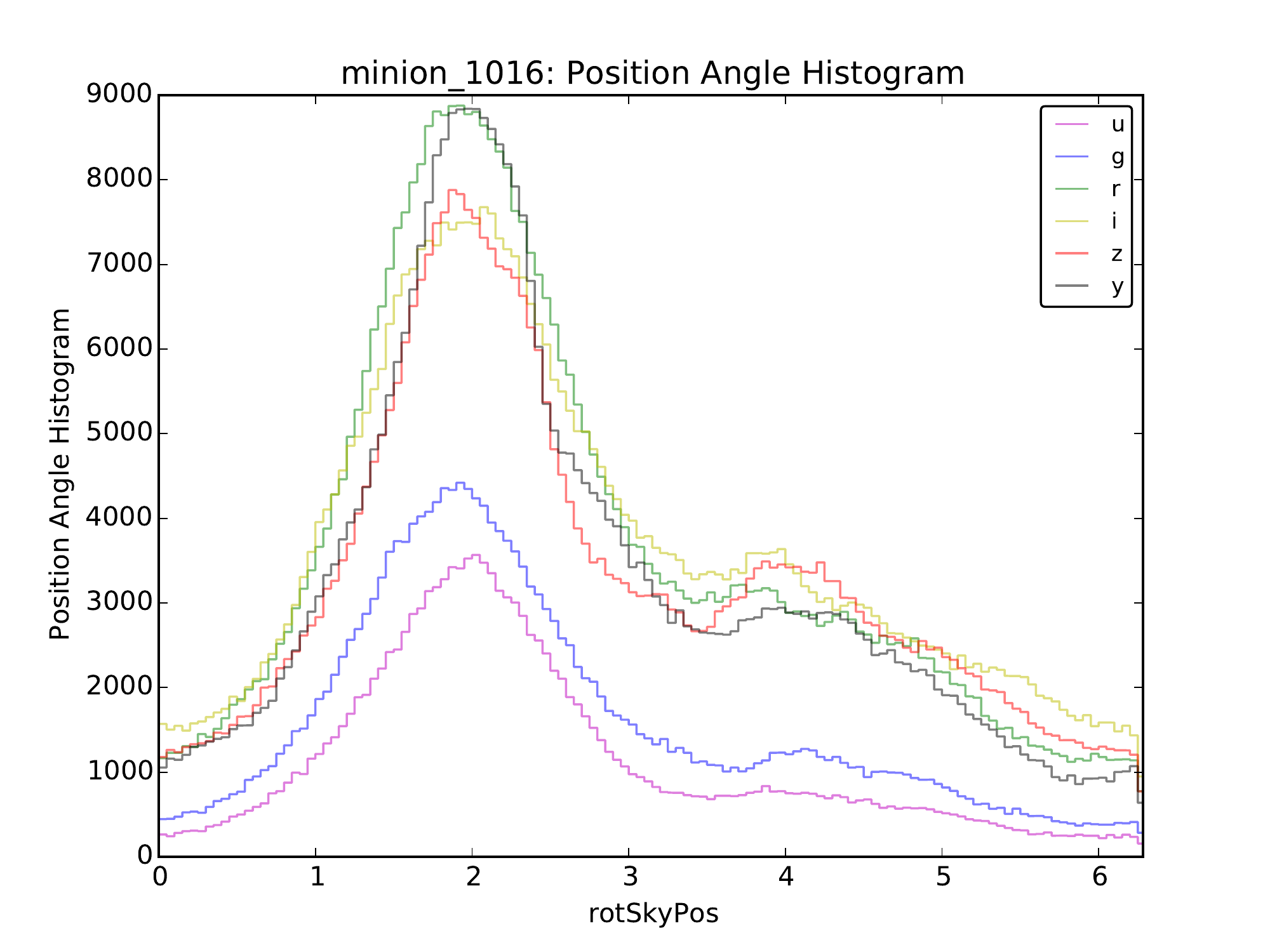}
\includegraphics[angle=0,width=0.49\hsize,clip]{figs/cadence/minion_1016_RmsAngle_rotSkyPos_r_band_WFD_OPSI_SkyMap.pdf}
\vskip -0.1in
\caption{The left panel shows the position angle distribution (in radians)  in each band for the
main survey fields in \opsimdbref{db:baseCadence}. The position angle is the angle between
``up'' in the image and North on the sky. The variation of the root-mean-square scatter of the
$r$ band distribution across the sky is shown in the right panel.}
\label{fig:rotator}
\end{figure}
%%%%%%%%%%%%%%%%%%%%%%%%%%%%%%%%%

Another potentially undesirable feature, seen in practically all
simulations analyzed here, is that up to about a quarter of visits in
the main survey area represents the third, the fourth and sometimes
even the fifth visit to a field in the same night. For a large number
of time-domain programs, these visits could be used instead to
decrease the field inter-night revisit time. For more details, see
\autoref{sec:solarsystem:discovery}. The position angle distributions for this simulation
are shown in \autoref{fig:rotator}.

\subsubsection{Time-domain Analysis}

\autoref{fig:enigmaGapAll} shows the median revisit time distribution
when all bands are considered, and \autoref{fig:enigmaGapr} shows the
median revisit time distribution in the band.  On average, fields in
the main survey get revisited about every 3 days using all filters,
and every 15 days when using only $r$ band visits (30 days when using
only $u$ band visits is the longest median revisit time).
\autoref{fig:enigmaMAXGapAll} shows the maximum inter-night gap, which
on average is about 5-6 months.

The temporal sampling for this simulation is sufficient to enable a
large recovery fraction for SNe. \autoref{fig:enigmaEarlySNe} shows
that a large fraction of LSST SNe will be detected before their
maximum brightness. Metrics for transient objects
are discussed in \autoref{chp:transients}, and the supernova section (\ref{sec:supernovae}) of \autoref{chp:cosmo}.
Intra-night revisit time distribution is discussed in more detail in
\autoref{sec:solarsystem:discovery}. The analysis of asteroid completeness is
discussed in \autoref{sec:solarsystem:discovery}.

%%%%%%%%%%%%%%%%%%%%%%%%%%%%%%%%%
\begin{figure}[t!]
\vskip -0.0in
\includegraphics[angle=0,width=0.49\hsize,clip]{figs/cadence/minion_1016_Median_Intra-Night_Gap_HEAL_SkyMap.pdf}
\includegraphics[angle=0,width=0.49\hsize,clip]{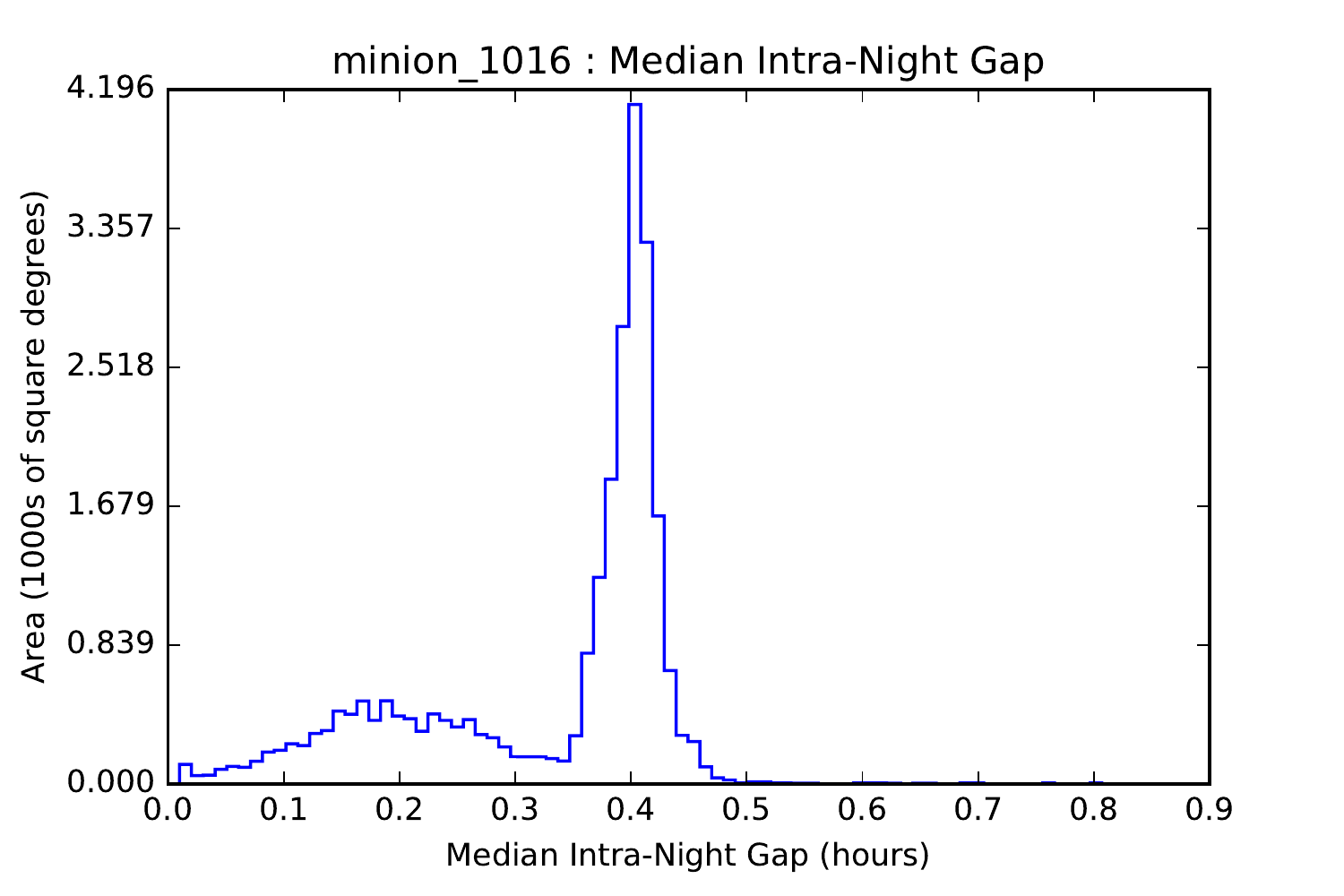}
\vskip -0.1in
\caption{The median intra-night gap (or revisit time) is shown in Aitoff projection
for all proposals and all filters for the Baseline Cadence \opsimdbref{db:baseCadence}.
On average, when a field is observed multiple times in a night there is a 25 minute gap between the observations.}
\label{fig:enigmaInterGapAll}
\end{figure}
%%%%%%%%%%%%%%%%%%%%%%%%%%%%%%%%%

%%%%%%%%%%%%%%%%%%%%%%%%%%%%%%%%%
\begin{figure}[t!]
\vskip -0.0in
\includegraphics[angle=0,width=0.49\hsize,clip]{figs/cadence/minion_1016_Median_Inter-Night_Gap_HEAL_SkyMap.pdf}
\includegraphics[angle=0,width=0.49\hsize,clip]{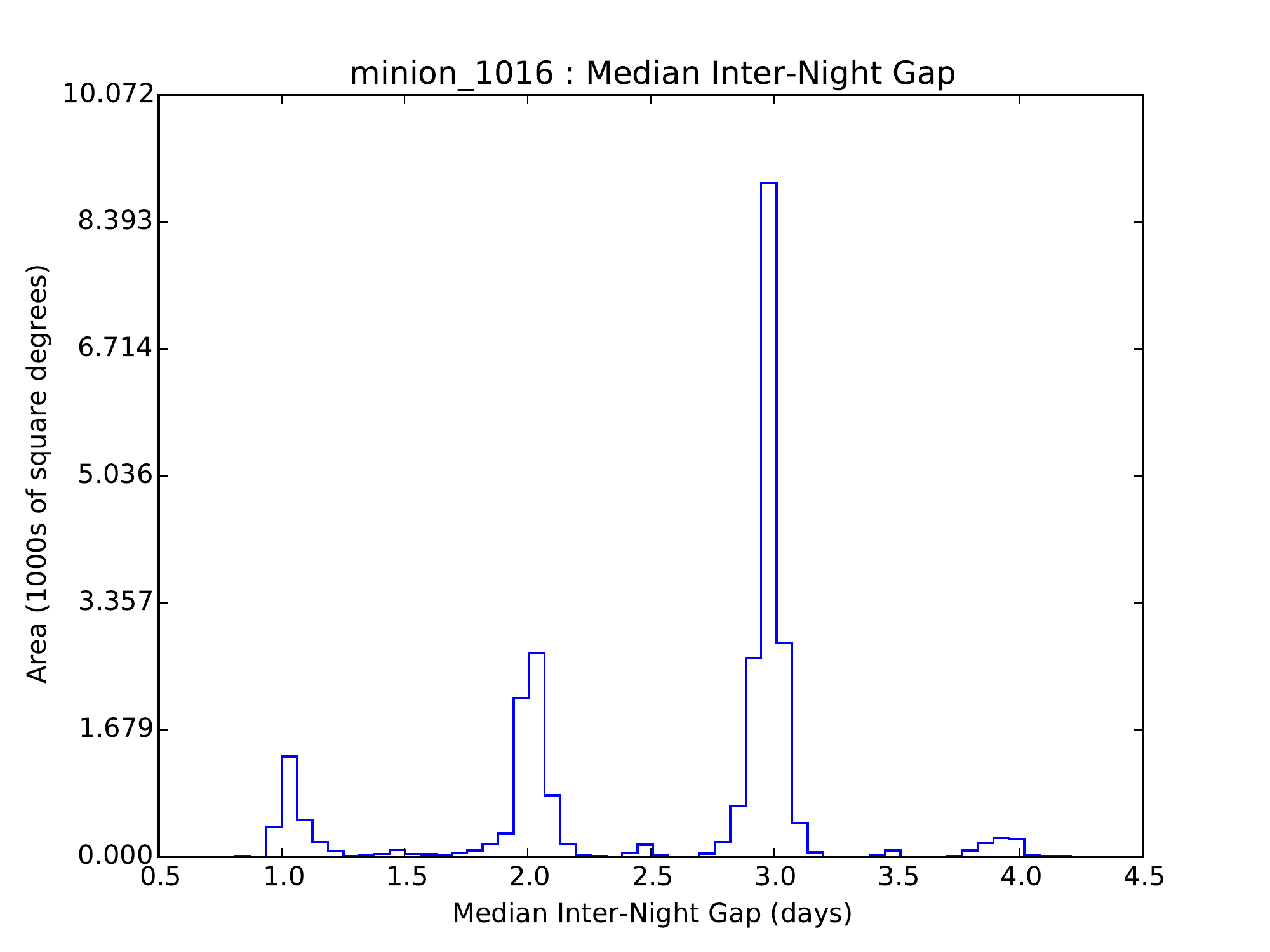}
\vskip -0.1in
\caption{The median inter-night gap (or revisit time) is shown in Aitoff projection
for all proposals and all filters for the Baseline Cadence \opsimdbref{db:baseCadence}.
On average, fields in the main survey get revisited about every 3 days.}
\label{fig:enigmaGapAll}
\end{figure}
%%%%%%%%%%%%%%%%%%%%%%%%%%%%%%%%%

%%%%%%%%%%%%%%%%%%%%%%%%%%%%%%%%%
\begin{figure}[h!]
\vskip -0.0in
\includegraphics[angle=0,width=0.49\hsize,clip]{figs/cadence/minion_1016_Median_Inter-Night_Gap_r_HEAL_SkyMap.pdf}
\includegraphics[angle=0,width=0.49\hsize,clip]{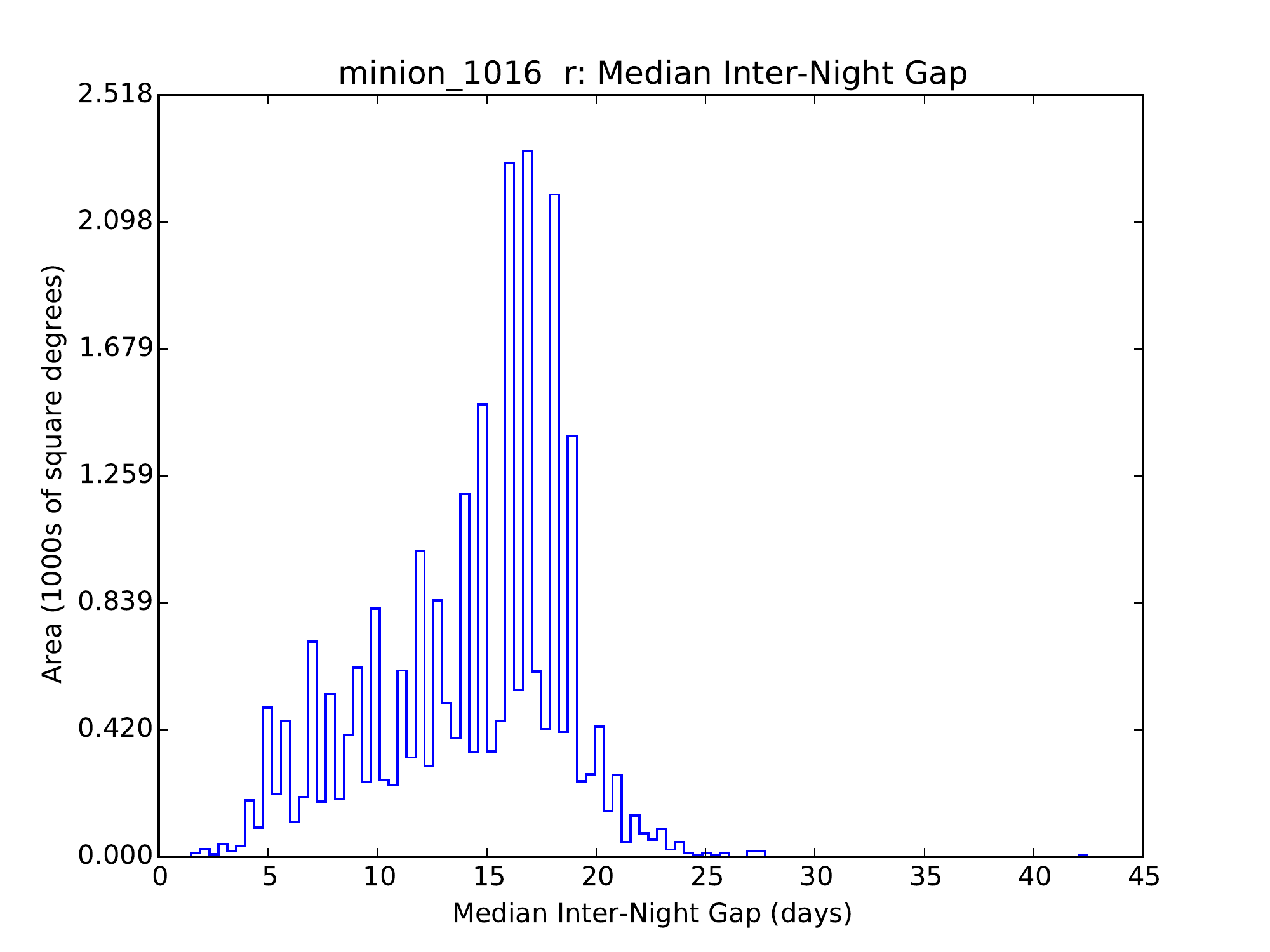}
\vskip -0.1in
\caption{The median inter-night gap for $r$ band visits is shown in Aitoff projection
for all proposals for the Baseline Cadence \opsimdbref{db:baseCadence}.
On average, fields in the main survey get revisited in the $r$ band about every two weeks.}
\label{fig:enigmaGapr}
\end{figure}
%%%%%%%%%%%%%%%%%%%%%%%%%%%%%%%%%

%%%%%%%%%%%%%%%%%%%%%%%%%%%%%%%%%
\begin{figure}[t!]
\vskip -0.0in
\includegraphics[angle=0,width=0.49\hsize,clip]{figs/cadence/minion_1016_Max_Inter-Night_Gap_HEAL_SkyMap.pdf}
\includegraphics[angle=0,width=0.49\hsize,clip]{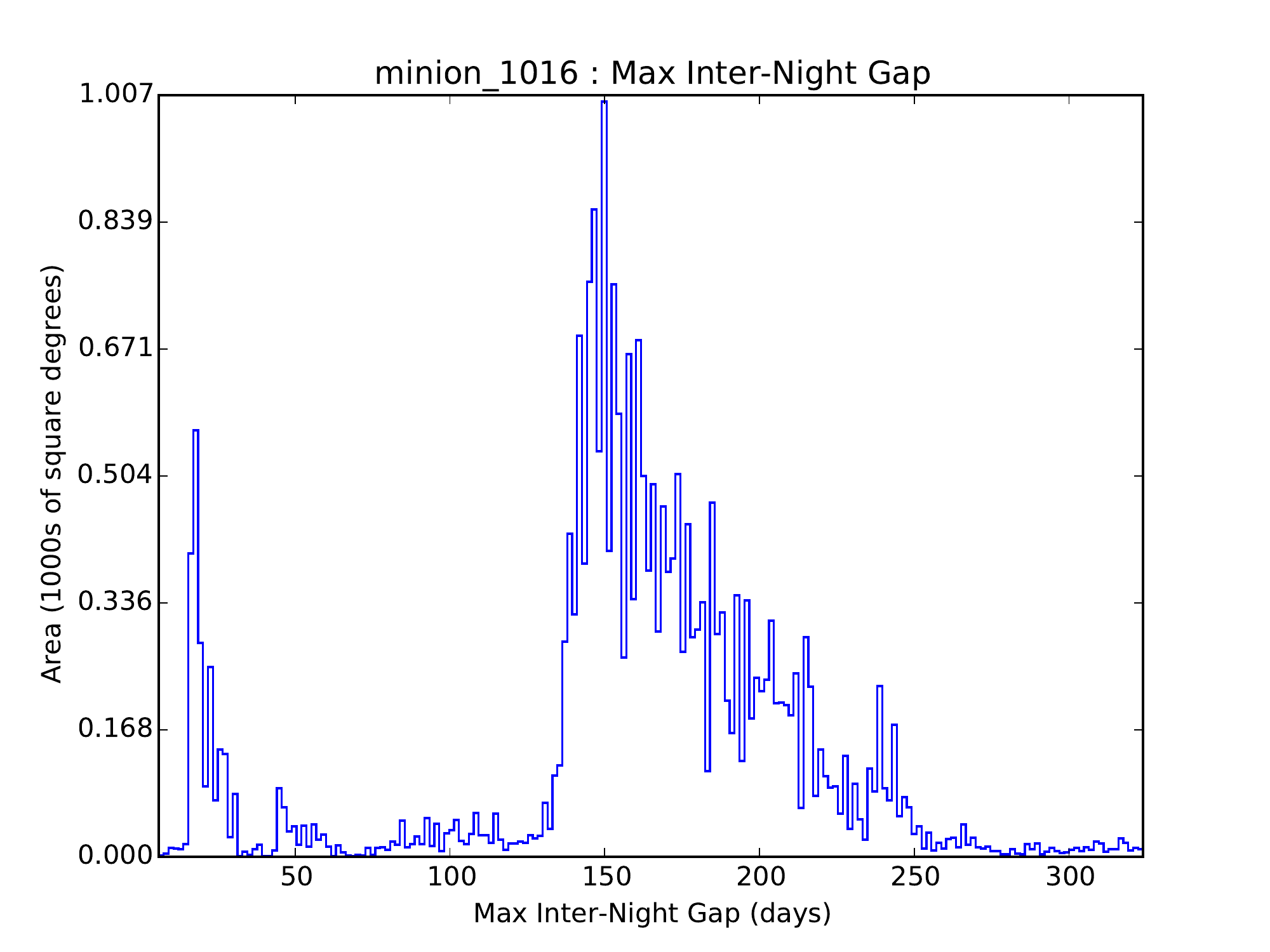}
\vskip -0.1in
\caption{The maximum inter-night gap (or revisit time) is shown in Aitoff projection
for all proposals and all filters for the Baseline Cadence \opsimdbref{db:baseCadence}.}
\label{fig:enigmaMAXGapAll}
\end{figure}
%%%%%%%%%%%%%%%%%%%%%%%%%%%%%%%%%

%%%%%%%%%%%%%%%%%%%%%%%%%%%%%%%%%
\begin{figure}[h!]
\vskip -0.0in
\includegraphics[angle=0,width=0.49\hsize,clip]{figs/cadence/minion_1016_SNAlert_HEAL_SkyMap.pdf}
\includegraphics[angle=0,width=0.49\hsize,clip]{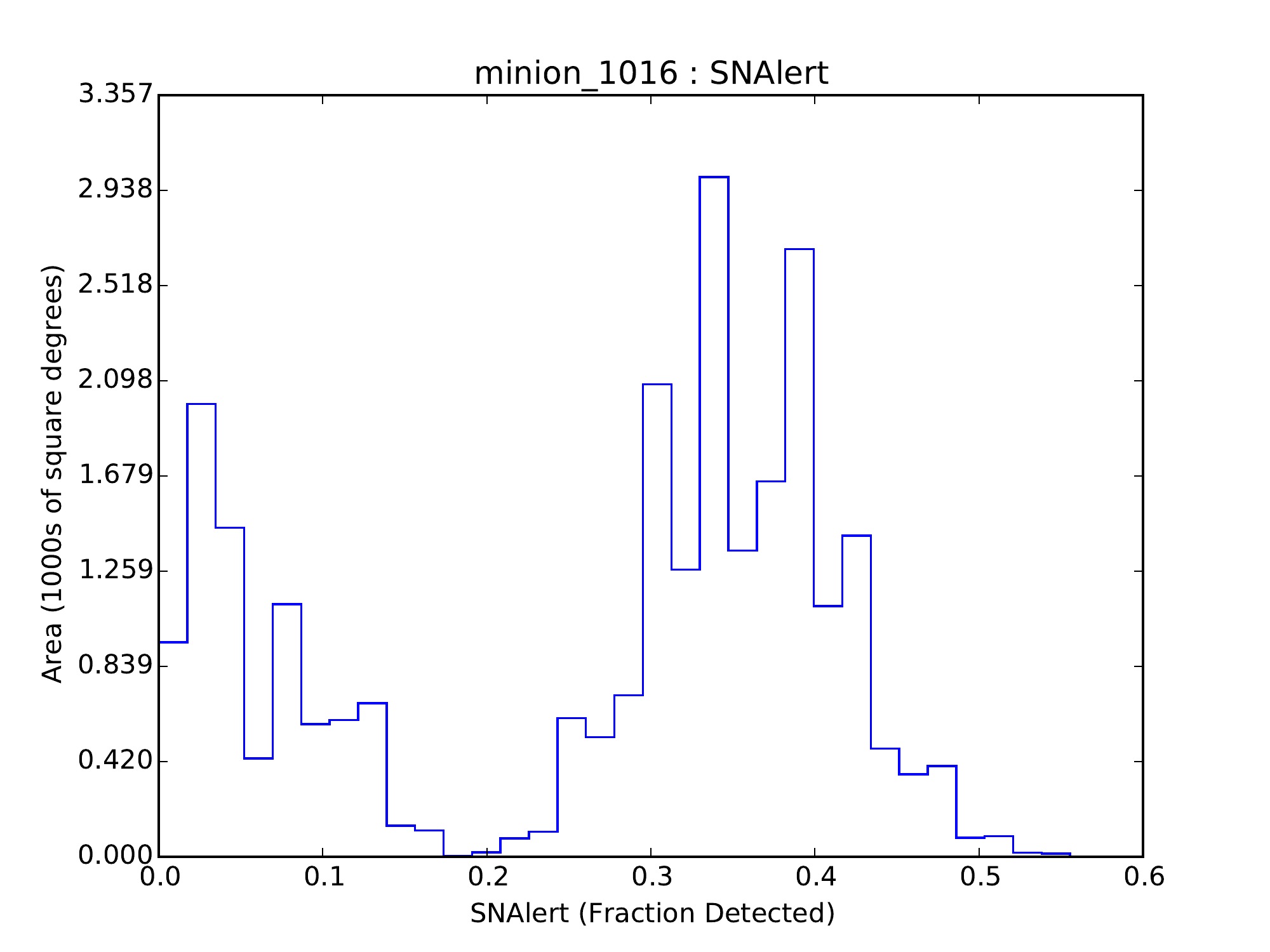}
\vskip -0.1in

\caption{The fraction of simulated Type Ia SNe at a redshift of 0.5 detected
pre-peak in any filter for the Baseline Cadence \opsimdbref{db:baseCadence}. About
40\% of all such SNe from the main survey will be detected before their
maximum brightness.}
\label{fig:enigmaEarlySNe}
\end{figure}
%%%%%%%%%%%%%%%%%%%%%%%%%%%%%%%%%

\subsubsection{Special Proposals}

Regarding the special proposals, here we only provide the basic
performance parameters. With the exception of the Deep Drilling
proposal, these proposals are essentially strawman placeholders. The
North Ecliptic Spur proposal (6.4\% of the observing time) obtained  300 visits per field, summed over $griz$ bands. These
fields are placed along the northern part of the Ecliptic. The
Galactic Plane proposal (1.7\%) obtained 30 visits per band in all six
bands, across the region extending in Galactic latitude 10 degrees
from the Galactic center, with the boundary approaching the Galactic
equator linearly with longitude, and the zone ending at $l=90$ deg.
and at $l=270$ deg. The South Celestial Pole proposal (2.1\%) obtained
30 visits per band in all six bands, for fields centers with Dec $<
-62.5$ deg. The Deep Drilling proposal (4.5\%) included 5
fields, with each obtaining several thousand visits per band as required for various cosmology investigations. The
coadded $5\sigma$ depths for these fields are much fainter than for
the main survey: the median values are (27.8, 28.4, 28.6, 28.0, 27.6,
26.1) in the $ugrizy$ bands, respectively.

\vskip 0.2in
{\bf Conclusions:}

The Baseline Cadence, \opsimdbref{db:baseCadence}, appears to
be an adequate replacement for the old Baseline Cadence
(\texttt{opsim3.61}). Based on this analysis, there are no
major problems with its performance. While there are patterns which
are not fully understood (most notably the observing bias towards
west),  or undesired (unnecessary revisits of the same field in the
same night), \opsimdbref{db:baseCadence} is used as a benchmark cadence,
and referred to as the ``Baseline Cadence'', in the rest of this
document. This simulation was proposed by the Project Science Team
and adopted as the new Baseline Cadence by the LSST Change Control Board in August 2016.

An important feature of \opsimdbref{db:baseCadence} simulation is that
the mean slew time of 6.8 sec (which includes filter change time) is
very close to the minimum possible slew time of about 4.5 sec. The
implication is that the surveying efficiency, assuming 30 sec exposure
time per visit, can be increased by at most about 6\% (that is, the
total open-shutter time is within about 6\% from its possible maximum,
given everything else unchanged).  Nevertheless, there are other
survey aspects, including sky coverage and temporal sampling
functions, that can be further optimized, as discussed in
\autoref{sec:cadexp:alternatives} below.

{\bf The main remaining known problems} with \opsimdbref{db:baseCadence} simulation include
\begin{itemize}
\item A strong bias towards observations west from the meridian for
  the main survey, see \autoref{fig:AltAzenigma}.  This bias
  significantly degrades the survey seeing and depth.
\item Several proposals complete after only 3-4 years, resulting in
  regions of sky where the proper motions are poorly constrained due
  to the short observing baseline.  See \autoref{fig:parapmenigma2}.
\item The sky brightness model has systematic errors, particularly in twilight time,
  resulting in estimates of limit depths ($m_5$ for single
  visits and coadded depths) that are too shallow by about 0.3 mag in
  the $u$ band, and 0.5-1.0 mag in the $z$ and $y$ bands.
\item The moon avoidance angle of 30 deg. allows too many $z$ band
  observations with elevated sky brightness due to moonshine,
  resulting in about 0.2-0.3 mag shallower depth.
\item There are too many unrequested and unnecessary revisits of the
  same field in the same night (that is, more than two visits to the
  same field in the same night).
\end{itemize}
Most of these issues are expected to be resolved by \OpSim version 4.

\navigationbar

% --------------------------------------------------------------------

\section{Some Simulated Alternative Observing Strategies}
\def\secname{cadexp:alternatives}\label{sec:\secname}

We now describe some alternatives to the Baseline Cadence that were
explored. These \OpSim databases are all available for further testing
with science-based MAF metrics.

% - - - - - - - - - - - - - - - - - - - - - - - - - - - - - - - - - -

%%%%%%%%%%%%%%%%%%%%%%%%%%%%%%%
\opsimdb[db:opstwo]{minion\_1012}{Only Wide, Fast, Deep Cadence, with pairs of visits.}
%%%%%%%%%%%%%%%%%%%%%%%%%%%%%%%

{\bf Motivation and description:} Formally, $\sim$85\% of observing
time is allocated to the main WFD Cadence program. The
remaining observing time is allocated to other programs, such as
``Deep Drilling'' programs (see Section 3.4 and Tables 22-26  in the
SRD). With this simulation, we wished to find out what would be the
effect of ignoring special programs and spending all of the observing
time on the main WFD Cadence program. \\

{\bf Expectations:} About 2.08 million visits (85\% of 2.44 million
visits) from the Baseline Cadence (\opsimdbref{db:baseCadence}) were allocated
to WFD Cadence. With \opsimdbref{db:opstwo} we expect that all of these 2.44 million visits
will be allocated to WFD Cadence. \\

{\bf Analysis Results:} This simulated cadence is named \opsimdbref{db:opstwo}.
Compared to the Baseline Cadence \opsimdbref{db:baseCadence}:
\begin{enumerate}
\item The total number of visits is close to the expected value: 2.42
  million.  The minimum number of visits per field for the 2,293 WFD
  fields in the Baseline Cadence is 965 for this simulation, compared to
  888 for the Baseline Cadence.
\item The median number of visits per night and the mean slew time are
  essentially the same as for the Baseline Cadence (807 vs. 816 and 7.2
  sec vs. 6.8 sec).
\item The median seeing, sky brightness and airmass in the $r$ and $i$ bands are
      essentially the same as for WFD fields in the Baseline Cadence.
\item The median trigonometric parallax and proper motion errors are
  improved by about 8\%, with improvements commensurate with the
  increase in the number of visits and the elimination of regions
  which are not observed for a full 10 years.
\item This simulation also shows observing bias towards west (that is,
  additional special programs in \opsimdbref{db:baseCadence} are not
  responsible for this bias).
\end{enumerate}

{\bf Conclusions:} \opsimdbref{db:opstwo}, using only the WFD Cadence
proposal, delivered 99.2\% of the number of visits obtained by the
Baseline Cadence. Therefore, {\it the ``filler'' aspect of other
proposals does not have a major impact on the surveying efficiency}.
The minimum number of visits per field for the 2,293 WFD fields in the
Baseline Cadence is 886 (the SRD design value is 825 and the stretch
goal value is 1000). Although the sky coverage of these 2293 fields is
about 18,000 sq.deg., that number of fields could cover 22,000 sq.deg if there was no field overlap. With
proper dithering (see \eg \autoref{sec:lss}), the effective number of visits could be increased to
$886\times22/18 = 1083$ (or the WFD area increased from 18,000 sq.deg.; see
analysis of \opsimdbref{db:opstwoPS} below). This increase is an improvement of 31\%
relative to the SRD design specification of 825 visits over 18,000
sq.deg. However, note again that there are no other programs in this
simulation (i.e., if other programs were allocated 10\% of the
observing time, the implied overall ``over-performance'' in the number
of  visits would be about 20\%).

% - - - - - - - - - - - - - - - - - - - - - - - - - - - - - - - - - -

%%%%%%%%%%%%%%%%%%%%%%%%%%%%%%%%%
\opsimdb[db:opstwoPS]{minion\_1020}{A Pan-STARRS-like observing strategy.}
%%%%%%%%%%%%%%%%%%%%%%%%%%%%%%%%%

{\bf Motivation and description:} ``Pan-STARRS-like cadence" attempts to apply a
uniform cadence strategy throughout the maximum size survey region, which is about 27,400~deg$^2$. This is similar to the Pan-
STARRS' 3PI survey which also tries to maximize sky area. The maximum acceptable
airmass is kept at its default value of 1.5; this excludes fields with Dec~$<-78$~deg and Dec~$> +18$~deg. The \opsimdbref{db:opstwoPS} simulations utilizes a maximum Dec of $+15$~deg, uniform cadence and no
other proposal, and requires pairs of visits as in the Baseline Cadence. \\

{\bf Expectations:} The total number of visits should be roughly the
same as in the Baseline Cadence, but spread over a 42\% larger sky area
(3,255 fields instead of 2,293), with fewer visits per field. \\

{\bf Analysis Results:}  This simulated cadence is named \opsimdbref{db:opstwoPS}.
Compared to the Baseline Cadence \opsimdbref{db:baseCadence}:
\begin{enumerate}
\item The total number of visits is 2.42 million, and essentially identical to the
number of visits in the Baseline cadence.
\item
The mean number of visits per field is 740, which is 81\% of the number of visits %minion_1016 median 912 visits per feild WFD
for WFD fields obtained by the Baseline Cadence (but here the sky area is 42\% larger). We note that this is below the number required by the SRD.
\item The median number of visits per night and the mean slew time are
  essentially the same as for the Baseline Cadence.
\item The median seeing, sky brightness and airmass in the $r$ and $i$
  bands for WFD fields are essentially the same as in the Baseline
  Cadence.
\item The median trigonometric parallax and proper motion errors show
  uniform behavior over the entire enlarged area (see
  \autoref{fig:parapmenigma2}), with the values similar to those
  obtained for the Baseline Cadence.
\item This simulation also shows observing bias towards west.
\end{enumerate}

Due to increased sky area, which samples regions that can never
achieve low airmass, the median coadded depth is about 0.15 mag
shallower for this simulation than for the Baseline Cadence. As a result,
the counts of galaxies per unit area down to a fixed SNR would
decrease by about 15-20\%. At the same time, the area outside the
Galactic plane is increased by about 30\%, and thus the total number
of galaxies would be increased by about 10\%, compared to WFD fields
in the Baseline Cadence. However, the increased median airmass also
results in larger seeing, especially for the borderline regions, as
illustrated in \autoref{fig:PS-seeing}. The increased median seeing
would decrease the number of galaxies effectively resolved for weak
lensing by about 3-5\%. In addition, the additional area has somewhat
larger extinction due to interstellar dust which further decreases the
galaxy counts (this impact of dust extinction on galaxy counts is not
yet implemented in MAF). As a result of these effects, the two
strategies result in similar weak lensing galaxy samples.

{\bf Conclusions:} When only the WFD Cadence proposal is
employed, the survey area could be increased by about 40\%, while
still delivering the mean number of fields at the level of 81\% of
that in the Baseline Cadence (or 90\% of the SRD design value of 825).
Hence, simulations \opsimdbref{db:opstwoPS} and \opsimdbref{db:opstwo}
demonstrate that this hypothetical ``survey reserve'', relative to the WFD
Cadence design specifications from the SRD, can be used to i) increase
the number of visits per field over the WFD area, or ii) increase the
surveyed area while keeping the number of visits per field
statistically unchanged, or iii) increase both area and the number of
visits, and/or iv) execute additional programs (the current baseline).
Of course, it must be remembered that this ``survey reserve'' is derived
using design system characteristics and thus unproven at the time of
writing.

%%%%%%%%%%%%%%%%%%%%%%%%%%%%%%%%%
\begin{figure}[t!]
\vskip -0.03in
\includegraphics[angle=0,width=0.49\hsize:,clip]{figs/cadence/minion_1020_Parallax_Normed_All_Visits_non-dithered_HEAL_SkyMap.pdf}
\includegraphics[angle=0,width=0.49\hsize:,clip]{figs/cadence/minion_1020_Proper_Motion_Normed_All_Visits_non-dithered_HEAL_SkyMap.pdf}
\includegraphics[angle=0,width=0.49\hsize:,clip]{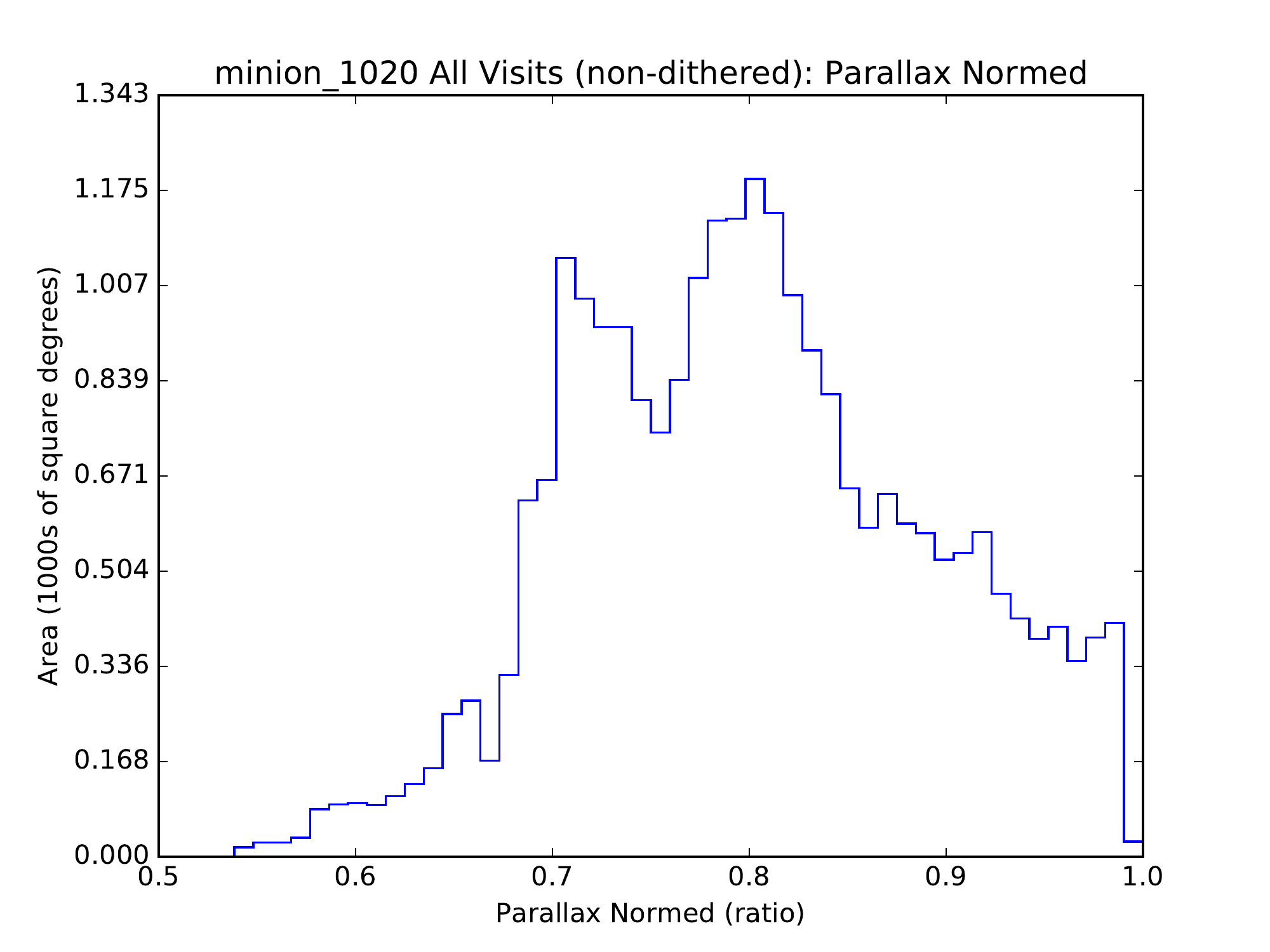}
\includegraphics[angle=0,width=0.49\hsize:,clip]{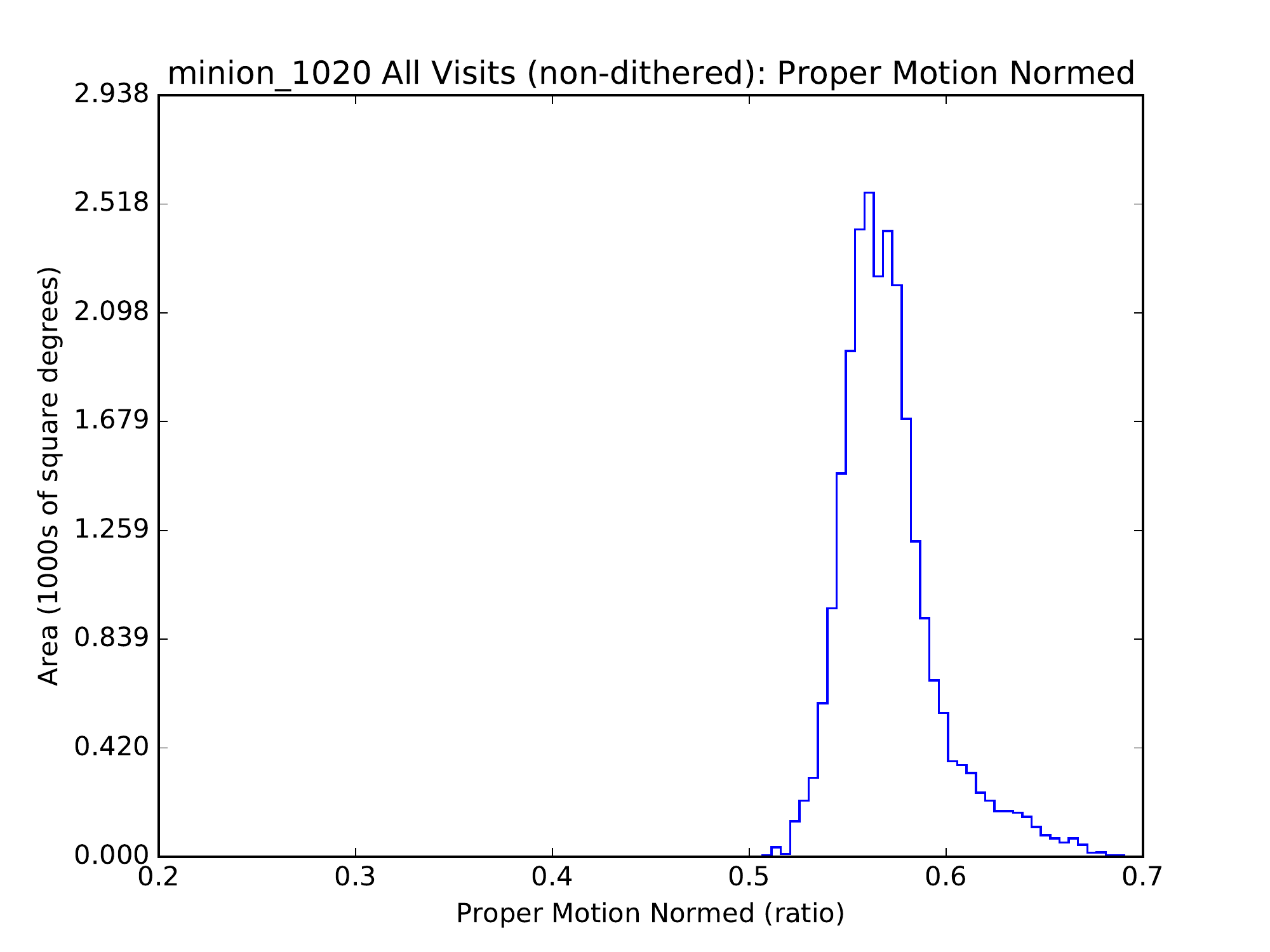}
\vskip -0.2in
\caption{The trigonometric parallax errors (left) and proper motion errors (right) for simulated cadence
minion\_1020 (``Pan-STARRS-like'' Cadence), normalized by the values for idealized perfectly optimized
cadence, are shown in Aitoff projection of equatorial coordinates (compare to \autoref{fig:parapmenigma}).}
\label{fig:parapmenigma2}
\end{figure}
%%%%%%%%%%%%%%%%%%%%%%%%%%%%%%%%%

%%%%%%%%%%%%%%%%%%%%%%%%%%%%%%%%%
\begin{figure}[t!]
\vskip -0.03in
\includegraphics[angle=0,width=0.49\hsize:,clip]{figs/cadence/minion_1020_Median_FWHMeff_r_band_all_props_OPSI_SkyMap.pdf}
\includegraphics[angle=0,width=0.49\hsize:,clip]{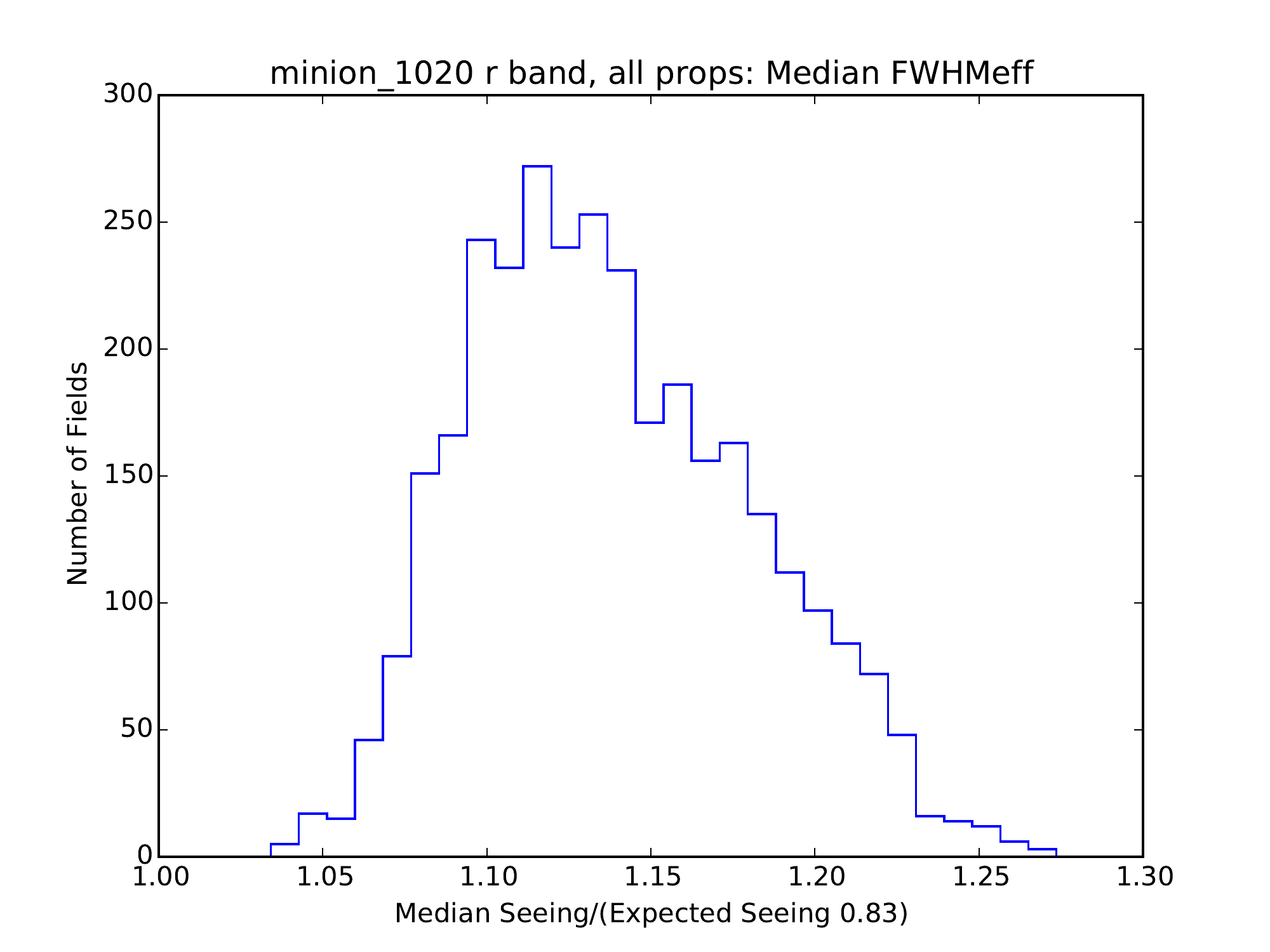}
\vskip -0.2in
\caption{The median seeing in $r$ filter, for simulated cadence minion\_1020 (``Pan-STARRS-like'' Cadence),
normalized by expected value (0.83$^{\prime\prime}$). Note that fields with the most positive and most negative
declination have on average larger values. For comparison, the median normalized seeing for WFD fields
in the Baseline Cadence is 1.08, with a negligible fraction of fields with values above 1.18.}
\label{fig:PS-seeing}
\end{figure}
%%%%%%%%%%%%%%%%%%%%%%%%%%%%%%%%%

% - - - - - - - - - - - - - - - - - - - - - - - - - - - - - - - - - -

%%%%%%%%%%%%%%%%%%%%%%%%%%%%%%%%%%%%%%%%%%%
\opsimdb[db:UConlyNoVisitPairs]{minion\_1013}{Only Wide, Fast, Deep Cadence, no visit pairs.}
%%%%%%%%%%%%%%%%%%%%%%%%%%%%%%%%%%%%%%%%%%%

{\bf Motivation and description:} The main goal of this simulation was
to assess the impact of the requirement for visit pairs on the survey
efficiency (the Baseline Cadence requests two visits per night to the same
field, separated in time by about an hour, and driven by asteroid
orbit determination). It is plausible that the removal of this
requirement could result in a more efficient survey. In order to allow
as simple analysis as possible, only the WFD Cadence proposal is
requested. Hence, this simulation should be directly compared to
simulation \opsimdbref{db:opstwo}. \\

{\bf Expectations:} If the requirement for visit pairs decreases
surveying efficiency, then this simulation should deliver more than
the 2.42~million visits delivered by \opsimdbref{db:opstwo}. \\

{\bf Analysis Results:} This simulated cadence is named \opsimdbref{db:UConlyNoVisitPairs}. Compared
to \opsimdbref{db:opstwo}:
\begin{enumerate}
\item The total number of visits is 2.42 million, identical to \opsimdbref{db:opstwo}.
\item The median slew time, and the median coadded depth and seeing in the $r$ band
are essentially identical, too.
\item The median airmass in the $r$ band of 1.25 is a bit higher than 1.18 obtained
for \opsimdbref{db:opstwo}.
\item The median fraction of revisits faster than 30 minutes of 0.32 is smaller than 0.38
for \opsimdbref{db:opstwo}, and is consistent with the absence of pair contributions (that is,
such revisits are due to field edge overlaps, and unintentional revisits, in case of \opsimdbref{db:UConlyNoVisitPairs}).
\end{enumerate}

{\bf Conclusions:} The comparison of this simulation and
\opsimdbref{db:opstwo} shows that requiring pairs of visits (in a
given observing night) does not result in an appreciable loss of
surveying efficiency. Indeed, pairs of visits result in a better
short-timescale coverage that would enhance many types of time-domain
science (and, of course, it's crucial for asteroid science).

% - - - - - - - - - - - - - - - - - - - - - - - - - - - - - - - - - -

%%%%%%%%%%%%%%%%%%%%%%%%%%%%%%%%%%%%%%%%%%%
\opsimdb[db:NoVisitPairs]{kraken\_1043}{Baseline Cadence, but with no visit pairs.}
%%%%%%%%%%%%%%%%%%%%%%%%%%%%%%%%%%%%%%%%%%%

{\bf Motivation and description:} The main goal of this simulation was
to assess the impact of the requirement for visit pairs on the survey
efficiency. Instead of the idealized case above which compared only
the WFD Cadence proposal fields, in this more realistic case
{\it all proposals from the Baseline Cadence are executed}. Hence, this
simulation should be compared to the Baseline Cadence
(\opsimdbref{db:baseCadence}). \\

{\bf Expectations:} A slight, or no, increase in surveying efficiency
and thus the total number of visits is expected when compared to the
Baseline Cadence. \\

{\bf Analysis Results:}  This simulated cadence is named
\opsimdbref{db:NoVisitPairs}. Compared to \opsimdbref{db:baseCadence},
\begin{enumerate}
\item The total number of visits is 2.51 million, or 2.4\% more than
in the Baseline Cadence.
\item The mean slew time is 5.8 sec, or 15\% shorter than for the Baseline
Cadence. This decrease in the mean slew time implies an efficiency
increase of 2.8\% and explains the actual 2.4\% improvement implied by
the total number of visits.  Note that this simulation has the
shortest mean slew time of all simulations investigated here (the
nominal shortest slew and settle time is about 4.5 sec).
\item The median airmass in the $r$ band is slightly larger for this
simulation than for the Baseline Cadence: 1.29 vs. 1.22.
\end{enumerate}

{\bf Conclusions:}
Unlike the comparison of \opsimdbref{db:UConlyNoVisitPairs} and
\opsimdbref{db:opstwo}, here the removal of visit pair requirement
results in a 15\% shorter mean slew time and consequently in 2.4\%
more visits.

% % - - - - - - - - - - - - - - - - - - - - - - - - - - - - - - - - - -
%
% %%%%%%%%%%%%%%%%%%%%%%%%%%%%%%%%%%%%%%%%%%%
\opsimdb[db:NEOswithVisitTriplets]{enigma\_1281}{NEO test: triplets of visits.}
\opsimdb[db:NEOwithVisitQuads]{enigma\_1282}{NEO test: quads of
  visits.}
% %%%%%%%%%%%%%%%%%%%%%%%%%%%%%%%%%%%%%%%%%%%

{\bf Motivation and description:} Many science programs can benefit
from having more than a pair of visits in a night; in particular,
Solar System science may critically depend on having more than just a
pair, depending on the performance of the Moving Object Pipeline
Software (MOPS). These two simulations were run to investigate the
effects of requiring more than just a pair of visits in each
night. The first, \opsimdbref{db:NEOswithVisitTriplets}, requests
sets of three visits (triplets) in each night. The second,
\opsimdbref{db:NEOwithVisitQuads}, requests sets of four visits (quads)
in each night. There is no constraint on the filter chosen for these
sets of visits -- it may be changed or it may remain the same. These
simulations should be compared to the Baseline Cadence,
\opsimdbref{db:baseCadence}, and to the \opsimdbref{db:NoVisitPairs},
which all keep the special surveys, but simply vary the sequences
requested in the WFD Cadence. \\

{\bf Expectations:} The general expectations are that science cases
which require many visits on timescales of a few hours will benefit
with these runs, while science cases which prefer visits to be spaced
more widely over time will see negative impacts.\\

{\bf Analysis Results:}
First, we emphasize that ``requested'' is not the same as
``delivered'': even the ``no pairs''
simulation \opsimdbref{db:NoVisitPairs} ends
up having multiple visits in a given night to the same fields, and
when multiple visits per night are requested, not all fields get to
have completed sequences. The statistics of how many fields are
combined into sequences of a given number of visits is shown in
\autoref{fig:NvisitStats}.  As evident, the highest peak is at the
requested number of visits in a sequence, but not all visits are
incorporated into requested sequences: some are in both shorter and
longer sequences. The ``no pairs'' simulation includes
multiple visits to some fields, because the current
version of the algorithm is not told not to do so. As illustrated in
\autoref{fig:intranightgapCompare}, such revisits typically happen
within 10 minutes from the first visit. This (unintended) behavior
implies that the naive expectation above is probably incorrect, or at
least softened.

The median inter-night revisit rate is affected by requesting more
visits within single night, as expected -- there are only so many
visits available, and if more occur in a particular night, it is
likely (without some kind of rolling cadence) that the result is
longer intervals between subsequent nights. This is demonstrated in
\autoref{fig:internightgapCompare}, where it can be seen that the
inter-night revisit rate increases by about 30\% from 3 nights to 4
nights if we go from pairs to triplets (or quads).

Details of the impacts on Solar System science is
left to \autoref{chp:solarsystem}, in particular the impact on
completeness is evaluated in \autoref{sec:solarsystem:discovery}.

The impact of requesting sequences with 3 or 4 visits to the same
field on other science programs is not yet analyzed in detail.  The
impact on static science should be minimal, except perhaps for a bit
worse behavior of various systematic errors (because fewer nights,
with their observing conditions, are sampled).

%%%%%%%%%%%%%%%%%%%%%%%%%%%
\begin{figure}[t!]
\vskip -0.03in
\includegraphics[angle=0,width=0.99\hsize,clip]{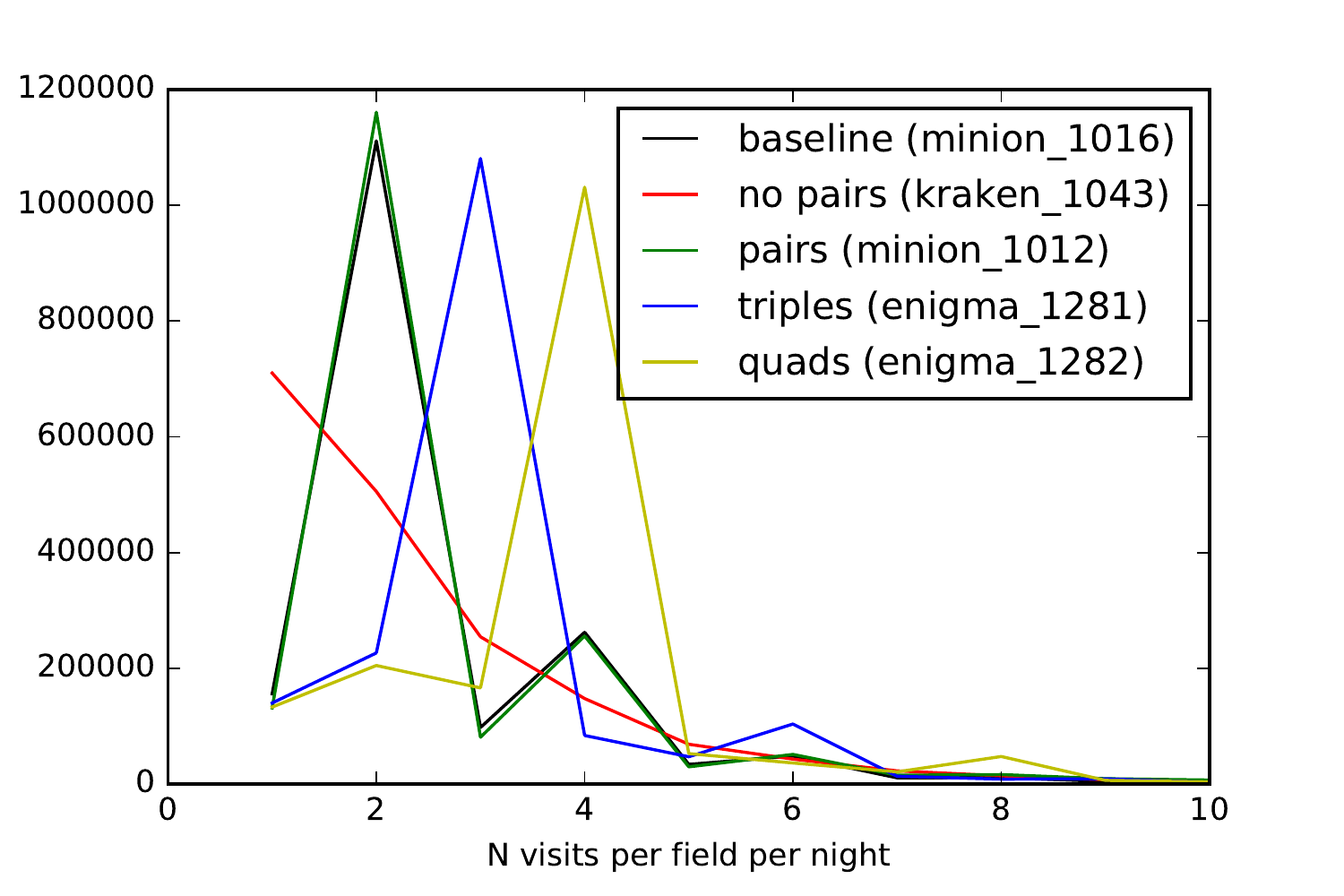}
\vskip -0.2in
\caption{The distribution of the number of visits used for nightly sequences of
length given on the horizontal axis. Only $griz$ bands are used. Note that even
``no pairs'' simulation (\opsimdbref{db:NoVisitPairs})
includes multiple visits. The highest peak is at the
requested number of visits in a sequence.}
\label{fig:NvisitStats}
\end{figure}
%%%%%%%%%%%%%%%%%%%%%%%%%%%

%%%%%%%%%%%%%%%%%%%%%%%%%%%
\begin{figure}[t!]
\vskip -1.2in
\includegraphics[angle=0,width=0.49\hsize:,clip]{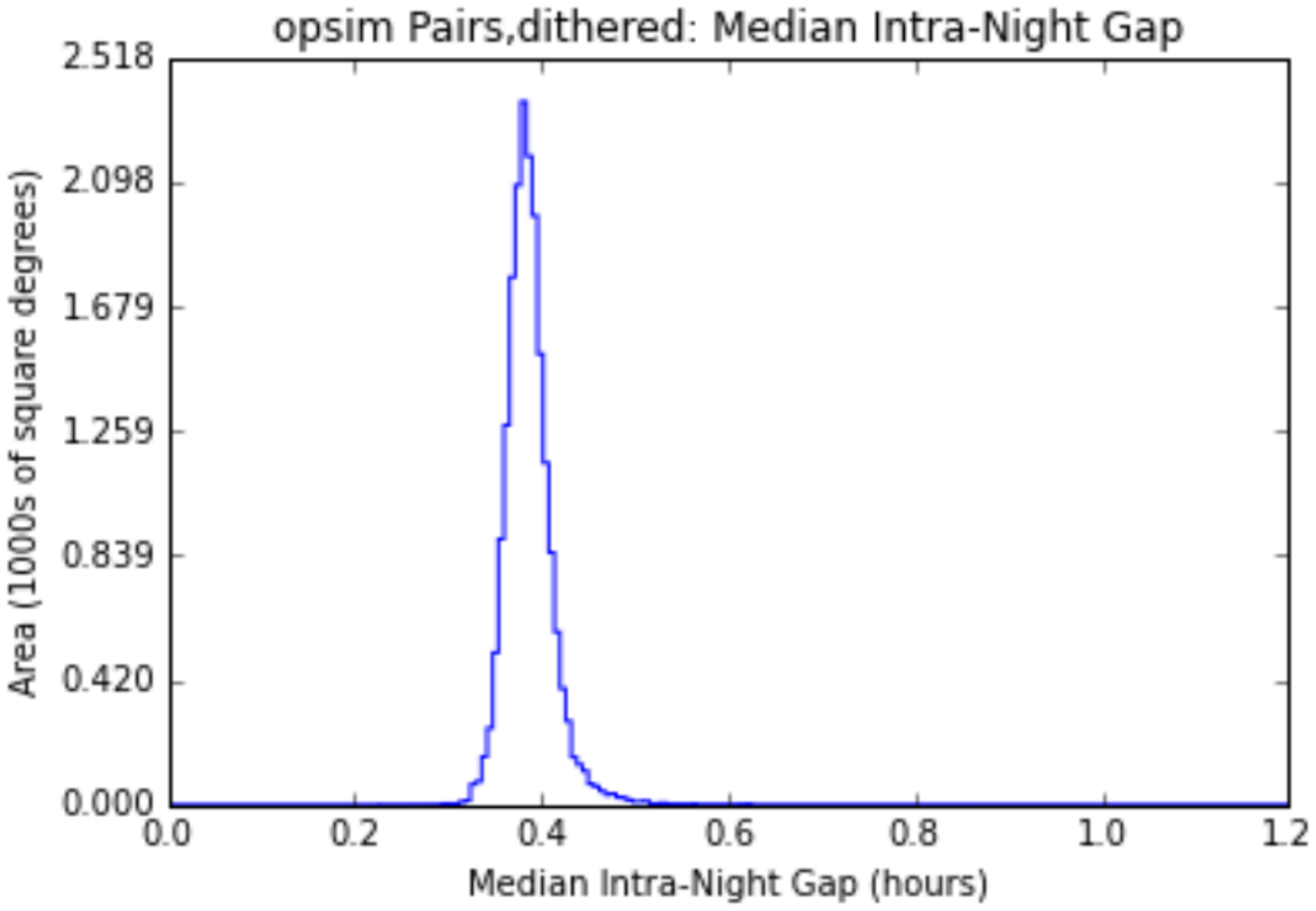}
\includegraphics[angle=0,width=0.49\hsize:,clip]{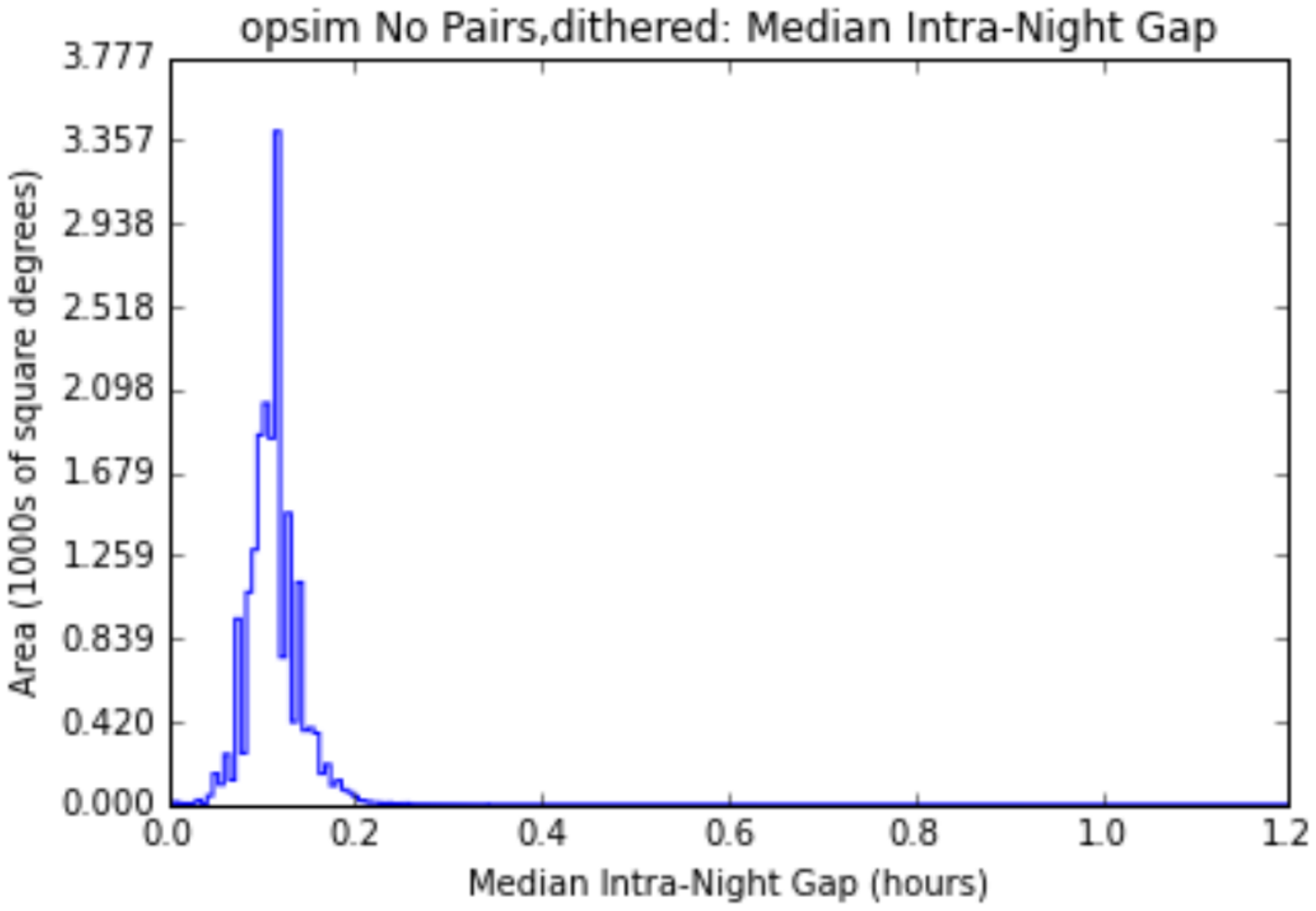}
\vskip -1.3in
\caption{%
The comparison of the median intra-night gap (per field) distributions for the
Baseline Cadence (left)
and simulation \opsimdbref{db:NoVisitPairs}, which did not request pairs of visits per night.
Despite no need for pairs, simulation \opsimdbref{db:NoVisitPairs} produced them ``spontaneously'',
as well as longer sequences (see \autoref{fig:NvisitStats}). The mean field revisit
time is much shorter (about 6 minutes, see the right panel) than for the Baseline Cadence
(22 minutes).}
\label{fig:intranightgapCompare}
\end{figure}
%%%%%%%%%%%%%%%%%%%%%%%%%%%

\begin{figure}[h]
%\vskip -2.5in
\includegraphics[angle=0,width=0.99\hsize:,clip]{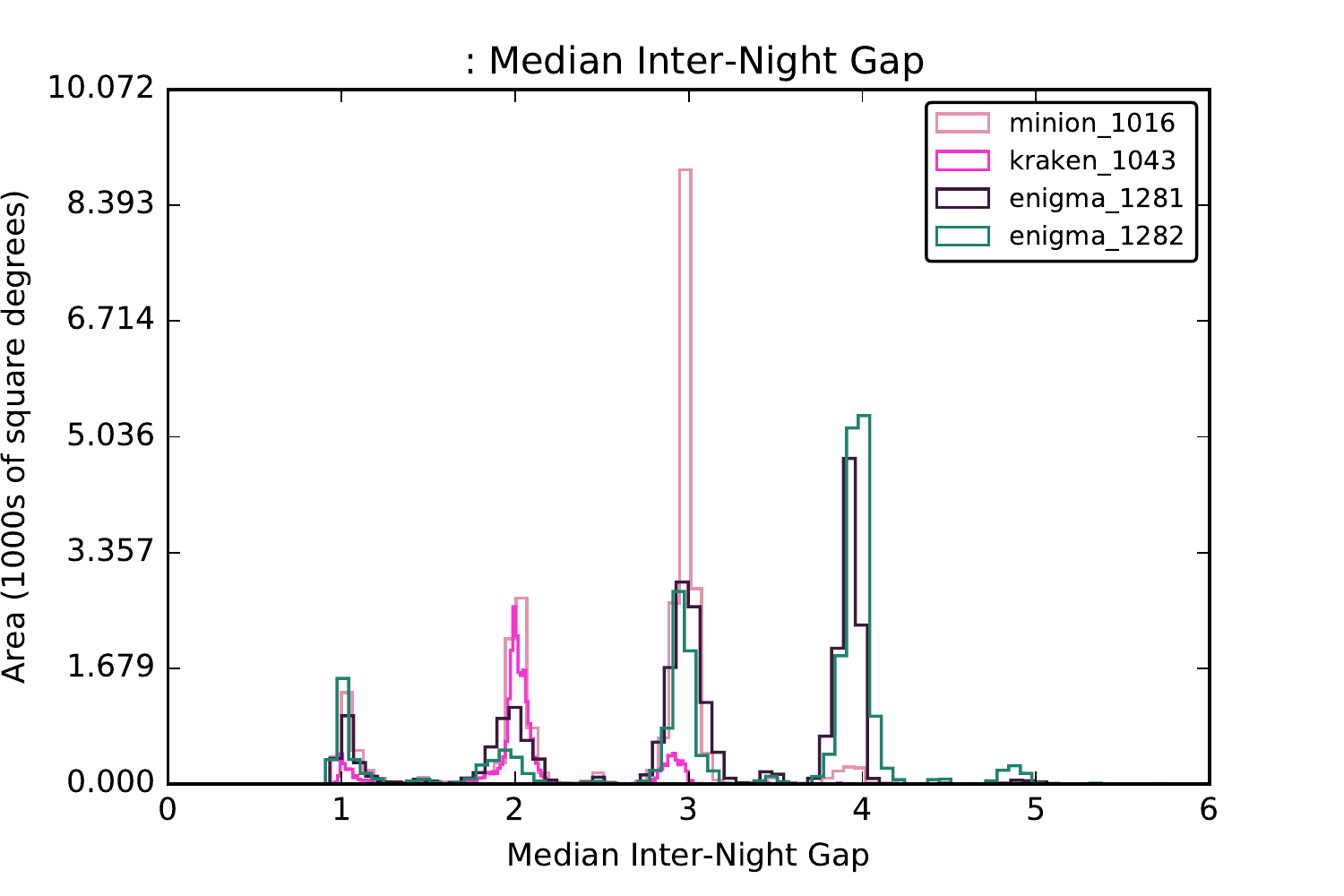}
%\vskip -2.7in
\caption{
Comparison of the median inter-night gap distribution for
\opsimdbref{db:baseCadence}, \opsimdbref{db:NoVisitPairs},
\opsimdbref{db:NEOswithVisitTriplets}, and
\opsimdbref{db:NEOwithVisitQuads}.
The peak median inter-night revisit rate (per field) is about 3 days for the
baseline cadence, \opsimdbref{db:baseCadence}, about 2 days for
\opsimdbref{db:NoVisitPairs}, and closer to 4 for both
\opsimdbref{db:NEOswithVisitTriplets} and
\opsimdbref{db:NEOwithVisitQuads}.}
\label{fig:internightgapCompare}
\end{figure}

{\bf Conclusions:}
The effect of requesting pairs, single visits, triplets or quads, is
softened with the current behavior of the scheduler, where it is not
uncommon to receive more than the requested number of visits within a
night. The inter-night revisit rates are affected, increasing the
inter-night revisit rate by about a night for triplets and quads and
reducing the inter-night revisit rate by about a night when no visit
pairs are requested (from a baseline value of about 3 nights).

% - - - - - - - - - - - - - - - - - - - - - - - - - - - - - - - - - -

%%%%%%%%%%%%%%%%%%%%%%%%%%%%%%%%%%%%%%%%%%%
\opsimdb[db:ShortExptime]{kraken\_1052}{Baseline Cadence, but with 33\% shorter exposure time.}
%%%%%%%%%%%%%%%%%%%%%%%%%%%%%%%%%%%%%%%%%%%

{\bf Motivation and description:} The optimal exposure time per visit
for the main survey, in the limit of a single value for all bands and
at all times, is in the range of about 20--60 seconds \citep[see Section
2.2.2 in the version 3.1 LSST overview paper]{IvezicEtal2008}. This
simulation investigates the effect of decreasing the exposure time per
visit to 20 seconds (from its nominal value of 30 seconds). The
shorter exposure time results in 0.22 mag shallower faint limit per
visit (the effect is larger in the $u$ band, see
\opsimdbref{db:DoubleUbandExptime}). \\

{\bf Expectations:} The total number of visits is expected to increase
by about 50\%, compared to \opsimdbref{db:baseCadence}, to 3.70 million,
for the same survey efficiency. However, the shorter exposure time
will have a significant impact on the survey efficiency: assuming a
slew time of 7 sec, the efficiency drops from 73\% to 65\% (comparing
30/(30+4+7) vs. 20/(20+4+7)). Therefore, the expected increase in the
number of visits is about 32\% and the expected number of visits is
3.2 million.  \\

{\bf Analysis Results:}  This simulated cadence is named
\opsimdbref{db:ShortExptime}. Compared to the Baseline Cadence:
\begin{enumerate}
\item The total number of visits is 3.23 million, representing an
increase of 31\% that is very close to the expected value of 32\%.
\item The median number of visits per night is 1079, or about 32\%
more than for the Baseline Cadence. The total open shutter time is 14\%
smaller for this simulation, and easily understood as due to expected
11\% decrease due to smaller surveying efficiency (the mean slew time
is the same as in the Baseline Cadence, 6.9 sec).
% 2.8 vs 2.1
\item The main survey (WFD, 18,000 sq.deg.) fields received 33\% more
visits than in the Baseline Cadence. The increase in the minimum number of
visits over that area is 9\% (from 886 to 963). In addition, another
1,000 sq.deg. (6\% of the nominal WFD) area has more than 961 visits.
\item Most other performance parameters are essentially unchanged: the
fraction of visits spent on the main survey (88\% vs. 85\%), the
median seeing in the $r$ band (0.93 arcsec vs. 0.94 arcsec), and the
median airmass (1.23 vs. 1.22).
\end{enumerate}

{\bf Conclusions:}
The comparison of \opsimdbref{db:ShortExptime} and
\opsimdbref{db:baseCadence} simulations demonstrates that the effect of
shorter exposures can be easily understood using simple efficiency
estimates. With the visit exposure time is decreased from 30 sec to 20
sec, the surveying efficiency and the total open shutter time drops by
$\sim$10\%, while the number of (shorter exposure time) visits (for
all proposals) increases by 32\%.

% - - - - - - - - - - - - - - - - - - - - - - - - - - - - - - - - - -

%%%%%%%%%%%%%%%%%%%%%%%%%%%%%%%%%%%%%%%%%%%
\opsimdb[db:LongExptime]{kraken\_1053}{Baseline Cadence, but 100\% longer exposure time.}
%%%%%%%%%%%%%%%%%%%%%%%%%%%%%%%%%%%%%%%%%%%

{\bf Motivation and description:} This simulation investigates the
effect of increasing the exposure time per visit to 60 seconds (from
its nominal value of 30 seconds). The longer exposure time results in
0.38 mag deeper faint limit per visit (the effect is larger in the
$u$ band, see \opsimdbref{db:DoubleUbandExptime}). The number of exposures per visit was kept at 2. \\

{\bf Expectations:} The total number of visits is expected to decrease by about
a factor of 2 in case of no significant impact on the survey efficiency.
However, the longer exposure time improves efficiency by a factor of
$2\times(34+7)/(64+7)-1=15\%$, and thus the expected total number of visits is
$0.5\times1.15 = 58\%$ of the number of visits in the Baseline Cadence (assuming
the same mean slew time of 7 seconds).

{\bf Analysis Results:} This simulated cadence is named \opsimdbref{db:LongExptime}.
Compared to the Baseline Cadence:
\begin{enumerate}
\item The total number of visits is 1.42 million or 58\% of the visits
obtained with the Baseline Cadence, and the total open-shutter time is
16\% higher than for the Baseline Cadence. Both results are in good
agreement with above expectations.
\item The median number of visits per night is 467, or 57\% of the
value obtained with the Baseline Cadence. The mean slew time is 0.5 sec
longer than that obtained with the Baseline Cadence.
\item This simulation has significantly different time allocation per
proposal, compared to the Baseline Cadence: 74\% spent on the WFD
proposal (vs. 85\%) and 11\%  spent on the North Ecliptic Spur proposal
(vs. 6\%)  (with smaller and less important differences for other
proposals). Because of these differences, {\it the results of this
test may not be very robust.}
\end{enumerate}

{\bf Conclusions:}
Simple estimates of the total number of visits and the
improvement in efficiency are in good agreement with delivered values. Of
course, the increased efficiency comes at the cost of fewer visits, which is
disadvantageous for time-domain science. There are other consequence of a longer
exposure time, including saturation at fainter mags, worse bleed trails and
scattered light from bright stars, and more trailing of fast-moving asteroids.

%{\bf Note to OpSim team: this simulation should be repeated} with the
%requested number of visits per field set to 60\% of the values used
%for Baseline Cadence for {\bf all} proposals. For example, instead of
%(75, 105, 240, 240, 210, 210) for Universal-18-0824B proposal, (45,
%63, 144, 144, 126, 126) should be used.  This way the additional
%observing time due to improved surveying efficiency will be allocated
%to all proposals, including WFD Cadence. {\it This simulation
%will be repeated with the same North Ecliptic Spur proposal as used
%for \opsimdbref{db:baseCadence}, and with the modified requested number of
%visits.}

% - - - - - - - - - - - - - - - - - - - - - - - - - - - - - - - - - -

%%%%%%%%%%%%%%%%%%%%%%%%%%%%%%%%%%%%%%%%%%%
\opsimdb[db:DoubleUbandExptime]{kraken\_1059}{Baseline Cadence, but with
doubled $u$-band exposure time and the number of visits halved.}
%%%%%%%%%%%%%%%%%%%%%%%%%%%%%%%%%%%%%%%%%%%

{\bf Motivation and description:} The read-out noise in the $u$ band is
not negligible compared to the background noise as in other bands, due
to darker $u$ band sky. The current best estimates for survey
performance \citep[see Table 2 in the version 3.1 LSST overview paper]{IvezicEtal2008} indicate that the {\it coadded} depth in the $u$ band
could be improved by 0.24 mag by increasing the exposure time per
visit from 30 seconds to 60 seconds
%\footnote{In the background-limited
%case, a factor of two increase of the exposure time results in 0.38
%mag deeper data. Since in the $u$ band the read-out noise is not
%negligible compared to the background noise, the total noise increases
%by less than a factor of $\sqrt{2}$ and there is an extra depth
%improvement of 0.24 mag (see eq.~7 and Table 2 the overview paper).
%Conversely, when exposure time is shorter than 30 seconds, there is an
%extra penalty of 0.16 mag, in addition to a loss of depth of 0.22 mag
%due to shorter exposure time in the limit of negligible read-out
%noise.}
(assuming the same total exposure time, which implies a
decrease in the number of visits by a factor of two). To keep the
total exposure time in the $u$ band unchanged, the requested number of
visits in this simulation is decreased by a factor of 2 relative to the
Baseline Cadence specification. \\

{\bf Expectations:} The {\it total} exposure time in the $u$ band should
remain unchanged. The single visits depth should be 0.62 mag deeper
due to increased exposure time, and the coadded depth should be
0.24 mag deeper (due to decreased impact of readout noise). \\

{\bf Analysis Results:} This simulated cadence is named \opsimdbref{db:DoubleUbandExptime}.  Compared
to the Baseline Cadence (\opsimdbref{db:baseCadence}):
\begin{enumerate}
\item The total number of visits is 2.32 million or 95\% of the
Baseline Cadence values. The fraction of time allocated to the main
survey is 84\% vs. 85\% for the Baseline Cadence, and for the NES
proposal 7\% vs. 6\%. %Given that the NE spur proposal was different
%than for \opsimdbref{db:baseCadence}, this simulation needs to be rerun.
\end{enumerate}

{\bf Conclusions:} The $u$ band exposure time can be increased from 30
seconds to 60 seconds without a significant impact on the survey
efficiency. This change would result in a gain of about 0.2 mag in the
coadded depth, with the same total exposure time allocated to the $u$ band.
However, the number of visits in the $u$ band would be decreased by about
a factor of two, with a negative impact on time-domain science.
% - - - - - - - - - - - - - - - - - - - - - - - - - - - - - - - - - -

%%%%%%%%%%%%%%%%%%%%%%%%%%%%%%%%%%%%%%%%%%%
% The simulation previously known as DoubleUbandExptimewithNESpur:
\opsimdb[db:DoubleUbandExptimeSameVisits]{kraken\_1045}{Baseline Cadence, but with
doubled $u$-band exposure time and the same number of visits.}
%%%%%%%%%%%%%%%%%%%%%%%%%%%%%%%%%%%%%%%%%%%

{\bf Motivation and description:} This simulation is similar to
\opsimdbref{db:DoubleUbandExptime}, which increased the exposure time
per visit in the $u$ band from 30 seconds to 60 seconds, with the
requested number of visits decreased by a factor of 2. That change
resulted in a gain of about 0.2 mag in the coadded depth, just from reducing overheads.
However, in order to keep the same total exposure time allocated to the $u$ band,
the number of $u$ band visits in \opsimdbref{db:DoubleUbandExptime} was
decreased by about a factor of two relative to Baseline Cadence, and this had a negative impact on
time-domain science. To further investigate this tradeoff, the \opsimdbref{db:DoubleUbandExptimeSameVisits} simulation was designed and carried out. \opsimdbref{db:DoubleUbandExptimeSameVisits} retains the Baseline Cadence
requested number of visits per field: hence, the coadded depth in the
$u$ band in this simulation could be improved by up to 0.6 mag. \\

{\bf Expectations:}  Given that about 7\% of all visits are allocated
to the $u$ band (56 out of 825), the total number of visits for other bands
will decrease when the number of the $u$ band is doubled, resulting in
about 0.04 mag shallower data in bands other than the $u$ band. \\

{\bf Analysis Results:}  This simulated cadence is named
\opsimdbref{db:DoubleUbandExptimeSameVisits}.  Compared to the Baseline
Cadence (\opsimdbref{db:baseCadence}):
\begin{enumerate}
\item The total number of visits is 2.36 million or 95.5\% of the
Baseline Cadence values. The fraction of time allocated to the main
survey is 85\% vs. 85\% for the Baseline Cadence, and for the NES
proposal 7\% vs. 6\%.

The median number of visits in the $grizy$ filters in the main survey
decreases by about 7\%.  The median single visit depth for main survey
is 0.53 mags deeper, and the median coadded $u$ band image is 0.50 mags
deeper. The other filters reach shallower coadded depth
than in the Baseline Cadence by about 0.05 mag.
\end{enumerate}

{\bf Conclusions:} When the $u$ band exposure time is increased from
30 seconds to 60 seconds, and the number of visits is kept unchanged,
the single-visit and coadded depths would be improved by about 0.5 mag.
This improvement would come at the expense of about 7\% fewer visits in
other bands (with about 0.05 mag shallower coadded depths). These
promising results should be considered when designing the next
incarnation of baseline cadence.

% - - - - - - - - - - - - - - - - - - - - - - - - - - - - - - - - - -

%%%%%%%%%%%%%%%%%%%%%%%%%%%%%%%%%%%%%%%%%%%
\opsimdb[db:UConlyRelaxedAirmass]{minion\_1022}{Only Wide, Fast, Deep Cadence, with relaxed airmass limit.}
%%%%%%%%%%%%%%%%%%%%%%%%%%%%%%%%%%%%%%%%%%%

{\bf Motivation and description:}  What is the effect of changing the
airmass limit from 1.5 to 2.0?  To avoid complicated analysis, we use
only the WFD cadence proposal and thus compare to
\opsimdbref{db:opstwo} (which has the same footprint on ths sky).

{\bf Analysis Results:}  This simulated cadence is named
\opsimdbref{db:UConlyRelaxedAirmass}.  Compared to
\opsimdbref{db:opstwo}, it collected 99.1\% visits. This fraction is
identical to the loss of efficiency due to slightly longer mean slew
time: 7.2 sec vs. 6.8 sec. In addition,
\opsimdbref{db:UConlyRelaxedAirmass} has much worse airmass
distributions than \opsimdbref{db:opstwo}, extending to the allowed
maximum of 2.0. For example, the median for the $r$ band and WFD fields
is 1.30, compared to 1.19 for \opsimdbref{db:opstwo}.

{\bf Conclusions:} This simulation confirms that it's a bad idea to
relax the airmass limit: as a result, the airmass distribution always
widens. In addition, relaxed airmass limit tends to result in a longer
mean slew time.  For a given proposal, the airmass limit has to be as
tight as possible, while still allowing observations of all requested
fields.

% - - - - - - - - - - - - - - - - - - - - - - - - - - - - - - - - - -

%%%%%%%%%%%%%%%%%%%%%%%%%%%%%%%%%%%%%%%%%%%%%%%
\opsimdb[db:UConlyStringentAirmass]{minion\_1017}{Only Wide, Fast, Deep Cadence, with stringent airmass limit.}
%%%%%%%%%%%%%%%%%%%%%%%%%%%%%%%%%%%%%%%%%%%%%%%

{\bf Motivation and description:} What is the effect of changing the airmass
limit from 1.5 to 1.3? To avoid complicated analysis, we use only the WFD
Cadence proposal and thus compare to \opsimdbref{db:opstwo} (i.e., we are not
including the North Ecliptic Spur, South Celestial Pole, or Deep Drilling Fields
in this simulation).

 {\bf Analysis Results:}  This simulated cadence is named
 \opsimdbref{db:UConlyStringentAirmass}. Compared to
 \opsimdbref{db:opstwo}, it collected essentially the same number of
 visits. The mean slew time is also essentially unchanged (7.4 sec vs.
 7.2 sec). The airmass distributions is improved compared to
 \opsimdbref{db:opstwo}. For example, the median for the $r$ band and
 WFD fields is 1.11, compared to 1.18 for \opsimdbref{db:opstwo}.  The
 limiting coadded depth in $u$ and $g$ bands is about 0.1 mag deeper than
 for the Baseline Cadence.

{\bf Conclusions:}  It is possible to achieve the same surveying
efficiency with much more stringent airmass limit than 1.5, which was
used in most simulations to date.  {\it Given this encouraging
behavior, an analogous experiment should be executed for the Baseline
Cadence (i.e.\ a simulation like \opsimdbref{db:baseCadence}, with airmass
limit for the main survey set to 1.3) -- after the ``Western bias'' is
fixed in \OpSim version 4.}

% - - - - - - - - - - - - - - - - - - - - - - - - - - - - - - - - - -

%%%%%%%%%%%%%%%%%%%%%%%%%%%%%%%%%%%%%%%%%%%%%%%
\opsimdb[db:NormalGalacticPlane]{astro\_lsst\_01\_1004}{Extend Wide, Fast, Deep
  Cadence to the Galactic Plane.}
%%%%%%%%%%%%%%%%%%%%%%%%%%%%%%%%%%%%%%%%%%%%%%%

{\bf Motivation and description:} What is the effect of extending the
`normal' WFD Cadence and number of exposures to the region of the
Galactic Plane within the WFD declination limits?  Fields which were
previously contained in the Galactic Plane proposal, but which fell
within the range of the WFD declination limits, were reassigned to the
standard WFD proposal instead of the much more limited Galactic Plane
proposal. The remaining proposals were unchanged, and thus we compare
this to the Baseline Cadence, \opsimdbref{db:baseCadence}.

 {\bf Analysis Results:}  This simulated cadence is named
 \opsimdbref{db:NormalGalacticPlane}. Compared with
 \opsimdbref{db:baseCadence}, there are small to no changes in
 proposals other than the WFD and Galactic Plane proposals
 themselves. The fields which remained in the Galactic Plane proposal
 obtained the same number of visits as expected (30 in each
 bandpass). There were differences observed in the overall status of
 observations taken under the WFD Cadence, in the WFD proposal,
 however. The number of fields included in the WFD proposal increased
 from 2293 to 2470 ($\sim8\%$), as 177 fields were moved from the Galactic Plane
 proposal into the WFD. The total number of observations obtained by
 the WFD proposal increased from 2085270 to 2116203; an increase of
 only $1.5\%$. As a result, the median number of visits per field was
 reduced in all bands, as well as the corresponding coadded depth.

{\bf Conclusions:} In both of these runs, the WFD proposal received
about 85\% of the total number of visits in the survey. Extending the
WFD Cadence to the Galactic Plane, without increasing the
overall priority of the WFD proposal, results in slightly fewer
visits per field as a result, by about 5\%, and a slightly lower
coadded depth, by about 0.04 magnitudes. It is worthwhile to note that
increasing the priority of the WFD proposal could increase the total
fraction of time devoted to WFD, returning the typical number of
visits and coadded depth to baseline levels while decreasing time
spent in other proposals. Metrics targeted for specific science cases,
explored in later chapters, will help determine whether this is a
worthwhile trade.

\navigationbar

% --------------------------------------------------------------------

% ====================================================================
%+
% NAME:
%    rollingcadence.tex
%
% CHAPTER:
%    cadexp2.tex
%
% ELEVATOR PITCH:
%
% AUTHORS:
%    Steve Ridgway (@StephenRidgway)
%-
% ====================================================================

\section{Future Work: Rolling Cadence}
\def\secname{rolling}\label{sec:\secname}

\credit{StephenRidgway}

With a total of $\sim 800$ visits spaced approximately uniformly over 10
years, and distributed among 6 filters, it is not clear that LSST can
offer the sufficiently dense sampling in time for study of transients
with typical durations less than or $\simeq 1$week. This is particularly
a concern for key science requiring well-sampled SNIa light curves.
``Rolling'' cadences stand out as a general solution that can
potentially enhance sampling rates by 2$\times$ or more, on some of the
sky all of the time and all of the sky some of the time, while
maintaining a sufficient uniformity for survey objectives that require
it. In this section we provide an introduction to the concept of rolling
cadence, and give some examples of ways in which it can be implemented.
% We also present three preliminary \OpSim experiments, that serve to
% illustrate some of the features and challenges of the rolling cadence
% strategy.

% --------------------------------------------------------------------

\subsection{The Uniform Cadence}

Current schedule simulations allocate visits as pairs separated by 30-60
minutes, for the purposes of identifying asteroids.  For most other science
purposes, the 30-60 minute spacing is too small to reveal temporal
information, and a pair will constitute effectively a single epoch of
measurement.  If the expected 824 (design value) LSST visits are
realized as 412 pairs, and distributed uniformly over 10 observing
seasons of 6 months each, the typical separation between epochs will be
4 days (typically in different filters).   The most numerous visits will be in the $r$ and $i$
filters, and the repeat visit rate in either of these will be $\simeq$
20 days.

The possibility is still open that, for asteroid identification, visits
might be required as triples or quadruples (which provide a more robust tracking), in which case the universal
temporal sampling will be further slowed by 1.5 or 2$\times$.

Under a strict universal cadence it is not possible to satisfy a need
for more frequent sample epochs.  This has led the LSST Project
simulations group to investigate the options opened up by reinterpreting
the concept of a universal cadence.  Instead of aiming for a strategy
which attempts to observe all fields ``equally'' all the time, it would
allow significant deviations from equal coverage during the survey,
returning to balance at the end of the survey.

Stronger divergence from a universal cadence, allowing significant
inhomogeneities to remain at the end of the survey, is of course
possible, but is not under investigation or discussed here.

There is currently considerable interest in the community in strategies
that provide enhanced sampling over a selected area of the sky, and
rotating the selected area in order to exercise enhanced sampling over
all of the survey area part of the time.  The class of cadences that
provides such intervals of enhanced visits, with the focus region
shifting from time to time, is termed here a ``rolling cadence.''  As a
point of terminology, observing a single sky area with enhanced cadence
for a period of time will be described as a ``roll''.

% --------------------------------------------------------------------

\subsection{Rolling Cadence Basics}

Assume a fixed number of observing epochs for each point on the sky,
nominally distributed uniformly over the 10-year survey duration.  A subset of
these can be reallocated to provide improved sampling of a given sky region in a given time interval.
This will have the inevitable effects of: (1) reducing the number of
epochs available for that sky region during the rest of the survey (affecting, for example, proper motion studies), and
(2) displace observations of other sky regions during the time of the
improved temporal sampling (affecting, for example, early large scale structure studies needing high depth uniformity).  In short, the cadence outside the enhanced
interval will be degraded.

The essential parameters of rolling cadence are: (1) the number of
samples taken from the uniform cadence, and (2) the enhancement factor
for the observing rate.
\href{https://project.lsst.org/meetings/ocw/sites/lsst.org.meetings.ocw/files/OpSim%20Rolling%20Cadence%20Stratgey-ver1.3.pdf}{LSST document 16370},
``A Rolling Cadence Strategy for the Operations Simulator'', by K. Cook
and S. Ridgway, contains more detailed discussion and analysis.

\subsection{Supernovae and Rolling Cadence}
\label{sec:rolling:supernovae}

% Supernovae as a science topic are addressed elsewhere.
% In this section, the demands of SN are used to directly constrain or
% orient the rolling cadence development.

Pending more quantitative guidance, the SN objective for rolling cadence
is to obtain multi-band time series that are significantly longer than
the typical SN duration, and that have a cadence significantly faster
than uniform. As an example we discuss the option of a rolling cadence
with the regular distribution of filters.

As a simple example, consider improving the cadence by a factor of 2 or
3.  If we accept that some regions of the sky will be enhanced every
year, and that uniform sky coverage will only be achieved at the end of 10
years, then we could use, e.g., 10\% of the total epochs in a single
roll.  If the enhancement is 2$\times$, each roll would last for
$\simeq$ 6 months, with high efficiency for capture of complete SN
events.  If the enhancement is 4$\times$, each roll would last for 2
months, with lower efficiency.

If it is important to achieve survey uniformity after 3 years, the
available visits for each roll would be reduced also.  With a 2$\times$
enhancement of epoch frequency, a roll would last 2 months.

Some leverage would be gained by using more than 10\% of the available
visits for a single roll.  However, this begins to impact the sampling
of slow variables reduce schedule flexibility and robustness, and should
be approached with caution.

From these examples, it appears that a 2$\times$ enhancement with
uniformity closure after 10 years is relatively feasible and promising.
Much higher gains, or more rapid closure, require additional
compromises.

% --------------------------------------------------------------------

\subsection{Fast Transients and Rolling Cadence}
\label{sec:rolling:transients}

Fast transients as a science topic are addressed in \autoref{chp:transients}. In this
section, the demands of fast transients are used to directly constrain
or orient the rolling cadence development.

By ``fast transients'', we are referring to events that are sufficiently
fast that they are not addressed by the rolling cadence designed for SN
observations, and slow enough that they are not covered in the cosmological ``deep
drilling field'' cadences (\autoref{sec:intro:baseline}, \autoref{sec:cadexp:opsim}).  For higher tempo rolls, it is quite
difficult to obtain full color data, because of the constraints on
filter selection.  For this example, we will examine a rolling cadence
utilizing only the {\it r} and {\it i} filters, as they are used for
most visits. They are close in wavelength, and we assume that sufficient
color information will be obtained by the ``background'' uniform survey
that continues during a roll.

Again using 10\% of the available visits from the full 10 year survey
for a single roll, we find that there would be enough epochs for each
roll to acquire 1 visit per day for 21 consecutive days, giving an
enhancement of 10$\times$.

Alternatively, the same epochs could be used to observe a target every
20 minutes for 12 hours during a single night (here it is assumed that
visit pairs are not required, doubling the available epochs) for an
enhancement of 300$\times$.

Several different possible redeployments of portions of a uniform survey
have been described, each using 10\% of available time.  Of course it is
possible in principle to implement multiple options, sequentially or
maybe in parallel in some cases. This may pose considerable challenges
to the scheduling strategy design by introducing incompatible boundary
conditions.

While rolling cadences are powerful, they have limitations.  For
example, sampling events that last longer than $\simeq$1 day and less
than $\simeq$ 1 week have the obvious problem of diurnal availability.
In this example, intermediate cadences could be implemented in the
circumpolar region, where diurnal access is much extended.  This is an
example of a case in which a mini-survey of a limited number of regions
could be considered as an alternative to a rolling cadence applied to
the entire main survey.

% --------------------------------------------------------------------

\subsection{Constraints, Trades and Compromises for Rolling Cadences}
\label{sec:rolling:trades}

While rolling cadences offer some attractive benefits, it is important
to realize that rolling cadences are very highly constrained, and that
they do bring disadvantages and compromises.

There are strong arguments against beginning a rolling cadence in the
first, or even the second year of the survey.  Early in the survey, it
is important to obtain for each field/filter combination, an adequate
number of good quality photometric images, and at least one image in
excellent seeing, to support closure of photometry reductions, generation of template images for differencing analysis, and to establish the baseline for proper motion studies.

Since many major science goals require a significant degree of survey
homogeneity, it may be advisable to implement a strategy that brings the
survey to nominal uniform depth at several times, e.g.\ after 3 or 5
years.  This would strongly constrain rolling cadences. We note that such a strategy would also allow a lot of science to be done without waiting the whole 10-years.

Some science objectives favor certain distributions of visits.  For
astrometry, visits early and late in the survey and at large parallax
factors, are beneficial.  Slow variables may benefit from uniform
spacing.  Rolling cadences might impact these constraints either
favorably or unfavorably.

Many objectives are served by randomization of observing conditions for
each field.  Some rolling cadences could tend to reduce this
randomization, for example by acquiring a large number of observations
during a meteorologically favorable or unfavorable season, or during a
period of instrument performance variance.

Dithering may prove challenging with a rolling cadence, since it reduces
temporal coverage at the boundaries of the selected sky region.  This is
negligible for small dithers, but important for large dithers, which are
under consideration; making the contiguous roll area as large as possible should mitigate this issue.

These cautions illustrate that evaluation of rolling cadences must be
based on the {\it full range of schedule performance metrics,} and not just
those targeted by rolling cadence development.

% --------------------------------------------------------------------

\subsection{Directions}
\label{sec:rolling:directions}

While preliminary experiments with rolling cadences have been carried
out with \OpSim, these experiments have significant deficiencies, and
are not suited for in-depth study as of this writing. Designing and
simulating a family of rolling cadences is one of the main goals for the
Project's ``SOCS and Scheduler'' team for \OpSim version 4.
% However, analysis of the cadences described above may guide
% development of objectives for enhancement by rolling cadence.

Rolling cadences will need to satisfy the basic survey science
requirements (including those on sky area, depth and visit count ), and
then be evaluated using the same set of metrics as for other cadences.
Of particular interest will be metrics that clearly distinguish the
gains available with rolling cadences; that is, metrics that measure
schedule performance for variable targets, and especially those with
strong sampling requirements, or more rapid variability.

\navigationbar

% % --------------------------------------------------------------------
%
% \section{Future Work}
% \def\secname{cadexp:future}\label{sec:\secname}
%
%
% \subsection{Ongoing: Extended time-domain metrics}
%
% A number of very sophisticated time-domain metrics have been
% implemented in recent MAF development cycle (and some were contributed
% by the community) but they have not been systematically run yet on all
% available simulations. Time-domain metrics, together with metrics for
% analyzing special programs (e.g. deep drilling programs), will be
% further expanded in the next development cycle.
%
%
% \subsection{Ongoing: Rolling Cadence experiments}
%
% Analysis of a few prototype runs (\texttt{ops2\_1102},
% \texttt{enigma\_1260}, \texttt{enigma\_1261}), which implemented the
% so-called ``swiss cheese rolling cadence'' is in progress.
%
%
%
% \navigationbar
%
% % --------------------------------------------------------------------

\section{Summary}
\def\secname{cadexp:summary}\label{sec:\secname}

The most important conclusion of this chapter is that the upper limit on
possible scheduling efficiency improvements for the Baseline Cadence is
close to 6\%. This conclusion is by and large based on the fact that
the mean slew time for the Baseline Cadence is 6.9 sec, and
thus only slightly larger than the design specifications for the
system slew and settle time of 4.5 sec.  Nevertheless, there are a
number of features to understand, and some to fix, and there is
substantial optimization potential in temporal sampling functions and
further optimization of the sky area and observing strategy details,
that can result in enhanced science even with the same integrated
open-shutter time (e.g. by obtaining deeper data through an improved
sampling of observing conditions).

\vskip 0.2in
The main other questions addressed here are:

\begin{enumerate}

\item {\it By what factor could we exceed the SRD design specification
for the number of visits if only the WFD Cadence proposal was
implemented?}

A simulation that only implemented the WFD Cadence proposal exceeded
the design specification for the number of visits by about 40\% (over
the design specification for the sky area of 18,000 sq.deg.)

\item {\it By what factor could we exceed the SRD design specification
for the sky coverage if only the WFD Cadence proposal was
implemented with the design specification for the number of visits?}

The Pan-STARRS-like strategy results in about 40\% larger sky
coverage (about 25,000 sq.deg.), with the mean number of visits at
92\% of the design specification. The total number of visits is the
same as for the Baseline Cadence, implying similar surveying efficiency.

Therefore, the available ``margin'' relative to the SRD design specifications
for the main survey is equivalent to about 30-40\% larger sky coverage, or
about 30-40\% more visits per field. The SRD assumes that 10\% margin
will be available for other programs. The implied ``survey reserve'',
relative to the WFD Cadence design specifications from the SRD, can
be used to:
  \begin{enumerate}
  \item increase the number of visits per field over the WFD area,  or
  \item increase the surveyed area while keeping the number of visits
  per field statistically unchanged, or
  \item increase both area and the number of visits, and/or
  \item execute additional programs (the current baseline).
  \end{enumerate}

\item {\it What is the effect of auxiliary proposals on surveying
efficiency?}

A comparison of simulations which only implemented the WFD Cadence
proposal to those that included all other programs did not show a
significant change of efficiency (older simulations, not analyzed
here, showed increases in surveying efficiency of up to about 3\% due
to shorter slewing time).

\item {\it What is the effect of visit pairs on surveying efficiency? }

Relinquishing the visit pair requirement results in up to 2-3\%
improvement of the surveying efficiency. The impact on some
time-domain science would be positive, while for NEO and main-belt
asteroid science it would be strongly negative.

\item {\it Can the effects of variations of the visit exposure time on
surveying efficiency be predicted using simple efficiency estimates?}

Simple estimates based on comparing exposure (open shutter) and total
visit times are in good agreement with simulations. Decreasing the
visit exposure time to 20 seconds decreases the total open shutter
time by 10\%, and increasing it to 60 seconds increases the total open
shutter time by 16\%, relative to the Baseline Cadence and standard
exposure time of 30 seconds. The number of visits changes by factors
of 1.35 and 0.58.

\item {\it What are the effects of doubling the exposure time only in
the $u$ band?}

The effect of doubling the exposure time only in the $u$ band, while
simultaneously halving the number of requested visits, has no
significant effect on the survey efficiency.

The effect of doubling the exposure time only in the $u$ band, with
the number of requested visits unchanged, is a decrease in the number
of visits in other bands by about 6\%.

\item {\it What is the impact of the hard airmass limit, $X<1.5$, on the
surveying efficiency?}

It is a very bad idea to relax the airmass limit! It is possible to
achieve the same surveying efficiency with a much more stringent airmass
limit than 1.5, which was used in most simulations to date.

\end{enumerate}

% --------------------------------------------------------------------

\navigationbar

%%  LocalWords:  STARRS cadexp Bremerton MAF opsim baseCadence Aitoff
%%  LocalWords:  airmass coadded WFD Msec airmassenigma ugrizy ivezic
%%  LocalWords:  SRD coaddm urz ury parapmenigma OpSim pre HAenigma
%%  LocalWords:  AltAzenigma radians enigmaGapAll enigmaGapr SNe griz
%%  LocalWords:  enigmaMAXGapAll enigmaEarlySNe Intra opstwoPS opstwo
%%  LocalWords:  UConlyNoVisitPairs NoVisitPairs kraken ShortExptime
%%  LocalWords:  DoubleUbandExptime LongExptime UConlyRelaxedAirmass
%%  LocalWords:  DoubleUbandExptimeSameVisits UConlyStringentAirmass
%%  LocalWords:  NEO PHA NEOs PHAs MOID Neptunian detections intra eq
%%  LocalWords:  differencing NEOsNoVisitPairs NEOswithVisitPairs LMC
%%  LocalWords:  NEOswithVisitTriplets NEOwithVisitQuads NvisitStats
%%  LocalWords:  enigmaNEO intranightgapCompare NEOquads Gould's SMC
%%  LocalWords:  Stubbs multi yoachim rhiannonlynne quartiles swiss
%%  LocalWords:  strawman unrequested

% --------------------------------------------------------------------

\chapter[Solar System]{Discovering and Characterizing Small Bodies in
  the Solar System}
\def\chpname{solarsystem}\label{chp:\chpname}

Chapter editors:
\credit{rhiannonlynne},
\credit{davidtrilling}.

Contributing authors:
\credit{ivezic},
\credit{migueldvb}.

% \section*{Summary}
% \addcontentsline{toc}{section}{~~~~~~~~~Summary}
%
% Executive summary goes here, highlighting the primary conclusions from
% the chapter's science cases. This should be abstract length, no more:
% say, 200 words.

% ====================================================================

\section{Introduction}
\label{sec:\chpname:intro}

LSST has tremendous potential as a discovery and characterization tool
for small bodies in the Solar System. With LSST, we have the
opportunity to increase our sample sizes of Potentially Hazardous
Asteroids (PHAs), Near Earth Objects (NEOs), Main Belt Asteroids
(MBAs), Jupiter Trojans, Centaurs, Trans-Neptunian Objects (TNOs),
Scattered Disk Objects (SDOs), comets and other small body populations
such as Earth mini-moons, irregular satellites, and other planetary
Trojan populations, by at least an order of
magnitude, often two orders of magnitude or more. In addition to
hundreds of astrometric measurements for most objects, LSST will also
provide precisely calibrated multiband photometry. With this
information, we can also characterize these populations -- deriving
colors, light curves, rotation periods, spin states, and even shape
models where possible.

The motivation behind studying these small body populations is
fundamentally to understand planet formation and evolution. The
orbital parameters of these populations record traces of the orbital
evolution of the giant planets. The migration of Jupiter, Saturn and
Neptune in particular have left marks on the orbital distribution of
MBAs, Jupiter Trojans, TNOs and SDOs. Rapid migration of
Jupiter and Saturn may have emplaced a large number of planetesimals
in the Scattered Disk; later slow migration of Neptune will affect the
number of TNOs in resonance and the details of their orbital parameters
within the resonance. Adding color information provides further
insights; colors roughly track composition, indicating formation
location and temperature or space weathering history. For example, the color
gradient of main belt asteroids, combined with their orbital
distribution, suggests that perhaps Jupiter migrated inwards,
mixing planetesimals from the outer Solar System into the outer parts
of the main belt, before eventually migrating outwards. Studying the
size distribution of each of the small body populations themselves
provides more constraints on planetesimal formation; this is
complicated by the effects of dynamical stirring from the giant
planets, which can increase the rate of erosion vs.\ growth during
collisions, and by the existence of the remnants of collisions such as
collisional families in the main belt. The presence of binaries and range
of spin states and shapes provides further constraints on the history
of each population. The location
of the planets before migration, the amount of migration, and the size
distribution of the small bodies themselves (after detangling the
dynamical evolution) all tell a deeper story about how the planets in
the Solar System formed, and how our formation history fits into the
range of observed extrasolar planetary systems.

These Solar System populations are unique when compared to other
objects which will be investigated by LSST, due to the simple fact
that they move across the sky. Metrics to evaluate
LSST's performance for moving objects need to be based on `per object'
measurements, rather than at a series of points on the sky or per
field pointing. For all metrics discussed in this chapter, the orbit
of each object is integrated over the time of the simulated opsim
survey and the detections of each object are recorded (using the
footprint of the focal plane and including
trailing losses and adjusting for the colors of the objects in each
filter to generate SNR and likelihood of detection); these
series of observations per object are then the basis for metric
evaluations.

\navigationbar

% ====================================================================

% ====================================================================
%+
% SECTION:
%    SolarSystem_Discovery.tex
%
% CHAPTER:
%    solarsystem.tex
%
% ELEVATOR PITCH:
%    Discovery of solar system objects, all objects.
%    Figure of merit is completeness.
%
%-
% ====================================================================

\section{Discovery: Linking Solar System Objects}
\def\secname{\chpname:discovery}\label{sec:\secname}

\credit{rhiannonlynne},
\credit{davidtrilling},
\credit{ivezic}

Discovering, rather than simply detecting, small objects throughout
the Solar System requires unambiguously linking a series of detections
together into an orbit. The orbit provides the information necessary
to scientifically characterize the object itself and to understand the
population as a whole. Without orbits, the detections of Solar System
Objects (SSOs) by LSST will be of limited use; objects discovered with
other facilities could be followed up by LSST, but almost the entire
science benefit to planetary astronomy would be lost. Linking and
orbit determination for Solar System objects is similar to source
association for non-moving objects; it provides the means to identify
multiple detections as coming from a single object.

Therefore, the first concern regarding the Solar System is related
to the question ``Can we accurately link individual detections of moving objects into
orbits?''.  This requirement poses varying levels of difficulty as we
move from Near Earth Objects (NEOs) through the Main Belt Asteroids
(MBAs) and to Trans-Neptunian Objects (TNOs) and Scattered Disk Objects
(SDOs), as well as for comets and for other unusual but very
interesting populations such as Earth minimoons
\citep[see \eg][]{2014Icar..241..280B, 2017Icar..285...83F}.
Due to their small
heliocentric and geocentric distances, NEOs appear to move with
relatively high velocities and are distributed over a large fraction
of the sky, including regions far from the ecliptic plane. MBAs are densely distributed,
primarily within about 30 degrees of the ecliptic. TNOs and SDOs move
slowly, however short time intervals between repeat visits in each night may make these difficult
to link. Comets and Earth mini-moons may require more complicated
orbit fitting to allow for non-gravitational or geocentric
orbits. The requirements of accurately linking individual detections
into orbits also implies that we do not create false objects by
incorrectly linking detections and/or noise.

Much of the answer to this question comes down to the performance of
various pieces of LSST Data Management software. In particular,
important questions are the
rate of false positive detections resulting from difference imaging, possible
limitations of the Moving Object Processing System (MOPS) to extend to high
apparent velocities, and the capability to unambiguously determine if
a linkage is `real' or not via orbit determination (done as part of
MOPS). Thus this question includes concerns beyond the limits of the OpSim simulated
surveys, although it still bears on the observing strategy requirements for
discovering Solar System Objects. An in-depth study of the performance
of difference imaging and MOPS is currently ongoing. However, we can
make a range of assumptions on how MOPS will perform and evaluate how
many and which objects can be linked under observational cadence, given those assumptions.

It is important to emphasize that for the vast majority of
LSST-observed moving objects, LSST will be both the discovery and the
recovery facility. Most LSST-observed objects will be too faint
to be observed by assets that are currently carrying out
Solar System small bodies observations, and the number of
detections will be so large that the existing infrastructure
could not observe more than a tiny fraction, even if they
were bright enough. Therefore, as described in this section,
LSST's performance for detecting and re-detecting minor
bodies is critical to the success of the project; no outside
contributions will be significant.

% --------------------------------------------------------------------

\subsection{Target measurements and discoveries}
\label{sec:\secname:targets}

The criteria for `discovery' with MOPS depends on the number
of observations of an object acquired per night within some time
window (creating `tracklets'), repeated over a number of nights within window of some
days (creating `tracks'), which are then linked into an orbit with a threshold on
astrometric residuals. The current assumptions are that we can link
detections into orbits with 2 detections per night within 90 minutes,
repeated for 3 nights within a window of 15 days. The additional assumptions are
that with these 6 observations, we will be able to create low-accuracy orbits that will suffice to link
additional observations obtained at later (or earlier, in the LSST
archive) times, and that the orbit fitting will enable rejection
of mislinkages.

We can also use other requirements for discovery. Requiring 4
detections in each night is a fairly common discovery criteria for
NEO surveys, as it reduces the number of mislinked tracklets to almost
zero. We could also require 4 nights of pairs within a window of 20 days, in order to improve the
initial orbit fitting and mislinkage rejection. We can also assume
MOPS will perform better than the current assumptions, and evaluate
discovery criteria of 3 pairs within a 30 day window.

With these discovery criteria, we can then evaluate the completeness
of an LSST simulated survey, for a given population. We can look at
this as a function of H magnitude and as a function of orbital
parameters.

For PHAs and NEOs there are special considerations in terms of
completion that arise from planetary defense concerns. For most other
populations, the general desire is simply to have a high level of
completeness, with no gaps in completeness that depend strongly on
orbital parameters. In particular, the desire is to be able to
calibrate any selection effects in discovery so that the survey completeness can
be used to debias the underlying population models.

Discovery opportunity, and thus the completeness of the underlying
population, is very sensitive to the time interval between
observations. Waiting longer between observations within a night means that objects
may move beyond a single LSST field of view. Longer times between
revisits means that observing in large contiguous blocks (rather than
narrow, disconnected strips) within a night becomes more important to make sure that
objects are followed between pointings, especially if the time
interval is much longer than 30 minutes. Because the objects must be
detected on several nights within the window, the inter-night revisit
rate for similar large contiguous blocks of sky is important.

An optimal discovery strategy for moving
objects could be ensuring a minimum (default: two) number of revisits
within a night within a short time window (default: 90 minutes),
preferably over a large block of sky, and
covering large contiguous amounts of sky several (default: 3) times within a
longer time window (default: 15 days).  Observations within a single
night do not need to be in the same filter, however we will be
constrained by the shallower limiting magnitude of the pair; {\it e.g.}
preferably $r$ band observations would be paired with $i$ rather than
$u$ observations. Finally, the fill factor of the camera is important;
in these simulations we have used the LSST focal plane, which has an
approximately 92\% fill factor.

% --------------------------------------------------------------------

\subsection{Metrics}
\label{sec:\secname:metrics}

The \MAFmetric{DiscoveryChancesMetric} can be used to identify sets of
detections of a particular object that meet the defined criteria for
discovery: X detections within T minutes in a night, Y nights within a
W day window; this describes the number of discovery opportunities for
each object. The results from the \MAFmetric{DiscoveryChancesMetric} can be fed to the
\MAFmetric{MoCompletenessMetric} summary metric, where if an object achieves
a user-defined requirement for the minimum number of discovery
opportunities (typically 1), then it is counted as `discovered'.  The
total number of objects discovered at each H magnitude is compared to
the total number of objects in the population at that H magnitude, in
order to evaluate `completeness' as a function of H. Discovery
opportunities can be evaluated as a function of orbital parameters, to
look for areas of orbital space that may be missed in a particular
survey strategy; completeness, since it marginalizes over the entire
population at a particular H value, loses this
capability. Completeness can be evaluated as a differential value
(completeness @ H=X) or integrated over the size distribution
(completeness @ H $\leq$ X).

The completeness can be parametrized by the completeness ($C_b$) at
some bright absolute magnitude ($H_b$), combined with the magnitude at
which this falls to 50\% ($H_f$). A draft set of requirements for
these parameters has been written up in the Solar System Object
Specifications document, although these requirements are still quite
preliminary. The current goal parameters are described in Table~\ref{ssoreqs},
balancing a desireable level of completeness with reasonable goals for
the standard LSST observing strategy.

\begin{table}[]
\centering
\caption{Solar System Object Differential Completeness Goals}
\label{ssoreqs}
\begin{tabular}{llll}
    & $C_b$ & $H_b$ & $H_f$ \\
NEA & 80\%  & 18.4  & 21.9  \\
MBA & 90\%  & 19.5  & 20.2  \\
TNO & 90\%  & 7.0   & 8.1
\end{tabular}
\end{table}

A further simplification of the completeness can be achieved by simply
measuring the completeness at a preset absolute magnitude. For
example, completeness for PHAs at H=22 is an important summary value,
and is discussed in its own section, \ref{sec:solarsystem:phas}.

% --------------------------------------------------------------------

\subsection{\OpSim Analysis}
\label{sec:\secname:analysis}

The basic output from the \MAFmetric{DiscoveryChancesMetric} is the number
of discovery opportunities (e.g.\ sets of observations that match the
required discovery criteria) available. For objects which have at
least a given number of discovery opportunities (here, we simply use
one required opportunity), then this object can be considered
``found'' and marked towards the completeness of the population at a
given H magnitude with the \MAFmetric{MoCompletenessMetric} summary metric.
Examining the \opsimdbref{db:baseCadence} Baseline Cadence with these metrics,
we find that most objects have many discovery opportunities. This is shown in
Figure~\ref{standard_discovery}.

Evaluating these metrics requires choosing an input solar system population;
for all tests here, we have chosen a random set of 2,000 objects of each type from
the Grav S3M model \citep{2011PASP..123..423G}. By cloning these orbits over a
range of $H$ values, we can rapidly generate detections and evaluate the discovery
metrics for a range of simulated surveys. The sample of 2,000 orbits per solar system
population provides a completeness estimate comparable with the estimate
resulting from evaluating a larger population, although with slightly more
statistical noise.

\begin{figure}
\includegraphics[width=3.3in]{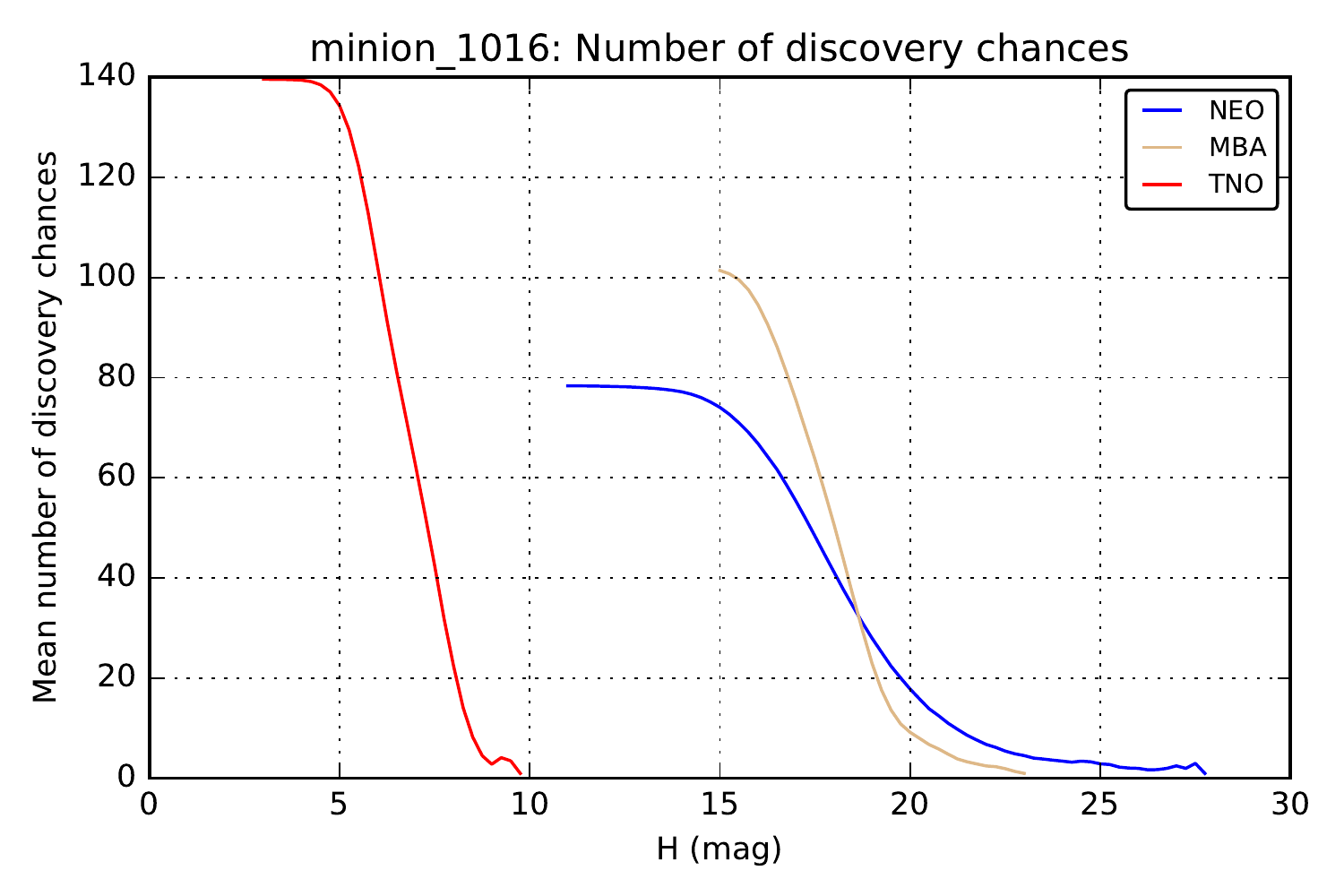}
% PJM: need shorter filenames for arxiv
% \includegraphics[width=3.3in]{figs/solarsystem/minion_1016_CumulativeCompleteness_tno_mba_neo_10_year_3_pairs_in_15_nights_MOOB_ComboMetricVsH.pdf}
\includegraphics[width=3.3in]{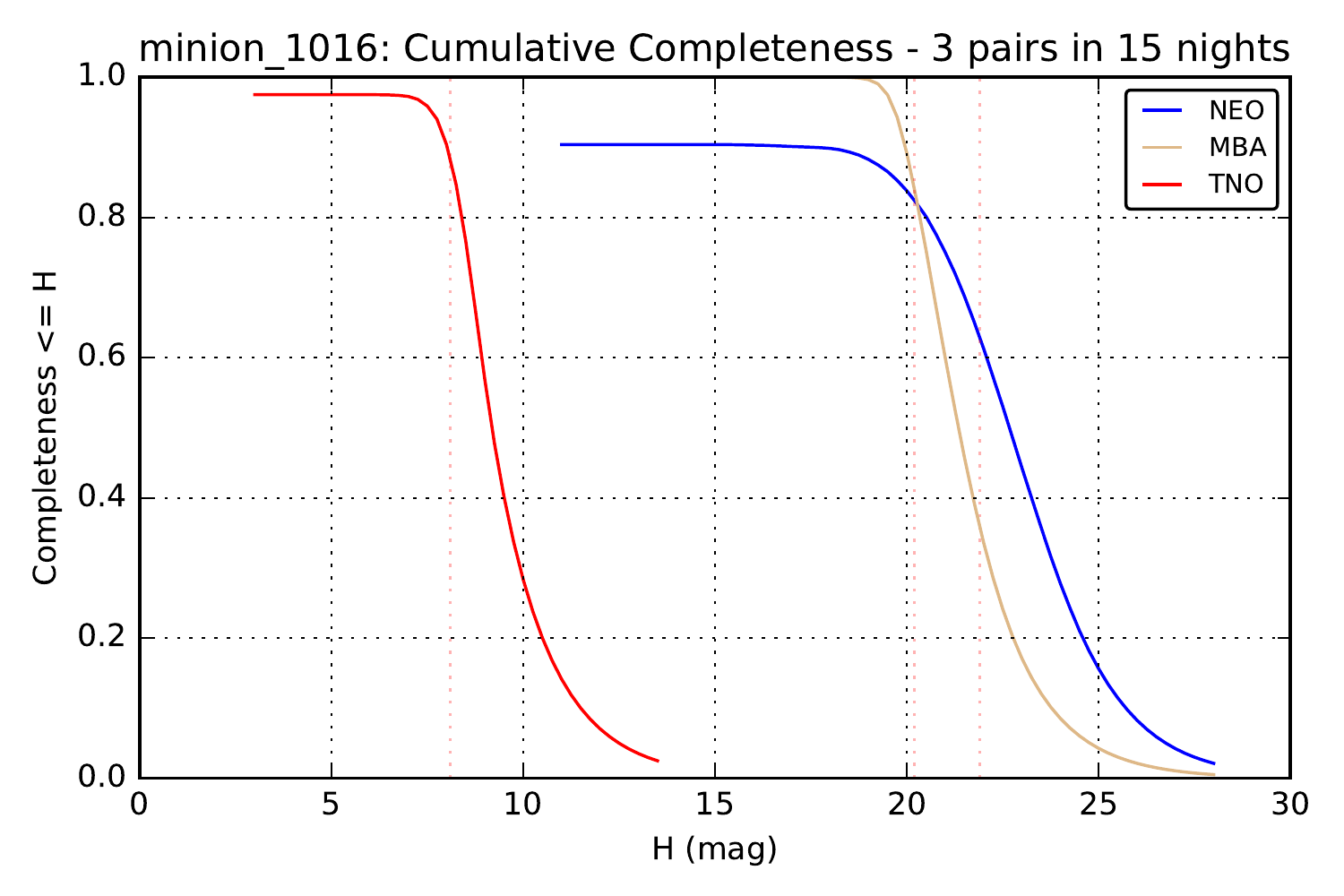}
\caption{Left: Number of discovery chances as a function of H
  (mean value for all objects at each $H$ value), assuming the minimum criteria for
  discovery - 2 visits per night within 90 minutes, repeated for 3
  nights within 15 days. Right: Resulting cumulative completeness for
  each population, assuming that only 1 discovery opportunity is
  required to `discover' each object.
\label{standard_discovery}}
\end{figure}

The runs \opsimdbref{db:baseCadence}, \opsimdbref{db:NoVisitPairs},
\opsimdbref{db:NEOswithVisitTriplets}, \opsimdbref{db:NEOwithVisitQuads}
are particularly interesting to evaluate in light of the different
sets of discovery criteria. Because OpSim does not currently require
only pairs (or singles, triplets or quads), but instead will sometimes
acquire more than the requested number of visits, changing the
discovery criteria from pairs within a night to triplets or quads,
does not automatically cause the completeness to plummet, although it
does decrease. Looking at the raw numbers of discovery chances offers some
enlightenment: the number of discovery opportunities does falls dramatically as we go from pairs to quads, however, there
are still some times when observations are obtained in triplets or
quads, so there are still some discovery chances. This behavior of the
scheduler (to frequently acquire more than the requested number of
visits) is likely to change in the future and make this effect more pronounced, but the completeness will
clearly be very sensitive to how observations are acquired. This effect is shown in
Figure~\ref{completeness_changes}.

\begin{figure}
% PJM: the following filenames are too long for the arxiv! Made copies:
%
% \includegraphics[width=3.3in]{figs/solarsystem/minion_1016_CumulativeCompleteness_pairs_20_4_quads_3_30_3_30_triplets_3_30_pairs_3_15_pairs_nights_in_neo_year_10_MOOB_ComboMetricVsH.pdf}
% \includegraphics[width=3.3in]{figs/solarsystem/kraken_1043_CumulativeCompleteness_pairs_20_4_quads_3_30_3_30_triplets_3_30_pairs_3_15_pairs_nights_in_neo_year_10_MOOB_ComboMetricVsH.pdf} \\
% \includegraphics[width=3.3in]{figs/solarsystem/enigma_1281_CumulativeCompleteness_pairs_20_4_quads_3_30_3_30_triplets_3_30_pairs_3_15_pairs_nights_in_neo_year_10_MOOB_ComboMetricVsH.pdf}
% \includegraphics[width=3.3in]{figs/solarsystem/enigma_1282_CumulativeCompleteness_pairs_20_4_quads_3_30_3_30_triplets_3_30_pairs_3_15_pairs_nights_in_neo_year_10_MOOB_ComboMetricVsH.pdf}
%
% Shorter filenames:
%
\includegraphics[width=3.3in]{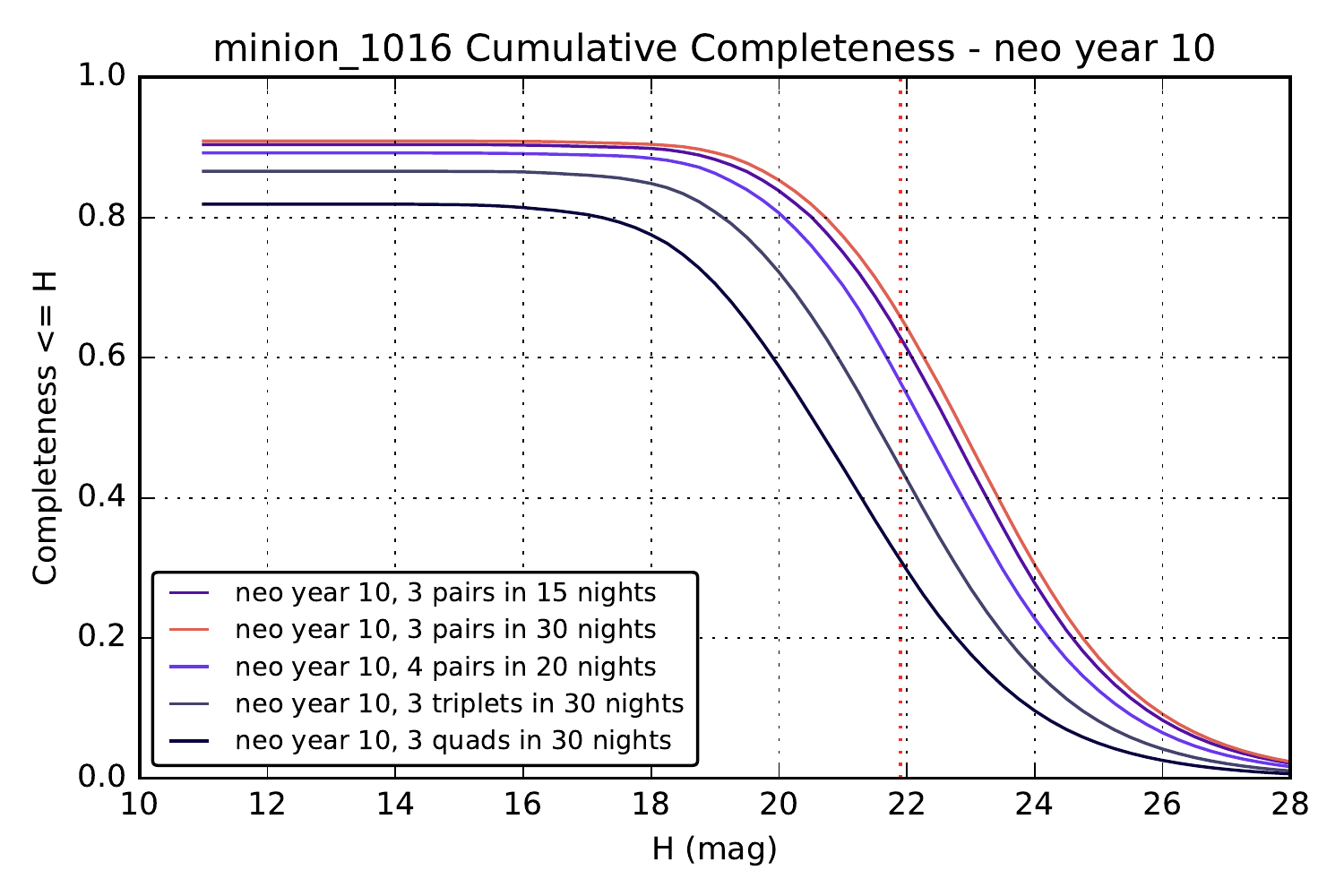}
\includegraphics[width=3.3in]{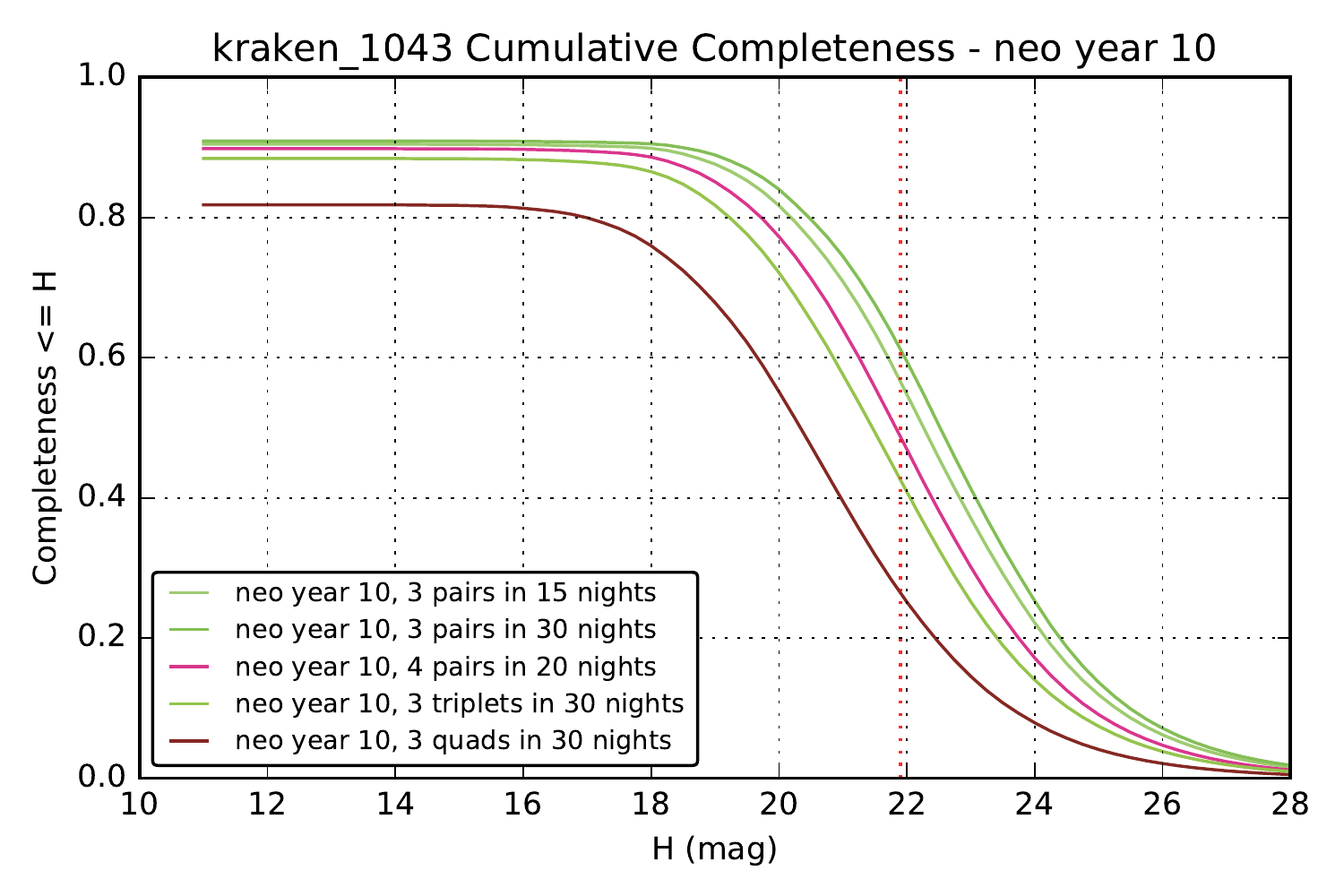} \\
\includegraphics[width=3.3in]{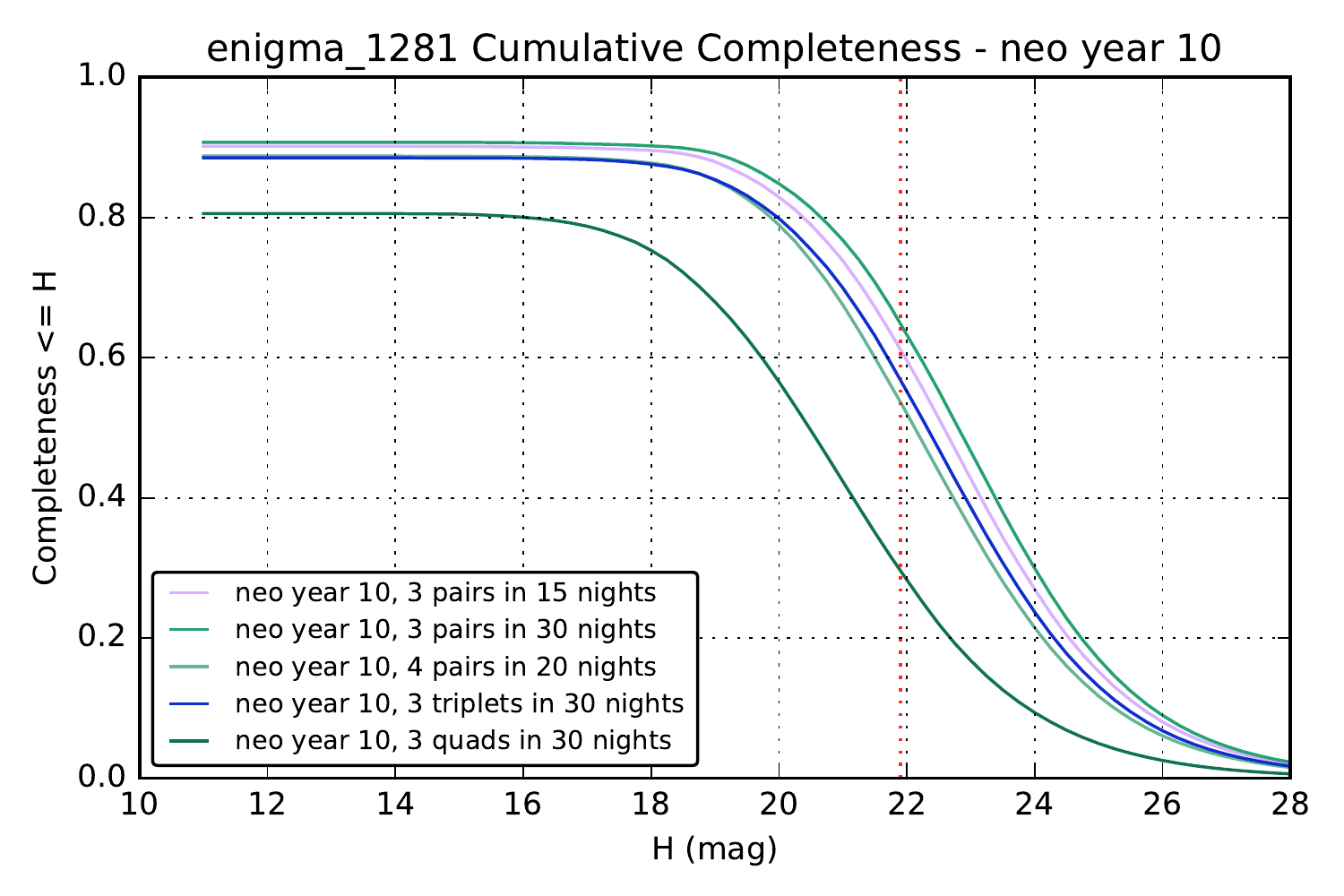}
\includegraphics[width=3.3in]{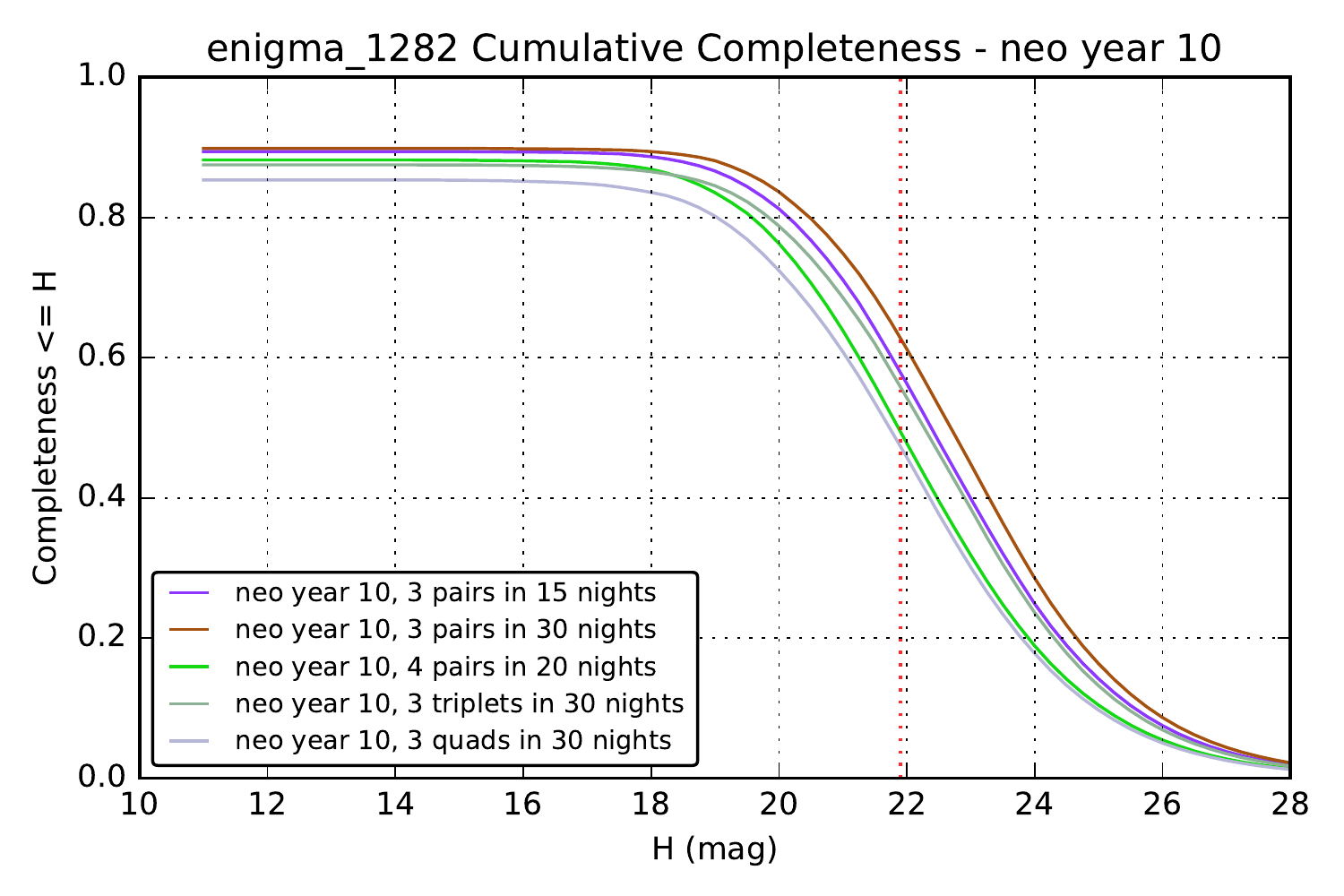}
\caption{Cumulative completeness for an NEO population, given
  different sets of discovery criteria, for the \opsimdbref{db:baseCadence}, \opsimdbref{db:NoVisitPairs},
\opsimdbref{db:NEOswithVisitTriplets},
\opsimdbref{db:NEOwithVisitQuads} simulated surveys. The results in the
lower right come from a simulated survey, \opsimdbref{db:NEOwithVisitQuads},
which attempted to obtain four visits to each field in each
night; the results on the upper left, come from the baseline simulated
survey, \opsimdbref{db:baseCadence}, which attempts to obtain pairs of
visits.
\label{completeness_changes}}
\end{figure}

Another aspect to consider is to look at how the completeness
increases over time. The completeness as a function of time is plotted
for particular $H$ values, depending on the
population, in Figure~\ref{completeness_time}.

\begin{figure}
\includegraphics[width=2in]{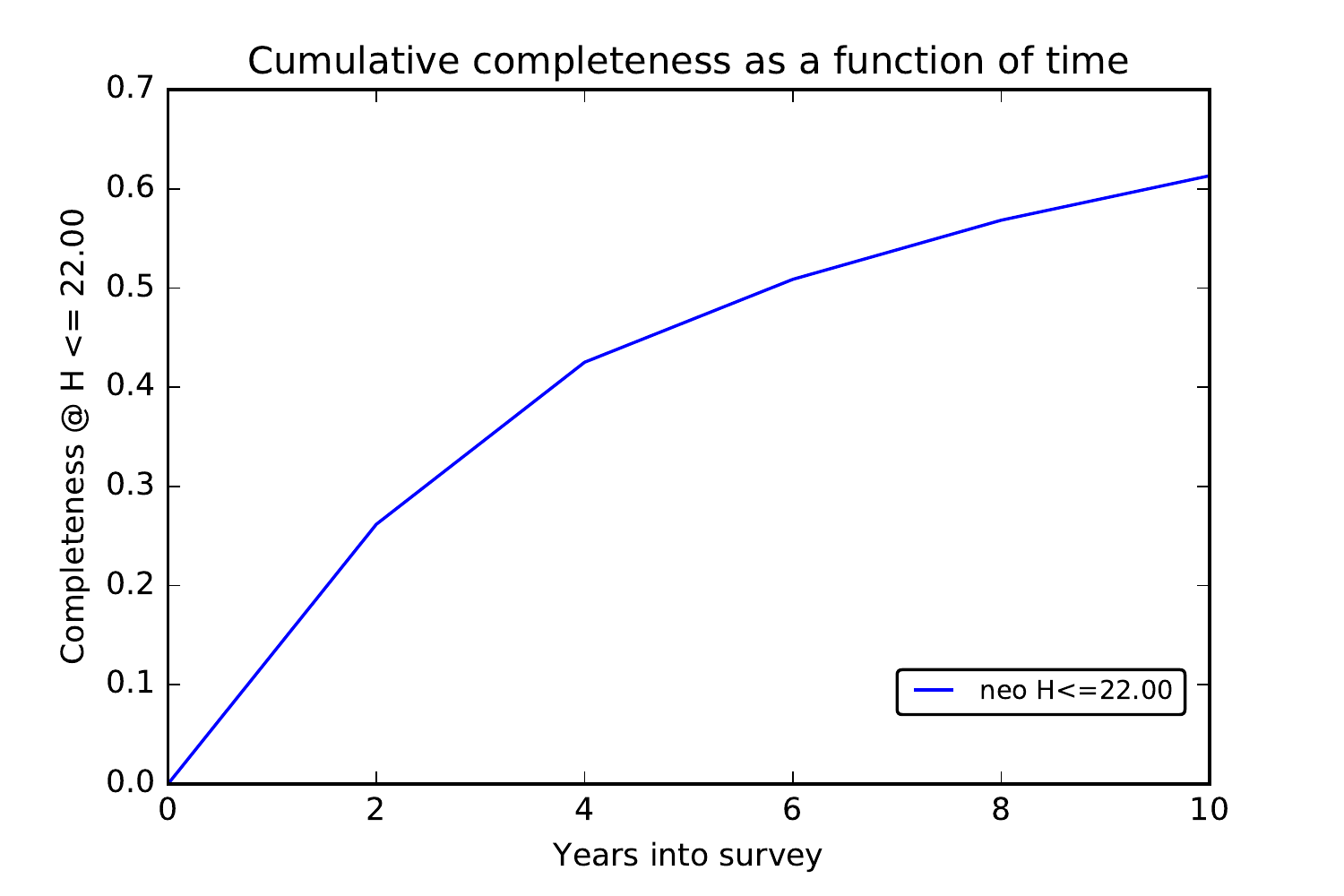}
\includegraphics[width=2in]{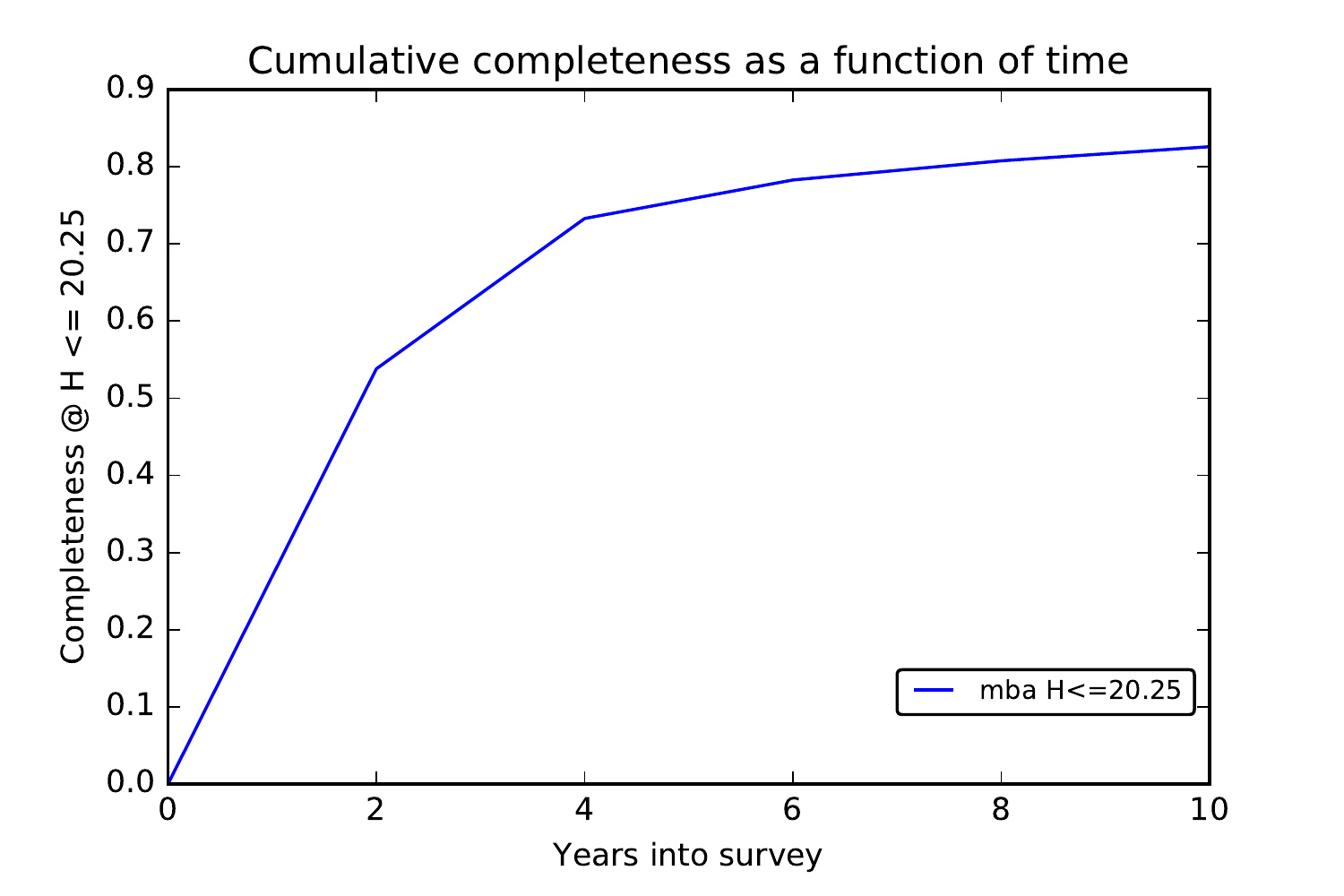}
\includegraphics[width=2in]{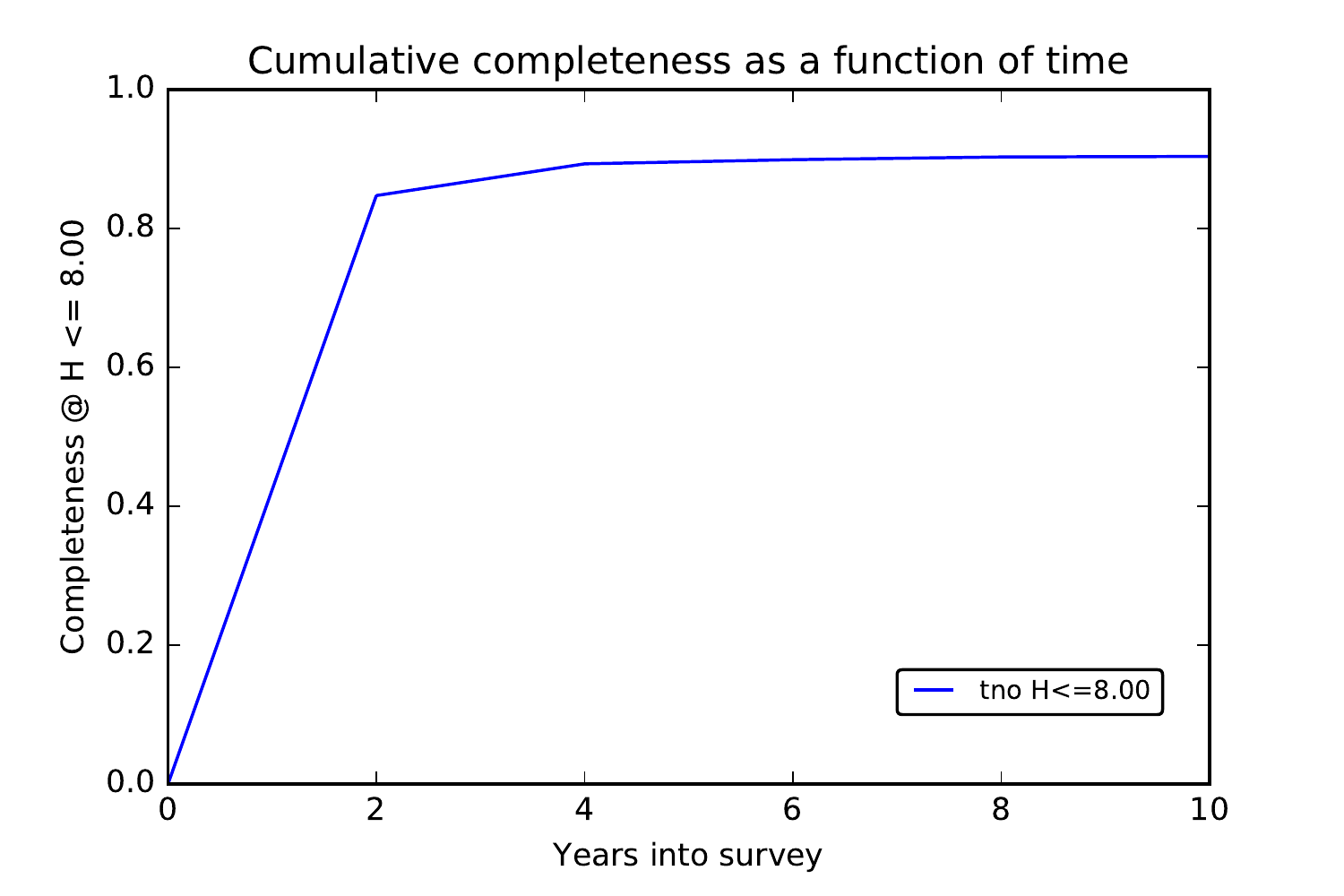}
\caption{Completeness as a function of time, for NEO, MBA and TNO
  populations. The completeness increases rapidly for the first few
  years, then increases more slowly. The NEO completeness rises more
  slowly than other populations, as more NEOs become available to
  discover due to changing their orbital positioning relative to Earth
  (becoming closer and brighter, or moving away from sightlines behind
  the Sun). The TNO completeness rises most rapidly with time, as
  these objects move slowly; we find most of these objects within the
  first two years and then improve their characterization over the
  rest of the survey (measuring better orbits and obtaining
  lightcurves and colors).
\label{completeness_time}}
\end{figure}

\begin{table}[]
\centering
\caption{Solar System Object Differential Completeness in \opsimdbref{db:baseCadence}}
\label{ssoperf}
\begin{tabular}{llll}
    & $C_b$ & $H_b$ & $H_f$ \\
NEA & 87.5\%  & 18.5  & 21.5  \\
MBA & 89\%  & 19.5  & 20.2  \\
TNO & 96\%  & 7.0   & 8.3
\end{tabular}
\end{table}

% --------------------------------------------------------------------

\subsection{Discussion}
\label{sec:\secname:discussion}

A large portion of the risk in being able to discover moving objects
lies in the currently uncertain performance of the MOPS
software. Figure~\ref{completeness_changes} clearly shows that with
the baseline cadence, if we must have triplets or pairs, or even just
require 4 pairs of observations over 20 nights, that the completeness
falls. The performance will likely fall even further if the scheduler
stops obtaining more than the minimum requested number of observations.

With the expected MOPS discovery requirements,
\opsimdbref{db:baseCadence} performs adequately for most solar system
objects, as seen in Table~\ref{ssoperf}, although completeness falls off more rapidly for faint
objects than desired for NEOs. To investigate this effect, more
metrics will have to be developed to discover why these fainter NEOs
are not being discovered (are they simply missing appropriate
sequences of observations due to the cadence or is something more
subtle occurring?).

\navigationbar

% ====================================================================

% ====================================================================
%+
% SECTION:
%    SolarSystem_PHA.tex
%
% CHAPTER:
%    solarsystem.tex
%
% ELEVATOR PITCH:
%    Discovery of PHAs in particular. Discussion of wider 'impacts'.
%
%-
% ====================================================================

\section{Discovery of Potentially Hazardous Asteroids}
\def\secname{\chpname:phas}\label{sec:\secname}

\credit{ivezic},
\credit{rhiannonlynne}.

The U.S. Congress has given a mandate to NASA to implement a
Near-Earth Object (NEO) Survey program to detect, track, catalog, and
characterize the physical characteristics of near-Earth objects equal
to or greater than 140 meters in diameter\footnote{See
\url{http://www.gpo.gov/fdsys/pkg/PLAW-109publ155/pdf/PLAW-109publ155.pdf}}. The
goal is to achieve a completeness of 90\%. In recent practice, adopted
here, the completeness is evaluated for a subset of NEOs called
Potentially Hazardous Asteroids\footnote{Potentially Hazardous
Asteroids (PHAs) are defined as asteroids with a minimum orbit
intersection distance (MOID) of 0.05 AU or less.} (PHA), with
H$\le$22, where H is the absolute magnitude\footnote{Absolute
magnitude is the magnitude that an asteroid would have at a distance
of 1 AU from the Sun and from the Earth, viewed at zero phase
angle. This is an impossible configuration, of course, but the
definition is motivated by desire to separate asteroid physical
characteristics from the observing configuration.} in the Johnson's V
band.

The discovery criteria for PHAs follows the same guidelines and metrics found in the previous
section, \ref{sec:solarsystem:discovery}, but is worth discussing
separately to focus on its main figure
of merit - completeness for PHAs with H$\le$22 magnitudes.

% --------------------------------------------------------------------

\subsection{Target measurements and discoveries}
\label{sec:\secname:targets}

Using the same range of discovery criteria as in the previous section,
\ref{sec:solarsystem:discovery}, we can look at the differential and
cumulative completeness for a population of PHAs. For this sample of
PHAs, we pulled the orbits of 2,000 objects with MOID~$<= 0.05$~AU from
the Grav S3M model \citep{2011PASP..123..423G}. These orbits were
then cloned over a range of $H$ values to evaluate the chances of
discovery for that orbit at each of those $H$ values. The differential
completeness as a function of $H$ is then simply the fraction of
objects which receive at least one set of observations which meet the
discovery criteria during the course of the survey. The cumulative
completeness is similar, but integrated over $H$ by assuming an $H$
distribution with a power-law index of $\alpha=0.3$. Both
differential and cumulative completeness are relevant metrics: the
former provides more insight in the behavior of a particular
simulation, while the latter is a metric given to NASA by the U.S.
Congress.

To match the NEO mandate, the cumulative completeness at $H$=22 can be
used as a figure of merit.

% --------------------------------------------------------------------

\subsection{Metrics}
\label{sec:\secname:metrics}

The metrics used here are the same as in
\ref{sec:solarsystem:discovery}, although run with different input populations.

% --------------------------------------------------------------------

\subsection{OpSim Analysis}
\label{sec:\secname:analysis}

The differential and cumulative completeness for the baseline survey,
\opsimdbref{db:baseCadence}, at a range of years is shown in
\autoref{fig:baselinePHA}. The baseline cadence achieves a cumulative completeness of 66\% for
H$\le$22 PHAs when requiring pairs of visits on 3 separate nights within 15 days.
The differential completeness at $H$=22 for the same
survey is 49\%, 17\% lower due to increasing completeness toward
smaller $H$ (larger objects).

%%%%%%%%%%%%%%%%%%%%%%%%%%%
\begin{figure}[th]
%\vskip -1.1in
\includegraphics[angle=0,width=0.49\hsize:,clip]{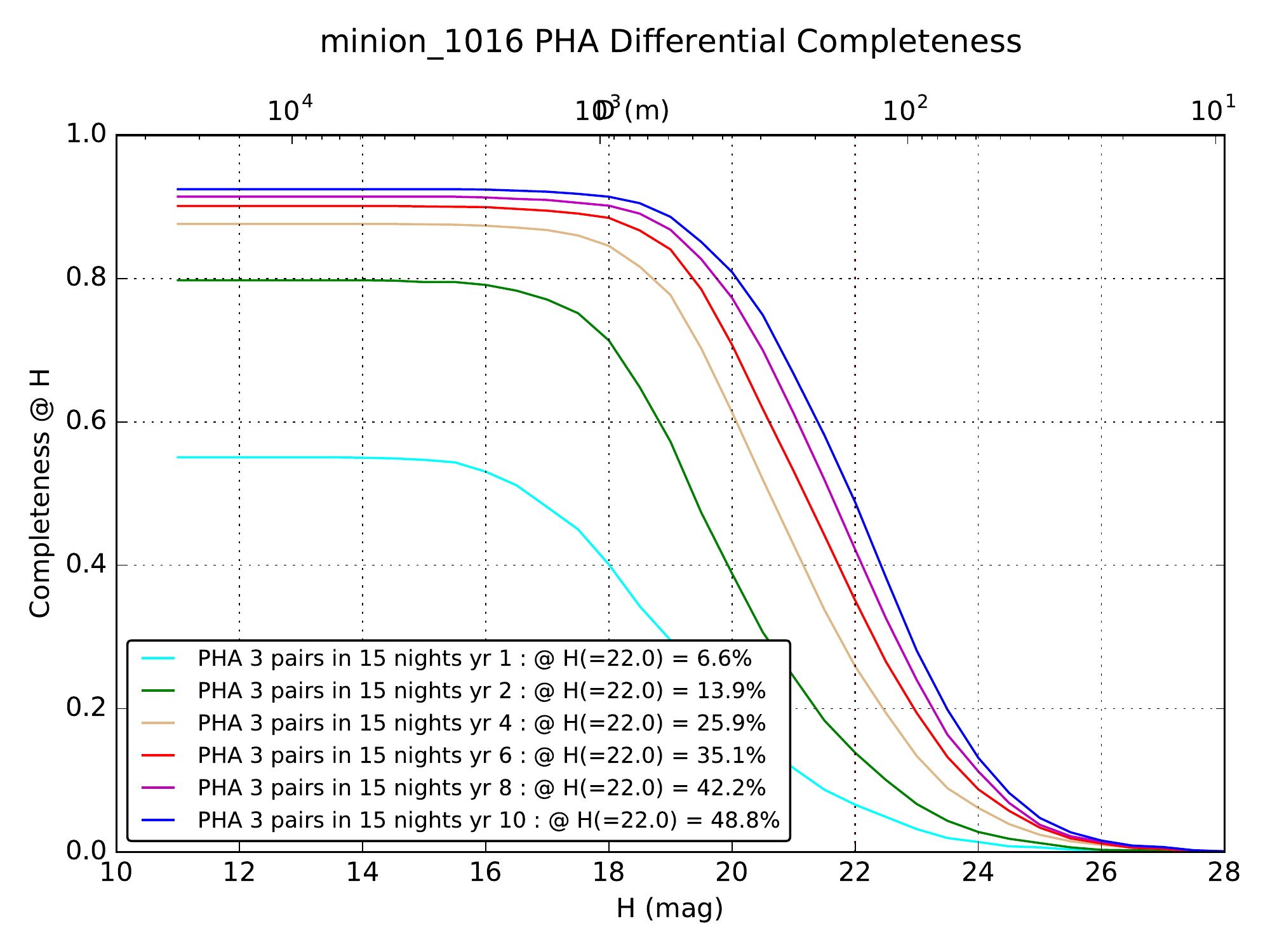}
% PJM: need shorter filenames for the arxiv:
% \includegraphics[angle=0,width=0.49\hsize:,clip]{figs/solarsystem/minion_1016_CumulativeCompleteness_PHA_3_pairs_in_15_nights_Years_1_to_10_MOOB_ComboMetricVsH.pdf}
\includegraphics[angle=0,width=0.49\hsize:,clip]{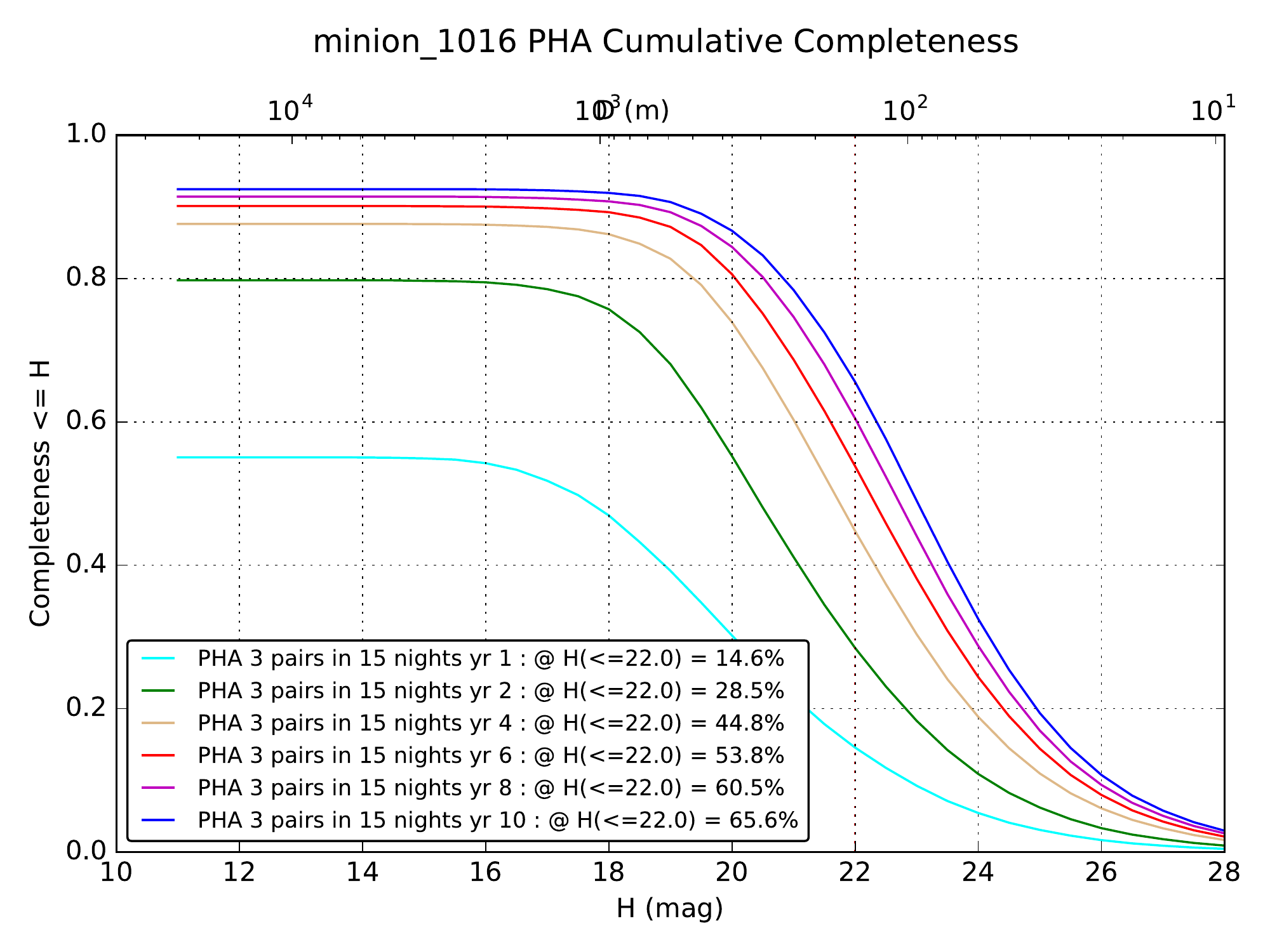}
%\vskip -1.2in
\caption{The PHA completeness for \opsimdbref{db:baseCadence}, as a function of the object's absolute
visual magnitude H on the horizontal axes (left: differential completeness at a given H;
right: cumulative completeness for all objects brighter than a given H), as it increases year over year.
The cumulative completeness for H$\le$22 NEOs (those with diameters larger than 140m)  for this
simulation is 66\% after 10 years.}
\label{fig:baselinePHA}
\end{figure}
%%%%%%%%%%%%%%%%%%%%%%%%%%%

We find that the PHA and NEO completeness are very similar for a given simulated survey and set of discovery criteria, as shown in \autoref{fig:neopha},
a reasonable result given that the input populations are very similar.
The analysis of the various observing run strategies (singles, pairs, triples or quads of visits) described in the previous section thus applies to PHAs as well.

%%%%%%%%%%%%%%%%%%%%%%%%%%%
\begin{figure}[bh]
%\vskip -1.2in
\includegraphics[angle=0,width=0.49\hsize:,clip]{figs/solarsystem/minion_1016_CumulativeCompleteness_NEO_and_PHA_Cumulative_Completeness.pdf}
%\vskip -1.3in
\caption{%
Comparison of the cumulative NEO and PHA completeness for the baseline cadence
\opsimdbref{db:baseCadence}.}
\label{fig:neopha}
\end{figure}

\navigationbar

% ====================================================================

% ====================================================================
%+
% SECTION:
%    SolarSystem_OrbitalAccuracy.tex
%
% CHAPTER:
%    solarsystem.tex
%
% ELEVATOR PITCH:
%    How secure is the orbit - is it going to hit us?
%    Libration amplitude distribution for TNOs?
%    Can we find it after X years for further study?
%    Can we identify the source region for NEOs within the main belt?
%-
% ====================================================================

\section{Orbital Accuracy}
\def\secname{\chpname:orbits}\label{sec:\secname}

\credit{rhiannonlynne},
\credit{davidtrilling}

A vast number of moving objects will appear in LSST images. Multiple
observations of a common object will be linked, and a preliminary orbit
derived. However, the orbital elements (semi-major axis, eccentricity,
etc.) will have some uncertainty. Short arcs --- that is, a small amount
of time between the first and last observation of a given object ---
produce orbits with large uncertainties on the orbital elements. As arc
length grows, the orbital uncertainties decrease.

A number of science cases require relatively small uncertainties on
orbital elements. Perhaps most importantly, small uncertainties can aid
in discriminating between Near Earth Objects that might and might not
impact the Earth. A more subtle example relates to the libration
amplitude distribution for TNOs, which can be compared to predictions
from Solar System formation models. Only with small uncertainties on
orbital elements can the libration amplitudes be determined to
sufficient precision to compare to the predictive models. Finally,
during and after the primary LSST survey additional measurements will be
desired for further characterization of many objects. Only if the
orbital elements are sufficiently well known can objects be studied later
with other facilities. For example, to carry out spectroscopy, the
position of the object must be known to approximately 1~arcsec (the
width of a typical slit). This places strong requirements on the
knowledge of the orbital elements.

% --------------------------------------------------------------------

\subsection{Target measurements and discoveries}
\label{sec:\secname:targets}

The relevant data here are positions as a function of time for a given
object (assuming that the linking of measurements to a given object is
satisfactory). Assuming that the accuracy and precision of each
measurement are approximately constant (likely, since all will be made
by the same observing system), the only significant factor that improves
the knowledge of the orbit is extending the observational arc. The
observing strategy employed by LSST must therefore have a cadence in
which objects are revisited with the largest possible arcs that still
allow linking of observations. In other words, if the observations of a
given object are too widely spaced, linking may not be possible, so,
even though the arc is long, the linking is poor and the object yield is
low. If the observations are made too densely in time, linking is likely
to be good, but the arc may not be very long. A middle ground is
desired.

% --------------------------------------------------------------------

\subsection{Metrics}
\label{sec:\secname:metrics}

The best metric here would be to take the actual series of observations
of each object, add appropriate astrometric noise to each observation
according to its SNR, cull observations which would not be `linkable' to
the rest (i.e.\ observations which occur on a single night far from other
nights in the arc, or even a series of observations which occur too many
years away from other observations of the same object), and then fit an
orbit to the remaining observations and determine the uncertainty in its
parameters. This is work for the future however; our first simple proxy
uses the \MAFmetric{ObsArcMetric} to just look at the time between the first
and last observation of an object. For many objects, this will be fairly
close to the actual arc length of the linkable observations, as most
objects receive many observations clumped together when they are
observable, so this simple proxy makes a reasonable starting point.

% --------------------------------------------------------------------

\subsection{OpSim Analysis}
\label{sec:\secname:analysis}

\begin{figure}
\includegraphics[width=6in]{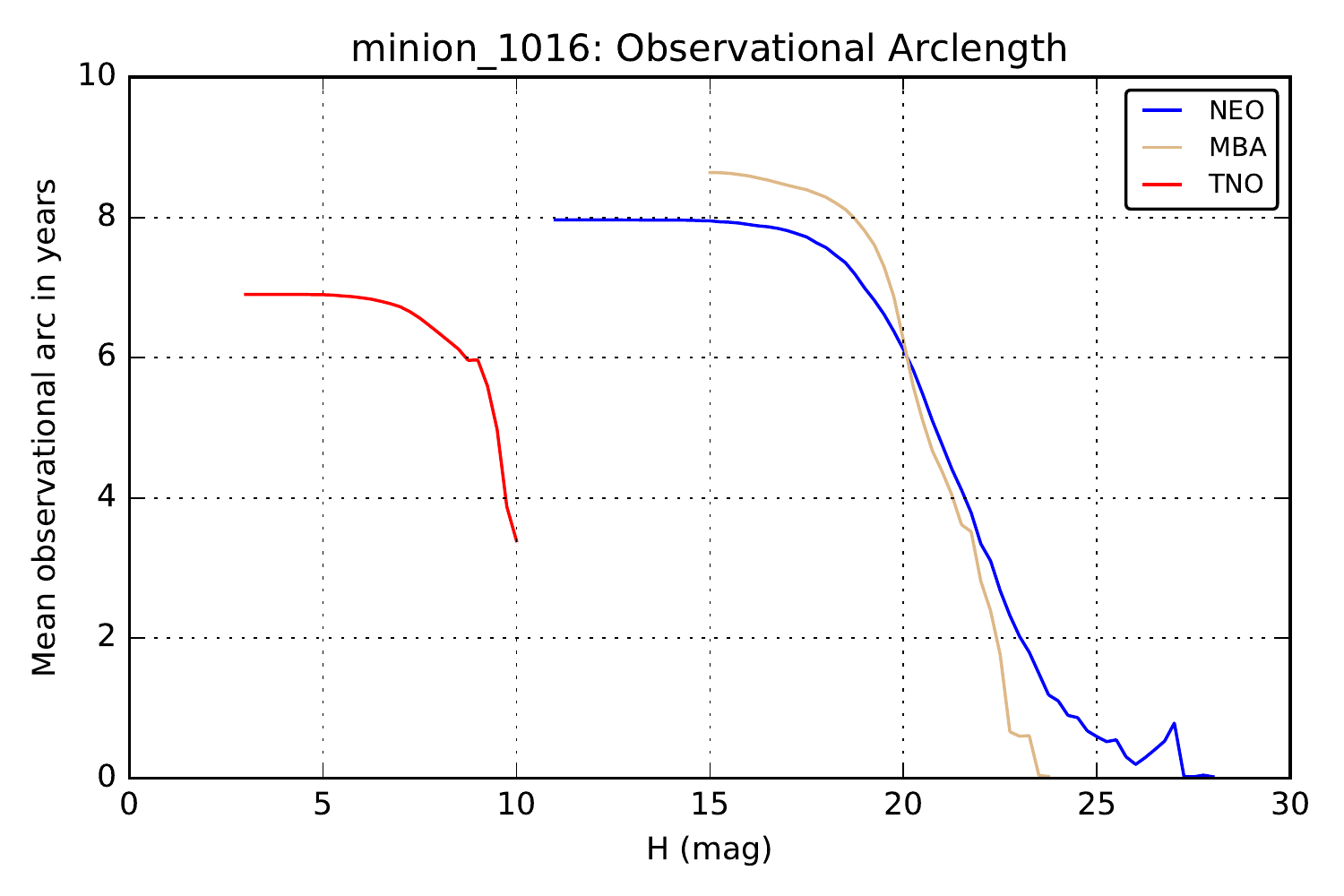}
\caption{Mean observational arc length, in years, for NEO, MBA and TNO
  populations as a function of $H$ magnitude.
\label{obsarc}}
\end{figure}

In \opsimdbref{db:baseCadence}, the mean observational arc length for
NEOs and MBAs is about 8~years for bodies larger than 1~km, and about
6~years for 300~m bodies. With these
orbital arcs, the orbits will be quite well known,
% xxx quantify -- how well! xxx,
meaning that the majority
of LSST-observed objects will have orbits that are sufficiently
well known that the above science cases can be carried out.
In some special cases --- for example, the case where
an NEO's orbit still presents a significant probability
of terrestrial impact --- additional non-LSST follow-up
may be needed, but this will be a small minority of cases.

% xxx what we probably want is fraction of objects
% with arcs longer than 3 years (say) as a fxn of
% H mag, for NEOs/MBAs/TNOs. xxx

% --------------------------------------------------------------------

\subsection{Discussion}
\label{sec:\secname:discussion}

The simple proxy metric above should be improved to account for
potential difficulties in linking observations, and to include actual
orbital fitting to determine orbital uncertainties. The timing of
observations effects the final orbital accuracy significantly,
particularly for TNOs, and having a good distribution on the times of
observations can improve orbital accuracy more quickly than would
naively be expected from a simple observational arclength scaling.

A figure of merit, including requirements on the orbital accuracy for
various classes of objects, should also be developed.

As an intermediate step, we have carried out two anecdotal studies
of orbital accuracy. In the first,
we took an arbitrary (real) NEO --- object 2016~DL ---
with a six year arc. This is representative
and typical of NEOs that will be observed
by LSST.
The maximum positional uncertainty for this object
over the next ten years
is 25~arcsec (3$\sigma$). This is small enough
that essentially any kind of follow-up observation would be
possible (presuming that pre-imaging is possible
for spectroscopy, for example, to locate the moving
object).
The uncertainties on the orbital parameters
of this object are on the order of
1 part in 10$^4$ or even smaller. This level of
precision should allow all the investigations described
above.

In the second experiment, we took an arbitrary
(real) TNO --- object 2015~SO20 ---
that has a five year arc.
For this object, the maximum positional uncertainty
over the next ten years is $\sim$20'', and its orbital
elements are known to
around 1 part in 10$^5$. Again, this level
of precision should enable all of the science
investigations described above.

\navigationbar

% ====================================================================

% ====================================================================
%+
% SECTION:
%    SolarSystem_CometActivity.tex
%
% CHAPTER:
%    solarsystem.tex
%
% ELEVATOR PITCH:
%-
% ====================================================================

\section{Detecting Comet Activity}
\def\secname{\chpname:activity}\label{sec:\secname}

\credit{rhiannonlynne},
\credit{davidtrilling},
\credit{migueldvb}

Comets are the remnant building blocks of the Solar System
that have been stored at cold temperatures beyond the ice
line, either in the Kuiper belt or the Oort cloud, since their
formation.  Measuring the evolution of cometary activity over
a range of heliocentric distances with LSST will allow us to
understand the overall comet activity and to link these
observations with the physical and chemical conditions in the
early solar nebula during planet formation.  Comets are
classified in two main dynamical families, Jupiter Family
comets (JFCs) that have low-inclination orbits with periods
less than 20 years, and Long-period comets (LPCs) that
originate in the Oort Cloud at a distance of more than 10000
AU and have large orbital eccentricities and nearly isotropic
distribution of inclinations.  Currently there are over 400
Jupiter-family comets known, most of which are faint compared
with the LPCs.  LSST will observe about $10^4$ individual
comets repeatedly including measurements of known objects over
its 10-year survey \citep{2010PhDT.......241S}. The
determination of their activity levels at various heliocentric
distances will be used to study the time evolution of each
object individually and to find the connection between comet
families and their formation region in the Solar System.

Several cometary volatiles result in strong emission bands
excited by solar radiation that emit by resonant fluorescence
at optical and near-ultraviolet wavelengths.  The LSST $u$
filter peaks near the CN (0--0) emission band at 3880 \r{A}.
Although CN is not the most abundant daughter species from
cometary volatiles and the OH (0--0) emission band at 3080
\r{A} is generally stronger, CN production rates provide an
excellent proxy of the level of overall gas activity in
comets. LSST will offer a unique opportunity to produce
a large database of CN production rates, vastly increasing our current
knowledge \citep[see
e.g.][]{1995Icar..118..223A, 2012ApJ...758...29A}.
 Other bands such as $r$, $i$, and $z$ will detect
continuum brightness that is produced by reflected radiation
from dust particles in the coma. Thus, it will be possible to
obtain the evolution of the gas-to-dust production ratio at
high cadence as a function of heliocentric distance in
different comet families. The greatly increased sample size
compared with previous catalogs \citep{1995Icar..118..223A}
will allow for statistical comparison of the comet families
and to link them to other small body populations in the Solar
System.

A recently discovered population of main-belt asteroids eject
dust and produce coma and tails giving them the appearance of
comets \citep{2012AJ....143...66J}.  This so-called main-belt
comets or active asteroids have the orbital characteristics of
asteroids with $T_J > 3$ and lose mass during part of their
orbits. The cometary activity observed in these objects may be
driven by primordial water ice that is trapped near the
surface and sublimates when it is exposed to sunlight.
Main-belt comets are important because they may have been able to
preserve water ice despite the effect of solar radiation and
heating from the decay of short-lived radioactive nuclei.  The
asteroids in the outer regions of the main belt can therefore
have a substantial fraction of water and other volatiles that
may have supplied the volatile content of terrestrial planets.
Most of the main-belt-comets are faint with very weak comae
that are active during part of their orbits. Given the
expected flux sensitivity of LSST, the transient cometary
activity of main-belt asteroids will be observable including
many objects that could be below the detection limits of
current photometric surveys.  The LSST observations will thus
help to understand the overlap between different populations
in the Solar System such as the relationship between comets
and asteroids.

% --------------------------------------------------------------------

\subsection{Target measurements and discoveries}
\label{sec:\secname:targets}

LSST will make an exceptionally large number of comet
observations.  About $10^4$ comets will be observed on average
of 50 times by LSST during its main survey, while a few objects
will be observed more than 1000 times
\citep{2010PhDT.......241S}.  Simulations of characteristic
comet orbits have shown that LSST will observe some Jupiter
Family comets (JFCs) hundreds of times over their full orbits
\citep{2010PhDT.......241S}.  Individual LPCs are predicted to
be observed by LSST with dozens of observations as they
approach or recede from the center of the Solar System or
during their perihelion passage.  Thus, these observations
will trace the onset of outgassing from quiescence at large
heliocentric distances and the decline of activity after
perihelion.

Ensuring that any activity or outgassing of a comet or active asteroid
is clearly identifiable with LSST DM or contributed Level 3 software is not a
solved problem, though there is ongoing work toward this goal
both within and outside the project.
In the meantime, it does not seem unreasonable
to assume that the main requirement, in terms of cadence, is to actually
have an observation at a time when activity is visible, as well as at
surrounding times to determine the start and end points of that activity.

Cometary activity and outgassing can last for various periods of time,
usually on the order of days to weeks. It can be transient, perhaps
due to a collision or other resurfacing event, or it can be periodic,
such as repeated activity when an object approaches perihelion. Thus,
in order to characterize the fraction of active asteroids, or to
understand the causes of their activity, or to understand cometary
activity as a function of source population (and thus presumably
composition) and heliocentric distance, the goal would be to have
repeated observations spread throughout the period when the object is visible.

% --------------------------------------------------------------------

\subsection{Metrics}
\label{sec:\secname:metrics}

A full exploration of the cadence effects on measuring activity rates
for comets and active asteroids would include understanding the
selection effects of when the object was not observed, as well as the
likelihood of detecting activity based on when it was observed. For
now, we have focused on the likelihood of being able to detect
activity lasting a given amount of time.

The metrics \MAFmetric{ActivityOverTimeMetric} and \MAFmetric{ActivityOverPeriodMetric} look at when an object was observed (with
a detection above a given SNR), and
split those observations into bins based on time or position in the
orbit (true anomaly), respectively. The first is relevant when looking
for transient activity that is not expected to repeat at the same
point in the orbit, while the second seems more appropriate for
activity that would repeat at the same point in the orbit. Each of
these metrics takes only a single time or true anomaly window, distributes
the observations of each object into bins based on those windows, and
counts the number of bins which received observations. The number of
bins with observations, compared to the overall number of bins,
determines the calculated likelihood of detecting activity for that
object.

To investigate the sensitivity of LSST to activity on a range of
timescales and lasting various fractions of the period, we ran these
metrics over a range of values and then plot the minimum, mean, and
maximum likelihoods of objects at a particular $H$ value, for various populations.

% --------------------------------------------------------------------

\subsection{OpSim Analysis}
\label{sec:\secname:analysis}

Running these metrics on a sample Main Belt Asteroid population
generates results illustrated in Figure~\ref{activity}.
In the baseline survey \opsimdbref{db:baseCadence}, the metric results
indicate that for bright asteroids with activity lasting more than
about 60 days, we have between about 18-60\% chance of obtaining at
least one observation that captures the event. If the activity is
periodic, and lasts around 10\% of the orbit, we have between
20-65\% chance of observing the activity. If the periods of activity
last longer, we have a higher chance of having an observation which
captures that activity, as expected.

That the chance of detecting activity is not significantly higher for
repeating events than for transient events is interesting. It's not
clear if this reflects a characteristic of the observing cadence
(e.g.\ perhaps the observations always are clustered near the same
point in the orbit, leaving many ``bins'' unwatched), but it's seems
likely that at  the very least, the metric should be tuned to account for the additional
likelihood of activity occurring near perihelion.

\begin{figure}
\includegraphics[width=3.3in]{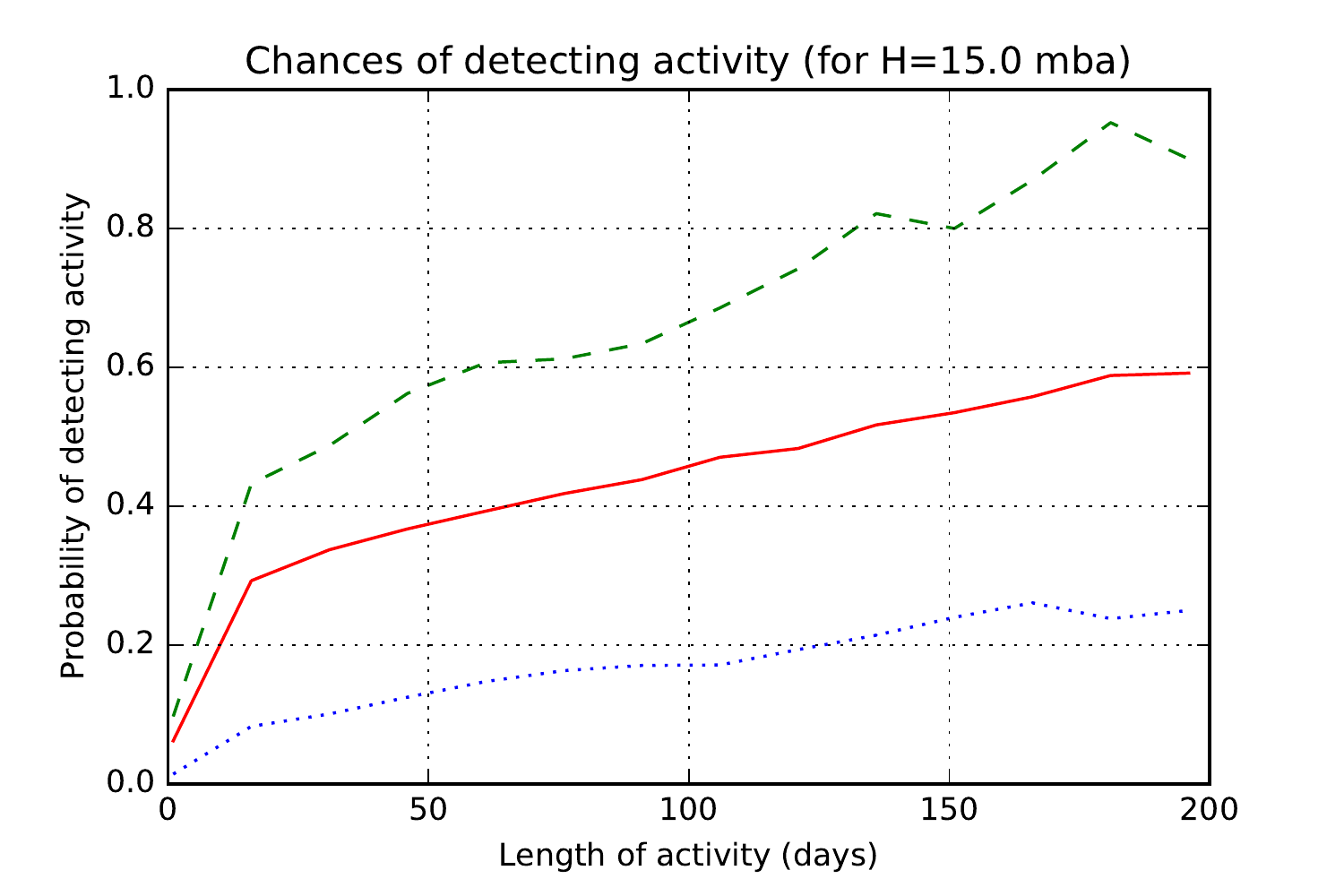}
\includegraphics[width=3.3in]{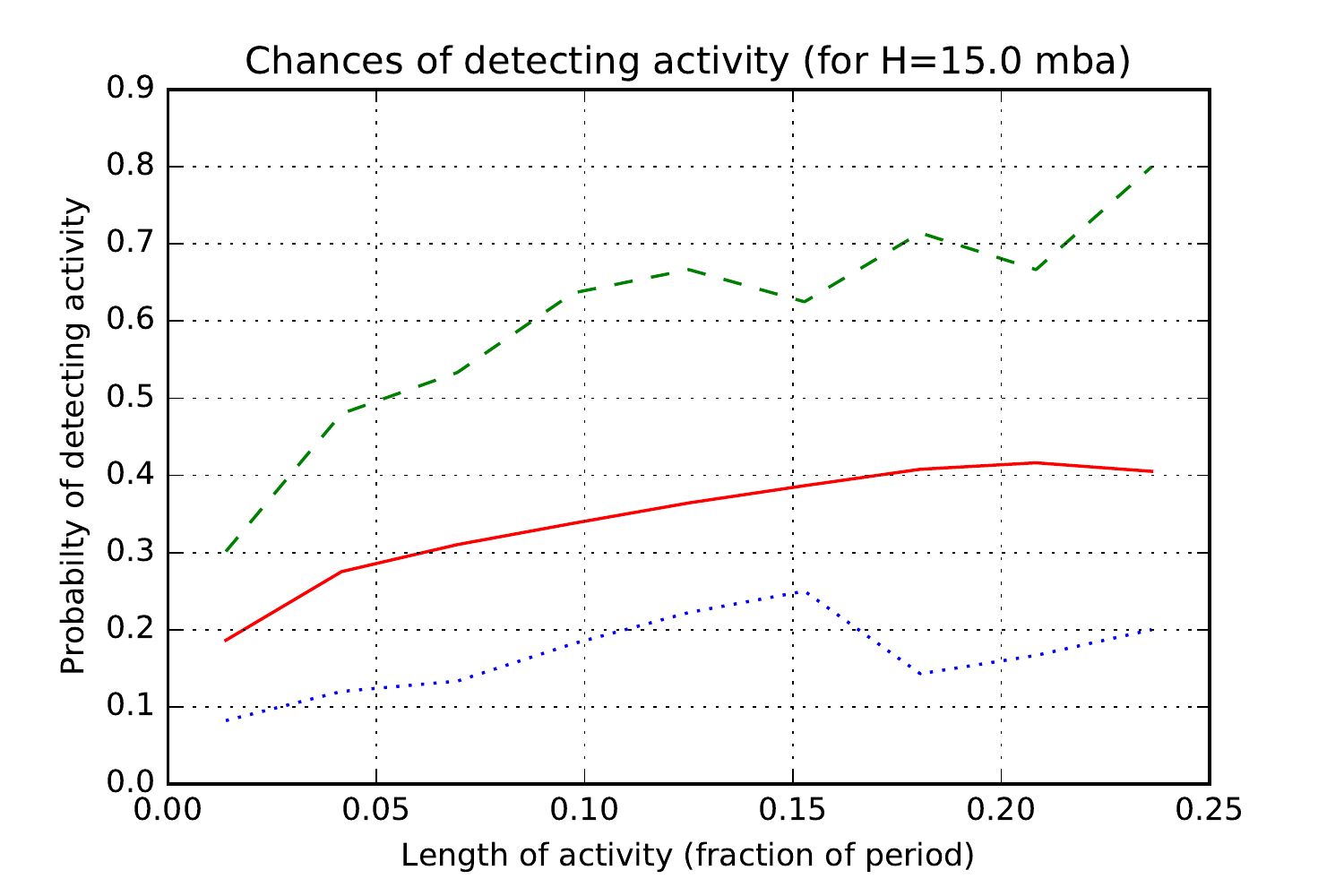}
\caption{Likelihood of detecting activity or outgassing for MBAs, for
  a single event lasting a given number of days (left) or an event
  potentially repeating for a given fraction of the period (right).
\label{activity}}
\end{figure}

% --------------------------------------------------------------------

\subsection{Discussion}
\label{sec:\secname:discussion}

The likelihood of detecting activity depends on if the observing
strategy is such that we observe an object when it is active, and the
capability to identify this activity in the acquired images.

In terms of observing strategy, if observations are clumped together
irregularly in time, we risk missing activity during the times we do
not observe the object, although with the possible benefit of being
more likely to detect short time-scale activity during the times we
have more frequent observations. Balancing these two tensions likely
requires more knowledge about the relevant timescales for activity on
active asteroids, as only a handful of active asteroids have currently
been identified. (comment on cometary activity?)

\navigationbar

% ====================================================================

% ====================================================================
%+
% SECTION:
%    SolarSystem_AsteroidLightCurves.tex
%
% CHAPTER:
%    solarsystem.tex
%
% ELEVATOR PITCH:
%-
% ====================================================================

\section{Measuring Asteroid Light Curves and Rotation Periods}
\def\secname{\chpname:lightcurves}\label{sec:\secname}

\credit{rhiannonlynne},
\credit{davidtrilling}

Two Solar System science projects require a series of photometric
measurements. These are (1) measuring lightcurves and therefore shapes
of minor bodies and (2) measuring the colors and therefore compositions
of minor bodies. This section and the next describe the science and the
metrics for these experiments.

% --------------------------------------------------------------------

\subsection{Target measurements and discoveries}
\label{sec:\secname:targets}

In general, minor bodies are aspherical, and therefore observations of
those bodies produce lightcurves with non-zero amplitudes. Constant
monitoring of such a body would reveal the detailed lightcurve, which
can be inverted to derive the effective observed shape at that epoch.
Observations over multiple epochs allow for observations at different
aspects, which can be used to determine the three dimensional shape and
pole orientation of the minor body. All of this information can be used
to understand, broadly, the orbital and physical evolution of minor
bodies in the Solar System.

LSST observations of minor bodies in the Solar System will not, however,
necessarily be dense in time (with the exception of observations made in
Deep Drilling Fields; see below). Therefore, lightcurves of minor bodies
must be combined across arbitrary rotational phase. Without knowing the
phase, the amplitude of the lightcurve (a proxy for asteroid shape) can
simply be determined. More complicated lightcurve inversion analysis
\citep[e.g.,][]{2016A&A...587A..48D}
can be carried out, given a sufficient number of points.

% --------------------------------------------------------------------

\subsection{Metrics}
\label{sec:\secname:metrics}

The general requirement for successful lightcurve inversion is to have
a large number of observations, at high SNR, over a wide range of
time. A guideline is that $\sim$100 measurements of an asteroid over
$\sim$years, calibrated with a photometric accuracy of
$\sim$5\% (SNR=20) or better, is sufficient to generate a coarse shape model.
This sparse data inversion gives correct results for both fast (0.2--2~h) and
slow ($>$24~h) rotators \citep{2007IAUS..236..191D}.

The metric \MAFmetric{LightcurveInversionMetric} simply checks to see if the
observations of a particular object meet these requirements, and if
so, identifies that object as having the potential for lightcurve
inversion.

% --------------------------------------------------------------------

\subsection{OpSim Analysis}
\label{sec:\secname:analysis}

\begin{figure}
\includegraphics[width=3.3in]{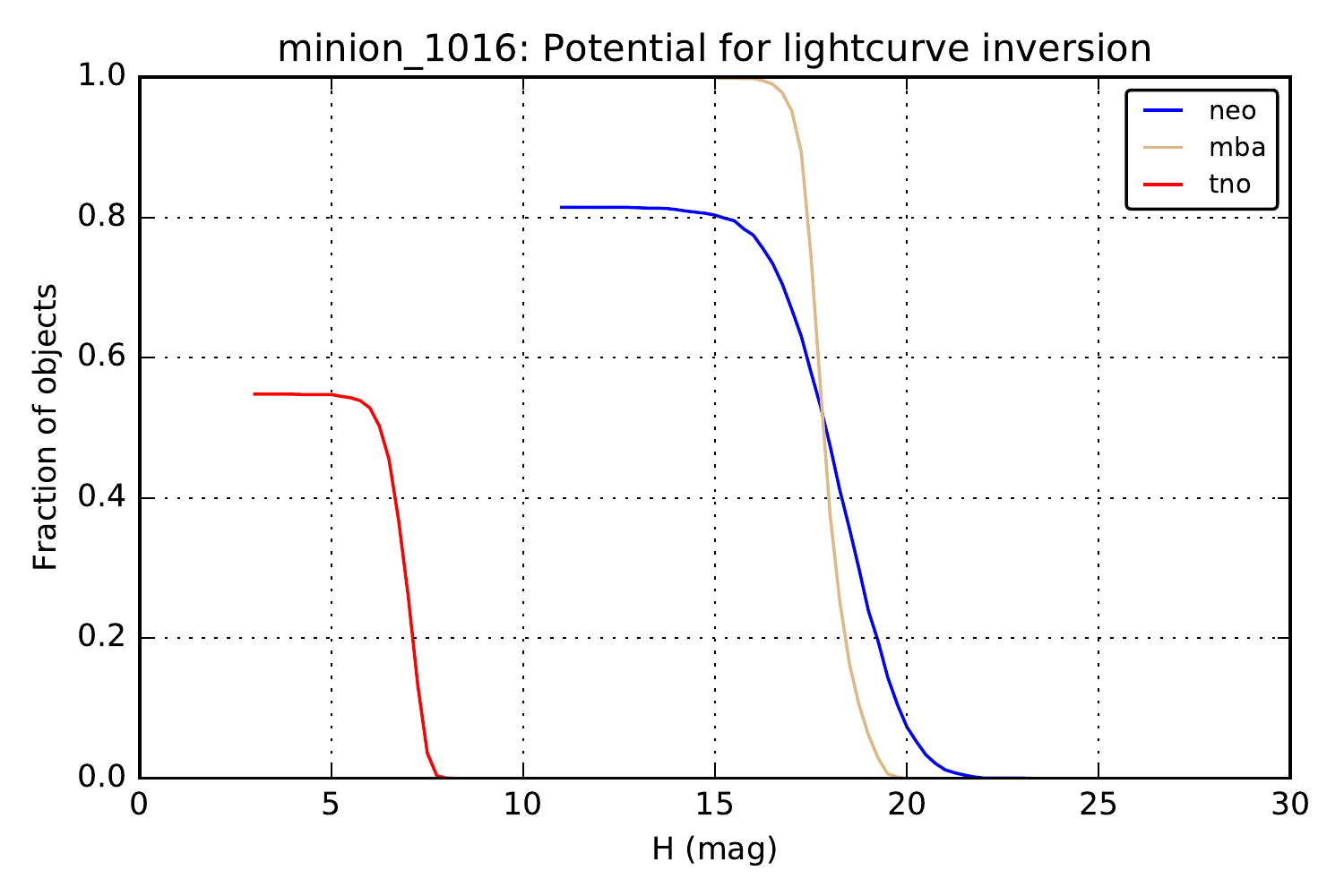}
\includegraphics[width=3.3in]{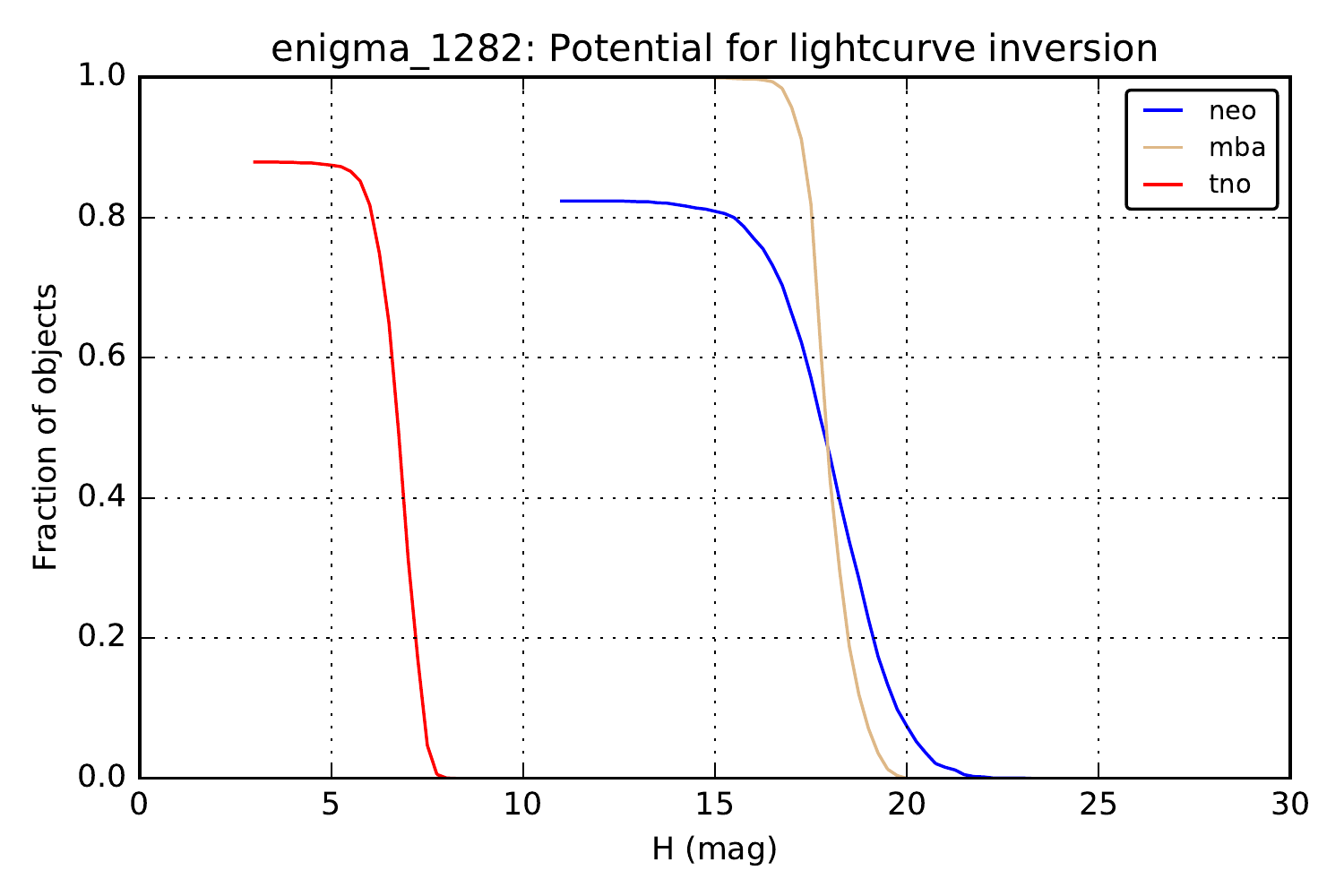}
\caption{Fraction of the sample population with the potential for
  lightcurve inversion, as a function of $H$ magnitude for NEOs, MBAs
  and TNOs, for simulated surveys \opsimdbref{db:baseCadence} and \opsimdbref{db:NEOwithVisitQuads}.
\label{lightcurveinversion}}
\end{figure}

Most solar system objects receive many observations, and bright
objects will have high SNR in most of those observations, so it is not
surprising that the baseline cadence, \opsimdbref{db:baseCadence},
performs reasonably well with this metric. Running the same metric on
other simulated surveys tends to show similar results, although
\opsimdbref{db:NEOwithVisitQuads} demonstrates that a higher fraction
of TNOs get suitable observations for lightcurve inversion, according
to this metric illustrated in Figure~\ref{lightcurveinversion}.
Of course, TNOs are unlikely to be good candidates for
full lightcurve inversion due to the limited range of phase angles the
observations cover (due to the large distance to the TNOs), and this indicates
one limitation of this simple metric. This does indicate however, that
it is likely we can significantly increase the numbers of TNOs for
which we can determine rotation periods (even if not full shape and spin
measurements).

% --------------------------------------------------------------------

\subsection{Discussion}
\label{sec:\secname:discussion}

A risk which is not captured by this simple metric is that the
observations included for the lightcurve inversion estimate here
could potentially occur far apart in time such that linking between
the observations (to determine that they belong to the same object) is
not possible.

Further work needs to be done to understand the necessary final figure
of merit, in particular, how many light curve inversion targets are
necessary, and how should they be spread among different sizes of
objects? Small objects have different shape and
rotation distributions than larger objects, so it is interesting
scientifically to understand objects at a range of $H$ magnitude. In
addition, this metric currently uses observations in any filter;
further work should be done to determine if this is sufficient, or if
observations must occur in a single filter.

A final point is that asteroid lightcurves can sometimes be
used to discover and characterize binary asteroid systems.
In general, sparse lightcurves, such as those that LSST
will produce, are not sufficient for binary studies, but
some sophisticated analysis and fortuitous observation
timing could reveal asteroid binarity.

\navigationbar

% ====================================================================

% ====================================================================
%+
% SECTION:
%    SolarSystem_AsteroidLightCurves.tex
%
% CHAPTER:
%    solarsystem.tex
%
% ELEVATOR PITCH:
%-
% ====================================================================

\section{Measuring Asteroid Colors}
\def\secname{\chpname:colors}\label{sec:\secname}

\credit{rhiannonlynne},
\credit{davidtrilling}

The varying compositions of asteroids result in a range of optical
colors. Sloan filters in general are sufficiently diagnostic to
discriminate among different compositional class
\citep[e.g.,][]{2008Icar..198..138P}.
Therefore, when a Solar System minor body is observed in $griz$ (Solar
System objects are generally quite faint in $u$ band and many fewer will
be detected),
% ; Y band xxx),
the color can be used to determine the
composition and, downstream, composition as a function of asteroid size,
family membership, orbital elements, or many other parameters.

% more motivation for color measurements?

One obstacle to determining asteroid colors is that asteroid rotation
periods are on the order of 2--20~hours, so that after an initial
measurement all further measurements (in the same filter, or other
filters) are obtained at an arbitrary rotational phase. Unless the
lightcurve is also known (perhaps determined from a large series of
measurements in the same bandpass, such as described in the section
above), multi-band measurements must occur at closely spaced times in
order to minimize the effects of the lightcurve on the measured color.

% --------------------------------------------------------------------

\subsection{Target measurements and discoveries}
\label{sec:\secname:targets}

Analysis of existing databases of TNO multi-band measurements indicate
that pairs of high SNR measurements (SNR$>10$), acquired within a
short time period ($<2$ hrs) can provide an accurate color measurement
\citep{2015A&A...577A..35P}. This is roughly consistent with
expectations based on the rotation periods.

% --------------------------------------------------------------------

\subsection{Metrics}
\label{sec:\secname:metrics}

The metric \MAFmetric{ColorDeterminationMetric} searches for pairs of
observations, taken within a given number of hours, where the pair
contains observations in each of two specified filters above a
specified SNR (e.g., $g$ and $r$ band observations taken within 2
hours of each other, with SNR$>10$). If an object receives a minimum
number of pairs of observations (currently set to just a single pair
of observations), then it is considered as having that color
``measured.'' Thus, this metric can measure the fraction of the sample
population which receives an adequate color measurement, for a series
of colors.

% --------------------------------------------------------------------

\subsection{OpSim Analysis}
\label{sec:\secname:analysis}

\begin{figure}
\includegraphics[width=3.3in]{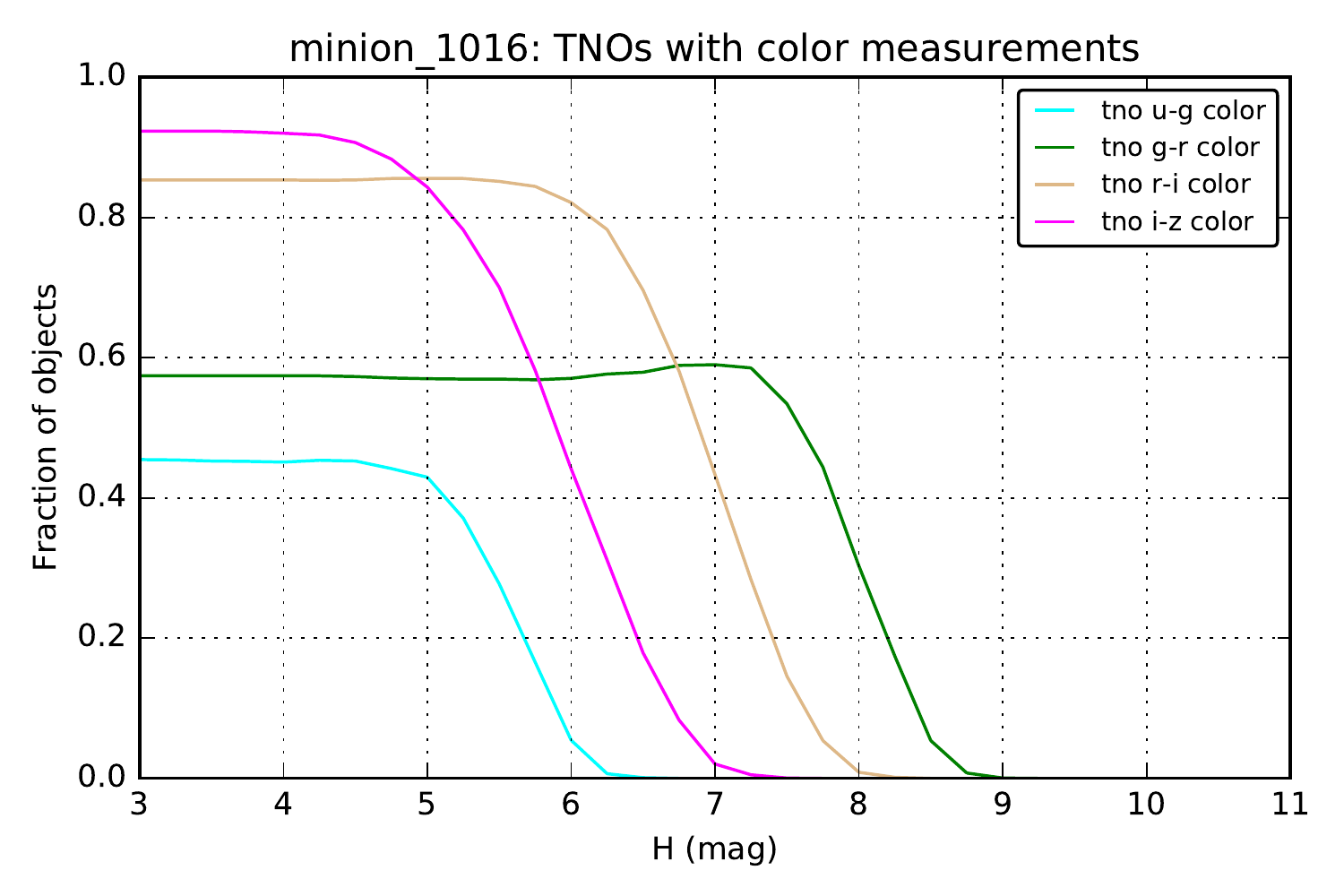}
\includegraphics[width=3.3in]{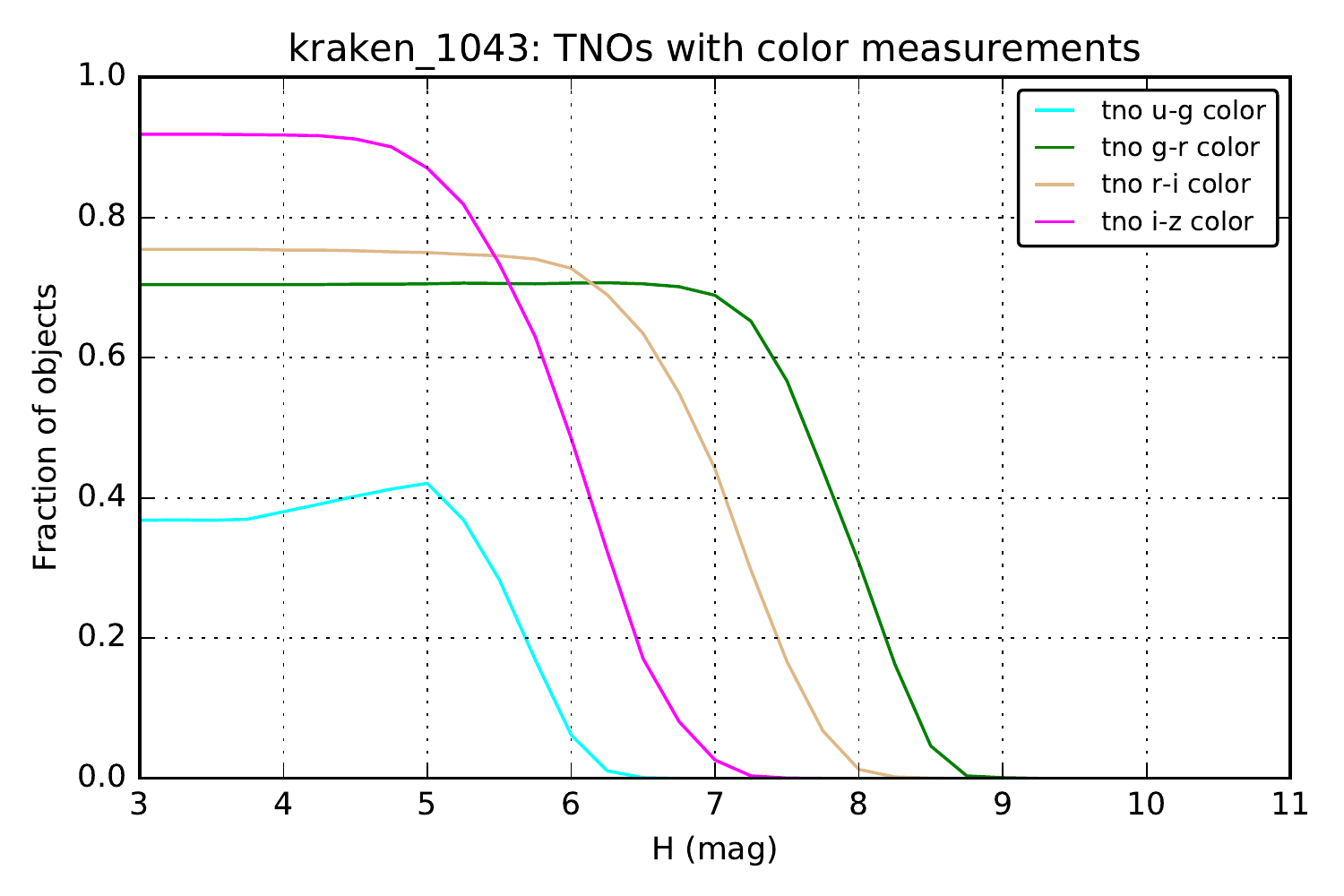}
\\
\includegraphics[width=3.3in]{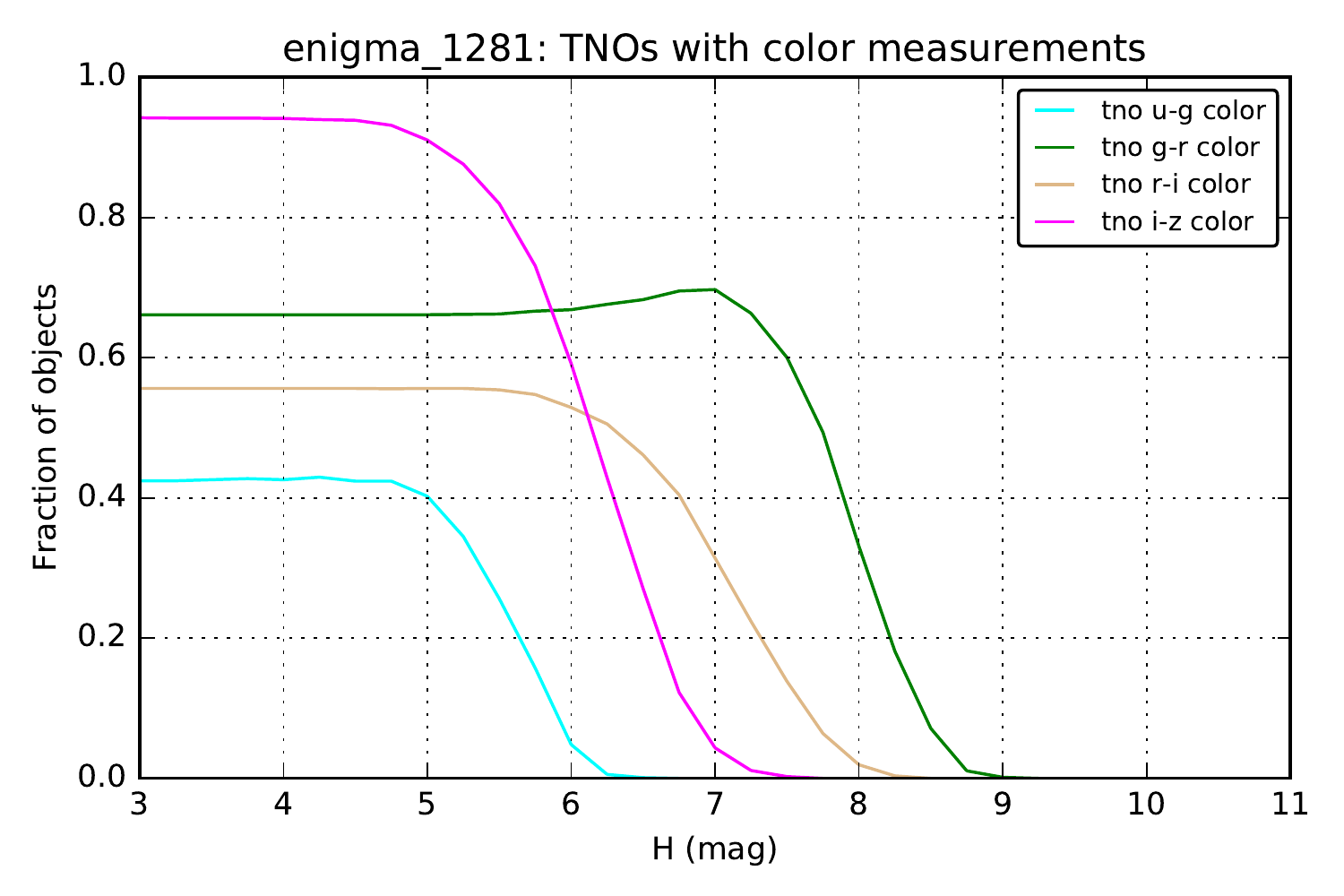}
\includegraphics[width=3.3in]{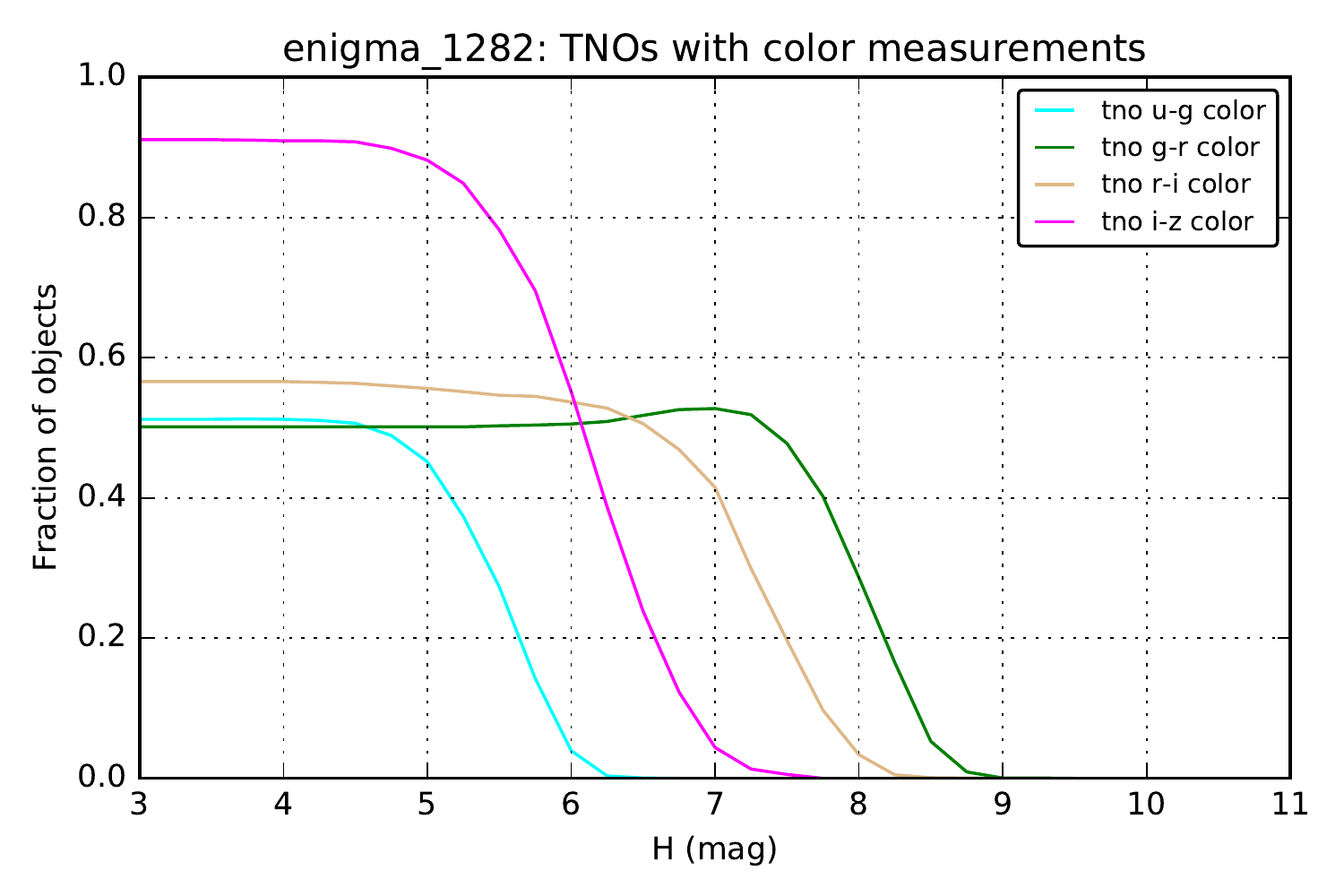}
\caption{Fraction of the sample population with the potential for
  color measurements in various bands, as a function of $H$ magnitude
  for TNOs for simulated surveys \opsimdbref{db:baseCadence},
  \opsimdbref{db:NoVisitPairs}, \opsimdbref{db:NEOswithVisitTriplets}
  and \opsimdbref{db:NEOwithVisitQuads}. The fraction of TNOs which
  could receive high accuracy color measurements, particularly in
  $r-i$, bounces around significantly.
\label{colordetermination}}
\end{figure}

The timing of repeat visits, and whether or not filter changes occur
between these repeat visits, affects the fraction of objects which can
receive color measurements with closely spaced observations
significantly. This can be seen in the metric results shown in
Figure~\ref{colordetermination}. The difference is most pronounced in
the $r-i$ color measurements. The baseline survey,
\opsimdbref{db:baseCadence}, seems to perform rather well --- and
indeed, the best out of this set of runs.

% --------------------------------------------------------------------

\subsection{Discussion}
\label{sec:\secname:discussion}

This metric does not yet account for potential linking difficulties
related to identifying that these two observations are part of a
particular object, and it should be developed further to determine
appropriate parameters for objects other than TNOs, as other objects
may have different ranges of rotation rates. In particular, the
timescales needed when obtaining observations in multiple filters to
determine colors need to be explored.

There is some tension between desiring to get observations in the same
bandpass on a given night (to maximize detection thresholds instead
of being limited by the shallower bandpass) and obtaining color
measurements by having observations in multiple bandpasses. The risk
here is that without at least some nights with multi-band observations
on short timescales, we may not be able to determine accurate colors.

\navigationbar

% ======================================================================

% ====================================================================
%+
% SECTION:
%    SolarSystem_FutureWork.tex
%
% CHAPTER:
%    solarsystem.tex
%
% ELEVATOR PITCH:
%    Ideas for future metric investigation, with quantitaive analysis
%    still pending.
%-
% ====================================================================

\section{Future Work}
\def\secname{\chpname:future}\label{sec:\secname}

In this section we provide a short compendium of science cases that
are either still being developed, or that are deserving of quantitative
MAF analysis at some point in the future.

% ====================================================================
%
% \section{Deep Drilling Observations}
\subsection{Deep Drilling Observations}
\def\secname{\chpname:dd}\label{sec:\secname}

\credit{rhiannonlynne},
\credit{davidtrilling}

Deep drilling observations provide the opportunity, via digital
shift-and-stack techniques, to discover Solar System Objects fainter
than the individual image limiting magnitude. These fainter objects
will be smaller, more distant, or lower albedo (or some combination of these)
than the general population found with individual images. Discovering smaller
objects is useful for constraining the size distribution to smaller
sizes; this provides constraints for collisional models and insights
into planetesimal formation. More distant objects are interesting in
terms of extending our understanding of each population over a wider
range of space; examples would be discovering very distant
Sedna-like objects or comets at larger distances from the Sun before
the onset of activity. Lower albedo objects may be useful to
understand the distribution of albedos, particularly to look for
trends with size.

Variations on the basic method of shift-and-stack have been used to
detect faint TNOs.
% XXX Allen, Bernstein, Gladman, Fuentes, ? XXX.
Computational limitations on these methods mean that, roughly and in
general for images taken at opposition, images taken over the timespan
of about an hour can be combined and searched for main belt asteroids,
and images taken over the timespan of about 3 days can be combined and
searched for more distant objects like TNOs.

With extragalactic deep drilling fields as in the baseline cadence,
where observations are taken in a series of filters ($g$, $r$, and $i$
would be useful for this purpose) each night, every three or four
days, we could use shift-and-stack to coadd the 50 images obtained in
$gri$ bandpasses in a single night. This would allow detection of
objects about 2 magnitudes fainter than in the regular survey, or
approximately $r=26.5$.

The Solar System Science Collaboration developed a deep drilling
proposal specifically targeted to search for very faint Main Belt
Asteroids (MBAs),  Jupiter Trojans,
and TNOs. This proposal can be summarized as
follows:
\begin{itemize}
\item 9 fields, in a 3x3 contiguous grid block centered on a spot
  where Jupiter Trojans and Neptune Trojans coincide (if possible,
  based on timing)
\item 8 sequences of $\sim1.5$ hour $r$-band exposures per field, in continuous
  observing blocks. Each of these blocks would have a coadded limiting
  magnitude of about $r=27$, letting us push to smaller sizes than
  possible with the general extragalactic deep drilling fields.
\item These 1.5 hour blocks would be spaced apart in time
   \begin{enumerate}
   \item Two blocks acquired on two nights, 1.5 months before the fields come to
     opposition
  \item  Two blocks acquired on two nights when the fields are at
    opposition
  \item Two blocks acquired on two nights, 1.5 months after the fields
    come to opposition
  \item Two blocks acquired on two nights when the fields are at
    opposition again, one year later. 
 \end{enumerate}
\item The location of the fields would be adjusted slightly to account
  for the bulk motion of TNOs in the field, thus letting us follow the
  majority of these very small objects over the course of a year,
  providing fairly accurate orbits. Most of the Jupiter Trojans and
  MBAs would diffuse out of the fields, however we would still have
  approximate sizes from the magnitude and distance estimates provided
  by two nights of observations. 
\end{itemize}

This proposal differs from the general extragalactic deep drilling
fields in that the field selection, observing cadence, and filter
choice is better suited for exploring faint Solar System Objects. More
details are available in the Solar System Collaboration Deep Drilling
whitepaper, \url{https://lsstcorp.org/sites/default/files/WP/Becker-solarsystem-01.pdf}.

% Describe how we will evaluate DD proposal, with TNO population +
% estimate on number of times objects observed (but not doing actual
% shift-and-stack). Need large-i populations to test if useful, probably.

% ====================================================================

\navigationbar

% ======================================================================

% --------------------------------------------------------------------

% --------------------------------------------------------------------

\chapter[The Milky Way Galaxy]{The Milky Way Galaxy}
\def\chpname{galaxy}\label{chp:\chpname}

Chapter editors:
\credit{willclarkson},
\credit{akvivas}

Contributing authors:

\credit{cbritt4},
\credit{DanaCD},
\credit{chomiuk},
\credit{vpdebattista},
\credit{jgizis},
\credit{ivezic},
\credit{mliu},
\credit{pmmcgehee},
\credit{dgmonet},
\credit{dnidever},
\credit{ctslater},
\credit{caprastro},
\credit{yoachim}
% {\it and others to follow}

\section*{Summary}
\addcontentsline{toc}{section}{~~~~~~~~~Summary}

LSST should produce significant contributions to most areas of
Galactic astronomy, with appropriate choices of observing strategy
enabling a huge diversity of investigations. Galactic science cases
fall roughly into two observational categories based on stellar
density and/or Galactic latitude. (i) Strategy assessment for the {\it
  high-density or low-latitude} cases is dominated by large variation
in total time allocation for the inner Plane (where most of the
Galaxy's stars are found), since the current strategy options tend to
complete their inner-Plane observations within the first $\sim
7$~months of the survey (see Chapter \ref{chp:cadexp}). Any science
case requiring coverage longer than a year in these regions will not
be well-served by most of the currently-run strategy simulations.
Quantitatively, Figures of Merit (FoMs)\footnote{See
  \autoref{sec:intro:evaluation} for definition of terms, and Chapter
  \ref{chp:cadexp} for more on the simulated observing strategies.  }
for these science cases therefore suggest the baseline cadence to be
factors $\sim 6-60$~worse than the two comparison strategies chosen
that allocate more time to the inner Plane (Tables
\ref{tab_SummaryMWDisk} \& \ref{tab_SummaryMWAstrometry}).  While
figures of merit for a number of science cases in this category have
been specified and are in the implementation phase (Section
\ref{sec:MW_Disk:MW_Disk_metrics}), at present it is at least as
important that a wider range of strategies be specified and run {\it
  with inner-Plane coverage spread over the full ten-year survey time
  baseline.} We encourage the reader to supply suggestions for
simulations meeting this need.

(ii). For science cases not restricted in latitude (including Solar
Neighborhood studies), basic figures of merit have been run for three choices of strategy (e.g. \autoref{sec:MW_Astrometry:MW_Astrometry_metrics}). For example: while some degree of
calibration is ongoing, at present roughly $12\%$~of fields show
problematic degeneracy against differential chromatic refraction for
all three strategies, impacting precision astrometry. Effort is now needed to implement the science
figures of merit in \autoref{sec:MW_Astrometry:MW_Astrometry_metrics} that are based on these
astrometry indications. Assessment of the science strategy impact on
the Halo (\autoref{sec:MW_Halo}) largely rests on the
implementation of a robust star/galaxy separation metric into \MAF.
This development is ongoing, led by one of the authors of this Chapter
(CTS).

\section{Introduction}

\def\secname{MW_Intro}\label{sec:\secname}

% WIC 2017-06-03 - the following commented part has been promoted to
% the executive summary to aid readability.

%LSST should produce significant contributions to most areas
%of Galactic astronomy.
LSST Milky Way science cases cover lengthscales
ranging from a few pc (such as sensitive surveys of low-mass objects in
the Solar Neighborhood), up to many tens of kpc (such as surveys for
low-mass satellite galaxies of the Milky Way and their post-disruption
remnant streams, and beyond this, investigations of resolved stellar
populations in the Local Volume). In this chapter, we investigate a limited set of representative Galactic science cases, with the aims of demonstrating the scientific trade-offs of different observing strategies and of motivating readers to contribute to consderations of LSST's observing strategy.  The LSST Science Book
(particularly chapters 6 and 7) and \citet[][in particular Sections
2.1.4 and 4.4]{IvezicEtal2008} present a broader treatment of both science questions and science cases relevant to Galactic science.\footnote{We do however provide
motivating details for certain science cases in this Chapter,
particularly in \autoref{sec:MW_Disk}, as those cases are not
emphasized in the LSST Science Book or the relevant sections of
\citet{IvezicEtal2008}.}

Concern about observations towards the inner Galactic Plane has for
several years been a common theme in feedback on LSST's observing
strategy, as the Baseline survey currently expends relatively little
observing time per field at low Galactic latitudes (30 visits per filter, closely spaced in time).  With few visits per filter and a condensed time sampling, populations found in scientifically interesting numbers only at low
Galactic latitudes might be detected with low efficiency
by LSST under this Baseline strategy. (An example is the probing of the
mass function of moderate-separation extrasolar planets via intra-disk
planetary microlensing, as argued forcefully in \citealt{gould13}).

%At the time of
%writing, observations towards the inner Plane still seem to be
%compromised further by the way OpSim arranges shorter programs to
%completion at early times in the simulated survey. This reduces the time
%baseline available for inner-Plane measurements by a significant factor
%compared to observations away from the Plane.

We explore the scientific
impact of shallow Plane observations by comparing Metrics and Figures of
Merit evaluated for the Baseline cadence (\opsimdbref{db:baseCadence}),
for the PanSTARRS-like strategy (\opsimdbref{db:opstwoPS}) which has
essentially uniform depth at all Galactic latitudes observed, and for
the \opsimdbref{db:NormalGalacticPlane} strategy in which the plane is part of
the Wide-Fast-Deep survey. The evaluation of
Figures of Merit generated by these three observing strategies will
quantify how a range of Milky Way science cases could be
substantially improved by selecting a strategy with increase Plane
coverage, without significant cost to the rest of LSST's scientific
investigations.

\subsection{Chapter terminology and structure}

To tame the diversity of science cases, we have picked
representative cases and grouped them within broad scientific areas,
devoting one Section of this chapter to each grouping of cases. A small
number of Figures of Merit (FoMs) have been described for each case.
At present, science cases are grouped in the following way:
\autoref{sec:MW_Disk} assesses the impact of observing strategy on
LSST's ability to map some representative astrophysically important
populations that are found mostly or exclusively in the Plane.
% \autoref{sec:MW_SFH} discusses the use of LSST to probe star formation
% histories through mostly young populations (see also Section 5.6.).
% \autoref{sec:MW_Dust} discusses the impact of observing strategy on
% ISM constraints.
Several observational challenges for LSST find their
sharpest expression in Milky Way science, including (but not limited to)
measurements of stellar parallax, absolute astrometry, and proper
motions (including the tie-in to the reference frame which will be
provided by the \textit{gaia} mission). For this reason, specific issues
relating to precision astrometry are developed in
\autoref{sec:MW_Astrometry}.
\autoref{sec:MW_Halo} assesses the degree to which structures in the Milky Way's halo can be discriminated and mapped, using tracer populations distinguished by variability and/or derived stellar parameters.
Finally, \autoref{sec:MW_future}
presents descriptions of investigations that are needed to properly
determine LSST's utility for Milky Way science, but which
% at this date (late-April 2016)
are as yet relatively incompletely developed.

Summary Tables are provided that present the figure of merit for each
science case within a given section (one row per figure of merit)
evaluated for each tested observing strategy (one column per
strategy). This summary information appears in
\autoref{tab_SummaryMWDisk} (the Disk),
% \autoref{tab_SummaryMWDust} (the ISM),
% \autoref{tab_SummaryMWHalo} (the Halo), and
\autoref{tab_SummaryMWAstrometry} (Astrometry).

\navigationbar

% Examples
% identified at this stage include the uses of LSST to set constraints on
% Galactic components (including the structure of the Bulge, and the
% impact of radial migration in the Disk) and the study of resolved
% stellar populations in the Local Group.

%\subsection{Summary tables for Figures of Merit}
%
%The tables below organize the comparison of Figures of Merit for all
%the science cases considered in this chapter:
%begin{itemize}
 % \item Mapping the Milky Way Halo: Table \ref{tab_SummaryMWHalo}
%  \item Mapping the Milky Way Disk: Table \ref{tab_SummaryMWDisk}
%    \item The ISM: Table \ref{tab_SummaryMWDust}
%  \item Astrometry with LSST: Table \ref{tab_SummaryMWAstrometry}
%\end{itemize}

%\subsection{Needed input}
%
%While many of the diagnostic Metrics are relatively well-developed,
%implementation is needed of Figures of Merit (FoMs) that depend on
%these metrics. In some Sections we have sketched out such figures of
%merit, in others the development of a practical FoM is still a topic
%of active development.

% ====================================================================
%+
% SECTION:
%    MW_Disk.tex
%
% CHAPTER:
%    galaxy.tex
%
% ELEVATOR PITCH:
%
%-
% ====================================================================

\section{Populations in the Milky Way Disk}
\def\secname{MW_Disk}\label{sec:\secname}

\credit{willclarkson}, \credit{caprastro}, \credit{cbritt4}, \credit{chomiuk}

Many populations of great importance to Astronomy exist predominantly
in or near the Galactic Plane, and yet are sufficiently
sparsely-distributed (and/or faint enough) that LSST is likely to be
the only facility in the forseeable future that will be able to
identify a statistically meaningful sample. Some (such as the novae
that allow detailed study of the route to Type Ia Supernovae) offer
unique laboratories to study processes of fundamental importance to
astrophysics at all scales. Others \citep[like slow-microlensing events
  heralding an unseen compact object population; e.g.][]{2016MNRAS.458.3012W} offer the {\it only}
probe of important populations.

We remind the reader that {\it variability} studies can be performed
in regions where stars are substantially more spatially crowded than could be
tolerated for aperture photometry. Microlensing studies provide a good
example, routinely using difference imaging \citep[e.g.][]{1998ApJ...503..325A, 2010MNRAS.409..247K} to detect variability in highly crowded regions, with observations by
facilities with higher spatial resolution used for follow-up
characterization.

%the rapid completion of inner-Plane observations in the
%Baseline strategy (run \opsimdbref{db:baseCadence}) dominates all
%other effects.

At present, \opsimdbref{db:baseCadence} allocates 30
exposures in all six filters to fields in the Galactic ``zone of
avoidance,'' typically completing these fields within the first 200
days of the survey (see Figure
\ref{fig_durationInGalacticCoords}). This rapid completion of inner-plane observations imprints a strong signal on
metrics and FoMs (see for example the right panel of Figure
\ref{fig:parapmenigma}).  In this section, we consider the scientific merit of \opsimdbref{db:baseCadence} and other observing strategies to study the relative impacts of the distribution and number of visits on Galactic plane science.

Before proceeding further, we point out that static science should not
be sacrificed completely to variability studies. LSST's adopted
observing strategy must retain at least a minimum total depth in each
of the $\left\{u,g,r,i,z,y\right\}$~filters, along all sight-lines
(and evaluated appropriately for each field), in order to constrain
and disentangle populations photometrically
\citep[e.g.][]{2012ApJ...757..166B, ivezic08}. If there are any
filters deemed unimportant to variability studies in any fields, it
will likely be the photometric confusion limit (for static science)
that sets the minimum total integration times in those cases.

%Comparison is now needed of multiple
%strategies that do allocate sufficient observations to inner-Plane
%observations, so that the scientific impact of the distribution (in
%time and across filters) of visits within surveys of similar total
%duration can be properly explored.

\begin{figure}
    \includegraphics[width=6.0in]{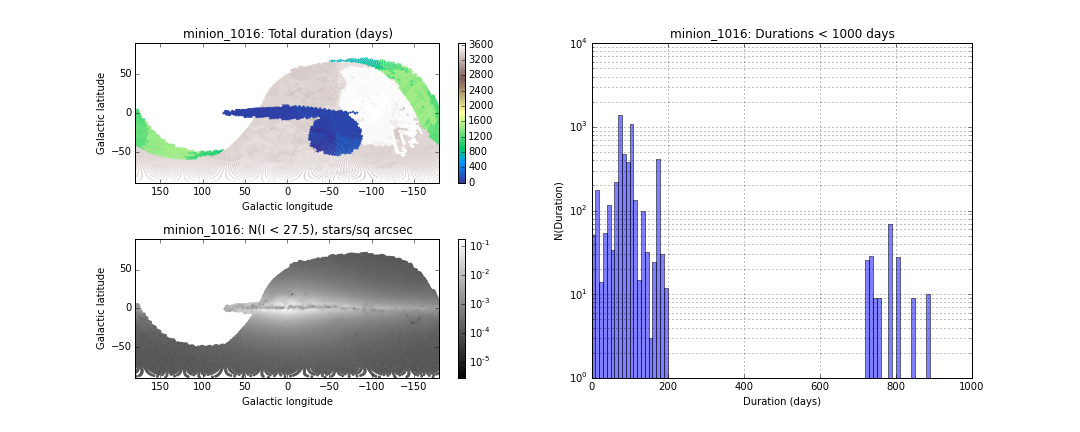}
  \caption{The first-order systematics that dominate transient metrics in the Galactic Plane. These panels are plotted for strategy \opsimdbref{db:baseCadence}. {\it Bottom-left:} stars per square arcsecond (at $i < 27.5$). {\it Top-left:} survey duration (defined as the last minus the first MJD for observations in $i$-band). {\it Right:} histogram of durations in $i$-band over the range 0-1000 days. The population below 200 days shows the Plane and South Polar Cap, the population at $\sim 800$~days is the shorter end of the population of Northern Ecliptic observations. Spatial maps are plotted in Galactic co-ordinates.}
\label{fig_durationInGalacticCoords}
\end{figure}

%An important collateral benefit of studies in the plane with an
%LSST-like facility, is improved mapping of the distribution and
%observational effects of the ISM (particularly dust), which is of
%importance to all IR/Optical/UV observational studies. \new{Due to its
%  importance for {\it all} Milky Way Astronomy, a separate Section is
%  devoted to the impact of observing strategy on the utility of LSST
% data to constrain the distribution of interstellar dust in three
%  dimensions (\autoref{sec:MW_Dust}).}

% --------------------------------------------------------------------

\subsection{Target measurements and discoveries}
\label{sec:\secname:MW_Disk_targets}

%Describe the discoveries and measurements you want to make.

%Now, describe their response to the observing strategy. Qualitatively,
%how will the science project be affected by the observing schedule and
%conditions? In broad terms, how would we expect the observing strategy
%to be optimized for this science?

Four Milky Way disk science cases that have complementary dependencies on
observing strategy (e.g. slow intrinsic variability vs fast intrinsic
variability vs no variability) are:

\begin{enumerate}
  \item Quantifying the large quiescent compact binary population via variability;
  \item New insights into the behavior of Novae and the route to Type Ia Superovae;
  \item The next Galactic Supernova;
  \item Measuring population parameters of planets outside the Snow Line with Microlensing;
  %\item A three-dimensional Dust map and improvements in the reddening law
\end{enumerate}

Below we provide more detail on these science cases, including
qualitative discussions of the expected impact of observing
strategy, as these science cases are not discussed in detail elsewhere
(in the LSST Science Book or the \citealt{IvezicEtal2008} summary paper).

%When the figures of merit have been
%computed for these science cases, the results will be summarized in a
%Table in \autoref{sec:\secname:MW_Disk_discussion}.

{\bf 1. Probing quiescent compact binaries via variability:} Of the
millions of stellar-mass black holes formed through the collapse of
massive stars over the lifetime of the Milky Way, only $\sim 20$ have
been dynamically confirmed through spectroscopic measurements
\citep[e.g.,][]{2015arXiv151008869C}.  Many questions central to modern
astrophysics can only be answered by enlarging this sample: which
stars produce neutron stars and which black holes; whether there is a
true gap in mass between neutron stars and black holes; whether
supernova explosions result in large black hole kicks.

There is expected to be a large population of black hole binaries in quiescence
with low X-ray luminosities from $\sim 10^{30}$--$10^{33}$ erg s${-1}$.
Such systems can be identified as optical variables that show unique,
double-humped ellipsoidal variations of typical amplitude $\sim 0.2$
mag due to the tidal deformation of the secondary star, which can be a
giant or main sequence star. In some cases analysis of the light curve
alone can point to a high mass ratio between the components,
suggesting a black hole primary; in other cases the accretion disk
will make a large contribution to the optical light which results in
intrinsic, random, and fast variations in the light curve. The disk
contribution to optical light can change over time, and several years
of data is necessary to properly subtract the accretion disk
contribution in order to properly fit ellipsoidal variations
\citep{2010ApJ...710.1127C}.
The brighter sources will be amenable to spectroscopy
with the current generation of 4-m to 10-m telescopes to dynamically
confirm new black holes; spectroscopy of all candidates should be
possible with the forthcoming generation of large telescopes. Thus,
LSST would trigger a rich variety of observational investigations of
the accretion/outflow process through studies of this large, dark
population.

While we have focused above on black hole binaries, we note that LSST
will, with suitable cadence, allow crucial measurements of the
populations of neutron star and white dwarf binaries. For example, the
total number of compact binaries is presently poorly
understood---Population models of neutron star X-ray binaries diverge
by orders of magnitude, largely due to uncertainties in the common
envelope phase of binary evolution
\citep[e.g.,][]{2003ApJ...597.1036P,2006MNRAS.369.1152K,2015A&A...579A..33V}.
This is poorly constrained but has a large impact on, for example,
LIGO event rates. A simple test case of common envelope evolution is
available in the number of dwarf novae (DNe; accretion disk
instability outbursts around white dwarfs), a population that does not
suffer from some of the complicating factors that neutron star and
black hole binaries do (e.g. supernova kicks).  Theoretical estimates
routinely yield a significantly higher number of DNe than are observed
in the solar neighborhood. Understanding the true specific frequency
of these systems provides a key check on common envelope evolution.
LSST will detect dwarf novae, which last at least several days with
typical amplitudes of 4--6 mag, out to kpc scales. This will allow a
test of not only the number of cataclysmic variables, but also of the
3D distribution within the Galaxy and dependence on metallicity
gradients \citep{2015MNRAS.448.3455B}.

{\it Response to observing strategy:} Since most black hole candidates
have been identified near the plane in the inner Milky Way (68\%~and 92\%
identified within $5^{\circ}$~and $10^{\circ}$~of the Plane, respectively), this science case {\it
    requires} that LSST observe the plane with sufficient cadence to
  detect the $\sim$hundreds of quiescent black-hole binaries by virtue
  of their variability. The natural choice for a survey for
  low-luminosity black hole binaries would be to extend the
  Wide-Fast-Deep survey throughout the Plane in the direction of the
  inner Milky Way. The orbital period of these systems is short (typically $<1$ day), so that a rolling cadence
  for at least parts of the Plane should be considered. For DNe, the cadence of observations
  is critical in obtaining an accurate measure of the population of
  cataclysmic variables, as a long baseline is necessary to recover systems with a low duty cycle, while widely-spaced observations
 would miss short outbursts.

%Describe the discoveries and measurements you want to make.

%Now, describe their response to the observing strategy. Qualitatively,
%how will the science project be affected by the observing schedule and
%conditions? In broad terms, how would we expect the observing strategy
%to be optimized for this science?

{\bf 2. Novae and the route to Type Ia Supernovae:} Only $\sim 15$
novae (explosions on the surfaces of white dwarfs) are discovered in
the Milky Way each year, while observations of external galaxies show
that the rate should be a factor of $\sim 3$ higher
\citep{2014ASPC..490...77S}.
Evidently, we are missing 50--75\% of novae due to their
location in crowded, extinguished regions, where they are not bright
enough to be discovered at the magnitude limits of existing transient
surveys. Fundamental facts about novae are unknown: how much mass is
ejected in typical explosions; whether white dwarfs undergoing novae
typically gain or lose mass; whether the binary companion is important
in shaping the observed properties of nova explosions. Novae can serve
as scaled-down models of supernova explosions that can be tested in
detail, e.g., in the interaction of the explosion with circumstellar
material \citep[e.g.,][]{2015arXiv151007662C}.  Further, since accreting white
dwarfs are prime candidates as progenitors of Type Ia supernovae, only
detailed study of novae can reveal whether particular systems are
increasing toward the Chandrasekhar mass as necessary in this
scenario.

{\it Response to observing strategy:} Most novae occur in the Galactic
Plane and Bulge, and therefore the inclusion of the Plane in a survey
of sufficient cadence to find these events promptly is of paramount
importance for this science. These events will trigger
multi-wavelength follow-up ranging from the radio to X-ray and
$\gamma$-rays; these data are necessary for accurate measurements of
the ejected mass.

{\bf 3. The First Galactic Supernova:} A supernova in the Milky Way
would be among the most important astronomical events of our lifetime,
with enormous impacts on stellar astrophysics, compact objects,
nucleosynthesis, and neutrino and gravitational wave astronomy. The
estimated rate of supernovae (both core-collapse and Type Ia) in the
Milky Way is about 1 per 20--25 years \citep{2013ApJ...778..164A}; hence there
is a 40--50\% chance that this would occur during the 10-year LSST
survey. If fortunate, such an event will be located relatively close
to the Sun and will be an easily observed (perhaps even naked-eye)
event. However, we must be cognizant of the likelihood that the
supernova could go off in the mid-Plane close to the Galactic Center
or on the other side of the Milky Way---both regions covered by
LSST. While any core-collapse event will produce a substantial
neutrino flux, alerting us to its existence, such observations will
not offer precise spatial localization. The models of \citet{2013ApJ...778..164A}
indicate that LSST is the \emph{only} planned facility that
can offer an optical transient alert of nearly all Galactic
supernovae.

{\it Response to observing strategy:} Even if the supernova is not too
faint, LSST will likely be the sole facility with synoptic
observations preceding the explosion, providing essential photometric
data leading up to the event---but only if LSST covers the Plane at a
frequent cadence. Just {\it how} frequent is open to exploration at
present, but the prospect of high-sensitivity observations of the
location of such a supernova {\it before} it takes place are clearly
of enormous scientific value.

% WIC 2016-06-03 - removed following feeback

%A secondary issue is the prospect that
%an easily-observed Milky Way supernova might be too bright for LSST to
%measure precisely with its planned exposure time, with a roughly 82\%
%chance of a core-collapse supernova reaching one or two magnitudes
%brighter than LSST's nominal saturation limit \citep[with a 1/3 chance that
%a ccSN would reach $m_V \sim 5$;][]{2013ApJ...778..164A}. For a Type Ia in
%the Milky Way,
%\citet{2013ApJ...778..164A} estimate $m_{V, max} \lesssim 13.5$~in 92\% of
%cases.

{\bf 4. Population parameters of planets beyond the Snow Line with
  Microlensing:} \citet{gould13}  shows that, LSST could contribute a
highly valuable survey for intra-disk microlensing (in which disk
stars are lensed by other objects in the disk, such as exoplanets,
brown dwarfs, or compact objects). The lower stellar density compared
to past bulge-focused microlensing surveys would be offset by the
larger area covered by LSST. The predicted rate of high magnification
microlensing events that are very sensitive to planets would be $\sim
25$ per year. This survey would be able to detect planets at moderate
distances from their host stars, a regime poorly probed by standard
Doppler and transit techniques. The LSST data alone would not be
sufficient: the detection of a slow ($\sim$ days) timescale increase
in brightness of a disk star would need to trigger intensive
photometric observations from small (1-m to 2-m class) telescopes that
would observe at high cadence for the 1--2 months of the microlensing
event. This would represent an excellent synergy between LSST and the
wider observing community, and would directly take advantage of the
capabilities unique to LSST.

{\it Response to observing strategy:} To catch lensing events as they
start to brighten, with sufficient fidelity to trigger the intensive
follow-up required, the models of \citet{gould13} suggest each field
should be observed once every few nights. With sparser coverage, the
survey would lose sensitivity to microlensing events in
progress. Comparison with a similar sample towards the inner Milky Way
would be highly useful, which would argue for observations of the
entire visible Plane with similar cadence.

Microlensing is also discussed elsewhere in this document, in the
context of exoplanets \autoref{sec:planets}, of the Magellanic Clouds
(\autoref{chp:MCs}), of AGN (\autoref{sec:agn:microlensing}), and of
WFIRST fields towards the Bulge (\autoref{sec:wfirst:microlensing}).
Although the WFIRST discussion in \autoref{sec:wfirst:microlensing}
assumes that Bulge fields will be observed at least at low cadence
($\sim 1$~observation per day) for the first {\it eight} years of the
survey, this is inconsistent with the Baseline strategy that puts all
the Galactic Plane observations into the first few years of the
survey.

% --------------------------------------------------------------------

\subsection{Figures of Merit}
\label{sec:\secname:MW_Disk_metrics}

%Quantifying the response via MAF metrics: definition of the metrics,
%and any derived overall figure of merit.

Here we describe the Figures of Merit (FoMs) we currently intend to implement
and evaluate for the candidate observing strategies of interest. Where
these FoM have already been evaluated, we provide the numerical
results in Table \ref{tab_SummaryMWDisk}. At present, only FoM 3.1 has been
implemented and run for three surveys, with the others in
development. The FoMs are:

\begin{itemize}
  \item FoM 1.1 - Fraction of quiescent black hole binaries detectable through ellipsoidal variability;
  \item FoM 1.2 - Uncertainty on the duty cycle distribution of Dwarf Novae;
  \item FoM 2.1 - Fraction of Novae detected by LSST (specific and total);
  \item FoM 2.2 - Fraction of Novae caught early enough by LSST to schedule followup observations;
  \item FoM 3.1 - Fraction of Galactic supernovae for which LSST would detect variability {\it before} the main Supernova event;
  \item FoM 4.1 - Fraction of accurately-triggered Microlens candidates;
  \item FoM 4.2 - Uncertainty in the mass function of intra-disk microlensed planets.
\end{itemize}

With the exception of FoM 3.1 above, all these Figures of Merit are
likely to be impacted by spatial confusion, as the populations of
interest tend to lie at low Galactic latitudes. Metrics for assessing
the impact of crowding have been developed (e.g. \href{https://github.com/LSST-nonproject/sims_maf_contrib/blob/master/science/static/CrowdingMetrics.ipynb}{{\tt
  CrowdingMetrics.ipynb} in {\tt sims\_maf\_contrib}}), and these should be
incorporated into all the FoMs described here. For the present,
however, we note that intrinsic source confusion is not a function of
observing strategy (assuming the confusion limit is well above the
formal limiting magnitude without it). Running the FoMs without
accounting for source confusion isolates the impact of strategy alone
on the science that can be performed, as FoMs can be compared in a
relative sense. Inclusion of crowding will later set the absolute
scale for each FoM.

In these FoMs, ``uncertainty'' can be taken to mean both random and
systematic uncertainty, likely recorded as separate numbers for each
FoM. We anticipate determining the FoMs that record population
parameter-uncertainty in a Monte Carlo sense. This is particularly
relevant for FoMs in which the event rate per pointing may be low
($\lesssim 1$~event per pointing per decade) but not so low that only
1-few events are expected over the whole sky over the lifetime of the
survey (as is the case for FoM 3.1, the First Galactic Supernova). In
very rare-event cases, the FoM can scale with the stellar density and
the recovery fraction of that particular transient, and need only be
evaluated once for the entire survey.

Since (at the time of writing) evaluating a Metric with relaxed SQL
constraints typically takes about 0.5-1.5 hours (on a reasonably
modern laptop), we do not expect to perform Monte Carlo population
simulations initially. In the medium-term, when Monte Carlo
experiments in the target populations are desired, the best strategy
may be to evaluate the run of a particular metric against a parameter
of interest (apparent magnitude, say, which is also expected to lead
to a turnover in the importance of confusion error), and the
investigator's preferred Monte Carlo framework for their population of
interest can interpolate the stored Metric values at the time of
trial-population generation. We have begun investigating the use of
these ``Vector Metrics'' for Figures of Merit, and describe the
anticipated FoMs in these cases below.

{\bf FoM 1.1 - Fraction of quiescent black hole binaries detectable
  through ellipsoidal variability:} Table
  \ref{table:pseudoFOM_1p1} outlines the steps to evaluate FoM
  1.1. Since the lightcurve {\it shape} matters in addition to the
  detection (i.e. we expect LSST data to be used to characterize
  ellipsoidal variations, not just to trigger followup by other
  observatories) the metric choice for detectability should take the
  lightcurve shape into account.

  We envisage two levels to implementing FoM 1.1. In the immediate
  future, the ellipsoidal lightcurve could be characterised as a
  sinusoid, with filter-dependent amplitude to match typical quiescent
  Low Mass X-ray Binaries (qLMXBs) from the literature. The next level
  of sophistication would be to input a more complicated shape that
  captures deviations from pure sine-wave behavior, likely using the \MAFmetric{TransientAsciiMetric}.

An open question is how best to meaningfully quantify the
  recovery fraction of a population with a wide range in binary
  parameters. However, as an initial FoM, evaluating once for an
  ``average'' population will allow direct comparison between
  observing strategies.

{\it Possible higher-order FoM:} A higher-order FoM for the fraction
of qLMXBs recovered might take the form of the uncertainty on the
population size (or physical population parameter like mass function
slope) derived from a survey under a given observing strategy. One can
imagine summing the ``recovered'' qLMXB population count and comparing
it to that simulated. Some white noise componant of varying strengths
could be added to the light curves to simulate various contributions
of the accretion disk to the continuum light.  Note that the survey
will necessarily be highly incomplete (due to inclination effects,
etc.), it is the likely {\it uncertainty} on the
completeness-correction that would be crucial in this case.

\begin{table}[h]
  \small
  \begin{tabular}{c p{12cm}}
    & {\it FoM 1.1 - Fraction of quiescent black hole binaries (qLMXB) detectable by LSST through ellipsoidal variability} \\
    \hline
  1. & Pick a typical binary mass ratio and separation for $qLMXB$ \\
  2. & Identify typical ellipsoidal variation amplitude and period, for each filter \\
  3a. & {\it Near-term:} Run \metric{periodicStarMetric} to determine the fraction of typical qLMXBs that would be recovered. Or; \\
  3b. & Run a variant of \href{https://github.com/LSST-nonproject/sims_maf_contrib/blob/master/science/periodicVariables/periodicStarFit.ipynb}{\tt periodicStarFit.ipynb} that allows the appropriate double-humped lightcurve shape. \\
  4. & {\bf Arrive at FoM 1.1:} Load the result from 4. and sum over the spatial region (to be determined: Galactic Latitude range? Comparison high-latitude clusters?) where the qLMXBs are expected.\\
\hline
    \end{tabular}
 \caption{Description of Figure of Merit 1.1.}
  \label{table:pseudoFOM_1p1}
\end{table}

%Dependencies:
%\begin{itemize}
%  \item Monte Carlo in period, phase and shape parameters (ASCII input?) for va%riables as measured in a particular OpSim run. Likely run Monte Carlo for a rep%resentative number (ten?) of well-chosen orbital periods within the 0.1-5d range;%
%  \item (Since these are short-period objects): the ``PeriodicMetric'' of Lund et al. (2015);
%  \item Will likely need reasonably high-spatial-resolution HEALPIX slices and a prescription for population density as a function of position on-sky (can be analytic).
%\end{itemize}
%Possible higher-order FoM: errors on the population size (mass
%function??) derived from a survey under a given observing
%strategy. Can imagine just adding up the ``recovered'' qLMXB
%population and comparing it to that simulated. Some white noise componant of va%rying strengths could be
%added to the light curves to simulate various contributions of the accretion di%sk to the continuum light.
%Note that the survey will necessarily be highly incomplete (inclination effects, etc.), it
%is the likely {\it uncertainty} on the completeness-correction that
%would be crucial in this case.

{\bf FoM 1.2 - Uncertainty in the duty cycle distribution of Dwarf
  Novae:} Table \ref{table:pseudoFOM_1p2} illustrates a version of
this FoM that could be run in the near future. This FoM could then be
adapted later in a more sophisticated analysis that estimates the
uncertainty in the population recovered (of a Dwarf Nova sub-class of interest,
perhaps). Dwarf Novae are a heterogeneous class; we imagine assigning
an average lightcurve to a population and determining the recovered vs
input duty cycle, under the assumption that the spatial distribution
of recurrent Dwarf Novae is uniform. This isolates the impact of
observing strategy alone due to gaps in coverage. Rather than a full
Monte Carlo simulation in Dwarf Novae populations, initially the
investigator might compute the FoM for a representative range of duty
cycles (since the error on timescale recovery may be expected to scale
with the duty cycle itself).

\begin{table}[h]
  \small
  \begin{tabular}{c p{12cm}}
    & {\it FoM 1.2 - Uncertainty in the Dwarf Nova duty cycle} \\
    \hline
  1. & Pick a typical lightcurve for the Dwarf Nova class of interest; \\
  2. & Assign a duty cycle (and thus typical outburst recurrence timescale); \\
  3. & Run \MAFmetric{TransientMetricASCII} using this lightcurve and duty cycle; \\
  4. & Combine the (spatially distributed) results of 3. into a median and formal random uncertainty estimate on the duty cycle estimated from each line of sight; \\
  5. & Compute the offset and its formal error, between the median from 4. and the input duty cycle from 2;\\
  6. & {\bf Arrive at FoM 1.2:} The four numbers from steps 4. and 5. are the characterization of the uncertainty in duty cycle required. \\
\hline
    \end{tabular}
 \caption{Description of Figure of Merit 1.2.}
  \label{table:pseudoFOM_1p2}
\end{table}

{\it Possible higher-order FoM:} The uncertainty in LIGO event rates
due to uncertainties in common envelope evolution, which drives
uncertainties in both LIGO event rates and DN population. While
Advanced LIGO is already starting to place limits on compact object
merger rates \citep[e.g.][]{2016PhRvX...6d1015A}, LSST observations
will provide important electromagnetic constraints on merger rate
results from direct gravitational detection.

%Dependencies:
%\begin{itemize}
%  \item Monte Carlo in distribution of maximum brightness and rise/decay timescale;
%    \item "Triples" without filter constraints (given a prior detection in each filter)- what fraction are recovered?
%    \item Histogram of duty cycle recovery efficiency versus duration of outburst and recurrence time.
%    \item Histogram of recovery efficiency of maximum brightness of dwarf nova versus duration and recurrence time (if assume subsequent outbursts have similar profiles).
%    \end{itemize}

{\bf FoMs 2.1 \& 2.2 - Fraction of Novae characterized by LSST; and the fraction detected early enough to schedule scientifically useful follow-up observations:} Since the set of Novae is so heterogeneous, one can imagine a two-stage process. In the near-future, a single run of \MAFmetric{TransientMetricASCII} using some sense of an ``average'' Nova as tracer to enable comparison between observing strategies. We present this in Table \ref{table:pseudoFOM_2p1}, which includes examples for the specific and total fraction of Novae recovered.

In the longer term, a Monte Carlo simulation could be run on a particular class of Novae depending on the parameters of the overall population whose constraints are desired. This latter effort would likely require further development of the Vector Metrics.

\begin{table}[h]
  \small
  \begin{tabular}{c p{12cm}}
    & {\it FoMs 2.1 \& 2.2 - Novae identified from LSST data} \\
    \hline
  1. & Pick a typical lightcurve for the Nova class of interest; \\
  2. & Run \MAFmetric{TransientMetricASCII} using this lightcurve; \\
  3. & {\bf Arrive at FoM 2.1a: Specific fraction of Novae discovered:} Sum the result from 2. over the spatial region of interest; \\
  4. & {\bf Arrive at FoM 2.1b:} Multiply the result of 2. by the result of the {\tt Starcounts} metric. Sum this to find the fraction of Novae recovered if their spatial density follows the stellar density in the Milky Way. \\
  \hline
  5. & From the typical lightcurve and a typical follow-up scenario, determine the time interval before outburst peak that would be required to schedule followu-up observations;\\
  6. & Use these to produce appropriate parameters for a transient metric that returns the fraction of events detected; \\
  7. & {\bf Arrive at FoM 2.2:} Sum the result of step 6. over the spatial region of interest.\\
\hline
    \end{tabular}
 \caption{Description of Figures of Merit 2.1. \& 2.2.}
  \label{table:pseudoFOM_2p1}
\end{table}

{\it Possible higher-order metrics:} Uncertainty on the specific rate
of Type Ia supernovae using LSST data taken under various observing
strategies.

%Dependencies:
%\begin{itemize}
%  \item Is the ``Triplets'' metric sufficient (i.e. is this ``just'' a case of supplying the metric the correct $\Delta t$~parameter values)?;
%    \item What is the maximum interval since initial rise that would be acceptable? (Is this a function of waveband for followup?)
%\end{itemize}
%Possible higher-order FoM: error on inferred rate of Type Ia supernovae?

%{\bf FoM 3.1 - Maximum time-interval {\it before} triggering of a SN in the Milky Way that LSST would h%ave taken precursor data.}
%Dependencies:
%\begin{itemize}
%  \item This FoM would probably be very easy to calculate (just estimate the mean time between observat%ions). However the acceptable limits still need thought:
%  \item How many colors are sufficient? Any observations at all before the SN goes off, or would a comp%lete set in all filters be needed to characterize the candidate progenitor?
%    \item What level of variability sensitivity is really needed? Would just an extremely deep image of% the SN field before the event be sufficient?
%\end{itemize}
%{\bf FoM 3.2 - Maximum time-interval {\it after} the SN event for triggering followup.}
%Dependencies:
%\begin{itemize}
%  \item Similar to FoM 3.1.
%    \item Is it important to know the discovery space for LSST? If a
%      supernova at $m_V~15$~goes off, other facilities are likely to
%      spot it...
%\end{itemize}

{\bf FoM 3.1 - Fraction of Galactic supernovae for which LSST would
  detect variability {\it before} the main Supernova event:} We have
implemented a simple FoM for the Galactic Supernova case, using the
parameters of SN2010mc as an example whose pre-SN outburst could be
discovered first by LSST. The FoM is defined as the density-weighted
average fraction of transient events recovered, where the average is
taken over the sight-lines within the simulated strategy:
\begin{equation}
  FoM_{preSN} \equiv \frac{ \sum^{sightlines}_{i} f_{var, i} N_{\ast, i} } {\sum^{sightlines}_{i} N_{\ast, i}}
\label{eqn:def_FOM_3p1}
\end{equation}
Here $f_{var, i}$~is the fraction of transient events that
LSST would detect for observing strategy including the $i$'th
sightline, $N_{\ast,i}$~the number of stars present along the $i$'th
sightline, and the FoM is normalized by the total number of stars
returned by the density model over all sightlines. (For the \OpSim
runs tested here, \opsimdbref{db:baseCadence} and
\opsimdbref{db:opstwoPS} and \opsimdbref{db:NormalGalacticPlane}, the normalization factors differ by $\sim
2\%$.) FoM values are in the range $0.0 \le FoM_{preSN} \le 1.0$.

We assume the Pre-SN variability similar to the pre-SN outburst of
SN2010mc \citep{2013Natur.494...65O}. The pre-SN variability is
modeled as a sawtooth lightcurve (in apparent magnitude). We assume
this transient event will always reach brightness sufficient for LSST
to observe, so opt for a very bright peak apparent magnitude in all
filters. We assume that the probability of a supernova going off is
proportional to the number of stars along a particular line of
sight.

In definition (\ref{eqn:def_FOM_3p1}), a lightly-modified version of
\MAFmetric{CountMetric} was used to determine $N_{\ast, i}$~with the output
summed over all sight-lines to produce $N_{\ast}$. Module \MAFmetric{TransientMetric} was used to determine $f_{var, i}$~for each
sight-line.\footnote{The notebooks used to evaluate this version of
  FoM 3.1 can be found in subdirectory {\tt notebooks} of the
  experimental github repository {\tt lsstScratchWIC}, available at
  this link: \url{https://github.com/willclarkson/lsstScratchWIC}}

% WIC 2016-04-26 - removed the old bullets here since they are
% obsolete now!

{\bf FoM 4.1 - Fraction of accurately-triggered Microlens candidates
  within a spatial region of interest:} Table
  \ref{table:pseudoFOM_4p1} lays out a possible FoM for the fraction
  of microlens candidates that LSST might catch sufficiently early
  that follow-up observations can be planned for other facilties. This
  low-level FoM should be straightforward to compute, for a microlens
  template lightcurve corresponding to some suitable average over the
  regime of interest.

\begin{table}
  \small
  \begin{tabular}{c p{12cm}}
    & {\it FoM 4.1 - Fraction of microlens events triggered from LSST observations} \\
    \hline
  1. & Decide on the typical microlens scenario of particular interest; \\
  2. & Produce a template ASCII lightcurve for this scenario; \\
  3. & Determine the characteristics for a trigger; \\
     & ~~~~ e.g. slow rise to 20\% flux above baseline at 7$\sigma$~significance;\\
     & ~~~~ e.g. must be at most $T$~days after the initial rise to schedule follow-up; \\
  3. & run the metric {\tt transientASCII} on this template; \\
  4. & {\bf Arrive at FoM 4.1:} Sum the fraction of detected candidates from 3. spatially over the region of interest.\\
\hline
    \end{tabular}
 \caption{Description of Figure of Merit 4.1}
  \label{table:pseudoFOM_4p1}
\end{table}

{\bf FoM 4.2 - Uncertainty in the mass function of microlensed planets
  past the Snow Line:} Table \ref{table:pseudoFOM_4p2}
  illustrates a possible FoM for a science case concerning uncertainty
  in the parameters of a particular planetary population of
  interest. As with FoM 4.1, in this scenario LSST is used as the
  initial trigger for follow-up observations by other facilities, but
  the observing strategy imposes uncertainty and bias on the eventual
  derived parameters through its removal of parts of the population
  from further study. The investigator could assume a particular
  uncertainty imposed on the mass determination from follow-up
  observations, but this should be fixed for all evaluations of the
  FoM so that LSST strategies can be compared. Strictly speaking, a
  Monte Carlo simulation over many realizations of the input planetary
  population should probably be run. However, formal errors would
  probably be acceptable in the near-term (i.e. formal errors on the
  determination of mass function parameters that are determined from
  the subset of objects that survive LSST's selection function for a
  particular strategy).

\begin{table}
  \small
  \begin{tabular}{c p{12cm}}
    & {\it FoM 4.2 - Uncertainty in the mass function for planets beyond the Snow Line through Microlensing}\\
    \hline
  1. & Parameterize the mass function of the population of interest;\\
  2. & Parameterize its distribution of lens amplitude and timescale;\\
  3. & Parameterize the scaling of event rate with local stellar density;\\
  4. & Generate a sample population over the sky; \\
     & ~~~ Scaling from 3. might be used in conjunction with {\tt maf\_contrib/starcounts}; \\
  5. & Run the transient-recovery metric for this population; \\
     & ~~~ Choice of metric needs to handle spatially varying lightcurve template; \\
     & ~~~ Or, the time-interval parameters should be allowed to spatially vary; \\
  6. & Determine the population of microlens planets that would have been triggered by LSST. Do not sum, but record the indices of the surviving objects; \\
  7. & Apply typical measurement uncertainty from a likely followup campaign; \\
  8. & Fit the determined mass function from this sample of survivors only; \\
  9. & {\bf Arrive at FoM 4.2:} Find the offset (from input) and formal uncertainty on the mass function parameters.\\
\hline
    \end{tabular}
 \caption{Description of Figure of Merit 4.2}
  \label{table:pseudoFOM_4p2}
\end{table}

%{\bf FoM 5.1 - Errors in derived $E(B-V)$, $n_H$~as a function of
%  location in the Plane.}
%Dependencies:
%\begin{itemize}
%  \item SNR scaling with apparent magnitude
%    \item For the M dwarf based technique, the relation between
%      reddening-invariant index $[Q_{gri}]$~and intrinsic $g-i$~are
%      expressed as polynomials, so expect non-linear relation with
%      photometric error.
%      \item This uses a 5th-order polynomial to describe the
%        $(Q_{gri}, g-i)$~stellar locus for M dwarfs $(g-i > 1.6)$.
%        \item Care must be taken to correctly propagate errors through
%          the various indices used - not trivial with so many choices
%          of flux ratio used.
%          \item Uncertainties in the parameterizations used for e.g. color-$M_V$~relationships.
%            \item The above are all for every location probed on the map.
%\end{itemize}

% --------------------------------------------------------------------

\subsection{OpSim Analysis}
\label{sec:\secname:MW_Disk_analysis}

{\bf FoM 3.1: the First Galactic Supernova:} This FoM is described in
Definition (\ref{eqn:def_FOM_3p1}) of
\autoref{sec:MW_Disk:MW_Disk_metrics}.

{\it Parameters used:} The lightcurve used has the following parameters:
rise slope $-2.4$; time to peak; $20$~days; decline slope: $0.08$; total
transient duration: 80 days. All filters are used in the detections, and
20 evenly-spaced phases are simulated for sensitivity to pathological
cases (parameter nPhaseCheck=20). Peak apparent magnitudes used: $\{
11,9,8,7,6,6\}$~in $\{u,g,r,i,z,y\}$. Then, $f_{var, i}$~is taken as the
``Sawtooth Alert'' quantity returned by \MAFmetric{TransientMetric}. For the
stellar density metric, distance limits ($10$pc $\le d \le 80$kpc) are
used to avoid biases in the FoM estimate by the Magellanic Clouds.

{\it \bf Results:} $FoM_{preSN}$(\opsimdbref{db:baseCadence})=0.13,
while
$FoM_{preSN}$(\opsimdbref{db:opstwoPS})=0.83.\footnote{2016-04-25: For
  comparison, when run on 2015-era \OpSim runs \opsimdbname{enigma\_1189}
  (Baseline strategy) and \opsimdbname{ops2\_1092} (PanSTARRS-like strategy)
  the results were 0.251 (Baseline) and 0.852 (PanSTARRS-like
  strategy). So the 2016-era \OpSim runs show a sharper disadvantage
  suffered by the Baseline cadence, than before the update.} This FoM
suggests that \opsimdbref{db:opstwoPS} is the best of the cadences we
have tested thus far. However, it is covers less total area on the sky
(spending no time at all on certain regions of community interest like
the South Polar regions) and thus might be unfairly advantaged
compared to \opsimdbref{db:baseCadence}. Exposures are spread over a
smaller area, thus regions including the inner Plane receive better
coverage than they might under a strategy that meets the needs of all
stakeholders).

A more direct comparison includes the recently-completed (at the time
of writing) \OpSim run \opsimdbref{db:NormalGalacticPlane}, which
covers the same regions on the sky as \opsimdbref{db:baseCadence} but
applies the Wide-Fast-Deep strategy to the inner Galactic Plane. This
survey therefore covers the same regions on the sky as the Baseline
cadence. The stragegy \opsimdbref{db:NormalGalacticPlane} still shows
a strong advantage compared to the Baseline survey, with
$FoM_{preSN}$(\opsimdbref{db:NormalGalacticPlane})=0.73, compared to
0.13 for Baseline cadence. See Table \ref{tab_SummaryMWDisk}. Figure
\ref{f_opSim_GalacticSN} presents a breakdown of this figure of merit
across sightlines, for the three observing strategies considered.

\begin{figure}
\begin{center}
  \includegraphics[width=5.25cm]{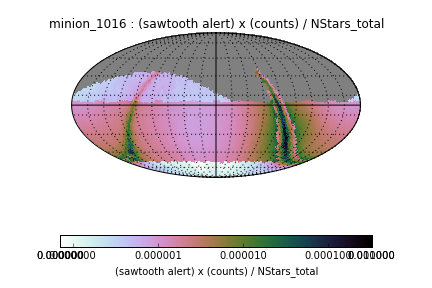}
  \includegraphics[width=5.25cm]{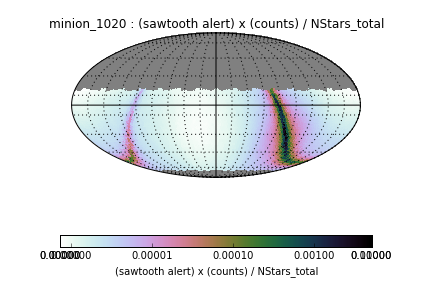}
  \includegraphics[width=5.25cm]{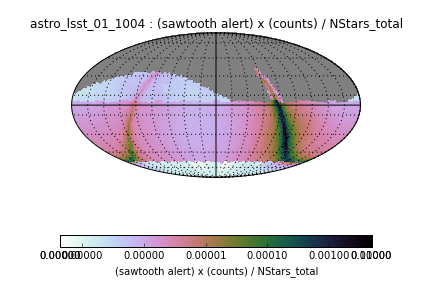}
  \caption{Figure of merit $FoM_{preSN}$~describing LSST's sensitivity
  to any pre-Supernova outburst for the Galactic Supernova science case,
  broken down by sightline. $FoM_{preSN}$~is estimated for
  three \OpSim runs (to-date); \opsimdbref{db:baseCadence} (left; Baseline
  cadence), \opsimdbref{db:opstwoPS} (center; PanSTARRS-like
  strategy), and \opsimdbref{db:NormalGalacticPlane} (which assigns Wide-Fast-Deep cadence to the inner Galactic Plane). The normalizing factors $N_{\ast, total}$ are $3.793 \times 10^{10}$~for both \opsimdbref{db:baseCadence} and \opsimdbref{db:NormalGalacticPlane} (that both strategies have the same $N_\ast$~is not a surprise since both cover the same area) and $3.692\times
  10^{10}$~for \opsimdbref{db:opstwoPS}. The imprint of reduced sampling towards
  the inner plane can be clearly seen for \opsimdbref{db:baseCadence}.
  Notice the difference in color scale between the panels. See \autoref{sec:MW_Disk:MW_Disk_analysis}}
\label{f_opSim_GalacticSN}
\end{center}
\end{figure}

% The Figures of Merit listed above must now be implemented and applied to the OpSim databases.

% The metrics listed above should be carefully compared between our proposed run and the baseline cadence.

% --------------------------------------------------------------------

\subsection{Discussion}
\label{sec:\secname:MW_Disk_discussion}

The Figures of Merit listed above must now be implemented within the
\MAF framework and applied to representative science cases.
See Table \ref{tab_SummaryMWDisk} at the end of this subsection for
initial efforts along these lines. We welcome input and volunteers for
this effort.

Qualitatively, however, we can note immediately that the current
baseline cadence (\opsimdbref{db:baseCadence}) partially excludes the
Galactic Plane from the deep-wide-fast survey and instead adopts a
nominal 30 visits per filter as part of a special proposal - which
also tends to cluster the visits in the inner Plane within the first
few years of the survey. This already seriously compromises the time
baseline (see \autoref{fig_astrom_ByTime_pmError} of
\autoref{sec:MW_Astrometry:MW_Astrometry_OpSim} for a demonstration applied to
proper motions).

Whether the Plane should be observed as a ``special survey'' or as
part of Wide-Fast-Deep, remains an open question that we expect the
Figures of Merit (FoM) to answer when fully implemented. For example,
one can imagine reducing observing time in $u$~(and possibly $g$)
bands towards regions of very high extinction at the lowest galactic
latitudes. The scientific impact of such a strategy (e.g. possibly
lower sensitivity to directly-detected compact objects versus improved
coverage for transients in general) should become clear when
implementation and evaluation of the FoMs described in this Section
are complete.

%We have proposed an OpSim run that includes the Galactic Plane in the
%deep-wide-fast survey:

%\url{https://github.com/LSSTScienceCollaborations/ObservingStrategy/blob/master/opsim/Proposal_GP.md}

\begin{table}
  \begin{tabular}{l|p{6cm}|c|c|c|c|p{5cm}}
    FoM & Brief description & {\rotatebox{90}{\opsimdbref{db:baseCadence}}} & {\rotatebox{90}{\opsimdbref{db:opstwoPS}}} & {\rotatebox{90}{\opsimdbref{db:NormalGalacticPlane}  }} &  {\rotatebox{90}{future run 2}} & Notes \\
    \hline
    1.1 & \footnotesize{LMXB ellipsoidal variations}      & - & - & - & - & - \\
    1.2 & \footnotesize{Uncertainty in DNe duty cycle}   & - & - & - & - &  \footnotesize{LSST as initial trigger} \\
    2.1 & \footnotesize{Fraction of Novae detected}       & - & - & - & - &  - \\
    2.2 & \footnotesize{Fraction of Nova alerts}       & - & - & - & - &  - \\
    3.1 & \footnotesize{Galactic Supernova pre-variability} & 0.13 & {\bf 0.83} & 0.73 & - & \footnotesize{Fraction of SN2010mc-like outbursts that LSST would detect; $FoM_{preSN} = f_{var} \times N_{\ast}$} \\
    4.1 & \footnotesize{Fraction of LSST-triggered microlens candidates} & - & - & - & - & - \\
    4.2 & \footnotesize{Uncertainty in derived planetary mass function} & - & - & - & - & \footnotesize{LSST as initial microlens trigger} \\
%    5.1a & \footnotesize{Median (over sight-lines) of the uncertainty in $E(B-V)$} & - & - & - & - & \footnotesize{(Most useful FoM probably a spatial map of the uncertainty.)} \\
%    5.1b & \footnotesize{Variance (over sight-lines) of the uncertainty in $E(B-V)$} & - & - & - & - & - \\
  \end{tabular}
\caption{Summary of figures-of-merit for the Galactic Disk science cases. The best value of each FoM is indicated in bold. Runs \opsimdbref{db:baseCadence} and \opsimdbref{db:opstwoPS} refer to the Baseline and PanSTARRS-like strategies, respectively. Column \opsimdbref{db:NormalGalacticPlane} refers to a recently-completed \OpSim run that includes the Plane in Wide-Fast-Deep observations. See \autoref{sec:MW_Disk:MW_Disk_analysis}. }
\label{tab_SummaryMWDisk}
\end{table}

%Discussion: what risks have been identified? What suggestions could be
%made to improve this science project's figure of merit, and mitigate
%the identified risks?

% ====================================================================
%
\subsection{Conclusions}

We answer below the ten questions posed in \autoref{sec:intro:evaluation:caseConclusions}.

While the science cases developed thus far all rest on
  variability sensitivity in one way or another, we should point out
  that static science should not be sacrificed completely to
  variability studies. This suggests that any strategy covering the
  inner Milky Way must retain at least a minimum total depth in each
  of the $\left\{u,g,r,i,z,y \right\}$~filters, along all sight-lines, in
  order to constrain stellar effective temperatures, metallicities,
  and interstellar extinction directly from LSST photometry \citep[see
    for example the presentation in][and associated
    papers]{ivezic08}. The implied figure of merit for static science
  is straightforward in principle - simply evaluate the photometric
  depth in each filter at which the confusion limit is reached at the
  required photometric precision - but in practice depends on the
  actual set stellar populations along each line of sight. This figure of
  merit is likely to set the minimum acceptable exposure time in each
  filter in each field, and will require a combination of modeling
  work and experience with existing photometric datasets.

\begin{description}

 \item[Q1:] {\it Does the science case place any constraints on the
 tradeoff between the sky coverage and coadded depth? For example, should
 the sky coverage be maximized (to $\sim$30,000 deg$^2$, as e.g., in
 Pan-STARRS) or the number of detected galaxies (the current baseline
 of 18,000 deg$^2$)?}

 \item[A1:] Figures of Merit addressing this question have been
   specified in this chapter, but implementation and execution are
   still in progress. Co-added depth is less important to this science
   than temporal coverage sufficient to measure the variability (at a
   range of timescales) central to the science cases in this chapter
   section. {\it Qualitatively,} we expect the FoMs will indicate that
   the sky coverage within the inner Disk should be maximized subject
   to the constraint that all fields experience LSST coverage
   throughout the full ten-year survey (a condition currently {\it
     not} satisfied by the current baseline cadence,
   \opsimdbref{db:baseCadence}). This is because some of the important
   transients are relatively rare, and/or the spatial distribution of
   the population within the Milky Way disk is of scientific
   importance. To our knowledge, only two \OpSim runs currently exist
   (\opsimdbref{db:opstwoPS} and \opsimdbref{db:NormalGalacticPlane})
   which subject the populations of interest in this section to
   coverage approaching that of Wide-Fast-Deep. A greater variety of
   strategies of inner-plane coverage is needed to explore any
   tradeoffs between time coverage and spatial coverage within the
   inner Disk. One example might be an \OpSim run with (say) 50\%
   spatial coverage at Wide-Fast-Deep-like levels, the other 50\% at
   coverage similar to the Plane mini-survey in
   \opsimdbref{db:baseCadence}.

 \item[Q2:] {\it Does the science case place any constraints on the
 tradeoff between uniformity of sampling and frequency of  sampling? For
 example, a rolling cadence can provide enhanced sample rates over a part
 of the survey or the entire survey for a designated time at the cost of
 reduced sample rate the rest of the time (while maintaining the nominal
 total visit counts).}

\item[A2:] The frequency of sampling is critical for the science
  cases in this section. Requirement on the uniformity of the sampling
  depends on the timescale of variability and is difficult to project
  without the FoMs being evaluated. A FoM to quantify the relative
  importance of each has been outlined in the chapter, but its
  implementation and execution are still in progress. A greater
  variety of observing strategy simulations that offer
  full-time-baseline coverage of the inner Plane are also needed on
  which to run the FoMs. We require suggestions for \OpSim runs with a
  variety of temporal distributions of exposures within the ten-year
  survey lifetime, towards the inner Plane.

 \item[Q3:] {\it Does the science case place any constraints on the
 tradeoff between the single-visit depth and the number of visits
 (especially in the $u$-band where longer exposures would minimize the
 impact of the readout noise)?}

\item[A3:] This depends strongly on the line of sight,
  especially in $u$-band where reddening reduces sensitivity in any
  individual visit (thus impacting variability-searches for very blue
  objects). In those lines of sight, however, large co-added depth in
  $u$~would still be useful for static population studies. For some
  science cases, single-visit photometric depth provides diminishing
  returns even in the redder filters (for example, once depth
  $i\gtrsim 24$~were reached, going deeper would likely not result in
  finding more dwarf novae, since this depth would already be
  sufficient to find such sources beyond the far side of the
  Bulge). However, \OpSim coverage is still relatively sparse in the
  spatial regions discussed. We suggest augmenting the \OpSim coverage
  towards the inner Milky Way with runs with a variety of u-band
  depths (see for example the discussion of strategies
  \opsimdbref{db:DoubleUbandExptime} and
  \opsimdbref{db:DoubleUbandExptimeSameVisits} in Chapter
  \ref{chp:cadexp}). To-date, sufficient \OpSim coverage to answer
  questions about preferred coverage vs. filter for the inner plane
  simply does not exist.

 \item[Q4:] {\it Does the science case place any constraints on the
 Galactic plane coverage (spatial coverage, temporal sampling, visits per
 band)?}

 \item[A4:] Yes. Galactic plane coverage (in most cases, the inner Plane)
   is crucial to all the science cases in this Section, because the
   populations of interest are not found outside the plane in
   sufficient numbers. In most cases, this coverage must be
   sufficiently spread out in time for discovery through variability
   to be possible over the entire ten-year survey.

 \item[Q5:] {\it Does the science case place any constraints on the
 fraction of observing time allocated to each band?}

\item[A5:] We do not have quantitative answers to this question
  yet since the FoMs are mostly still under development and await a
  sufficient range of \OpSim runs to answer them. {\it Qualitatively,}
  however: LSST will likely be crucial in providing prompt color
  information for the rare-but-important events in the inner plane
  (including unusual or scientifically critical microlensing events
  alerted by LSST or other facilities), so we urge nonzero coverage in
  all six filters. Taking the specific case of outbursts from compact
  objects: having two red filters (e.g. $r-i$) would help tremendously
  in breaking degeneracies between distance and luminosity since the
  spectrum of the outbursts discussed here is typically vega-like (so
  any unusual $r-i$~color should be the result largely of interstellar
  reddening). The minimum observing time allocated to each band is likely to be set by the requirement to retain static science goals such as population dissection \citep[see, e.g.][]{ivezic08}; at the present date, however, figures of merit for static science in the inner plane have not yet been evaluated.
%To be most generally useful for static population
%  studies, such as those constraining reddening, $T_{eff}$~and
%  $[Fe/H]$~photometrically following the manner of the SDSS tomography
%  studies \citep[see, e.g.][and references therein]{ivezic08}, coverage in at
%  least $ugriz$~will be required that affords precise photometry at
%  least down to the confusion limit. We do not as-yet have
%  quantitative limits on what this would mean for the distribution of observing time across filters, however.}

\item[Q6:] {\it Does the science case place any constraints on the cadence for deep drilling fields?}

\item[A6:] The science cases in this chapter do not place any
  constraints on the deep-drilling fields already identified. Were
  deep-drilling-like cadence to be possible for a few fields within
  the inner plane, however, it would be hugely useful. For example, to
  recover ellipsoidal variables at short periods, the shortest gap in
  observations should be $\sim$20 minutes. Not all observations need
  to be so close together, but it is important to have at least a few
  baselines that short in order to reliably recover periods between 80
  minutes and 2 hours, where the majority of CVs (and potentially a
  new population of X-ray faint short period XRB outbursts) should
  be. These considerations also apply to any LSST observational campaign that will complement WFIRST observations of the Bulge (Section \ref{sec:wfirst:microlensing}).

 \item[Q7:] {\it Assuming two visits per night, would the science case
 benefit if they are obtained in the same band or not?}

\item[A7:] Qualitatively, visits in two different filters are weakly preferred. Science cases requiring discrimination of
   the class of object would benefit from color information afforded
   by observing in different filters during the same night, and for
   most of the science cases in this section the variation plays out
   in days rather than hours. However, for a subset of the science
   cases (e.g. ellipsoidal variations of counterparts to eRosita
   sources), observations at the same filter would be preferred. If
   the color variation with phase is smaller than the orbital
   variations, then having the colors could give the best of both
   worlds.

 \item[Q8:] {\it Will the case science benefit from a special cadence
 prescription during commissioning or early in the survey, such as:
 acquiring a full 10-year count of visits for a small area (either in all
 the bands or in a  selected set); a greatly enhanced cadence for a small
 area?}

\item[A8:] No, as long as all fields obtain at least some
  coverage in all filters throughout the full 10-year survey
  lifetime. Our expectation is that a given field will experience
  nonuniform cadence over the ten-year survey, with some intervals of
  high cadence followed by some intervals of low cadence. It would be
  useful for some fields to experience their high-cadence intervals
  during the first year in order to better understand detectability
  and systematics, and to update any population simulations as LSST's
  performance is better constrained.

 \item[Q9:] {\it Does the science case place any constraints on the
 sampling of observing conditions (e.g., seeing, dark sky, airmass),
 possibly as a function of band, etc.?}

\item[A9:] At this stage we do not believe that any special
   conditions are required. The inner plane is somewhat unusual in
   that it will likely be confusion-limited rather than
   sky-limited. Thus better seeing would obviously be better for the
   science, however we do not at this stage have any quantitative
   constraints.

 \item[Q10:] {\it Does the case have science drivers that would require
 real-time exposure time optimization to obtain nearly constant
 single-visit limiting depth?}

\item[A10:] Not as long as the achieved limiting depth is accurately
   understood once observations have been taken. In these
   confusion-limited regions, this question is probably constrained
   more by the performance of the photometry software than by
   real-time observing considerations.

 \end{description}

 % ====================================================================

\navigationbar

% PJM: Perry's content is to be merged with Pat's, in Section 5.6
% \input{MilkyWay/MW_SFH.tex}

% PJM: moved the following to FutureWork, while the metric(s) is/are being implemented
% \input{MilkyWay/MW_Dust.tex}

% ====================================================================
%+
% SECTION:
%    MW_Astrometry.tex
%
% CHAPTER:
%    galaxy.tex
%
% ELEVATOR PITCH:
%
%-
% ====================================================================

\section{Astrometry with LSST: Positions, Proper Motions, and Parallax}
\def\secname{MW_Astrometry}\label{sec:\secname}

\credit{willclarkson}, \credit{caprastro}, \credit{dgmonet}, \credit{DanaCD}, \credit{jgizis}, \credit{mliu}, \credit{yoachim}

A number of Milky Way science cases of interest to the Astronomical
community will depend critically on the astrometric accuracy LSST will
deliver. While ``astrometry'' is not a science case in the framework
of this white paper, LSST's astrometric performance will be sensitive
to the particular choice of observing strategy.
%While astrometry is not a science case, high astrometric accuracy enables
%a large number of science cases.
Hence, the LSST Observing Strategy needs to be examined for systematic
trends that might limit or even preclude precise measures of
stellar positions, proper motions, parallaxes, and perturbations that
arise from unseen companions.

Tying together the LSST and Gaia optical reference frames will be
critical to many of the science cases discussed here (and in the LSST
Science book), but the final as-delivered performance of Gaia is of
course not yet known. For example, the apparent magnitude range needed
for reference frame tie-in is not yet constrained for crowded fields
(where bright stars will likely form the essential LSST-Gaia reference
frame tracer population), but this will impact the useful LSST
exposure time for these regions, with corresponding observing strategy
implications.

It will be vital for the interaction between LSST and Gaia to be
studied in detail, in terms of astrometry, photometry, spatial
crowding, and instrumental issues, which will be greatly facilitated
by the second Gaia data-release
\citep[e.g.][]{2016AA...595A...1G}. With Gaia-DR2 still in the future
at the time of writing, here we outline some candidate astrometry
science cases, and perform the relative comparison between candidate
observing strategies. As we show, stark differences due to
time-sampling are already apparent, particularly for inner-Plane
regions of the Milky Way.

\autoref{sec:\secname:MW_Astrometry_measurements} highlights two
science cases at opposite scales of distance from the Sun that require
accurate and precise astrometry and/or proper motion
measurements. \autoref{sec:\secname:MW_Astrometry_metrics} presents
Metrics for LSST's astrometric performance, and discusses Figures of
Merit for the two highlighted science cases. These metrics are applied to two example \OpSim runs in
\autoref{sec:\secname:MW_Astrometry_OpSim}. Finally in Section
\ref{sec:\secname:MW_Astrometry_furtherwork}, the work that is still
needed is discussed, both in terms of the Metrics and the Figures of
Merit that depend on them.

%Each of these cases stresses different aspects of the LSST hardware, software,and observing strategies.

%, here we highlight three representative science cases.
%that illustrate the various impacts of the observing strategy might
%be:
%To highlight the
%various astrometric impacts of the strategy, three science cases have
%been chosen for particular attention:

%\subsection{Introduction: Astrometry as a special case}
%\label{sec:\secname:MW_Astrometry_intro}

\subsection{Target Measurements and Discoveries}
\label{sec:\secname:MW_Astrometry_measurements}

%\begin{itemize}
%\item[1.] Identification of Streams in the Galactic Halo using proper motions.
%\item[2.] A complete sample of stars in the solar neighborhood.
%\end{itemize}
%\item The tie between the Radio and Optical realizations of the International Celestial Reference System.
%\item The specific and ensemble agreement between LSST and Gaia parallaxes.

{\bf 1. Identification of Streams in the Galactic Halo Using Proper Motions}

Much of the Milky Way's stellar halo was built by the accretion of smaller galaxies. Given that these galaxies
were generally of low mass, their tidal debris should still form coherent structures in phase space, especially
in the outer Galaxy where dynamical times are long. The identification of these streams would allow
a reconstruction of the accretion history of the Milky Way. Tides also lead to the dissolution of globular clusters,
leaving notably thin streams that serve as sensitive tracers both of the Galactic potential and of the presence of dark
subhalos.

A relatively small number of streams, originating from both dwarfs and globular clusters, have been identified via photometry
of individual stars in large surveys such as SDSS\@. However, only the highest surface brightness structures can be found
in this manner, and it is often difficult to trace the streams over their full extent. LSST will enable streams to be identified
by stellar proper motions, and combined with targeted follow-up spectroscopy, will yield full 6-D position and velocity measurements suitable for dynamical modeling.
Further, it will allow the discovery of tidal debris that is no longer spatially coherent but which can be unambiguously identified in phase space.

Finally, streams and other kinematically-distinct halo substructure
can be identified and characterized by combining proper motions and
photometry in reduced proper-motion diagrams \citep[e.g.,][]{carlin12},
and by analyzing proper-motions of tracers such as
RR Lyrae and giants over large portions of the sky \citep[e.g.,][]{casettidinescu15}.

{\it Response to observing strategy:} Most stars in streams will be main-sequence stars, and the old main sequence turnoff  is located at $r\sim24$ at a distance of 100 kpc.
The nominal LSST proper motion precision at this magnitude is 1 mas yr$^{-1}$, corresponding to about 475 km s$^{-1}$ at this distance. The proper motion
measurements will be better for brighter stars, but in general ensembles of stars will be necessary for accurate measurements. To make accurate proper motion measurements for faint stars, several key components are required. First, a zero point must be established, possibly via background galaxies located in each field. Next, the observations must cover a sufficient range of epochs to reliably detect linear proper motions.

In principle, streams can be detected using proper motions alone, as
kinematically colder populations than the Galactic field through which
they are observed. On the assumption that intrinsic proper motion
dispersion in a stream is negligible compared to the field dispersion,
the faintness limit for proper motion-only detection of streams is set
approximately by the apparent magnitude at which LSST's proper motion
error becomes comparable to or larger than the field proper motion
dispersion. Based on LSST's nominal astrometric performance (as
communicated in the LSST Science Book), this limit will be reached at
$r \sim 22.5$~(Casetti-Dinescu et al. 2017, forthcoming).

However, {\it characterization} of these streams requires their
  identification over their full lengths of many degrees of the sky, for which relative astrometry over small fields will not be
sufficient. Over most of the main-survey area, asbolute proper
  motion calibration may be achieved using the large number of
  background galaxies expected \citep[e.g.][]{IvezicEtal2008} to set
  the proper motion zero point. By using Gaia stars at the
  bright end as absolute proper-motion calibrators we can quantify the
  precision and accuracy of background galaxies as a link to an
  inertial reference system, and thus improve the calibration at the
  faint end of the survey.

Matching LSST's astrometry to the radio International Celestial Reference System (ICRS) - and thus calibrating LSST's absolute astrometry - is highly likely to proceed via the Gaia optical reference frame. Therefore, tying LSST's position to Gaia's final delivered reference frame will be essential. The accuracy with which this tie-in can be achieved, will likely vary spatially and needs further exploration.

Secondary links to the radio ICRS might proceed by matching LSST positions directly to the radio frame traced by background QSOs. These considerations should be developed by the user community, and will include issues including detectable optical or radio structures that degrade the positions or suggest a displacement between the location of the sources of the radio and optical radiation.

{\bf 2. The Most Complete Sample of Stars in the Solar Neighborhood}

The direct solar neighborhood offers our only chance to get make a complete sample of stars, brown dwarfs, and stellar remnants that encompass the entire formation and dynamical history of the Milky Way. While Gaia will offer parallax measurements for perhaps billions of stars, its faint magnitude limit of $G\sim 20$ will limit its measurements of the lowest-mass objects
and remnants to nearby objects, much less than the thin disk scale height of $\sim 300$ pc. For example, Gaia can only measure parallaxes for $0.2 M_{\odot}$ M dwarfs to about 100 pc
and $0.1 M_{\odot}$ M dwarfs to only \emph{10 pc}, showing that Gaia is ill-suited for studies of the coolest dwarfs. By contrast, LSST can measure parallaxes for $> 10^5$ M dwarfs and thousands of L/T brown dwarfs (the coolest Y dwarfs are too faint even for LSST; little contribution is likely here beyond the sample provided by WISE). Gaia will likewise be limited to cool white dwarfs within $\sim 100$ pc with which to estimate the age of the disk, and the thick disk and halo will be out of reach. LSST can directly compare white dwarf luminosity functions to determine precise differential ages for the thin disk, thick disk, and halo.

{\it Response to observing strategy:} Successfully completing this project will require parallax measurements much fainter than possible with Gaia as well as a verification that the LSST and Gaia parallax measurements are consistent in the overlapping magnitude range.

The measurement of stellar parallax puts the substantial constraints on the observing cadence. There are two major issues: the need to sample a wide range of parallax factor (related to time of year), and breaking the correlation between differential color refraction and parallax factor.

``Parallax factors" characterize the ellipse of the star's apparent motion as seen over the course of a year. The shape of the ellipse is given by the Earth's orbit and is not a free parameter in the astrometric solution. The amplitude of the right ascension parallax factor is close to unity while the amplitude of the declination parallax factor is dominated by the sine of ecliptic latitude.
The right ascension parallax factor has maximum amplitude when the star is approximately six hours from the Sun, so the optimum time for parallax observing is when the
star is on the meridian near evening or morning twilight. Atmospheric refraction displaces the star's apparent position in the direction of the zenith by an amount dependent on both the wavelength of the light and the distance to the zenith. Whereas the measured position of star is a function of the total refraction, the measurement of parallax
and proper motion depends on the differences in the refraction as a function of the color of each star and the circumstances of the observations.  This
dependence is called differential color refraction. The combination of parallax factor and differential color refraction leads to two rules: (i) Observations need to cover the widest possible range in parallax
factor, and (ii) The correlation between parallax factor and hour angle in the observations needs to be minimized.

%with respect to the meanmotion of the reference frame.

%The search for faint proper motion stars has two key components.  The first is the need to identify stars that move from the ensemble of other image features that can cause confusion.  For example, a compact group of stars that contains one or more stars of variable brightness can confuse the catalog correlation algorithm.  The other is the need to establish the zero point. For the case of relative astrometry, meaning the measurement of relative positions in an image, the question remains on how to remove the mean motion of the reference frame.  For example, astrometry on certain classes of galaxies might produce a zero point of sufficient accuracy.  This leads to a third constraint on the observing cadence.
% \begin{itemize}
%\item [3)] Observations must cover a sufficient range of epochs so that stars with
%linear or periodic motions can be identified at a high level of confidence.
%\end{itemize}

%\subsection{Sensitivity of parallax measurements to observing strategy}
%\label{sec:\secname:MW_Astrometry_cadence}

%\medskip

\subsection{Metrics and Figures of Merit for LSST's delivered astrometric accuracy}
\label{sec:\secname:MW_Astrometry_metrics}

%\medskip

First we discuss metrics for the observing strategy that affect all of
LSST's astrometric measurements, then discuss figures of merit for the
two science cases. (The three general metrics were identified years
ago and are already in the suite of \MAF utilities, and they should be
reviewed prior to making final decisions. For this reason, in addition
to the Figures of Merit later in the chapter, we present spatial maps
and histograms for the metrics themselves in Section
\ref{sec:\secname:MW_Astrometry_OpSim}, for representative \OpSim
strategies.)

\begin{itemize}
\item[A)] For each LSST field, the parallax factors at each epoch of
observation need to be computed.  The ensemble of these must be checked for
sufficient coverage of the parallactic ellipse.  In particular, the number of
measures with RA parallax factor less than --0.5 and greater than +0.5
needs to be tallied because these carry the most weight in the solution
for the amplitude (parallax).
\item[B)] For each LSST field,
%the hour angle of the observation needs to be
%computed, and
the correlation between hour angle and parallax factor
needs to be examined for significance.  The observing strategy must minimize
the number of fields with this correlation.
\item[C)] The epochs of observation for each field must be checked for a
reasonable coverage over the duration of the survey and to avoid
collections of too many visits during a few short intervals.
\end{itemize}

Within \MAF, metrics A (parallax factor distribution) and B
  (hour angle and parallax correlation) are implemented in a slightly
  different manner from the prescription above. We describe the
  implemented metrics here.

{\it Parallax factor coverage:} This is \MAFmetric{ParallaxCoverageMetric} in \MAF. For
  each pointing, the parallax factor coverage is parameterized as the
  weighted mean radius of arc $\langle l \rangle$~from the center of
  motion due to parallax, scaled to the range $0 \le \langle l \rangle
  \le 1$. Inverse-variance weighting is used both for $\langle l \rangle$~and for the center of motion due to parallax.
%The
%  inverse-variance weighted mean parallax offset is subtracted from
%  the set of parallax offsets for an object at a given location with
%  unit parallax amplitude, and the inverse-variance weighted mean
%  $\langle l \rangle$~of the resulting residuals is returned,
%The reported $\langle l \rangle$~is scaled
%  to the range $0 \le \langle r \rangle \le 1$.
For each measurement,
  the variance used in the weighting is the estimate of the
  (uncrowded) astrometric uncertainty returned by \OpSim for a star of
  specified fiducial magnitude $r$~at the center of the HEALPIX of
  interest. What constitutes a ``good'' value for $\langle l \rangle$~depends on the location
  of the star in ecliptic co-ordinates. Near either ecliptic pole a
  star with uniform parallax coverage would have $\langle l \rangle
  \approx 1.0$~while on the ecliptic uniform coverage would produce
  $\langle l \rangle \approx 0.5$. For any location, $\langle l
  \rangle \approx 0$~would mean all the observations were taken with
  identical parallax factor and therefore any attempt to fit the
  parallax amplitude would be completely degenerate with the object's
  position.

{\it Parallax-Hour angle correlation:} This is \MAF metric \MAFmetric{ParallaxDcrDegenMetric}. At the level of tens
  of milliarcsec, Differential Chromatic Refraction (DCR) shifts the
  apparent location of the star in a color-dependent manner. Depending
  on the hour-angle distribution of observations throughout the year,
  motion due to parallax can become degenerate with motion due to the
  pattern of DCR values sampled. This metric returns the Pearson
  correlation coefficient $\rho$~between the best-fit parallax
  amplitude and DCR amplitude, returning values in the range $-1.0 \le
  \rho \le +1.0$. The range of acceptable values for this metric is
  still under investigation; Monte Carlo simulation by one of us (DGM)
  suggests the parallax error becomes independent of
  $\rho$~(i.e. other effects dominate) for values $|\rho| \lesssim
  0.7$.

For the stream project discussed above, a simple to state (but perhaps complex to implement) figure of merit
is the number of streams that can be discovered in LSST via their proper motions. As a first
attempt, it would be reasonable to assume about 100 halo streams from old, metal-poor dwarf galaxies with
stellar masses $10^5-10^7 M_{\odot}$ distributed as $r^{-3.5}$. The stream widths and internal velocity
dispersions can be set from galaxy scaling relations, and their 3-D velocities consistent with a simple Galactic mass
model at their radii. Setting the stream lengths is more complicated, but should cover a large range from a few to many kpc.
Over a given area, the stream ``S/N" can roughly be taken as the number of stream stars (identified via proper motion, color, and magnitude)
divided by the square root of the number of field stars. For globular clusters, a similar number of streams could be included, but these should have much smaller widths (10s of pc)
and typical masses $10^4-10^5 M_{\odot}$. Eventually it would be desirable to use actual simulated stream parameters taken from cosmological models of the Milky Way (e.g.,
from the Aquarius simulation).

Solar neighborhood projects will be sensitive to the general parallax and proper motion metrics discussed above. More specific science figures of merit are {\it required} at this stage.  For example, the precision of the differential age measurement between the thin disk and halo, which would depend on the number of white dwarfs that can be isolated
from each population.

\subsection{OpSim Analysis}
\label{sec:\secname:MW_Astrometry_OpSim}

Here we present initial analysis of LSST's astrometric
performance. Two example strategies are assessed: the current baseline
strategy, \opsimdbref{db:baseCadence}, and the more recently-evaluated cadence
\opsimdbref{db:NormalGalacticPlane}, which extends the Wide-Fast-Deep
survey to the Galactic Plane (see Section \autoref{sec:cadexp:alternatives}
for more detail on this run).

%the PanSTARRS-like cadence,
%\opsimdbref{db:opstwoPS}, which greater spatial uniformity and
%superior coverage of the Galactic Plane.

\subsubsection{Metrics: Parallax and proper motion precision}

Here we present the expected astrometric performance of LSST as a function of
location on-sky, for two main cuts on the survey strategies:
\begin{itemize}
  \item By time: objects detected in $\{g,r,i,z\}$, after years 1, 2 and 10
    of the survey (Figures \ref{fig_astrom_ByTime_PACoverage} -
    \ref{fig_astrom_ByTime_paError});
\item By filter: objects detected in $\{g,r,i,z\}$, or in $u$ only, or $y$ only, over the full 10 years of the survey (Figures~\ref{fig_astrom_ByFilter_PACoverage} - \ref{fig_astrom_ByFilter_paError}).
\end{itemize}

Astrometric performance for parallax is quantified using the following
metrics:
\begin{itemize}
  \item[1.] Parallax factor coverage (following metric A of \autoref{sec:\secname:MW_Astrometry_metrics}); values farther from 0 are better). See Figures \ref{fig_astrom_ByTime_PACoverage} \&  \ref{fig_astrom_ByFilter_PACoverage};
    \item[2.] Parallax-Hour angle correlation (metric B of \autoref{sec:\secname:MW_Astrometry_metrics}; values closer to 0 are better). See Figures \ref{fig_astrom_ByTime_PADegen} \& \ref{fig_astrom_ByFilter_PADegen};
      \item[3.] Proper motion error, for a star at apparent magnitude 21.0 in the filter specified (this addresses the distribution of measurement epochs, as recommended in Metric C in \autoref{sec:\secname:MW_Astrometry_metrics}; smaller values are better). See Figures \ref{fig_astrom_ByTime_pmError} \& \ref{fig_astrom_ByFilter_pmError};
        \item[4.] Parallax error, for a star at apparent magnitude 21.0 in the filter specified (smaller values are better). See Figures \ref{fig_astrom_ByTime_paError} \& \ref{fig_astrom_ByFilter_paError}.
\end{itemize}

{\it Limitations of the results presented in Figures \ref{fig_astrom_ByTime_PACoverage} to \ref{fig_astrom_ByFilter_paError}.:}
\begin{itemize}
  \item[i.] The spatial maps are clipped at $95\%$~in order to keep
    the color-scale at a sensible range; in some cases this has had
    the side effect of removing parts of the spatial coverage in the
    \opsimdbref{db:baseCadence} maps.

  \item[ii.] This analysis neglected spatial confusion in high-density regions. While this
    confusion would be the same whatever observing strategy was
    chosen, the measurement uncertainties for proper motion and parallax uncertainty
    should be regarded as lower limits.

    \item[iii.] The choice of fiducial apparent magnitude $r = u = y =
      21.0$~is arbitrary. It
      would be informative to repeat the analysis for a range of
      target apparent magnitudes that are better-matched to the
      specific science cases.

      \item[iv.] The comparison between single-filter and $griz$
        detections likely overestimates the measurement precision for
        the $u$-only and $y$-only detections, as an object only
        detected in a single filter may well not be detected in all
        images taken in that filter. While the comparison between
        filter subsets for a given strategy may therefore be highly
        approximate, the comparison between strategies for the same
        filter should be more reliable.

% WIC 2016-06-01 - item below removed, now that we are comparing two strategies with
% similar sky coverage.

%  \item[v.] We have not yet subdivided the samples by a meaningful
%    spatial co-ordinate (galactic latitude would be the obvious
%    choice). A large part of the breadth of the various metric values in
%    \opsimdbref{db:baseCadence} as compared to \opsimdbref{db:opstwoPS} may be
%      due to spatial nonuniformity of the sampling; replotting the
%      histograms coded by galactic latitude would be highly informative in this context.

\end{itemize}

{\it Indications at this date:} Despite these limitations, we note the following:

% WIC 2016-06-01 - updated for comparing wfdPlane to Baseline, not PanSTARRS-1 as was the case.

\begin{itemize}

  \item[I1.] To first order, proper motion and parallax error are dominated by the total time coverage, as might be expected. For some of the mini-survey regions, particularly (but not limited to) the inner-Plane, the baseline strategy \opsimdbref{db:baseCadence} is {\it dramatically} worse for astrometry than the two alternative candidate strategies tested here. When the tendency of \OpSim to cluster mini-surveys within the first year is corrected, detailed comparison of strategies can proceed given the full spread of observing epochs.

  \item[I2.] Taking snapshots of the survey at various stages of completion (Figures \ref{fig_astrom_ByTime_PACoverage} -  \ref{fig_astrom_ByTime_paError}), strategy \opsimdbref{db:NormalGalacticPlane} is not significantly worse than the current baseline strategy \opsimdbref{db:baseCadence};

\item[I3.] For the extremes of object color (objects detected
  only in the bluest or only in the reddest filter), most of the
  astrometry metrics suggest worse performance with additional
  structure in the spatial variation of astrometric performance (see
  Figures \ref{fig_astrom_ByFilter_PACoverage} -
  \ref{fig_astrom_ByFilter_paError}), compared to objects detected in all filters. Further exploration is strongly suggested.

\item[I4.] The two strategies offering enhanced inner-Plane coverage, \opsimdbref{db:NormalGalacticPlane} and \opsimdbref{db:opstwoPS}, do not meaningfully differ from each other purely in terms of delivered astrometric precision.\footnote{However, the absence of the Northern Ecliptic Spur from \opsimdbref{db:opstwoPS} makes it unlikely that this candidate strategy will score highly for Solar System studies.}

\end{itemize}

%% In the current incarnation, these will be big figures on the page. Consider
%% finding a way to summarize them!
\begin{figure}[ht]
  \begin{center}
  \includegraphics[width=2.0in]{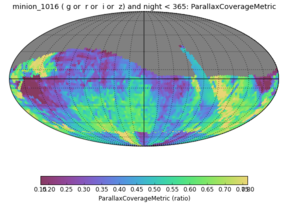}
  \includegraphics[width=2.0in]{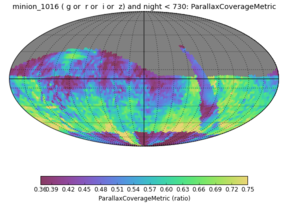}
  \includegraphics[width=2.0in]{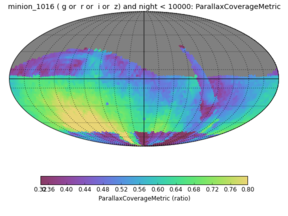}
  \end{center}
  \begin{center}
  \includegraphics[width=2.0in]{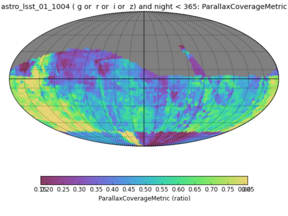}
  \includegraphics[width=2.0in]{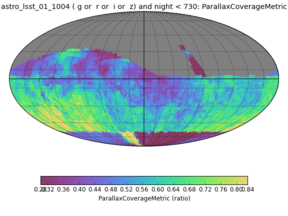}
  \includegraphics[width=2.0in]{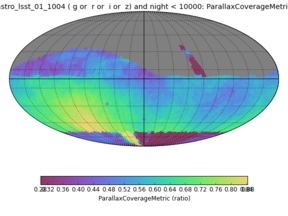}
  \end{center}

  \begin{center}
  \includegraphics[width=2.0in]{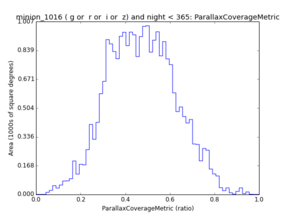}
  \includegraphics[width=2.0in]{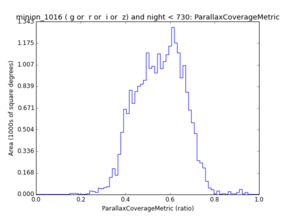}
  \includegraphics[width=2.0in]{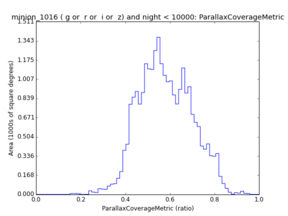}
  \end{center}
  \begin{center}
  \includegraphics[width=2.0in]{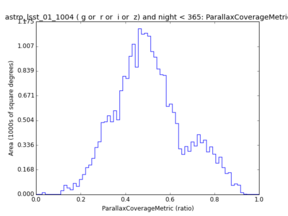}
  \includegraphics[width=2.0in]{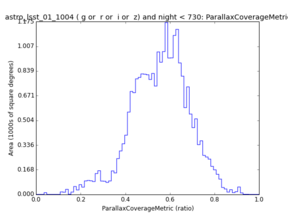}
  \includegraphics[width=2.0in]{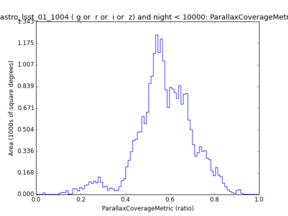}
  \end{center}
  \caption{Parallax coverage achieved at different epochs within the survey. {\it Top and Third row:} \OpSim run \opsimdbref{db:baseCadence}. {\it Second and bottom row:} \OpSim run \opsimdbref{db:NormalGalacticPlane} (wide-fast-deep extended to much of the inner Plane). Reading left-right, columns represent: {\it Left column:} all observations within the first 365 days of operation; {\it Middle column:} first two years; {\it right column:} the full 10-year survey. Spatial maps are clipped at 95\%, with histogram horizontal limits (0.0 - 1.0).}
  \label{fig_astrom_ByTime_PACoverage}
\end{figure}

\begin{figure}[ht]
  \begin{center}
  \includegraphics[width=2.0in]{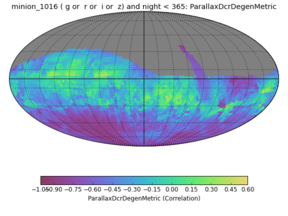}
  \includegraphics[width=2.0in]{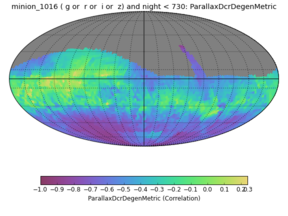}
  \includegraphics[width=2.0in]{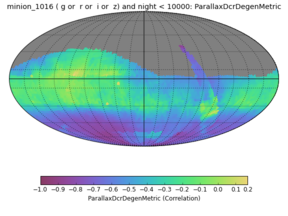}
  \end{center}
  \begin{center}
  \includegraphics[width=2.0in]{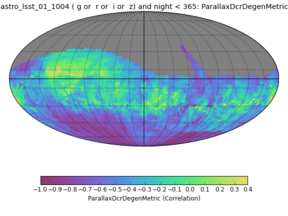}
  \includegraphics[width=2.0in]{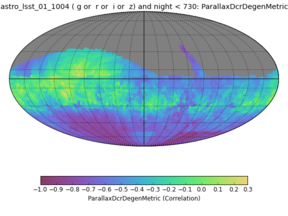}
  \includegraphics[width=2.0in]{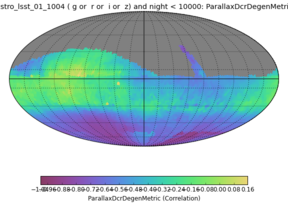}
  \end{center}

  \begin{center}
  \includegraphics[width=2.0in]{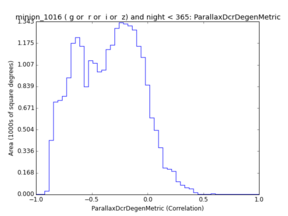}
  \includegraphics[width=2.0in]{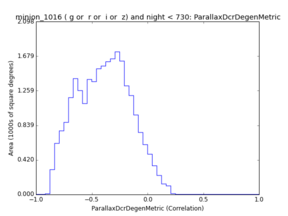}
  \includegraphics[width=2.0in]{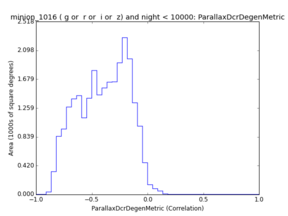}
  \end{center}
  \begin{center}
  \includegraphics[width=2.0in]{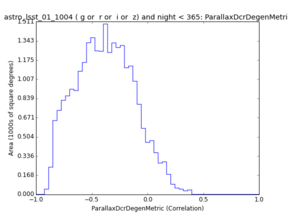}
  \includegraphics[width=2.0in]{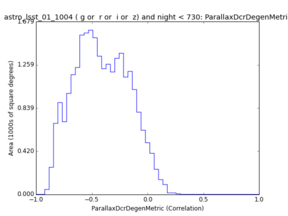}
  \includegraphics[width=2.0in]{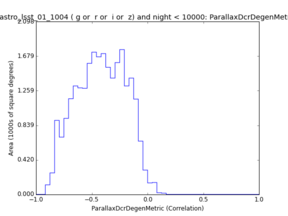}
  \end{center}
  \caption{Correlation coefficient $\rho$~between parallax and Differential Chromatic Refraction (DCR) up to different epochs within the survey. {\it Top and Third row:} \OpSim run \opsimdbref{db:baseCadence}. {\it Second and bottom row:} \OpSim run \opsimdbref{db:NormalGalacticPlane} (wide-fast-deep extended to much of the inner Plane). Reading left-right, columns represent: {\it Left column:} all observations within the first 365 days of operation; {\it Middle column:} first two years; {\it right column:} the full 10-year survey. Spatial maps are clipped at 95\%, with histogram horizontal scale set to the range $-1.0 \le \rho \le +1.0$.}
  \label{fig_astrom_ByTime_PADegen}
\end{figure}

\begin{figure}[ht]
  \begin{center}
  \includegraphics[width=2.0in]{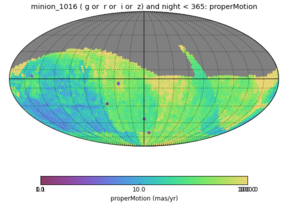}
  \includegraphics[width=2.0in]{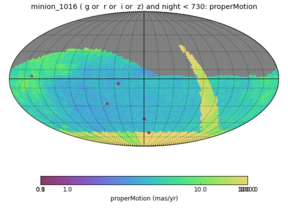}
  \includegraphics[width=2.0in]{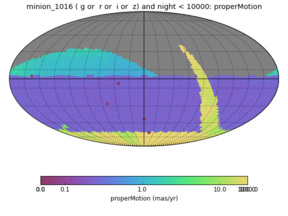}
  \end{center}
  \begin{center}
  \includegraphics[width=2.0in]{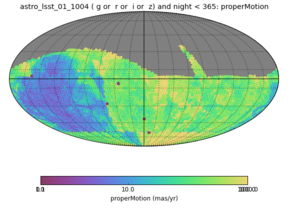}
  \includegraphics[width=2.0in]{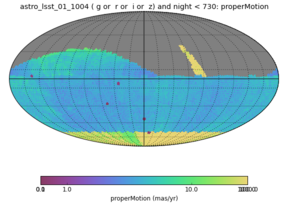}
  \includegraphics[width=2.0in]{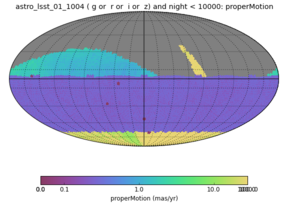}
  \end{center}

  \begin{center}
  \includegraphics[width=2.0in]{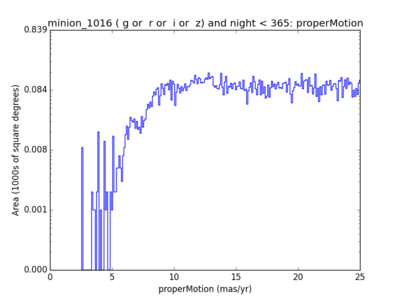}
  \includegraphics[width=2.0in]{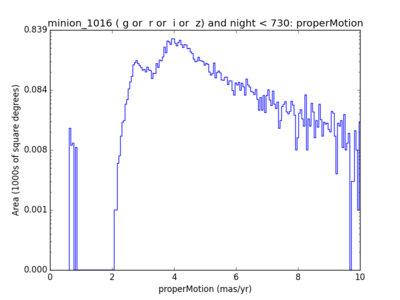}
  \includegraphics[width=2.0in]{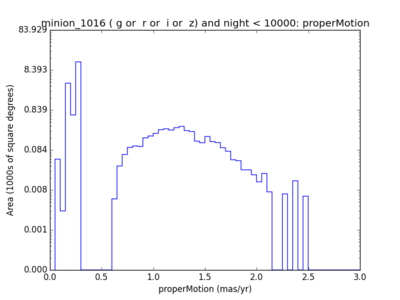}
  \end{center}
  \begin{center}
  \includegraphics[width=2.0in]{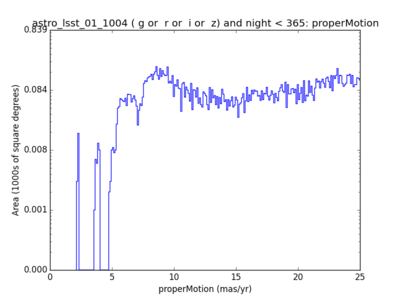}
  \includegraphics[width=2.0in]{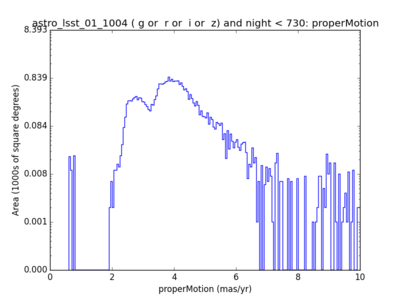}
  \includegraphics[width=2.0in]{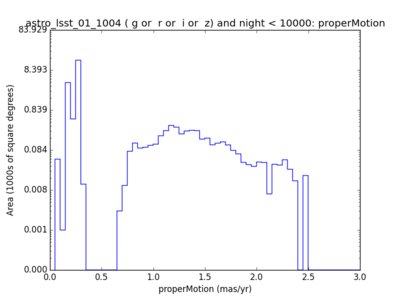}
  \end{center}
  \caption{Proper motion error for a star at $r=21.0$, for different epochs within the survey. Crowding errors are ignored. {\it Top and Third row:} \OpSim run \opsimdbref{db:baseCadence}.  {\it Second and bottom row:} \OpSim run \opsimdbref{db:NormalGalacticPlane} (wide-fast-deep extended to much of the inner Plane). Reading left-right, columns represent: {\it Left column:} all observations within the first 365 days of operation; {\it Middle column:} first two years; {\it right column:} the full 10-year survey. Spatial maps are clipped at 95\% and a log-scale is used for the maps and histograms. Reading left-right, the horizontal upper limits on the histograms are (25, 10, 3.0) mas yr$^{-1}$, respectively. Note that the histograms do not include the full range of values reported in the maps.}
  \label{fig_astrom_ByTime_pmError}
\end{figure}

\begin{figure}[ht]
  \begin{center}
  \includegraphics[width=2.0in]{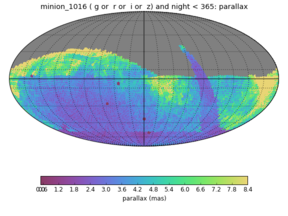}
  \includegraphics[width=2.0in]{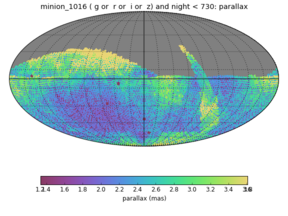}
  \includegraphics[width=2.0in]{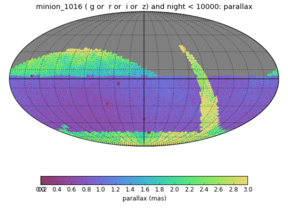}
  \end{center}
  \begin{center}
  \includegraphics[width=2.0in]{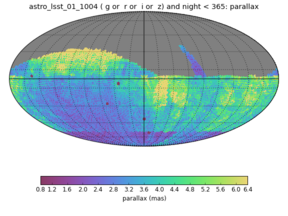}
  \includegraphics[width=2.0in]{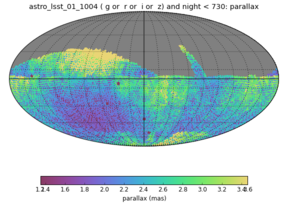}
  \includegraphics[width=2.0in]{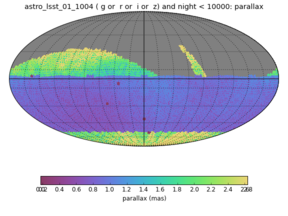}
  \end{center}

  \begin{center}
  \includegraphics[width=2.0in]{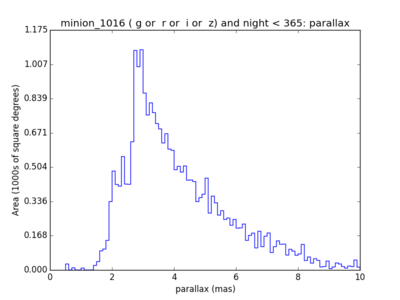}
  \includegraphics[width=2.0in]{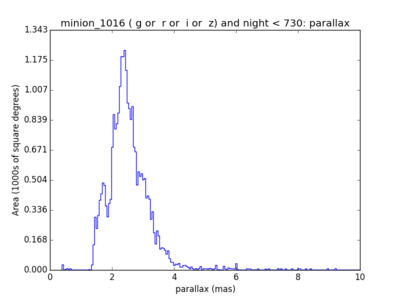}
  \includegraphics[width=2.0in]{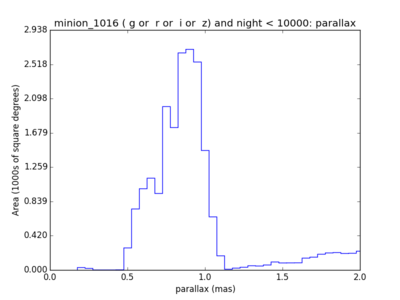}
  \end{center}
  \begin{center}
  \includegraphics[width=2.0in]{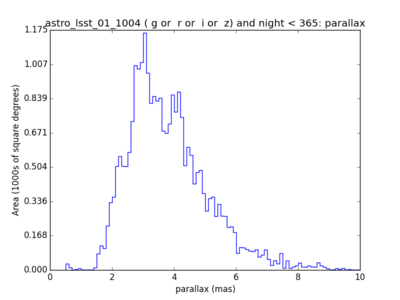}
  \includegraphics[width=2.0in]{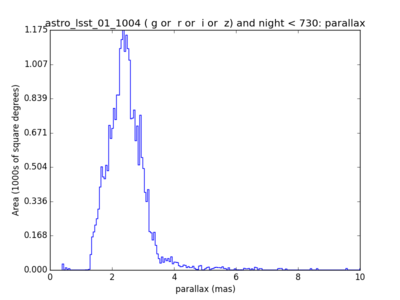}
  \includegraphics[width=2.0in]{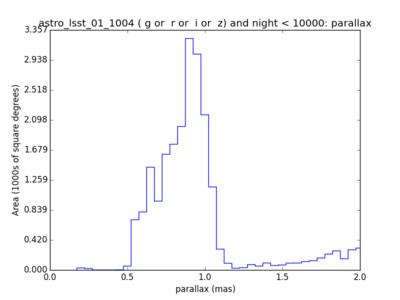}
  \end{center}
  \caption{Parallax error for a star at $r=21.0$, for different epochs within the survey. Crowding errors are ignored. {\it Top and Third row:} \OpSim run \opsimdbref{db:baseCadence}. {\it Second and bottom row:} \OpSim run \opsimdbref{db:NormalGalacticPlane} (wide-fast-deep extended to much of the inner Plane). Reading left-right, columns represent: {\it Left column:} all observations within the first 365 days of operation; {\it Middle column:} first two years; {\it right column:} the full 10-year survey. Spatial maps are clipped at 95\%.  Reading left-right, the horizontal upper limits on the histograms are (10, 10, 2.0) mas, respectively.}
  \label{fig_astrom_ByTime_paError}
\end{figure}

%%% Now for the metrics by filter.
\begin{figure}[ht]
  \begin{center}
  \includegraphics[width=2.0in]{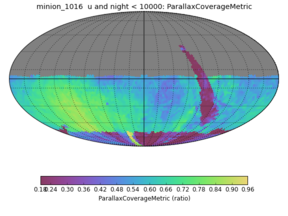}
  \includegraphics[width=2.0in]{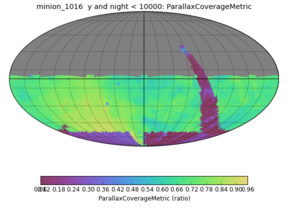}
  \includegraphics[width=2.0in]{./figs/milkyway/astromPanels/MW_Astrom_paCovge_Baseline_10y_map.png}
  \end{center}
  \begin{center}
  \includegraphics[width=2.0in]{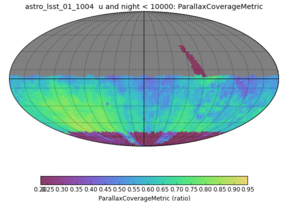}
  \includegraphics[width=2.0in]{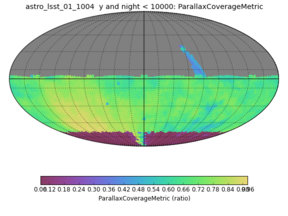}
  \includegraphics[width=2.0in]{./figs/milkyway/astromPanels/MW_Astrom_paCovge_wfdPlane_10y_map.png}
  \end{center}

  \begin{center}
  \includegraphics[width=2.0in]{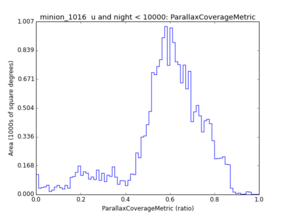}
  \includegraphics[width=2.0in]{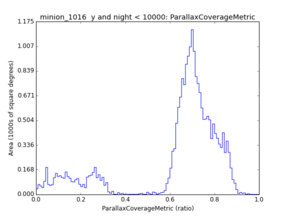}
  \includegraphics[width=2.0in]{./figs/milkyway/astromPanels/MW_Astrom_paCovge_Baseline_10y_hst.png}
  \end{center}
  \begin{center}
  \includegraphics[width=2.0in]{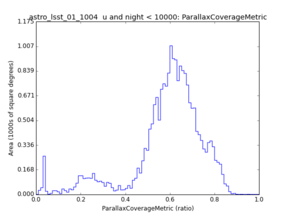}
  \includegraphics[width=2.0in]{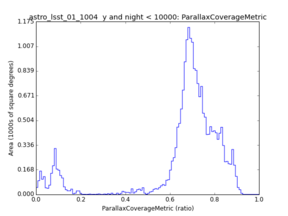}
  \includegraphics[width=2.0in]{./figs/milkyway/astromPanels/MW_Astrom_paCovge_wfdPlane_10y_hst.png}
  \end{center}
  \caption{Parallax coverage achieved for three extremes of object color, over the full 10-year survey. {\it Top and Third row:} \OpSim run \opsimdbref{db:baseCadence}. {\it Second and bottom row:} \OpSim run \opsimdbref{db:NormalGalacticPlane} (wide-fast-deep extended to much of the inner Plane). Reading left-right, columns represent: {\it Left column:} Objects detected only in the bluest filter; {\it Middle column:} objects detected only in the reddest filter; {\it Right column:} objects detected in all filters. Spatial maps are clipped at 95\%, with histogram horizontal limits (0.0 - 1.0).}
  \label{fig_astrom_ByFilter_PACoverage}
\end{figure}

\begin{figure}[ht]
  \begin{center}
  \includegraphics[width=2.0in]{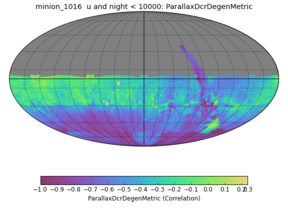}
  \includegraphics[width=2.0in]{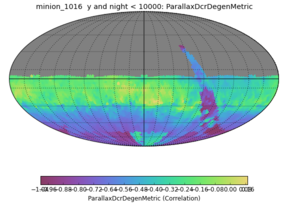}
  \includegraphics[width=2.0in]{./figs/milkyway/astromPanels/MW_Astrom_paDcrDegen_Baseline_10y_map.png}
  \end{center}
  \begin{center}
  \includegraphics[width=2.0in]{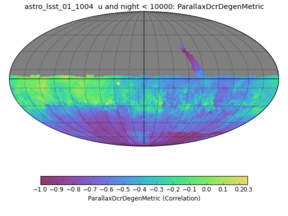}
  \includegraphics[width=2.0in]{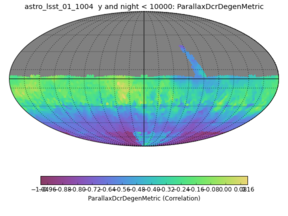}
  \includegraphics[width=2.0in]{./figs/milkyway/astromPanels/MW_Astrom_paDcrDegen_wfdPlane_10y_map.png}
  \end{center}

  \begin{center}
  \includegraphics[width=2.0in]{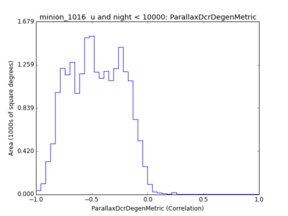}
  \includegraphics[width=2.0in]{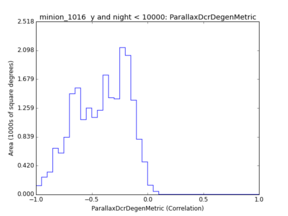}
  \includegraphics[width=2.0in]{./figs/milkyway/astromPanels/MW_Astrom_paDcrDegen_Baseline_10y_hst.png}
  \end{center}
  \begin{center}
  \includegraphics[width=2.0in]{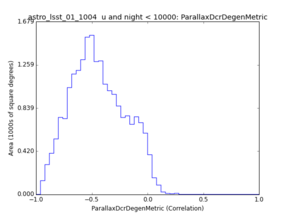}
  \includegraphics[width=2.0in]{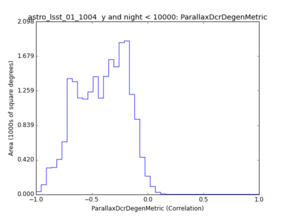}
  \includegraphics[width=2.0in]{./figs/milkyway/astromPanels/MW_Astrom_paDcrDegen_wfdPlane_10y_hst.png}
  \end{center}
  \caption{Correlation coefficient $\rho$~between parallax and Differential Chromatic Refraction (DCR), selecting filters for three extremes of object color, over the full 10-year survey. {\it Top and Third row:} \OpSim run \opsimdbref{db:baseCadence}. {\it Second and bottom row:} \OpSim run \opsimdbref{db:NormalGalacticPlane} (wide-fast-deep extended to much of the inner Plane). Reading left-right, columns represent: {\it Left column:} Objects detected only in the bluest filter; {\it Middle column:} objects detected only in the reddest filter; {\it Right column:} objects detected in all filters. Spatial maps are clipped at 95\%, with histogram horizontal scale set to the range $-1.0 \le \rho \le +1.0$.}
  \label{fig_astrom_ByFilter_PADegen}
\end{figure}

\begin{figure}[ht]
  \begin{center}
  \includegraphics[width=2.0in]{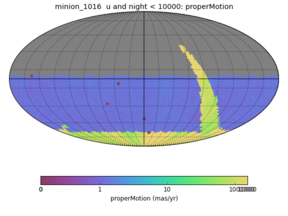}
  \includegraphics[width=2.0in]{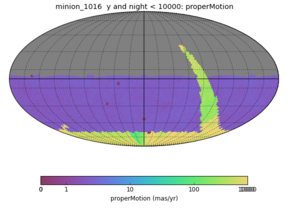}
  \includegraphics[width=2.0in]{./figs/milkyway/astromPanels/MW_Astrom_pmError_Baseline_10y_map.png}
  \end{center}
  \begin{center}
  \includegraphics[width=2.0in]{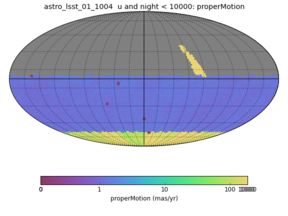}
  \includegraphics[width=2.0in]{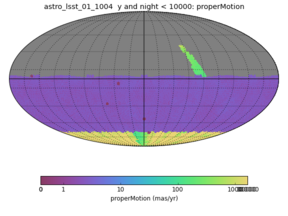}
  \includegraphics[width=2.0in]{./figs/milkyway/astromPanels/MW_Astrom_pmError_wfdPlane_10y_map.png}
  \end{center}

  \begin{center}
  \includegraphics[width=2.0in]{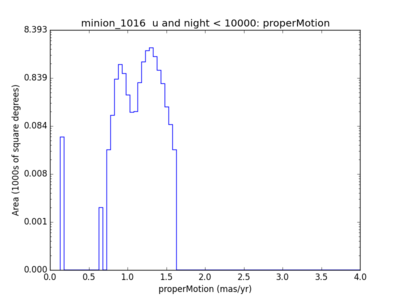}
  \includegraphics[width=2.0in]{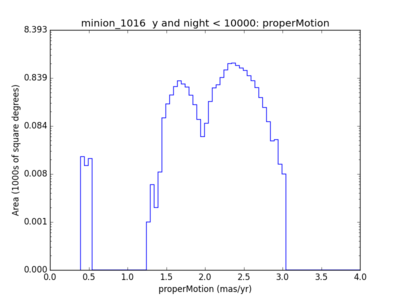}
  \includegraphics[width=2.0in]{./figs/milkyway/astromPanels/MW_Astrom_pmError_Baseline_10y_hst.png}
  \end{center}
  \begin{center}
  \includegraphics[width=2.0in]{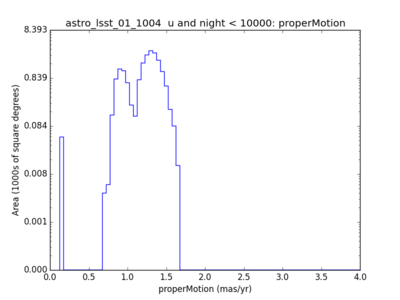}
  \includegraphics[width=2.0in]{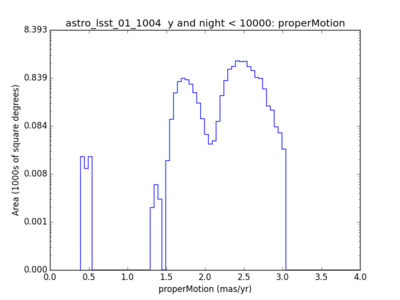}
  \includegraphics[width=2.0in]{./figs/milkyway/astromPanels/MW_Astrom_pmError_wfdPlane_10y_hst.png}
  \end{center}
  \caption{Proper motion error for a star at apparent magnitude $m=21.0$, for three extremes of object color and assessed over the full survey. Crowding errors are ignored. {\it Top and Third row:} \OpSim run \opsimdbref{db:baseCadence}. {\it Second and bottom row:} \OpSim run \opsimdbref{db:NormalGalacticPlane} (wide-fast-deep extended to much of the inner Plane). Reading left-right, columns represent: {\it Left column:} Objects detected only in the bluest filter; the fiducial object has apparent magnitude $u=21.0$; {\it Middle column:} objects detected only in the reddest filter (so $y = 21.0$); {\it Right column:} objects detected in all filters (using $r=21.0$~and a ``flat'' spectrum within \MAF). Spatial maps are clipped at 95\% and a log-scale is used for both the spatial maps and histograms. Reading left-right, the horizontal upper limits on the histograms are (4.0, 4.0, 3.0) mas yr$^{-1}$, respectively. Note that the histograms do not include the full range of values reported in the maps.}
  \label{fig_astrom_ByFilter_pmError}
\end{figure}

\begin{figure}[ht]
  \begin{center}
  \includegraphics[width=2.0in]{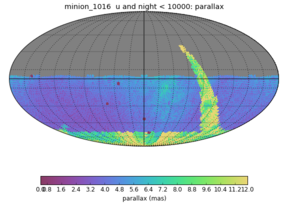}
  \includegraphics[width=2.0in]{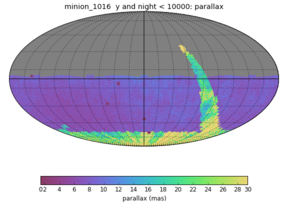}
  \includegraphics[width=2.0in]{./figs/milkyway/astromPanels/MW_Astrom_paError_Baseline_10y_map.png}
  \end{center}
  \begin{center}
  \includegraphics[width=2.0in]{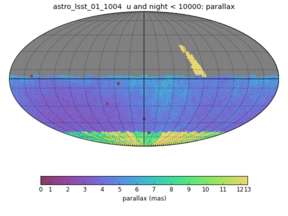}
  \includegraphics[width=2.0in]{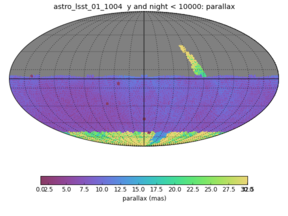}
  \includegraphics[width=2.0in]{./figs/milkyway/astromPanels/MW_Astrom_paError_wfdPlane_10y_map.png}
  \end{center}

  \begin{center}
  \includegraphics[width=2.0in]{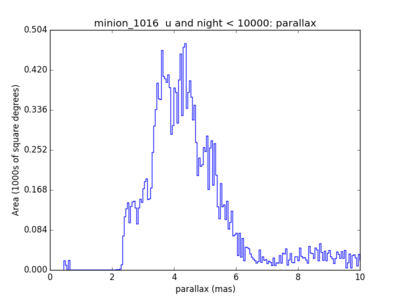}
  \includegraphics[width=2.0in]{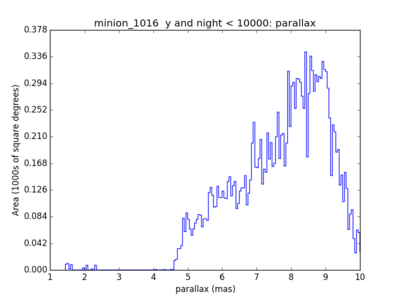}
  \includegraphics[width=2.0in]{./figs/milkyway/astromPanels/MW_Astrom_paError_Baseline_10y_hst.png}
  \end{center}
  \begin{center}
  \includegraphics[width=2.0in]{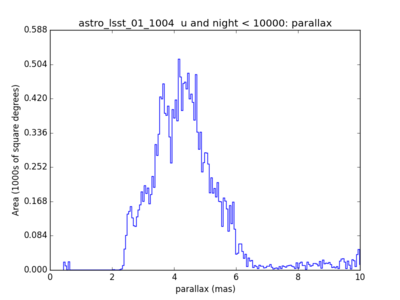}
  \includegraphics[width=2.0in]{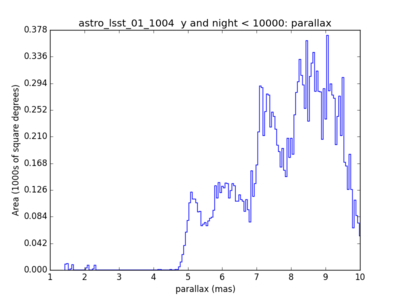}
  \includegraphics[width=2.0in]{./figs/milkyway/astromPanels/MW_Astrom_paError_wfdPlane_10y_hst.png}
  \end{center}
  \caption{Parallax error for a star at apparent magnitude $m=21.0$, for three extremes of object color and assessed over the full survey. Crowding errors are ignored. {\it Top and Third row:} \OpSim run \opsimdbref{db:baseCadence}. {\it Second and bottom row:} \OpSim run \opsimdbref{db:NormalGalacticPlane} (wide-fast-deep extended to much of the inner Plane). Reading left-right, columns represent: {\it Left column:} Objects detected only in the bluest filter; the fiducial object has apparent magnitude $u=21.0$; {\it Middle column:} objects detected only in the reddest filter (so $y = 21.0$); {\it Right column:} objects detected in all filters (using $r=21.0$~and a ``flat'' spectrum within \MAF). Spatial maps are clipped at 95\%. Reading left-right, the horizontal upper limits on the histograms are (10, 10, 2.0) mas, respectively. Note that the histograms do not include the full range of values reported in the maps.}
  \label{fig_astrom_ByFilter_paError}
\end{figure}

\subsubsection{Figures of Merit depending on the Metrics}

Building on the first-order metrics above, this subsection communicates scientific figures of merit for the cases identified in \autoref{sec:\secname:MW_Astrometry_measurements} above.

Table \ref{tab_SummaryMWAstrometry} summarizes the Figures of Merit
(FoMs) for Astrometry science cases. At the time of writing, FoMs have
been implemented to summarize the random uncertainty in proper motion
and parallax, for two regions experiencing extreme values of these
quantities: the inner Plane (conservatively defined in this section as
$|b| \lesssim 7^o$~and $|l| \lesssim 80^o$), and the main survey
(excluding the inner plane and the Southern Polar region, taken
here as $\delta_{2000.0} < -60.0^o$). Figure
\ref{fig_astrom_RegionSelKey} illustrates these selection-regions on
the sky. These form FoM 1.1-1.4, and have to-date been run for the
\OpSim runs \opsimdbref{db:baseCadence} (Baseline cadence),
\opsimdbref{db:opstwoPS} (similar to PanSTARRS-1), and the
recently-completed \opsimdbref{db:NormalGalacticPlane} (which applies
Wide-Fast-Deep cadence to much of the inner Galactic Plane).

From the point of view of parallax and proper motion, the latter two
strategies do not negatively impact the non-plane regions, but they
{\it substantially} improve the sampling for proper motions and
parallax (again, neglecting the effects of spatial crowding).

FoM 1.5 in Table \ref{tab_SummaryMWAstrometry} reports the total number of fields with Parallax/Hour-angle correlation $|\rho| < 0.7$.

At the time of writing, FoMs 2-5 in Table
\ref{tab_SummaryMWAstrometry} are still at the specification stage,
and are described in Section
\ref{sec:\secname:MW_Astrometry_furtherwork}.

%%%% Figures used as ``key'' for the astrometry FoMs:

\begin{figure}[h]
  \begin{center}
    \includegraphics[width=2.0in]{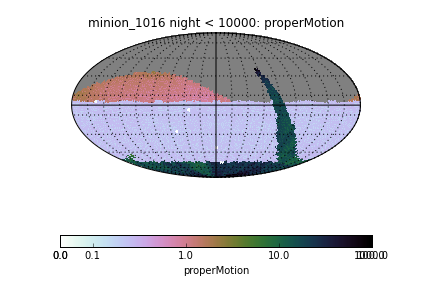}
  \includegraphics[width=2.0in]{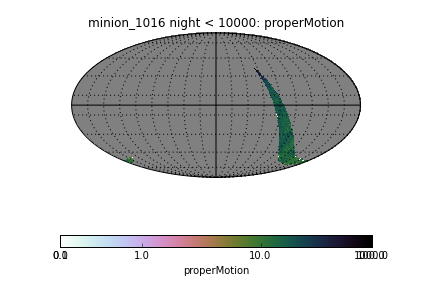}
  \includegraphics[width=2.0in]{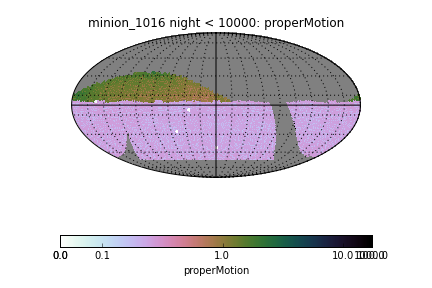}
    \end{center}
  \caption{Selection regions for the Astrometry Figures of Merit (FoMs) 1.1-1.4. Figures of Merit 1.1 and 1.3 refer to the ``main survey'' region shown in the middle panel (which for the FoM also avoids the region of the South Galactic Pole). The right panel shows the inner Plane region to which FoMs 1.2 \& 1.4 refer. The left-hand panel shows the entire survey region for reference. This example shows run \opsimdbref{db:baseCadence}. See Table \ref{tab_SummaryMWAstrometry} and Section \ref{sec:\secname:MW_Astrometry_metrics}.}
  \label{fig_astrom_RegionSelKey}
\end{figure}

\subsection{Topics that will need to be addressed}
\label{sec:\secname:MW_Astrometry_furtherwork}

Here we present suggestions for further work, first on figures of
merit for the science cases, and then on additional Metrics for LSST's
astrometric performance.

\subsubsection{Further work on science Figures of Merit}
%\medskip

At the time of writing, the Figures of Merit for both the highlighted
Science cases need to be implemented and applied to \OpSim output,
preferably in a format that can be summarized in a single Table in
this section. These figures of merit are discussed above in Section
\ref{sec:\secname:MW_Astrometry_metrics} (particularly for the Halo
Streams science project). Figures of merit for the two science cases
might be:
\begin{itemize}
  \item[1.] Number of streams that LSST can discover via their proper motions;
\item[2.] Uncertainty and bias in the thin and thick disk differential age measurement when using white dwarfs from each population as tracers.
\end{itemize}

Given the diversity of science cases that use local Solar Neighborhood
populations as tracers, it may be advantageous to subdivide the Solar
Neighborhood projects into further figures of merit. Two further example
figures of merit might then be:
\begin{itemize}
  \item[3.] Uncertainty and bias in the Brown Dwarf mass function using Solar Neighborhood tracers;
   \item[4.] Uncertainty and bias in the thickness in the main sequence of M-dwarfs within 25pc from the Sun, once variability has been characterized and removed.
\end{itemize}

%%%% SUMMARY TABLE FOR THIS SECTION

\begin{table}
  \begin{tabular}{l|p{4.8cm}|p{1.1cm}|p{1.1cm}|p{1.1cm}|c|p{3.5cm}}
    FoM & Brief description & {\rotatebox{90}{\opsimdbref{db:baseCadence} }} & {\rotatebox{90}{\opsimdbref{db:opstwoPS} }} & {\rotatebox{90}{\opsimdbref{db:NormalGalacticPlane}   }} &  {\rotatebox{90}{future run 2}} & Notes \\
    \hline
    1.1 & \footnotesize{Median parallax error at $r=21$ (main survey)}      & 0.69  & 0.72 & 0.69 & - &
%\footnotesize{Summarize the presentation in Figures \ref{fig_astrom_ByTime_PACoverage}-\ref{fig_astrom_ByFilter_paError} }
\footnotesize{See region definitions in Figure \ref{fig_astrom_RegionSelKey}.}
\\
    1.2. & \footnotesize{Median parallax error at $r=21$ (plane)}   & 2.68 & {\bf 0.91} & {\bf 0.89} & - &
\footnotesize{Smaller values are better.}\\
    1.3. & \footnotesize{Median proper motion error at $r=21$ (main survey)}  & 0.19 & 0.19 & 0.19 & - &
%\footnotesize{Take median of Figure \ref{fig_astrom_ByTime_pmError} over the ``plane'' region.}
\\
    1.4. & \footnotesize{Median proper motion error at $r=21$ (plane)} & 16.7
%$^\dagger$
& {\bf 0.26} & {\bf 0.25} & - &
%\footnotesize{$^\dagger$no, this is not a typo.}
\\
1.5. & \footnotesize{Fields with Parallax-DCR correlation coefficient $\rho \ge 0.7$~/ total fields} & \footnotesize{ \bf{3486} / \bf {31116} } & \footnotesize{3586 / 30107} & \footnotesize{3690 / 31116} & - & \footnotesize{Smaller is better. Value reported after full 10 years of survey for $griz$~detections.}  \\
    \hline
    2.1. & \footnotesize{Number of streams LSST can discover via proper motions} & - & - & - & - &  - \\
    3.1. & \footnotesize{Uncertainty and bias in thin- and thick-disk differential age measurement via white dwarfs} & - & - & - & - &  - \\
    4.1. & \footnotesize{Uncertainty and bias in brown dwarf mass function from the Solar Neighborhood}  & - & - & - & - & \footnotesize{Using astrometry metrics for objects detected only in the reddest filter(s)} \\
    4.2. & \footnotesize{Uncertainty and bias in white dwarf mass function from the Solar Neighborhood}  & - & - & - & - & \footnotesize{Using astrometry metrics for objects detected only in the bluest filter(s)} \\
    5.1. & \footnotesize{Uncertainty and bias in Solar Neighborhood M-dwarf thickness on the MS}  & - & - & - & - &  - \\
\end{tabular}
\caption{Summary of Figures of Merit for the Milky Way Astrometry science cases. The best value of each FoM is indicated in bold. Runs \opsimdbref{db:baseCadence} and \opsimdbref{db:opstwoPS} refer to the Baseline and PanSTARRS-like strategies, respectively. Column \opsimdbref{db:NormalGalacticPlane} refers to a recently-completed \OpSim run that includes the Plane in Wide-Fast-Deep observations. See \autoref{sec:MW_Astrometry}.}
\label{tab_SummaryMWAstrometry}
\end{table}

%%%% SUMMARY TABLE FINISHES HERE

\subsubsection{Further work on Astrometry Metrics}

The MAF metrics presented in Sections \ref{sec:\secname:MW_Astrometry_OpSim} and \ref{sec:\secname:MW_Astrometry_metrics} are only part of the
study of LSST's predicted astrometric performance.  Detailed simulations
and studies need to be done in many other areas as part of the
prediction and verification of LSST's astrometric performance.  Among
the most important are the following.
\begin{itemize}
\item How well do galaxies perform as astrometric reference objects? Are certain shapes or colors better than others? What is the
surface density of ``good" astrometric reference galaxies as a function of filter?
% 2017-06-03 commented out
%\item How well can we identify optically point-like QSOs that will be useful in matching the optical reference frame to the ICRS?
%\item Given the LSST exposure time, site, and physical characteristics, how can we mitigate the limitations on astrometric accuracy imposed by the seeing and local atmospheric turbulence?
\item How does the astrometric performance depend on stellar density? If there are fields in which photometry is only possible via difference imaging, what are the limitations
on astrometry in these fields?
%\item What tools do we need to compare the general and specific agreement between the {\it Gaia} results and the LSST results?
\item Does the ``brighter-wider" effect in the deep-depletion CCDs introduce a magnitude term into the centroid positions?

\item Deep-driling fields will likely be essential to fully understand LSST's astrometric behavior over the entire survey. How should Deep-Drilling fields be specified to fully characterize LSST's delivered astrometric performance?

\end{itemize}

% ====================================================================

\subsection{Conclusions}

Here we answer the ten questions posed in
\autoref{sec:intro:evaluation:caseConclusions}:

\begin{description}

\item[Q1:] {\it Does the science case place any constraints on the
tradeoff between the sky coverage and coadded depth? For example, should
the sky coverage be maximized (to $\sim$30,000 deg$^2$, as e.g., in
Pan-STARRS) or the number of detected galaxies (the current baseline
of 18,000 deg$^2$)?}

\item[A1:] We do expect tradeoffs between depth and sky
  coverage, but we do not yet have the FoM evaluations to set
  quantitative constraints. For example, we expect some combination of
  depth and survey volume would optimize the completeness to objects
  among the populations in the Solar Neighborhood. More generally,
  perhaps, in fields away from the galactic mid-plane, the
  lengthscales over which the proper motion zeropoints can be
  accurately constrained will depend on the spatial density of
  well-measured background galaxies (finer lengthscale corresponding
  to greater co-added depth). The depth must therefore be sufficient
  to sample enough of these galaxies to constrain variations of
  astrometric zeropoint on lengthscales at least as fine as those
  imposed by the LSST system itself (or the atmosphere, whichever is
  finer). We anticipate that this tradeoff can be informed by
  simulation under a set of assumptions for these variations.

\item[Q2:] {\it Does the science case place any constraints on the
tradeoff between uniformity of sampling and frequency of  sampling? For
example, a rolling cadence can provide enhanced sample rates over a part
of the survey or the entire survey for a designated time at the cost of
reduced sample rate the rest of the time (while maintaining the nominal
total visit counts).}

\item[A2:] Yes, although for astrometry the language of these
  constraints is slightly different. The parallactic ellipse must be
  sufficiently covered, the correlation between hour angle and
  parallax factor must be minimized, and the visits must be
  sufficiently distributed (both within a year and over the ten-year
  time baseline) to produce the best precision in both proper motion and parallax. See Section \ref{sec:MW_Astrometry:MW_Astrometry_metrics}.

\item[Q3:] {\it Does the science case place any constraints on the
tradeoff between the single-visit depth and the number of visits
(especially in the $u$-band where longer exposures would minimize the
impact of the readout noise)?}

\item[A3:] More visits at the standard exposure time are
  generally preferred to a few visits with longer exposures, in order
  to achieve as broad a temporal coverage as possible (see Section
  \ref{sec:MW_Astrometry:MW_Astrometry_metrics}). The $u$-band itself
  is likely to be of limited use for astrometry (except possibly for
  extremely blue objects with little signal in any of the other
  filters) due to differential chromatic refraction (DCR), however of
  course $u$-band will still be useful for photometric constraints.

\item[Q4:] {\it Does the science case place any constraints on the
Galactic plane coverage (spatial coverage, temporal sampling, visits per
band)?}

\item[A4:] Not for the example of detecting Galactic Halo
  streams via proper motions (Section
  \ref{sec:MW_Astrometry:MW_Astrometry_measurements}). For the Solar
  Neighborhood populations, avoiding the inner Galactic mid-plane
  would obviously reduce the completeness of the census of nearby
  objects with parallax determinations due to the reduction in total
  area surveyed. However, this reduction may be incremental rather
  than serious. The impact of an inner-plane zone of avoidance on the
  recovery of the parameters describing these constituent populations
  has not yet been evaluated. Of course, this all changes for
  astrometry of objects of interest that lie in the inner plane (see
  also Section \ref{sec:MW_Disk}), where (for example) the reduced
  proper motion will be a useful diagnostic. At present, however, the
  performance of the LSST software stack towards crowded fields is
  as-yet unknown - as is the performance of Gaia in these regions. As this performance becomes better understood, it
  will be possible to quantitatively compare strategies for astrometry
  towards the inner plane.

\item[Q5:] {\it Does the science case place any constraints on the
fraction of observing time allocated to each band?}

\item[A5:] Yes, but indirectly through the requirement to
  measure all the populations of interest in the Solar
  Neighborhood. Making the assumption that this science case requires
  parallax measurements for extremely blue objects as well as
  extremely red objects, which might each be measurable only in a
  single very red or blue filter, would suggest at a minimum that the
  coverage considerations of Section
  \ref{sec:MW_Astrometry:MW_Astrometry_metrics} be applied to
  observations in $u$~and $Y$ filters separately, as well as at least
  one mid-range filter. However the quantitative impact on population
  recovery from various filter-distributions has yet to be assessed at
  this date. Further work is needed to determine if the increased
  sensitivity of $u$-band astrometry to DCR relative to $g$-band would
  prevent its use for astrometry.

\item[Q6:] {\it Does the science case place any constraints on the
cadence for deep drilling fields?}

\item[A6:] The deep-drilling fields will provide essential characterization of LSST's delivered astrometric performance, providing at least a sanity check on the characterization of astrometric accuracy for {\it all} observed fields. The observational design of the deep-drilling fields needs to be carefully considered in terms of their utility for calibrating the non-deep-drilling fields, as well as in terms of maximizing the astrometric precision of the deep-drilling fields in their own right (for which the considerations of Section \ref{sec:MW_Astrometry:MW_Astrometry_metrics} will apply).

%If precision astrometry is desired for the deep drilling fields, then the considerations of Section \ref{sec:MW_Astrometry:MW_Astrometry_metrics} apply to those fields as well.

\item[Q7:] {\it Assuming two visits per night, would the science case
benefit if they are obtained in the same band or not?}

\item[A7:] While detailed investigation is still pending, we
  expect that using different filters within the same night would be
  preferred to allow better constraint of DCR effects. Doing different
  filters on the same night might reduce the number of free parameters
  (like seeing and parallax factor) and give more pairs for direct
  filter-A vs. filter-B astrometry.

\item[Q8:] {\it Will the case science benefit from a special cadence
prescription during commissioning or early in the survey, such as:
acquiring a full 10-year count of visits for a small area (either in all
the bands or in a  selected set); a greatly enhanced cadence for a small
area?}

\item[A8:] It is vital for astrometry that at least a few fields
  be observed with both sufficient parallax factor coverage and
  sufficient number of visits, early in the survey, to demonstrate
  parallax precision specified in the Science Requirements
  Document. In these fields, sufficient exposures must be reserved for
  the entire 10-year survey baseline so that proper motion precision
  is not too badly compromised in these fields. This combination of
  factors may require dedicated commissioning observations of these
  fields in addition to the 10-year survey operations. In addition,
  however, at least a few fields must be observed at a variety of
  values of single-visit achieved depth, and FWHM, in order to
  constrain the degree to which FWHM will actually predict the
  achieved astrometric precision (see also the answer to Q10
  below). This second set of requirements may also be best served by
  dedicated commissioning observations.

\item[Q9:] {\it Does the science case place any constraints on the
sampling of observing conditions (e.g., seeing, dark sky, airmass),
possibly as a function of band, etc.?}

\item[A9:] The observations need to be planned in such a way
  that the correlation between parallax and hour-angle is minimized,
  to avoid deneracies between the motion due to atmospheric refraction and the motion that is sought due to parallax. See
  Section \ref{sec:MW_Astrometry:MW_Astrometry_metrics}.

\item[Q10:] {\it Does the case have science drivers that would require
real-time exposure time optimization to obtain nearly constant
single-visit limiting depth?}

\item[A10:] While optimization on an exposure-to-exposure basis
  is perhaps unlikely, {\it selection} between observations in
  response to conditions (on a timescale of perhaps 10 minutes) will
  be crucial to maximize achieved astrometric precision. The rules by
  which this selection would proceed, still need to be charted. For
  example, while maintaining limiting depth might suggest shorter
  exposure times when the FWHM is narrow, this may not translate to
  improved astrometric error across an LSST chip, because the
  lengthscales of the turbulence driving the FWHM is not the same as
  that of the turbulence driving astrometric error across an LSST chip.

\end{description}

\navigationbar

% --------------------------------------------------------------------

% PJM: moved the following to FutureWork, while the metric(s) is/are being implemented
% \input{MilkyWay/MW_Halo.tex}

% Under development:
% \input{MilkyWay/MW_Bulge.tex}

% Under development:
% \input{MilkyWay/MW_LocalVolume}

% ====================================================================
%+
% SECTION:
%    MW_Halo.tex
%
% CHAPTER:
%    galaxy.tex
%
% ELEVATOR PITCH:
%
%-
% ====================================================================

\section{Mapping the Milky Way Halo}
%\subsection{Mapping the Milky Way Halo}
\def\secname{MW_Halo}\label{sec:\secname}

\credit{akvivas}, \credit{ctslater}, \credit{dnidever}

The study of the halo of the Milky Way is of the highest importance,
not only to understand the formation and early evolution of our own
galaxy, but also to test current models of hierarchical galaxy
formation. LSST will provide an unprecedented combination of area,
depth, wavelength range and long time-baseline for imaging data,
allowing detailed studies of the present-day structure of this old
Galactic component.  Here we
focus our attention on halo investigations using three tracer
populations. While we anticipate more cases will be developed and
compared between strategy choices, we have selected populations here
that illustrate many of the most important challenges.
%We focus here on three investigations of the Halo to be
%pursued with LSST data.
We describe the figures of merit (and the
diagnostic metrics on which they depend) that will allow quantitative
assessment of the impact of the choice of observing strategy on the
constraints LSST will afford.
%We expect more projects will join later.  % WIC - not sure what this meant.
We first briefly discuss the use of three main population tracers to
chart the halo population. More detail on each of these tracer
populations, and the general scientific motivations for studies of the
Milky Way halo with LSST, can be found in \citet{2009arXiv0912.0201L}.

{\it 1. RR Lyrae stars} have been known for several decades as
excellent tracers of the halo population. They are not only old stars
($>10$ Gyrs) but they are also excellent standard candles that allow
construction of three-dimensional maps. RR Lyrae stars have been used
to survey Milky Way halo populations extending out to
%The halo of the Milky Way has
%been now surveyed in a very large extension up to
$\sim 60-80$ kpc from the Galactic center \citep[][among
  others]{drake13a,drake13b,zinn14,torrealba15}. Beyond $\sim 80$~kpc,
the halo is mostly uncharted territory.

The RR Lyrae surveys suggest the halo is filled with substructures
(clumps of elevated stellar density) which are usually interpreted as
debris from destroyed satellite galaxies. This substructure overlies a
smooth component in the distribution of RR Lyrae stars, whose number density is well-described by a power law in galactocentric distance,
%which is well described
%with a power-law in the mean number density of RR Lyrae stars as a
%function of galactocentric distance,
steepening at radii $\gtrsim 30$ kpc \citep{zinn14}.  Thus, beyond
$\sim 60$ kpc, few field RR Lyrae stars are expected. However, we
presume that any RR Lyrae star beyond this distance may be part of
either debris material or distant low-luminosity satellite galaxies
% of low luminosity
that have been escaped detection until now \citep{sesar14,baker15}.
LCDM models predict debris as far as $0.5$~Mpc from the galactic
center. This is the territory that will be explored by LSST.

{\it 2. Red giant stars} can similarly be used to trace the structure
of the halo up to large distances. They have the advantage of being
bright and are numerous compared to the RR Lyrae stars but not as good
distance indicators.

{\it 3. Main sequence stars}, although less luminous than RR Lyraes or
Red Giants, are so much more numerous that statistical studies can be
pursued in a manner not generally possible for those populations.
%Fainter than these two tracers, main
%sequence stars stand up as a tool for studying the Halo. They are the
%most numerous type of stars available and statistical studies are
%possible.
Using the technique of photometric metallicities \citep{ivezic08}, the
Sloan Digital Sky Survey (SDSS) provided unprecedented maps of the
metallicity distribution up to $\sim 10$ kpc from the Galactic center,
unveiling not only the mean metallicity distribution of the halo but
also, sub-structures within the halo. LSST will extend these studies
all the way to the outermost parts of the Galaxy.
%This kind of works will be extended to the outermost parts of the
%Galaxy with LSST data.

% --------------------------------------------------------------------

\subsection{Target measurements and discoveries}
%\subsubsection{Target measurements and discoveries}
\label{sec:\secname:MW_Halo_targets}

Accurate measurement of these three tracer populations implies the following requirements:

%The three projects just described require the discovery and/or measurement of t%he following
%type of objects:

\begin{itemize}

\item[1.] RR Lyrae stars: These are bright horizontal-branch variable
  stars with periods between 0.2 to 1.0 days and large amplitudes,
  particularly in the bluer bandpasses (g amplitudes $0.5 -
  1.5$~mag). \citet{2012AJ....144....9O} made an intensive search for
  RR Lyrae stars in simulated LSST data and reached to the conclusion
  that this type of stars can be recovered to distances $\sim 600$
  kpc. A similar procedure can now be performed using MAF to directly
  compare LSST cadence scenarios to each other.
  %and current cadence scenarios.
  Chapter \ref{chp:variables} discusses the
  discovery metrics for variable stars including RR Lyrae
  stars. However, optimal recovery may involve more complex metrics
  involving the simultaneous use of multi-band time series
  \citep{vanderplas15,vivas16}. Besides recovery of variable stars,
  red-wavelength mean magnitudes $z$~and $y$ are particularly
  valuable since they provide the most accurate distance indicators.
  %valuable measurement to track for studies in the halo is the
  %infrared mean magnitudes z and y
\citep{caceres08}.

\item[2.] Main sequence stars: lacking any distinguishable variability, the
challenge in selecting a large and clean sample of main sequence stars comes
from tremendous number of small and nearly-unresolved galaxies present at
faint magnitudes. Precise star/galaxy separation is thus the limiting factor
on the useful depth of the main sequence sample. In addition to identifying
dwarfs, using dwarfs to map the metallicity distribution of the halo requires
precise $u$-band data, since it exhibits the strongest metallicity dependence of
the LSST filters.

\item[3.] Red Giants: due to their intrinsic luminosity, the Red Giants will sample
a far larger volume than main sequence stars at similar apparent magnitudes. However, they must first be identified and separated from the very numerous main
sequence stars present in the foreground. A gravity-sensitive photometric index can
be used for separating efficiently giants from dwarfs. The $u$-band magnitude is essential for such an index, so
%is
%an essential ingredient in this process, and
the behavior of the $u$-band limiting magnitude must therefore be charted under the various observational strategies under consideration.
% and it is necessary to follow-up
%the behavior of the u limiting magnitude under different observational
%strategies.
\autoref{fig-MW-giants} shows the distance that can be reached
by M-giants of different metallicities assuming limiting magnitude $u = 26.0$.
%-band limiting magnitude of 26.0..

\end{itemize}

\begin{figure}[h]
\begin{center}
  \includegraphics[scale=0.5]{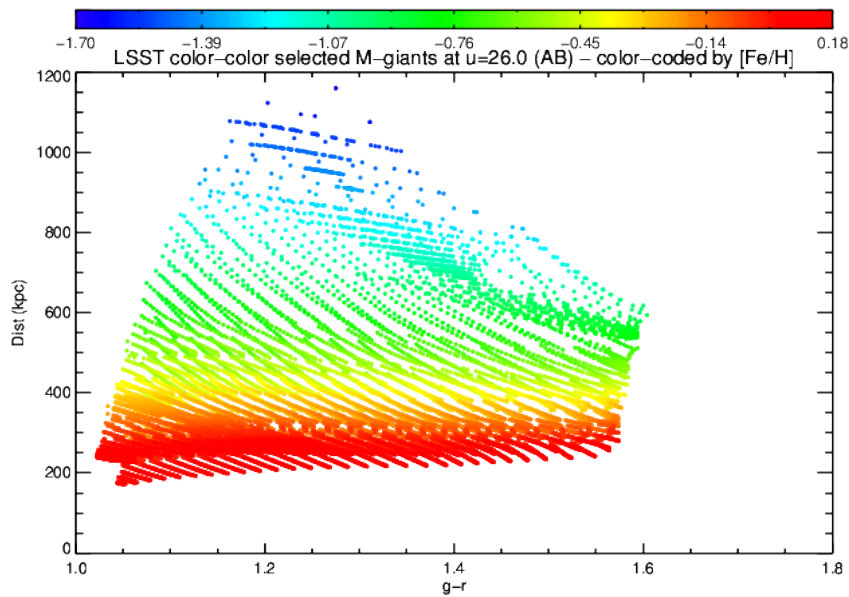}
  \caption{Distance to which red giant stars can be identified in the galactic halo assuming a limiting magnitude
  of u=26.0. The color code scales with the metallicity of the stars. More metal-poor stars can be
  detected to farther distances. \label{fig-MW-giants}}
\end{center}
\end{figure}

% --------------------------------------------------------------------

\subsection{Metrics}
% \subsubsection{Metrics}
\label{sec:\secname:MW_Halo_metrics}

\textbf{Star-Galaxy Separation:} For main sequence stars, the useful depth of
the survey will likely not be the photometric detection limit, but will instead
be set by the ability to differentiate stars from unresolved background
galaxies. Towards faint magnitudes the contamination by galaxies worsens
significantly for several reasons: the number of galaxies is rising
substantially, the angular size of galaxies is shrinking, and our ability to
distinguish stars from marginally resolved galaxies diminishes for faint
sources simply due to photon statistics. While the fundamental properties of
the contaminant sources are beyond our control, our ability to reject these
sources depends on survey parameters which do vary with the choice of observing strategy, such as the distribution of seeing across
visits and the depth of these visits.

We are currently in the process of developing a metric that will estimate our
ability to separate stars and galaxies for any observation depth and seeing
conditions. This requires both an understanding of how images of a source are
measured and classified as either a star or galaxy, and how the population of
stars and galaxies vary in number and size (for galaxies) with depth. Our model
uses the distribution of galaxies in size and number, derived from HST COSMOS
observations, along with a fully Bayesian model decision formalism to compute
the expected completeness and contamination in star-galaxy separation.
Computationally, for each position in the survey footprint we interpolate the
results from that work on a grid in seeing, galaxy size, and coadd depth, then
integrate over the distribution of galaxy sizes. This modeling process is
currently being verified against existing surveys, and will be incorporated into
the observing strategy study at a later date.

%This is a diagnostic metric and some of the higher level metrics
%described below will depend on it.
Some of the higher-level figures of merit described below will depend on this star/galaxy separation diagnostic metric.

\textbf{Distance to the farthest RR Lyrae stars:} This metric charts our ability to
recover an RR Lyrae star from LSST data as a function of its distance. An RR Lyrae star may be
considered as recovered if its period and amplitude are within 10\% of the intrinsic values.
The procedure followed by \citet{2012AJ....144....9O} is a good example on how this can be
achieved. They built a large number of synthetic light curves spanning the properties of
known RR Lyrae stars and ``observed'' them with the cadence given by the \OpSim runs
available at that time. Anticipated improvements over this previous work include the use
of simultaneous multi-band information to recover periods \citep[e.g.,][]{vanderplas15,vivas16}.

However, a first look into this problem using MAF can be achieved
by simplifying the procedure and only test if a star with period 0.55 days (the mean period for
RR Lyrae stars) can be recovered by metrics already available in MAF.
Then, distance can be calculated using the mean magnitude of the recovered RR Lyrae stars
(in the reddest bands available to LSST) and the interstellar extinction at that point of the sky (maps are available
now in MAF).  This metric should compute the largest distance that can be measured with a 10\% precision
at which certain percentage of RR
Lyrae stars (eg. 80\%) can be recovered by LSST. It is expected that the results
of this metric at low galactic latitudes will be largely dependent on the chosen observational
strategy (through variations in cadence towards the Plane).
%(how sparse will the cadence be  in the galactic plane).

A reasonable Figure of Merit for this sub-project is the volume of the
halo within RR Lyrae stars  can be recovered. Similarly, another Figure
of Merit would be the fraction of the Galactic thick disk's volume that
can be traced by RR Lyrae stars.

\textbf{Distance to the farthest main sequence stars and giant stars:}
Since variability is not the signal property for these tracer
populations, metrics are somewhat simpler than for the RR Lyrae.
%Being non-variable objects,
%the metrics for these objects are somewhat simpler than for the RR Lyrae. and
Here the distance metric requires the determination of the limiting $u$-band
magnitude
%(in u band)
for which galaxy/star separation is reliable to a certain level. In
these cases, distances depend on metallicity. Then, a figure of merit is
the volume of the halo mapped with stars within a specified metallicity
range.

\textbf{WFIRST Synergy:} The extended optical-IR baseline that will be
provided by WFIRST \citep{2015arXiv150303757S} in $\sim 2,000$ sq degrees of the sky at high galactic latitudes
will provide additional ways to improve both the star-galaxy separation \citep[e.g.][]{banerji15} and the disentangling of stellar tracers in the halo
of the Milky Way \citep[e.g.][]{dalcanton12}. Chapter~\ref{chp:wfirst} discusses some areas of interaction between both
surveys. Metrics related to stellar populations in the halo using the overlap between both surveys are 
still to be developed.

\subsection{Conclusions}

 Here we answer the ten questions posed in
 \autoref{sec:intro:evaluation:caseConclusions}:

 \begin{description}

 \item[Q1:] {\it Does the science case place any constraints on the
 tradeoff between the sky coverage and coadded depth? For example, should
 the sky coverage be maximized (to $\sim$30,000 deg$^2$, as e.g., in
 Pan-STARRS) or the number of detected galaxies (the current baseline 
 of 18,000 deg$^2$)?}

\item[A1:] Yes - depth is generally preferred over increased
  sky-coverage. Fields with insufficient depth for star-galaxy
  separation to the required level of reliability will not be useful
  for the science in this section. While we expect this will lead to
  preference of achieving minimum depth vs expanding the sky-coverage,
  at this date we do not yet have quantitative limits.

 \item[Q2:] {\it Does the science case place any constraints on the
 tradeoff between uniformity of sampling and frequency of  sampling? For
 example, a rolling cadence can provide enhanced sample rates over a part
 of the survey or the entire survey for a designated time at the cost of
 reduced sample rate the rest of the time (while maintaining the nominal
 total visit counts).}

\item[A2:] Uniformity is generally preferred over sampling frequency. Coverage over the full ten-year survey must be maintained at sufficient cadence to be sensitive to RR Lyrae and $\delta$~Scuti populations, which are identified by variability. For the other tracers (where co-added depth is the key requirement), sampling considerations are less important.

 \item[Q3:] {\it Does the science case place any constraints on the
 tradeoff between the single-visit depth and the number of visits
 (especially in the $u$-band where longer exposures would minimize the
 impact of the readout noise)?}

\item[A3:] Single-visit depth is not a strong requirement for Halo science. Depth in $u$-band is dependent on the choice of tracer. For example, selection of red giants depends on accurate $u$-band photometry, which is not the case for turnoff stars (Section \ref{sec:MW_Halo:MW_Halo_targets}).

 \item[Q4:] {\it Does the science case place any constraints on the
 Galactic plane coverage (spatial coverage, temporal sampling, visits per
 band)?}

\item[A4:] No.

 \item[Q5:] {\it Does the science case place any constraints on the
 fraction of observing time allocated to each band?}

 \item[A5:] Yes, depending on the tracer population (Section \ref{sec:MW_Halo:MW_Halo_targets}).

 \item[Q6:] {\it Does the science case place any constraints on the
 cadence for deep drilling fields?}

\item[A6:] Not strongly. However, achieving huge co-added depth in fields containing Halo structure would greatly aid the understanding of observations of Halo structure throughout the main survey region.

 \item[Q7:] {\it Assuming two visits per night, would the science case
 benefit if they are obtained in the same band or not?}

\item[A7:] Halo science does not constrain the intra-night filter choice for a given field.

 \item[Q8:] {\it Will the case science benefit from a special cadence
 prescription during commissioning or early in the survey, such as:
 acquiring a full 10-year count of visits for a small area (either in all
 the bands or in a  selected set); a greatly enhanced cadence for a small
 area?}

\item[A8:] It would be good to observe at least a few fields towards known Halo structure, with a high fraction of observations in the first year or so, to best characterize LSST's performance for Halo science and make predictions for the rest of the survey.

 \item[Q9:] {\it Does the science case place any constraints on the
 sampling of observing conditions (e.g., seeing, dark sky, airmass),
 possibly as a function of band, etc.?}

\item[A9:] Generally best-seeing is preferred to enable star/galaxy separation. We do not at this date have quantitiative evaluations of a Figure of Merit for this, however.

 \item[Q10:] {\it Does the case have science drivers that would require
 real-time exposure time optimization to obtain nearly constant
 single-visit limiting depth?}

\item[A10:] No.

\end{description}

% ====================================================================

\navigationbar

% ====================================================================
%+
% SECTION:
%    MW_FutureWork.tex
%
% CHAPTER:
%    galaxy.tex
%
% ELEVATOR PITCH:
%    Ideas for future metric investigation, with quantitaive analysis
%    still pending.
%-
% ====================================================================

\section{Future Work}
\def\secname{MW_future}\label{sec:\secname}

% ====================================================================

% \input{MilkyWay/MW_Dust.tex}

% ====================================================================

% WIC - promoted this back to MW Halo section

% \input{MilkyWay/MW_Halo.tex}

% ====================================================================

% \subsection{Other Ideas}

\credit{willclarkson}, \credit{akvivas}, \credit{vpdebattista}

In this final section we provide an extremely brief list of important science
cases that are still in an early stage of development, but that are
deserving of quantitative \MAF analysis in the future.

\subsection{Further considerations for Milky Way static science}

  One important area of Milky Way science on which further
  community input is still sorely needed, is {\it static science} (a
  category that includes population disentanglement through deep,
  multicolor photometry), particularly in regions outside the main
  ``Wide-Fast-Deep'' (WFD) survey (Sections \ref{sec:MW_Astrometry} \&
  \ref{sec:MW_Halo} include discussion of static science in WFD regions). Since
  static science depends on depth (for, e.g., precise colors near the
  main sequence turn-off of some population) and uniformity over the
  survey (to aid characterization of strong selection functions),
  static science observing requirements may be in tension with (or at
  least not explicitly addressed by) requirements communicated
  elsewhere in this chapter.

  For example, probing deep within spatially crowded
  populations may lead to a sharper requirement on the selection of
  observations in good seeing conditions towards crowded regions, than
  has been apparent to-date. This needs quantification. 

  To pick another example, while we have indicated that co-added
  depth is a lower priority than temporal coverage for
  variability-driven studies in the Galactic Disk (conclusion A.1 in
  Section \ref{sec:MW_Disk}), co-added depth will likely be crucial for
  population disentanglement through photometry. While a
  judiciously-chosen observing strategy should be able to support both
  static and variable science, at this date quantitative trade-offs have not yet been specified.

  In many cases the implementation of figures of merit for static
  science in the Milky Way is complicated by the requirement to
  interface custom population simulations with the observational
  characterizations produced by the \MAF framework. For many
  investigators, the preferred method may be to use \MAF to produce
  parameterizations of the observational quantities of interest - for
  example, the run of photometric uncertainty against apparent
  magnitude, for each location on the sky, and including spatial
  confusion (all of which \MAF can currently produce\footnote{See the tutorials at \url{https://github.com/LSST-nonproject/sims_maf_contrib} for more information.}) - and then use
  these characteristics as input to their own population simulations,
  on which the investigator may have invested substantial time and
  effort.

  To provoke progress, we specify in Table
  \ref{table:strawmanMWstaticScience} a possible Figure of Merit for
  static science in terms of capabilities mostly already provided by
  the \MAF framework, which does not require custom simulation. This
  Figure of Merit - which asks what fraction of fields in a spatial
  region of interest, are sufficiently well-observed to permit
  population disentanglement to some desired level of precision -
  could form the basis of several science FoM's (for example, the
  fraction of fields in which photometric age determinations of
  bulge/bar populations might be attempted). We encourage community
  development and implementation of this and other FoMs for Milky Way
  static science.

\begin{table}[h]
  \small
  \begin{tabular}{c p{12cm}}
    & {\it FoM innerMW-Static: fraction of fields in inner Galactic plane adequately covered for population discrimination} \\
    \hline
    1. & Produce absolute magnitudes $M_{u,g,r,i,z,y}$~of the population of interest, using \MAF's spectral libraries; \\
    2. & For each HEALPIX (i.e. pointing):\\
       & 2.1. Place the fiducial star at appropriate line-of-sight distance, produce apparent magnitudes $m_{u,g,r,i,z,y}$;\\
       & 2.2. Modify $m_{u,g,r,i,z,y}$~for extinction using \MAF's extinction model;\\
    & 2.3. Compute the photometric and astrometric uncertainties due to sampling and random error (from \MAF's ``m52snr'' method);\\
    & 2.4. Convert exposure-by-exposure estimates of photometric and astrometric uncertainty due to spatial confusion, into co-added uncertainties;\\
    & 2.5. Combine the random and confusion uncertainties into final measurement uncertainties on the photometry and astrometry;\\
    & 2.6. Use uncertainty propagation to estimate color uncertainties $u-g, g-r, r-i, i-z$;\\
    3. & Count the fraction of sight-lines for which the color below the threshold needed for "sufficient" accuracy in parameter determination \citep{ivezic08}. {\bf This is the figure of merit.} \\
\hline
    \end{tabular}
 \caption{Description of Figure of merit ``innerMW-Static''}
  \label{table:strawmanMWstaticScience}
\end{table}

Finally, we provide an extremely brief list of important science
cases that are still in an early stage of development, which fall into the ``Static science'' category of Milky Way science:
\begin{itemize}
  \item {\it Formation history of the Bulge and present-day balance of
  populations:} Sensitivity to metallicity and age distribution of Bulge
  objects near the Main Sequence Turn-off;
  \item {\it Migration and heating in the Milky Way disk:} Error and
  bias in the determination of components in the (velocity dispersion vs
  metallicity) diagram, for disk populations along various lines of
  sight \citep[e.g.][]{2016ApJ...818L...6L}.

% WIC 2017-05-24: commented this out in favor of a dedicated local volume subsection.
%
%\item Fraction of Local-Volume objects discovered as a function of
%  survey strategy.
\end{itemize}

\subsection{Short exposures}

  Populations near, or brighter than, LSST's nominal saturation
  limit ($r \sim 16$~with 15s exposures) are likely to be crucial to a
  number of investigations for Milky Way investigations, whether as
  science tracers in their own right, or as contaminants that might
  interfere with measurements of fainter program objects (due, for
  example, to charge bleeds of bright, foreground disk objects).

  Quantitative exploration of these issues now requires involvement from
  the community (e.g. to determine LSST's discovery space for bright
  tracers in context with other facilities and surveys like ZTF, Gaia
  and VVV), and the project (e.g. to determine the parameters of short
  exposures that might be supported by the facility). To provoke
  development, here we list a few example questions regarding short
  exposures that still need resolution:

\begin{itemize}
  \item{What level of bright-object charge-bleeding can be tolerated? For example, is some minimum distribution of position-angles required in order to spread the bleeds azimuthally over the set of observations of a particular line of sight (so that different pixels fall under bleeds in different exposures)?\footnote{Note that with $\sim 30$~exposures per filter per field over ten years towards some regions, charge bleeds on the detector for a given object might not be spread over a large range of orientations. The spread of charge bleed angles on the detector - as a spatially-varying metric - should be implemented and included in figures of merit.} What is this minimum?}
    \item{What is the science impact of restricting targets to $r \gtrsim 16$?}
      \item{Will the proposed twilight survey of short exposures be adequate for science cases requiring short exposures? Are there enough observations from other surveys (e.g. DES or other DECam surveys) to cover the bright end of the entire LSST footprint, and are {\it new} observations of very bright objects required scientifically in any case?}
        \item{What is the bright limit required for adequate astrometric cross-calibration against Gaia?}
          \item{To what extent would a given bright cutoff hamper the combination of LSST photometry or astrometry with that from other surveys (like VVV)?}
          \item{How short an exposure time can the facility support? Does the OpSim framework already include all the operational limitations on scheduling short exposures?}
\end{itemize}

\subsection{The Local Volume}

  Finally, we remind the reader that substantial opportunity
  remains to develop science figures of merit for {\it Local Volume}
  science cases. Figures of Merit for Local Volume science likely
  share much common ground with those for the Halo (discussed in
  Section \ref{sec:MW_Halo}), and substantial prior expertise exists
  \citep[e.g.][]{2014ApJ...795L..13H}. One straightforward Figure of
  Merit (FoM) might be the fraction of dwarf galaxies in the Local
  Volume that are correctly identified as a function of survey
  strategy.

% ====================================================================

\navigationbar

% --------------------------------------------------------------------

% --------------------------------------------------------------------

\chapter[Variable Objects]{Variable Objects}
\def\chpname{variables}\label{chp:\chpname}

Chapter editors:
\credit{AshishMahabal},
\credit{lmwalkowicz}.

Contributing authors:
\credit{lundmb},
\credit{StephenRidgway},
\credit{keatonb},
\credit{phartigan},
\credit{CJohnsKrull},
\credit{pmmcgehee},
\credit{ShashiKanbur}

% \section*{Summary}
% \addcontentsline{toc}{section}{~~~~~~~~~Summary}
%
% Executive summary goes here, highlighting the primary conclusions from
% the chapter's science cases. This should be abstract length, no more:
% say, 200 words.

% --------------------------------------------------------------------

\section{Introduction}

Variable objects are defined as those that exhibit brightness changes,
either periodic or non-periodic, which are detected in quiescence and
non-destructive to the object itself. Variable objects span a wide range
in timescale-of-interest (sometimes even within a single class of
objects), and so different science cases benefit from different sampling
strategies. These strategies may be significantly disparate from one
another, sometimes even mutually exclusive; competing objectives
described in this chapter and the next are therefore at the heart of
LSST observing strategy and cadence design.

Below we develop a number of key science cases for LSST studies of
variable objects, associating them with related metrics that can be used
within the Metrics Analysis Framework (MAF) to understand the impact of
a given survey strategy realization on the scientific results for that
case. The science cases outlined are by no means exhaustive, but rather
are motivated by providing key quantitative examples of LSST's
performance given any particular deployment of survey strategy. The
authors encourage community contribution of similar cases, where the
scientific outcome can be quantified using specific metrics.

%When evaluating a particular observation or series of observations in
%light of how they perform for a specific science case, it may be
%helpful to think of metrics as lying along a continuum between
%discovery and characterization. Discovery requires a minimum amount of
%information to recognize an event or object as a candidate of
%interest, which necessarily involves some level of bare-bones
%characterization (upon which said recognition is based); rich
%characterization, on the other hand, implies that an event may not
%only be recognized as a candidate of interest, but basic properties of
%the event or object may be determined from the observation (e.g.
%including but not limited to classification of the event). The
%interpretation of a given metric along this continuum has implications
%for the subsequent action and analysis required, particularly as
%regards possible follow-up observations with other facilities.

%Target types are here grouped in subsections by variability
%characteristics, but as will be seen, this does not mean that all
%targets in a group require a common cadence, since the times scales
%may vary dramatically.  Acquiring suitable data for a wide range of
%time scales presents a fundamental problem for LSST, since the
%available $~$800 visits to a field over the survey cannot be deployed
%so as to usefully sample all time scales at all times.  This fact
%leads to the concept of a non-uniform survey, in which parts of the
%sky are visited more frequently part of the time.  The merits of such
%options must be traded against the benefits of a more uniform survey
%strategy.

\begin{center}
\begin{tabular}{| l | p{8cm} |l | l |}
\hline Periodic Variable Type & Examples of target science & Amplitude & Timescale\\
\hline
RR Lyrae & Galactic structure, distance ladder, RR Lyrae properties&  large &  day \\
Cepheids & Distance ladder, cepheid properties&  large &  day \\
Long Period Variables & Distance ladder, LPV properties & large  &  weeks \\
Short period pulsators & Instability strip, white dwarf interior properties, evolution&  small & min  \\
Periodic binaries & Eclipses, physical properties of stars, distances, ages, evolution, apsidal precession, mass transfer induced period changes, Applegate effect &  small &  hr-day \\
Rotational Modulation & Gyrochronology, stellar activity & small  &  days \\
Young stellar populations & Star and planet formation, accretion physics & small  &  min-days \\
 \hline \end{tabular}
 \end{center}

 \navigationbar

% --------------------------------------------------------------------

% ====================================================================
%+
% SECTION:
%    cepheids.tex
%
% CHAPTER:
%    variables.tex
%
% ELEVATOR PITCH:
%
%-
% ====================================================================

\section{The Cepheid Mass-Luminosity Relation}
\def\secname{cepheids}\label{sec:\secname}

Classical Cepheids begin to pulsate once they evolve up the giant branch and
execute blueward loops on the HR diagram that take them into the Instability
Strip. Cepheid masses and luminosities in the instability strip are connected
through the Mass-Luminosity (ML) relation.  This ML relation is strongly
dependent on stellar evolution physics. Canonical/Non-canonical ML relations
arise from varying treatments of convective core overshoot and mass loss
\citep{1992A&A...258..397B,2000ApJ...529..293B,2013ApJ...768L...6M}.

Stellar pulsation models adopt a given ML relation and then compute a
theoretical light curve for a range of different temperatures and
metallicities. This theoretical light curve can then be transformed into LSST
wavebands using stellar atmospheres (\citep{Bono2000} and references therein) and
quantitatively compared to observed LSST light curves through Fourier
decomposition of the form
$$V = A_0 + \sum_{k=1}^{k=N}A_k cos(k\omega t + {\phi}_k),$$
where $\omega = 2\pi/P,$ with $P$ the period, $N$ is the order of the fit. The
coefficients $R_{k1}=A_k/A_1$ and ${\phi}_{k1}={\phi}_k - k{\phi}_1$ can be
computed for both theoretical and LSST observed light curves. These
coefficients are sensitive to the adopted $ML$ relation and other global
parameters such as metallicity and effective temperature.  By utilizing such a
decomposition, the multiwavelength light curves that the LSST will produce for
both Cepheids and RR Lyraes can rigorously constrain Cepheid and RR Lyrae
global stellar parameters such as the ML relation. Of course, knowledge of the
ML relation through this approach can then lead to another ``theoretical''
distance scale using both Cepheids and RR Lyraes.  However, in the case of
Cepheids, given good enough cadence in LSST bands and thus an accurate Fourier
decomposition with precisely known Fourier parameters, it will be possible to
discriminate between canonical and non-canonical ML relations and thus provide
constraints for stellar evolution physics. {\it A good Figure of Merit for
this science case would be one based on the precision with which
we can infer the parameters of these relations.}

\citet{2014MNRAS.445.2655B} describes in detail the way the quantitative structure of
Cepheid and RR Lyrae light curves vary with period and optical band. Given
appropriate cadence, LSST light curves will provide accurate Fourier
decompositions at multiple wavelengths that can significantly augment these
results and provide an important database with which to connect quantitative
aspects of Cepheid and RR Lyrae light curve structure to pulsation envelope
physics. Two examples are the following.
\begin{itemize}
\item{1)} RRab stars found in stripe 82 of the
SDSS exhibit a flat Period-Color (PC) relation at minimum light at certain SDSS colors but not at others. LSST observations of RRab stars will be able
to augment this result and investigate if there are any links to the structural properties of observed light curves.
\item{2)} Short period ($\log P \le 0.4$) FU Cepheids in the SMC exhibit a noticeable break in their $(V-I)$ PC relation at certain phases of pulsation
phases. At the same period, the Fourier parameter $R_{21}$ displays a strong turnover. LSST data will be important in seeing if this result extends
to LSST colors.
\end{itemize}

\citet[and references therein]{2014MNRAS.445.2655B,2015MNRAS.447.3342B} have
demonstrated how Cepheid and RR Lyrae
Period-Color(PC)/Period-Luminosity (PL)/
Period-Wesenheit(PW)/Period-Luminosity-Color(PLC) relations vary significantly
both as a function pulsation phase and period and observation band.  The LSST
database on Cepheids and RR Lyraes will provide an excellent database to further
investigate the variation of these relations with pulsation phase with a view
to understanding pulsation physics and constraining theoeretical models.
Currently, the literature only discusses these relations at mean light, that
is the average over pulsation phase. Yet this averaging process masks
some dependencies: there are pulsation phases with very high/low PL/PC
dispersion and there are some phases where the relation is highly nonlinear.
In the era of precision cosmology, it is important to understand the tools that
we use to construct a distance scale. The LSST database on Cepheids and RR
Lyraes will be an important database with which to investigate the multiphase
properties of PL/PC/PW/PLC relations.

What cadence is required for ``accurate multiwavelength Fourier
decomposition''? Cepheids and RR Lyraes are strictly periodic. Then for known variables, whose periods are already known, such
Fourier decompositions can be carried out on folded light curves. In such
situations given a cadence, one possible measure of the success of this Fourier
decomposition could be the maximum phase gap in the folded light curve. Another
measure could be the error on the Fourier parameters
\citep{1986A&A...170...59P}. One way forward is to carry out a detailed
Fourier analysis of any one simulated schedule.

\subsection{Description of Relevant Metrics}
\label{sec:\chpname:variablemetrics}

Despite the range in scientific motivation for the cases presented
in this chapter,
there are some common metrics that are widely applicable (or may
be combined in a variety of ways with other metrics to suit a variety of
applications).
\new{The present science case provides as good a venue as
any in which to introduce these common metrics.}

%\subsection{Metrics}
%\label{sec:\chpname:metrics}

\begin{center}
\begin{tabular}{| p{5cm} |p{10cm} |}
\hline Metric & Description\\
\hline
Eclipsing/transiting system discovery & Fraction of discoveries vs fractional duration of eclipse\\
Lightcurve shape recovery & ... \\
%Transiting exoplanets (depth dependent) & Fraction of discoveries vs fractional duration of eclipse\\
Phase gap & Histogram vs period of the median and maximum phase gaps achieved in all fields\\
Period determination (period dependent) & Fraction of targets vs survey duration, for which the period can be determined to 5-sigma confidence\\
Period variability (period dependent) & Fraction of targets vs survey duration, for which a period change of 1\% can be determined with 5-sigma confidence\\
  \hline \end{tabular}
 \end{center}

The ability to identify that an object is periodic, and to correctly
determine that object's period, are widely applicable measures of
discovery. In the case of regular variables (as outlined below), these
two measures together can uniquely identify a population. Other kinds of
periodic systems (transiting planets for example) also require a
measurement of periodicity, but have a much wider range of relevant
periods, and looser requirements on the strictness of that periodicity.

\citet{LundEtal2016} discuss three {\it diagnostic} metrics that have been
incorporated into the MAF. Two of these metrics deal explicitly with time
variable behavior: a) observational triplets, and b) detection of periodic
variability.

%%%%%%%%%%%%%%%%%%%%%%%%%%%%%%%%%
\begin{figure}[tbh!]
%\vskip -4.1in
%\hskip -0.5in
%\includegraphics[angle=0,width=1.19\hsize:,clip]{figs/enigma1189_earlySNe.pdf}
%\vskip -4.0in
\centering\includegraphics[width=0.95\linewidth]{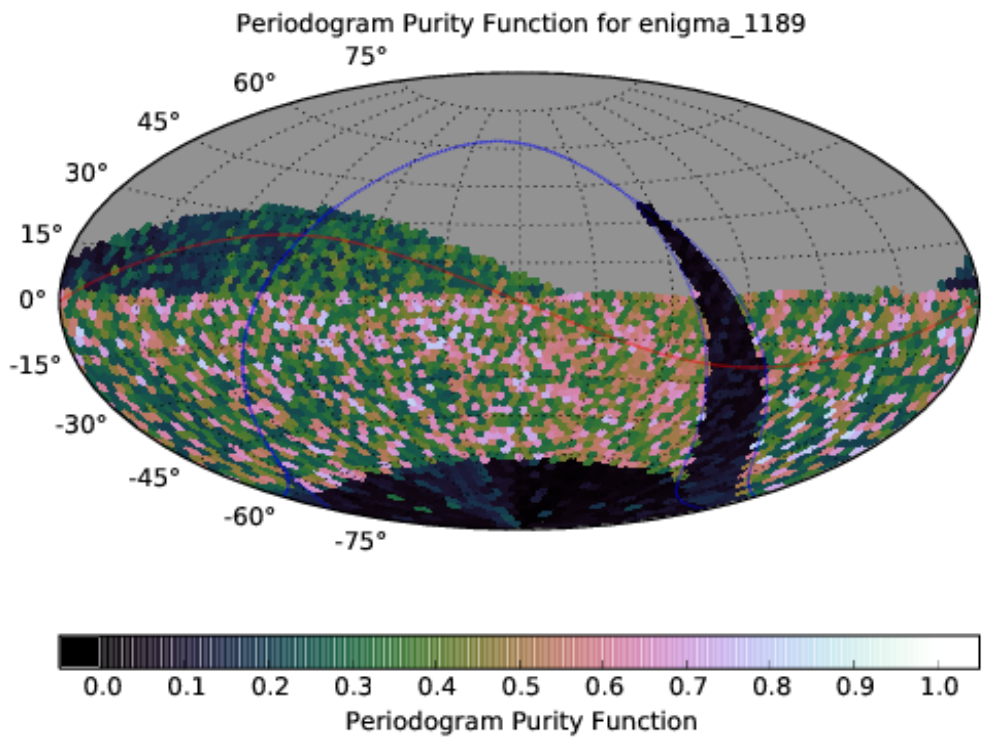}
\caption{The value for the Periodogram Purity Function for candidate
Baseline Cadence \opsimdbref{db:baseCadence}. The Periodogram Purity
Function provides a measure of the power lost due to aliasing.}
\label{fig:enigmaPeriodogramPurity}
\end{figure}
%%%%%%%%%%%%%%%%%%%%%%%%%%%%%%%%%

%%%%%%%%%%%%%%%%%%%%%%%%%%%%%%%%%
\begin{figure}[tbh!]
%\vskip -4.1in
%\hskip -0.5in
%\includegraphics[angle=0,width=1.19\hsize:,clip]{figs/enigma1189_earlySNe.pdf}
%\vskip -4.0in
\centering\includegraphics[width=0.95\linewidth]{figs/variables/enigma_1189_Phase_Gap_MedianGap_OPSI_SkyMap.pdf}
\caption{The median phase gap for candidate Baseline Cadence \opsimdbref{db:baseCadence}.
The \MAF \MAFmetric{PhaseGapMetric} looks at periods between 3 and 35 days by default.}
\label{fig:enigmaMedianGap}
\end{figure}
%%%%%%%%%%%%%%%%%%%%%%%%%%%%%%%%%

%%%%%%%%%%%%%%%%%%%%%%%%%%%%%%%%%
\begin{figure}[tbh!]
%\vskip -4.1in
%\hskip -0.5in
%\includegraphics[angle=0,width=1.19\hsize:,clip]{figs/enigma1189_earlySNe.pdf}
%\vskip -4.0in
\centering\includegraphics[width=0.95\linewidth]{figs/variables/enigma_1189_Phase_Gap_LargestGap_OPSI_SkyMap.pdf}
\caption{The maximum phase gap for candidate Baseline Cadence \opsimdbref{db:baseCadence}.
The \MAFmetric{PhaseGapMetric} looks at periods between 3 and 35 days by default.}
\label{fig:enigmaMaxGap}
\end{figure}
%%%%%%%%%%%%%%%%%%%%%%%%%%%%%%%%%

\subsubsection{Periodogram purity function (PeriodicMetric)}

This metric calculates the Fourier power spectral window function of each field
\citep{1987AJ.....93..968R} as a means of quantifying the completeness of phase
coverage for a given periodic variable. The periodogram purity is defined as 1
minus the Fourier power spectral window function. in the perfect case, all
power in the window function is concentrated in a delta function at zero, and
is zero at all other frequencies. As power ``leaks'' away from the correct
frequency as a consequence of discrete, non-ideal data sampling, the
periodogram becomes more structured. For the purposes of MAF metrics, which are
designed to quantify performance as a single number, the periodogram purity is
quantified as the minimum value of the periodogram purity function at non-zero
frequency shifts; the ideal case would be a periodogram purity metric value of
1.

\subsubsection{Phase Gap Metric \MAFmetric{PhaseGapMetric}}

The Phase Gap Metric is designed to examine the largest phase gaps in
the observing schedule. For a given point in the sky, a series of
periods are randomly selected (by default, 5 periods), with a default
minimum of 3 days and maximum of 35 days. The largest phase gap for each
period is calculated, and the metric plots the median
(\autoref{fig:enigmaMedianGap}) and maximum (\autoref{fig:enigmaMaxGap})
of this subset of values that contains the maximum phase gap per period.
The Phase Gap Metric is part of varMetrics.

\subsubsection{Period Deviation Metric \MAFmetric{PeriodDeviationMetric}}

The Period Deviation Metric calculates the error in recovering the
correct period of a sinusoid using a given observing schedule and a
Lomb-Scargle periodogram. For a given point in the sky, a series of
periods are randomly selected (by default, 5 periods), and the metric
returns the worst period deviation, and the period at which this
occurred. The Period Deviation Metric is part of varMetrics.

%\subsection{Proposed Metrics}

%The following is a raw list of metric ideas; these need specificity and further description.

%FWHM of the window function (to quantify sampling)

%Maximum hour angle difference

%Fraction of discoveries vs fractional duration of eclipse

%Fraction of targets vs survey duration, for which the period can be determined to 5-sigma confidence

%Fraction of targets vs survey duration, for which a period change of 1$\%$ can be determined with 5-sigma confidence

% ====================================================================
%
% \subsection{OpSim Analysis}
%
% ====================================================================
%
 \subsection{Conclusions}

 Here we answer the ten questions posed in
 \autoref{sec:intro:evaluation:caseConclusions}:

 \begin{description}

 \item[Q1:] {\it Does the science case place any constraints on the
 tradeoff between the sky coverage and coadded depth? For example, should
 the sky coverage be maximized (to $\sim$30,000 deg$^2$, as e.g., in
 Pan-STARRS) or the number of detected galaxies (the current baseline
 of 18,000 deg$^2$)?}

 \item[A1:] Cepheid light curves will be well sampled with far less than the ~800 visits acquired in the baseline survey, hence increased sky coverage would measure cepheids in a larger number of galaxies, which would be valuable.

 \item[Q2:] {\it Does the science case place any constraints on the
 tradeoff between uniformity of sampling and frequency of  sampling? For
 example, a rolling cadence can provide enhanced sample rates over a part
 of the survey or the entire survey for a designated time at the cost of
 reduced sample rate the rest of the time (while maintaining the nominal
 total visit counts).}

 \item[A2:] An enhanced cadence could provide earlier discovery and period confirmation for a subset of targets, but this is
not a high priority.

 \item[Q3:] {\it Does the science case place any constraints on the
 tradeoff between the single-visit depth and the number of visits
 (especially in the $u$-band where longer exposures would minimize the
 impact of the readout noise)?}

 \item[A3:] No strong constraint.

 \item[Q4:] {\it Does the science case place any constraints on the
 Galactic plane coverage (spatial coverage, temporal sampling, visits per
 band)?}

 \item[A4:] LSST observations of galactic cepheids are not high priority, since they are bright for LSST.

 \item[Q5:] {\it Does the science case place any constraints on the
 fraction of observing time allocated to each band?}

 \item[A5:] No strong constraint as long as good representation of all.

 \item[Q6:] {\it Does the science case place any constraints on the
 cadence for deep drilling fields?}

 \item[A6:]  This would only apply in deep drilling fields that were chosen to study relatively near-by galaxies, for which cepheid
optimization would no doubt be considered explicitly.

 \item[Q7:] {\it Assuming two visits per night, would the science case
 benefit if they are obtained in the same band or not?}

 \item[A7:] Different bands would provide more rapid characterization of targets, but this is not a strong benefit.

 \item[Q8:] {\it Will the case science benefit from a special cadence
 prescription during commissioning or early in the survey, such as:
 acquiring a full 10-year count of visits for a small area (either in all
 the bands or in a  selected set); a greatly enhanced cadence for a small
 area?}

 \item[A8:]  Enhanced cadences could provide earlier science, but
cepheids are not a strong driver for this.

 \item[Q9:] {\it Does the science case place any constraints on the
 sampling of observing conditions (e.g., seeing, dark sky, airmass),
 possibly as a function of band, etc.?}

 \item[A9:] No constraints that are particular to cepheids, but good seeing will aid the study of stars in the crowded fields of
external galaxies.

 \item[Q10:] {\it Does the case have science drivers that would require
 real-time exposure time optimization to obtain nearly constant
 single-visit limiting depth?}

 \item[A10:] No.

 \end{description}

\navigationbar

% --------------------------------------------------------------------

% PJM: moved to Future Work while MAF analysis is pending:
% \input{Variables/periodicpulsators.tex}

% --------------------------------------------------------------------

% ====================================================================
%+
% NAME:
%    multiperiodicvariables.tex
%
% CHAPTER:
%    variables.tex
%
% ELEVATOR PITCH:
%-
% ====================================================================

\section{Characterizing Multiperiodic, Short-Period Pulsating Variables}
\def\secname{multiperiodicvariables}\label{sec:\secname}

\credit{keatonb}

Most pulsating variable stars exhibit a superposition of multiple
simultaneous pulsation modes.  While these multiperiodic pulsators are
observationally more complex than classical pulsators, they communicate
a greater wealth of information about the stars.  Global nonradial
pulsations pass through and are affected by the interiors of stars, and
the measured frequencies of photometric variability are eigenfrequencies
of stars as physical systems.  Therefore, the observation and study of
photometric variability in multiperiodic pulsating stars is the most
powerful method by which we can probe stellar interiors.

The current state and modern tools of the field of asteroseismology are
thoroughly discussed by \citet{2010aste.book.....A}.  The locations of
the known classes of pulsating variable in the H-R Diagram are indicated
in their Figure 1.12, and a summary of their general
properties---including pulsation amplitudes and timescales---is provided
in their Appendix A.

Sections 8.6 and 8.7 of the Science Book express the anticipated
contributions that LSST will make to the study of pulsating variables.
While the sparseness of observations (low duty cycle) over the LSST
survey lifetime will make exact period solutions difficult, if not impossible for
most multiperiodic, short-period pulsating variables, the Science Book
correctly emphasizes that the photometric precision and multi-color
information will yield detections of variability that can be treated
statistically to determine the ensemble properties of different classes
of pulsating star.  The dependence of pulsational power on a star's
position in 5-color parameter space, as well as the relative pulsation
amplitudes in different passbands, informs the least understood aspects
of pulsation theory: mode selection and amplitude limiting mechanisms. A
thorough statistical view of pulsating stars requires on the order of
thousands of objects per type at a minimum, with robust detections of
variability in multiple passbands.

\subsection{The Case for ZZ Cetis}
\label{sec:\secname:targets}

Of the known classes of pulsating variable star, the ZZ Cetis
(pulsating hydrogen-atmosphere white dwarfs) are the faintest, and among
those with the lowest pulsation amplitudes.  These will be the most
difficult for LSST to detect and characterize, and they therefore
provide an important benchmark for assessing the effectiveness of
different proposed observing strategies for the study of pulsating
stars.  If LSST is well suited for ZZ Ceti science, it will be a useful
tool for all classes of pulsator.

Most details of ZZ Ceti stars are not strictly relevant to this
analysis, so we direct the interested reader to recent reviews by
\citet{2008ARA&A..46..157W}, \citet{2008PASP..120.1043F}, and
\citet{2010A&ARv..18..471A}.

As white dwarfs with atmospheres spectroscopically dominated by hydrogen
cool to between roughly 12,500 and 10,600\,K, they are observed as the
photometrically variable ZZ Cetis.  The square root of the total
observed pulsational power (intrinsic root-mean-squared signal) in these
objects is on the order of 1\% and the mean pulsation periods are
$\sim$10\,min \citep[e.g.,][]{2006ApJ...640..956M}.  Because the pulsation
periods are $\ll$ the typical LSST revisit time, the survey will essentially
sample the pulsations randomly in phase.

The ability to detect pulsations relies on the recognition that scatter
in the flux measurements significantly exceeds what can be attributed to
noise.  LSST's sensitivity to these pulsations depends primarily on two things:
the photometric precision and the total number of measurements.  By
affecting these survey characteristics, the choice of observing strategy
impacts LSST's success in its goal of exploring the variable universe.
Strategies that maximize photometric precision and the total number of
visits in all filters optimize the survey toward this goal, but there are
tradeoffs between these dual requirements related primarily to exposure
time that are complicated and must be explored in MAF to be understood.

While consideration of observations across all filters together provides
the greatest sensitivity to detecting overall variability, the measurement
of pulsational power in individual filters serves the science needs for
pulsating stars best.  Since sparse temporal sampling will make measuring
the individual periods of complex, multi-modal pulsating stars challenging,
LSST's greatest contributions to this field may lie
in its ability to measure relative pulsational power across many
passbands.  For ZZ Cetis in particular, there is a dependence of
the relative amplitudes measured in different filters on the geometry of
the pulsations---specifically the spherical degrees, $\ell$, of the
spherical harmonic wave patterns associated with the pulsation modes.
Determining the $\ell$ values associated with individual modes is
essential for comparing measured pulsation frequencies with those
calculated for asteroseismic stellar models.  The difficulty in
determining $\ell$ is currently the greatest limitation on white dwarf
asteroseismology. LSST has the potential to statistically constrain the
relative contributions of modes of different $\ell$ to the
overall photometric variations.  The calculations by
\citet{1995ApJS...96..545B} of relative pulsation amplitudes in
different filters show that measuring the amplitude in the $u$ band is
crucial for gaining leverage on this problem.

% --------------------------------------------------------------------

\subsection{Metrics}
\label{sec:\secname:metrics}

We have developed a custom MAF metric that calculates a ``variability
depth'' for every point on the sky equal to the magnitude limit for
detecting a population of photometric variables with a given
disk-integrated root-mean-squared (r.m.s.)\ underlying signal to a
desired level of completeness and a tolerable level of contamination.
The metric makes the simplifying
assumptions that the typical revisit time for a field is longer than the
pulsation periods (appropriate for many pulsators, including ZZ Cetis)
and that the intrinsic variability takes the form of a Gaussian (which,
for multi-periodic pulsators, is supported by the central limit
theorem).  The metric relies on the total number of visits and
signal-to-noise per visit (scaled from the 5$\sigma$-depth, with
Gaussian errors assumed) for the calculation, and is included in {\tt
sims\_maf\_contrib} as {\tt VarDepth}.  Example output of this metric
is displayed in Figure~\ref{fig:vardepth}.

\begin{figure}
  \centering
  \includegraphics[width=0.76\columnwidth]{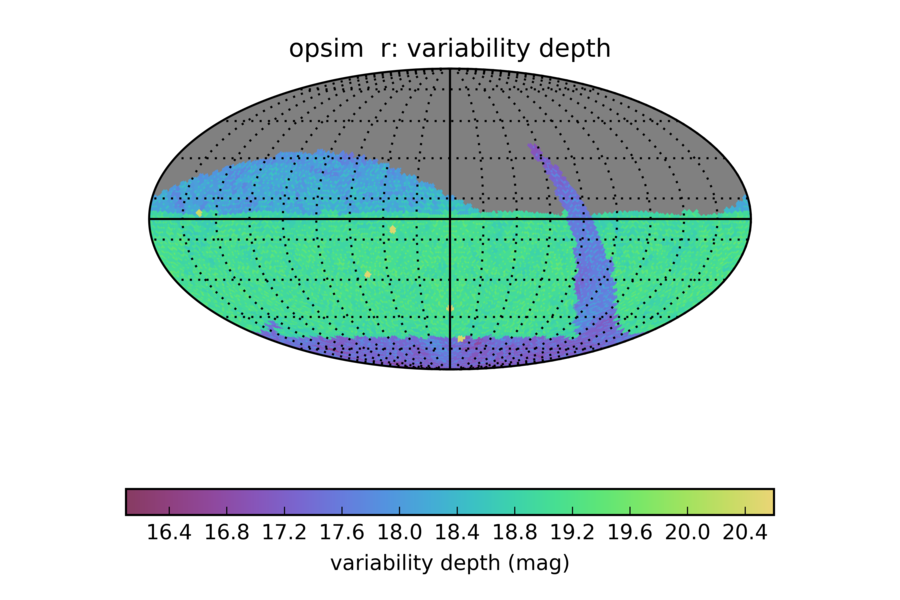}
  \caption{Example output of the {\tt VarDepth} MAF metric run on the
  current baseline cadence, \opsimdbref{db:baseCadence}, after 10 years of survey
  operations. Input parameters and SQL queries were set to calculate the
  magnitude limit for detecting 90\% of pulsators with 1\% r.m.s.\
  variability from a cut on the measured variance in the $r$ band
  (allowing contamination from 10\% of nonvariable sources).}
  \label{fig:vardepth}
\end{figure}

We specialize this metric for the case of ZZ Ceti pulsators by including
maps of the expected distribution of ZZ Cetis in the Galaxy. These maps
were precomputed for the $u$ and $r$ filters by querying the CatSim
database for white dwarfs with an SQL constraint to include only those
with hydrogen atmospheres and with effective temperatures between 10,600
and 12,500 K (i.e., inside the ZZ Ceti instability strip).  While the
boundaries of the instability strip depend slightly on surface gravity,
the width of the strip does not change much, so these temperature cuts
yield representative counts.  The white dwarf spectral energy
distributions were calculated for CatSim by Pierre Bergeron et
al.\footnote{\url{http://www.astro.umontreal.ca/~bergeron/CoolingModels/}}\
The {\tt ZZCetiCounts} metric calculates the number of ZZ Cetis that we
expect to detect in each part of the sky, and the sum of the results
gives the total number of ZZ Cetis with variability detected by LSST.

The measured r.m.s.\ scatter from pulsations in ZZ Cetis is typically of
order $\sim$1\%, with a mean value around 3\%
\citep{2006ApJ...640..956M}.  Since we aim to statistically determine
the ensemble amplitude properties in multiple filters for a large
population of ZZ Cetis (particularly in the $u$ filter), we adopt the
following figure of merit for LSST's ability to study pulsating stars:
that significant variability should be detected in the $u$ band for at
least 1000 ZZ Cetis by the end of
the 10\,yr survey operations.  A sample of this size can be binned to track
changes in pulsational power as white dwarfs cool across the ZZ Ceti
instability strip, without the results being unduly influenced by random
sampling of inclination angle or the white dwarf mass distribution. It also
ensures that LSST contributes an order-of-magnitude improvement to the
number of ZZ Cetis known.

% --------------------------------------------------------------------

\subsection{\OpSim Analysis}
\label{sec:\secname:analysis}

For comparison purposes, we calculate the number of ZZ Cetis detected in
both the $u$ and $r$ filters, assuming intrinsic r.m.s.\ variability of
1\% for two of the currently available \OpSim runs:
\opsimdbref{db:baseCadence}, the current baseline cadence, and
\opsimdbref{db:DoubleUbandExptimeSameVisits}, with doubled $u$-band exposure
times. We require that 90\% of ZZ Cetis with 1\% r.m.s.\  variability
are detected to the computed ``variability depth,'' with a tolerance for
up to 10\% of nonvariables with the same $u$ and $r$ magnitudes to yield
false detections.  The total number of ZZ Cetis detected for each of
these analyses, \emph{if all ZZ Cetis exhibit exactly the assumed level
of r.m.s.\ variability}, is provided in \autoref{tab:zz1pertab}.

\begin{table}[h]
\begin{center}
    \caption{ZZ Ceti Recovery for 1\% R.M.S.\ Variability}\label{tab:zz1pertab}
    \begin{tabular}{| l | l | l |}
    \hline
    \OpSim Run & Filter & \# ZZ Cetis \\ \hline
      \opsimdbref{db:baseCadence} & $u$ & 9  \\
      & $r$ & 127 \\ \hline
      \opsimdbref{db:DoubleUbandExptimeSameVisits} & $u$ & 17\\
      & $r$ & 123  \\ \hline
    \end{tabular}
\end{center}
\end{table}

Clearly LSST appears to fall very short of the proposed figure of merit
by this measure.   However, the 1\% level of variability assumed for ZZ
Cetis in this treatment was chosen to assess how well the
lowest-amplitude ZZ Cetis are recovered by LSST and doesn't fairly
represent the more typical, higher amplitude variables.  If we relax
this constraint and repeat the analysis with ZZ Cetis all modeled as 3\%
r.m.s.\ variables (the mean r.m.s.\ variation observed), we get the
results shown in \autoref{tab:zz3pertab}. This better represents a
total number of ZZ Cetis expected, allowing for incompleteness to low
amplitude.  While we still do not detect $\sim$1000 ZZ Cetis in $u$,
overall we see an increase in the total number of detected ZZ Cetis by
roughly an order of magnitude (in $r$), with the
\opsimdbref{db:DoubleUbandExptimeSameVisits} simulation reaching markedly closer
to the $u$-band goal than the \opsimdbref{db:baseCadence} strategy.

\begin{table}[h]
\begin{center}
    \caption{ZZ Ceti Recovery for 3\% R.M.S.\ Variability}\label{tab:zz3pertab}
    \begin{tabular}{| l | l | l |}
    \hline
    \OpSim Run & Filter & \# ZZ Cetis \\ \hline
     \opsimdbref{db:baseCadence} & $u$ & 197  \\
      & $r$ & 1601 \\ \hline
     \opsimdbref{db:DoubleUbandExptimeSameVisits}  & $u$ & 325\\
    & $r$ & 1534  \\ \hline
    \end{tabular}
\end{center}
\end{table}

% --------------------------------------------------------------------

\subsection{Discussion}
\label{sec:\secname:discussion}

This simplistic analysis ignores most subtleties of ZZ Ceti variables,
but Table~\ref{tab:zz3pertab} captures the order of magnitude of
expected ZZ Ceti detections \emph{from excess scatter alone} for
the \OpSim runs considered.  More sophisticated analysis
of the LSST data will recover additional variables, especially since the
information about the timing of visits has been completely ignored here.
The statistical treatment of large populations of stars will also improve the
detection of variability over this star-by-star treatment.
Still, it appears that the science results for the lowest amplitude ZZ Cetis
will be severely limited, and therefore LSST will capture a less-than-complete
census of pulsating variable stars.

We emphasize again that low-amplitude ZZ Cetis are the most difficult
pulsating stars to observe. LSST will capture variability in higher
amplitude ZZ Cetis and other types of variable star more completely, and
LSST observations of all candidate pulsating white dwarfs will certainly
still be worthy of careful analysis.  We note that the
\opsimdbref{db:DoubleUbandExptimeSameVisits} strategy with 60\,s exposures in $u$
does a much better job of measuring stellar variability overall, at the
expense of a practically negligible decrease in $r$-band detections.  We
argue that any observing strategy that increases the typical {\tt VarDepth}
limit in the $u$ band will improve LSST's scientific yield for
\emph{all} stellar pulsation studies.

% ====================================================================

 \subsection{Conclusions}

 Here we answer the ten questions posed in
 \autoref{sec:intro:evaluation:caseConclusions}:

 \begin{description}

 \item[Q1:] {\it Does the science case place any constraints on the
 tradeoff between the sky coverage and coadded depth? For example, should
 the sky coverage be maximized (to $\sim$30,000 deg$^2$, as e.g., in
 Pan-STARRS) or the number of detected galaxies (the current baseline 
 of 18,000 deg$^2$)?}

 \item[A1:] Coadded depth is not directly a limitation on detecting variability,
 but both improve with the number of revisits to the same field, $N$,
 as roughly $\sqrt{N}$.  A larger field of view also increases the potential population
 of variable stars discovered, but pushing to higher airmass where data quality is
 poorer does not make for a good trade unless the survey area is extended to strategically
 target more variable stars, e.g., over more of the Galactic plane.

 \item[Q2:] {\it Does the science case place any constraints on the
 tradeoff between uniformity of sampling and frequency of  sampling? For
 example, a rolling cadence can provide enhanced sample rates over a part
 of the survey or the entire survey for a designated time at the cost of
 reduced sample rate the rest of the time (while maintaining the nominal
 total visit counts).}

 \item[A2:] While the time distribution of observations has not been considered
 here, this will significantly impact variable star science with LSST.
 Future development of MAF metrics should address specifically how
 the distribution of visits (rather than only number, area, and signal-to-noise as addressed
 above) ease the challenges of amplitude and \emph{period} recovery in sparely sampled
 observations.  Revisit times much shorter than pulsation timescales ($< \sim$5\,min)
 can help to constrain the periods.  Perhaps more importantly, cadences that
 produce the least complicated spectral windows and the highest
 ``effective Nyquist frequencies'' will best facilitate period recovery \citep[e.g.,][]{1999A&AS..135....1E}.

 \item[Q3:] {\it Does the science case place any constraints on the
 tradeoff between the single-visit depth and the number of visits
 (especially in the $u$-band where longer exposures would minimize the
 impact of the readout noise)?}

 \item[A3:] We specifically support increased time spent in the
 $u$-band, with the goal of measuring the relative pulsation amplitudes
 in multiple filters. Increased signal-to-noise per $u$-band exposure enables
 more measurements of pulsational power in $u$, though pulsating star
 science would be hurt by a corresponding decrease in the total number of
 $u$-band visits per field .

 \item[Q4:] {\it Does the science case place any constraints on the
 Galactic plane coverage (spatial coverage, temporal sampling, visits per
 band)?}

 \item[A4:] The Galactic plane contains a higher concentration of multi-periodic
 pulsating stars that LSST will be able to characterize in the time domain.
 Pulsating star science benefits from more survey time spent in the Galactic
 plane.

 \item[Q5:] {\it Does the science case place any constraints on the
 fraction of observing time allocated to each band?}

 \item[A5:] For the ZZ Ceti case study, $u$-band observations are particularly
 important for potentially constraining pulsation geometries. The metric
 developed in this section measures the expected number of ZZ Cetis
 detected in the $u$-band, and we advocate for observing strategies
 that increase this number.

 For classical pulsators, such as $\delta$ Scuti stars (P$\sim$hrs) and $\gamma$ Dor stars (P$\sim$days), multi-band photometry is important for mode identification. Enough data in each band is required to determine the pulsation's amplitude and phase. For these pulsators, the $u$-band is the least informative \citep[e.g.,][]{1990A&A...234..262G}.

 \item[Q6:] {\it Does the science case place any constraints on the
 cadence for deep drilling fields?}

 \item[A6:] The results of the {\tt VarDepth} MAF metric run on the
 current baseline cadence, \opsimdbref{db:baseCadence}, shows that after
 10 years, deep drilling fields are sensitive to pulsation
 \emph{amplitude} measurements in the $r$-band $\approx$1.5\,mag fainter
 than for identical objects in the main survey.  The improvement to
 period recoverability could be far greater in deep drilling fields, but
 new MAF metrics must be developed to demonstrate this and to help
 select between specific proposed cadences.

 The possibility to obtain data at a high cadence for extended periods of time in the deep drilling fields will also facilitate the asteroseismology of solar-like oscillators ($\sim$minutes for main-sequence stars and $\sim$hours for red giants). However, a MAF metric is required to determine the specific requirements.

 \item[Q7:] {\it Assuming two visits per night, would the science case
 benefit if they are obtained in the same band or not?}

 \item[A7:] Analyses that will successfully extract the maximum pulsation
 science from LSST will have to address data across all passbands together,
 and there is no current indication of how or if the nightly revisit filter selection
 will affect this.

 \item[Q8:] {\it Will the case science benefit from a special cadence
 prescription during commissioning or early in the survey, such as:
 acquiring a full 10-year count of visits for a small area (either in all
 the bands or in a  selected set); a greatly enhanced cadence for a small
 area?}

 \item[A8:] A few hours of continuous sampling of a few fields during
 commissioning would produce immediate science results for many
 multi-modal pulsating stars, as well as help accelerate LSST's ultimate
 characterization of variable stars in these fields.

 Furthermore, by mimicking the 10-yr survey for a small area, the scientific community can obtain evidence-based estimates of the yield of different variable classes and example power spectra. This will also provide proof-of-concept data that will facilitate the recruitment of scientists to work on LSST.

 \item[Q9:] {\it Does the science case place any constraints on the
 sampling of observing conditions (e.g., seeing, dark sky, airmass),
 possibly as a function of band, etc.?}

 \item[A9:] The {\tt VarDepth} MAF metric can help assess how the
 adjustments made by the observing strategy to observing conditions
 affect the variable star science output.

 \item[Q10:] {\it Does the case have science drivers that would require
 real-time exposure time optimization to obtain nearly constant
 single-visit limiting depth?}

 \item[A10:] No. Consistent exposure times are simpler to interpret as they
 smooth over any underlying variability by a consistent amount.

 \end{description}

% ====================================================================

\navigationbar

% --------------------------------------------------------------------

% PJM: moved to Future Work while MAF analysis is pending:
% \input{Variables/planets.tex}

% --------------------------------------------------------------------

% PJM: Commenting out placeholder:
% \input{Variables/rotationalvariables.tex}

% --------------------------------------------------------------------

% ====================================================================
%+
% NAME:
%    youngstars.tex
%
% CHAPTER:
%    variables.tex
%
% ELEVATOR PITCH:
%-
% ====================================================================

\section{Discovery and Characterization of Young Stellar Populations}
\def\secname{youngstars}\label{sec:\secname}

\credit{phartigan},
\credit{CJohnsKrull},
\credit{pmmcgehee}

\subsection{Introduction}

All young stars exhibit some form of
photometric variability, and these variations hold the key
to understanding the diverse physical processes present at starbirth
such as mass accretion events from circumstellar disks, presence of
warps in envelopes, creation of new knots in stellar jets,
evolution of stellar angular momenta, starspot longevity and cycles,
and the frequency and strength of flares.
With the proper cadences and filter choices,
LSST will make a significant impact in our understanding of all
these phenomena simply by providing large enough samples to allow
us to relate these aspects of the young stars and their environments
to stellar properties such as mass, age, binarity,
and their location within their nascent dark clouds in a statistically
significant manner.

Low-mass ($\lessim 1.5 M_{\odot}$) pre-main-sequence stars separate
into two main categories, depending upon whether or not an optically thick
dusty circumstellar disk exists in the system: young stars without dusty
inner disks are known as `weak-lined T Tauri stars' (wTTs), while those
that have inner dusty disks are called `classical T Tauri stars'.
The nature of the variability in young stars changes with evolutionary status.
In cTTs, variability is primarily caused by unsteady mass infall from circumstellar disks
onto their stars, and from periodic extinction events that unfold as dense
clumps or disk warps circle the star in Keplerian orbits.
Once the disks become optically thin, variability in the wTTs phase
is dominated by small-amplitude (typically $\sim$ 0.1 mag)
quasi-periodic variations that arise
from cool star spots, though active regions that generate X-rays in
these objects undoubtedly produce optical flares as well.

At the extreme end of cTTs phenomena, rare massive
outbursts of up to 6 magnitudes with decay times between several months
to over a hundred years in pre-main-sequence stars (EX Ori's \citep{herbig01}
and FU Ori's \citep{hartmann96})
are of great interest to studies of disk accretion because they indicate the onset of
a major disk instability. Only a handful of such systems have been found, and LSST
will easily detect any new ones.  In addition to triggering
follow-up observations, LSST will define the first population
constraints on the duration of high states, particularly for the shorter-lived
versions of these eruptive variables.

Obtaining rotation periods for large numbers of pre-main-sequence stars is the only way to
quantify how angular momenta of young stars vary with age. For both wTTs and cTTs,
phase-coherence in the rotation periods defines the longevity of star spots, while the
amplitude of the periodic component of the lightcurve constrains the spot coverage (and
temperature if multiple filters are used).  As described in section 8.10.2 in the
LSST Science Book (p298--299), irregular flaring in young stars exists across the entire
$ugrizy$~bandpass of LSST. Flaring can last from minutes to years, with
amplitudes from a few tenths to several magnitudes.  In cTTs, $u$-band fluxes rise during
accretion events, which we can distinguish from extinction events if red magnitudes are
also available. A large accretion event is a signal to observe the system
in the future with other instrumentation to look for evidence of a newly-created
jet knot. In wTTs, Flares also occur in wTTs as a consequence of high chromospheric activity.
Flaring in wTTs is also easiest to monitor at $u$, though the rapid decline of
chromospheric flares requires a rapid cadence to capture correctly.

LSST also provides a potential means to discover new young stars by way of
their variability and colors. One of the challenges in this regard will be to
distinguish young stars from other low-amplitude variable stars in the field.
In that regard we expect that machine-learning techniques that incorporate knowledge of
fluxes in other wavebands as well as the LSST lightcurves and colors will
be an ongoing effort. It is possible that X-ray detections will be more
reliable for detection of new young stars, but LSST will at minimum assist
by identifying non-YSO X-ray sources, and should be a means for discovery
for older pre-main-sequence stars (10 $-$ 30 My range) that have had time
to wander away from the well-known sites of star forming activity that are
typical targets for deep, pointed X-ray surveys.

Galactic star formation regions are largely found at low Galactic
latitudes or within the Gould Belt structure. As such study of young
stars with LSST is closely tied to other science goals concerning the
Milky Way Disk and is subject to the concerns of both crowded field
photometry and the observing cadence along the Milky Way. However, recent
DECam observations at the CTIO 4-m that reached depths similar to those
proposed for LSST show negligible crowding in the optical, and $\lessim$
5\%\ crowding at $z$ in Carina. In this case, extinction in the molecular
clouds helps by significantly lowering the frequency of background contamination.

% --------------------------------------------------------------------

\subsection{Analysis}
\label{sec:\secname:analysis}

{\bf Target Regions}

LSST will allow us to survey the outstanding collection of
star formation regions in the Southern hemisphere, including the closest
such regions ($\rho$ Oph, CrA, Cha~I and Lupus), and the most famous intermediate-mass
(Orion) and massive (Carina) examples.
The closest star forming regions have only low-mass molecular clouds, and each contain
only about 100 young stars. More massive molecular clouds produce both higher
mass stars and more low mass stars. In the Orion Nebula Cluster (d = 414 pc), the number of
identified YSOs is $\sim$ 3000, and we expect $\gtrsim$ 30000 pre-main-sequence
stars in the famous southern star formation region in Carina (2.3 kpc).
LSST will make its greatest impact
when observing the more massive star formation regions, where the amount
of young stars is much higher.

Owing to extinction in the dark clouds,
source confusion will generally not be an issue (as evidenced by typical
deep optical images of such regions), though the large fraction
of pre-main-sequence binaries at all separations ensures that many
lightcurves will be composites of the primaries and secondaries.
More distant star-forming regions will suffer from enhanced foreground
contamination, though it should be possible to eliminate most contaminating
variable stars by combining close inspection of their lightcurves with
colors.

{\bf Metrics:}

{\bf A. Magnitude Limits, Filter Choices}

To quantify YSO studies with LSST, we consider V~927 Tau,
a rather faint, moderately-reddened 0.2 M$_\odot$ young star in the Taurus cloud
as a target goal. Extrapolating this star
to the distance of Carina we have $u$=24.0, $g$=23.0, $r$=20.8, $i$=19.4, and $z$=18.0. For reference,
a typical young star in the Carina X-ray catalog has an $i$-magnitude of 18.
Objects that suffer larger extinctions along the line of sight will
be easiest to observe in the red. The universal cadence option of 2$\times$15 sec
exposures will yield $\sigma$ = 0.02 mag for $r$=21.8, a magnitude fainter than
V~927 Tau would be in Carina. We show below that this photometric uncertainty
suffices to recover a typical period from such an object. The mass function
of young stars peaks around 0.3 M$_\odot$, so {\it LSST will
determine periods to near the hydrogen burning limit with nominal r-band measurements
for a region like Carina}. Of course, several additional magnitudes of extinction will
exist towards many embedded sources. For example, if we assume an additional five magnitudes
of extinction at V for the V~927 Tau-like example above,
then $r$=25.2 and $\sigma$ = 0.41 mag per visit with universal cadence,
so no usable lightcurve will be possible at $r$.
However, $z$=20.5 in this case, where $\sigma$ $<$ 0.01 mag and precision
lightcurves are again possible.

{\bf B. Period Recovery for wTTs}

In order to assess how well LSST will recover periods, we created the following
model for wTTs variability.  Based on current surveys of wTTs, the periods are
distributed approximately as a Poisson distribution with a mean of 3.5 days
\citep{Affer2013}.
Amplitudes are typically 0.1 magnitude \citep{Grankin2008},
so we adopt a Poisson distribution
that has a mean amplitude of 0.05 mag, and then add 0.05 mag to ensure that
the mean variability is 0.1 mag. Shapes of T-Tauri lightcurves can be sinusoidal,
but many are `bowl-shaped', influenced by the distribution of large dark starspots
\citep{Alencar2010}.
For the bowl lightcurves we assumed a Gaussian shape with a FWHM in a uniform
distribution of extent between 0 and 0.75 in phase.
Our simulations cover both of these shapes.
Errors for each point were taken to be 0.02 magnitude, corresponding to
about r $\sim$ 21.8 for a universal cadence.

One set of simulations assumed a cadence of
one observation every three days over the course of a year.
If we define a successful period recovery to be better than a 1\%\ error, then
using the standard Scargle method \citep{Horne1986}
we are able to find the correct period in
98\%\ of the the sinusoidal, and 86\%\ of the bowl lightcurves, with the most
difficult challenges being at the short end of the period distribution. If we change
the cadence to once every 7 days, the ability to recover periods drops to
82\%\ and 59\% , respectively, for the two shapes. Interestingly, restricting the
sample to the highest-amplitude sources ($\gtrsim$ 0.1 magnitude) does little
to aid period recovery. The main issue remains the short-period systems where
P $\lesssim$ 2 days.

Overall, standard cadences of once every few days should suffice
to find most periodic T-Tauri stars that have periods $\gtrsim$ 3 days.
A dedicated campaign to observe star-forming regions
at time intervals of an hour or less is required to capture the shorter-period systems.
The r-filter should suffice for most objects, though some benefit will be had
by going to z to allow the more heavily-extincted sources to be observed.

\begin{figure}[h]
\centering\includegraphics[width=0.95\linewidth]{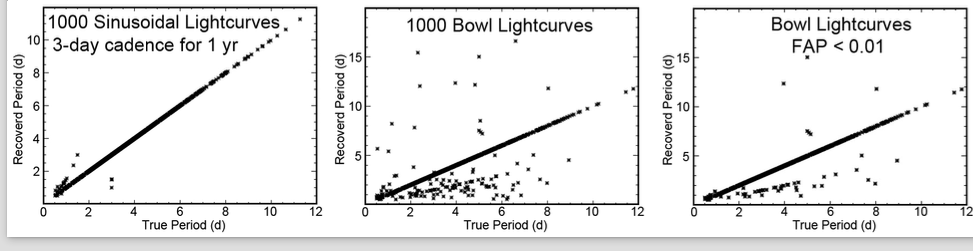}
\caption{Recovered period vs. true period for a sample of sinusoidal (left),
bowl-shaped (middle) and bowl-shaped with False Alarm Probability $<$ 0.01 (right),
assuming a 3-day cadence and one year of observing. What appears as a solid line are the
individual points with periods that are recovered correctly. The bowl-shaped curves are
more difficult to recover than the sinusoids, but the method is highly successful in both cases.}
\label{tts1}
\end{figure}

\begin{figure}[h]
\centering\includegraphics[width=0.95\linewidth]{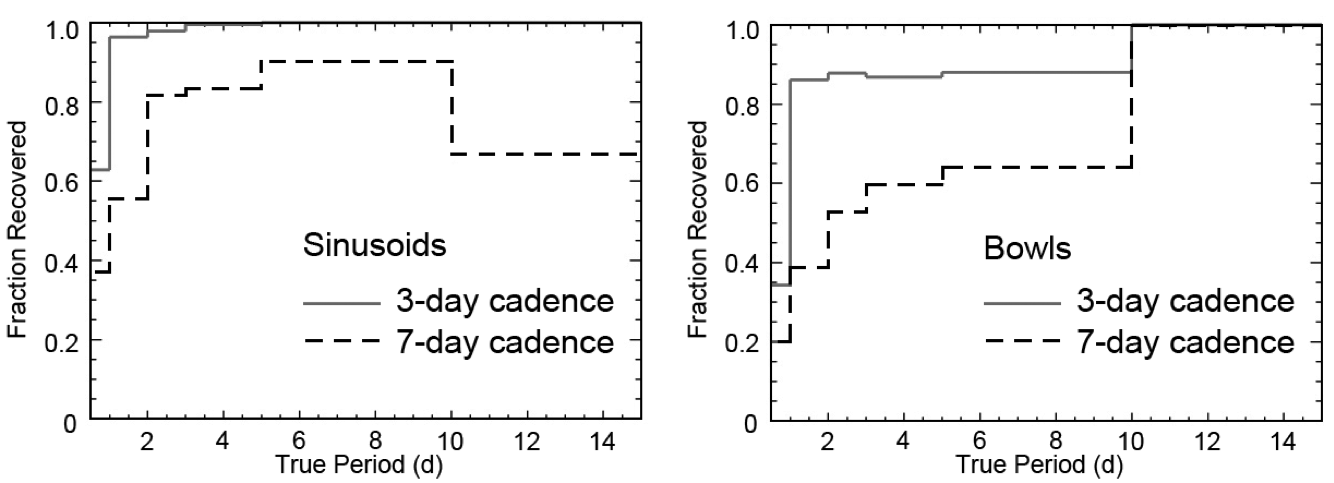}
\caption{ Fraction of periods recovered correctly for sinusoidal (left) and bowl light curves (right) for
3-day (solid line) and 7-day (dashed line) cadences over an observing period of one year.
A 3-day cadence is significantly better than a 7-day one. Over 98\%\ of sinusoidal, and 86\%\ of bowl
light curve periods are recovered successfully with the 3-day cadence. The percentages drop to about
82\%\ and 59\%\, respectively, for the 7-day cadence.
}
\label{tts2}
\end{figure}

\begin{figure}[h]
\centering\includegraphics[width=0.95\linewidth]{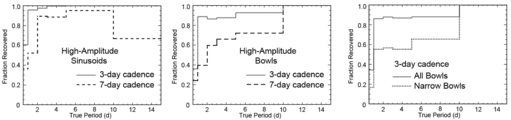}
\caption{Left and center: Same as Fig 2 but restricting the sample to amplitudes greater than 0.1 mag. The
method is only marginally more successful with the larger amplitude objects than it is with the entire sample.
Right: The narrowest 278 bowls have a significantly higher error rate than the entire sample does.
}\label{tts3}
\end{figure}

{\bf C. Period Recovery for cTTs}

Complex irregular variations in cTTs lightcurves make it much more difficult,
and in many cases impossible to
recover periods in these systems. While sparse coverage of one observation every
few days is adequate for identifying sudden changes from accretion events, these
events to a large degree overwhelm low-amplitude periodic signatures. Even when
period searches yield a low false-alarm probability, the results are not necessarily
reliable. Results from Palomar Transient Factory surveys in the North American
Nebula \citep{Findeisen2013} and with Spitzer
\citep{Cody2014}
reveal several types of both short- and long-term variations including both bursting and
fading.  These observations emphasize how important it will be to have some
dense phase coverage as a reality check to ensure the reliability of
any periods recovered from sparse data in these objects, as well as to
follow the short-term variations that characterize accreting systems.

{\bf D. Discovery, Accretion and Extinction Events}

As we indicated above, any
cadence will uncover FU Ori and EX Ori events in all filters.
Periodic extinction events follow the same restrictions and
have the same requirements as rotational periods described in subsection B.

In order to assess the ability of LSST to identify and classify
eruptive variables (FUor/EXor), we construct
\autoref{table:pseudoForExor}, which shows a possible Figure of Merit for the
recovery by LSST of the distribution of EXor high-state duration in
outburst.

\begin{table}
\small
\begin{tabular}{c p{12cm}}
& {\it Figure of Merit for recovery of EXor high--state duration distribution}\\
\hline
1.  & Produce ASCII lightcurve for eruptive outburst \\
2.  & Initialise large array to store the maps of fraction detected as a function of duration and amplitude. \\
2.  & for {\it duration T} in range \{min, max\}:  \\
3.  & ~~~~ for {\it amplitude A} in range \{min, max\}: \\
4.  & ~~~~~~~~~~ run {\tt mafContrib/transientAsciiMetric} \\
5.  & ~~~~~~~~~~ store the spatial map of the fraction detected for this (A, T) pair \\
6.  & Initialise master arrays to hold the run of duration distribution measurements.\\
7. & Produce distribution of high--state durations and amplitudes from which the simulations will be drawn. \\
8.  & for {\it iDraw} in range \{1, nDraws\}:\\
9.  & ~~~~ construct model population with input duration distribution \\
10.  & ~~~~ Apply the stored metrics from 2-5 to measure fraction recovered \\
11.  & ~~~~ Characterize the duration distribution for this draw \\
12. & ~~~~ Fill the {\it iDraw}'th entry in the master arrays. \\
13. & {\bf FoM 1:} Compute the median and variance of the upper/lower quintiles. \\
14. & {\bf FoM 2:} Evaluate the bias between recovered and input high-state duration. \\
\hline
\end{tabular}
\caption{Steps for Figure of Merit recovering the distribution
  for the duration of EXor high states.}
\label{table:pseudoForExor}
\end{table}

% --------------------------------------------------------------------

\subsection{Summary and Recommendations}
\label{sec:\secname:discussion}

{\bf Performance for Nominal Cadences}

Nominal cadences that return to a star-forming region every 3-4 days
will suffice to determine rotation periods for $\sim$ 90\%\ of the
young stars within the magnitude limits of LSST.
These cadences are also adequate to detect major episodic
accretion events like FU Ori's and EX Ori's. However, a more focused
annual campaign of about a week duration is necessary to optimize
period recovery and angular momentum studies of young stars.

{\bf The Need for Annual Dense Coverage of a Few Selected Regions}

Occasional dense coverage of targeted regions is the only way to
get quantitative information on short-term accretion and flare
activity.  Dense coverage also removes degeneracies
for periodic variables that have periods less than a day, and is
the only way to provide a sanity check on any periods recovered for
cTTs, which have complex irregular light variations.
Comparing longevities of starspots across the mass ranges of young
stars requires two well-sampled lightcurves separated by large
time intervals.  The embedded and Classical T Tauri stars also undergo
significant and rapid color changes due to both accretion processes
and extinction variations, so it is important
to include multiple filters in any dense coverage campaign.

These goals can be accomplished by having a week every year where
one or more selected fields are observed once every 30 minutes in u, r and z.
A young star with a 2-day period sampled every 30 minutes provides a
data point every 0.01 in phase. For the best-case scenario, observing for 7 nights
and 10 hours per night would yield 140 photometric points in each filter.
Depending on the period aliasing, this coverage should populate the
phases well enough to identify most of the large starspots on the stellar photospheres.

At the beginning of LSST operations we argue that a targeted test field
be observed in this manner to illustrate what can be done with LSST in this mode.
Combining a densely-packed short-interval
dataset with a sparse but long baseline study maxmizes the scientific return
for both methods, and allows LSST to address all of the accretion and
rotational variability associated with young stars.

\subsection{Conclusions}
\begin{description}
\item[Q1:] {\it Does the science case place any constraints on the
tradeoff between the sky coverage and coadded depth? For example, should
the sky coverage be maximized (to $\sim$30,000 deg$^2$, as e.g., in
Pan-STARRS) or the number of detected galaxies (the current baseline
of 18,000 deg$^2$)?}
\item[A1:]
Most young stars congregate into clusters in specific regions, though there is an older
population that is more distributed. The vast majority are within $\sim$ 25 degrees of the
galactic plane. As long as that swath of sky is covered to the degree possible from Chile,
the survey will provide the young star community with the monitoring capability needed to
identify transient outbursts and to study periodic and aperiodic phenomena in young stars.
While it may be useful to have deep coadded images of some regions, for example, to identify
optical counterparts to X-ray point sources, dust extinction typically limits such efforts
in the optical towards most regions of interest.  Hence, deep coadded frames are a secondary
priority.

\item[Q2:] {\it Does the science case place any constraints on the
tradeoff between uniformity of sampling and frequency of  sampling? For
example, a rolling cadence can provide enhanced sample rates over a part
of the survey or the entire survey for a designated time at the cost of
reduced sample rate the rest of the time (while maintaining the nominal
total visit counts).}

\item[A2:]
We discussed cadences in some detail above. To obtain the best constraints on periods, a time-intensive
($\sim$ one week/year) campaign on a few selected regions is warranted. Nominal coverage over the
galactic plane will suffice to identify eruptive variables.

\item[Q3:] {\it Does the science case place any constraints on the
tradeoff between the single-visit depth and the number of visits
(especially in the $u$-band where longer exposures would minimize the
impact of the readout noise)?}

\item[A3:]
This item is discussed in section A above. In young stars, the u-band is particularly useful
as a measure of accretion. At the same time, for period determinations, denser cadences produce
fewer problems with aliasing for the typical young star variable with a rotation period between
a day and two weeks.

\item[Q4:] {\it Does the science case place any constraints on the
Galactic plane coverage (spatial coverage, temporal sampling, visits per
band)?}

\item[A4:]
A nominal sampling of once every few days will suffice to identify interesting
eruptive variables as long as the galactic plane coverage is good (item A1).
However, a sparse temporal sampling such as this will make it difficult to interpret
the lightcurves of the tens of thousands of T Tauri stars observed by LSST. A single
week's campaign of dense sampling, e.g., 2-3 times per night, of targeted regions
will greatly improve our ability to separate periodic variables from aperiodic ones.
Knowing the rotation periods of thousands of young stars within a given star forming
region will be a major contribution that LSST makes to this field of research. Such data
will allow us to learn how angular momentum is distributed among newborn
stars, whether it changes with mass and location in the cloud,
and how it varies as the stars age.

\item[Q5:] {\it Does the science case place any constraints on the
fraction of observing time allocated to each band?}

\item[A5:]
In the case we made above for dense sampling, we argued for z, r, and u. The
u-band allows us to follow mass accretion variability, while r and z (for most
objects) will be dominated by photospheric flux. The r-z color is an important
constraint for models of star spots. More colors are always useful, but having
a photospheric color index plus one accretion measure are the science drivers
for filter choices. Once u, r, and z are observed, having higher cadences
is preferable to having more bands.
The exposure times and magnitudes for typical objects are described in section A
above.

\item[Q6:] {\it Does the science case place any constraints on the
cadence for deep drilling fields?}

\item[A6:]
Absolutely. We argue in the `Summary and Recommendations' section above
for the advantages of a single week of denser monitoring for specific
regions, with the goal of having several observations per night, separated
in time by at least an hour from one another. The goal here is to provide
some basis in reality for interpreting the irregular lightcurves of young
stars, which also typically have a periodic component. Depending on the longevity of
the star spots, a period might be obvious in high-cadence data taken over
a week, but disappear over a year if some spots vanish and others form.

\item[Q7:] {\it Assuming two visits per night, would the science case
benefit if they are obtained in the same band or not?}

\item[A7:]
If the data are taken in the same band
it means better sampling for periods. With different bands it means a lightcurve in two bands,
which is also useful. It's probably a wash, and we can follow whatever the drivers are for the
other science cases of LSST.

\item[Q8:] {\it Will the case science benefit from a special cadence
prescription during commissioning or early in the survey, such as:
acquiring a full 10-year count of visits for a small area (either in all
the bands or in a  selected set); a greatly enhanced cadence for a small
area?}

\item[A8:]
Yes, definitely. We ought to see what we can get out of a dedicated week-long
LSST cadence on specific regions. If the results are impressive we follow up
on different regions every year (or even return to the same regions).

\item[Q9:] {\it Does the science case place any constraints on the
sampling of observing conditions (e.g., seeing, dark sky, airmass),
possibly as a function of band, etc.?}

\item[A9:]

Better seeing helps with unresolved binaries and for regions where contamination comes
into play, for example, in the plane but away from dark clouds. Some of the fainter objects
will be affected if the Moon is very bright and close. However generally these constraints
are probably in a ``typical'' category and will not affect the design of the survey.

\item[Q10:] {\it Does the case have science drivers that would require
real-time exposure time optimization to obtain nearly constant
single-visit limiting depth?}

\item[A10:]
No.

\end{description}

% ====================================================================

% bibtems need pushing into the relevant file!

%Hartmann \& Kenyon 1996, ARA\&A, 34, 207 \\
%Herbig et al. 2001, PASP, 113, 1547 \\
%Herbig 1977, ApJ, 217, 693 \\
%Aspin et al. 2009, ApJ, 692L, 67 \\
%Hodapp et al. 1996, ApJ, 468, 861 \\
%McGehee et al. 2004, ApJ, 616, 1058 \\

%\bibitem[Affer et al. (2013)]{Affer13}
%{Affer, L., Micela, G., Favata, F., Flaccomio, E., \& Bouvier, J.} 2013
%\textit{MNRAS}, 430, 1433

%\bibitem[Alencar al. (2010)]{CoRoT}
%{Alencar, S. H. P. et al.} 2010,
%\textit{A\&A}, 519, A88

%\bibitem[Cody al. (2014)]{cody14}
%Cody, A., et al.
%{Cody, A. et al.} 2014,
%\textit{AJ}, 147, 82

%\bibitem[Grankin al. (2008)]{ROTOR}
%\bibitem[Findeisen al. (2013)]{findeisen13}
%Findeisen, K., Hillenbrand, L., Ofek, E., Levitan, D., Sesar, B., Laher, R., \& Surace, J.
%{Findeisen, K. et al.} 2013,
%\textit{ApJ}, 768, 93

%\bibitem[Grankin al. (2008)]{ROTOR}
%{Grankin, K. N., Bouvier, J., Herbst, W., \& Melnikov, S. Yu.} 2008,
%\textit{A\&A}, 479, 827

%\bibitem[Horne \& Baliunas (1986)]{Scargle}
%{Horne, J. H. \& Baliunas, S.} 1986, \textit{ApJ}, 302, 757

\navigationbar

% --------------------------------------------------------------------

% ====================================================================
%+
% SECTION:
%    future.tex
%
% CHAPTER:
%    variables.tex
%
% ELEVATOR PITCH:
%    Ideas for future metric investigation, with quantitaive analysis
%    still pending.
%-
% ====================================================================

\section{Future Work}
\def\secname{\chpname:future}\label{sec:\secname}

In this section we provide a short compendium of science cases that
are either still being developed, or that are deserving of quantitative
MAF analysis at some point in the future.

% ====================================================================

% ====================================================================
%+
% SECTION:
%    periodicpulsators.tex
%
% CHAPTER:
%    variables.tex
%
% ELEVATOR PITCH:
%
%-
% ====================================================================

% \section{Discovery of Periodic Pulsating Variables}
\subsection{Discovery of Periodic Pulsating Variables}
\def\secname{periodicvariables}\label{sec:\secname}

\credit{lmwalkowicz},
\credit{StephenRidgway}

Regular variables, such as Cepheids and RR Lyraes, are valuable tracers
of Galactic structure and cosmic distance. In this case of these and
other strictly (or nearly-strictly) periodic variables, data from
different cycles of observation can be phase-folded to create a more
fully sampled lightcurve as LSST visits will occur effectively at random
phases. In a 10-year survey, most periodic stars of almost any period
will benefit from excellent phase coverage in all filters (only a very
small period range close to the sidereal day will be poorly observed).
Therefore, most implementations of the LSST observing strategy will
provide good sampling of periodic variables.

However, different implementations of the survey may result in different
resulting sample sizes of these periodic variables, and may also affect
the environments in which these stars are discovered. In this section,
we create a framework for understanding how current implementations of
the observing strategy influence (or even bias) the resultant sample
size and environments where these important tracers may be identified.

\subsubsection{Tracing Galactic Structure with RR Lyrae}

RR Lyrae variables are crucial tracers of structure in the Galaxy and beyond
into the Local Group. The incredible sample of RR Lyrae anticipated from LSST
observations will enable discovery of Galactic tidal stream and neighboring
dwarf galaxies throughout much of the Local Group
\citep{IvezicEtal2008}.  LSST also creates the possibility
of detecting and studying RRL variables in the Magellenic Clouds; see Chapter
[CHAPTER] for discussion.

\citet{2012AJ....144....9O} carried out an extensive simulation of period
and lightcurve shape recovery of RR Lyrae variables using an early OpSim run
opsim1$\_$29. Correctly identifying the period aids in building the sample of
interest, whereas fitting the lightcurve shape makes it possible to measure the
metallicity of the star. In their simulation, they employed both a Fourier
analysis and template matching to recover the lightcurve shape, finding that
template matching yielded a more accurate lightcurve shape measurement in the
presence of sparse data. The results of this simulation showed that the vast
majority of RR Lyrae will be discovered by the baseline observing strategy (as
deployed in opsim1$\_$29) within 5 years of survey operations. Half of both
RRLab and RRLc stars will be found out to $\sim$600 kpc and $\sim$250 kpc
(respectively) by the end of the 10-year main survey, and template matching
techniques for lightcurve shape recovery will provide metallicities to
$\sim$0.15dex.

% --------------------------------------------------------------------

\subsubsection{The Cepheid Cosmic Distance Ladder}

Classical cepheids remain an essential step in the cosmic distance
ladder. Their calibration is based largely on LMC cepheids and known
(assumed) distance of the LMC.  The associated errors, while uncertain,
are believed to be of $\>\sim$7\%. (Madore, Barry F.; Freedman, Wendy L.
(2009). "Concerning the Slope of the Cepheid Period?Luminosity
Relation". The Astrophysical Journal 696 (2): 1498. arXiv:0902.3747.
Bibcode:2009ApJ...696.1498M. doi:10.1088/0004-637X/696/2/1498.) New
developments in galactic studies are poised to support substantially
improved descriptive information concerning nearby galactic cepheids,
with possible substantial reductions in this error, by accurately
securing the PL slope and zero point.

Cepheid calibration errors are associated in part with uncertainties in
extinction, both interstellar and in some cases circumstellar, and in
metalicity.  At present, the direct, local calibration of cepheids is limited
by the availability of a few direct distance measurements, obtained with HST,
with errors $\sim$10\%.  The GAIA mission is expected to return $\sim$9000
Galactic cepheids, of all periods, colors and metallicities, with distance
errors less than 10\% (many of them much less) - Windmark, F.; Lindegren, L.;
Hobbs, D., 2011A\&A...530A..76W. It is expected to deliver at least 1000
cepheids in the LMC with expected mean distance error $\sim$7-8\% (Clementini
(2010) - 011EAS....45..267C).  GAIA, as well as other methods, will also
support determination of the 3-d map of galactic interstellar extinction -
including possible variations in the extinction law. These rich data sets will
be supported with direct measurements of cepheid diameters (A. Merand et al,
A\&A in press) and advances in stellar hydrodynamics
\citep{2013MNRAS.435.3191M} which will provide theoretical and empirical basis
for calibrations to reconcile known physics with observational correction
factors.

Galactic cepheids will generally be too bright for LSST, but cepheids in
the local group are sufficiently bright that LSST photometry will be
limited by calibration errors rather than by brightness.  This dataset
will provide superb support for integration of GAIA-based galactic
cepheid studies with extra-galactic cepheid studies.

GAIA will provide similar precision data with the potential to identify
or support distance determinations from many other galactic star types.
LSST photometric catalogs will represent a uniquely extensive and
complete database for such investigations.

% --------------------------------------------------------------------

% \subsection{Metric Analysis}
\subsubsection{Metric Analysis}
\label{sec:\secname:analysis}

Several metrics currently exist in the MAF for evaluating how LSST
survey strategy affect the recovery of periodic sources.
For example,
the PeriodicMetric makes use of the periodogram purity function, which effectively
quantifies aliasing introduced into periodogram analysis from the
sampling of the lightcurve.
%
% and period deviation metric (PeriodDeviationMetric) all return relevant
% information
%
Similarly, the
phase gap metric (PhaseGapMetric),
evaluates the periodicity of the source lightcurve and its coverage
in phase space (the latter being relevant for shape recovery).

Recreating the template matching results of the
\citet{2012AJ....144....9O} simulation requires sampling specific input
lightcurves and comparing with the library of available shapes; this
necessarily requires a step outside of the MAF, but can easily be
enabled using the lightcurve simulation tools under development.
%  [NAME OF FED'S LIGHTCURVE
% TOOL].

Current simulations of the main survey show a broad uniformity of
visits, with thorough randomization of visit phase per period, giving
very good phase coverage with minimum phase gaps.

% % --------------------------------------------------------------------
%
% \subsection{Discussion}
% \label{sec:\secname:discussion}

For periodic variable science, two cadence characteristics should be avoided:
\begin{itemize}
\item an exactly uniform spacing of visits (which is anyway virtually impossible); \
\item a very non-uniform distribution, such as most visits concentrated in a few survey years.
 \end{itemize}

A metric for maximum phase gap will guard against the possibility that a
very unusual cadence might compromise the random sampling of periodic
variables.

In each case, it would help to jump-start science programs if some
fraction of targets had more complete measurements early in the survey.

% ====================================================================

 \subsection{Conclusions}

 Here we answer the ten questions posed in
 \autoref{sec:intro:evaluation:caseConclusions}:

 \begin{description}

 \item[Q1:] {\it Does the science case place any constraints on the
 tradeoff between the sky coverage and coadded depth? For example, should
 the sky coverage be maximized (to $\sim$30,000 deg$^2$, as e.g., in
 Pan-STARRS) or the number of detected galaxies (the current baseline 
 of 18,000 deg$^2$)?}

 \item[A1:] No strong constraint - but see Q4.

 \item[Q2:] {\it Does the science case place any constraints on the
 tradeoff between uniformity of sampling and frequency of  sampling? For
 example, a rolling cadence can provide enhanced sample rates over a part
 of the survey or the entire survey for a designated time at the cost of
 reduced sample rate the rest of the time (while maintaining the nominal
 total visit counts).}

 \item[A2:] An enhanced cadence could provide earlier discovery and period confirmation for a subset of targets, but this is not a high priority.

 \item[Q3:] {\it Does the science case place any constraints on the
 tradeoff between the single-visit depth and the number of visits
 (especially in the $u$-band where longer exposures would minimize the
 impact of the readout noise)?}

 \item[A3:] No strong constraint.

 \item[Q4:] {\it Does the science case place any constraints on the
 Galactic plane coverage (spatial coverage, temporal sampling, visits per
 band)?}

 \item[A4:] Most stars are in the galactic plane. For periodic pulsators, it is desirable to obtain multiple visits
in all filters, with at least 20-50 epochs in several filters. Obtaining this coverage is more important to periodic 
pulsator science than greater sky coverage.

 \item[Q5:] {\it Does the science case place any constraints on the
 fraction of observing time allocated to each band?}

 \item[A5:] No strong constraint, as long as there is a good representation of each, including u.

 \item[Q6:] {\it Does the science case place any constraints on the
 cadence for deep drilling fields?}

 \item[A6:] For deep drilling on the galaxy or nearby galaxies, the best cadences would cover a range of periods from
minutes to days - longer periods would be satisfactorily sampled by the main survey.

 \item[Q7:] {\it Assuming two visits per night, would the science case
 benefit if they are obtained in the same band or not?}

 \item[A7:] Different bands would provide more rapid characterization of targets, but this is not a strong benefit.

 \item[Q8:] {\it Will the case science benefit from a special cadence
 prescription during commissioning or early in the survey, such as:
 acquiring a full 10-year count of visits for a small area (either in all
 the bands or in a  selected set); a greatly enhanced cadence for a small
 area?}

 \item[A8:] Enhanced cadences could provide earlier science, or somewhat deeper science, depending on details, but 
periodic variables are not a strong driver for this.

 \item[Q9:] {\it Does the science case place any constraints on the
 sampling of observing conditions (e.g., seeing, dark sky, airmass),
 possibly as a function of band, etc.?}

 \item[A9:] No constraints that are particular to variable stars.

 \item[Q10:] {\it Does the case have science drivers that would require
 real-time exposure time optimization to obtain nearly constant
 single-visit limiting depth?}

 \item[A10:] No.

 \end{description}

% ====================================================================

\navigationbar

% ====================================================================

% ====================================================================
%+
% NAME:
%    planets.tex
%
% CHAPTER:
%    variables.tex
%
% ELEVATOR PITCH:
%-
% ====================================================================

% \section{Probing Planet Populations with LSST}
\subsection{Probing Planet Populations with LSST}
\def\secname{planets}\label{sec:\secname}

\credit{lundmb},
\credit{shporer},
\credit{stassun}

This section describes the unique discovery space for
extrasolar planets with LSST, namely,
planets in relatively unexplored environments.

% \subsubsection{Planets In Relatively Unexplored Environments}

A large number of exoplanets have been discovered over the past few
decades, with over 1500 exoplanets now confirmed. These discoveries are
primarily the result of two detection methods: The radial velocity (RV)
method where the planet's minimum mass is measured, and the transit
method where the planet radius is measured and RV follow-up allows the
measurement of the planet's mass and hence mean density. Other methods
are currently being developed and use to discover an increasing number of
planets, including the microlensing method and direct imaging. In
addition, the Gaia mission is expected to discover a large number of
planets using astrometry \citep{2014exha.book.....P}.

The {\it Kepler} mission has an additional almost 4000 planet
candidates. While these planet candidates have not been confirmed, the
sample is significant enough that planet characteristics can be studied
statistically, including radius and period distributions and planet
occurrence rates. LSST will extend previous transiting planet searches
by observing stellar populations that have generally not been
well-studied by previous transiting planet searches, including star
clusters, the galactic bulge, red dwarfs, white dwarfs (see below), and
the Magellanic Clouds (see \autoref{sec:MCs:MC_exoplanets}). Most known
exoplanets have been found relatively nearby, as exoplanet systems with
measured distances have a median distance of around 80~pc, and 80\% of these
systems are within 320~pc (exoplanets.org). LSST is able to recover transiting
exoplanets at much larger distances, including in the galactic bulge and the
Large Magellanic Cloud, allowing for measurements of planet occurrence rates in
these other stellar environments
\citep{2015AJ....149...16L,2015AJ....150...34J}.
Red dwarfs have often been
underrepresented in searches that have focused on solar-mass stars, however red
dwarfs are plentiful, and better than 1 in 7 are expected to host earth-sized
planets in the habitable zone \citep{2015ApJ...807...45D}.

Another currently unexplored environment where LSST will be able to
probe the exoplanet population is planets orbiting white dwarfs (WDs).
Such systems teach us about the future evolution of planetary systems
with main-sequence primaries, including that of the Solar System. When a
WD is eclipsed by a planet (or any other faint low-mass object,
including a brown dwarf or a small star) the radius and temperature
ratios lead to a very deep eclipse, possibly a complete occultation,
where during eclipse the target can drop below the detection threshold.
The existence of planets orbiting WDs has been suggested
observationally
\citep[e.g.,][]{2009ApJ...694..805F,2009AJ....137.3191J,2010ApJ...722..725Z,2012ApJ...747..148D}.
and theoretically \citep[e.g.,][]{2010MNRAS.408..631N}.
A few brown dwarf companions were already discovered
\citep[e.g.,][]{2006Natur.442..543M,2012ApJ...759L..34C,2006Sci...314.1578L,2014MNRAS.445.2106L},
and \citet{2015Natur.526..546V}
recently discovered a disintegrating planetary body orbiting a WD
\citep[see also][]{2015arXiv151006434C,2016ApJ...818L...7G,2016MNRAS.458.3904R}.

While most of the sky that LSST will survey will be at much lower
cadences than transiting planet searches employ, a sufficient
understanding of the LSST efficiency for detecting planets combined with
the large number of targets may still provide significant results.
Additionally, the multiband nature of LSST provides an extra benefit, as
exoplanet transits are achromatic while many potential astrophysical
false positives, such as binary stars, are not.
Indeed, as demonstrated by \citet{2015AJ....149...16L}, the multi-band LSST light curves
can likely be combined to create merged light curves with denser sampling and effectively
higher cadence, enabling detection of transiting exoplanets. The deep-drilling fields in
particular should prove to be a rich trove of transiting exoplanet detections, with
transit-period recoverability rates as high as $\sim$50\% or more among Hot Jupiters around
solar-type stars out to distances of many kpc and even the Magellanic Clouds in some cases
\citep{2015AJ....149...16L,2015AJ....150...34J}.
Yields may be expected perhaps as early as the third year of LSST operations
(Jacklin et al., in prep).
The ability to detect transiting planets outside of the deep-drilling fields is less certain;
here the details of the cadence among the various passbands will likely be particularly
important to assess carefully.

\subsection{Metrics}
\label{sec:\secname:metrics}
The detection of transiting planets will be dependent on having observations
that will provide sufficient phase coverage for transiting planets, with periods
that can range from less than one day up to tens of days. In order to address
this range of periods, an initial metric that can be used to address the detection
of transiting planets is the Periodogram Purity Function, discussed more thoroughly
in Section~5.2.1.

\subsection{Discussion}
\label{sec:\secname:discussion}
In general, the detection of transiting exoplanets with LSST will rely on
a small subset of potentially detectable planets that can be sufficiently
separated from statistical noise, rather than a clear threshold in a planet's
properties that would distinguish detectable planets vs. nondetectable planets.
This will mean that the best calculation of planet yields will have to come
from simulations of light curves for large numbers of stellar systems in order
to characterize LSST. The computation time involved in this process is sufficiently
prohibitive to prevent a metric being developed based directly on these
simulated light curves, however future work may be able to map relationships
between metric values for individual fields and the corresponding numbers
of planets that can be detected.

%
% % --------------------------------------------------------------------
%
% \subsection{OpSim Analysis}
% \label{sec:\secname:analysis}
%
% % --------------------------------------------------------------------
%
% \subsection{Discussion}
% \label{sec:\secname:discussion}
%
% ====================================================================
%
 \subsection{Conclusions}
%
% Here we answer the ten questions posed in
% \autoref{sec:intro:evaluation:caseConclusions}:
%
 \begin{description}
 \item[Q1:] {\it Does the science case place any constraints on the
 tradeoff between the sky coverage and coadded depth? For example, should
the sky coverage be maximized (to $\sim$30,000 deg$^2$, as e.g., in
 Pan-STARRS) or the number of detected galaxies (the current baseline 
of 18,000 deg$^2$)?}
\item[A1:] Longer time series and more data points per star are more valuable than greater sky coverage.
\item[Q2:] {\it Does the science case place any constraints on the
 tradeoff between uniformity of sampling and frequency of  sampling? For
 example, a rolling cadence can provide enhanced sample rates over a part
 of the survey or the entire survey for a designated time at the cost of
 reduced sample rate the rest of the time (while maintaining the nominal
 total visit counts).}
\item[A2:] Cadences suitable for transit timescales (hours-days) are very useful, and complementary to 
standard longer timescale cadence.
 \item[Q3:] {\it Does the science case place any constraints on the
 tradeoff between the single-visit depth and the number of visits
 (especially in the $u$-band where longer exposures would minimize the
 impact of the readout noise)?}
 \item[A3:] Number of visits is most valuable.
 \item[Q4:] {\it Does the science case place any constraints on the
 Galactic plane coverage (spatial coverage, temporal sampling, visits per
 band)?}
 \item[A4:] Most planets will be identified in galactic plane fields. Long visit sequences to
rich stellar fields may be productive, if crowding and flux contamination can be managed.
 \item[Q5:] {\it Does the science case place any constraints on the
 fraction of observing time allocated to each band?}
 \item[A5:] No strong constraint, but multiple bands care needed to test for achromatism.
 \item[Q6:] {\it Does the science case place any constraints on the
 cadence for deep drilling fields?}
 \item[A6:] Only applicable for deep drilling in the galaxy or local group - cadences that sample transit timescales (hours-days)
are complementary to long, sparse time series from the main survey.
 \item[Q7:] {\it Assuming two visits per night, would the science case
 benefit if they are obtained in the same band or not?}
 \item[A7:] No strong preference.
 \item[Q8:] {\it Will the case science benefit from a special cadence
prescription during commissioning or early in the survey, such as:
 acquiring a full 10-year count of visits for a small area (either in all
 the bands or in a  selected set); a greatly enhanced cadence for a small
 area?}
 \item[A8:] A mix of cadence types, including a greatly enhance cadence, in a rich stellar field, would serve to evaluate
different cadences and to test the photometric performance in crowded fields.
 \item[Q9:] {\it Does the science case place any constraints on the
 sampling of observing conditions (e.g., seeing, dark sky, airmass),
 possibly as a function of band, etc.?}
 \item[A9:] None unique to planets.
 \item[Q10:] {\it Does the case have science drivers that would require
 real-time exposure time optimization to obtain nearly constant
 single-visit limiting depth?}
 \item[A10:] No.
 \end{description}
%
% ====================================================================

\navigationbar

% ====================================================================

% --------------------------------------------------------------------

% --------------------------------------------------------------------

% --------------------------------------------------------------------

\chapter[Transients]{Eruptive and Explosive Transients}
\def\chpname{transients}\label{chp:\chpname}

Chapter editors:
\credit{ebellm},
\credit{fedhere}

Contributing Authors:

\credit{arcavi},
\credit{chomiuk},
\credit{Doctor},
\credit{Fong},
\credit{Haiman},
\credit{Kalogera},
\credit{AshishMahabal},
\credit{raffaellamargutti},
\credit{tmatheson},
\credit{StephenRidgway},
\credit{ohadshemmer},
\credit{nathansmith},
\credit{paulaszkody},
\credit{Trimble},
\credit{svalenti},
\credit{Zauderer}

% \section*{Summary}
% \addcontentsline{toc}{section}{~~~~~~~~~Summary}
%
% Executive summary goes here, highlighting the primary conclusions from
% the chapter's science cases. This should be abstract length, no more:
% say, 200 words.

% --------------------------------------------------------------------

\section{Introduction}

Explosive and eruptive transients are physically and
phenomenologically diverse.   What these events share
are rare, large-amplitude deviations from a quiescent state.  These
outbursts are typically unpredictable and of limited duration, and so their
discovery and characterization are sensitive to the detailed observing
strategy.  Often, followup observations with other facilities can provide
significant additional scientific value, but this creates a challenge to
identify candidate events while they are still visible.

Transients such as novae, supernovae (SNe), and long gamma-ray bursts (GRBs)
probe the final stages of stellar evolution. Tidal Disruption Events
(TDEs), short GRBs, and
Cataclysmic Variables (CVs) give us the opportunity to study
compact and binary objects. Massive star eruptions allow us to understand
mass loss mechanisms and chemical enrichment.
The brightest transients (GRBs, TDEs,
SNe) are light beams that can be seen over cosmic distances, and some
transients---most notably Type Ia SNe---are cosmological tracers.
In this chapter we focus on LSST's potential to advance the astrophysics of
eruptive and explosive transients; the use of SNe for cosmology is
discussed in \autoref{chp:cosmo}. Transients in the Milky Way Disk are
discussed in more detail in \autoref{chp:galaxy}.

Cadence choices will determine LSST's ability to discover, classify, and
characterize these events. However, due to their different time scales,
different phenomena will benefit from different sampling
strategies---sometimes
significantly different, and at times orthogonal.  Competing objectives
described in this chapter are at the heart of LSST's observing strategy and
cadence design.

When evaluating a particular observation or series of observations in
light of how they perform for a specific science case, it may be
helpful to think of metrics as lying along a continuum between
discovery and characterization. Discovery requires a minimum amount of
information to recognize an event or object as a candidate of
interest.  It is particulary relevant for science cases that require
triggering followup resources in real time from the live event stream.

Characterization, on the other hand, implies
that basic properties of the event may be determined from the
LSST observations, including but not limited to the classification of
the event.
It is particular relevant for science cases requiring analysis of large
samples of completed lightcurves.

Characterization and classification of transient events
benefits from substantial temporal sampling over the finite duration of the
event along with color information (perhaps contemporaneous).
Transient events slower than $\sim$ weeks may be adequately sampled by
a uniform LSST cadence.  Obtaining adequate sampling for faster-evolving
events may require special scheduling
strategies.  For some event types, LSST can only be expected to
provide a discovery service, and followup will necessarily be
performed elsewhere---so long as the cadence is sufficient to identify the
event type.
For some events, such as detecting electromagnetic counterparts to
gravitational wave events (GWs),
serendipitous discoveries are unlikely, but
enabling a ToO program would provide the opportunity for LSST to
contribute significantly to this science.

%The interpretation of a given metric along this continuum has
%implications for the subsequent action and analysis required,
%particularly as regards possible follow-up observations with other
%facilities.

We consider a non-exhaustive set of ``astronomical transients'' in the
paragraphs that follow. For a few of these transients, we quantify the
ability of LSST cadences to produce data useful for various science
goals. These case studies include SNe, GRBs, and GWs. For other
transient families (Novae, LBVs, TDEs) we provide more general
information in \autoref{sec:\chpname:future}, and we invite the
community to further develop the ideas proposed here, as well as
further other related goals, into quantified science cases.

\subsection{Targets and Measurements}
\label{sec:\chpname:targets}

%The class of transients includes a heterogeneous assortment of objects
%and phenomena.
\autoref{tab:transient_types} is a \emph{non-exhaustive} list of
phenomena to which we are referring as \emph{eruptive and explosive
  transients} in this document.

  \begin{table}
\begin{center}
  \begin{tabular}{| p{4.0cm} | p{4.0cm} | l | l | p{1.5cm}|}
    \hline

    Transient Type & Science drivers & Amplitude & Time Scale & Event Rate\\
\hline

Flare stars & Flare frequency, energy, stellar age, space weather & large &
	  min & very common\\

X-ray Novae & Interacting binaries, stellar evolution, SN progenitors,
nuclear physics & large & weeks & rare\\

Cataclysmic variables (\ref{sec:\chpname:CVtransients})& Interacting binaries, stellar evolution, compact
objects & large & min - days & common\\

LBV variability (\ref{sec:\chpname:LBVs})& Late stages stellar evolution, Mass loss, SN progenitors
& large & weeks-years & rare \\

Massive star eruptions (\ref{sec:\chpname:LBVs})& Late stages stellar evolution, Mass loss, SN progenitors & extreme & weeks-years & rare\\

Supernovae (\ref{sec:\chpname:SNtransients})& stellar evolution, feedback, chemical enrichment, cosmology & extreme & days - months & very common\\

GRBs (\ref{sec:\chpname:grbs})& jet physics, SN connection, stellar evolution & extreme & min - days
	  & rare (optical discovery) \\

TDEs (\ref{sec:\chpname:tdes})& Massive BH demographics, accretion physics & large & weeks-months & very rare\\

LIGO detections (GW, \ref{sec:\chpname:gw}) & EM characterization & unknown
	  & unknown & very rare\\

\emph{Unknown} & Discovery & unknown & unknown & rare\\

 \hline \end{tabular}
\end{center}
\caption{Overiew of major types of optical transients.  Common events have
hundreds or thousands of exemplars, while rare event classes have only a
	  few or tens.
\label{tab:transient_types}}
\end{table}

\begin{figure}[hbt]
\centerline{
\includegraphics[width=0.6\textwidth]{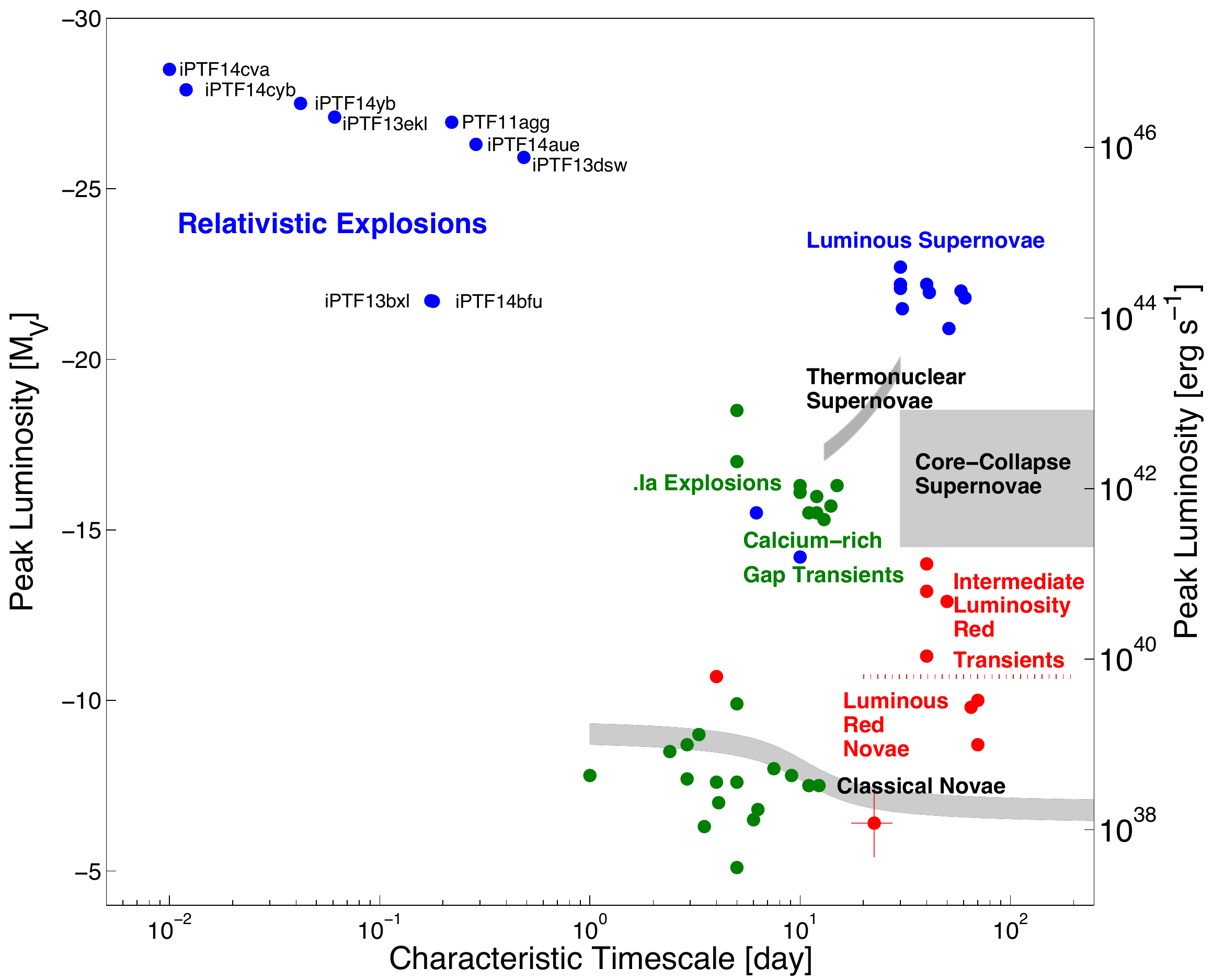}
}
\caption{
Peak luminosity-characteristic decay timescale plot for explosive
transients \citep[adapted from][]{2011PhDT........35K}.
}
\label{fig:transient_phase_space}
\end{figure}

%templates and stacks

%confusion

%

% --------------------------------------------------------------------

\subsection{Transient time scales}

Optical transients display a wide range of intrinsic timescales
(\autoref{fig:transient_phase_space}), and even some long-lived events
have short-duration features of interest.

For very short-lived phenomena (stellar flares, GRBs),
the main function of LSST will be to provide discoveries and/or simple
characterization.  Followup to discovery/identification, if required,
must take place elsewhere. This implies that the LSST observations
must be sufficient to recognize in real time that an event is fast-evolving
in order to to trigger followup. However, assessing the rise
slope is best done with a single filter,
because when observing with different
filters it is very difficult to separate lightcurve evolution from
color variations. Colors are informative when considering statistical
samples (\autoref{sec:\chpname:SNtransients}) as long as the epoch of peak
can be reliably determined.

SNe fall in an intermediate time range.  LSST will provide
multiple visits in multiple filters during the typical SN duration
(months).  This sampling may still be insufficient for many science
objectives, such as photometric classification of SN subtypes.
However, moderate changes to LSST
observing strategy may enhance the sampling for part of the sky part
of the time, greatly improving the usefulness of SN observations.
Metrics that assess the discovery rate of SN are included in
\autoref{chp:cosmo}.
Here we are interested in assessing the ability of
LSST to discriminate SN from other transients, SN subtypes from one
another, and to identify particularly interesting SNe: for example
those that show signature of shock break-out, companion interaction
in the early light curve, or would be candidates for \emph{flash
spectroscopy} follow-up \citep[e.g.,][]{2014Natur.509..471G}.
In addition, metrics that
quantify LSST's ability to constrain SN physics in a statistically
large sample of SN are needed.

TDEs have only recently started to be characterized in the optical
bands. The current sample of events show relatively long
time-scales (rise and decline of the light curve is over months). We
would like to assess through metrics how well TDEs can be
distinguished from supernovae based on their light curve shape and
color (in real-time so that followup observations can be triggered)
and how well the LSST light curves themselves can be used to model the
TDE emission and deduce the black hole properties.
%Ref. Science Book:
%10.6.1, \citet{Gezari2012, Chornock2014, Arcavi2014, Holoien2014,
%Holoien2015, Holoien2016}.

Large amplitude flares from AGN may mimic
explosive transients; they are discussed in \autoref{chp:agn}.

In addition we hope that LSST will provide a wealth of serendipitous
discoveries of yet-to-be-observed transients.  An ideal transient
discovery survey would include balanced coverage of all time scales. LSST
will cover longer time periods well, but will have to make some
choices of emphasis in coverage of shorter time-scales.

In the sections that follow we will use several case studies to assess
LSST's performance for a range of time-domain science:

\begin{itemize}
\item
  The ability of a given LSST cadence to discriminate truly young
  transients from those only first detected well after their explosion date (\autoref{sec:\chpname:transientsAge}).
  This ability is a crucial input to follow-up strategy design.
  We identify a region of lightcurve slope
  % the phase-space of rising speed and color
  that is characteristic of a
  variety of transients in their early phases and assess the ability of LSST's
  cadences to place transients within this phase-space.
  More sophisticated classification algorithms will likely be necessary, but
  are beyond the scope of this whitepaper.
\item
  The statistical constraints to a transient class that can be obtained
  over the course of the LSST survey, from the LSST survey data alone
  (assuming a successful classification). We discuss SN Ia early interaction
  signatures and IIb shock break-out (\autoref{sec:\chpname:SNtransients}).
\item
  The ability to identify in real-time a rapidly-evolving
  object of interest and
  trigger prompt follow-up observations. For this topic GRBs are used as
  a case study (\autoref{sec:\chpname:grbs}).
\item
  The value of triggered Target-of-Opportunity observations for
  following up very rare, fast-evolving events.   Here the kilonova
  counterparts expected from Advanced LIGO triggers are used as a
  case study (\autoref{sec:\chpname:gw}).
\item
  The insight that a cadence gives into single transient classes. We
  discuss CVs, massive star eruptions, and TDEs (\autoref{sec:\chpname:future}).

\end{itemize}

% --------------------------------------------------------------------

\subsection{Metrics}
\label{sec:\chpname:metrics}

Two metrics were developed and are used in this chapter specifically for transient phenomena:
\begin{itemize}
  \item{\metric{transientAsciiMetric}: accepts an ASCII file in input, so that realistic transient shapes can be used, with different shapes for different filters. The output can be the series of LSST observations (magnitude and error), or the fraction of transients detected (with user-specified constraints). This metric is used in \autoref{sec:\chpname:SNtransients}.}
  \item{\metric{GRBTransientMetric}: calculates the fraction of GRB-like transients detected (with user-specified constraints) using an $F(t) \propto t^{-\alpha}$
    lightcurve. This metric is used in \autoref{sec:\chpname:grbs}}.
\end{itemize}

Additionally, the standard MAF metrics
that quantify the gaps between consecutive visits to
a field within a night and over multiple nights
(\MAFmetric{IntraNightGapsMetric} and \MAFmetric{InterNightGapsMetric})
are of great value and are
heavily used throughout this chapter, in \autoref{sec:\chpname:transientsAge},
~\autoref{sec:\chpname:SNtransients}, and~\autoref{sec:\chpname:gw} for example.
These quantify the median times between consecutive visits to a field
within one night and over multiple nights, respectively.

Further metrics relevant to transient science are discused in \autoref{chp:galaxy},~\autoref{chp:cosmo}, and ~\autoref{chp:variables}.

Many science cases can be developed and tested with these metrics, and we
encourage users to do so. In addition, we are collecting a library of representative transient lightcurves in a separate GitHub repository\footnote{\url{https://github.com/LSSTTVS/LSST_TVS_RoadMap}} and we encourage readers to contribute their transient models or observations.

% --------------------------------------------------------------------

\subsection{OpSim Analysis}
\label{sec:\chpname:analysis}

The current set of simulated cadences
provide poor coverage in any one
filter for transient events longer than a visit pair ($\sim$30
minutes) and shorter than $\sim$ weeks (\autoref{fig:tgaps} and
\autoref{fig:tgaps_r}; \autoref{tab:visitgaps}).

As discussed in the subsequent sections, this
gap in the sampling hinders characterization of fast-evolving
transients.  A cadence with two visits separated by an hour or two rather
than 20 minutes would provide better discrimination.  A third visit in the
same night in a different filter would provide
color information valuable for realtime classification.
If a subset of those third
visits were in the same filter as the first two, it would improve the shape
characterization of the fastest-evolving transients.

\begin{table}
  \begin{tabular}{l|p{6cm}|c|c|c|c|p{5cm}}
    FoM & Brief description & {\rotatebox{90}{\opsimdbref{db:baseCadence}}}
          & {\rotatebox{90}{\opsimdbref{db:NEOswithVisitTriplets}}} &
          {\rotatebox{90}{\opsimdbref{db:NoVisitPairs}}} &
          {\rotatebox{90}{\opsimdbref{db:opstwoPS}}} & Notes \\
    \hline

    \thesection-1 & \footnotesize{\MAFmetric{IntraNightGapsMetric},
    any filter}      & 0.39 & 0.42 & 0.18 & 0.40 &
    \footnotesize{Median gap (hours) between consecutive observations of a field
	    in any pair
    of filters in a single night.} \\

    \thesection-2 & \footnotesize{\MAFmetric{IntraNightGapsMetric},
    $r$ band}      & 0.40 & 0.44 & 0.17 & 0.41 &
    \footnotesize{Median gap (hours) between consecutive $r$-band observations of a
	    field in a single night.} \\

    \thesection-3 & \footnotesize{\MAFmetric{InterNightGapsMetric},
    any filter}      & 3.0 & 3.9 & 2.0 & 3.0 &
	    \footnotesize{Median gap (days) between consecutive observations of a field
	    in any pair
    of filters over multiple nights.} \\

    \thesection-4 & \footnotesize{\MAFmetric{InterNightGapsMetric},
    $r$ band}      & 15.0 & 22.8 & 11.0 & 21.9 &
    \footnotesize{Median gap (days) between consecutive $r$-band observations
    of a field over multiple nights.} \\

\end{tabular}
\caption{
Inter- and intra-night revisit metrics in any filter and in $r$-band for
several simulated surveys.
}
\label{tab:visitgaps}
\end{table}

\begin{figure}[hbt]
\centerline{
	\includegraphics[width=0.45\textwidth]{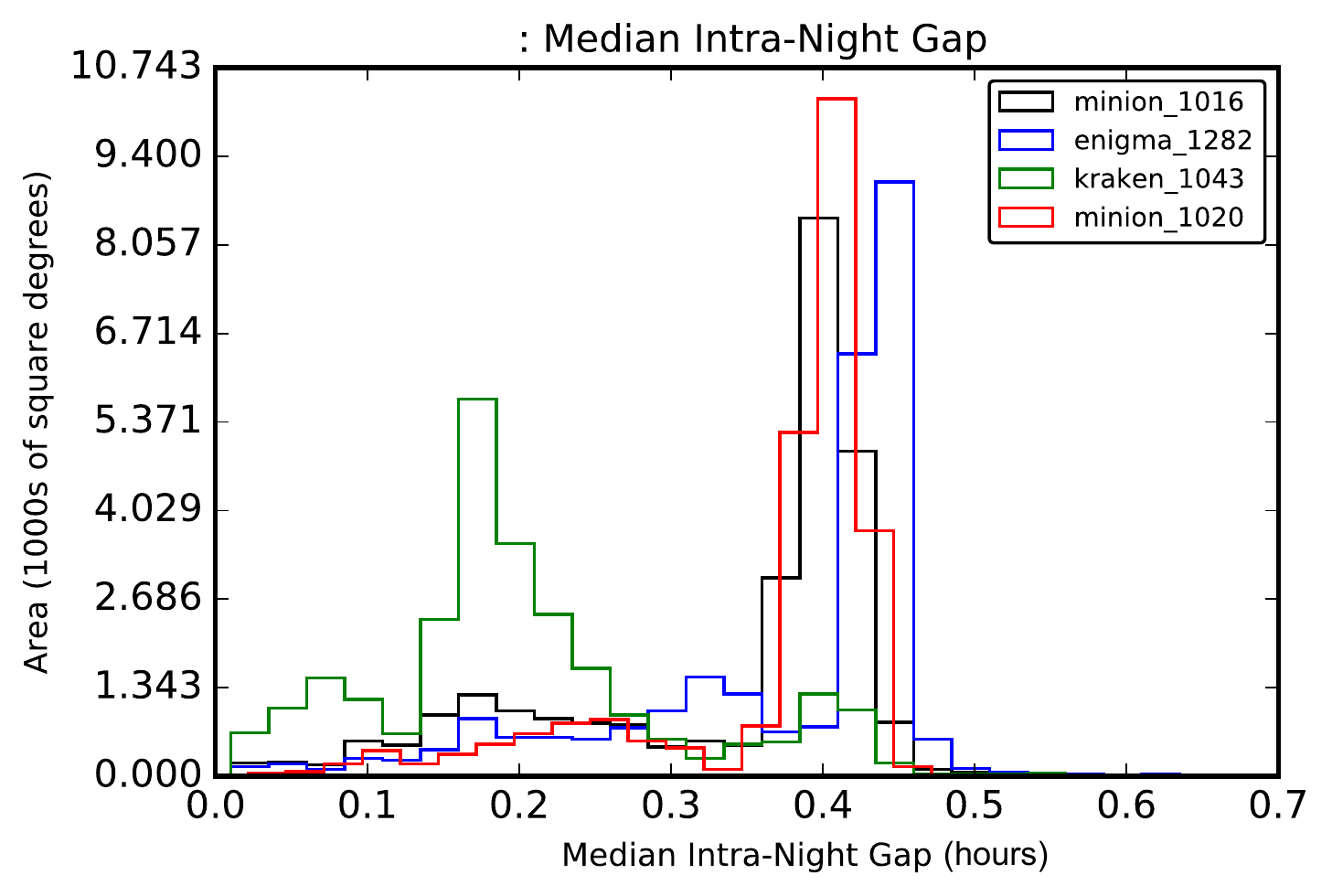}
	\includegraphics[width=0.45\textwidth]{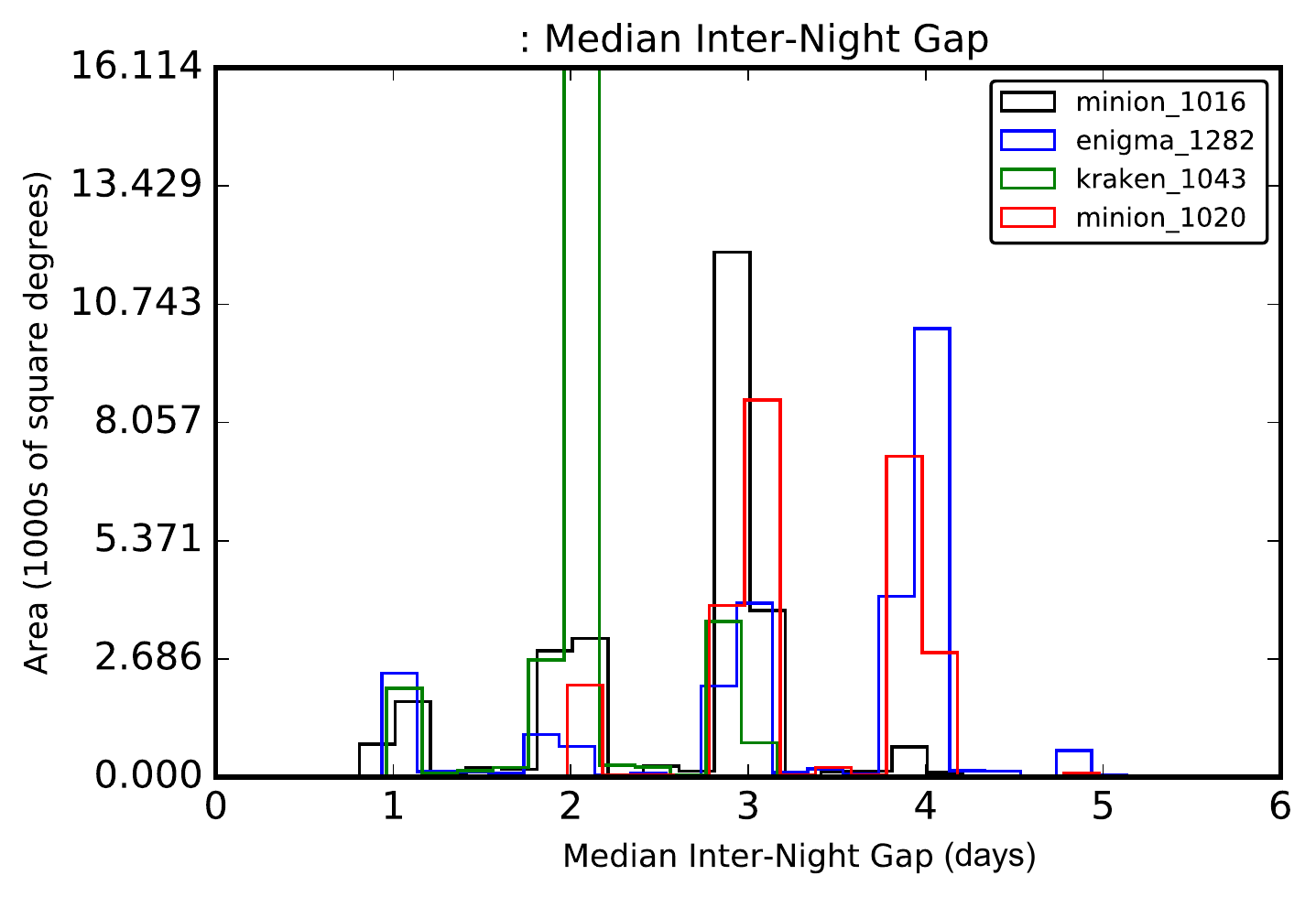}
}
\caption{ Histograms of median intra-night visit gaps (hours, left) and
	inter-night visit gaps (days, right)
for any band for several OpSim runs.  Current simulations provide little
	temporal coverage for transient timescales between 30 minutes and
	2--3 days.}
\label{fig:tgaps}
\end{figure}

\begin{figure}[hbt]
\centerline{
	\includegraphics[width=0.45\textwidth]{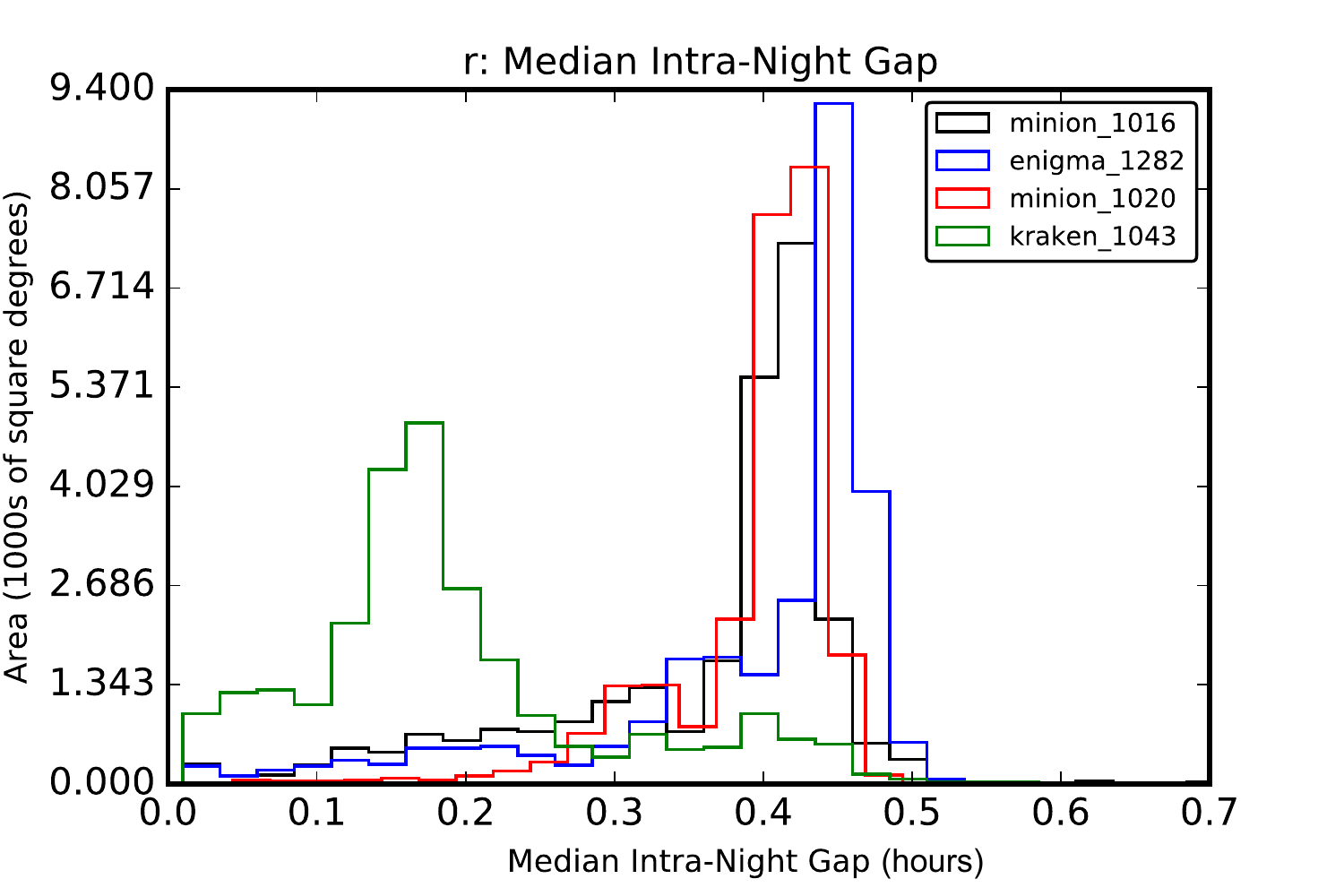}
	\includegraphics[width=0.45\textwidth]{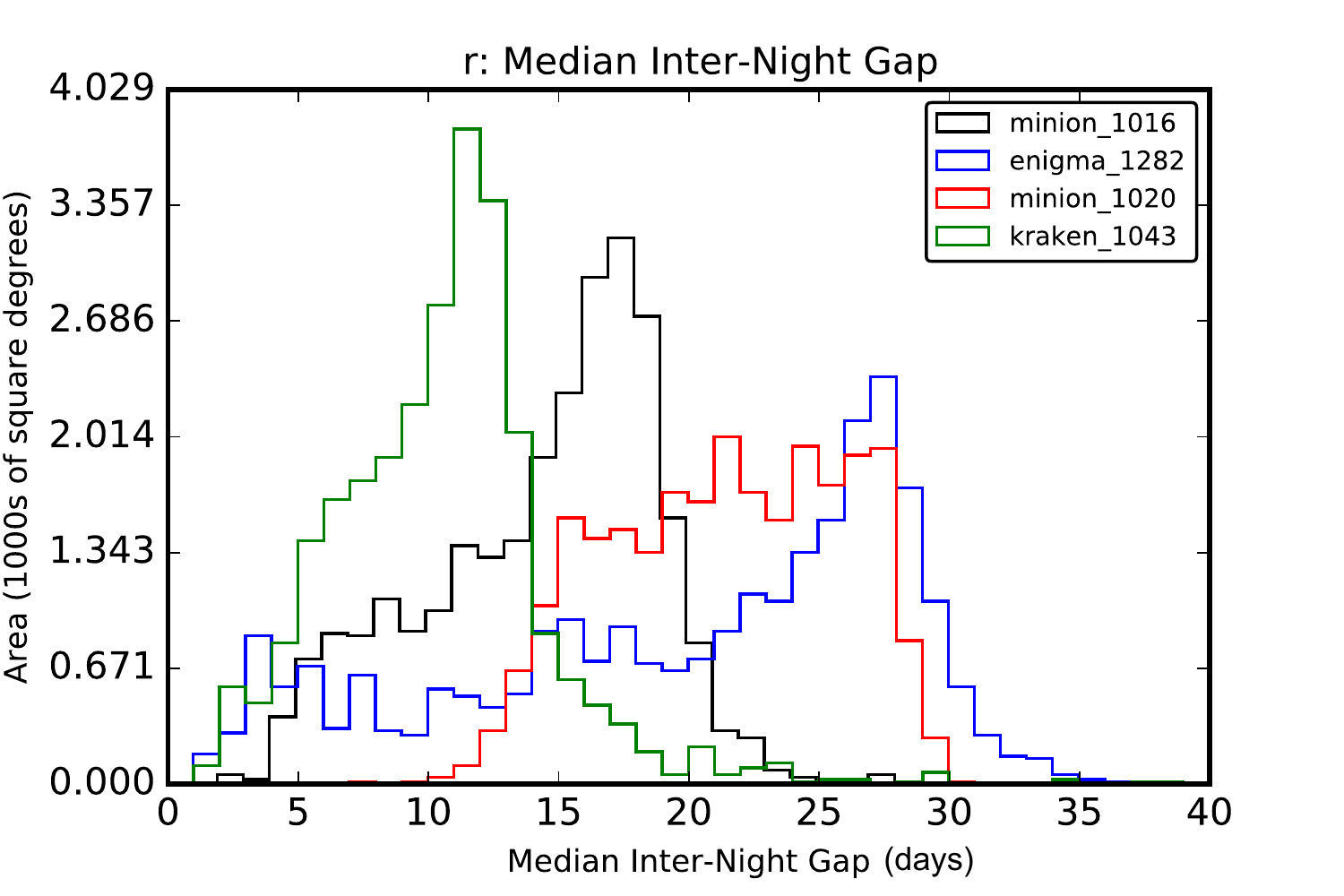}
}
\caption{ Histograms of median $r$-band intra-night visit gaps (hours,
	left) and inter-night visit gaps (days, right)
for several OpSim runs.  Current simulations don't revisit a field in the
	same filter for a period of weeks after a 30-minute visit pair.}
\label{fig:tgaps_r}
\end{figure}

% don't just care about gap between consecutive obs, though; first-last obs
% in a night
\emph{In fact, if the transient community were to design an optimal strategy for short and intermediate duration transients it would likely include 2 visits at a short time interval in different filter, and a third visit at a later time, but within the same night, with one of the two filters already used.}

% --------------------------------------------------------------------

\subsection{Discussion}
\label{sec:\chpname:discussion}

LSST's currently simulated cadences have significant cadence gaps for
timescales between nightly visit pairs and intra-night revisits.  For
many transient science cases, rolling-type cadences that improve the
sampling of a subset of events may be helpful in maximizing the
transient science that can be done with LSST: the minimum lightcurve
sampling required to adequately discover or characterize them may
still be larger than that provided by baseline cadences.  However,
even moderate adjustments (e.g., lengthening the visit pair spacing,
or optimizing the deep drilling filter strategy) may yield
improvements.

The metrics presented in these sections are initial efforts towards
quantifying these goals, and they suggest specific directions for new
OpSIM experiments.  More detailed efforts to understand and model the
challenging problem of transient classification with sparse
lightcurves will be needed in order to best guide LSST's time-domain
observing strategy.

\navigationbar

% --------------------------------------------------------------------

% SECTION:
%    transient.tex
%
% CHAPTER:
%    transients.tex
%
% ELEVATOR PITCH:
%    Explain in a few sentences what the relevant discovery or
%    measurement is going to be discussed, and what will be important
%    about it. This is for the browsing reader to get a quick feel
%    for what this section is about.
%
% COMMENTS:
%
%
% BUGS:
%
%
% AUTHORS:
%  Stefano Valenti , Federica Bianco (@fedhere)
%
% ====================================================================

\section{Realtime Identification of Young Transients}
\def\secname{\chpname:transientsAge}\label{sec:\secname}

\credit{svalenti}, \credit{fedhere} % (Writing team)

For many transients, the first few hours after event beginning reveal a tremendous amount of fundamental information. A large number of resources in the transient community are devoted to the study of the very early phases of transients (e.g. SNe, GRBs). Since real-time discrimination is a very hard task, it is then important to be able to select, among the large number of transients discovered by LSST, the youngest objects, in order to devise follow-up plans and best distribute precious follow-up resources. In this section we investigate the feasibility of identification of young transients (identified within few hours after the event occurs) from the LSST data alone, using the intra-night visits.

The Baseline Cadence~\opsimdbref{db:baseCadence} predicts that, on average,
fields in the main survey are revisited every $\sim3$ days in any filter
(\autoref{fig:enigmaGapAll}), and every $\sim15$ days when using only
$r$ band visits (\autoref{fig:enigmaGapr}).  Hence, we are most likely
to discover faint transients that are within 3 days of peak brightness.
However, for the small subset of nearby events, we can hope to discover
them within a few days of explosion.  The challenge is to discriminate
these truly young events from newly-discovered SN that are near peak
brightness.
Within the~\opsimdbref{db:baseCadence} cadence, and most cadences
realized thus far, the second intra-night visit occurs around 30 minutes (left panel of \autoref{fig:tgaps_r}).
after the first visit (to maximize the Solar System moving objects recovery,~\autoref{chp:solarsystem}). We want to understand {\emph{how the intra-night gap enables, affects, and can be used to maximize the identification of new transients as young}}, where, by young, we mean within a day of outburst/explosion.

To begin to answer this question, we limit our investigation to light curve shape in just the $r$ band, and specifically to what can be done in $r$ band. We have selected a representative set of transients with good photometric coverage in the first week after the the outburst/explosion (left panel of \autoref{fig:earlyslope}) and computed the light curve slope as a function of time in magnitudes per day (right panel of \autoref{fig:earlyslope}). In \autoref{fig:earlyrise} we report the change in $r$ brightness between the first and the second visit for the same set of transients as function of phase. The similarity matrices in \autoref{fig:simmatrix} represent the distance in this quantity for each transient pair for time gaps between observations ranging between 30 minutes and 24 hours.

Despite the heterogeneity in light curve shapes, most of the transients show a similar change in brightness on short time scales.
This confirms that early classification of the transient sub-type is a
major challenge. However, since in general young transients show a fast
increase in brightness, it is much easier to assess whether a transient is
\emph{young}.  Simply put, young transients will show a much larger
brightness change between visits than old events.
This discrimination is aided by a larger time gap between visits (e.g. 2 hours).
Within 30 minutes the change in brightness is of the order of $1\%$, or
even less, for most transients even within the first $\sim3$ days from the
start of the outburst/explosion (\autoref{fig:earlyslope}, left). Thus a
measurement of the change would require a $SNR\gtrsim500$ on each
measurement. Longer gaps give us more leverage: with a time gap of
$\sim2$~hours after the first visit, the change of brightness will increase
to $\sim5\%$. However the breadth of the gap is not unlimited: a gap of 24
hours imposes a significant delay in triggering follow up for these
fast-evolving events.

A natural metric to compare cadences for this purpose
is the median time difference between
the first and last observations of a field each night.  This differs from
the \MAFmetric{IntraNightGapsMetric}, as the latter only compares consecutive
observations of a field and hence underestimates the nightly time baseline
when there are three or more observations of a field in a night.

The classification of interesting transients, at an early stage, can be
aided by using supplementary information, such as historical information
from previous visits, and by color information about the transient. But to
properly assess the color of an evolving transient, the
gap between observations in different filters should not exceed a few hours
(\autoref{sec:\chpname:SNtransients}).

Finally, we stress that the quality and completeness of early
multiwavelengh data available at this time is limited. The sample of
astronomical transients used here is not comprehensive, and a uniform
set of homogeneous data of different transients is still needed in
order to further investigate the ideal separation between
observations, the need for color information, and the tension between
the two.

{\emph{In the light of these considerations, we recommend the
    simulation of a cadence with three visits per field, per night,
    two in the same band, but spaced by two hours or more, and a third
    in a different band. This criteria could be limited to the
    extragalactic sky, away from both the ecliptic plane and the
    galactic plane, where recovery of Solar System objects puts less
    strain on the cadence requirements.  The current 3-visit \OpSim run
    (\opsimdbref{db:NEOswithVisitTriplets}) is inadequate since it
    does not include the constraint of one visit being in a different
    filter.}}

{\emph{Furthermore, we note that the currently envisioned deep
    drilling cadence prioritizes depth per visit at the expense of a
    higher cadence. One hour per night on a deep drilling field
    reaches a depth that not required for almost all transient science
    cases, and by cycling to a different field each night, the time
    between visits for a particular field (4-5 days) is too long for
    many important science cases. Even with higher overhead, a more
    useful approach for nearly all transient science, that the deep
    drilling fields are designed to facilitate, would be to observe
    three or four of the available fields each night for 15 or 20
    minutes.}}

\begin{figure}[hbt]
\centerline{
\includegraphics[width=\textwidth]{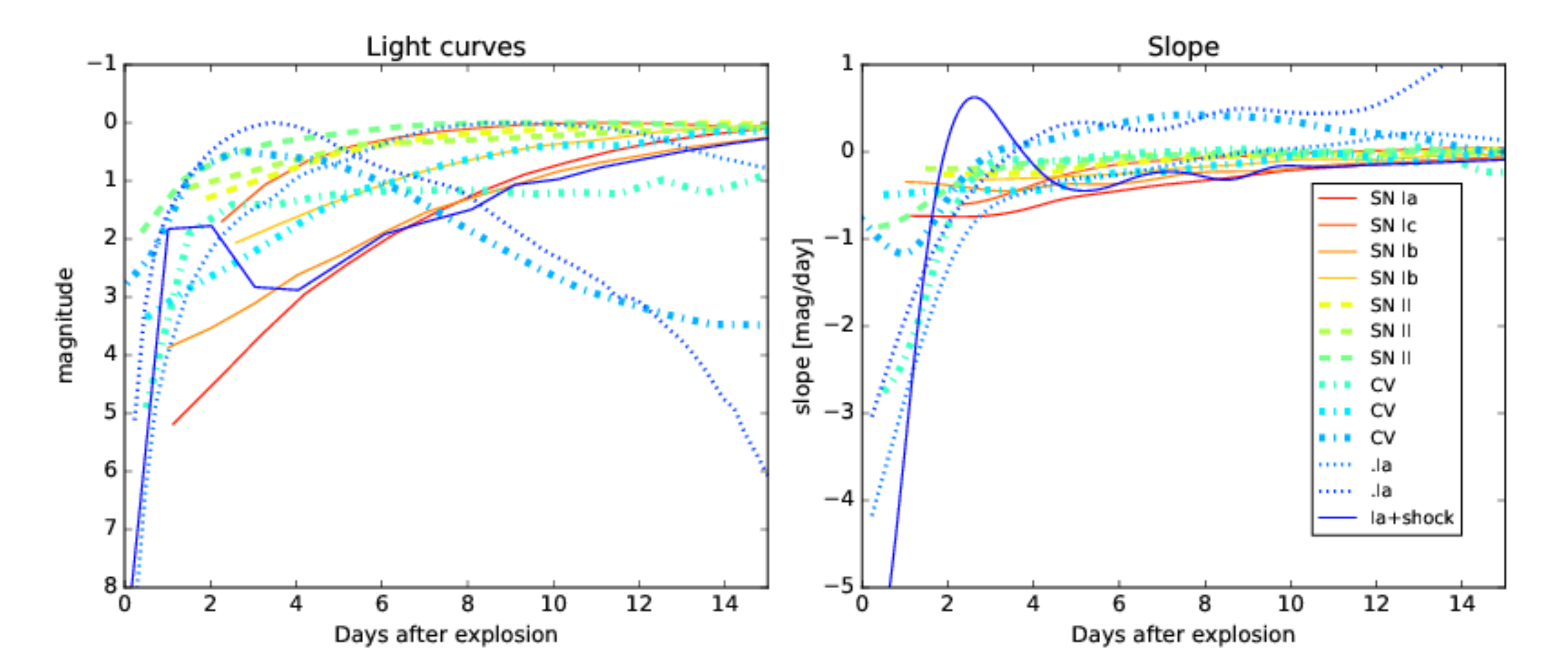}
}
\caption{\emph{Left}: $r'$-band light curve for representative transients as function of the phase from the beginning of the transient outburst/explosion for the first few days of the transient life. \emph{Right}: slope of the transient evolution. Data from: SN~Ia,~\citet{Olling15}; SNII,~\citet{Rubin16}; SN~.Ia,~\citet{Shen10}; SN~Ib,~\citet{Valenti11},~\citet{Cao13}; SN~Ic,~\citet{Mazzali02}; CV, ~\citet{Sokoloski13}, Finzell et al. (in prep), SN~Ia+interaction (see~\autoref{sec:\chpname:SNtransients}).}
\label{fig:earlyslope}
\end{figure}

\begin{figure}[hbt]
\centerline{
\includegraphics[width=\textwidth]{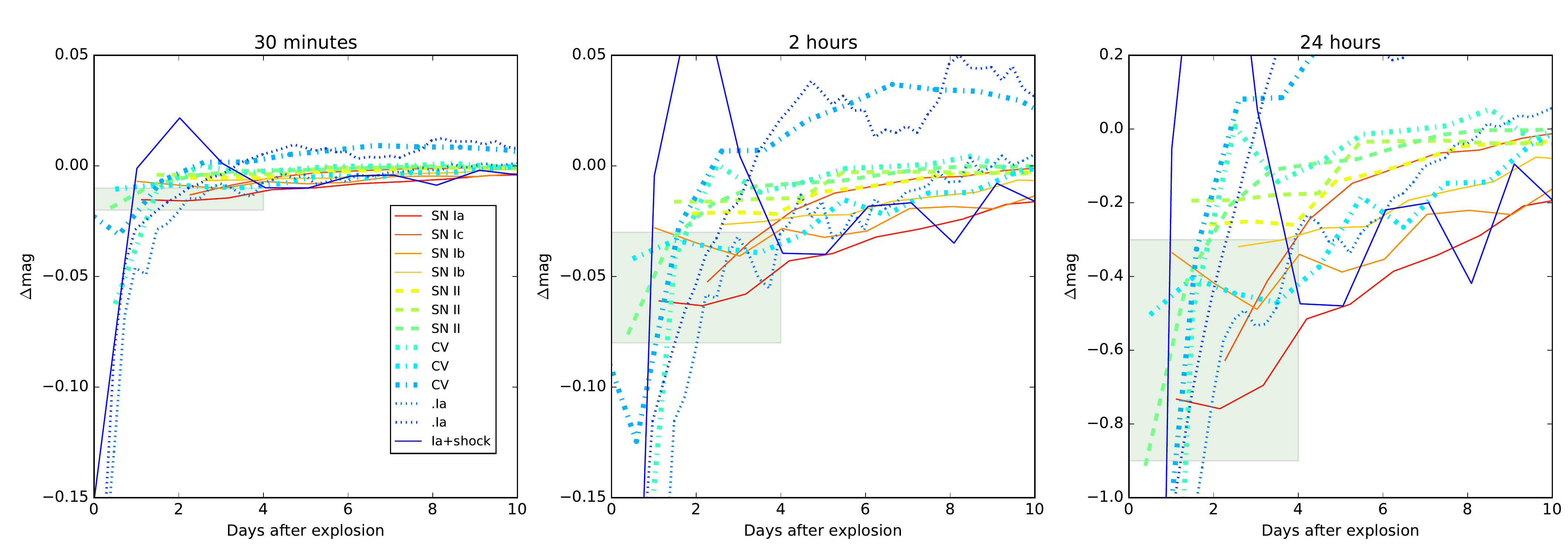}
}
\caption{Observed magnitude change between two consecutive observations for a representative set of astronomical transients, as a function of the phase. We consider observation gaps of 30 minutes  (left panel), 2 hours (central panel) and 24 hours (right panel).
}
\label{fig:earlyrise}
\end{figure}
\begin{figure}[hbt]
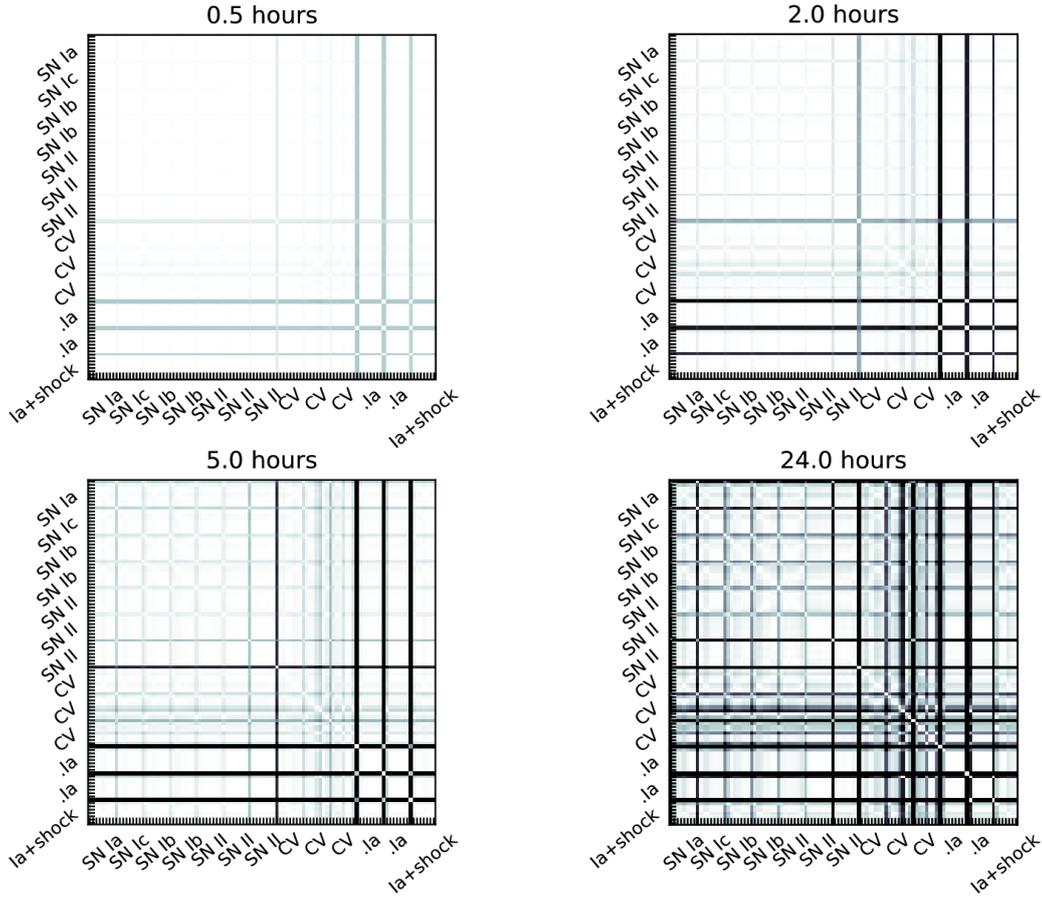

\centering
    \begin{subfigure}[t]{0.45\textwidth}
        \centering
        \includegraphics[width=0.8\textwidth]{figs/transients/TransientsAgeSimilarity1.pdf}
    \end{subfigure}%
    ~
    \begin{subfigure}[t]{0.45\textwidth}
        \centering
        \includegraphics[width=0.8\textwidth]{figs/transients/TransientsAgeSimilarity2.pdf}
    \end{subfigure}

    \begin{subfigure}[t]{0.45\textwidth}
        \centering
        \includegraphics[width=0.8\textwidth]{figs/transients/TransientsAgeSimilarity3.pdf}
    \end{subfigure}%
    ~
    \begin{subfigure}[t]{0.45\textwidth}
        \centering
        \includegraphics[width=0.8\textwidth]{figs/transients/TransientsAgeSimilarity4.pdf}
    \end{subfigure}

    \caption{Similarity matrix of the transients in \autoref{fig:earlyslope} and \autoref{fig:earlyrise} in $r$ filter for time gaps (starting with the top left panel) of  0.5, 2, 5, and 24 hours.  The similarity is measured as the Euclidean distance between the magnitude change of each transient's pair and it is represented in log scale, with darker colors indicating a larger distance (to a maximum difference $\lvert{\Delta\mathrm{Mag}_1 - \Delta\mathrm{Mag}_2}\rvert ~\sim~8$). For each transient's pair the similarity is calculated for 8 phases within the life of the transients, with the first observation starting at explosion, and as late as 3.5 days after explosion, in 12 hour increments, thus 8 values are associated to each transient pair (as indicated by the tick markers).}
  \label{fig:simmatrix}
\end{figure}

% ====================================================================

 \subsection{Conclusions}

 Here we answer the ten questions posed in
 \autoref{sec:intro:evaluation:caseConclusions}:

 \begin{description}

 \item[Q1:] {\it Does the science case place any constraints on the
 tradeoff between the sky coverage and coadded depth? For example, should
 the sky coverage be maximized (to $\sim$30,000 deg$^2$, as e.g., in
 Pan-STARRS) or the number of detected galaxies (the current baseline
 of 18,000 deg$^2$)?}

 \item[A1:] No strong constraint, as long as larger sky coverage does not compete with dense cadences
required for fast transients.

 \item[Q2:] {\it Does the science case place any constraints on the
 tradeoff between uniformity of sampling and frequency of  sampling? For
 example, a rolling cadence can provide enhanced sample rates over a part
 of the survey or the entire survey for a designated time at the cost of
 reduced sample rate the rest of the time (while maintaining the nominal
 total visit counts).}

 \item[A2:] Frequency of sampling is far more important than uniformity of sampling for early classification of interesting cadence. The rolling cadence is definitely the first step to take, but it is still not enough to identify young transients. A longer than .4 hours intra-night gap or, even better a one day cadence would allow young transients to vary enough to be identified. If the second visit occurs on 0.4-hour time scale, a different filter would be preferred (color information can be alternatively used to identify young transients)

 \item[Q3:] {\it Does the science case place any constraints on the
 tradeoff between the single-visit depth and the number of visits
 (especially in the $u$-band where longer exposures would minimize the
 impact of the readout noise)?}

 \item[A3:] Anything that reduces the number of visits per field potentially compromises the objectives.

 \item[Q4:] {\it Does the science case place any constraints on the
 Galactic plane coverage (spatial coverage, temporal sampling, visits per
 band)?}

 \item[A4:] The requirement for multiple visits/filters per night applies to all fields.

 \item[Q5:] {\it Does the science case place any constraints on the
 fraction of observing time allocated to each band?}

 \item[A5:] No.

 \item[Q6:] {\it Does the science case place any constraints on the
 cadence for deep drilling fields?}

 \item[A6:] As discussed in detail above, short bursts of visits separated by several days is not
satisfactory - deep drilling cadences can be devised to provide excellent sampling.

 \item[Q7:] {\it Assuming two visits per night, would the science case
 benefit if they are obtained in the same band or not?}

 \item[A7:] If two visits only, different filters may be the most effective. If more than two visits, the most closely spaced pair should be in the same filter, and the other visits should include at least one other filter.

 \item[Q8:] {\it Will the case science benefit from a special cadence
 prescription during commissioning or early in the survey, such as:
 acquiring a full 10-year count of visits for a small area (either in all
 the bands or in a  selected set); a greatly enhanced cadence for a small
 area?}

 \item[A8:] A greatly enhanced cadence would provide a strong test of the methodology and would jump-start the science.

 \item[Q9:] {\it Does the science case place any constraints on the
 sampling of observing conditions (e.g., seeing, dark sky, airmass),
 possibly as a function of band, etc.?}

 \item[A9:] None unique to transients.

 \item[Q10:] {\it Does the case have science drivers that would require
 real-time exposure time optimization to obtain nearly constant
 single-visit limiting depth?}

 \item[A10:] No.

 \end{description}

 \navigationbar

% --------------------------------------------------------------------

% ====================================================================
%+
% SECTION:
%    sn.tex
%
% CHAPTER:
%    transients.tex
%
% ELEVATOR PITCH:
%
%-
% ====================================================================

\section{Supernovae as Transients}
\def\secname{\chpname:SNtransients}\label{sec:\secname}

\credit{fedhere}

Supernovae (SNe) represent the final dramatic stages of the life of many
stars. The term SN covers a diverse set of phenomena: explosion of
low mass stars in binary systems, thermonuclear SN or SN Ia (also
discussed in \autoref{sec:supernovae}), explosions of high mass stars,
core collapse (CC) SNe, and even terminal explosions of more exotic
systems, yet to be understood, like Super Luminous SNe
(SLSNe). Phenomenologically, the observables of the explosion are
also diverse. The transient duration ranges from weeks to
months and even years. The electromagnetic energy radiated ranges between
$\sim0.1$ (faintest CC SNe), to $\sim1$ (SN Ia) and $\sim100$ (SLSNe)
$\times 10^{49}$ erg, corresponding to absolute magnitudes at peak
ranging between $\sim-19$ and $\sim-22$.

LSST's contribution to SNe studies can be substantial. Synoptic
surveys such as SDSS, SNLS, PTF, PanSTARRS have revolutionized our
understanding of SN time and again, exposing their diversity,
and revealing different progenitor channels. LSST's first crucial
input will be discovery: the normal type Ia SN rate out to redshift
$z=1$ is estimated to be $\sim200 ~(\mathrm{sq. deg.})^{-1}$ per
year\footnote{\url{http://www.lsst.org/sites/default/files/docs/Wood-Vasey_086.11.pdf}},
and SN~Ia represent only about 1/4 of all SN
events~\citep{Li11b}: tens of millions of stars will explode
within the LSST footprint every year. The main factors affecting
LSST SN science concern:
\begin{enumerate}
\item
LSST's SN discovery power,
\item
LSST's discrimination power,
\item
the quality of the statistical sample over time.
\end{enumerate}
Items 1 and 2 are \emph{time sensitive}, while the latter
is not, although it is interesting to understand the pace at which a
science question can be advanced in the lifetime of LSST.

{\bf \emph{Discovery:}} the SN~Ia discovery is rate is a standard LSST time-domain metric: a
fraction of $\sim40\%$ SN~Ia $z\lesssim0.5$ are expected to be discovered pre-peak
luminosity within the standard LSST survey
(e.g. \opsimdbref{db:baseCadence}~Figure~\ref{fig:enigmaEarlySNe}). The
topic of SNe discovery is discussed in further detail in
\autoref{sec:supernovae}.

The next step is then {\emph{discrimination}}, and the question we need
to answer, for SNe as well as for most other transients, is: will LSST
photometry allow us to distinguish SN from other transients, and to
distinguish the different types of SN? And further: will this be
achievable in time to appropriately direct follow-up efforts? This is
particularly difficult considering that photometric classification
schemes have only achieved modest performances in distinguishing, for
example, SN~Ic from SN~Ia.

When a large statistical sample of SNe is generated, LSST's photometry
may allow setting constraints on the diversity of the sample, even as a
standalone survey, without the aid of follow-up efforts.  {\emph Thus
  LSST \bf{alone} can shed light on the diversity within the
  population of SN}, which in turn may constrain the genesis of the
explosion.\footnote{Reliable typing of a SN and redshift determination
  would still require auxiliary data.} For SN~Ia, where the exploding
star is a carbon-oxygen White Dwarf (WD), a major outstanding question
that can be answered by an LSST photometric sample is
what is the percentage of SN Ia that arise from a
\emph{double-degenerate} (DD) progenitor system (a carbon-oxygen WD-WD
binary), from a \emph {single-degenerate} (SD) system (a WD-Main Sequence
 or WD-Red Giant (RG) binary), or from a \emph{merger} (a WD-WD
 binary with a He and a carbon-oxygen WD).
 Answering this question would reduce the
scatter in the Hubble diagram if SNe from different progenitors are
shown to require different standardization~\citep{Scolnic2014}. On the
CC~SN side: the diversity of SN sub-classes, and the relationship
between them (is there a phenomenological continuum or are they actually
distinct classes, e.g. between IIp and IIL, or Ib and IIb?) is yet to
be understood. Exceptionally well-studied objects may answer these
questions: individual SN Ia with tight constraints on the progenitor
system show, for example, that both single and double degenerate
progenitors exist (e.g. SN 2011fe, \citealt{Li11}, ~\citealt{Olling15}
and PTF 11kx, \citealt{Dilday12}). However, a statistical sample is
needed to set constraints on populations~\citep{Hayden2010, Bianco11}.

Thus the technical question to be answered is: how much detail can be
sacrificed in favor of sample size without compromising diagnostic
power? And the diagnostic power relies on color and sampling: thus
what is the trade-off between cadence in the same filter, and
observations in different filters. Specifically, transients can be
distinguished early from two photometric characteristics: rise time
and color. There is a tension between these observables, as discussed
in Section~\ref{sec:\chpname:transientsAge}. Obtaining colors relies
of course on obtaining photometry in different bands as close as
possible to \emph{simultaneously}.  However, assessing the rise slope
is best done with a single filter, so prompt characterization also
needs multiple epochs within a night, although separated by at least a
few hours, in the same filter, as observing with different
filters it is impossible (or very hard) to separate shape from
color. Colors are an important diagnostic for the
statistical sample: as long as the epoch of peak is reliably assessed
coadded light curves can be studied, which is the goal of the analysis
that follows.

\subsection{Distinguishing progenitor scenarios}

In this chapter we envision and design a SN related metric that works
on a large sample (months-to-years of LSST data) and assesses the
ability to characterize the contribution of SNe with specific features
to the global population: as a test case we will use the identification of
an early blue excess for SN type Ia, a signature of interaction with a
companion, and thus of a SD progenitor. Equivalently, the presence of
an early blue excess in CC~SNe could be the signature of
shock breakout which directly measures the radius of the progenitor
star. We perform the simulation on SN~Ia since statistical studies of
samples that set constraints on progenitor fractions (fraction of DD
vs SD progenitors) exist and can be used as a
benchmark~\citep{Hayden2010, Bianco11}.  What we evaluate as a
\emph{figure of merit} (FoM) for this science deviates from the
guidelines for figures of merit, since LSST will surely be able to
answer this question \emph{at some point} and we measure \emph{how
  fast} LSST can answer this question. Our FoM is time
within the survey required to achieve a sufficiently large sample of
SNe to enable us to distinguish
populations with different contribution from DD and SD progenitors.
We rely on simulations of the observables of the population for
different sample sizes, and on the \MAFmetric{transientAsciiMetric} to
determine the detectability of interacting vs non-interacting
SNe. We are developing a metric (\texttt{colorGapMetric}) to assess
the gap between of detections in 2 filters. In the meantime, we rely on
the estimated of the gap between observations in a single filter, and
in any filters (see~\autoref{sec:\chpname:analysis}).

We simulate interacting SNe from the Nugent templates \citep{Nugent02}
injecting the angle-dependent effects of interaction with a companion
as simulated by \citep{Kasen10}, for a $2~M_\odot$ and a $6~M_\odot$
MS companion stars, and a $1~M_\odot$ RG companion, following the
procedure designed in ~\citep{Bianco11}. We create synthetic progenitor
populations with a fraction of single degenerate progenitor systems
$0.05 \leq f_\mathrm{SD} \leq 0.6 $ in 0.05 intervals, and random lines of
sight with respect to the binary's geometry. One such lightcurve, with
maximal interaction effects, is shown in \autoref{fig:kasenlc}, also
indicating how it may be observed by LSST. For each population, we
simulate the observation of colors by selecting random epochs with a
granularity of 1 day within the first 10 days after explosion, and
subtracting the magnitude in different filters at the same epoch
$\pm~1$~day for each SN, and we include the effects of observational
noise by generating datapoints from a draw within a Gaussian
distribution centered at the color measured in the previous step and
with standard deviation $\sigma_\mathrm{pop} = 0.1$, 0.3, and 0.5.
The SNR requirement is
translated into a requirement on each
observation of $\mathrm{SNR} >
\frac{1.0}{\sqrt{2.0}~\sigma_\mathrm{pop}}$.
We generate populations of $N_\mathrm{pop}=100$,~1000,~10000 $z~=~0.5$ SNe,
observed in $g'-r'$, as a representative case. Because the effect is
heavily chromatic and becomes essentially negligible
by $r$ band, $u'-i'$ gives the most leverage. However $g'$ and $r'$
are the best observed LSST bands in most cadences. An extension of
this work should then consider $g'-r'$, $u'-r'$, $g'-i'$, and $u'-i'$.

\begin{figure}[hbt]
\centerline{
\includegraphics[width=0.6\textwidth]{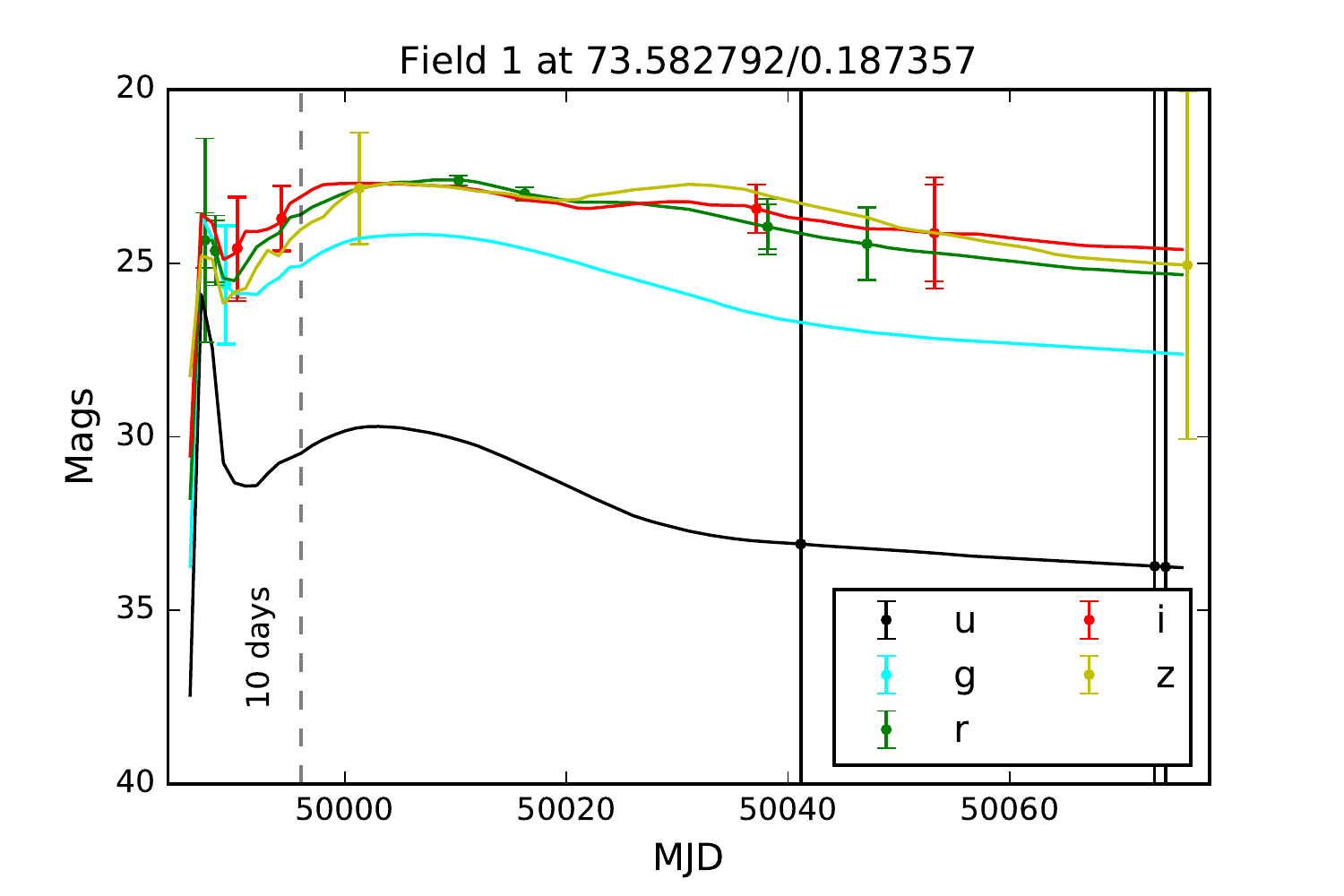}
}
\caption{ A normal SN Ia lightcure at z=0.5 showing interaction with a
  RG companion as seen from the most favorable viewing angle: the
  effect of interaction as simulated by \citet{Kasen10} is added on
  top of a lightcurve simulated from the \citealt{Nugent02}
  templates. The data points represent one possible set of LSST
  observations of this transient, obtained by running the
  \MAFmetric{transientAsciiMetric}.  This particular event is detected in
  $g'$, $r'$, and $i'$ within the first 10 days.}
\label{fig:kasenlc}
\end{figure}

We perform Kolmogorov-Smirnoff ($KS$) and Anderson-Darling ($AD$) tests
to evaluate our diagnostic power as a function of sample quality,
$SNR$, and sample size, $N_\mathrm{pop}$.  In
\autoref{tab:SNprogenitors} we report the ability to distinguish a
population with a $f_\mathrm{SD} > x$ from $f_\mathrm{SD}=0.05$; \emph{the
number reported is the SN~Ia fraction from SD progenitors that can be
distinguished at a $p\mathrm{-value}~\leq ~0.05$}.

\begin{table}
\begin{center}
  %\begin{tabular}{ c | c| c| c | c | c| c| c | }
  \begin{tabular}{ c | c| c| c |  }
$g-r$&\bf{$N_\mathrm{pop}$=100}&\bf{$N_\mathrm{pop}$=1,000}&\bf{$N_\mathrm{pop}$=10,000}\\%& $g-i$&\bf{$N$=100}&\bf{$N$=1,000}&\bf{$N$=10,000} \\
  \hline
  {\bf $\mathrm{SNR}~\geq~1.4$}&  -  & 0.2 & 0.1 \\%& &  -  & -  &  0.2 \\
  {\bf $\mathrm{SNR}~\geq~2.0$}&  - & 0.2 & 0.1 \\%& &  -  & 0.2 & 0.1 \\
  {\bf $\mathrm{SNR}~\geq~7.0$}& 0.2 & 0.1 & 0.1 \\%& & 0.4 & 0.1 & 0.1 \\
%&&&&&&&\\ $u-i$&\bf{$N$=100}&\bf{$N$=1,000}&\bf{$N$=10,000} & $u-r$&\bf{$N$=100}&\bf{$N$=1,000}&\bf{$N$=10,000}\\
% \bf{$\sigma_\mathrm{pop} = 2.0$}&  -  & -  &  0.2 & &  -  & -  &  0.2 \\
% \bf{$\sigma_\mathrm{pop} = 1.0$}&  -  & 0.2 & 0.1 & &  -  & 0.2 & 0.1 \\
% \bf{$\sigma_\mathrm{pop} = 0.5$}& 0.4 & 0.1 & 0.1 & & 0.4 & 0.1 & 0.1 \\

 \hline
  \end{tabular}
  \caption{Minimum fraction of single-degenerate (SD) SN~Ia in a sample of size $N_\mathrm{pop}$ of $z~=~0.5$ SNe that can be distinguished from a population with a fraction of 0.95 double-degenerate (DD) and 0.05 SD SNe~Ia, for a given quality cut on each observed datapoint ($\sigma_\mathrm{pop}$).}
\label{tab:SNprogenitors}
\end{center}
\end{table}

Now we can evaluate how long it will take for a given LSST
cadence to obtain a sufficient number of observations in the 2 desired
bands, separated by less than 1 day, that pass the SNR requirements.
This should be done in a full Monte Carlo simulation, injecting
light curves with the proper light curve shape at the proper rate.  Note
that, because the early light curves of interacting SD SN~Ia are
brighter, they should be more easily detected. However, at this stage we
can take some shortcuts. \emph{First shortcut}: we evaluate the
relative observability of SNe with excess, and SNe without excess at
$z~=~0.5$ and adjust the number of detections according to
the injected ratio.  The relative detectability can be assessed with
the \MAFmetric{transientAsciiMetric}, which allows us to see how OpSim
recovers observations of transients with realistic shapes. We conclude
that for RG-WD progenitors the detectability is enhanced by $\sim50\%$ in
 $g'$ compared to SD progenitors, and slightly less in $r'$.
Then we extract from the \MAFmetric{transientAsciiMetric}, the number of
\emph{color observations}, i.e. observations in 2 bands within 1 day
of each other, each fulfilling our SNR requirement for the color for
3-, 6-, and 12 months of survey in year 1.

With the goal of distinguishing a SD contribution of 10\% to the SN Ia
population from a 5\% contribution to a three-sigma level ($p$-value
$<0.05$) we need more than 1000 detections within 1 day in 2 filters
at a $\mathrm{SNR}~\geq~7$: \autoref{tab:SNprogenitors}. But the pairs
of observations we recovered in the previous steps are within the
first 10 days but with any gap in time. \emph{Second shortcut}: To
include the constraint that the detections should be within 24 hours
we use to the \MAFmetric{InterNightGapsMetric}, which is plotted in
~\autoref{fig:enigmaGapAll}.  For the \opsimdbref{db:baseCadence} we
estimate $~\sim10\%$ of the observation are revisited within a
night. With the assumption that this is likely to happen in two
different filters, which is \emph{non-conservative}, but neglecting
intra-night observations that may happen in the two different filters,
which is a \emph{conservative} assumption our numbers drop by a factor
10. The lightcurves are injected with an event rate designed to be
consistent with the discovery rate measured in \ref{sec:supernovae}.

With all these assumptions standing, we find that that only 3 months
of survey are sufficient to provide a sufficiently large and
sufficiently high SNR sample for our purpose, and improve on the
findings on this topic that were achieved with
SDSS~II~\citep{Hayden2010}, and 3 years of SNLS data~\citep{Bianco11}
with \opsimdbref{db:baseCadence} or the
\opsimdbref{db:NEOswithVisitTriplets}. The
\opsimdbref{db:NEOswithVisitTriplets} requires three visits, thus
increasing the timeline for inter-night observations. Although it does not
require the observations to be in any specific filters, and with the
addition of the third visit within the same night, it increases the
typical intra-night gap, it outperforms \opsimdbref{db:baseCadence} slightly.
It is possible that a detailed investigation
of the true \emph{inter-night gap between different filters}, or the
addition of a requirement in the cadence that one of the night filters
be different than the others (possibly requiring an increased gap
between two of the three images to minimize filter changes) would
provide valuable data for this kind of studies even faster.

\emph{This exercise demonstrates the power of LSST in collecting large high SNR samples of transients, but we must remind the reader that these conclusions, and generally large sample analysis, rely on having properly identified both the transient class (normal SN~Ia) and the date of maximum! This, once more, highlights the importance of prompt identification and classification: for SN~Ia this likely will limit this work to objects that could be identified spectroscopically, enhancing the importance of follow-up.}

\autoref{fig:sndetect} shows the detection rate for SN~Ia at $z=0.5$
in absence of shock interaction as a function of SNR (obtained by summing in quadrature the errors on $g'$
and $r'$) for 3, 6 months, and a year of
\opsimdbref{db:baseCadence} and \opsimdbref{db:NEOswithVisitTriplets}.
% (ideally I will plot it for the other survey as well tonight)

\begin{table}
  \begin{tabular}{l|p{8cm}|c|c|p{3cm}}
    FoM & Brief description & {\rotatebox{90}{\opsimdbref{db:baseCadence}}}
	  & {\rotatebox{90}{\opsimdbref{db:NEOswithVisitTriplets}}} & Notes \\
    \hline
    \thesection-1 & \footnotesize{\texttt{SNIaprojenitorMetric},
    \nolinebreak{\texttt{1,000 detections}}}      & 3 & $<3$ &
    \footnotesize{Time in month to collect 1,000 relevant observations to distinguish a 5\% from a 10\% SD contribution} \\
    \thesection-2     & \footnotesize{\texttt{SNIaprojenitorMetric},
    \texttt{10,000 detections}}      & $>12$ & $>12$ &
    \footnotesize{Time in month to collect 10,000 relevant observations in months  to distinguish a 5\% from a 10\% SD contribution.}\\
\end{tabular}
\caption{FoMs for statistical SN Ia progenitor studies to assess the contribution of SD progenitors to the SN Ia population.
}
\label{tab:SummarySNprojs}
\end{table}

\begin{figure}[hbt]
  \centerline{
    \includegraphics[width=0.6\textwidth]{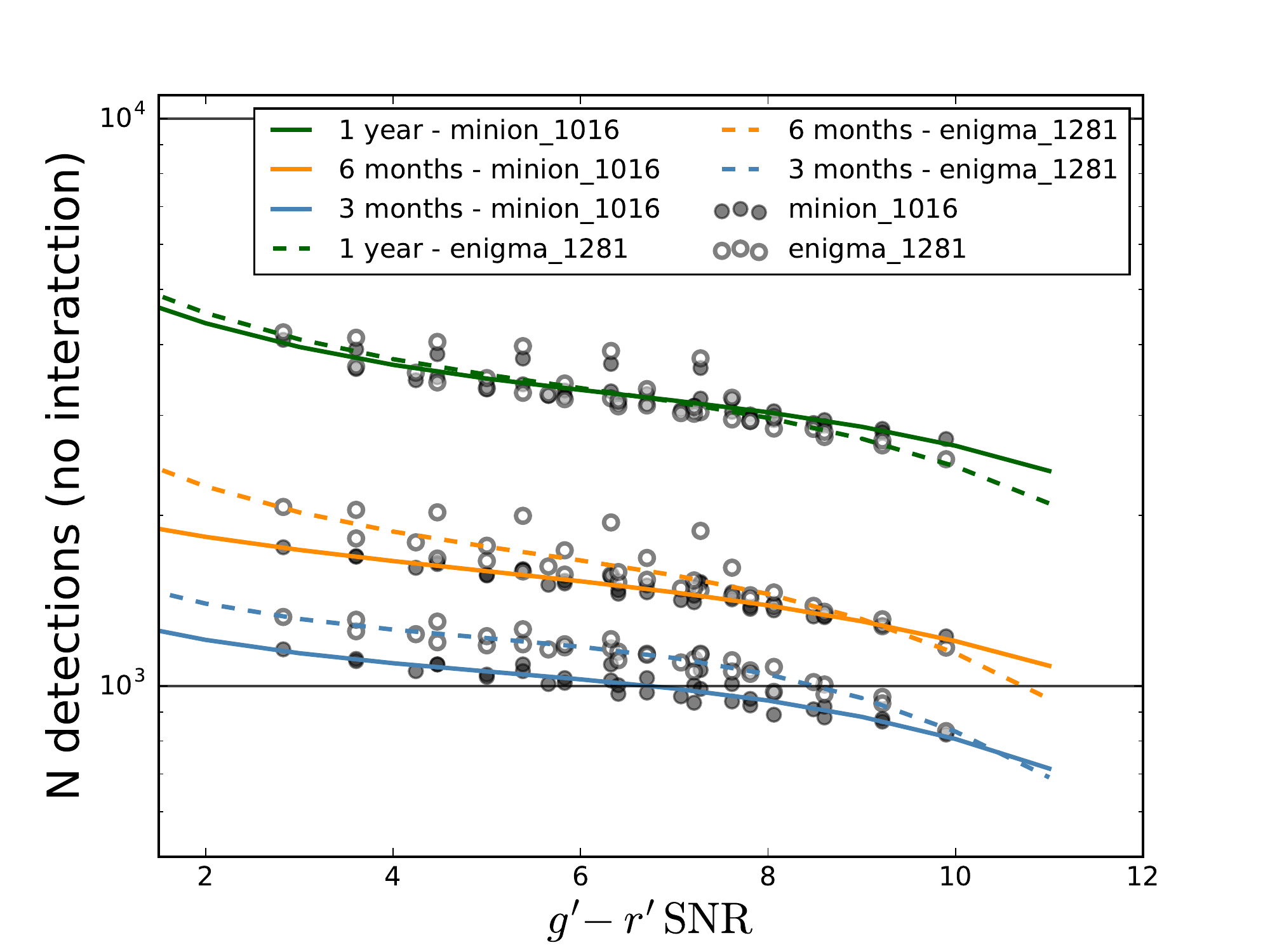}
  }
  \caption{
    Normal SN Ia light cure at z=0.5 detected by the \opsimdbref{db:baseCadence} (solid lines) and \opsimdbref{db:NEOswithVisitTriplets} cadence (dashed line) in 3 months, 6 months, and 1 year, that provide color information useful to constrain the progenitor distribution. Line are third-degree polynomial fits.}
  \label{fig:sndetect}
\end{figure}

% ====================================================================
%
 \subsection{Conclusions}

 Here we answer the ten questions posed in
 \autoref{sec:intro:evaluation:caseConclusions}:

 \begin{description}

 \item[Q1:] {\it Does the science case place any constraints on the
 tradeoff between the sky coverage and coadded depth? For example,
 should the sky coverage be maximized (to $\sim$30,000 deg$^2$, as e.g.,
 in Pan-STARRS) or the number of detected galaxies (the current baseline
 but with 18,000 deg$^2$)?}

 \item[A1:] Yes, although this question can be answered with a full simulation that includes a coordinate dependent event rate while the current simulation assumes uniform probability of a transient in the field-of-view and cannot evaluate this trade-off

 \item[Q2:] {\it Does the science case place any constraints on the
 tradeoff between uniformity of sampling and frequency of sampling? For
 example, a rolling cadence can provide enhanced sample rates over a
 part of the survey or the entire survey for a designated time at the
 cost of reduced sample rate the rest of the time (while maintaining the
 nominal total visit counts).}

 \item[A2:] Yes: this science case is sensitive to both. A more sophisticated simulation which also measures the ability to correctly identify the epoch of maximum would be more powerful to answer the question.

 \item[Q3:] {\it Does the science case place any constraints on the
 tradeoff between the single-visit depth and the number of visits
 (especially in the $u$-band where longer exposures would minimize the
 impact of the readout noise)?}

 \item[A3:] Yes, because the diagnostic power depends on both the SNR of each observation and the gap between observations.

 \item[Q4:] {\it Does the science case place any constraints on the
 Galactic plane coverage (spatial coverage, temporal sampling, visits
 per band)?}

 \item[A4:] No.

 \item[Q5:] {\it Does the science case place any constraints on the
 fraction of observing time allocated to each band?}

 \item[A5:] Yes, since it relies on obtaining observations in at least 2 filters.

 \item[Q6:] {\it Does the science case place any constraints on the
 cadence for deep drilling fields?}

 \item[A6:] Yes, although the results have not been analyzed separately for WFD and DD fields.

 \item[Q7:] {\it Assuming two visits per night, would the science case
 benefit if they are obtained in the same band or not?}

 \item[A7:] No. Although  we would benefit greatly from 2 visits in the same filters, and one visit in a different filter to constrain simultaneously shape and color."

 \item[Q8:] {\it Will the case science benefit from a special cadence
 prescription during commissioning or early in the survey, such as:
 acquiring a full 10-year count of visits for a small area (either in
 all the bands or in a  selected set); a greatly enhanced cadence for a
 small area?}

 \item[A8:] A full event rate needs to be included in the simulation to answer this question.

 \item[Q9:] {\it Does the science case place any constraints on the
 sampling of observing conditions (e.g., seeing, dark sky, airmass),
 possibly as a function of band, etc.?}

 \item[A9:] Indirectly, since detection efficiency depends on SNR.

 \item[Q10:] {\it Does the case have science drivers that would require
 real-time exposure time optimization to obtain nearly constant
 single-visit limiting depth?}

 \item[A10:] No.

 \end{description}

 \navigationbar

% --------------------------------------------------------------------

% ====================================================================
%+
% SECTION:
%    grb.tex
%
% CHAPTER:
%    transients.tex
%
% ELEVATOR PITCH:
%-
% ====================================================================

\section{Gamma-Ray Burst Afterglows}
\def\secname{\chpname:grbs}\label{sec:\secname}

\credit{ebellm}

Gamma-ray bursts (GRBs) are relativistic explosions typically classified
by the temporal duration of their initial gamma-ray emission: Long GRBs,
that mark the endpoint of the lives of some massive stars, and short
GRBs, believed to originate from the merger of binary neutron stars. GRB
emission is known to be beamed: the initial prompt gamma-ray emission is
seen only for observers looking at the jet axis. The longer-wavelength
X-ray, optical, and radio afterglow may be seen both by on- and off-axis
observers.  The latter case is known as an orphan afterglow, due to the
absence of gamma-ray emission. On- and off-axis afterglows are predicted
to have different temporal signatures in the optical: On-axis events
decay as a power-law until a jet break, while off-axis events should be
fainter and show an initial rise (Figure \ref{fig:afterglow_lcs}).
Despite systematic searches, no convincing orphan afterglow candidates
have yet been discovered, limiting our knowledge of the beaming fraction
of GRBs and hence their true rates. Well-sampled orphan afterglow
lightcurves would also permit study of the GRB jet structure.

\begin{figure}[hbt]
\centerline{
\includegraphics[width=0.6\textwidth]{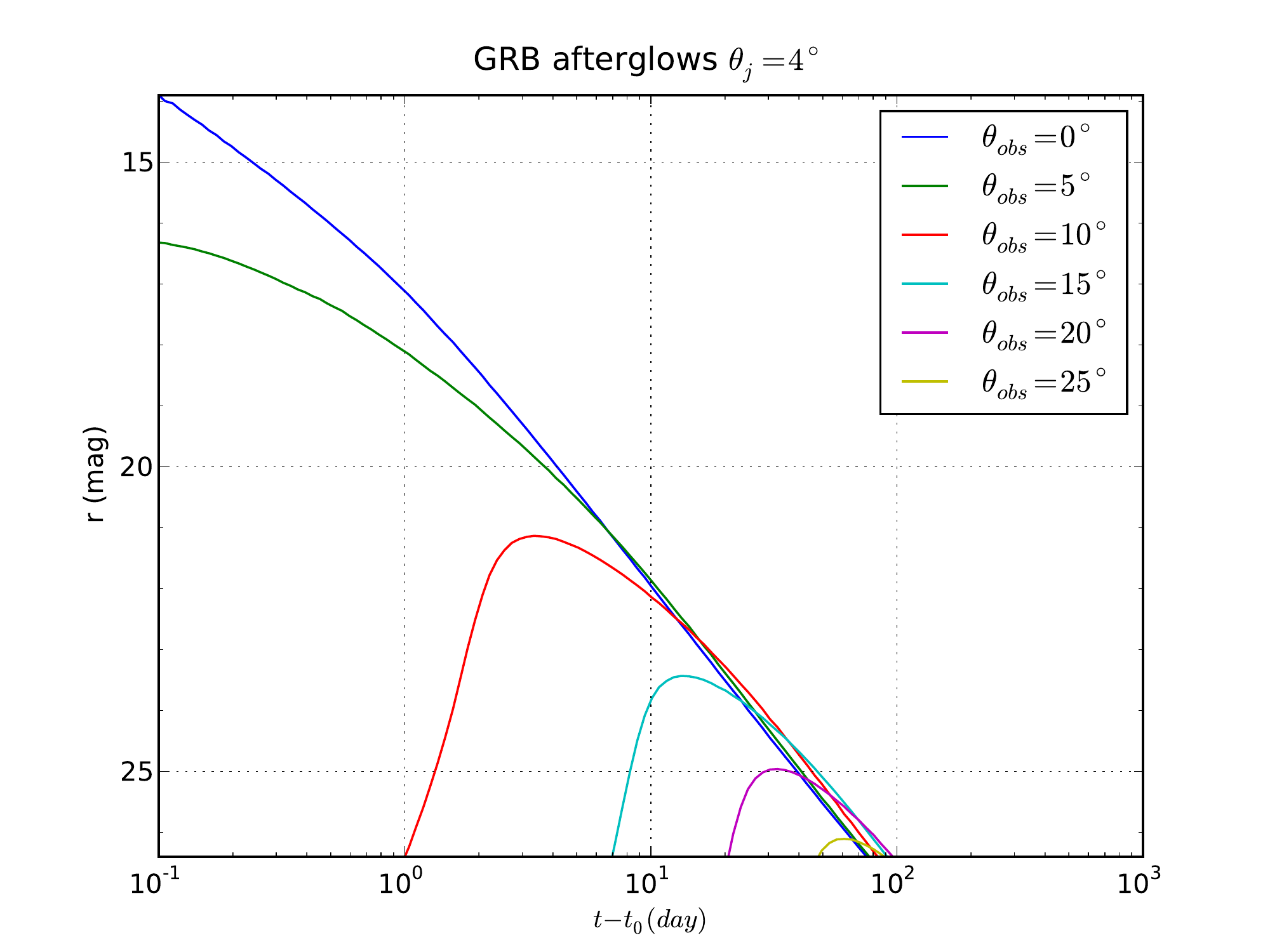}
}
\caption{ Predicted light curves of GRB afterglows by off-axis angle
with respect to the jet axis $\theta_{\rm obs}$ \citep[Figure
8.8,][]{2009arXiv0912.0201L}. The forward shock model is derived from
\citet{2002ApJ...576..120T} and assumes a jet half opening angle
$\theta_j = 4^\circ$, the isotropic equivalent energy of $E_{\rm iso} =
5\times10^{53} \rm erg$, ambient medium density $n = 1$ g cm$^{-3}$, and
the slope of the electron energy distribution $\rm p = 2.1$. The
apparent AB $r$-band magnitudes assume a source redshift $z = 1$. }
\label{fig:afterglow_lcs}
\end{figure}

Because of their rarity, in all but one case \citep{2015ApJ...803L..24C}
to date GRBs have been discovered using their prompt emission by hard
X-ray or gamma-ray all-sky monitors. This selection imposes biases on
the population of relativistic explosions we observe. Baryon-loading in
the GRB jet---a ``dirty fireball'' \citep{2003ApJ...591.1097R}---can
lead to on-axis events without gamma-ray emission.  Only one plausible
candidate has been identified to date \citep{2013ApJ...769..130C}.
Discovery of new dirty fireballs---if distinguished from off-axis
events--would clarify the rates of these events and enhance our
understanding of the diversity of stellar death.

LSST is the survey most capable of resolving these decades-old
questions.  Due to its large aperture and etendue, LSST can detect
faint, fast-fading, and rare cosmological events, potentially enabling
population studies of the high-redshift universe.
\citet{2015A&A...578A..71G} estimated LSST could detect 50 orphan
afterglows each year, more than any other planned survey.

%deep survey helps due to time dilation

%beaming fraction and true rates; jet structure; dirty fireballs?
%GRB-SN connection; probe high-z star formation?

%other fast transients: Fast transients and SN shock breakout?  flash spectroscopy

The challenge of detecting and recognizing GRB afterglows in the LSST data in
real time makes this science case a useful proxy for other fast transient
science cases that benefit from $N > 2$ visits per night.  In particular, this
includes discovering supernovae soon after explosion for flash spectroscopy or
shock breakout searches.

% need appropriate cadences to support value of realtime alert stream

% --------------------------------------------------------------------

\subsection{Target measurements and discoveries}
\label{sec:\secname:targets}

GRB afterglow discovery is among the science cases that places the
greatest stress on the LSST cadence.  Because afterglows fade
rapidly---dropping several magnitudes in the first few hours---high
cadence observations are required to detect the fast fading. If an
afterglow candidate can be recognized in real time, it will be possible
to trigger TOO spectroscopy (to measure a redshift and confirm the event
is cosmological), X-ray and radio observations (to detect a high-energy
counterpart and the presence of a jet), and additional photometry (to
characterize the lightcurve evolution).  If there is no source at the
location of the transient in the coadded reference image, two
consecutive observations in the same filter separated by an hour or two
are the minimum required to potentially trigger followup of a
fast-fading event. However, a third observation within a night or
two---ideally in the same filter---would improve the purity of the
sample and reduce the reliance on triggered followup. Observations in
other bands at high cadence are less useful because they require
assumptions about the event's SED and its evolution to determine if a
source is truly fading.

Distinguishing orphan afterglows from on-axis events (whether conventional
GRBs or dirty fireballs) will also require more than two detections.
Orphan events may prove harder to recognize in real time, because they are
intrinsically fainter than on-axis events and show an initial rise rather
than a rapid decay (Figure \ref{fig:afterglow_lcs}).  Additionally, because
of relativistic time dilation, high redshift events are easier to detect,
but these events will be fainter and more difficult to follow up.
Accordingly, population studies of orphan afterglow candidates may by
necessity be conducted with LSST photometry alone.  Such studies may only
be productive if LSST has sufficiently frequent revisits to a field in a
single filter.

% --------------------------------------------------------------------

\subsection{Metrics}
\label{sec:\secname:metrics}

The core figure of merit for GRB afterglows is simply the raw number of
on- and off-axis events detectable in two, three, or more observations,
preferably in a single filter.

The appropriate way to derive these detections is to conduct a Monte
Carlo simulation of a cosmological population of GRBs and fold it
through the LSST observing cadence \citep[cf.][]{2011PASP..123.1034J}.
We are developing this infrastructure for the MAF framework.

In the meantime, simplified metrics can give us a general idea of how well
a given cadence can characterize fast-evolving transients such as GRBs.  We
have created a new metric, \metric{GRBTransientMetric}, that replaces the
linearly rising and decaying lightcurve in \MAFmetric{TransientMetric} with
the $F \sim t^{-\alpha}$ decay characteristic of on-axis afterglows.  (For
the time being, we neglect the jet break that steepens the rate of decay;
this implies that our detectability estimates are optimistic.)

We simulate on-axis afterglows at random sky positions using the parameters of
\citet{2011PASP..123.1034J}. The R-band apparent magnitude at 1 minute
after explosion is randomly drawn from a Gaussian with $\mu=15.35$,
$\sigma=1.59$ and decays with $\alpha=1.0$.  For these estimates we
simply assume zero color difference between in all LSST bands.
There are roughly 300 on-axis GRBs per year with these parameters;
we calculate the average fraction of these events which have at least one,
two, or three detections in any single filter.

% Can use https://github.com/lsst/sims_maf/blob/master/python/lsst/sims/maf/metrics/tgaps.py or https://github.com/lsst/sims_maf/blob/master/python/lsst/sims/maf/metrics/cadenceMetrics.py (Inter/Intra-night) to get histograms.  Would be nice to extend to single-band, N-offset

% --------------------------------------------------------------------

\subsection{OpSim Analysis}
\label{sec:\secname:analysis}

We ran \metric{GRBTransientMetric} on several OpSim v3.3.5 runs with a range of
characteristics:  \opsimdbref{db:baseCadence}, the baseline cadence;
\opsimdbref{db:NEOswithVisitTriplets}, with three visits per WFD field;
\opsimdbref{db:NoVisitPairs}, with no visit pairs; and
\opsimdbref{db:opstwoPS}, a PanSTARRS-like cadence.

\autoref{tab:SummaryGRBs} lists the fraction of on-axis afterglows
detected in at least one, two, and three visits in a single filter.

Because of its wider areal coverage, the PanSTARRS-like cadence of
\opsimdbref{db:opstwoPS} maximizes the fraction of events detected in
one and two epochs.  Not surprisingly, the triplet-visit WFD cadence of
\opsimdbref{db:NEOswithVisitTriplets} maximizes the three-epoch detection
rate.

\begin{table}
  \begin{tabular}{l|p{6cm}|c|c|c|c|p{5cm}}
    FoM & Brief description & {\rotatebox{90}{\opsimdbref{db:baseCadence}}}
	  & {\rotatebox{90}{\opsimdbref{db:NEOswithVisitTriplets}}} &
	  {\rotatebox{90}{\opsimdbref{db:NoVisitPairs}}} &
	  {\rotatebox{90}{\opsimdbref{db:opstwoPS}}} & Notes \\
    \hline
    \thesection-1 & \footnotesize{\metric{GRBTransientMetric},
    \texttt{nPerFilter}\,$=1$}      & 0.17 & 0.16 & 0.20 & \textbf{0.21} &
    \footnotesize{Fraction of GRB-like transients detected in at least one
    epoch.} \\
    \thesection-2     & \footnotesize{\metric{GRBTransientMetric},
    \texttt{nPerFilter}\,$=2$}      & 0.12 & 0.10 & 0.09 & \textbf{0.14} &
    \footnotesize{Fraction of GRB-like transients detected in at least two
    epochs in any single filter.} \\
    \thesection-3     & \footnotesize{\metric{GRBTransientMetric},
    \texttt{nPerFilter}\,$=3$}      & 0.05 & \textbf{0.08} & 0.04 & 0.04 &
    \footnotesize{Fraction of GRB-like transients detected in at least
	    three epochs in any single filter.}
\end{tabular}
\caption{Mean figures-of-merit (FoMs) for on-axis Gamma-Ray Bursts for one,
two, and three detections in a filter.
The best value of each FoM is indicated in bold.
The wider areal coverage of \opsimdbref{db:opstwoPS} improves its detection
rate of GRBs in one and two epochs, while the triplet visits
in \opsimdbref{db:NEOswithVisitTriplets} naturally improve the
three-detection efficiency.
}
\label{tab:SummaryGRBs}
\end{table}

% --------------------------------------------------------------------

\subsection{Discussion}
\label{sec:\secname:discussion}

An LSST cadence purely designed for discovering GRB afterglows would
include three or more visits to each field every night, with the visits
separated by an hour or two. Moreover, it would be conducted in a single
filter in order to best identify the lightcurve shape of off-axis
events.

While the current surveys simulated are far from this ideal
(usually just two closely spaced visits, with subsequent revisits days
later), nonetheless an appreciable number of GRBs are detectable.
\opsimdbref{db:NEOswithVisitTriplets} would detect about 25 events each
year in three epochs, already potentially the largest sample of untriggered
afterglows.

However, some care is required in interpreting these values:
while the GRB afterglow fades rapidly over the first day of the explosion
(Figure \ref{fig:afterglow_lcs}), at later times a 30 minute visit
separation is not enough to reveal significant evolution in the lightcurve.
We intend to enhance our metric to require that detections are counted only
if significant evolution is statistically distinguishable with 1\%
photometry.

In future work we intend to simulate cosmological populations of on- and
off-axis in order to better determine how many events could be discovered
in time to trigger real-time followup\footnote{or conversely,
constrain the area
over which high-cadence observations are required to detect a meaningful
population of GRBs.}.

\begin{figure}[hbt]
\centerline{
\includegraphics[width=0.6\textwidth]{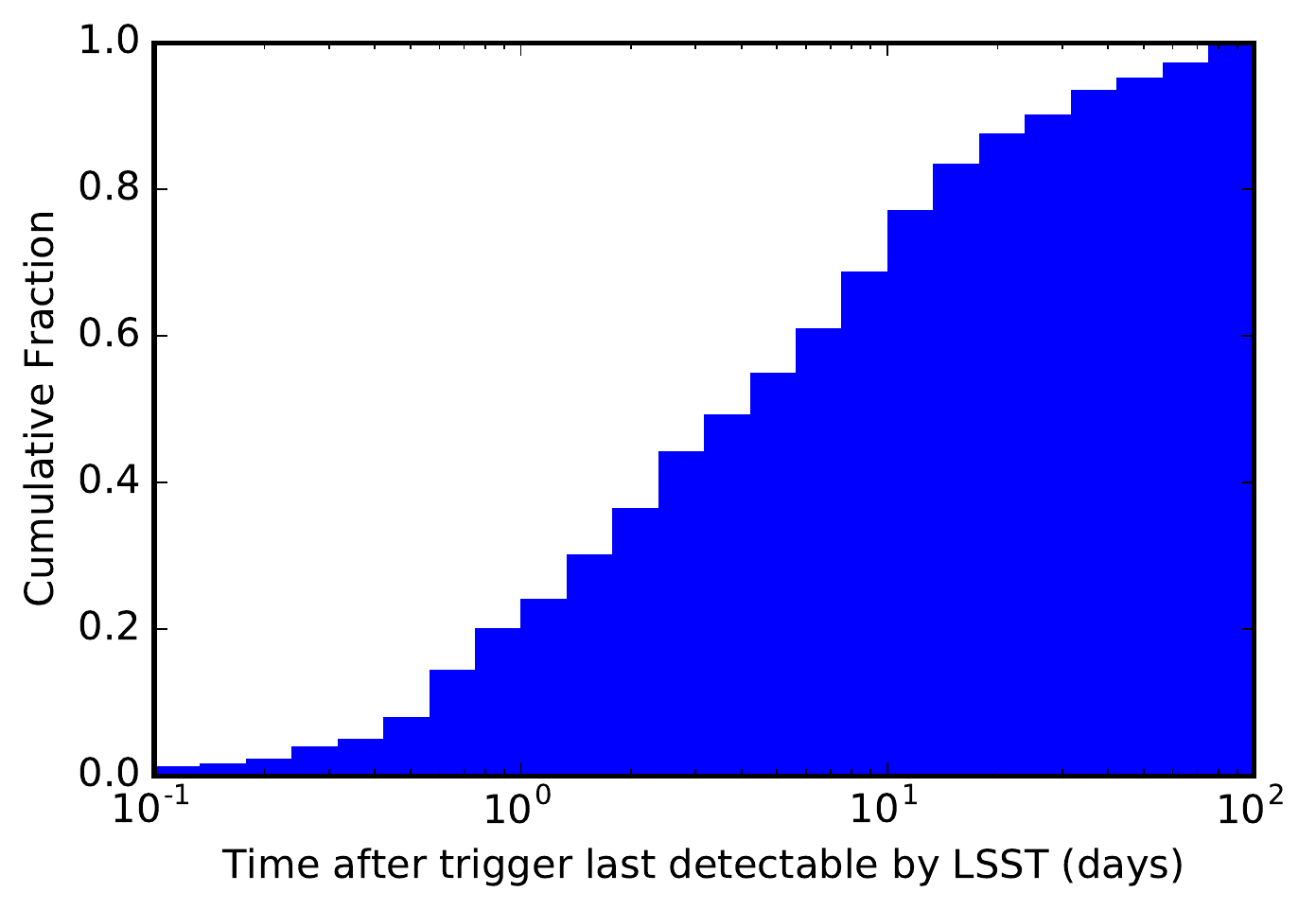}
}
\caption{ Cumulative fraction of GRB on-axis afterglows fainter than
magnitude 24.7 at a given time after the burst. We use an $\alpha=1$
decay with no jet breaks and the brightness parameters of
\citet{2011PASP..123.1034J}. }
\label{fig:afterglow_visibility}
\end{figure}

Thanks to LSST's depth, GRBs can be visible for weeks (Figure
\ref{fig:afterglow_visibility}).  Accordingly,
modest enhancements to the intra- and inter-night revisit rate with
single-filter rolling cadences should substantially improve LSST's
discovery and characterization of relativistic explosions.

\subsection{Conclusions}

Here we answer the ten questions posed in
\autoref{sec:intro:evaluation:caseConclusions}:

\begin{description}

\item[Q1:] {\it Does the science case place any constraints on the
tradeoff between the sky coverage and coadded depth? For example, should
the sky coverage be maximized (to $\sim$30,000 deg$^2$, as e.g., in
Pan-STARRS) or the number of detected galaxies (the current baseline
of 18,000 deg$^2$)?}

\item[A1:] No strong constraint, although on average
	larger sky coverage provides fewer epochs for the dense time sampling
	required to detect fast-fading events like GRBs.

\item[Q2:] {\it Does the science case place any constraints on the
tradeoff between uniformity of sampling and frequency of  sampling? For
example, a rolling cadence can provide enhanced sample rates over a part
of the survey or the entire survey for a designated time at the cost of
reduced sample rate the rest of the time (while maintaining the nominal
total visit counts).}

\item[A2:]  Frequency of sampling is far more important than uniformity of
	sampling for fast transients like GRBs.  Rolling cadences with
		three or more epochs per night or two are needed for
		realtime discovery of young events.

\item[Q3:] {\it Does the science case place any constraints on the
tradeoff between the single-visit depth and the number of visits
(especially in the $u$-band where longer exposures would minimize the
impact of the readout noise)?}

\item[A3:]  While greater single visit depth probes a greater cosmological
	volume within which to detect GRBs and other fast transients,
		our efficiency at discovering
		GRBs with LSST is driven entirely by the time sampling.
		Accordingly we prefer a larger number of visits per field
		to greater single-exposure depth, independent of band..

\item[Q4:] {\it Does the science case place any constraints on the
Galactic plane coverage (spatial coverage, temporal sampling, visits per
band)?}

\item[A4:] As GRBs are cosmological events, we expect to detect very few at
	low Galactic latitudes due to extinction.  Accordingly this
		science case is insensitive to observing plans in the Plane
		except insofar as they limit the number and cadence of
		exposures at higher latitudes.

\item[Q5:] {\it Does the science case place any constraints on the
fraction of observing time allocated to each band?}

\item[A5:] No strong constraints, although detection of faint afterglows
	will benefit from exposures taken in the bands with deepest
		single-exposure depth.

\item[Q6:] {\it Does the science case place any constraints on the
cadence for deep drilling fields?}

\item[A6:] Deep drilling fields provide an excellent opportunity to achieve
	the high time sampling required to discover GRBs while they are
		young.  As discussed above, a good fast transient cadence
		might have a \textit{minimum} of three visits
		in a night separated by an hour or two, preferably in a
		single filter, with revisits every night.

\item[Q7:] {\it Assuming two visits per night, would the science case
benefit if they are obtained in the same band or not?}

\item[A7:] Due to the need to constrain the lightcurve shape, it is best if
	the observations are in the same filter--especially with only two
	visits per night.  Otherwise there is a degeneracy between the
	(evolving) color of the event and its lightcurve shape.

\item[Q8:] {\it Will the case science benefit from a special cadence
prescription during commissioning or early in the survey, such as:
acquiring a full 10-year count of visits for a small area (either in all
the bands or in a  selected set); a greatly enhanced cadence for a small
area?}

\item[A8:] A dedicated experiment providing enhanced cadences over a small
	area  (as described above) would provide an ideal
	experiment to determine the rate of fast transients.  Such an
	observing strategy would also
	facilitate organizing necessary followup resources
	(spectroscopy, X-ray followup) because the observing period would
	be known in advance.

\item[Q9:] {\it Does the science case place any constraints on the
sampling of observing conditions (e.g., seeing, dark sky, airmass),
possibly as a function of band, etc.?}

\item[A9:] None unique to the science.

\item[Q10:] {\it Does the case have science drivers that would require
real-time exposure time optimization to obtain nearly constant
single-visit limiting depth?}

\item[A10:] No.

\end{description}

\navigationbar

% --------------------------------------------------------------------

% PJM: moved to Future Work while MAF analysis is pending:
% \input{Transients/tde.tex}

% --------------------------------------------------------------------

% PJM: moved to Future Work while MAF analysis is pending:
% \input{Transients/cv.tex}

% --------------------------------------------------------------------

% PJM: moved to Future Work while MAF analysis is pending:
% \input{Transients/eruptive.tex}

% --------------------------------------------------------------------

% ====================================================================
%+
% SECTION:
%    gw.tex
%
% CHAPTER:
%    transients.tex
%
% ELEVATOR PITCH:
%-
% ====================================================================

\section{Gravitational Wave Sources}
\def\secname{\chpname:gw}\label{sec:\secname}

\credit{raffaellamargutti},
\credit{Doctor},
\credit{Fong},
\credit{Haiman},
\credit{Kalogera},
\credit{Trimble},
\credit{Zauderer}

The first detection of Gravitational Waves (GW) by the advanced
LIGO/Virgo collaboration \citep{Abbott16, Abbott09, Acernese08} has
recently opened a new window of exploration into our Universe. The
amount of information that can be revealed by the properties of the GW
emission is immense and holds promises for revolutionary insights,
including accurate masses and spins of neutron stars and black holes,
tests of General Relativity and an accurate census of the neutron star
(NS) and black hole (BH) populations that might challenge our current
understanding of massive stellar evolution. However, GW events are
poorly localized (10-100 deg$^2$ at the time of LSST operations). The
identification of EM counterparts would provide precise localization and
distance measurements, in addition to the necessary astrophysical
context (e.g. host galaxy properties, connection to specific stellar
populations) to fully exploit the revolutionary power of this new GW
era.

% --------------------------------------------------------------------

\subsection{Target measurements and discoveries}
\label{sec:\secname:targets}

The first GW event was found to be associated with the merger of two
black holes \citep{Abbott16,Abbott16b}. Although no EM counterpart was
expected to accompany a black-hole black-hole (BBH) merger, it seems now
possible that even BBH mergers  might produce short GRB-like EM emission
\citep{Connaughton16,Loeb16,Zhang16,Perna16,Stone16}. Indeed, in
analogy with supermassive BH mergers, shocks might develop in the
just-formed circumbinary accretion disk (if a disk forms), which can
produce a bright afterglow following the BBH merger (e.g.
\citealt{Lippai08,Corrales10,Schnittman13}). Albeit speculative in
nature, it is advisable to keep an open mind about the possibility of EM
counterparts to BBH mergers.

The most promising and better understood EM counterparts to GW events
are ``kilonovae" \citep{Li98,Metzger10,Metzger12,Kasen13,Barnes13}.
Kilonovae are short-lived (typical time scale of one week), apparently
faint ($z\sim21$ mag at peak at 120 Mpc), red ($i-z\approx1$ mag),
isotropic transients (\autoref{Fig:kilonova}) due to the radioactive
decay of r-process elements synthesized in the merger ejecta of a NS-NS
or NS-BH system. These merging systems are the favored progenitors of
short GRBs. Indeed, a signature consistent with kilonova emission has been
recently found following the short GRB\,130603B
\citep{Berger13,Tanvir13}. The key piece of information that enabled the
discovery of kilonova-like emission associated with  this short GRB was
its sub-arcsecond localization enabled by the detection of the optical
afterglow, which allowed for an effective kilonova search with the
Hubble Space Telescope (\autoref{Fig:kilonova}). In contrast, the
typical localization region of GW events in the LSST era is expected to
be of the order of a few tens of square degrees \citep{aaa+13}. It is
thus clear that the major challenges faced by the optical follow-up of
GW events is represented by the combination of poor localizations with
faint and fast evolving red electromagnetic counterparts.

The detection of an optical counterpart in conjunction with a GW event
will significantly leverage the GW signal. LSST, with its the wide FOV,
wavelength coverage and exquisite sensitivity is uniquely poised to
identify and characterize counterparts to GW events.

\begin{figure}
\vskip -0.0 true cm
\centering
\includegraphics[scale=0.85]{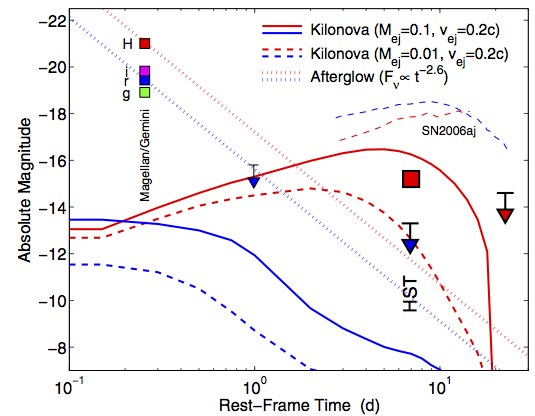}
\caption{Kilonova signature in the short GRB\,130603B as revealed by the
Hubble Space Telescope (HST). The Magellan and Gemini telescopes sampled
the optical afterglow of the GRB (dotted lines). The kilonova light
starts to dominate the emission in the H band around a few days after
the merger. Thick and dashed lines: theoretical kilonova models from
\citet{Barnes13} showing that kilonovae are fast-evolving, faint and red
transients. The light-curve of the SN\,2006aj associated with the long
GRB\,060218  is also shown for comparison. From \citet{Berger13}.}
\label{Fig:kilonova}
\end{figure}

% --------------------------------------------------------------------

\subsection{OpSim Analysis and Discussion}
\label{sec:\secname:analysis}

Effective follow up of GW triggers relies on the capability to sample a
relatively large portion of the sky, repeatedly, over a time scale $<1$
week, with different filters \citep{Cowperthwaite15}. In the optical
band, the kilonova signature is expected to be more prominent in the
$i$, $z$  and $y$ filters, which we identify as the most promising
filters for the kilonova search. We emphasize however that another set
of contemporaneous observations in  a ``bluer" filter is necessary to
acquire color information and distinguish kilonovae from other
fast-evolving transients.

We use the median inter-night gap  for visits in the same filter derived
from the candidate Baseline Cadence \opsimdbref{db:baseCadence} to show that,
in the absence of a Target of Opportunity (ToO) capability, it is
\emph{not} possible for LSST  to play a major role in the identification
of EM counterparts of GW triggers.

To identify kilonova candidates we need at least 2 observations acquired
within $\sim 1$ week  of the GW event \citep{Cowperthwaite15}. Using the
inter-night gap distribution for visits in the $y$ filter (which is the
most promising filter for a kilonova search), the area of the sky
covered with cadence  $\Delta t<7$ days at any given time, is
$A_{sky}\sim 3000$ deg$^2$ (including deep drilling fields).  This is
the area that can be searched for fast evolving transients.  Two
important considerations follow:

\begin{itemize}
\item[(1)] $A_{sky}$ only covers $P\sim7$\% of the sky. The  probability
that the \emph{entire} GW localization region is contained, by chance,
within $A_{sky}$ is thus very small.
\item[(2)] Even if LSST is able to cover a meaningful portion of the GW
region, we would still not have color information, and we would thus be
unable to filter out contaminating transients.
\end{itemize}

\textit{We conclude that relying on the serendipitous alignment of the
LSST fields with the GW localization map is not an effective strategy to
follow up GW triggers and identify their EM counterparts. We thus
strongly recommend a ToO capability as part of the baseline LSST
operations strategy.}

Ideally, the ToO capability will allow for imaging of the GW
localization map at least twice over $\Delta t\lesssim$1 week with a
``red" filter ($i$, $z$  or $y$),  and  will include the possibility to
designate a desired set of filters to obtain color information. By the
time of LSST operation the typical size of the GW localization region is
expected to be 10-100 deg$^2$, which would require a small number of
LSST re-pointings. We thus do \emph{not} anticipate a significantly
disruptive impact on other LSST campaigns (especially if only the GW
triggers with the best localizations in the southern sky are selected
for LSST ToOs).

\textit{At the price of re-shuffling a reasonably small number of
fields, \textbf{if} equipped with ToO capabilities, LSST can be the
premier player in the era of EM follow up to GW sources.}

\navigationbar

% --------------------------------------------------------------------

% ====================================================================
%+
% SECTION:
%    transientFuture.tex
%
% CHAPTER:
%    transients.tex
%
% ELEVATOR PITCH:
%    Ideas for future metric investigation, with quantitaive analysis
%    still pending.
%-
% ====================================================================

\section{Future Work}
\def\secname{\chpname:future}\label{sec:\secname}

In this section we provide a short compendium of science cases that
are either still being developed, or that are deserving of quantitative
MAF analysis at some point in the future.

% ====================================================================

% ====================================================================
%+
% SECTION:
%    tde.tex
%
% CHAPTER:
%    transients.tex
%
% ELEVATOR PITCH:
%    Tidal disruption events (TDEs) are the disruptions of stars by supermassive black holes.
%    They can produce flares in the optical and UV (sometimes accompanied by X-ray and radio emission as well).
%    These flares can be used to reveal the properties of otherwise quiescent SMBHs and to study accretion physics.
%    TDEs are rare, LSST will allow the first statistical sample of such events.
%
% AUTHORS:
%    Iair Arcavi (@arcavi)
%
% ====================================================================

% \section{Tidal Disruption Events}
\subsection{Tidal Disruption Events}
\def\secname{\chpname:tdes}\label{sec:\secname}

\credit{arcavi}

A star passing close to a supermassive black hole (SMBH;
$M\gtrsim10^{6}M_{\odot}$) will be torn apart by tidal forces. For
certain ($\lesssim10^{8}M_{\odot}$) black hole masses, the disruption
will occur outside the event horizon and will be accompanied by an
observable flare \citep{Hills1975, Rees1988}. Such flares can be used to
study inactive SMBHs, which are otherwise inaccessible beyond the nearby
($\lesssim100$ Mpc) universe.

We are now building our understanding of how observational properties of
TDEs are affected by the SMBH. Theory claims to provide such a
connection \citep[e.g.][]{Lodato2009, Guillochon2014}, but uncertainties
in the physics of the disruption, subsequent accretion and emission
mechanisms are currently topics of debate \citep[e.g.][]{Strubbe2015,
Guillochon2014, Roth2015}, and new models are vigorously being developed
\citep[e.g.][]{Piran2015, Hayasaki2015, Svirski2015, Bonnerot2015}.

TDEs are rare ($\sim10^{-5}-10^{-4}$ events per galaxy per year;
\citealp{Wang2004, Stone2015}), and until recently, TDE candidates were
discovered mostly in archival data \citep[e.g.][]{Donley2002,
Gezari2006, Esquej2007}. Now, however, wide-field transient surveys have
started discovering TDEs in real time.

Generally, two types of TDE candidates have been identified:
\begin{enumerate}
	\item \textit{High energy TDEs}. The prototype is Swift J1644
\citep{Bloom2011, Burrows2011, Levan2011, Zauderer2011}, with two other
events known \citep{Cenko2012, Brown2015}. These events display emission
in $\gamma$-rays and X-rays as well as in the radio, but are not
detected in the optical.
\item \textit{Optical-UV TDEs}.  The prototype is PS1-10jh
(Figure \ref{fig:tde}; \citealp{Gezari2012}).
		About $8$ other events are known
\citep{Chornock2014, Arcavi2014, Holoien2014, Holoien2015, Holoien2016}.
Some events were detected also in the X-rays and radio (in addition to
the optical and UV), but the X-ray and radio signatures are different
than those of the high energy TDE candidates.
\end{enumerate}

\begin{figure}[hbt]
\centerline{
\includegraphics[width=0.6\textwidth]{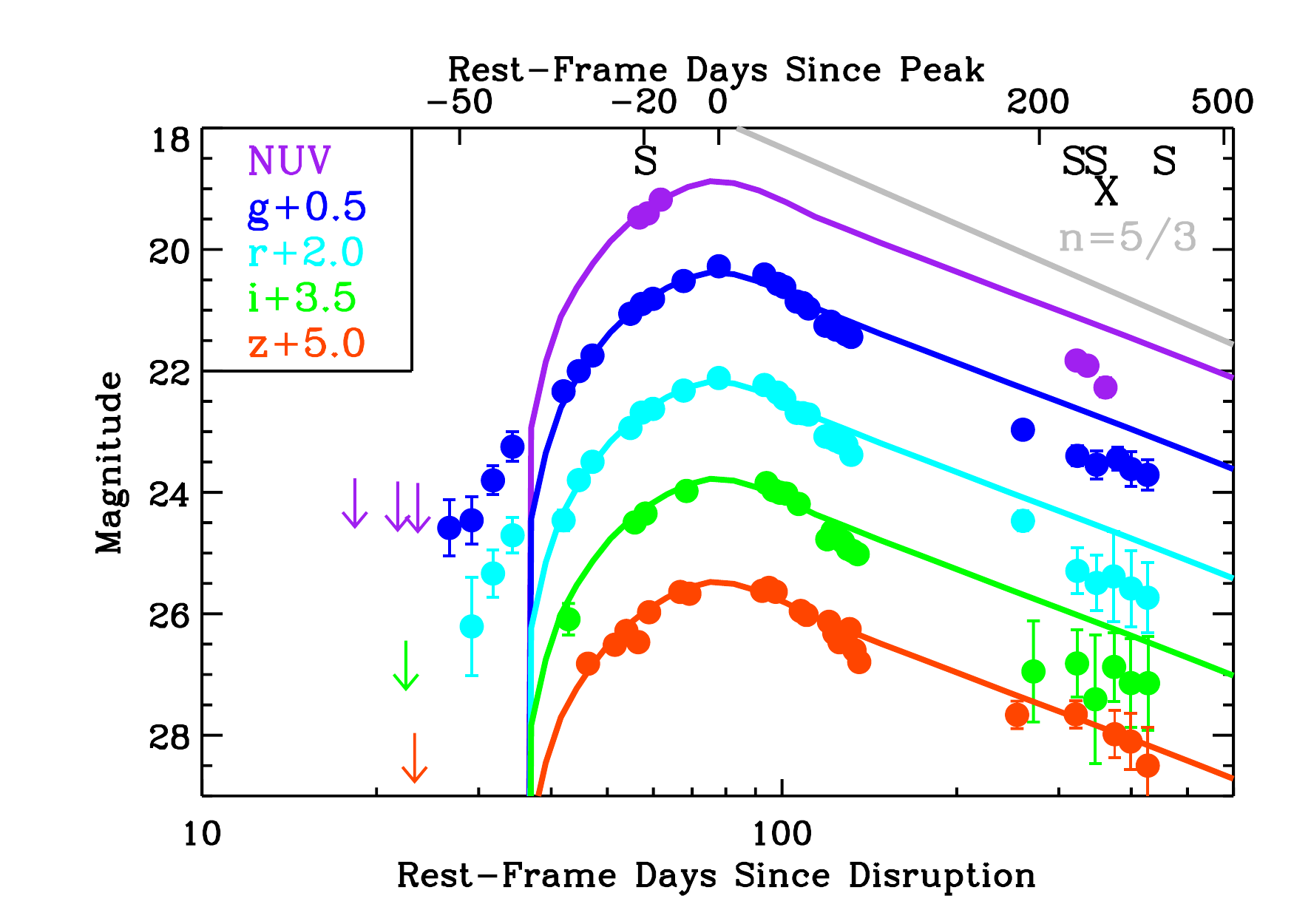}
}
\caption{
Optical and near-UV lightcurve of the TDE PS1-10jh \citep{Gezari2012}.
}
\label{fig:tde}
\end{figure}

It is still not clear whether both of these classes of transients are TDEs,
and if so, why they are so different form each other. One option raised is
that some TDEs may launch jets, which when directed towards us, appear as
the high energy events, but otherwise appear as the optical-UV events.
It is still not clear if this is indeed the case \citep[e.g.][]{VanVelzen2013}.

Here we focus on the second type of TDE candidates, which is the
relevant class for LSST, since they can be discovered in the optical.
However, multi-wavelength coordinated observations of
optically-discovered events are required in order to better understand
the connection between the two types of candidates.

The first well-sampled TDE of the optical+UV class was PS1-10jh (discovered by Pan-STARRS;
\citealt{Gezari2012}). \citet{Arcavi2014} later presented three new TDE candidates
from PTF and one discovered by ASAS-SN, all with similar properties
as PS1-10jh. These events exhibit blue colors, broad light curves,
peak absolute magnitudes of $\sim-20$ and a $\sim t^{-5/3}$ decay
at late times. This decay law has been suggested as a unique signature
of accretion-powered TDE light curves \citep{Rees1988, Evans1989, Phinney1989}.
Early-time deviations from the $t^{-5/3}$ rate
can be used to constrain the density profile of the disrupted star
\citep{Lodato2009, Gezari2012}. Late-time deviations would
test the accretion power-source hypothesis altogether.

The spectral signatures of these TDEs are still a puzzle. PS1-10jh displayed
only He II emission lines, lacking any signs of H. Some of the \citet{Arcavi2014}
sample, however, do display H emission.
In fact, a continuum of H / He emission ratios for this class of transients
is being revealed, and is now a focus of theoretical modelling \citep{Strubbe2015, Roth2015}.

The second recent discovery relating to this new sample concerns their
host galaxies \citep{Arcavi2014, French2016}, most of which are
post-starburst galaxies. These galaxies show little or no signs of on-going star formation, but their significant A stellar populations indicate that star formation ceased abruptly a few hundred Myr to a Gyr ago \citep{Dressler1983}. Galaxies with these characteristics often show signs of recent galaxy-galaxy mergers \citep{Zabludoff1996}, which produced the starburst and evolved the bulge. Optical-UV TDEs are intrinsically over abundant in post-starburst galaxies by a factor of $\sim30-200$ (depending on the characteristics of the galaxy; \citealt{French2016}). The reason for the strong preference of TDEs for post-starburst galaxies has still not been determined.

LSST's contribution to TDE studies will be substantial.
\citet{VanVelzen2011} estimate that LSST could discover approximately
4000 TDEs per year. The main drivers for studying TDEs with LSST are:
\begin{itemize}
\item Measuring black hole masses: This involves fitting models to TDE
light curves. It is also relevant to correlate these measurements with
host galaxy properties (mass, bulge/disk decomposition).
\item Constraining galactic dynamics by measuring the TDE rates as
functions of black hole mass and galaxy types.
\item Characterizing TDE emission signatures.
\end{itemize}

A metric is required for measuring how well TDEs can be identified and
distinguished from supernovae and active galactic nuclei. In general, we
expect TDEs to:
\begin{itemize}
\item Be located in the center of their host.
\item Display approximately constant blue (few $10^4$K) colors.
\item Evolve slowly (weeks-months).
\item Not show past AGN-like variability.
\item Preferentially peak around mag -20.
\item Preferentially be hosted in a post-starburst galaxy.
\end{itemize}
These criteria are based on our current knowledge of optical TDEs, which
is still in its early stages. The field is rapidly evolving, and it is
possible that new observations will change the current picture of TDE
emission. This metric is probably best combined with those discussed in
 \autoref{sec:\chpname:SNtransients} for identifying supernovae, though the
luminosity function of TDEs (or what constitutes a ``typical'' TDE light
curve) is not yet known.

A second metric is required to asses the accuracy with which the black
hole mass can be constrained from the TDE light curves. This metric can
be based on existing theoretical models to fit simulated TDE light
curves (such as TDEFit; \citealt{Guillochon2014}).

% ====================================================================
%
 \subsection{Conclusions}

 Here we answer the ten questions posed in
 \autoref{sec:intro:evaluation:caseConclusions}:

 \begin{description}

 \item[Q1:] {\it Does the science case place any constraints on the
 tradeoff between the sky coverage and coadded depth? For example, should
 the sky coverage be maximized (to $\sim$30,000 deg$^2$, as e.g., in
 Pan-STARRS) or the number of detected galaxies (the current baseline
 of 18,000 deg$^2$)?}

 \item[A1:] The number of events discovered will be approximately proportional to the sky area surveyed, multiplied by the
 average length of coverage, so added sky area is beneficial, as long as the observing season per field is not shortened.

 \item[Q2:] {\it Does the science case place any constraints on the
 tradeoff between uniformity of sampling and frequency of  sampling? For
 example, a rolling cadence can provide enhanced sample rates over a part
 of the survey or the entire survey for a designated time at the cost of
 reduced sample rate the rest of the time (while maintaining the nominal
 total visit counts).}

 \item[A2:] Uniform sampling is optimum.

 \item[Q3:] {\it Does the science case place any constraints on the
 tradeoff between the single-visit depth and the number of visits
 (especially in the $u$-band where longer exposures would minimize the
 impact of the readout noise)?}

 \item[A3:] No, as long as all filters are well represented.

 \item[Q4:] {\it Does the science case place any constraints on the
 Galactic plane coverage (spatial coverage, temporal sampling, visits per
 band)?}

 \item[A4:] This science will be accomplished away from the galactic plane.

 \item[Q5:] {\it Does the science case place any constraints on the
 fraction of observing time allocated to each band?}

 \item[A5:] Increased $u$-filter cadence valuable for TDE.

 \item[Q6:] {\it Does the science case place any constraints on the
 cadence for deep drilling fields?}

 \item[A6:] Only cadences that sample timescales of weeks and greater will be useful.

 \item[Q7:] {\it Assuming two visits per night, would the science case
 benefit if they are obtained in the same band or not?}

 \item[A7:] Different bands would be strongly preferred.

 \item[Q8:] {\it Will the case science benefit from a special cadence
 prescription during commissioning or early in the survey, such as:
 acquiring a full 10-year count of visits for a small area (either in all
 the bands or in a  selected set); a greatly enhanced cadence for a small
 area?}

 \item[A8:] A sparse cadence applied to a large sky area would be scientifically productive, but this is not a strong constraint.

 \item[Q9:] {\it Does the science case place any constraints on the
 sampling of observing conditions (e.g., seeing, dark sky, airmass),
 possibly as a function of band, etc.?}

 \item[A9:] No.

 \item[Q10:] {\it Does the case have science drivers that would require
 real-time exposure time optimization to obtain nearly constant
 single-visit limiting depth?}

 \item[A10:] No.

 \end{description}

 \navigationbar

% ====================================================================

% ====================================================================
%+
% SECTION:
%    cv.tex
%
% CHAPTER:
%    transients.tex
%
% ELEVATOR PITCH:
%    Explain in a few sentences what the relevant discovery or
%    measurement is going to be discussed, and what will be important
%    about it. This is for the browsing reader to get a quick feel
%    for what this section is about.
%
% AUTHORS:
%    Federica Bianco (@fedhere)
%
% ====================================================================

% \section{Cataclismic Variables}
\subsection{Cataclismic Variables}
\def\secname{\chpname:CVtransients}\label{sec:\secname}

\credit{paulaszkody},
\credit{fedhere}

Cataclysmic Variables (CVs) encompass a broad group of objects
including novae, dwarf novae, novalikes, and AM CVn systems, all with different
amplitudes and rate of variability. The one thing they all have in
common is active mass transfer from a late type companion to a
white dwarf. These create variability on a wide range of timescales:

\begin{itemize}
	\item \textit{minutes} flickering in
dwarf novae and novalikes, pulsations in accreting white dwarfs in
the instability strip, orbital periods of AM CVn systems
\item \textit{hours} orbital periods of novae, dwarf novae and novalikes
\item \textit{days} normal outburst lengths of dwarf novae
\item \textit{weeks} outburst length of superoutbursts in short orbital period
dwarf novae, outburst recurrence time of normal outbursts in short
orbital period dwarf novae
\item \textit{months} outburst recurrence time of
longer period dwarf novae, various state changes in novalikes, declines
in novae
\item \textit{years} for the outburst recurrence timescales of the
shortest period dwarf novae and the recurrence times in recurrent novae
\end{itemize}
The
amplitudes range from tenths of mags for flickering and pulsations to 4 mags
for normal dwarf novae and changes in novalike states up to 9-15 mags for the
largest amplitude dwarf novae and classical novae.

These large differences make correct classification with LSST difficult
but necessary in order to reach goals of assessing the correct number
of types of objects for population studies of the end points of
binary evolution. Multiple filters (especially the blue $u$ and $g$)
along with amplitude and recurrence of variation provide the best
discrimination, as all CVs are bluer during outburst and high states of
accretion. Long term, evenly sampled observations can provide indications
of the low amplitude random variability and catch some of the more frequent
outbursts, but higher sampling is needed to determine whether an object
has a normal or superoutburst, to catch a rise to outburst or to a
different accretion state or to follow a nova. Novae typically
have rise times of a few days, while the decline time and shape provide
information as to the mass, distance and composition. The time to decline
by 2-3 magnitudes is correlated with composition,
%
% FED: what is a range of time scales for this decline? days? months?
%
WD mass and location in
the galaxy, thus enabling a study of Galactic chemical evolution.  As with SN,
the diagnostic power for all these systems rests on color and sampling.

Metrics to be developed would assess the abilities of  a given observing
strategy to distinguish between new novae and dwarf novae outbursts and
identify high and low states.  This discriminiation is provided by
measurement of the shapes and recurrence times of large variations as well
as blue colors to distinguish low amplitude variability that would indicate
new pulsators or novalikes. Population studies rely on the numbers of long
orbital period (low amplitude, wide outbursts) vs. short orbital period
(patterns of short outbursts followed by larger, longer superoutbursts)
dwarf novae at different places in the galaxy, as well as the numbers of
recurrent (1-10 yrs) vs. normal novae (10,000 yrs, about 35/galaxy/yr).
Objects particulary worthy of later followup are containing highly magnetic
white dwarfs. These objects can be identified in a large sample when the
magnitude for a majority of the years is a faint (low) state and a small
percentage of time is a bright (high) state, combined with a red color (due
to cyclotron emission from the magnetic accretion column).

% ====================================================================
%
\subsection{Conclusions}
%
% Here we answer the ten questions posed in
% \autoref{sec:intro:evaluation:caseConclusions}:
%
 \begin{description}

 \item[Q1:] {\it Does the science case place any constraints on the
 tradeoff between the sky coverage and coadded depth? For example, should
 the sky coverage be maximized (to $\sim$30,000 deg$^2$, as e.g., in
 Pan-STARRS) or the number of detected galaxies (the current baseline
 of 18,000 deg$^2$)?}

 \item[A1:] Extending the sky area is not a priority for this science.

 \item[Q2:] {\it Does the science case place any constraints on the
 tradeoff between uniformity of sampling and frequency of  sampling? For
 example, a rolling cadence can provide enhanced sample rates over a part
 of the survey or the entire survey for a designated time at the cost of
 reduced sample rate the rest of the time (while maintaining the nominal
 total visit counts).}

 \item[A2:] Intervals of higher cadence are extremely valuable and a rolling cadence is a
satisfactory approach, subject to cadence details.

 \item[Q3:] {\it Does the science case place any constraints on the
 tradeoff between the single-visit depth and the number of visits
 (especially in the $u$-band where longer exposures would minimize the
 impact of the readout noise)?}

 \item[A3:] Increasing the number of $u$-band visits will improve the characterization of CV phenomena.

 \item[Q4:] {\it Does the science case place any constraints on the
 Galactic plane coverage (spatial coverage, temporal sampling, visits per
 band)?}

 \item[A4:] Most CVs will be detected in the galactic plane, and a long, rich series of visits is needed.

 \item[Q5:] {\it Does the science case place any constraints on the
 fraction of observing time allocated to each band?}

 \item[A5:] $u$-band is diagnostic, especially $u$-$g$.

 \item[Q6:] {\it Does the science case place any constraints on the
 cadence for deep drilling fields?}

 \item[A6:] For deep drilling in the galactic plane or for local group galaxies, the best cadences would obtain several epochs  per night in
 each filter, rather than concentrating all acquisition with a filter in a single rapid burst.

 \item[Q7:] {\it Assuming two visits per night, would the science case
 benefit if they are obtained in the same band or not?}

 \item[A7:] Same filter and different filter each offer valuable information, and a mix of these two options would be preferred
pending test of both cadences.

 \item[Q8:] {\it Will the case science benefit from a special cadence
 prescription during commissioning or early in the survey, such as:
 acquiring a full 10-year count of visits for a small area (either in all
 the bands or in a  selected set); a greatly enhanced cadence for a small
 area?}

 \item[A8:] It would be very helpful to CV studies - and many other areas of transient science - to understand variability across all timescales.
 Especially valuable would be a cadence that would cover one (cluster, rich star field) with all the timescales that will not be
strongly represented  in the main survey, starting at 15 seconds.

 \item[Q9:] {\it Does the science case place any constraints on the
 sampling of observing conditions (e.g., seeing, dark sky, airmass),
 possibly as a function of band, etc.?}

 \item[A9:] No.

 \item[Q10:] {\it Does the case have science drivers that would require
 real-time exposure time optimization to obtain nearly constant
 single-visit limiting depth?}

 \item[A10:] No.

 \end{description}

 \navigationbar

% ====================================================================

% ====================================================================
%+
% SECTION:
%    eruptive.tex
%
% CHAPTER:
%    transients.tex
%
% ELEVATOR PITCH:
%-
% ====================================================================

% \section{LBVs and related non-supernova transients}
\subsection{LBVs and related non-supernova transients}
\def\secname{\chpname:LBVs}\label{sec:\secname}

\credit{nathansmith}

There is a large and diverse class of visible-wavelength transient
sources recognized in nearby galaxies that appear to be distinct from
traditional novae and from SNe, and have often been associated with
the giant eruptions of luminous blue varibles (LBV), such as the 19th
century outburst of $\eta$ Carinae.  Broadly speaking, members of this
class of transients share the common properties that they have peak
luminosities below those of most core-collapse SNe and more luminous
than novae and CVs (absolute magnitudes of roughly $-$9 to $-$15 mag).
They also have H-rich spectra (usually) with relatively narrow lines
that indicate modest bulk outflow velocities of 10$^2$ to 10$^3$ km
s$^{-1}$ (although some have exhibited small amounts of material at
faster speeds).  They tend to evolve on fairly long timescales of
weeks to years (although sometimes they exhibit a quick rise to peak
similar to SNe II-P). This group of transients has gone by many names,
such as LBV eruptions, SN impostors, Type V supernovae,
intermediate-luminosity optical (or red) transients, as well as others
that often include a physical interpretation.
%For brevity, these are
%often collectively referred to as ``LBVs'', although many of them may
%not actually be LBVs.
%This may be largely for historical reasons,
%since LBVs were the first of these to be recognized as a class.  Some
%of the subgroups may be very different from objects like $\eta$
%Carinae, however.

Observationally, these eruptions are understood to represent important
and dramatic mass-loss episodes in the lives of massive stars, based
on empirical estimates of the amount of ejected matter.
%Guided
%largely by nearby LBVs with resolved shells, t
These eruptions are
expected to instigate mass loss that is comparable to or more
important than metallicity-dependent winds of massive stars.  This
mode of mass loss, regardless of the mechanism, may be a very
important ingredient in the evolution of massive stars that is
currently not included in stellar evolution models.  Correcting this
is one of the key science drivers in trying to understand the physics
these eruptions.

%The degeneracy
%arises because when the objects are
%fully obscured by dust, one cannot actually meaure the star's
%temperature, and the bolometric luminosities of super-AGB and red and
%blue supergiants overlap.  Unfortunately,
%cases when we have strong constraints on the quiescent progenitor are
%rare, and once they reach their peak luminosity, there is a great deal
%of overlap in observed properties.

Theoretically, these eruptions are not understood.  %There are many
%ideas, but few if any confirmed mechanisms tied to observed
%objects.
Some
%previously discussed
theoretical ideas involve (1)
winds driven by super-Eddington instabilities (although the root cause
for suddenly exceeding the Eddington limit remains unexplained), (2)
hydrodynamic explosions caused by deep-seated energy deposition, such
as unsteady nuclear burning, (3) accretion onto companion stars in
binary systems (degenerate or not), (4) mergers in binary and triple
systems, (5) electron-capture SNe, and (6) ``failed SNe'' associated
with a weak explosion and envelope ejection that results from black
hole formation during core collapse.
Because of the relatively low total energy indicated by
radiative luminosities and outflow speeds, these are usually discussed
as non-terminal eruptions, however, the last two are terminal events
that are less luminous and lower energy than normal SNe, and the last
3 should only occur once for a given source.
%Together with several well-studied examples that indicate
%repeating eruptions, there are indeed many
%cases where only one such transient has been seen at the same
%position, and some cases where late-time observations suggest that no
%source has survived with a luminosity comparable to its progenitor.
%However, there are also several well-studied examples that indicate
%repeating eruptions (multiple repeating transients, multiple nebular
%shells with different ages, etc).
All these theoretical mechanisms
may lead to similar observed phenomena: weak explosions, moderate
luminosities, slow expansion, dusty aftermath, but this class of objects
may represent a mixed-bag of different mechanisms that get lumped
together by default as ``other'' because they are not traditional SNe.

Rates for these LBV-like eruptions are still very poorly constrained;
%largely because most previous SN and transient searches with small
%telescopes have been optimized for finding more luminous SNe in a
%larger volume.  This field begun to change with recent surveys, and will
%be revolutionized with LSST.  F
%from discovered examples we have,
numbers are very roughly consistent with a volumetric rate comparable
to that of core-collapse SNe or larger.  %, but with a large error bar.
Limited information often makes classification into various
subgroups difficult or highly subjective, thus subclass rates are even
less well constrained.  %The ``rate'' also depends on how
%faint the lower limit of inclusions is; e
Evidence suggests that the brightest events occurr less frequently and that
numbers increase as one moves to lower luminosity.  At the faint end,
it becomes difficult to distinguish between eruptions and regular
variability of LBVs, or between massive star eruptions and CVs.  \emph{With
deep LSST stacks identifying faint CV in quiescent states this will
 change dramatically in the LSST age, with the unvailing of
detailed progenitor information}.  Having deep, pre-eruption
characterization of sources at the positions of these eruptive
transients (as well as SN precursors) will likely be a major
contribution of LSST.
An important empirical discriminant of subgroups in this class comes
from their progenitor stars.  Some are indeed seen to be very
luminous, blue supergiant stars consistent with traditional LBVs.
Some, however, have somewhat less luminous, heavily dust-obscured
progenitor stars that have been associated with either dust-enshrouded
blue or red supergiants, or alternatively, with super-AGB stars of
8-10 M$_{\odot}$%, with uncertainty.

An area of recent interest is that eruptive non-terminal transients
have been observed, in some cases, to precede much more powerful
explosions that are seen as Type IIn supernovae.  \emph{Detectability of SN
precursors eruptions is discussed in ~\autoref{chp:galaxy}.
LSST can provide a large enough sample of these events to enable the study
or rates.} %There may also be
SN
precursors have observed or inferred properties that are very similar
to LBVs and related transients,
%.  This may suggest some link between them,
but then again, most of the LBVs and other SN impostors have been
observed for decades and have not gone SN (yet).  \emph{Being able to
  distinguish which of these optical transients are SN precursors and
  which are not is a major science driver.}  The amount of mass lost
in a precursor eruption may dramatically alter the type of SN that is
observed.  Even if the pre-SN transients are not observed directly,
pre-SN eruptive mass loss can be inferred and constrained with
persistent observations of the detected eruptions' lightcurves (and
spectra) through circumstellar interaction diagnostocs of
the bright eruptions and explosions.
%a continuum of energies in pre-SN outbursts, extending down to more
%normal classes of core-collapse SNe, but these may often go
%unrecognized unless the SN is caught very early after explosion.

In terms of timescales, many of the eruptive transients exhibit rise
and decline timescales similar to normal SNe~II-P or II-L, but with
fainter peak luminosity.  For these, observational cadence
requirements will be the same as SNe.  For some eruptive transients,
however, the rise timescales can be very long (rising a few magnitudes
in years).  While LSST's cadence will certainly be fast enough, being
able to discover slowly rising transients that do not change much from
night to night will be an important metric, and the edge effects
should be investigated.
For the faster-rising transients, just like for SNe, spectroscopic
follow-up is needed to discriminate these from normal SNe, and also
contextual information about the host galaxy (and hence, the absolute
magnitude) is needed to differentiate these non-terminal eruptions
from Type IIn supernovae (their spectra look similar, although LBVs do
tend to have narrower lines).  %Spectral and color evolution, as well
%as information about the progenitor, is needed to distinguish among
%subgroups within the class.
Multiwavelength follow-up is often extremely valuable or even
essential; i.e. mid-IR tells us if an optically invisible source is
cloaked in a dust shell but still quite luminous; Xrays and radio tell
us if an expanding shock wave is the likely source of persistent
luminosity.  For these reasons, nearby cases will continue to be the
most valuable in deciphering the physics of subclasses, whereas the
increased volume in which LSST discovers these fainter transients will
drastically improve our understanding of their rates.  Armed with both
a better understanding of their underlying physics and
characterization, as well as their rates and duty cycles, these
eruptive events can then be incorporated into stellar evolution models
and population synthesis/feedback models.

\navigationbar

% ====================================================================

\navigationbar

% --------------------------------------------------------------------

% ====================================================================

\chapter{The Magellanic Clouds}
\def\chpname{MCs}\label{chp:\chpname}

Chapter Editors:
\credit{dnidever},
\credit{knutago}

% \section*{Summary}
% \addcontentsline{toc}{section}{~~~~~~~~~Summary}
%
% Executive summary goes here, highlighting the primary conclusions from
% the chapter's science cases. This should be abstract length, no more:
% say, 200 words.

\section{Introduction}

The Magellanic Clouds have always had outsized importance for
astrophysics.  They are critical steps in the cosmological distance
ladder, they are a binary galaxy system with a unique interaction
history, and they are laboratories for studying all manner of
astrophysical phenomena.  They are often used as jumping-off points
for investigations of much larger scope and scale; examples are the
searches for extragalactic supernova prompted by the explosion of
SN1987A and the dark matter searches through the technique of
gravitational microlensing.  More than 17,000 papers in the NASA ADS
include the words ``Magellanic Clouds'' in their abstracts or as part
of their keywords, highlighting their importance for a wide variety of
astronomical studies.

An LSST survey that did not include coverage of the Magellanic Clouds
and their periphery would be tragically incomplete.  LSST has a unique
role to play in surveys of the Clouds.  First, its large $A\Omega$
will allow us to probe the thousands of square degrees that comprise
the extended periphery of the Magellanic Clouds with unprecedented
completeness and depth, allowing us to detect and map their extended
disks, stellar halos, and debris from interactions that we already
have strong evidence must exist.  Second, the ability of LSST
to map the entire main bodies in only a few pointings will allow us to
identify and classify their extensive variable source populations with
unprecedented time and areal coverage, discovering, for example,
extragalactic planets, rare variables and transients, and light echoes
from explosive events that occurred thousands of years ago.
Finally, the large number of observing opportunities that the LSST
10-year survey will provide will enable us to produce a static imaging
mosaic of the main bodies of the Clouds with extraordinary image
quality, an invaluable legacy product of LSST.

We have several important scientific questions that can be grouped into two themes, as follows.

\noindent{\bf Galaxy Formation and Evolution}

The study of the formation and
evolution of the Large and Small Magellanic Clouds (LMC and SMC,
respectively), especially their interaction with each other and the
Milky Way. The Magellanic Clouds (MCs) are a unique local laboratory
for studying the formation and evolution of dwarf galaxies in
exquisite detail.  LSST's large FOV will be able to map out the
three-dimensional structure, metallicity and kinematics in great
detail. Within this theme we have three main science questions:
\begin{enumerate}

\item What are the stellar and dark matter mass profiles of the
Magellanic Clouds?  To answer this we need to map their extended disks, halos, debris, and streams.  We can use
streams and RR Lyrae stars as probes of the 3D mass profile.

\item What is the satellite population of the Magellanic Clouds? The
discovery of dwarf satellites by the Dark Energy Survey and other
surveys hint at LSST's potential here.

\item What are the internal dynamics of the Magellanic Clouds?  Proper
motions from HST and from the ground have measured the bulk
motions of the Clouds and have, in combination with spectroscopy,
begun to unravel the three dimensional internal dynamics of the
Clouds.

\end{enumerate}

\noindent{\bf Stellar Astrophysics and Exoplanets}

The MCs have been
used for decades to study stellar astrophysics, microlensing and other
processes.  The fact that the objects are effectively all at a single
known distance makes it much easier to study them than in, for
example, the Milky Way, while the MCs' especial proximity allows us to explore deeper into the luminosity function of the stellar populations. LSST will extend these MC studies to fainter
magnitudes, higher cadence, and larger area. Within this theme we have three main science questions:
\begin{enumerate}

\item How do exoplanet statistics in the Magellanic Clouds compare to
those in the Milky Way?  The calculations in the next section show that
LSST can measure transits of Jupiter-like planets, an intriguing
prospect given the Clouds' lower metallicity environment.

\item What are the variable star and transient population of the Clouds?
LSST will enable population studies, linking star formation and chemical
enrichment histories.

\item What can we learn about supernovae and other explosive events from
their light echoes?  Echoes can give view of such events unavailable by
any other means.

\end{enumerate}

Many different types of objects and measurements with their own
cadence ``requirements'' will fall into these two broad categories
(with some overlap).
% These will be outlined in the next section.
A very important aspect of the ``galaxy evolution'' science theme is
not just the cadence but also the sky coverage of the Magellanic
Clouds ``mini-survey.''  A common misunderstanding is that the MCs
only cover a few degrees on the sky.  That is, however, just the
central regions of the MCs akin to the thinking of the Milky Way as
the just the bulge.  The full galaxies are actually much larger with
LMC stars detected at $\sim$21$^{\circ}$ ($\sim$18 kpc) and SMC stars
at $\sim$10$^{\circ}$ ($\sim$11 kpc) from their respective centers.
The extended stellar debris from their interaction likely extends to
even larger distances.  Therefore, to get a complete picture of the
complex structure of the MCs will require a mini-survey that covers
$\sim$2000 deg$^2$.
% At this point, it not entirely clear how to
% include this into the metrics.
Note, that for the second science case
this is not as much of an issue since the large majority of the
relevant objects will be located in the high-density, central regions
of the MCs.

Our investigation of how the LSST observing strategy will affect the
science outline here is still in its infancy. Some of the disgnostic and
figure of merit metrics developed elsewhere in this paper may be useful
for assessing the Magellanic Cloud science cases as well. In the
meantime we present below two science cases in the early stages of development that show some of the promise LSST shows in this area.

\navigationbar

% ====================================================================
%+
% SECTION:
%    MCs_FutureWork.tex
%
% CHAPTER:
%    transients.tex
%
% ELEVATOR PITCH:
%    Ideas for future metric investigation, with quantitaive analysis
%    still pending.
%-
% ====================================================================

\section{Future Work}
\def\secname{\chpname:future}\label{sec:\secname}

In this section we provide a short compendium of science cases that
are either still being developed, or that are deserving of quantitative
MAF analysis at some point in the future.

% ====================================================================

% ====================================================================
%+
% SECTION:
%    MCs_ProperMotion.tex
%
% CHAPTER:
%    magclouds.tex
%
% ELEVATOR PITCH:
%-
% ====================================================================

% \section{The Proper Motion of the LMC and SMC}
\subsection{The Proper Motion of the LMC and SMC}
\def\secname{\chpname:propermotion}\label{sec:\secname}

\credit{dnidever},
\credit{knutago}

% SAC Review from Jason Kalirai: This opening paragraph is missing the science hook. There is a clear explanation for how/why LSST proper motions are going to be better than anything before, but the text doesn't actually say what we will learn from such measurements. Is it the case that the resulting constraints on the orbit or past accretion history break some current uncertainty in models of the MC evolution?

In the last decade work with $HST$ has been able to measure the bulk
tangential (in the plane of the sky) velocities ($\sim$300 km/s) of
the Magellanic Clouds (Kallivayalil et al.\ 2016a,b,2013) and even the
rotation of the LMC disk \citep{2014ApJ...781..121V}. Gaia
will measure precise proper motions of stars to $\sim$20th magnitude
which will include the Magellanic red giant branch stars. LSST will be
complementary to Gaia and measure proper motions of stars in the
$\sim$20--24 mag range that includes Magellanic main-sequence stars
which are far more numerous than giants, and, therefore, more useful
for mapping extended stellar structures. The LSST 10-year survey
proper motion precision will be $\sim$0.3--0.4 mas/yr at LMC
main-sequence turnoff at r$\approx$22.5--23.  This will allow for
accurate measurement of proper motions of individual stars at the
$\sim$5$\sigma$ level.

% SAC Review from Jason Kalirai: it's not clear why individual stars are needed. Shouldn't the proper motion precision being referenced here be for the population as a whole?

Besides measuring kinematics, the LSST proper motions can be used to
produce clean samples of Magellanic stars.
In addition, LSST proper motions can be used to improve star/galaxy
separation which is quite significant for faint, blue Magellanic
main-sequence stars.

% streaming motions

% can we do individual LMC stars with LSST, or small groups?

% SRD says want 0.2 mas/yr accuracy over the course of the survey
% 0.2 mas/yr at r=20.5 (similar to Gaia)
% ~0.25 mas/yr at r=22
% ~0.3 mas/yr at r=22.5 LMC turnoff
% ~0.4 mas/yr at r=23
% 1 mas/yr for r=24
% See Figure 21 from Ivezic et al. (2012) or slide 46 of overview-sci-reqs.pdf
%have the astrometric precision to measure the proper motions of individual
%Proper motion cleaning to find the giants?? gaia does that already
%lsst can use proper motion cleaning to do star/galaxy separation as well
%
%The
%The Magellanic Clouds have a large tangential velocity (in the plane of the sky) that has been
%Gaia will be able to see the bright stuff, need lsst to get the MSTO
%gaia/lsst synergy
%
%metric that calculates the proper motion accuracy of LMC MSTO stars at r=23 and calculates the sigma-level of
%the proper motion measurement (2.0 mas/yr / sigma_pm ).

% --------------------------------------------------------------------

% \subsection{Metrics}
\subsubsection{Metrics}
\label{sec:\chpname:metrics}

The natural Figure of Merit for this science case is the precision
with which the proper motion of the Magellanic Clouds can be measured.
This is likely to depend on the following diagnostic metrics:

% metric on surface brightness limit in different parts of the sky
% to MC structure
% could use metric for how much of the Besla stellar debris we can detect
%  even just that region of the sky covered

\begin{itemize}

\item  Single star proper motion precision, possibly quantified as.
the significance level
($\sigma$-level) of the proper motion measurement of one Magellanic MSTO
star (r=23 mag). We would expect this to take values of
$\sim$2.0 mas yr$^{-1}$ / $\sigma_{\rm pm}$.
%  the precision proper motion metric that
%  Metric that calculates the proper motion accuracy of LMC MSTO stars at r=23 and calculates the sigma-level of
%the proper motion measurement (2.0 mas/yr / sigma_pm ).

\item Another useful diagnostic metric would be the surface
brightness limit of the Magellanic structures, using MSTO stars.

\item A metric quantifying how much of the expected Magellanic debris/structure
\citet[from][]{2012MNRAS.421.2109B} model) we can detect would allow the
proper motion science case to be extended to peripheral structures.
This would depend
mostly on the area covered, but we could also use the surface
brightness limit (calculated above) directly.

\end{itemize}

\subsection{Exoplanets in the LMC and SMC}
\def\secname{\chpname:MC_exoplanets}\label{sec:\secname}

\credit{lundmb},
\credit{migueldvb}

% SAC Review by Jason Kalirai: I was confused about the part of this section related to finding transiting exoplanets in LMC stars. First, it would be nice to show some of the analysis in the paper itself, hopefully backed up by simulations of LSST's performance. Second, if the motivation is to tackle this in the Clouds due to their low metallicity, why not simply propose for such an experiment in a more nearby metal-poor system (with or without LSST).

While exoplanets are discussed in greater depth in \autoref{sec:planets}, it is
also worth noting here the unique circumstance of exoplanets in the
Magellanic Clouds. To date, all detected exoplanets have been found around
host stars within the Milky Way. Any constraints that could be applied to
planet occurrence rates in such a different stellar population as is found
in the Magellanic Clouds would provide a fresh insight into the limits
that are to be placed on planet formation rates.

The transit method of exoplanet detection is constrained by sufficient
period coverage in the observations taken, and in the dimming caused by
the star's transit being large enough with respect to the noise in
observations that the periodic signal of the transit can be recovered.
The relatively small chance of a planet being present and properly
aligned is offset by observing a large number of stars simultaneously.
Simulations have already shown that LSST has the capability to recover
the correct periods for large exoplanets around stars at the distance of
the LMC \citet{2015AJ....149...16L}.  We note that it would be unlikely
to be able to conduct follow-up observations of the discovered
candidates to confirm their planetary nature at a distance of $\sim$
50 kpc.  Further work is needed to characterize the ability to detect
these planets with sufficiently significant power to determine the
planet yield that could be expected from the LMC (Lund et al. in prep).

% --------------------------------------------------------------------

% \subsection{Metrics}
\subsubsection{Metrics}
\label{sec:\secname:metrics}

The case of transiting exoplanets in the Magellanic Clouds will benefit
from the same metrics that are used by transiting exoplanets within the
Milky Way, and are addressed in \autoref{sec:variables:variablemetrics}
and \autoref{sec:planets}. The key properties of the OpSim to be
measured will be those that relate to the number of observations that
will be made during planetary transits, and the overall phase coverage
of observations.  Unlike the general case of transiting planets in LSST,
transiting planets in the Magellanic Clouds specifically will likely
only have any meaning in deep-drilling fields, or some other comparable
cadence.

\navigationbar

% ====================================================================

% --------------------------------------------------------------------

% --------------------------------------------------------------------

\chapter[AGN]{Active Galactic Nuclei}
\def\chpname{agn}\label{chp:\chpname}

Chapter editors:
\credit{ohadshemmer},
\credit{tanguita}.

Contributing authors:
\credit{AstroVPK},
\credit{tinapeters},
\credit{nielbrandt},
\credit{GordonRichards},
\credit{ScottAnderson},
\credit{mattodowd},
{\it Robert Wagoner}

\section*{Summary}
\addcontentsline{toc}{section}{~~~~~~~~~Summary}

To zeroth order, AGN science with LSST will benefit from the
longest temporal baseline (to aid both selection and variability studies), the most
uniform cadence in terms of even sampling for each band, and uniform sky
coverage while maximizing the area, but excluding the Galactic plane. It
is also expected that any reasonable perturbation to the nominal LSST
observing strategy will not have a major effect on AGN science. While
denser sampling at shorter wavelengths will aid investigations of the
size and structure of the AGN central engine via intrinsic continuum
variability and microlensing, care must be taken not to compromise the
coadded $Y$-band depth which is crucial for detecting the distant-most
quasars. Assuming two visits per night, two different bands are
preferred. Science cases related to intrinsic continuum and
broad-emission line variability will benefit from the denser sampling
offered in the DDFs. These fields will provide powerful ``truth tables''
that are crucial for AGN selection algorithms, enable construction of
high-quality power spectral density functions, and enable measurements
of continuum-continuum and line-continuum time lags. The benefits and
tradeoffs with respect to the main survey involve high-quality light
curves but for only a small fraction ($\sim1$\%) of all sources,
preferentially those at lower luminosities. Certain science cases will
benefit greatly from even denser sampling, i.e., $\sim1 -
1000$~d$^{-1}$, of a smaller area, perhaps during the commissioning
phase, as long as the temporal baseline will extend over the ten years
of the project. Another justification to this strategy is the fact that
very few AGNs, or transient AGNs, have been monitored at these
frequencies on such a long baseline, leaving room for discovery.

% --------------------------------------------------------------------

\section{Introduction}
\label{sec:\chpname:intro}

% Introduce, with a very broad brush, this chapter's science projects,
% and why it makes sense for them to be considered together.

The purpose of this chapter is to identify AGN science cases that may be
affected by the LSST observing strategy and to specify the metrics that can be
used to quantify any potential effects. Since the total number of metrics that
can be quantified is quite large, and the potential effects are not likely to be
significant in most cases, the goal of this chapter is to identify potential
``show stoppers'' that may undermine key AGN research areas. For example,
certain perturbations may reduce significantly the number of ``interesting'' AGNs,
such as $z > 6$ quasars, lensed quasars, or transient AGNs. Another example is
photometric reverberation mapping which is one of LSST's greatest potential
advantages for AGN research but is also very sensitive to the cadence; care must
be taken to ensure that the observing strategy does not undermine the ability to
make the best use of this method.

%This chapter discusses the potential effects of the LSST observing
%strategy on AGN science. In short, there appears to be a consensus
%among the AGN and galaxies communities that AGN science will benefit
%from the most uniform cadence in terms of even sampling for each band
%and uniform sky coverage. It is also expected that any reasonable
%perturbation to the nominal LSST observing strategy will not have a major
%effect on AGN science. This chapter attempts to identify all
%the areas of AGN science that may be affected by the observing strategy
%and to point out the metrics that can be used to quantify any potential
%effects. Since the total number of metrics that can be quantified is
%quite large, and the potential effects are not likely to be significant in
%most cases, the goal of this chapter is to identify potential ``show
%stoppers'' that may undermine key AGN research areas. For example, certain
%perturbations may reduce significantly the number of ``interesting'' AGNs,
%such as $z>6$ quasars, lensed quasars, or transient AGNs. Another example
%is photometric reverberation mapping which is one of LSST's greatest
%potential advantages for AGN research but is also very sensitive to the
%cadence; care must be taken to ensure that the observing strategy does
%not undermine the ability to make the best use of this method.

This chapter is organized as follows. Section~\ref{sec:AGNCensus}
describes potential effects the LSST Observing Strategy may have on the
selection of LSST AGNs in the entire survey, hereafter, the AGN
census. Subsamples of this census are then used throughout this chapter for
discussing particular science cases. Section~\ref{sec:AGNContinuum}
discusses potential effects of the Observing Strategy on studying AGN continuum
(or disk) variability. Section~\ref{sec:AGNMicrolensing} describes how the
LSST cadence may affect estimates or constraints on the size and structure of
the AGN central engine from observations of microlensing events.
Section~\ref{sec:AGNFuture} presents science cases that are still being
developed quantitatively, including sampling effects on measurements related
to the broad emission line region in AGNs. Finally,
section~\ref{sec:AGNDiscussion} discusses additional AGN science aspects that
may be affected significantly by the LSST cadence.

Note: Transient AGN and tidal disruption events are discussed in
detail in the transients chapter
(\autoref{chp:transients}), while gravitationally-lensed AGN are
covered in the cosmology chapter (\autoref{sec:lenstimedelays}).

% --------------------------------------------------------------------

% PJM: moved to Future Work while MAF analysis is pending:
% \input{AGN/AGN_Census}

% --------------------------------------------------------------------

% PJM: moved to Future Work while MAF analysis is pending:
% \input{AGN/AGN_Clustering}

% --------------------------------------------------------------------

% PJM: moved to Future Work while MAF analysis is pending:
% \input{AGN/AGN_BELR}

% --------------------------------------------------------------------

% ====================================================================
%+
% SECTION:
%    AGN_Census.tex
%
% CHAPTER:
%    agn.tex
%
% ELEVATOR PITCH:
%-
% ====================================================================

% \section{AGN Selection and Census}
\section{AGN Selection and Census}\label{sec:AGNCensus}
\def\secname{\chpname:census}\label{sec:\secname}

\credit{ohadshemmer},
\credit{nielbrandt},
\credit{GordonRichards},
\credit{AstroVPK},
\credit{ScottAnderson}

One basic figure of merit for AGN science is the total number of AGNs
discovered in the entire LSST survey, as a function of luminosity and
redshift. The main goal is therefore to adjust the Observing Strategy
in order to maximize this number.
%The primary goal for AGN science is to maximize the discovery of AGN
%with the LSST and construct the largest possible inventory of sources
%spanning the widest possible ranges in the redshift-luminosity parameter
%space. This, in turn, will provide
Doing so will provide tighter constraints within the context
of various cosmological science cases, such as quasar clustering,
$z>6$ quasars and reionization, and strong gravitational lensing.

% --------------------------------------------------------------------

% \subsection{Target measurements and discoveries}
\subsection{Target measurements and discoveries}
\label{sec:\secname:targets}

It is expected that $\approx 10^7 - 10^8$ AGNs will be selected in the
main LSST survey using a combination of criteria, split broadly into
four categories: colors, astrometry, variability, and multiwavelength
matching with other surveys \citep{2009arXiv0912.0201L, 2013AAS...22124710S}.
The LSST Observing strategy will mostly affect the first three of these
categories as described further below.

{\bf Colors:} The LSST observing strategy will determine the depth in each band,
as a function of position on the sky, and will thus affect the color selection
of AGNs. Additionally, it will affect the reliability of the actual
determination of the color, due to the non-negligible time gaps between
observations using two different filters for a particular LSST field. This will
eventually determine the AGN $L-z$ distribution and, in particular, may affect
the identification of quasars at $z\gtsim 6$ if, for example, $Y$-band exposures
are not sufficiently deep.

{\bf Variability:} some AGNs can be effectively distinguished from (variable)
stars, and from quiescent galaxies, by exhibiting certain characteristic
variability patterns (e.g., \citealt{ButlerandBloom2011}). Picking the
right cadence can increase the effectiveness of AGN selection. Ultimately,
hybrid color and variability algorithms will be employed to enhance
the selection process (e.g., \citealt{Petersetal2015}); this may be
particularly important for selecting obscured sources which comprise a
significant fraction of the entire AGN population.

%Non-uniform
%sampling may ``contaminate'' the variability signal of AGN candidates.

{\bf Astrometry:} In cases where selection by color and variability is
insufficient for a reliable identification, AGNs can be further selected
among sources having zero proper motion, within the uncertainties. The
LSST cadence may affect the level of this uncertainty in each band, and certainly the temporal baseline for proper motion measurement, and
may therefore affect the ability to identify (mostly fainter) AGN.
Differential chromatic refraction (DCR), making use of the astrometric offset a
source with emission lines has with respect to a source with a featureless
power-law spectrum, can help in the selection of AGNs and in confirming their
photometric redshifts \citep{KaczmarczikEtal2009}. The DCR effect is more
pronounced at higher airmasses. Therefore, it could be advantageous to have at
least one visit, per source, at airmass greater than about 1.4 (though of course
there is a trade-off with the additional extinction, for faint sources). AGN
selection and photometric redshift confirmation may be affected since the LSST
cadence will affect the airmass distribution, in each band, for each AGN
candidate.
The deep drilling fields (DDFs) will provide a truth table for determining
the predictive power of the DCR method as a function of the airmass
distribution of the observations.

The most critical measurement for the AGN census is having a reliable
and precise redshift for each source, obtained both from a photometric
and an `astrometric' redshift from DCR.

% --------------------------------------------------------------------

% \subsection{Metrics}
\subsection{Metrics}
\label{sec:\secname:metrics}

% Ideas for Metrics:
% detection - how many can LSST detect based on the luminosity function
% (depends on the depth in each band for single epoch and coadd)
% (how will this change with each DR)? @ohadshemmer

% classification - How many of these will we actually classify as quasars?
% non-simultaneous colors. variability of QSOs (how does depend on
% cadence/baseline/seasonal gaps?)

The following are most important for the AGN census:

1) Determine the mean (averaged across the sky) {\bf uncertainty on astrometric
redshifts derived from DCR} as a function of airmass, image quality, and
limiting magnitude. These uncertainties should be compared to the
corresponding uncertainties on the photomteric redshift.

2) Estimate the {\bf number of quasars at $z>6$ that LSST can discover}
during a single visit, as well as in the entire survey, and verify that
these numbers do not fall short of the original predictions. To first order this
simply requires computing the maximum depth in the $Y$-band (for both
single visits and the coadd), averaged across the sky for the nominal
OpSim, as well as assessing the ability to reject L and T dwarfs via astrometry.

3) Assess the effect of {\bf non-simultaneous colors on AGN selection.}
First, the term color should be clearly defined. Potential definitions
include the difference between the co-adds in two bands for the entire
survey (or at a certain point in time during the survey), the difference
between the mdian magnitude in each band during the survey, or the
difference between observations in two bands that are closely spaced in time.
Next, each source would be represented as an ellipse in color-color space.
The aim is to assess the sizes of the ellipses and how these sizes could be
minimized by perturbing the current cadence.

4) Assess how the sampling affects the selection of AGN by variability (e.g.,
interactions with red-noise power spectrum).

5) Check how overall survey length affects proper motion measurements and consequently AGN selection.

%4) Estimate the number of low-luminosity AGN (LLAGN) that can be
%identified during the entire survey.

% --------------------------------------------------------------------

% \subsection{OpSim Analysis}
\subsection{OpSim Analysis}
\label{sec:\secname:analysis}

% OpSim analysis: how good would the default observing strategy be, at
% the time of writing for this science project?

\begin{figure}
\centering\includegraphics[width=0.9\linewidth]{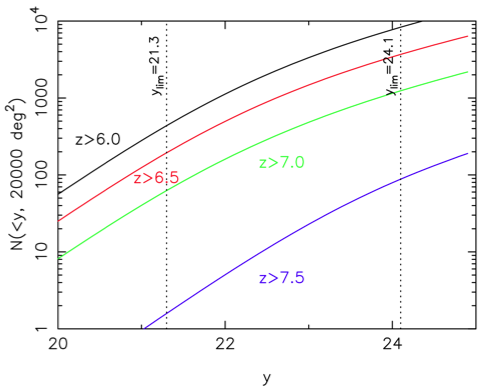}
\caption{Number of quasars at $z>6$ that LSST is expected to discover
based on $Y$-band limiting magnitude in a single epoch (entire survey)
marked by the first (second) dotted line from the left.}
\label{fig:zgt6}
\end{figure}

% LSST Review from Niel Brandt: check for updates needed to this figure, as it is over a decade old. Also, add a plot comparing LSST and WFIRST for high-z AGN selection

For assessing the limitations of DCR on the $L-z$ plane of LSST AGNs,
one needs to obtain from OpSim the current maximal airmasses for each band,
and the associated image quality and limiting magnitude. The current maximal
airmasses, per band, averaged across the sky are: 1.41, 1.50, 1.51, 1.51, 1.51,
and 1.51 for $u, g, r, i, z, Y$, respectively. Need to convolve this with the
seeing in each band to obtain the dependence on airmass and image
quality. This output should be converted into the mean and spread
of the uncertainty on the astrometric redshifts. The best way
to obtain this is to fold the astrometric redshift estimation from
DCR into MAF. Ultimately, one needs to check the implications of higher
airmasses and limiting magnitudes on the ability to obtain more accurate
and precise astrometric redshifts.

For predicting the number of detected $z>6$ quasars
%Compare this magnitude to the
%one required for identifying $\geq1000$ quasars at $z\geq6$.
the
% current enigma\_1189{}
% aquila\_{}1110
\opsimdbname{aquila\_1110}
\OpSim database for the main, i.e., WFD survey, gives a
single-visit $5\sigma$-depth
%old number computed in Bremerton: $Y=22.36$ mag,
%$Y=21.51$ enigma_\_{}1189
{\it Y} $= 21.45$ mag (as compared to {\it Y} $= 21.51$ for the older \opsimdbname{enigma\_1189} run).
For the final co-added $5\sigma$-depth the median magnitude from the \opsimdbname{aquila\_1110}
\OpSim run is
%$Y=24.36$. Enigma\_{}1189
{\it Y}$ = 23.07$ which is more than a magnitude shallower than the older \opsimdbname{enigma\_1189}
depth of {\it Y}$ = 24.36$.
{\bf The lower {\it Y}-band depth may reduce the total number of quasars at $z > 6$ discovered by LSST by
a factor of $\sim 2$
%correspondingly deeper (i.e., better) than enigma\_{}1189
%comparable to
%the original predictions
(see Fig.~\ref{fig:zgt6}).}
% (See the AAS poster from 2013: http://www.lsst.org/sites/default/files/221-RC-247.10-AAS_shemmer.pptx.pdf).

As for general AGN selection, the effects of the sampling on variability selection
should be assessed, and the amplitudes of the uncertainties in color-color space
and how these depend on the cadence should be simulated.
The combination of LSST photometry with that from WFIRST and/or Euclid data should also be considered, both for extending the upper limit in detectable redshift to $\sim10$, but also improving the completeness and purity of the sample at lower redshifts.

% % --------------------------------------------------------------------
%
% \subsection{Discussion}
% \label{sec:\secname:discussion}
%
% Discussion: what risks have been identified? What suggestions could be
% made to improve this science project's figure of merit, and mitigate
% the identified risks?
%
%
% ====================================================================

\subsection{Conclusions}

Here we answer the ten questions posed in
\autoref{sec:intro:evaluation:caseConclusions}:

\begin{description}

\item[Q1:] {\it Does the science case place any constraints on the
tradeoff between the sky coverage and coadded depth? For example, should
the sky coverage be maximized (to $\sim$30,000 deg$^2$, as e.g., in
Pan-STARRS) or the number of detected galaxies (the current baseline
of 18,000 deg$^2$)?}

\item[A1:] The main FoM for AGN science is maximizing the number of AGNs
detected.  Since the co-added depth is already pushing well into the
regime of low-luminosity AGNs, that suggests that area is preferred over
depth.

\item[Q2,3,5:] {\it Does the science case place any constraints on the
tradeoff between uniformity of sampling and frequency of  sampling? Does
the science case place any constraints on the tradeoff between the
single-visit depth and the number of visits (especially in the $u$-band
where longer exposures would minimize the impact of the readout noise)?
Does the science case place any constraints on the fraction of observing
time allocated to each band?}

\item[A2,3,5:] It should be possible to do a MAF analysis to determine
the relative selection completeness and efficiency for OpSim outputs
with different uniformity/frequency of sampling. The same can be said
for constraining the fraction of observing time in each band (where $u$
is the most important for $z\lesssim3$ and $Y$ is most important for $z\gtrsim6$)
and for determining whether nightly
visits should be in the same band or not, and for the trade-off of
single-visit depth and number of visits. However, the AGN census is
unlikely to be the driver in these decisions.

\item[Q4:] {\it Does the science case place any constraints on the
Galactic plane coverage (spatial coverage, temporal sampling, visits per
band)?}

\item[A4:] Given the desire for maximal extragalactic area, added Galactic plane
coverage would be detrimental to AGN science.

\item[Q6:] {\it Does the science case place any constraints on the
cadence for deep drilling fields?}

\item[A6:] Nearly any cadence discussed for the deep drilling fields
will be more than adequate for AGN selection (as opposed to AGN
physics, which will provide constraints); however added depth during
commissioning would enable more robust truth tables.

\item[Q7:] {\it Assuming two visits per night, would the science case
benefit if they are obtained in the same band or not?}

\item[A7:] No preference.

\item[Q8:] {\it Will the case science benefit from a special cadence
prescription during commissioning or early in the survey, such as:
acquiring a full 10-year count of visits for a small area (either in all
the bands or in a  selected set); a greatly enhanced cadence for a small
area?}

\item[A8:] No preference.

\item[Q9:] {\it Does the science case place any constraints on the
sampling of observing conditions (e.g., seeing, dark sky, airmass),
possibly as a function of band, etc.?}

\item[A9:] No.

\item[Q10:] {\it Does the case have science drivers that would require
real-time exposure time optimization to obtain nearly constant
single-visit limiting depth?}

\item[A10:] The census of AGNs is unlikely to be the determining factor
in terms of observing conditions and does not require nearly constant
single-visit depths.

\end{description}

% LSST Review from Niel Brandt: survey must span full 10 years to enable good astrometry / propoer motion measurements to aid in selection. PJM: doesn't fit in the 10 questions, but leaving a note here for the future!

\navigationbar

% ====================================================================
%+
% SECTION:
%    AGN_Disk_Intrinsic.tex
%
% CHAPTER:
%    agn.tex
%
% ELEVATOR PITCH:
%-
% ====================================================================

\section{Disc Intrinsic AGN Variability}\label{sec:AGNContinuum}
\def\secname{\chpname:variability}\label{sec:\secname}

\credit{ohadshemmer},
\credit{AstroVPK},
{\it Robert Wagoner}

A variety of AGN variability studies will be enabled by LSST. These are
intended to probe the physical properties of the unresolved inner regions
of the central engine. Relations will be sought between variability amplitude
and timescale vs. $L$, $z$, $\lambda_{\rm eff}$, color, multiwavelength and
spectroscopic properties, when available. For example, LSST AGNs exhibiting excess
variability over that expected from their luminosities will be further scrutinized
as candidates for lensed systems having unresolved images with the excess
(extrinsic) variability being attributed mainly to microlensing.

Measuring time-delayed responses between variations in the continuum flux in one
LSST band to the continuum flux in another, will be one of the main themes of
AGN science in the LSST era (e.g., \citealt{Chelouche2013};
\citealt{CheloucheandZucker2013}; \citealt{EdelsonEtal2015};
\citealt{FausnaughEtal2015}).
% LSST Review by Niel Brandt: Add Jiang et al? Sentences below are vague: what can be learned _specifically_ about accretion disk models?
Such measurements can test accretion disk models
in a robust manner for a considerably larger number of AGNs than is currently
feasible with microlensing (see section~\ref{sec:agn:microlensing}).

Theories of the hierarchical merger of dark matter halos over cosmic time
predict that galaxy-galaxy mergers should result in the formation of a large
number of binary SMBHs. This population is predicted to be a strong, stochastic
contributor to the overall gravitational wave background
\citep{2015arXiv151105564T}. The inspiral of gravitationally bound pairs of
SMBHs formed by a major merger may `stall', reducing the gravitational wave
signal \citep{2014SSRv..183..189C}. Potentially periodic AGN variability,
leading to tentative discoveries of binary SMBHs (e.g.,
\citealt{2015Natur.525..351D}; \citealt{GrahamEtal2015}; \citealt{LiuEtal2015}),
may be feasible for LSST for periods ranging from a few days up to $\sim3$~yr
over the entire survey. The fraction of close SMBHs, tentatively detected by
LSST, may provide strong constraints on the strength of the graviational wave
signal expected from the inspiral.

In the deep-drilling fields (DDFs), the LSST sampling is expected to provide
high-quality power spectral density (PSD) functions for $\approx10^{5} - 10^{6}$
AGNs across wide ranges of $L$, $z$, and $\lambda_{\rm eff}$ down to frequencies of
$\sim1$~d$^{-1}$. These PSDs can enhance AGN selection, and can be used to
constrain the SMBH mass and accretion rate/mode, as well as enable searching for
periodic or quasi-periodic oscillations (QPOs).

%
%Potentially periodic AGN variability, leading to tentative discoveries
%of binary SMBHs (e.g., \citet{GrahamEtal2015}), may also be
%measurable.
%

%Photometric reverberation mapping (PRM), measuring the time-delayed
%response of either the flux of the broad emission line region (BELR)
%lines to the flux of the AGN continuum or between the continuum flux
%in one (longer wavelength) band to the continuum flux in another (band
%with shorter wavelength), will be one of the cornerstones of AGN
%research in the LSST era (e.g., \citet{Chelouche2013};
%\citet{CheloucheandZucker2013}; \citet{CheloucheEtal2014};
%\citet{EdelsonEtal2015}; \citet{FausnaughEtal2015}). For example, LSST
%is expected to deliver BELR line-continuum time delays in
%$\sim10^5-10^6$ sources, which is unprecedented when compared to
%$\sim50-100$ such measurements conducted via the traditional, yet much
%more expensive (per source) spectroscopic method. Sources in the
%deep-drilling fields (DDFs) will benefit from the highest quality PRM
%time-delay measurements given the factor of $\sim10$ denser sampling.

The high-frequency QPO (HFQPO) periods expected from the inner accretion disk
(which provide stable clocks located closer to the horizon as the BH spin increases)
can be estimated from those of the fundamental $g$-mode, which agree with
the observed HFQPO frequencies in stellar-mass BH binaries. Utilizing the
theoretical upper bounds for BH spin and $L/L_{\rm Edd}$, and the lower
bound to the $k$- and bolometric correction $B_n(z)$, one obtains
$\log P({\rm hr}) > 0.4(1-m_n) + \log[(1+z)D_{\rm L}(z)^2]$ for magnitude
$m_n$ in a particular band $n$ and luminosity distance $D_{\rm L}(z)$.
The $k$- and bolometric correction $B_n(z)$ is a decreasing function of BH mass,
but an increasing function of BH spin. The Lyman-alpha forest limits the redshift
range to $z < 2.7$ for $g$-band observations. The HFQPOs will be weaker within longer
wavelength bands.
%The $\sim 80$ visits in the $g$-band proposed for the main survey appears
%insufficient to produce a useful PSD. The expected HFQPO periods are longer than a few hour visit in a DDF.
For instance, for $m_g  <  23$ and the optimal $z =  2.7$, the HFQPO period is $P > 4$~hr.
%
%QPO search will be most relevant for the DDFs,
Searching for HFQPOs in the DDFs will be most effective if the sampling frequency
in those fields for the $u$ and $g$ bands is at least nightly, i.e., $\gtsim3000$
visits, per band, during the entire survey. Given that the period of a typical HFQPO is
related to the SMBH mass by $P({\rm hr}) \approx (1.1-4.0)(1+z)(M/10^{7}M_{\odot})$,
LSST will be sensitive to probing SMBHs with $M < 10^{7}~M_{\odot}$ using HFQPOs.

%In addition, there's a need for an "ultradeep" field, e.g., the MCs, that will
%be monitored, during commissioning phase, with frequencies in the range ~$1$-$10^4$ min. (i.e., from
%minutes to weeks/months).

% --------------------------------------------------------------------

\subsection{Target measurements and discoveries}
\label{sec:\secname:targets}

%We will measure the power spectral density (PSD) of AGN light
%curves across $L$, $z$, and $\lambda_{\rm eff}$. Specifically, we will
%measure the short timescale ($\leq 5$~d) spectral index of the PSD and
%the locations of `features' such as QPOs
%and breaks in the PSD.

In the main survey, standard time-series analysis techniques will be used for
measuring time delays between pairs of continuum bands and for detecting
periodic AGNs. Correlation analyses will search for relations between AGN
variability properties and their basic physical parameters. In the DDFs, such
analyses will enable probing deeper and more frequently, resulting in
higher-quality data that will provide stronger constraints on AGN variability propertie; the only drawback is
the relatively smaller number of sources available at the high-luminosity end.

A key measurement enabled by the DDFs is a high-quality PSD, in six bands,
for the largest number of AGNs to date. These PSDs, which are rich
in diagnostic power, will be used to search for ``features'' such as QPOs
and breaks, as well as power-law slopes, that can help constrain SMBH masses
and accretion rates.
% LSST Review from Niel Brandt: describe current optical/NIR PSD results, add refs.
Additionally, the PSDs can serve as selection
tools, to more effectively distinguish AGNs from variable stars, as
well as a basis to propose cadence perturbations to further enhance
AGN selection.

A high-quality PSD, extending to high frequencies
% (reaching $\sim 1$ min timescales for stacked PSDs),
% PJM: commented out followin LSST Review by Niel Brandt: is 1 min really needed?
can effectively distinguish AGNs from other
variable sources, assuming AGN light curves are described by a particular
continuous-time autoregressive moving average model (C-ARMA; \citet{KellyEtal14}),
i.e., C-ARMA(2,1), corresponding to a damped harmonic oscillator.
Determination of the parameters that describe the PSD requires light curve
sampling at least as frequent as $\sim1$~d$^{-1}$. Figure~\ref{PSDvsFreq} shows
the frequency dependence of the spectral index of the PSD for one particular AGN,
Zw 229-15, observed with {\em Kepler}. The light curve of this source is
well-described by a C-ARMA(2,1) model. The C-ARMA(2,1) model is a
% higher order random walk
damped harmonic oscillator rather
than the damped random walk (DRW) model of \citet{Kelly09}, which
corresponds to a C-ARMA(1,0) model. Recent variability studies indicate that
the simple C-ARMA(1,0) model is insufficient to model AGN variability because
the spectral index of its PSD is mathematically constrained to be 2
\citep{KellyEtal14,Kasliwal15,Simm15}. Insufficient sampling of an AGN light
curve (e.g., a few times a month), can therefore result in the erroneous conclusion
that a DRW model adequately characterizes the variability.

\begin{figure}
  \begin{subfigure}[t]{0.5\textwidth}
    \centering\includegraphics[width=0.9\linewidth]{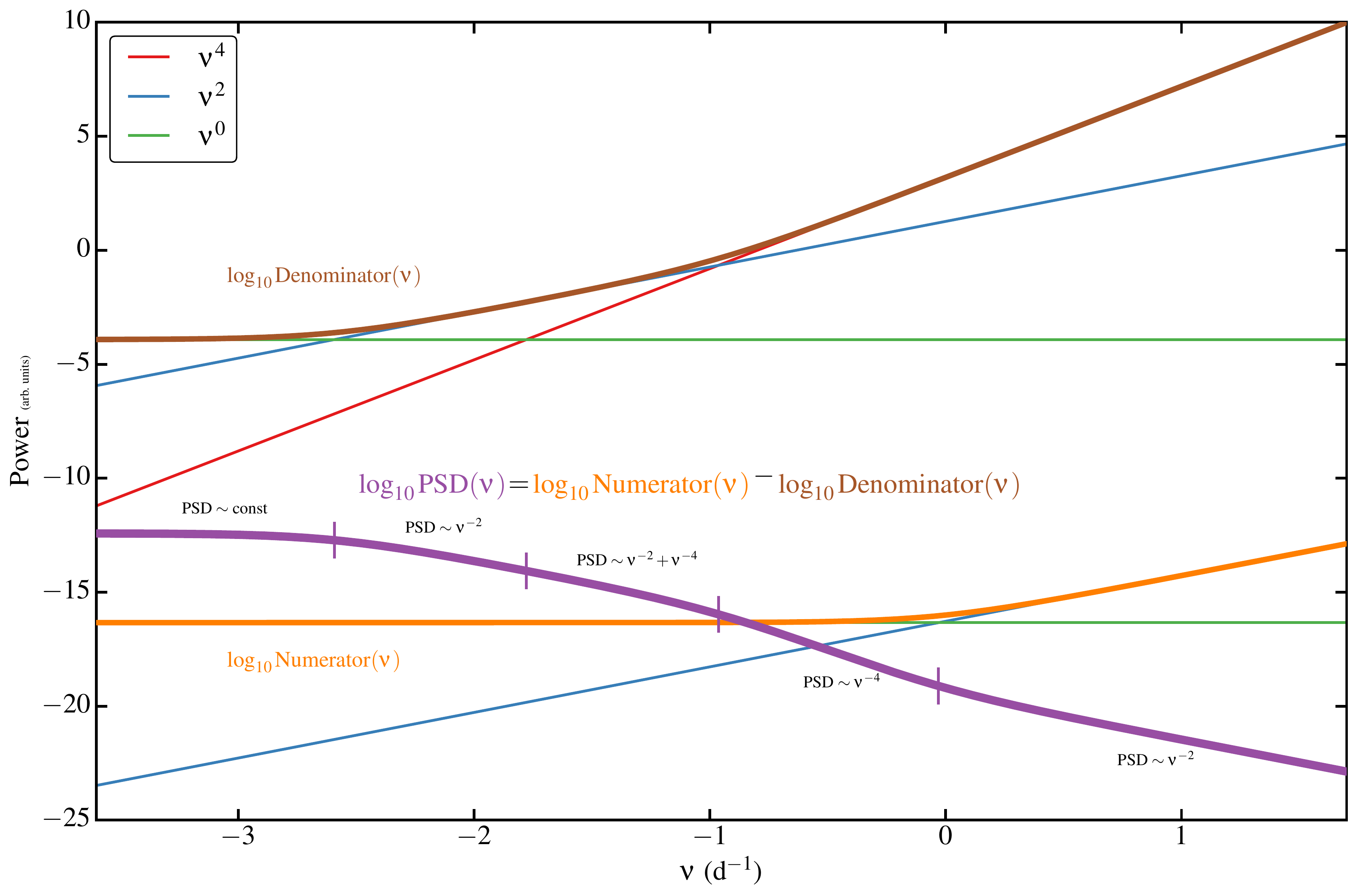}
    \centering
    \caption{}
    \label{fig:PSDvsFreq}
  \end{subfigure}
  %\medskip
  \begin{subfigure}[t]{0.5\textwidth}
    \centering\includegraphics[width=0.9\linewidth]{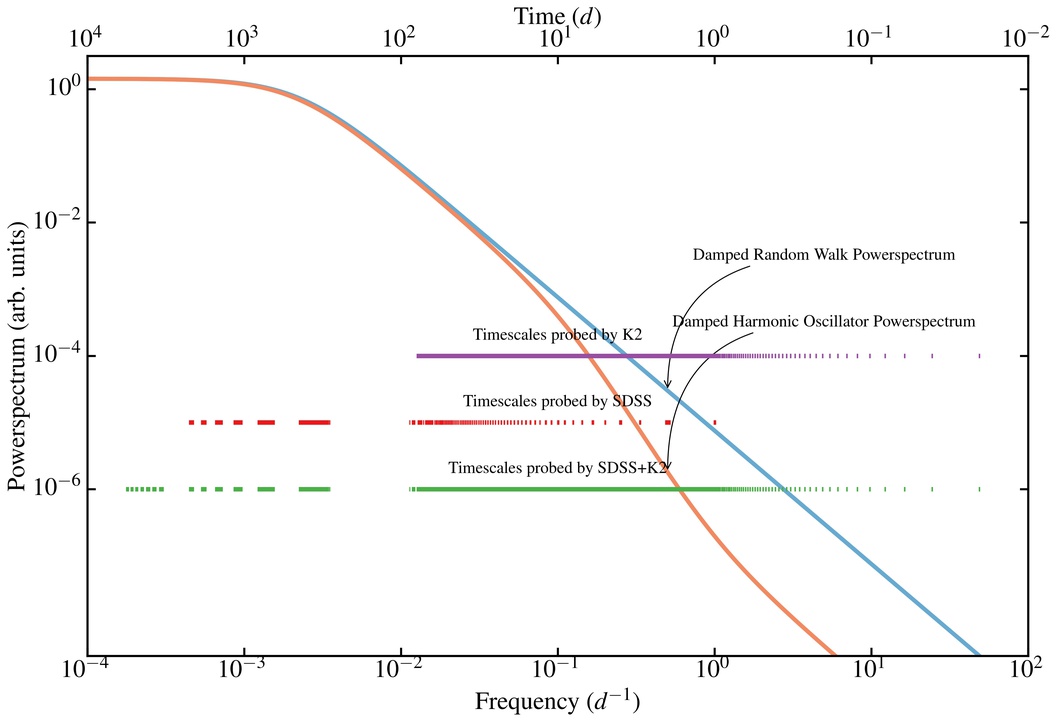}
    \centering
    \caption{}
    \label{fig:SDSSK2Power}
  \end{subfigure}
  \caption{(\subref{fig:PSDvsFreq}) shows the PSD of Zw 229-15 as a function of frequency,
  obtained from {\em Kepler} photometry. The PSD (purple) is the ratio of an even
  polynomial numerator (orange) to an even polynomial denominator (brown).
  This AGN is well-described by a C-ARMA(2,1) model; different powers of frequency
  dominate its PSD at different frequencies depending on the hyperparameters of
  this model. The wide frequency range enables detection of PSD spectral index variations
  ranging between 0 and 4. Clearly, the light curve of this AGN must be sampled on
  timescales {\em shorter} than $1-5$ days in order to observe the $\nu^{-4}$ behavior
  characteristic of a higher order random walk. This is illustrated in (\subref{fig:SDSSK2Power})
  where we see the frequencies and time intervals probed by SDSS, Kepler and a light curve
  constructed by combining the two datasets (SDSS+K2). Each vertical dashed line corresponds to
  a pair of observations seperated by the indicated $\delta t$ (top axis). We plot (for
  illustration), two C-ARMA models with the same overall power - a damped random walk, i.e. a
  C-ARMA(1.0) process, and a damped harmonic oscillator, i.e a C-ARMA(2,1) process.
  It is clear that SDSS (Kepler) cannot probe the highest (lowest) frequencies. However the
  combination of the two can cover the full frequency range. The LSST cadence should be chosen
  to provide similar temporal coverage in the DDFs.}
  \label{PSDvsFreq}
\end{figure}

Accurate recovery of the PSD parameters can be greatly enhanced by increasing the
sampling frequency. To illustrate the effects of the cadence, Figure \ref{CadenceEffect}
shows how the inferred joint distribution of two hyperparameters of the C-ARMA(2,1)
model, the oscillator timescale and the damping ratio, depend on the sampling frequency.
Degrading the sampling frequency from $1/$($30$~min), corresponding to {\em Kepler}
light curves, to $1/$($3$~d), corresponding to the nominal DDF cadence, changes both
the size and the shape of the joint distribution, degrading both the accuracy and
correlation of the inferred hyperparameters.
Furthermore, the C-ARMA formalism may enable adjusting the cadence of the DDFs once
the LSST survey begins to determine the sampling pattern in real time.

\begin{figure}
\centering\includegraphics[width=0.9\linewidth]{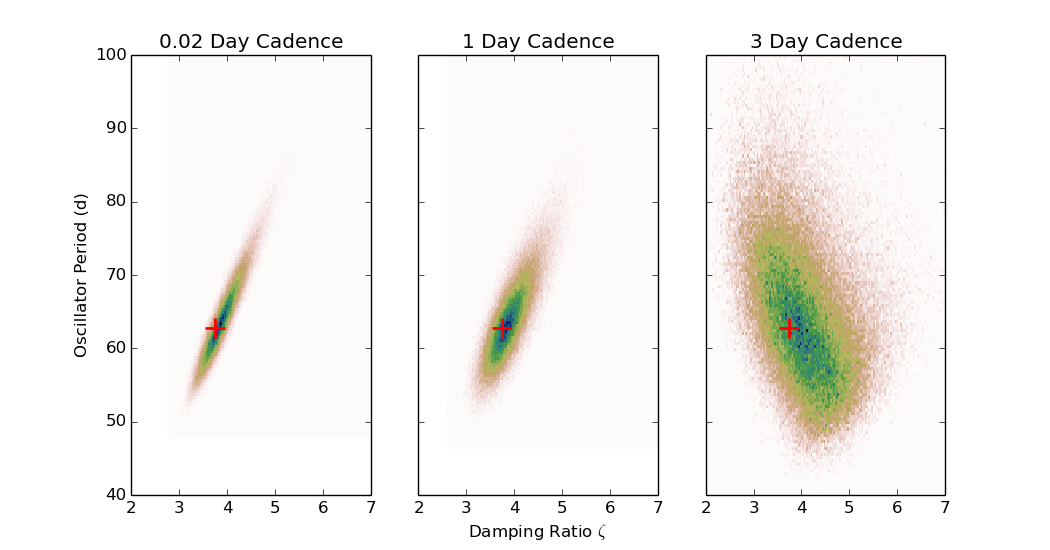}
\caption{The effect of sampling frequency on hyperparameter estimation (courtesy of
J.~Moreno). Light curves were generated using a C-ARMA(2,1) model using the best-fit
parameters for Zw 229-15, observed with {\em Kepler}, indicated by the red cross in
each panel. The light curve was then down-sampled to simulate the effect of observational
cadence. Constraints on the oscillator period and damping ratio begin to widen noticeably
at 3-day sampling. At 1 week and longer cadence (not shown), one does not recover the
correct model order. This strongly indicates the importance of further study to refine
the cadence requirements for LSST.
}
\label{CadenceEffect}
\end{figure}

%The PRM measurements will probe the size and structure of the
%accretion disk and BELR, in a statistical sense, and may provide
%improved SMBH mass estimates for sources that have at least
%single-epoch spectra. \new{Our goal is to understand the population of
%AGN broad line regions, including their geometry. We expect to do this
%via  a model that connects the BH mass, BLR geometry and AGN
%photometric variability properties via a set of scaling relations. A
%simple version of this is could be something like...\newline\newline
%So, our target measurements are of $a$ and $b$, the X parameters.
%Before we derive a metric that quantifies our ability to measure these
%parameters, we can anticipate some of the sensitivity of the
%photometric RM method to observing strategy.}

%\new{We focus on the PSD function as a way of characterizing AGN
%variability in various ways. What do we expect the AGN population to
%look like in PSD parameter space? The hyper-parameters that govern the
%relationships between PSD parameters and  AGN and host galaxy
%properties are probably of greatest scintific interest.}

% --------------------------------------------------------------------

\subsection{Metrics}
\label{sec:\secname:metrics}

% Quantifying the response via MAF metrics: definition of the metrics,
% and any derived overall figure of merit.

%\new{In lieu of a simulated AGN population, we focus on a few
%particular {\it diagnostic} metrics that capture  our likely ability
%to measure the PSD across the population. These include: the
%uniformity of the sampling pattern in log time lag?}

%Assess the number of meaningful BELR-continuum time delays that can be obtained
%with the nominal OpSim, and point out potential perturbations in the
%cadence to improve the number and quality of such time delays.

While additional quantitative work is required for determining the optimal cadence for
fully capturing AGN accretion physics and to enhance AGN selection, it is clear
that even the nominal DDF sampling (\eg in \opsimdbname{db:baseCadence}) is barely sufficient, and
more frequent sampling would be ideal.
% LSST Review from Niel Brandt: Haven't connected the hyperparameters to science impacts clearly. What exactly is lost going from 1 day to 3 day cadence?
The ability to detect HFQPOs
should also improve by increasing the sampling frequency, the amplitudes of such
features are quite uncertain, as are the (short) duty cycles. Observations,
theory and numerical simulations have only suggested that the fractional
modulation should be small (less than a few percent). Thus it is not obvious how
to choose a metric and observing strategy to maximize it, other than increasing
the sampling frequency to at least nightly samplings in the $g$ and $u$ bands
(i.e., increasing the sampling frequency in the DDFs at least by a factor of
$\sim 3$).

Specific metrics include:

1) LSST can make a significant contribution using
the C-ARMA formalism in the selection of low-luminosity AGN (LLAGN), i.e.,
sources with $L \ltsim 10^{42}$~erg~s$^{-1}$, in the DDFs. Such sources are
likely to be missed by traditional color-variability selection algorithms due to
a strong host contribution. The metric to be developed should assess how the
number of selected LLAGN depends on the sampling frequency in each band, and take into account the host galaxy light contamination.

2) Assessing the standard deviation of the error in recovered time-lag between
bands, $\tau$, using a cross-correlation analysis. The goal is to minimize
$\sigma_{\Delta \tau}$. Additionally, one should assess the worst case estimate of
the time-lag between bands, i.e., minimizing $\max \vert \Delta \tau \vert$.

3) Determining the fractional error on the slopes and features of the PSD.
Assuming that AGN variability is parametrized by a C-ARMA process with
autoregressive roots $\{\rho_{i}:1 \leq i \leq p\}$, the damping timescales and
QPO centers are given by $\tau_{i} = 1/|\Re(\rho_{i})|$ and ${\rm QPO}_{i} =
2\pi/|\Im(\rho_{i})|$, respectively. One needs to investigate how well different
sampling strategies recover each damping timescale and QPO center for a range of
assumed models. The choice of appropriate models can be guided by using
variability data from K2 observations of SDSS quasars.

% % --------------------------------------------------------------------
%
% \subsection{OpSim Analysis}
% \label{sec:\secname:analysis}
%
% OpSim analysis: how good would the default observing strategy be, at
% the time of writing for this science project?
%
%
% % --------------------------------------------------------------------

\subsection{Discussion}
% \subsubsection{Discussion}
\label{sec:\secname:discussion}

% Discussion: what risks have been identified? What suggestions could be
% made to improve this science project's figure of merit, and mitigate
% the identified risks?

While science-driven metric analysis is still to be performed, we expect the key requirement emerging from such an analysis to be to increase the nominal sampling
frequency in the DDFs by at least a factor of 3, i.e., having at least
3000 visits, per band, during the entire survey. Alternatively, if this
sampling is not feasible for all the DDFs, it would be beneficial to
identify a subset of ``special'' DDFs which would be sampled by this
frequency. Such DDFs would also benefit from being circumpolar,
% e.g., the Magellanic Clouds,
enabling a more uniform sampling to produce the
highest quality PSDs.

% ====================================================================
%
\subsection{Conclusions}

Here we answer the ten questions posed in
\autoref{sec:intro:evaluation:caseConclusions}:

\begin{description}

\item[Q1:] {\it Does the science case place any constraints on the
tradeoff between the sky coverage and coadded depth? For example, should
the sky coverage be maximized (to $\sim$30,000 deg$^2$, as e.g., in
Pan-STARRS) or the number of detected galaxies (the current baseline
of 18,000 deg$^2$)?}

\item[A1:] The disc-intrinsic variability science case places no direct
constraints on the tradeoff between the sky coverage and the coadded depth.

\item[Q2:] {\it Does the science case place any constraints on the
tradeoff between uniformity of sampling and frequency of sampling? For
example, a rolling cadence can provide enhanced sample rates over a part
of the survey or the entire survey for a designated time at the cost of
reduced sample rate the rest of the time (while maintaining the nominal
total visit counts).}

\item[A2:] Preliminary studies of the impact of the survey strategy on
the disc-intrinsic variability science case indicate that
%\begin {enumerate}
%\item
having the longest-possible temporal baseline, i.e., the interval between
the first and last observation, is mandatory. The temporal baseline
\emph{must} be at least 2-3 $\times$ the longest timescale ($\tau$) built
into the light curve \citep{2017A&A...597A.128K}. \citet{2010ApJ...721.1014M}
find that $\max \tau \sim 1000$ d suggesting that survey strategies
that do not visit every field at both the beginning and end of the survey will
be sub-optimal for this science case.
%\item
%having intervals of time during which the survey is performed with
%higher sampling frequency is mandatory.
Furthermore, Fig. \ref{fig:PSDvsFreq} demonstrates
that in order to accurately distinguish between various stochastic models of
AGN variability, the high frequency behavior of the light curve must be
sufficiently explored by utilizing the DDFs.
%\end{enumerate}

%In summary, non-uniform sampling is preferred. Some variant of the rolling-cadences
%strategies that provides a long temporal baseline but also periods of very high
%frequency observing is ideal.

\item[Q3:] {\it Does the science case place any constraints on the
tradeoff between the single-visit depth and the number of visits
(especially in the $u$-band where longer exposures would minimize the
impact of the readout noise)?}

\item[A3:] The science cases would benefit from maximizing the number of visits.

\item[Q4:] {\it Does the science case place any constraints on the
Galactic plane coverage (spatial coverage, temporal sampling, visits per
band)?}

\item[A4:] AGN science cases would benefit from minimizing coverage of
the Galactic plane.

\item[Q5:] {\it Does the science case place any constraints on the
fraction of observing time allocated to each band?}

\item[A5:] In general, the science cases place no constraints on
the fraction of observing time allocated to each band.
For QPOs, however, shorter wavelengths are preferred.

\item[Q6:] {\it Does the science case place any constraints on the
cadence for deep drilling fields?}

\item[A6:] The science cases would benefit from maximizing the sampling
rate in the DDFs to at least one visit per day in each band.

\item[Q7:] {\it Assuming two visits per night, would the science case
benefit if they are obtained in the same band or not?}

\item[A7:] Both continuum-continuum lag studies as well as color-variability studies
of AGN would be strongly benefited by having the two (or more) visits per night be in
\emph{different} bands.

\item[Q8:] {\it Will the case science benefit from a special cadence
prescription during commissioning or early in the survey, such as:
acquiring a full 10-year count of visits for a small area (either in all
the bands or in a  selected set); a greatly enhanced cadence for a small
area?}

\item[A8:] The disc-intrinsic variability science case would be benefited by having a
greatly enhanced cadence for a small commissioning area of the sky as long as this area
is followed up with the normal revisit schedule later during the survey in order to
ensure that the temporal baseline for this region is as long as possible.

\item[Q9:] {\it Does the science case place any constraints on the
sampling of observing conditions (e.g., seeing, dark sky, airmass),
possibly as a function of band, etc.?}

\item[A9:] The disc-intrinsic variability science case places no constraints on
the sampling of observing conditions.

\item[Q10:] {\it Does the case have science drivers that would require
real-time exposure time optimization to obtain nearly constant
single-visit limiting depth?}

\item[A10:] No

\end{description}

% ====================================================================

\navigationbar

% ====================================================================
%+
% SECTION:
%    AGN_Disk_Extrinsic.tex
%
% CHAPTER:
%    AGN.tex
%
% ELEVATOR PITCH:
%    Using AGN microlensing to measure the size and structure of
%    accretion disks. Depends on well-sampled multi-filter light curves,
%    and a large sample of detected strongly-lensed AGN.
%
% AUTHORS:
%    Timo Anguita (@tanguita), Matthew O'Dowd
%-
% ====================================================================

\section{AGN Size and Structure with Microlensing}\label{sec:AGNMicrolensing}
\def\secname{\chpname:microlensing}\label{sec:\secname}

\credit{tanguita},
\credit{mattodowd}

Microlensing due to stars projected on top of individual
gravitationally-lensed quasar images produces additional magnification which can be used to probe the structure of the background high-redshift AGN.

The microlensing method is based on the fact that the variability amplitude depends on the quasar size relative to the Einstein radius of a star (projected into the source plane). By comparing the variability amplitudes at different wavelengths, we can determine the relative source sizes and test the predicted wavelength scaling: Assuming a thermal profile for accretion disks, sizes in different emission
wavelengths can be probed and as such, constraints on the slope of this
thermal profile can be obtained. Given the sheer number of lensed systems that LSST is expected to discover ($\sim8000$), this will allow us to stack systems for better
constraints and hopefully determine the {\it luminosity and redshift evolution
of the disk size and profile.} Due to the typical relative velocities of lenses,
microlenses, observers (Earth) and source AGN, the microlensing variation
timescales are between months to a few decades.

%Using the fact that the Einstein radii of stars in lensing galaxies
%closely match the scales of different emission regions in
%high-redshift AGNs (micro-arcseconds), analyzing microlensing induced
%flux variations statistically on individual systems allows us to
%measure ``sizes'' of AGN regions.

% --------------------------------------------------------------------

\subsection{Target measurements and discoveries}
\label{sec:\secname:targets}

Analysis of microlensing induced variability will allow the measurement of
accretion disk sizes $R_\lambda$ and their thermal profile slope $\alpha$,
which together strongly contrain the physics of accretion.
This needs to be done per system discovered. Assuming $\sim$1000 lensed quasars with
high-quality light curves (i.e. that allow time-delay measurements, see
\autoref{sec:lenstimedelays}), a relationship between the size, thermal profile
slope, and luminosity of the accretion disk and the mass of the black hole
will likely be derived.

How precisely are we going to be able to measure these parameters for a given
survey strategy? This is not a simple question to answer due to the significant
degeneracies that plague the phenomenon. This is what our MAF metric will
quantify. Before we design this, we need to predict and statistically quantify
the degeneracies and sensitivities.

% \new{Our goal is to understand the population of AGN accretion disk sizes
% and profiles. We anticipate doing this via a hierarchical model where
% these properties are related to each other in some way, perhaps via
% power law scaling relations. A very simple version of this is the following...
% \newline\newline
% So, our targets are the parameters $a$ and $b$, that describe this
% simple population. How well will we be able to measure these, for a
% given survey strategy? This is what our MAF metric will quantify.
% Before we design this, we can predict the likely sensitivities of this
% measurement.}

The quasar microlensing optical depth is $\sim1$, so every lensed quasar should
be affected by microlensing at any given point in time to a different extent.
This continuous, low-level microlensing can provide useful statistical constraints on
size and geometry for both accretion disks (e.g., \citealt{kochanek2004}; \citealt{bate2008}; \citealt{floyd2009}; \citealt{blackburne2011,blackburne2014}; \citealt{jimenez2014}) and
broad emission line regions  (e.g. \citealt{kochanek2004}; \citealt{richards2004}; \citealt{wayth2005}; \citealt{sluse2011}; \citealt{odowd2011}; \citealt{guerras2013}). Over LSST's 10-year life,
the resulting library of light curves will enable a statistical analysis of this
low-level microlensing that will far exceed previous efforts.

Note, however, that the larger the apparent magnification, the more stringent are
the constraints on the geometric properties of the source.
\citet{MosqueraandKochanek2011} studied the expected microlensing
timescales for all known lensed quasars at the time. They found that the median
Einstein crossing time scales, which can statistically be interpreted as the
time between high-magnification events, in the observed $I$-band, is of the order
of $\sim20$~yr (with a distribution between 10 and 40~yr). Additionally, the source
crossing time (duration of a high-magnification event) is $\sim7.3$~months (with
a distribution tail up to 3~yr). This basically means that out of all the lensed
quasar {\em images} (microlensing between images is completely uncorrelated)
about half of them will undergo strong microlensing events during the 10~yr
baseline of LSST. However,
since the typical number of lensed images is either two or four, this means
that, statistically, in every system, one (for doubles) or two (for quads) high
magnification events should be observed in 10~yr of LSST monitoring.

Strong lensing events also provide our best chance of investigating anomalies in accretion
disk geometries. For example, warps due to multiple accretion events or magnetic
fields, fragmentation due to gravitational instability, and hot spots due to
embedded star formation can all result in deviations from smooth temperature profiles.
Such anomalies are expected to have an effect on the light curves of strong microlensing events.

Note that, the important cadence parameter is the source crossing time. Ideally,
high-magnification events would need to be as uniformly sampled as possible. The
$\sim 7.3$ months crossing time is the median for the observed $i$-band, but this time
would be significantly shorter for bluer bands: for a thermal profile with slope
$\alpha: R_\lambda \propto \lambda^\alpha$ implies source crossing time $t_{\rm
s} \propto \lambda^{1/\alpha} \rightarrow t_u=t_i \times (\lambda_{\rm u} /
\lambda_{\rm i})^{1/\alpha}$. For a Shakura-Sunyaev slope of $\alpha=0.75$ this
would correspond to $\sim 7.3 \times (3600/8140)^{4/3}$ months which is $\approx 2.5$
months in the $u$-band.

In terms of the cadence, at least three evenly sampled data points per band
within two to three months would be preferred to be able to map the constraining
high-magnification event, and these would hopefully be uniformly spaced.
Additionally, LSST can trigger imaging of high-magnification events with dedicated
facilities to enhance these constraints. More frequent sampling (e.g., in the DDFs)
would increase such constraints significantly. However, since lensed quasars are not
that common, this smaller area would mean that only a modest number ($<100$)
of suitable systems will be monitored in the DDFs.

Regarding the season length, the ``months'' timescale of high
magnification events very likely means that we can/will miss high
magnification events in the season gaps, at least in the bluer bands.

``Show stopper'': observations spread on timescales larger than 3 months.
This would likely miss the high-magnification events. In those cases
we could perhaps consider close consecutive photometric bands as
equivalent accretion disk regions, however this would mean weaker
constraints on the thermal profile.

Note that the discussion above is centered on high-magnification events. Even
though these produce the most valuable information on short timescales (e.g., \citealt{Anguita2008, eigenbrod2008}), low magnification events or
\emph{no} magnification events can set constraints on the structure of high
redshift AGN as well as the lensing galaxies (e.g. \citealt{gilmerino2005}). In particular, since isolated high-magnification events are rare, accretion disk size studies with microlensing are done by a Bayesian analysis of long (several years) light curves of microlensing in every lensed image simultaneously compared to source plane microlensing magnification patterns (e.g., \citealt{kochanek2004,blackburne2014}). Constraints using this method rely on the fact that even when not strong, several uncorrelated microlensing induced variations in all lensed images as well as anomalous (single epoch) flux ratios (e.g., \citealt{bate2008,rojas2014}) in all available bands need to be consistent with the underlying geometrical structure of the background accretion disk. It is important to take into consideration that the timescales directly depend on the projected velocities of the three-plane system: the redshifts of the lens
and source as well as their respective peculiar velocites along the CMB dipole velocity in the direction towards the lensing galaxies (observer's peculiar velocity).

Long-timescale high-accuracy multi-band data as will be delivered by
LSST have never been obtained to date for any lensed system. Coupling this fact
to a factor $\sim$10 increase in the number of lensed quasars known, LSST will
enable totally new and unprecedented perspectives for microlensing studies.

%
% Important Note: all this science needs to be done on lensed quasars
% with measured or very short time delays to remove the intrinsic
% variability signal, which might significantly reduce the sample.

{\bf Microlensing Aided Reverberation Mapping:} Quasar broad emission lines exhibit a very wide range of ionization energies. This implies differences in the incident ionizing flux, and suggests that we may expect a radial stratification of ionization species This is indeed observed in both reverberation mapping (see review by \citealt{gaskell2009}) and gravitational microlensing (\citealt{guerras2013}). However proximity to the ionizing source alone isn't enough to explain the wide range of ionizations observed in AGN broad-line flows; the details of self-shielding, gas density, and overall flow geometry will define where a given line will be efficiently produced within the flow. As such, measurement of size as a function of ionization potential can be highly constraining for broad line models. LSST's bands will provide reasonable isolation of prominent broad emission lines or line pairs at all redshifts. In particular, two or more of H$\beta$, H$\gamma$ + H$\delta$, MgII, CIII], and CIV + SiIV dominate broad line emission in single bands at most redshifts. These emission line species or pairs span a broad range of ionization potentials and so will measure its relationship to emission region size. Photometric Reverberation Mapping (PRM) may produce this measurement over a fraction of the LSST quasar sample. Now, given that microlensing mostly affects continuum emission rather than  broad line emission, microlensing can enable the disentangling of the broad line emission plus the continuum emission in single photometric bands, allowing the use of even single broad band PRM measurements \citep{SluseandTewes2014} in the lensed subsample. Note that at all redshifts at least one and often two bands are relatively free of broad line emission and so provide a rough measure of continuum strength for both variability and relative microlensing magnification measurement. As with the two-band PRM method discussed in section \ref{sec:AGNBELR}, the
denser (and the longer) the sampling, the more accurate are the constraints that
can be obtained for the time delays. This method allows constraining both the
accretion disk structure as explained above and the Broad Emission Line Region (BELR). The only additional
requirement is one spectroscopic observation to constrain the ``macro''
magnification ratios from narrow emission lines.

\begin{center}
	\begin{figure}[hbt]
		\includegraphics[width=\textwidth]{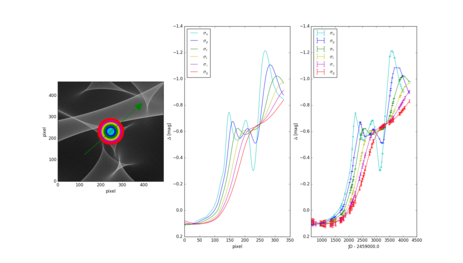}
		\caption{Ten year long simulated light curve for image A of RXJ1131-1231 using $\sigma_0$=2.0 light days at 2000\AA{}, $\alpha$=0.75, $\kappa$=0.494, $\gamma$=0.562 and 40\% of matter in compact form (stars). The left panel shows the concentric Gaussian emission regions observed by the LSST filters projected on top of the magnification pattern. The center panel shows the light curves with a ``perfect'' cadence. The right panel shows the magnitude values interpolated at epochs observed by LSST at the location of the system according the the ``Baseline Cadence'' (\opsimdbref{db:baseCadence})) Opsim output. The quasar was assumed to have the same intrinsic brightness in each band for easier comparison of variations between bands.}
		\label{microsimcurve}
	\end{figure}
\end{center}

% --------------------------------------------------------------------

\subsection{Metrics}
\label{sec:\secname:metrics}

Metrics for these section need to be defined by using simulated light curves
that take into account the several parameters that come into play in quasar
microlensing. These include: the time gap between visits in the same band,
projected CMB velocity, simulated peculiar velocities and redshifts of lenses
and sources as well as ``macro'' lens model parameters (i.e., surface mass
density and shear projected on top of lensed quasar images). Two metrics are
currently in consideration:

High Magnification Events recovery metric: This metric will measure the
number of high-magnification events recovered/missed considering the
cadence and season length in every LSST band and as the precision of the
brightness measurement.

Accretion disk size and slope metric: This metric will do a full
analysis of the ``pure'' microlensing light curves to recover these two
physical AGN parameters. The figure of merit would be the accuracy of
the measurement.

Since the microlensing signal can only be obtained after time delays between images
have been measured, both metrics need close interaction with time delay
measurements. As such, the ``Time Delays Challenges'' (see
\autoref{sec:lenstimedelays}) will include complete microlensing signal
simulations which also take into account the aforementioned parameters. Note
that given
the dependence on individual filter cadence and season length as well as
projected CMB velocities, every region on the sky needs to be considered
independently. Time Delay Challenge submissions will thus include recovered
``pure'' microlensing
light curves in addition to measured time-delays. By doing the reverse
procedure, i.e. using these ``pure'' microlensing light curves to statistically
re-obtain the input accretion disk sizes and thermal slopes, we will be able to
quantitatively measure the accuracy of the intrinsic accretion disk parameter
estimations for a given survey strategy.

We have build a preliminary tool that extracts micro lensing light curves from source plane magnification patterns as will be observed by LSST, taking into account:
\begin{itemize}
\item Projected peculiar velocities of the observer (projected CMB dipole in the direction of the system), lens and source.
\item Bulk motion of stars in the lensing galaxy (from the stellar velocity dispersion of the lens).
\item Specific LSST observing epochs in the direction of the system (from Opsim outputs).
\item Thermal profile slope of the accretion disk $\alpha$.
\item Scale size of the accretion disk $\sigma_{0}(\lambda)$ at a given wavelength $\lambda$.
\item Smooth (dark matter) to compact (stars) matter ratio on top of lensed quasar images.
\item ��Macro'' lens model parameters on top of lensed quasar images: Surface mass density $\kappa$ and shear $\gamma$.
\end{itemize}

\noindent In its current state, the tool assumes simple face-on concentric Gaussian emission regions for the accretion disk. An example of such a curve is shown in figure \ref{microsimcurve}. To recover the figure of merit, (measurement accuracy of $\alpha$ and $\sigma_0$), light curves generated with this tool for a given realistic lensed quasar system ($\kappa$, $\gamma$, $s$ and velocity dispersion of the lensing galaxy as well as time delays between lensed images) for every region in the sky need to analyzed using the above mentioned statistical analyses to recover the input accretion disk parameters.

% microlensing - convolve microlensing timescales for QSOs we already know
% about. how many of the high magnification events do we get? How bright?
% @tanguita

% --------------------------------------------------------------------

\subsection{OpSim Analysis}
\label{sec:\secname:analysis}

Much like the cosmology with lensed quasar time delays, we expect a strong
dependence of the proposed metrics with night-to-night cadence, uniformity and
season length. Maximizing these will maximize the likelihood of recovering high
magnification events, which in turn will provide the most stringent constraints
on
accretion disk structure. As mentioned above, since shorter wavelengths show
faster and stronger magnification events, in an ideal scenario, bluer bands would have
tighter night-to-night cadence.

\begin{center}
	\begin{figure}[hbt]
		\includegraphics[width=\textwidth]{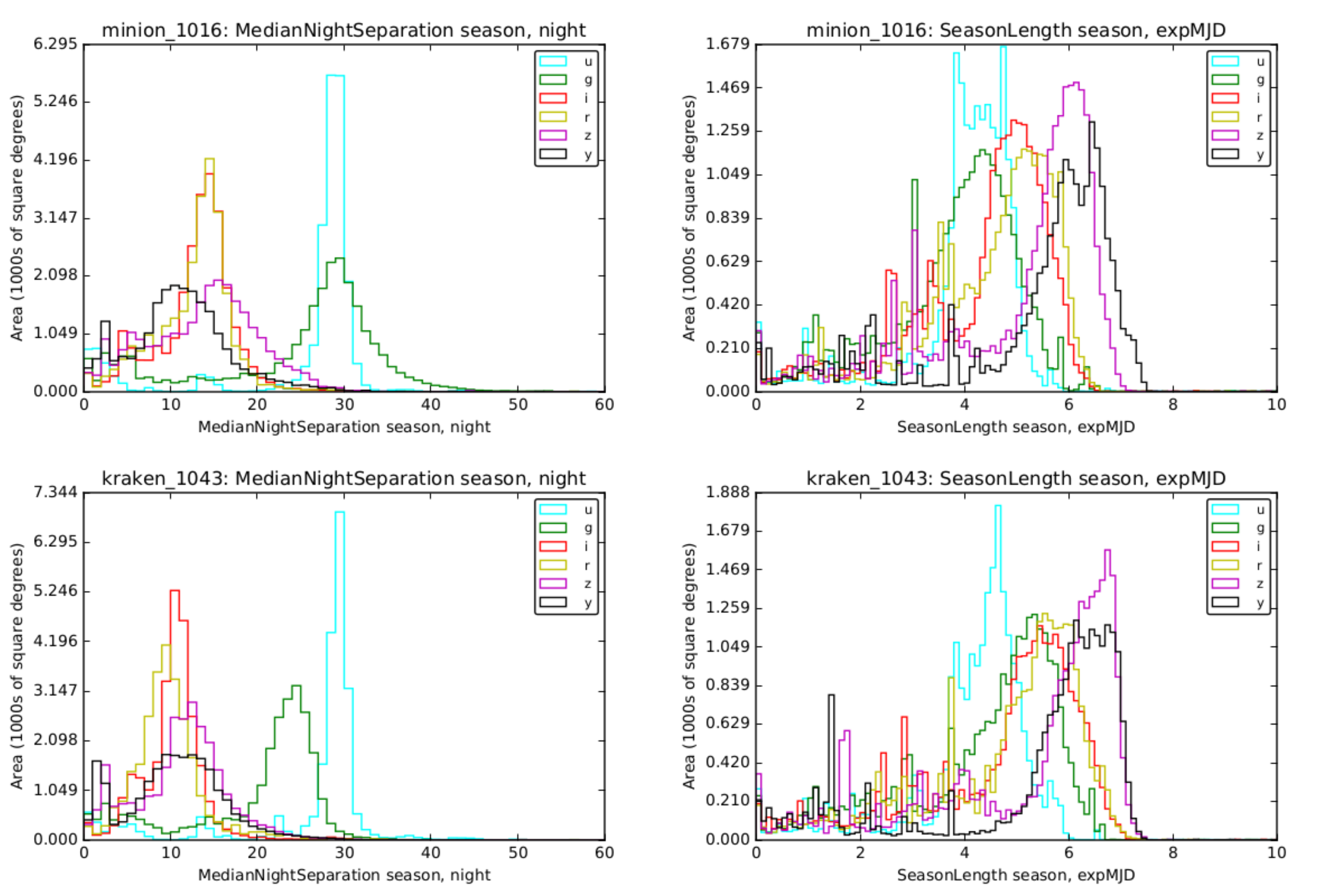}
		\caption{Median Night Separation in days (left) and median season length in months (right) for all bands in the current ``Baseline Cadence'' (\opsimdbref{db:baseCadence}, top) and ``No Visit Pairs''	(\opsimdbref{db:NoVisitPairs}, bottom) opsim outputs.}
		\label{microfig}
	\end{figure}
\end{center}

As shown in figure \ref{microfig}, it seems there is a slightly better prospect
for the AGN structure with microlensing science case using the ``No Visit Pairs'' observing
strategy in comparison to the baseline strategy due to the smaller inter-night
gaps and longer season lengths in the g band. In both cases the night-to-night
cadence in the longer wavebands are compatible with the detection of most
microlensing events. On the other hand, in the u and g bands in both survey strategies it might compromise the results. Furthermore, in all LSST bands the spread in the night-to-night cadence (uniformity) and season length will likely dominate the uncertainties.

\begin{center}
	\begin{figure}[hbt]
		\includegraphics[width=\textwidth]{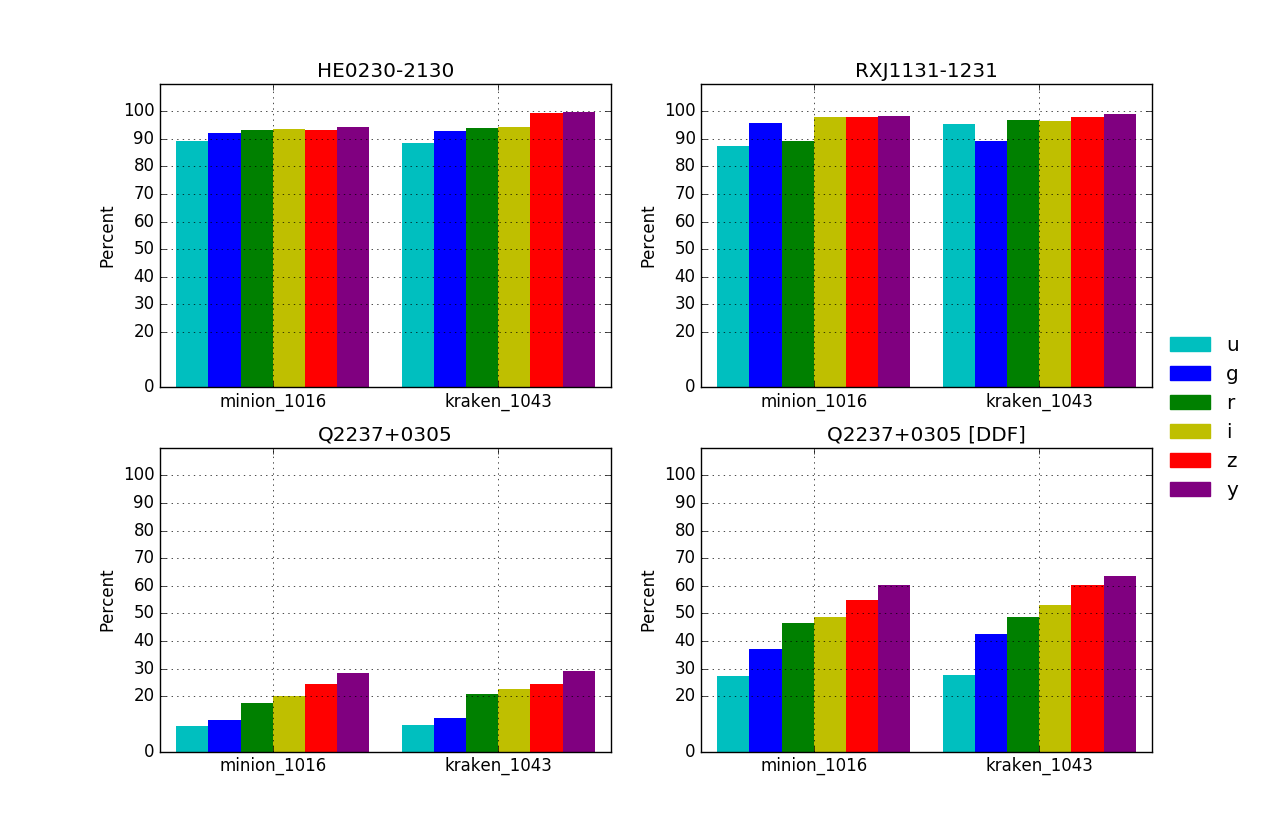}
		\caption{Relative brightness fluctuations seen by LSST compared to ``perfect'' cadence for four systems as described in the text. The current ``Baseline Cadence'' (\opsimdbref{db:baseCadence}) and ``No Visit Pairs''	(\opsimdbref{db:NoVisitPairs}) opsim outputs are shown.All systems assumed to have accretion disk parameters of $\sigma_0$=2.0 light days and $\alpha$=0.75. Ratios are the median of 10000 simulated light curves.}
		\label{ulens_sim}
	\end{figure}
\end{center}

Given the nature of the commonly used statistical analyses to constrain the structure of accretion disk (accretion disk and slope metric), the accuracy of the recovery of the parameters will depend on the amount of brightness variations seen in the light curves even if no high-magnification events are seen. As a first assessment, we have used the light curve simulation tool outlined above to generate 10000 light curves for images A of three well known lensed systems where microlensing has been studied: RXJ1131-1231, HE0230-2130 and Q2237+0305. Additionally, given the very fast time scales expected for Q2237+0305 due to the relative redshifts of the source and lens, we also generate 10000 light curves for this system as if it was located in one of the LSST Deep Drilling Fields. All systems where assumed to have a size ($\sigma_0$) and slope ($\alpha$) parameters of 2.0 light days at $\lambda$=2000\AA{} and 0.75, respectively. Even when this experiment does not yield our proposed figure of merit, it does allow the comparison of the relative performance of different observing strategy regarding microlensing signal recovery.

The results of this experiment are shown in Figure \ref{ulens_sim}. For this analysis, at least for the four systems studied, we can see that even when the ``No Visit Pairs'' observing strategy shows a slightly better prospect compared to the ``Baseline Cadence'' for our science case, the difference is negligible. Note also that with the exception of Q2227+0305 (due to its very short time scales), both observing strategies will be able to map the expected microlensing induced brightness variations at a very high rate. It is important to consider that this is an optimistic analysis because the photometric uncertainties (including the dominating ones that come from the intrinsic variability correction from time delay measurements) have not been taken into account.

\subsection{Conclusions}
\label{sec:\secname:questions}

Here we answer the ten questions posed in
\autoref{sec:intro:evaluation:caseConclusions}:

\begin{description}

\item[Q1:] {\it Does the science case place any constraints on the
tradeoff between the sky coverage and coadded depth? For example, should
the sky coverage be maximized (to $\sim$30,000 deg$^2$, as e.g., in
Pan-STARRS) or the number of detected galaxies (the current baseline
of 18,000 deg$^2$)?}

\item[A1:] Given that lensed quasars are rare, maximizing the area
coverage would increase the number of systems to study. Even if this
comes at a cadence cost given that in most cases the studied OpSim
experiments provide sufficient temporal constraints.

\item[Q2:] {\it Does the science case place any constraints on the
tradeoff between uniformity of sampling and frequency of  sampling? For
example, a rolling cadence can provide enhanced sample rates over a part
of the survey or the entire survey for a designated time at the cost of
reduced sample rate the rest of the time (while maintaining the nominal
total visit counts).}

\item[A2:] Given the long timescales of microlensing events and the inability of predicting high-magnification events, uniform sampling would be better.

\item[Q3:] {\it Does the science case place any constraints on the
tradeoff between the single-visit depth and the number of visits
(especially in the $u$-band where longer exposures would minimize the
impact of the readout noise)?}

\item[A3:] No.

\item[Q4:] {\it Does the science case place any constraints on the
Galactic plane coverage (spatial coverage, temporal sampling, visits per
band)?}

\item[A4:] Not really, however, minimizing the coverage of the Galactic would benefit all AGN science cases.

\item[Q5:] {\it Does the science case place any constraints on the
fraction of observing time allocated to each band?}

\item[A5:] Ideally, shorter wavelengths should get denser (uniform) sampling.

\item[Q6:] {\it Does the science case place any constraints on the
cadence for deep drilling fields?}

\item[A6:] As in the main survey, AGN microlensing would benefit if the deep drilling fields observations would span long timescales with uniform coverage.

\item[Q7:] {\it Assuming two visits per night, would the science case
benefit if they are obtained in the same band or not?}

\item[A7:] Not very important for our since case. Given a choice, different bands would be better.

\item[Q8:] {\it Will the case science benefit from a special cadence
prescription during commissioning or early in the survey, such as:
acquiring a full 10-year count of visits for a small area (either in all
the bands or in a  selected set); a greatly enhanced cadence for a small
area?}

\item[A8:] No.

\item[Q9:] {\it Does the science case place any constraints on the
sampling of observing conditions (e.g., seeing, dark sky, airmass),
possibly as a function of band, etc.?}

\item[A9:] Seeing and Airmass are directly related to the ability to
accurately separate the flux from possibly blended lensed images. As
with most transient science cases, excellent seeing images are required
as templates for image substraction.

\item[Q10:] {\it Does the case have science drivers that would require
real-time exposure time optimization to obtain nearly constant
single-visit limiting depth?}

\item[A10:] No.

\end{description}

% % --------------------------------------------------------------------
%
% \subsection{Discussion}
% \label{sec:\secname:discussion}
%
% %Discussion: what risks have been identified? What suggestions could be
% %made to improve this science project's figure of merit, and mitigate
% %the identified risks?
%
%
% ====================================================================

\navigationbar

% --------------------------------------------------------------------

% ====================================================================
%+
% SECTION:
%    AGN_Future_Work.tex
%
% CHAPTER:
%    agn.tex
%
% ELEVATOR PITCH:
%    Ideas for future metric investigation, with quantitaive analysis
%    still pending.
%-
% ====================================================================

\section{Future Work}\label{sec:AGNFuture}
\def\secname{\chpname:future}\label{sec:\secname}

In this section we provide a short compendium of science cases that
are either still being developed, or that are deserving of quantitative
MAF analysis at some point in the future.

% ====================================================================

%\input{AGN/AGN_Clustering.tex}

% --------------------------------------------------------------------

% ====================================================================
%+
% SECTION:
%    AGN_BELR.tex
%
% CHAPTER:
%    agn.tex
%
% ELEVATOR PITCH:
%
%-
% ====================================================================

% \section{The Size and Structure of the Broad Emission Line Region}
\subsection{The Size and Structure of the Broad Emission Line Region}\label{sec:AGNBELR}
\def\secname{\chpname:photoRM}\label{sec:\secname}

\credit{ohadshemmer}

LSST may provide estimates of the size and structure of the broad
emission line region (BELR) using the photometric reverberation
mapping (PRM) method. This method enables one to measure the
time-delayed response of the flux in one band to the flux
in another by using cross-correlation techniques on AGN light
curves (e.g.,
%\citet{Chelouche2013}; \citet{CheloucheandZucker2013};
\citealt{CheloucheEtal2014}).

%; \citet{EdelsonEtal2015}; \citet{FausnaughEtal2015}).
The main challenge of PRM is to detect,
with high confidence, the time lag between the variations of a BELR
line-rich band with respect to variations in a line-poor band, given
the relatively small flux contributions ($\sim10$\%)~of BELR lines to each
LSST band. Nevertheless, LSST is expected to deliver BELR line-continuum
time delays in $\sim10^5-10^6$ sources, which is unprecedented when
compared to $\sim60-80$ such measurements conducted, to date, via the
traditional, yet much more expensive (per source) spectroscopic method.
Sources in the DDFs will benefit from the highest
quality PRM time-delay measurements given the factor of $\sim10$ denser
sampling \citep{CheloucheEtal2014}.

% --------------------------------------------------------------------

% \subsection{Target measurements and discoveries}
\subsubsection{Target measurements and discoveries}
\label{sec:\secname:targets}

The PRM measurements will probe the size and structure of the BELR,
in a statistical sense, and may provide improved SMBH mass estimates
for sources that have at least single-epoch spectra. PRM will also be
used to trigger follow-up spectrophotometric monitoring of ``promising''
cases depending on their variability properties. The goal is to obtain
$R_{\rm BELR}$ measures for different BELR lines in certain luminosity
and redshift bins; for example, PRM may provide mean $R_{\rm BELR}$ for
Ly$\alpha$ in quasars at $2.1\ltsim z \ltsim 2.2$ with
$45 \ltsim \log L ({\rm erg~s}^{-1}) \ltsim 46$, or mean $R_{\rm BELR}$
for C~{\sc iv}~$\lambda 1549$ in quasars at $1.6\ltsim z \ltsim 1.7$
with $44 \ltsim \log L ({\rm erg~s}^{-1}) \ltsim 45$.
%\new{Our goal is to understand the population of
%AGN broad line regions, including their geometry. We expect to do this
%via  a model that connects the BH mass, BLR geometry and AGN
%photometric variability properties via a set of scaling relations. A
%simple version of this is could be something like...\newline\newline
%So, our target measurements are of $a$ and $b$, the X parameters.
%Before we derive a metric that quantifies our ability to measure these
%parameters, we can anticipate some of the sensitivity of the
%photometric RM method to observing strategy.}

% LSST Review by Niel Brandt: how to address selection bias towards high line strength systems?

The PRM method is very sensitive to the sampling in each band,
therefore the ability to derive reliable time delays can be affected
significantly by the LSST cadence. The best results will be obtained
by having the most uniform sampling equally for each band.
Since the observed line-continuum lags scale with luminosity and redshift,
PRM with the LSST will be limited by the average time gaps between successive
observations in a particular band.
Additionally, there is a trade-off between the number of DDFs and the
number of time delays that PRM can obtain \citep{CheloucheEtal2014}.
For example, an increase in the number of DDFs, with similarly dense
sampling in each field, can yield a proportionately larger number of
high-quality time delays, down to somewhat lower luminosities (to the
extent that host-galaxy contamination can be neglected), but at the
expense of far fewer time delays (for relatively high luminosity
sources) in the main survey.

% --------------------------------------------------------------------

% \subsection{Metrics}
\subsubsection{Metrics}
\label{sec:\secname:metrics}

The average and the dispersion in the number of visits, per band, across
the sky for the nominal OpSim (during the entire survey) should be computed.
Since PRM works best for uniform sampling, one should compare the distributions
of the number of visits in each band, averaged across the sky, and identify
ways to minimize any potential differences between these distributions. By
running PRM simulations, one should identify the 1) minimum number of visits
(in any band) that can yield any meaningful BELR-continuum lag estimates, and
2) the largest difference in the number of visits between two different bands
that can yield any meaningful BELR-continuum lag estimates. These simulations
should be repeated by doubling the nominal number of DDFs. Finally, the
%number of meaningful BELR-continuum time delays that can be obtained
uncertainties on $R_{\rm BELR}$ values achieved with the nominal OpSim
should be assessed, and potential perturbations to the cadence should be
pointed out to reduce these uncertainties.

Another metric is the accuracy of the slope $\alpha$, $\Delta \alpha$, in the
$R_{\rm BELR} \propto L^{\alpha}$ relation. Spectrophotometric monitoring
typically yields $\alpha \simeq 0.50 \pm 0.05$.

\navigationbar

% ====================================================================

% --------------------------------------------------------------------

\section{Discussion}\label{sec:AGNDiscussion}
\label{sec:\chpname:discussion}

%Some additional considerations/thoughts that came up during the Bremerton
%workshop:

The goals of this Chapter were 1) to define and quantify key metrics for AGN
science that can be measured and tested using the current LSST observing
strategy, and 2) to identify reasonable perturbations that may be beneficial
for AGN science.
Undoubtedly, additional work is required to refine many of these metrics
and perform more rigorous tests and simulations.
The following list identifies additional cadence-related aspects for further
consideration. Together with the metrics discussed in this Chapter, this may
be regarded as a road map for further investigations that should be performed
during the planning stage.

1) Assuming a total of ten DDFs, it would be beneficial if one of those fields
could be sampled more heavily than the others and would be visited nightly (or
even more frequently, e.g., from $\sim1$ to $\sim1000$ min) per band.
This can be justified by the fact that a) very few AGNs or transient AGNs have
been monitored at these frequencies on such a long baseline, leaving room for
discovery, and b) this may probe intermediate-mass black holes
($\sim10^4 - 10^5$~$M_{\odot}$) via PRM or PSDs. Good candidate fields are
the Magellanic Clouds and the Chandra Deep Field-South. An observational
strategy should be developed and implemented either in a new OpSim, or
during commissioning.

%2) An assessment of the effects of the LSST cadence on the ability to
%detect periodic AGNs and quasi-periodic oscillations (QPOs) in AGNs
%should be performed.

2) In order to have more informative metrics, accurate model light
curves are needed that can reproduce fiducial light curves in different
bands, at different inherent luminosities, and at different redshifts.
This may be developed together with the Strong Lensing Science Collaboration.

% DDF - we need longer duration OpSims in these fields. (Bob Wagoner)
% (https://github.com/LSSTScienceCollaborations/ObservingStrategy/blob/master/opsim/README.md)

% commissioning opportunity - one field, one night, one filter (u or g).
% 15/30 sec exposures. (Bob Wagoner)
% (https://github.com/LSSTScienceCollaborations/ObservingStrategy/blob/master/commissioning/README.md)

%\todo{}{
3) There is a need to compare the science content in this Chapter
with the AGN chapter in the LSST Science Book as well as with the
Ivezic et al. overview paper (http://arxiv.org/pdf/0805.2366v4.pdf)
to ensure that no key science aspect is overlooked or compromised
by the nominal cadence.
%}

%\todo{}{Compare the $Y$-band (and other bands) depths, single
%epoch and final co-added, from  enigma\_1189 with other OpSims.}

%\todo{}{Assess the effect of non-simultaneous colors on AGN selection.}

%\todo{}{Based on the current OpSim, need to specify the magnitude limits
%at the highest airmass and assess the limitations of the DCR method in
%the $L-z$ plane. Should check this with MAF and if indeed AM <= 1.4, need
%to add a request in:
%https://github.com/LSSTScienceCollaborations/ObservingStrategy/tree/master/opsim
%}

%\todo{}{Assess whether, e.g., a pair of $\sim2.5$ min exposures (i.e., $\sim10$
%times longer than the standard exposures) at airmass $\sim2$ would get deep
%enough for useful DCR constraints for a large fraction of the AGNs. This may
%be a non-negligible perturbation of the expected 56-184 visits per band, and
%may even be impossible given current upper limits on exposure times, but
%this would help improve photo-z's for galaxies and SNe too.}

%\todo{}{For PRM and microlensing: obtain distributions (mean and dispersion)
%of the number of visits, per band, across the sky for the nominal OpSim
%(during the entire survey).}

%\todo{}{
4) It is worth checking whether
%Add a discussion about blazars and LSST cadence (e.g., Isler+15?). Would
any aspect of blazar science might be compromised by the nominal cadence, or
would benefit from a specific cadence requirement.
%}

%\todo{}{
5) The effects of the cadence on the overall LSST astrometry accuracy and precision
should be assessed in terms of potential effects on AGN selection. For example,
AGN selection may benefit from very good depth at least in the 1st and 10th year of the
survey.
%}

%\todo{}{
%6) It is necessary to identify the frequency range and sampling for obtaining
%optimal PSDs required for QPO detection (given $M_{\rm BH}$, spin, and $L/L_{\rm Edd}$).
%Based on the nominal cadence, it is necessary to assess the potential of discovering
%QPOs in the main survey and in the DDFs.
%}

Beyond this, we anticipate more AGN science cases emerging, and needing to be evaluated via science-driven \MAF analysis. For example, regular (non-quasi) periodicity, possibly to probe the population of binary SMBHs, is an observing mode we have not yet focused much attention on.  What would be a good cadence to do that science? We encourage more questions like this as we build up the metric analysis outlined in the chapter to date.

\navigationbar

% --------------------------------------------------------------------

% --------------------------------------------------------------------

\chapter[Cosmology]{Accurate Cosmological Measurements on the Largest Scales}
\def\chpname{cosmo}\label{chp:\chpname}

Chapter editors:
\credit{egawiser},
\credit{MichelleLochner}.

Contributing authors:
\credit{humnaawan},
\credit{egawiser},
\credit{pkurczynski},
\credit{rhiannonlynne},
\credit{jmeyers314},
\credit{tonytyson},
\credit{MelissaGraham},
\credit{SamSchmidt},
\credit{connolly},
\credit{ivezic},
\credit{jhrlsst},
\credit{rbiswas4},
\credit{sethdigel},
\credit{astrofoley},
\credit{lgalbany},
\credit{pgris},
\credit{ReneeHlozek},
\credit{saurabhwjha},
\credit{RickKessler},
\credit{AlexGKim},
\credit{aimalz},
\credit{jasonmcewen},
\credit{janewman-pitt-edu},
\credit{hiranyapeiris},
\credit{kponder},
\credit{rlschuhmann},
\credit{astrostubbs},
\credit{wmwv},
\credit{drphilmarshall},
\credit{tanguita}

% \section*{Summary}
% \addcontentsline{toc}{section}{~~~~~~~~~Summary}
%
% Executive summary goes here, highlighting the primary conclusions from
% the chapter's science cases. This should be abstract length, no more:
% say, 200 words.

% --------------------------------------------------------------------

\section{Introduction}
\label{sec:\chpname:intro}

Cosmology is one of the key science themes for which LSST was
designed. Our goal is to measure cosmological parameters, such as the
equation of state of dark energy or departures from General
Relativity, with sufficient accuracy to distinguish one model from
another and hence drive our theoretical understanding of how the
universe works as a whole. To do this will necessarily involve a
variety of different measurements that can act as cross-checks and break
any parameter degeneracies.

The  Dark Energy Science Collaboration (DESC) has identified five
different cosmological probes enabled by the LSST: weak lensing (WL),
large-scale structure (LSS), Type Ia supernovae (SN), strong lensing
(SL), and clusters of galaxies (CL). In all these cases, the primary concern
is the residual systematic error: the shapes and photometric redshifts of
galaxies and the properties of supernova and lensed quasar light
curves will all need to be measured with extraordinary accuracy in
order to properly harness the statistical power available through LSST. This
accuracy will come from the abundance and heterogeneity of the
individual measurements and the degree to which they can be
modeled and understood. This latter point implies a need for uniformity
in the survey, which enables powerful simplifying assumptions to be made
when calibrating on the largest, cosmologically most important scales.
The need for heterogeneity in the measurements also requires uniformity in
the sense that the nuisance parameters that describe the systematic effects
need to be sampled over as wide a range as possible (e.g., the need to
sample a wide range of roll angles to minimize shape error;
observing conditions to understand photometric errors due to the
changing atmosphere).

In this chapter we look at some of the key measurements planned by DESC
and how they depend on the Observing Strategy.

%---------------------------------------------------------------------

% ====================================================================
%+
% SECTION NAME:
%    lss.tex
%
% CHAPTER:
%    cosmology.tex
%
% ELEVATOR PITCH:
%    Large Scale Structure, Weak Lensing, and Clusters all require
%    survey uniformity in the static 10-year survey.  A key contributor
%    to this is the pattern of dithers adopted.
%
% AUTHORS:
%   Eric Gawiser (@egawiser)
%-
% ====================================================================
\newcommand{\sigmaOS}[0]{$\sigma_{\mathrm{C_{\ell, {OS}}}}$}
\newcommand{\CellOS}[0]{$C_{\ell, \rm{OS}}$}
\newcommand{\statFloor}[0]{$\Delta C_\ell$}
\newcommand{\delobs}[0]{\delta_{\mathrm{obs},i}}
\newcommand{\dellss}[0]{\delta_{\mathrm{LSS},i}}
\newcommand{\delos}[0]{\delta_{\mathrm{OS},i}}
\newcommand{\ev}[1]{\left < {#1} \right >}

\section{Large-Scale Structure: Dithering to Improve Survey Uniformity}
\def\secname{lss}\label{sec:\secname}

\credit{humnaawan},
\credit{egawiser},
\credit{pkurczynski},
\credit{rhiannonlynne}

Three of the key cosmology probes available with LSST represent ``static science'', i.e., insensitive to time-domain concerns.  These are Weak Lensing, Large-Scale Structure, and Galaxy Clusters.  Nonetheless, due to the need to track and correct for the survey window function for these probes, cosmology with LSST will benefit from achieving survey depth as uniform as possible over the WFD area.  \OpSim tiles the sky in hexagons inscribed within the nearly-circular LSST FOV. It has been shown in \citet{CarrollEtal2014} that the default LSST survey strategy implemented in \OpSim leads to a strongly non-uniform honeycomb pattern due to overlapping regions on the edges of the hexagons receiving nearly double the observing time.  A pattern of large dithers, i.e. telescope-pointing offsets on the scale of the LSST FOV, greatly reduces these overlaps, leading to an increase in the median survey depth in each filter by 0.08 magnitudes.

In this section, we report the results from an investigation into different types of dithers, varying both in terms of the timescales on which dithers are implemented as well as the geometry of the dither positions. The results discussed largely follow the analysis in \citet{AwanEtal2016}, except that here we use the $i$-band-relative mock catalogs and magnitude cuts (as opposed to the $r$-band), and analyze the impacts of different observing strategies using the new baseline cadence \opsimdbref{db:baseCadence} and two other cadences. We also discuss the quantification of the effectiveness of a given observing strategy as a Figure of Merit.

% ====================================================================
% Subsection: Dither Patterns and Timescales
% ====================================================================
\subsection{Dither Patterns and Timescales}
\label{sec:\secname:strategies}
As in \citet{AwanEtal2016}, we consider three timescales to capture the range of time intervals on which dithers can be implemented: by visit, by night, and by season. Both visit and night timescales are well-defined in \OpSim \citep[see]{IvezicEtal2008}. For the seasons, we use the \href{https://github.com/lsst/sims_maf/blob/master/python/lsst/sims/maf/stackers/generalStackers.py}{SeasonStacker}, which assigns a season to every observation by tracking when each field's RA is overhead in the middle of the day. The season assignment leads to 11 seasons, and for our purposes, we treat the 0th and 10th seasons the same by assigning them the same dither position.

Another variation in the implementation timescale is added by dithering each field independently as opposed to dithering all fields collectively. For instance, FieldPerNight timescale assigns a new dither position to a field when it is observed on a new night, while PerNight implementation assigns a new dither to all fields every night regardless of whether a particular field is observed or not.

In \citet{AwanEtal2016}, we study five different geometries for the dither positions, where one geometry is specifically for PerSeason assignment and the rest are implemented on three timescales, namely FieldPerVisit, FieldPerNight, and PerNight. Here, we focus only on three combinations of the different geometries and timescales as an illustration of the impacts of these combinations: RepulsiveRandomDitherFieldPerVisit, FermatSpiralDitherPerNight,  and PentagonDitherPerSeason. These geometries are shown in \autoref{fig: dithGeometries}, adapted from Figure 1 in \citet{AwanEtal2016}.

\begin{figure*}[!htb]
      \centering\includegraphics[width=\linewidth, trim={25 570 25 50},clip=true]{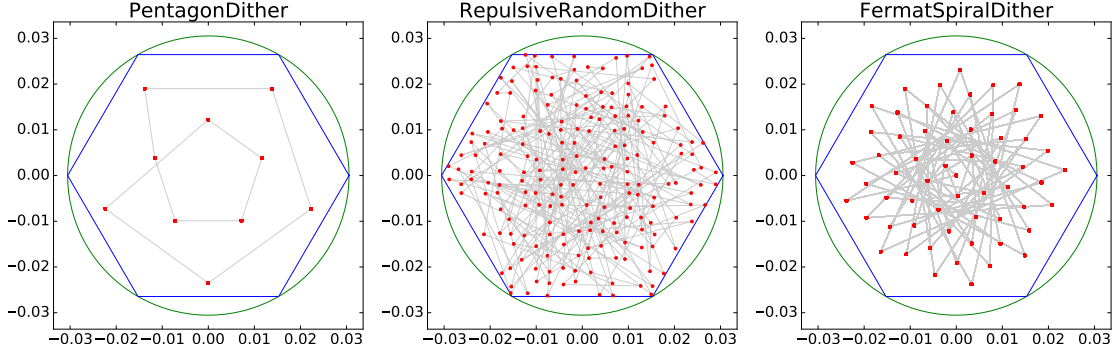}
\caption{Dither geometries implemented for various timescales. PentagonDither is implemented only on PerSeason timescale, while the rest are implemented on FieldPerVisit, FieldPerNight, and PerNight timescales. Here the green curve is the LSST FOV of radius 0.305 radians; the blue hexagon represents the hexagonal tiling of the sky originally adopted for the undithered observations; and the red points are the dithers. The axes are in radians.}
\label{fig: dithGeometries}
\end{figure*}

Here we note that all the dithers are restricted to lie within the hexagons inscribed in the 3.5$^\circ$ LSST FOV, and that we continue with the naming scheme [Geometry]Dither[Field]Per[Timescale], where the absence of the tag `Field' implies that all fields are assigned the same dither.

% ====================================================================
% Subsection: Metrics
% ====================================================================
\subsection{Metrics}
\label{sec:\secname:metrics}
Our first metric is the \href{https://github.com/lsst/sims_maf/blob/master/python/lsst/sims/maf/metrics/simpleMetrics.py}{CoaddM5Metric}, which we use to calculate the coadded 5$\sigma$ depth resulting from various observing strategies. This allows us to compare the artifacts in the coadded depth induced by the observing strategy. Then we account for the dust extinction, using \href{https://github.com/lsst/sims_maf/blob/master/python/lsst/sims/maf/metrics/exgalM5.py}{ExGalM5Metric}, before estimating the number of galaxies in each pixel at a particular depth. For this purpose, we use a mock LSST catalog \citep{MunozEtal2015} to estimate the power law coefficients for each redshift bin, converting the depth into an estimated number of galaxies for each pixel \citep[Eq. 2]{AwanEtal2016}:
\begin{equation}
	N_{\mathrm{gal}}= 0.5\int_{-\infty}^{m_\mathrm{{max}}} {\mathrm{erfc}[a(m-5\sigma_{\mathrm{stack}})] 10^{c_1m + c_2}dm}
	\label{eq: Ngal}
\end{equation}

Here the \texttt{erfc} function accounts for incompleteness while the constants c$_1$ and c$_2$ are determined from the mock catalog. $m_{\mathrm{max}}$ specifies the magnitude cut, and we modify both $m_\mathrm{{max}}$ and c$_2$ to account for colors (assumed to be $u-g= g-r= r-i= 0.4$). This calculation is carried out using the \href{https://github.com/humnaawan/sims_maf_contrib/blob/master/mafContrib/galaxyCountsMetric_extended.py}{GalaxyCountMetric}.

We then account for the effects of photometric calibrations in our estimated number of galaxies. As discussed in \citet{AwanEtal2016}, we model the calibration uncertainties using a simple ansatz relating the systematic errors in each pixel to the seeing in that pixel (relative to the average seeing across the survey region; $\Delta s_{i}$)  and number of observations $N_{\mathrm{obs},i}$:
\begin{equation}
	\Delta_{i}= \frac{k \Delta s_{i}}{\sqrt{N_{\mathrm{obs}, i}}}
\end{equation}
where $k$ is a constant such that $\sigma_{\Delta_i}$ matches the LSST photometric calibration goal of 1$\%$  \citep{2009arXiv0912.0201L}. Since the calibration uncertainties are pixel-dependent, we use the \href{https://github.com/humnaawan/sims_maf_contrib/blob/master/mafContrib/galaxyCounts_withPixelCalibration.py}{pseudo-GalaxyCountsMetric}, which handles pixel-by-pixel calculation to modify the upper limit in the integral in \autoref{eq: Ngal} to be $m_{\mathrm{max}} + \Delta_{i}$, thereby accounting for the fluctuations in the galaxy counts due to the calibration errors. We then calculate the fluctuations in the galaxy counts in each pixel, $\delta_i= \Delta N_i/\overline{N}$.

% ====================================================================
% Subsection: Figure of Metric
% ====================================================================
\subsection{Figure of Merit}
\label{sec:\secname:FoM}
As derived and discussed in \citet{AwanEtal2016}, the spurious power from the artificial fluctuations in the galaxy counts induced by the observing strategy (OS) represents a bias in our measurement of the LSS. Hence, the uncertainty in this bias becomes the limiting factor in our ability to correct for the structure induced by the observing strategy. More quantitatively, for an optimized LSS study, the uncertainties induced by the observing strategy, \sigmaOS, must be subdominant to the statistical uncertainty \statFloor\ inherent to the measured power spectrum due to ``cosmic variance'' \citep{Dodelson}:
\begin{equation}
	\Delta C_\ell= C_{\ell,\mathrm{LSS}} \sqrt{\frac{2}{f_{\mathrm{sky}} (2\ell + 1)}}
	\label{eq: statFloor}
\end{equation}
where $f_{\rm{sky}}$ is the fraction of the sky observed, accounting for the reduction in the observed information due to incomplete sky coverage.

Since we do not include any input LSS in our pipeline, the power spectrum we measure for any given band is due to the power induced by the observing strategy, \CellOS, for that band. Modeling the overall bias induced by the observing strategy as an average across $ugri$ bands, we calculate the uncertainties in the bias \sigmaOS\ as the standard deviation across \CellOS\ for $ugri$ bands to account for the effects of detecting the galaxy catalog through various bands. We then compare these uncertainties with the statistical floor for various redshift bins, where the statistical floor is based on the galaxy power spectra calculated using the code from \citet{Zhan2006}, which we pixelize to match the HEALPix resolution to account for the finite angular resolution of our simulations.

To quantify the effectiveness of each observing strategy in minimizing \sigmaOS, we construct a Figure of Merit (FoM) as the ratio of the ideal-case uncertainty in the measured power spectrum and the uncertainty arising from shot noise and the structure induced by the observing strategy:
\begin{equation}
	\mathrm{FoM} = \sqrt{\frac{\sum\limits_\ell{\left({\sqrt{\frac{2}{f_{\mathrm{sky, max}} (2\ell + 1)}}C_{\ell, \mathrm{LSS}}} \right)}^2}{\sum\limits_\ell \left[{ \left( { \sqrt{\frac{2}{f_{\mathrm{sky}} (2\ell + 1)}}\left\{{C_{\ell, \mathrm{LSS}} + \frac{1}{\bar{\eta}}} \right\}  } \right ) ^2 + \sigma_{C_{\ell,\mathrm{OS}}}^2  }\right]}}
\label{eq: FoM}
\end{equation}
Here, $\bar{\eta}$ is the surface number density in steradians$^{-1}$, and the term containing it accounts for the contribution from the shot noise to the measured signal \citep{HutererEtal2001,Jing2005}. This FoM measures the percentage of ideal-case information that can be measured in the presence of systematics. We note that the shot noise is negligible even for the shallowest (10-year) surveys we consider.

We define the ideal-case as being based on the largest coverage of the sky with LSST, i.e., $f_{\rm{sky, max}}$ is the largest WFD coverage with the baseline cadence. For \opsimdbref{db:baseCadence}, the observing strategy with RepulsiveRandomDitherFieldPerVisit dithers leads to the largest $f_{\rm{sky}}$ ($\sim 39.5\%$). Note that this fraction is calculated after masking the shallow borders of the main survey; for details, see \citet{AwanEtal2016}. %Therefore, we fix the $f_{\rm{sky, best}}$ to be $39.5\%$ and compare the the different observing strategies and cadences relative to it.

% ====================================================================
% Subsection: A Comment on Terminology
% ====================================================================
\subsection{A Comment on Terminology}
For clarity, we make a note on the terminology we have introduced. Strictly speaking, the bias caused by the observing strategy is a window function bias, as the survey window function ($W_i$) accounts for the effective survey geometry which scales the fluctuations in the galaxy counts in each pixel: $1+\delobs= W_i(1+\dellss)$. Comparing this with Equation 4 in \citet{AwanEtal2016}, $1+\delobs= (1+\delos)(1+\dellss)$, we see that the bias induced by the observing strategy is directly related to the window function: $1 + \delos= W_i$

Then, for the total power, we have
\begin{equation}
\ev{\delobs^2}=\ev{\dellss^2}\ev{(1 + \delos)^2}+ \ev{\delos^2}=\ev{\dellss^2}\ev{W_i^2}+ \ev{(W_i-1)^2}
\end{equation}
where the first equality is based on Equation 6 in \citet{AwanEtal2016} and the second one holds given the relation between $\delos$ and $W_i$. Since the bias induced by the observing strategy $\delos^2=  (W_i-1)^2$, the uncertainties in the bias are the window function uncertainties.

Generally the window function is assumed to be known perfectly and its uncertainties are not explicitly identified as such. To avoid confusion and focus on the window function uncertainties arising from the observing strategy, we continue using the terms bias induced by the observing strategy and its uncertainties in favor of window function and its uncertainties.

% ====================================================================
% Subsection: OpSim Analysis and Results
% ====================================================================
\subsection{\OpSim Analysis and Results}
\label{sec:\secname: analysis}
For the purposes of our analysis, we use HEALPix resolution of $N_\mathrm{_{side}}= 256$, effectively tiling each $3.5^\circ$ FOV with about 190 HEALPix pixels. Using the metrics discussed in \autoref{sec:\secname:metrics}, we analyze \sigmaOS\ from various observing strategies. First we present the results for the baseline cadence, \opsimdbref{db:baseCadence}.

\begin{figure*}[!htb]
      \centering\hspace*{-3em}\includegraphics[width=0.6\linewidth]{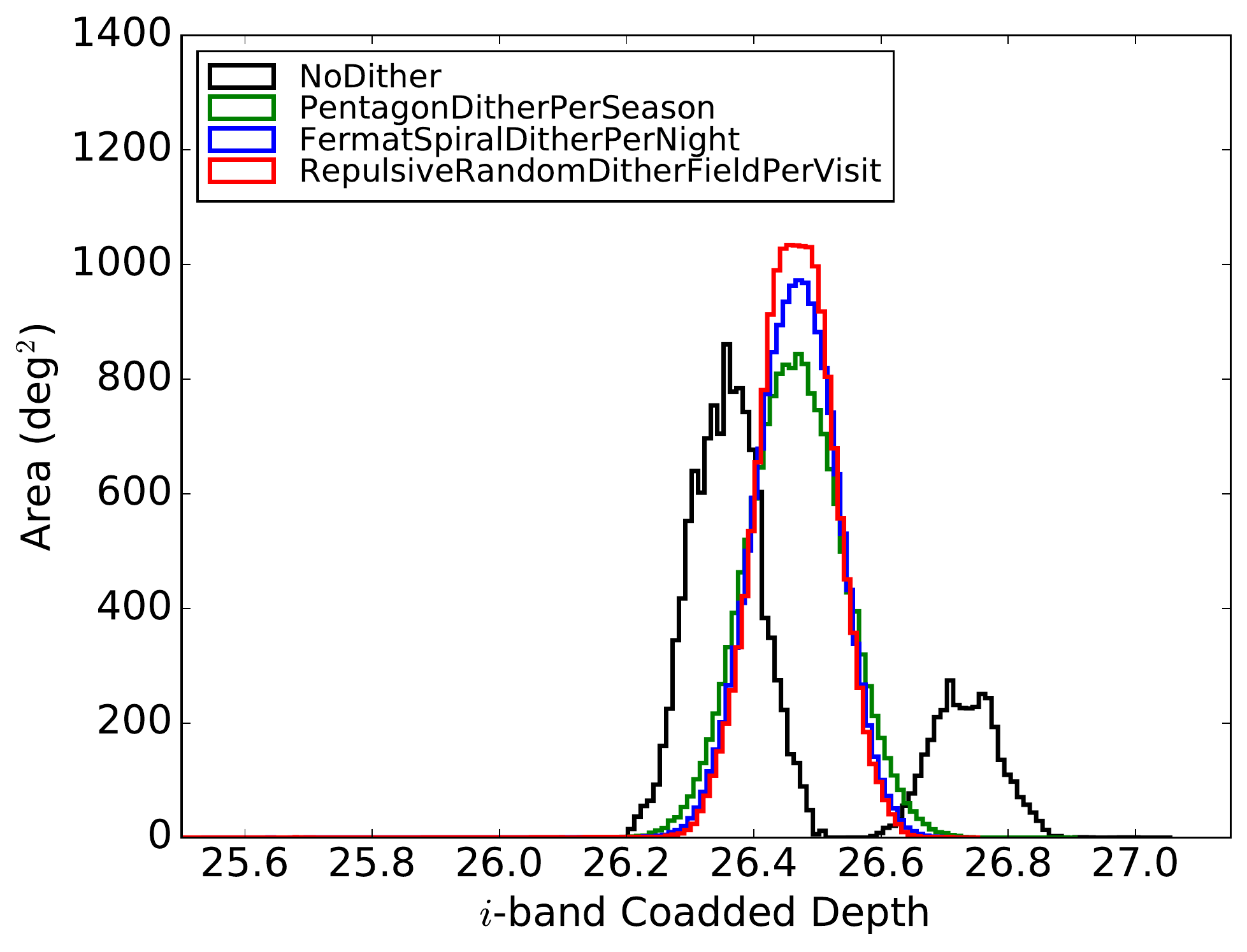}
      \vspace*{-1em}
\caption{Histogram for the $i$-band coadded 5$\sigma$ depth after the full, 10-year survey.}
\label{fig: coaddHistogram}
\end{figure*}

\autoref{fig: coaddHistogram} shows the histogram for the $i$-band coadded 5$\sigma$ depth from \opsimdbref{db:baseCadence} for the four observing strategies. We observe a bimodal distribution for the undithered survey -- the deeper depth mode corresponds to the overlapping regions between the hexagons, while the rest of the survey contributes to the shallower mode. In contrast, all dithered surveys lead to unimodal distributions as the overlapping regions between the fields change frequently, leading to more uniformity. We also note that frequent dithering leads to deeper regions as we observe more peaked histograms for FieldPerVisit and PerNight strategies.

\autoref{fig: coaddSkymaps} shows the plots for the $i$-band coadded 5$\sigma$ depth for the observing strategies. As in \citet{AwanEtal2016}, we find that the undithered survey leads to a strong honeycomb pattern which is much weaker in all of the dithered surveys. We again observe that the dithered surveys are deeper than the undithered survey in terms of the median depth across the survey region.

\begin{figure*}[!htb]
      \centering\includegraphics[width=\linewidth, trim={50 470 55 70},clip=true]{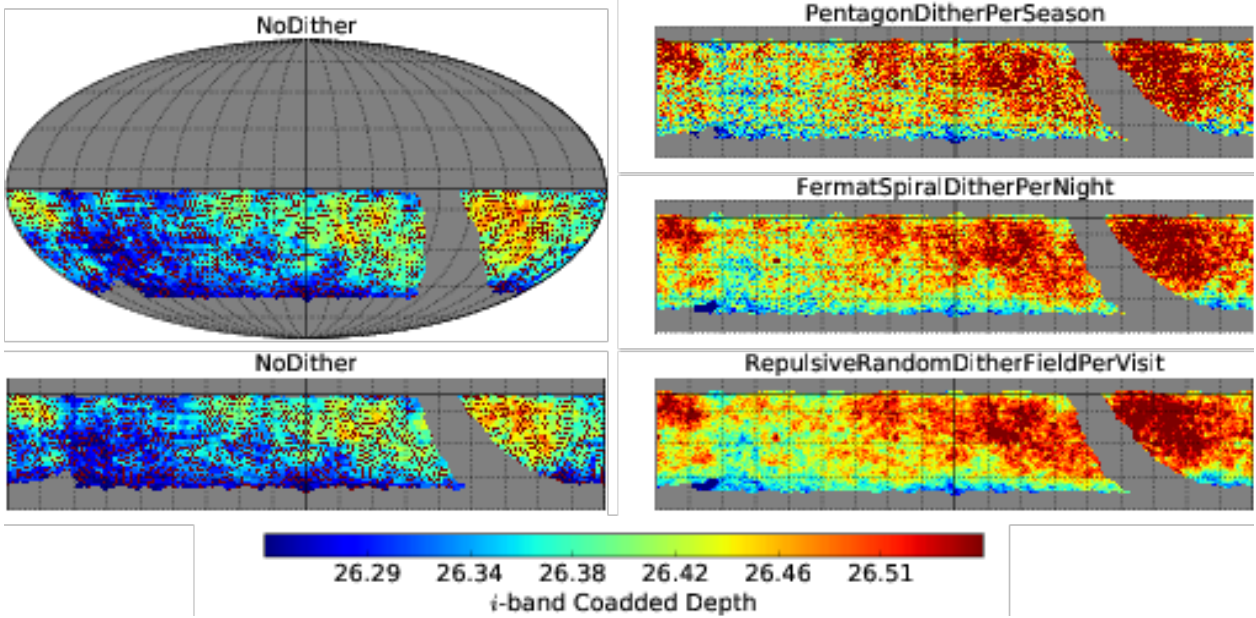}
\caption{Plots for the $i$-band coadded 5$\sigma$ depth based on \opsimdbref{db:baseCadence} for various observing strategies. The top left plot shows the Mollweide projection for NoDither while the bottom left shows the corresponding Cartesian projection, restricted to $180^\circ>$RA$>-180^\circ$ (left-right), $-70^\circ<$Dec$<10^\circ$ (bottom-top). Only the latter is shown for the rest of the strategies. }
\label{fig: coaddSkymaps}
\end{figure*}

In order to quantify the angular characteristics observed in the skymaps, we calculate the angular power spectra corresponding to the skymaps for the $i$-band coadded 5$\sigma$ depth. \autoref{fig: coaddPowerSpectrum} shows these spectra for the four observing strategies. We observe a sharp reduction in the artificial power in the dithered surveys when compared to the undithered one: the strong honeycomb pattern in the undithered survey leads to a large peak around $\ell\sim150$, while the peak is about 10 times weaker in the dithered surveys. We do, however, observe variations amongst the various dither strategies: while RepulsiveRandom dithers lead to small power for all timescales, PerSeason dithers lead to large power on larger angular scales, and both PerSeason and FermatSpiral lead to large power around $\ell\sim150$ (which still is $<10\times$ the corresponding peak from the undithered survey).

\begin{figure*}[!htb]
      \centering\includegraphics[width=\linewidth]{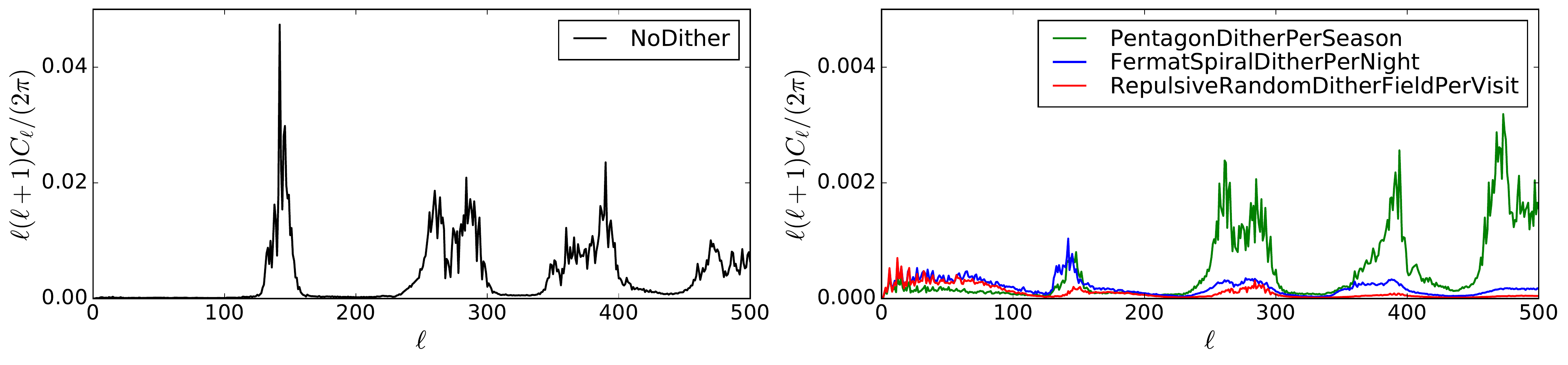}
      \vspace*{-2em}
\caption{Angular power spectra for the  $i$-band coadded 5$\sigma$ depth from \opsimdbref{db:baseCadence} for various observing strategies. We note that dithering reduces the spurious power by over 10$\times$.}
\label{fig: coaddPowerSpectrum}
\end{figure*}

We then proceed to calculate the bias induced by the observing strategy and its uncertainty from the different observing strategies. First, we examine simulated results after only one year of survey. \autoref{fig: minion1016: 1yr} shows the comparison between \sigmaOS\ and \statFloor\ for $0.66<z<1.0$ after the 1-year survey for two magnitude cuts: $i<24.0$ and $i<25.3$. We observe that the undithered survey leads to \sigmaOS\ 1-5$\times$ the statistical floor around $\ell\sim150$; PerSeason timescale does only slightly better. However, we see an improvement with frequent dithers: both FieldPerVisit and PerNight implementations lead to uncertainties 0.5-1$\times$ the statistical floor, although FermatSpiral dithers on PerNight timescale lead to a peak around $\ell\sim150$ more pronounced than the one from RepulsiveRandom dithers on FieldPerVisit timescale.

\begin{figure*}[!htb]
      \centering\hspace*{1em}\includegraphics[width=\linewidth]{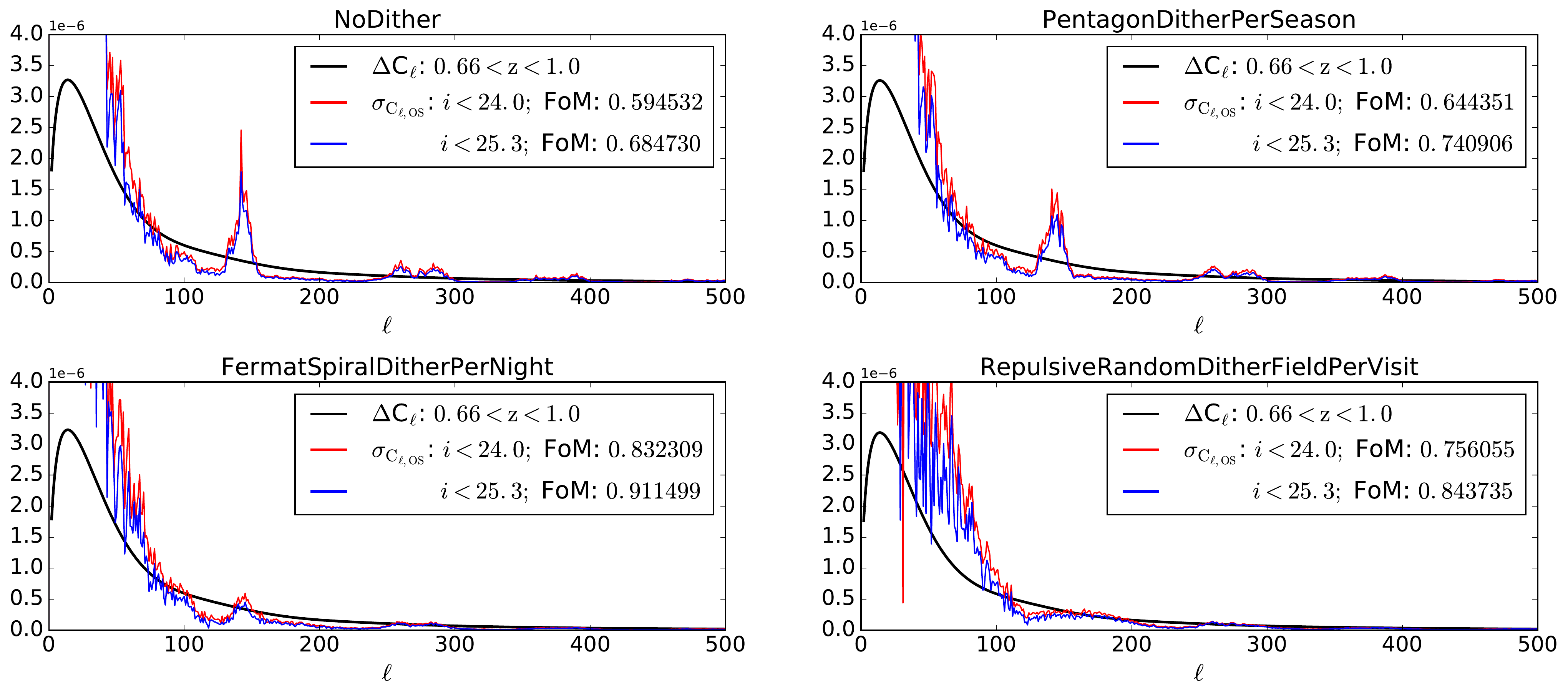}
       \vspace*{-2em}
\caption{\sigmaOS\ comparison with the minimum statistical uncertainty \statFloor\ for $0.66<z<1.0$ for different magnitude cuts after only one year of survey based on \opsimdbref{db:baseCadence}.}
\label{fig: minion1016: 1yr}
\end{figure*}

The trends are captured in the Figure of Merit, which we calculate using \autoref{eq: FoM} over the range $100<\ell<300$. We observe a smaller FoM for the shallower survey -- realistic given that although there is less structure and therefore weaker artifacts induced by the observing strategy, the shot noise becomes significant and makes the FoM smaller. For the deeper survey, we find that FermatSpiralDitherPerNight outperforms all others with the highest FoM, while RepulsiveRandomDitherFieldPerVisit is more effective than PerSeason dithers. The undithered survey, as expected, performs the worst.

In \autoref{fig: minion1016: 10yr}, we show simulated results after the full, 10-year survey for $0.66<z<1.0$ for three different magnitude cuts: $i<24.0$, $i<25.3$ and $i<27.5$. We observe stark differences between the undithered and dithered surveys: the former leads to large uncertainties in the bias induced by the observing strategy while the latter is effective in bringing \sigmaOS\ well below the statistical floor. The effectiveness of all three dithered surveys in minimizing the uncertainties implies more flexibility in choosing the dither strategy for years 2-10.

%For 10yr, 3 magnitude cuts with minion1016, NoDither was always bad but PerSeason dithers were giving us comparable FoM to PerNight and FieldPerVisit dithers. That is not the case anymore; PerSeason dithers are now doing slightly better than before though still not as good as PerNight and FieldPerVisit dithers.

Analyzing the FoM more closely, we observe that the gold sample leads to smaller FoM than both the shallower and deeper catalogs. The larger FoM for shallower catalog is realistic, given less structure with shallow depth leads to weaker artifacts and the shot noise is negligible over the full ten-year survey, but the out-of-trend behavior of gold sample hints at a peculiarity of the variance across the $ugri$ bands at that depth for the baseline cadence. We investigate this behavior briefly and find that the $u$-band-induced artifacts add the most to the uncertainties in the bias induced by the observing strategy, as the gold sample $u$-band cadence in the \opsimdbref{db:baseCadence} is different from $gri$ cadences. This issue still needs to be further investigated, with potentially incorporating the importance of each band to calculate an overall bias induced by the observing strategy. We note, however, that this peculiarity is particularly enhanced for the undithered survey.

\begin{figure*}[!htb]
      \centering\hspace*{1em}\includegraphics[width=\linewidth]{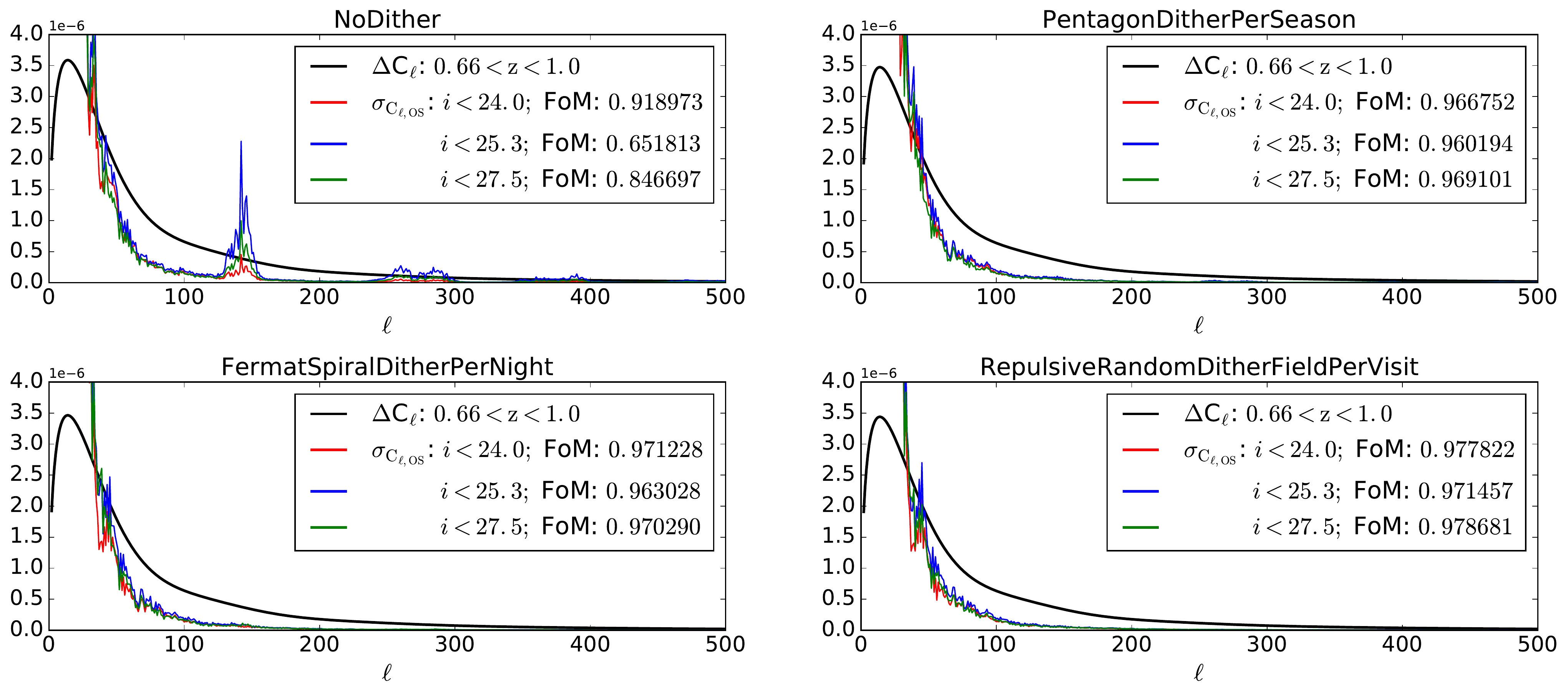}
       \vspace*{-2em}
\caption{\sigmaOS\ comparison with the minimum statistical uncertainty \statFloor\ for $0.66<z<1.0$ for different magnitude cuts after the full, 10-year survey based on \opsimdbref{db:baseCadence}.}
\label{fig: minion1016: 10yr}
\end{figure*}

The trends observed here remain consistent for all five redshift bins. We note that our choice of dithers is particularly important for the one-year survey as only one of the three dither strategies leads to a large FoM. Therefore, in the absence of effective dithers, systematics correction methods will become necessary after the one-year survey. However, these methods may not lead to significant improvements for a dithered 10-year survey as dithers of most kinds are effective in reducing the uncertainties well below the minimum statistical limit.

To further probe the effects of dithers, we run the 1-year and 10-year analyses for two cadences besides the baseline cadence: \opsimdbref{db:NoVisitPairs} which does not require visit pairs, and \opsimdbref{db:opstwoPS} which implements a Pan-STARRS-like observing strategy offering a larger area coverage. In \autoref{fig: cadences: 1yr}, we compare the results from these two cadences with those from \opsimdbref{db:baseCadence} for $0.66<z<1.0$ for the  $i<25.3$ galaxy sample after only one year of survey. We see that the undithered survey leads to large uncertainties in the bias induced by the observing strategy with all three cadences, with the peak uncertainty 5-15$\times$ the statistical floor. As expected, the undithered survey with the wider coverage \opsimdbref{db:opstwoPS} cadence leads to stronger artifacts and a much smaller FoM (by $\sim33\%$ in comparison with \opsimdbref{db:baseCadence}), while not requiring visit-pairs is slightly more effective than the baseline (FoM increases by about 6$\%$). We see very similar trends for the three cadences for PerSeason dithers although the peak \sigmaOS\ ranges between 3-9$\times$ the statistical floor; FoM based on \opsimdbref{db:opstwoPS} is worse than that from \opsimdbref{db:baseCadence} by about 25$\%$ and  \opsimdbref{db:NoVisitPairs} improves on the baseline FoM by $\sim5\%$.

% For different cadences, 1yr results, we now have FoM really high for the wider minion1020 for both PerNight and FieldPerVisit dithers while NoDither and PerSeason dithers are still performing poorly. RepRandom dithers with the wider survey gets us F0M=0.99 while FermatSpiral dithers perform better than RepRandom for the other two cadences.

As before, \sigmaOS\ improves with more frequent dithering. It is only about 1-3$\times$ the statistical floor for FermatSpiral dithers on PerNight timescale. In contrast to NoDither and PerSeason dithers, both \opsimdbref{db:opstwoPS} and \opsimdbref{db:NoVisitPairs} perform better than baseline\opsimdbref{db:baseCadence} with PerNight dithers: FoM from the wider coverage cadence is about $4.5\%$ better than for the baseline cadence, while we see a $4\%$ better FoM with \opsimdbref{db:NoVisitPairs}.

For RepulsiveRandom dithers on FieldPerVisit timescale, we find that the uncertainties in the bias induced by the observing strategy are on the same scale as the statistical floor. The wider coverage cadence outperforms the baseline cadence significantly as  the wider survey FoM is about $18\%$ better than the baseline FoM while the improvement is about 3$\%$ when not requiring visit-pairs. We emphasize that the differences between results with different cadences is highly dependent on the observing strategy: the wider coverage with no or infrequent dithers performs quite poorly while it significantly improves the FoM when large, frequent dithers are implemented. On the other hand, not requiring visit-pairs leads to comparatively larger improvement for infrequent dithers than frequent ones (compared to the baseline).

\begin{figure*}[!htb]
      \centering\includegraphics[width=\linewidth]{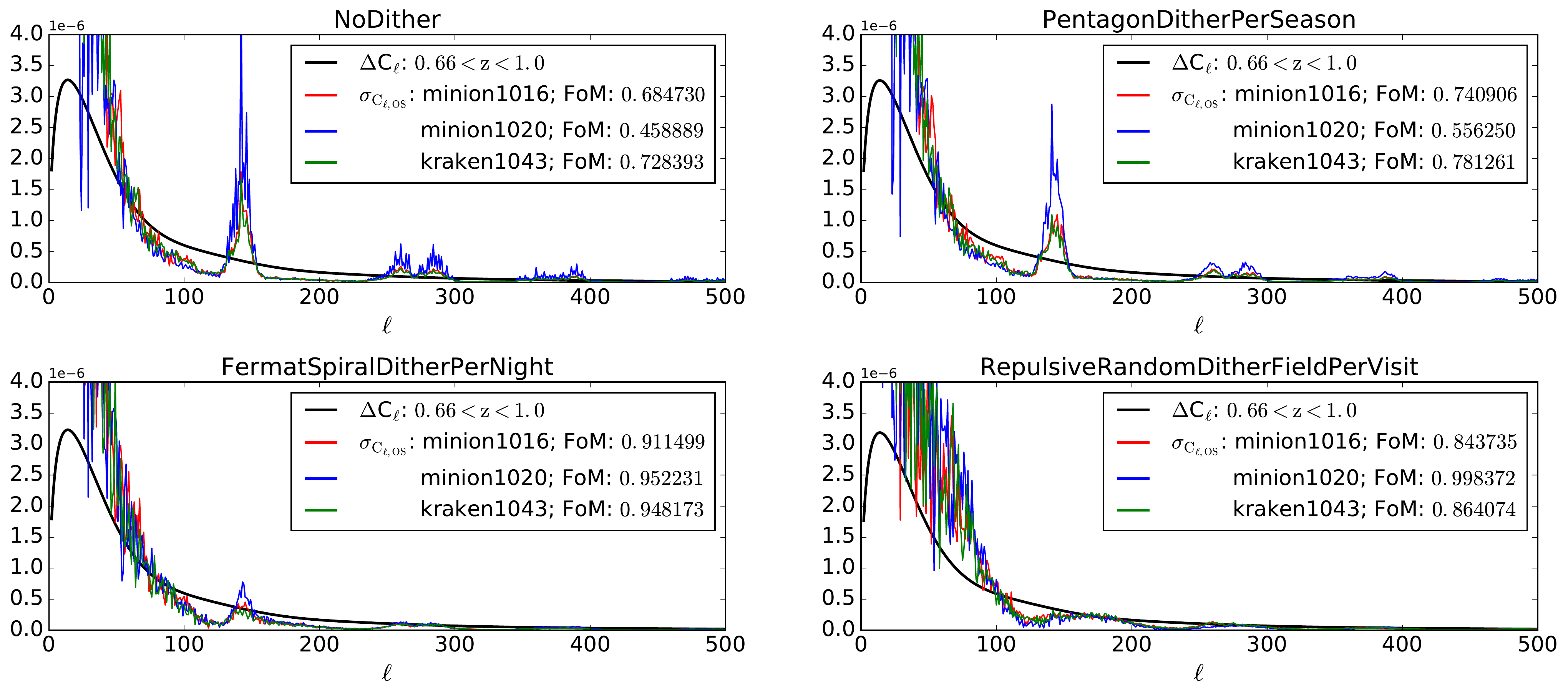}
       \vspace*{-2em}
\caption{\sigmaOS\ comparison with the minimum statistical uncertainty \statFloor\ for $0.66<z<1.0$ for three different cadences for $i<25.3$ after only one year of survey.}
\label{fig: cadences: 1yr}
\end{figure*}

Finally, we show the simulated results for different cadences after the 10-year survey in \autoref{fig: cadences: 10yr}. As in \autoref{fig: minion1016: 10yr}, we see that all the dithered surveys effectively minimize the uncertainties, regardless of the cadence. We do observe, however, that the wider coverage \opsimdbref{db:opstwoPS} still underperforms significantly for the undithered survey (FoM about 30$\%$ less than baseline FoM)  while all the dithered surveys see a stark improvement (FoM $>$ 1 for all; $\sim 20\%$ improvement on the baseline FoM). The improvement from \opsimdbref{db:NoVisitPairs} is comparable among the four observing strategies. Based on these results, we note than wider coverage offers significant improvements with large dithers on any implementation timescale.

\begin{figure*}[!htb]
      \centering\includegraphics[width=\linewidth]{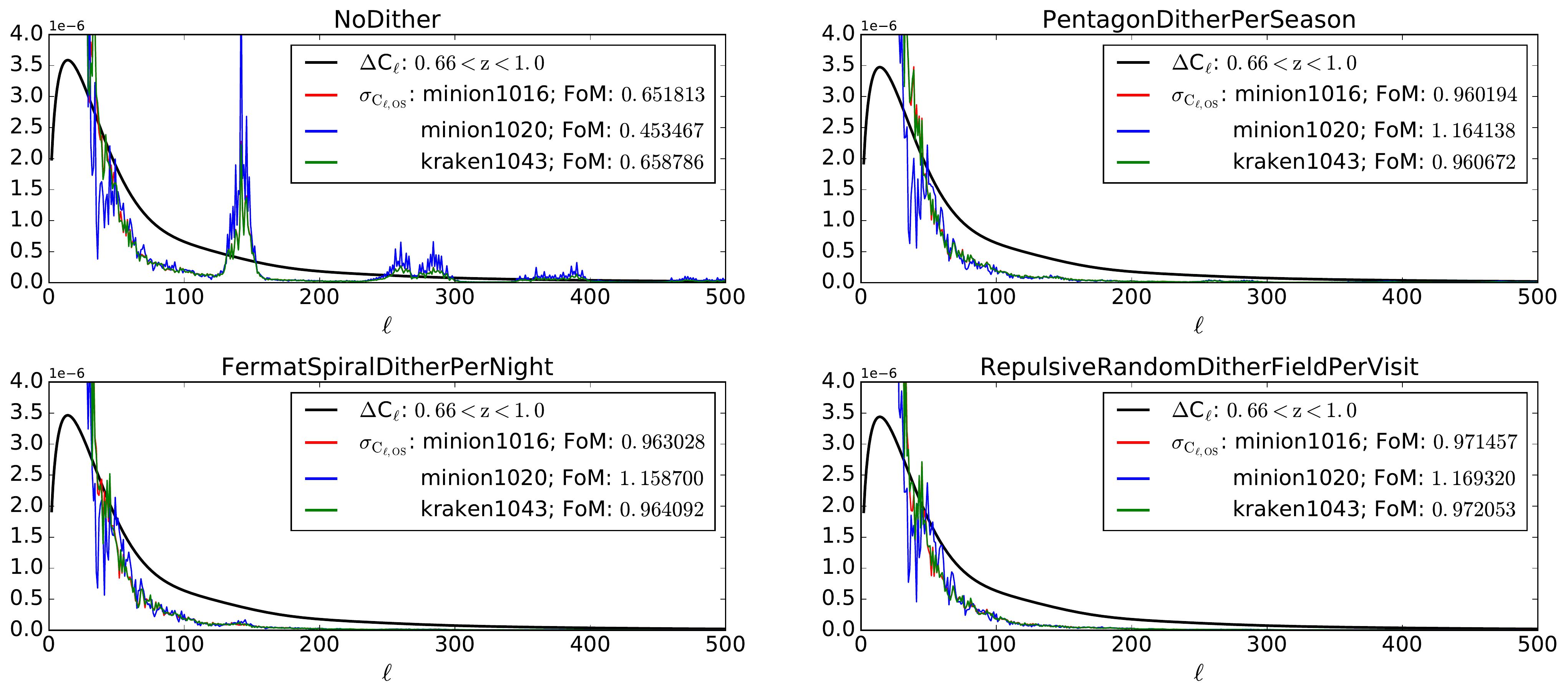}
       \vspace*{-2em}
\caption{\sigmaOS\ comparison with the minimum statistical uncertainty \statFloor\ for $0.66<z<1.0$ for three different cadences for $i<25.3$ after the full, 10-year survey.}
\label{fig: cadences: 10yr}
\end{figure*}

% ====================================================================
% Science Case Conclusions
% ====================================================================
\subsection{Conclusions}

Here we answer the ten questions posed in
\autoref{sec:intro:evaluation:caseConclusions}:

\begin{description}

\item[Q1:] {\it Does the science case place any constraints on the
tradeoff between the sky coverage and coadded depth? For example, should
the sky coverage be maximized (to $\sim$30,000 deg$^2$, as e.g., in
Pan-STARRS) or the number of detected galaxies (the current baseline
of 18,000 deg$^2$)?}

\item[A1:] As we see in \autoref{sec:\secname: analysis}, a deeper
catalog is more effective, though it makes the choice of the dither
strategy more important, especially in the first year of survey. We also see
that the wider-coverage cadence \opsimdbref{db:opstwoPS} performs
significantly better for LSS systematics with large, frequent dithers
while it performs much poorly with no or infrequent dithers; this
trend is consistent for both one-year and the full, ten-year surveys. We
note here that one year of the wider coverage (for gold sample) with frequent large
dithers leads to better systematics (as quantized here) than ten years of
the standard WFD  footprint, strongly supporting the effectiveness of wider area
coverage in the first year of survey for LSS systematics. We are definitely area-
limited more than depth-limited for LSS studies.

\item[Q2:] {\it Does the science case place any constraints on the
tradeoff between uniformity of sampling and frequency of  sampling? For
example, a rolling cadence can provide enhanced sample rates over a part
of the survey or the entire survey for a designated time at the cost of
reduced sample rate the rest of the time (while maintaining the nominal
total visit counts).}

\item[A2:] Depth uniformity is critical for LSS systematics. As we
demonstrated in \autoref{sec:\secname: analysis}, LSS studies will benefit
strongly from large dithers and wide area coverage. We do not have constraints
on the cadence.

\item[Q3:] {\it Does the science case place any constraints on the
tradeoff between the single-visit depth and the number of visits
(especially in the $u$-band where longer exposures would minimize the
impact of the readout noise)?}

\item[A3:] From our investigation into the large uncertainties in the bias induced by
the observing strategy observed in the gold sample in the baseline cadence, in
comparison with the shallower and deeper catalogs, we find that
$u$-band-induced artifacts add the most to the uncertainties in the
bias. Hence there could be a significant penalty from reducing the
number of $u$-band visit. At minimum, doing so would make the choice of
dither pattern more important. This issue still needs to be further
investigated.

\item[Q4:] {\it Does the science case place any constraints on the
Galactic plane coverage (spatial coverage, temporal sampling, visits per
band)?}

\item[A4:] LSS systematics do not place any constraints on the Galactic
plane coverage.

\item[Q5:] {\it Does the science case place any constraints on the
fraction of observing time allocated to each band?}

\item[A5:] Increasing the number of visits leads to greater survey
uniformity. At present, this is worst (among $ugri$) in the $u$-band, so
increasing the fraction of $u$-band observing time would likely help.

\item[Q6:] {\it Does the science case place any constraints on the
cadence for deep drilling fields?}

\item[A6:] LSS systematics do not constrain the cadence for deep
drilling fields as long as the main survey dithers are not affected.

\item[Q7:] {\it Assuming two visits per night, would the science case
benefit if they are obtained in the same band or not?}

\item[A7:] We do not see significant difference between obtaining two
visits per night in the same band or not, although we do see a mild
benefit in not obtaining the visits in the same band as it allows
greater variation in atmospheric conditions in each band.

\item[Q8:] {\it Will the case science benefit from a special cadence
prescription during commissioning or early in the survey, such as:
acquiring a full 10-year count of visits for a small area (either in all
the bands or in a  selected set); a greatly enhanced cadence for a small
area?}

\item[A8:] We will request full, 10-year depth during commissioning to
validate our choice of dither pattern.

\item[Q9:] {\it Does the science case place any constraints on the
sampling of observing conditions (e.g., seeing, dark sky, airmass),
possibly as a function of band, etc.?}

\item[A9:] Seeing will play a role in the photometric calibration
errors. However, these errors appear to be subdominant to the artifacts
induced by the observing strategy.

\item[Q10:] {\it Does the case have science drivers that would require
real-time exposure time optimization to obtain nearly constant
single-visit limiting depth?}

\item[A10:] We do not require any real-time exposure time optimization.

\end{description}

% ====================================================================
% Discussion
% ====================================================================
\subsection{Discussion}
\label{sec:\secname:discussion}

In this section, we presented results for the impacts of LSST observing
strategy on LSS studies. Using the \OpSim cadence baseline
\opsimdbref{db:baseCadence}, we demonstrate that dithers are necessary
for both 1-year and 10-year surveys. We find that of the three dither strategies
discussed here, FermatSpiral dithers on PerNight  timescale are the most
effective for the gold sample after one-year of survey while  dithers of all kinds
are effective after the ten-year survey. These results imply the need for a very careful
choice of the observing strategy in the first year while there is quite a range of choice
for years 2-10.

We also analyze two other cadences and find  that frequent dithering with
maximum  sky coverage could allow a significant  fraction LSST-enabled LSS
science after  one-year. Assuming that the quality of photometric redshifts is
fixed (when it actually improves with depth) and that our ansatz for window
function uncertainties is representative,  our results can go as far as implying
that the wider coverage for a few years is far more important than more years of the
baseline  WFD coverage.

Future work will entail improving our analysis to better constrain the
artifacts induced by the observing strategy by, e.g., including
uncertainties in the dust extinction, using improved models for the
photometric calibration uncertainties, more realistic galaxy colors,
incorporating improved mock catalogs to better estimate the galaxy
counts as well as its uncertainties, and a better estimate of the
uncertainties in the bias induced by the observing strategy by a more thorough accounting of the
effects of each band. Also, the development and the analysis of the
effectiveness of various systematics correction methods needs to be
carried out, especially for the 1-year survey, as only a few observing
strategies reduce the artifacts. Finally, the effectiveness
of various dithers still needs to be assessed for other science probes.

A baseline dithering pattern ``Hexdither'' is generated for all \OpSim runs with the dithered field pointings output as part of the Sqlite database (using the ``ditheredRA'' and ``ditheredDec'' parameters).  The dithering strategies discussed in this section are incorporated within the \MAF analysis framework and can be applied as a post-processing step to each simulated \OpSim survey. Integration of dithering within the scheduling of the telescope, as opposed to a post-processing step, is expected to be delivered with the v1.5 release of \OpSim v4 (currently scheduled for June 2019).

\navigationbar

% --------------------------------------------------------------------

% ====================================================================
%+
% SECTION NAME:
%    wl.tex
%
% CHAPTER:
%    cosmology.tex
%
% ELEVATOR PITCH:
%-
% ====================================================================

\newcommand{\red}[1]{\textcolor{red}{#1}}

\clearpage
\section{Weak Lensing}
\def\secname{wl}\label{sec:\secname}

\credit{jmeyers314},
\credit{tonytyson}.

Much of LSST cosmology may be limited by systematic rather than statistical
errors.  This is especially true of weak gravitational lensing, which relies on
very accurate (\ie low bias), estimates of the shear of large ensembles of
galaxies. Measurements of the noisy shapes of many galaxies, and high
signal-to-noise measurements of PSF calibration stars are made.   Even though
the shot noise of the shape of an individual galaxy is very large, any small
shear bias could accumulate over many such galaxies.  As outlined in the SRD,
uniformity of seeing in the bands used for weak lensing and special observing
strategies are required in order to reduce additive and multiplicative shear
systematics.

Achieving the ultimate sensitivity of the LSST to weak lensing science places
stringent requirements on our ability to accurately estimate galaxy shapes and
redshifts, which in turn demands precise and accurate knowledge of the point
spread function, astrometry, and photometry.  These measurements are influenced
by the interaction of light with the Earth's atmosphere, the telescope optics,
and the CCD sensors.  Systematics in the shear are introduced in each case.
Observing strategies have been developed for suppressing these systematics in
current lensing surveys, such as the Deep Lens
Survey\footnote{\url{dls.physics.ucdavis.edu}}.  These and new methods will be
applied to the LSST survey.

To leading order, we can express the effect of shear systematics in the observed
shear $\gamma^\mathrm{obs}_i$ as a small linear perturbation of the true shear
$\gamma_i$,

$$ \gamma_i^\mathrm{obs} = (1+m_i) \gamma_i + c_i, $$

where $m_i$ is the multiplicative and $c_i$ is the additive systematic in the
ith shear component.  These systematics have contributions from the atmosphere
and the detector+optics.  Systematic errors in modeling the PSF from images of
stars and in interpolating the PSF from the positions of stars to the positions of
galaxies propagate to systematic errors in the galaxy shear.  To leading order, the
PSF contributions to the additive systematic are a linear function of the PSF
ellipticity.  The best observing strategies cause the average PSF ellipticity at a
given point (over all exposures) to average towards zero.

From the LSST SRD requirements on residual systematics in the galaxy shear-shear
correlation function one can specify the level of residual shear systematics at
which statistical uncertainties become subdominant.  Over the sample of 3-4
billion galaxies, the shear systematics must be below 3 parts in 10,000 for
additive shear $|c|$, and 3 parts in 1000 for multiplicative shear $|m|$.  Each
visit to a sky patch encounters these systematics.  In particular, each re-visit
to a given field generates the same CCD-based additive shear systematic.  Some
observing strategies can effectively randomize these over all visits to a field.
It is important to note that the full survey shear-shear correlation error due
to these systematics is expected to be no better than the corresponding
systematic in any given field after all re-visits to that field. This is because
the useful angular scales in cosmic shear are less than a field radius of
several degrees, and the systematics floor in shear-shear correlation is set
therefore by the floor in any one typical field.  Below we discuss the observing
strategies for suppressing shear systematics and metrics for their success.

\subsection{Target Selection}

Image quality must be uniformly good in the bands used for weak lensing shear.
These will be mainly the $r$ and $i$ bands, though it is possible that the $z$
band will also be used for shear measurement.  The decision on which field to
observe next must be based mainly on its weak lensing priority \citep[Sec 3.1
and Figure 14.4]{2009arXiv0912.0201L}.  Depending on the current weather and
seeing, the scheduler will have a list of priorities for next-field, based on
prior history of coverage.  The relevant parameters are seeing, depth, and
camera rotation angle with respect to North and to zenith. Nearby fields in need
of coverage in these bands should be given high priority if the seeing is better
than some specified value, likely 0.7 arcsec FWHM \citep[Sec
14.5.2]{2009arXiv0912.0201L}.

\subsection{Target Measurements}

It is expected that even after optimization of camera optics and electronics,
systematic image shape errors will be associated with the orientation of the
camera focal plane.  Using data from vendor CCDs, simulations of LSST observing
have shown that a combination of x-y dithering on the sky and pipeline
processing with pixel re-map (to cancel much of the CCD frame fixed distortions)
can get well within a factor of ten of the goal for shear systematics residuals.
Simulations which add camera angle dithering show that the residual shear
systematics goal can be achieved in fields with relatively uniform seeing
history \citep{Jee&Tyson2011}.  To average down the PSF systematics over many
re-visits we benefit from uniformly distributed image quality over the ensemble.
A non-uniform history in some field can be addressed by the scheduler taking
that into account for the offending rotation angle(s) in the history of prior
visits.

Thus shear systematics will be reduced by randomization of the orientation of
the camera with respect to the sky.  This is represented by the parameter
RotSkyPos, defined as the angle between the $+y$ camera direction and North.  We
can construct diagnostic metrics that quantify the uniformity of its
distribution at each sky position.  Given the spin 2 symmetry of shear, the
optimal strategy for shear systematics will be to aim for uniformity of
RotSkyPos mod $\pi$, since angles separated by $\pi$ radians are degenerate.

Similarly, the telescope optics may harbor systematic aberrations, and these
also could be mitigated by recording images with a uniform distribution of
parallactic angle, which is the angle between North and zenith for a given field
observation.  Differential chromatic refraction (DCR) effects will also be
mitigated by varying the parallactic angle, though the exact relationship is
complex since the parallactic angle sets both the DCR magnitude and direction
for a given observation.  Re-visits to a given field should be distributed over
parallactic angles (or equivalently, hour angles), consistent with airmass and
seeing limits.  Note that wide hour angle coverage for a given field will also
be helpful in order to efficiently achieve full 180 deg coverage in CCD sky
angle.

As argued below, survey depth is more important than survey area early in the
survey.  Uniformity of depth is important, but less so than uniformity in camera
rotator shear suppression.  Simulations have shown that for the Gold sample of
galaxies, uniformity at the 0.2 mag level in limiting magnitude produces little
shear bias.  The largest effect comes from bias in weak lensing magnification
tomography \citet{Morrison2012}.  This is important in the joint analysis of
LSST multi probes of cosmology.  Trends in survey depth can also propagate to
trends in photometric redshift.  These trends must be understood, if not
minimized.

\subsection{Metrics}

For characterizing the isotropy of rotational sampling, both for rotSkyPos and
the parallactic angle, we investigate two metrics: the AngularSpreadMetric and
the KuiperMetric.  The AngularSpreadMetric characterizes the balance of a set of
angular values, in the sense that opposing angles, those separated by $\pi$
radians, have zero contribution to the AngularSpread.  The Kuiper statistic,
which is related to the well known Kolmogorov-Smirnov statistic, characterizes
the departure of a distribution from uniform, but with the added quality of
being invariant under cyclic transformations of the input set of angles.

The AngularSpread metric is computed as follows:  Given a set of angles
$\{\theta\}_{i=1, ..., N}$, map these angles onto a unit circle: $(x_i, y_i) =
(\cos \theta_i, \sin \theta_i)$, and find the 2D centroid: $(\bar{x}, \bar{y}) =
\frac{1}{N} (\sum_i x_i, \sum_i y_i)$.  The AngularSpread is the distance of the
2D centroid from the unit circle: $\mathrm{AngularSpread} = 1 - \sqrt{\bar{x}^2 +
\bar{y}^2}$.  An AngularSpread of 1 therefore corresponds to a perfectly
balanced distribution, in which the averages of both $\cos \theta$ and $\sin
\theta$ are zero, while an AngularSpread of 0 indicates a maximally anisotropic
distribution in which every angle is identical: $\theta_i = \mathrm{const}$.  As
mentioned above, weak lensing shear systematics cancel to first order when those
systematics are separated not by an angle of $\pi$ radians on the sky, but by an
angle of $\pi/2$ radians (i.e., the difference in shear \emph{phase} is $\pi$
radians).  To incorporate this spin-2 nature of shear systematics is simple, we
just multiply each angle $\theta_i$ (either RotSkyPos or the parallactic angle)
by 2 before applying the AngularSpread metric, so that, for example, pairs of
angles separated by $\pi/2$ radians on the sky are separated by $\pi$ radians in
shear phase and correctly cancel.  See \autoref{fig:angularSpread} and caption for
an example of how the AngularSpread metric is used for weak lensing.

\begin{figure}
\centering
\includegraphics[width=\linewidth]{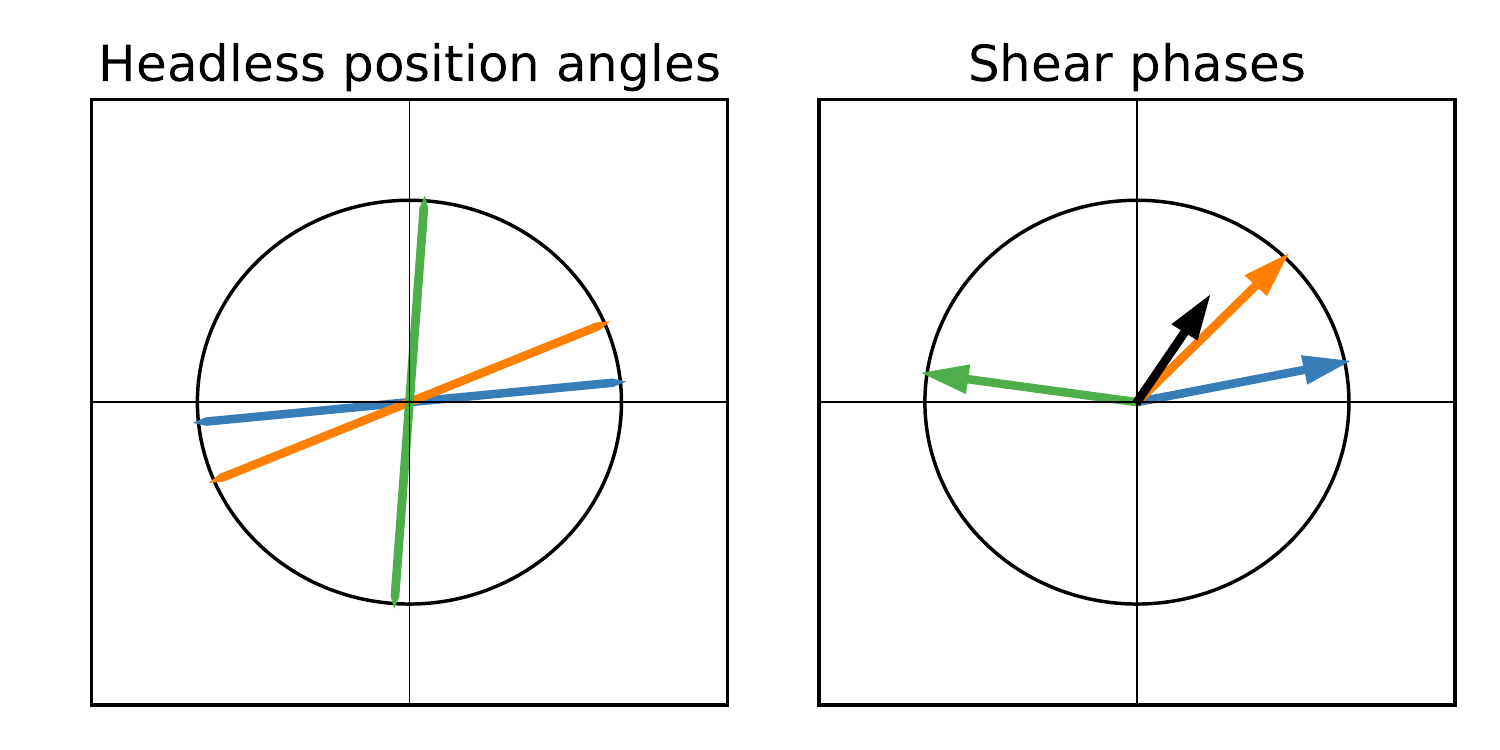}
\caption{Demonstration of AngularSpread metric.  \textbf{Left:} Three position
angles are indicated on a unit circle, using headless vectors since shear
systematics are invariant to 180-degree rotations.  \textbf{Right:} The
corresponding shear phases, which are twice the position angles, on a unit circle.
The black arrow indicates the 2D centroid of the 2D shear phases.  The
AngularSpread metric is the distance from the tip of this black arrow to the unit
circle.  Note that the blue and green position angle symbols, which are separated
by nearly 90-degrees and hence produce nearly opposing shear systematics, correspond
to blue and green shear phase vectors separated by nearly 180-degrees and hence
nearly cancel geometrically in the computation of the AngularSpread.}
\label{fig:angularSpread}
\end{figure}

% \begin{figure}
% \centering\includegraphics[width=\linewidth]{figs/enigma1189RmsAnglerotSkyPosugrizybandallpropsOPSIComboHistogram.png}
% \caption{The relative angle of the detector plane with respect to the sky, RotSkyPos, as a histogram showing the number of fields vs. rms of the parameter.}
% \label{RotSkyPos}
% \end{figure}

While the AngularSpread metric does a good job at characterizing the balance of
a distribution defined on a circle, it does not directly address the {\emph
{uniformity}} of said distribution.  For instance, the AngularSpread of the
angles $\{0, 0, 0, 0, \pi, \pi, \pi, \pi\}$ is zero, but the distribution is far
from uniform.  The Kolmogorov-Smirnov (KS) test is well known for investigating
whether a set of data are consistent with a given distribution.  The KS
statistic, off which the KS test is based, is defined as the maximum absolute
difference in the empirical cumulative distribution function (CDF) of the data
and the CDF of the distribution being tested.  The Kuiper statistic is a slight
modification of the KS statistic, defined as the sum of the maximum difference
and absolute minimum (maximally negative) difference between the empirical and
test CDFs.  This modification is convenient for characterizing distributions
defined on a circle, since it makes the statistic invariant under rotations of
the data.  The larger the Kuiper test statistic (which ranges between 0 and 1),
the larger the difference between the empirical distribution and the test
distribution.  To incorporate the spin-2 nature of shear systematics in the
Kuiper statistic, we map the values $\theta_i \rightarrow \theta_i \mod \pi$ and
compare to the uniform distribution between 0 and $\pi$.

The above metrics focus on the suppression of additive shear systematics.
Minimizing multiplicative shear systematics, and in particular estimating its
spatial variation is also important.  If the depth or average seeing are very
heterogeneous in their variation across the sky, then any multiplicative biases
that have to be corrected in the data analysis will have severe spatial
dependence. Homogeneity of depth and seeing certainly helps, and that is an
observing strategy question as discussed above.  A specific metric for
multiplicative systematics would be useful in our next \OpSim runs.  A useful
metric to be applied to full observing simulations is the mean ratio (and
spatial variance) of observed vs simulated shear amplitudes over a large sample
vs z-bins.  Since {\it Multi-Fit} joint star-galaxy fits to individual visits
will be used for data analysis, these PSF effects will be inherited for each
exposure and over that full field in a properly weighted fashion.  Other science
cases value uniformity of image quality as well, and this type of metric may be
applied during observing.

\subsection{\OpSim Analysis}

The distribution of AngularSpread for $2 \times$ rotSkyPos is shown in
\autoref{fig:WL_AngularSpread_rotSkyPos} for the latest baseline \OpSim run,
\opsimdbref{db:baseCadence}.  The left panel shows a sky map for the i-band (in this and the
following figures, the sky maps vary only minimally between the two principal
lensing filters, $r$ and $i$), while the right panel shows a histogram of values
for each LSST filter.  The distribution of the Kuiper statistic for rotSkyPos
mod $\pi$ is similarly shown in \autoref{fig:WL_Kuiper_rotSkyPos}.

While we do not currently have a method to quantitatively connect the
distribution of rotSkyPos to cosmological systematics, these figures appear to
indicate that rotSkyPos is already being well sampled in current simulations due
to the rotator tracking the sky during exposures, being subject to cable wrap
limits, and occasionally resetting to 0-degrees for filter changes.

\citet{Jee&Tyson2011} did a study of the shear residual systematics due to known
LSST CCD brighter-fatter anisotropy in 100 revisits to a single field with
random angular orientations and seeing sampled from the expected distribution.
The Data Management (DM) pipeline will use a model of the charge transport in
the CCD to re-map pixel shapes, sizes, and areas in pixel level data processing.
The needed factor of 10 suppression of the CCD-based shear systematic residuals
(post pixel remap pipeline correction) was obtained, reaching the SRD floor on
cosmic shear systematics (presented at weak lensing systematics workshop, Dec
2015)\footnote{\url{https://indico.bnl.gov/conferenceDisplay.py?confId=1604}}.
Of course, the requirements for spatial dithering for shear systematics
residuals depend on the precision of the pixel processing for removal of the CCD
based additive shear systematic.  We assume that this pixel level remap in the
DM pipeline cannot correct to better than 3 times the rms errors in the lab
tests for dynamic and static CCD systematics.

The distribution of parallactic angles is similarly shown in
\autoref{fig:WL_AngularSpread_ParallacticAngle} and
\autoref{fig:WL_Kuiper_ParallacticAngle}.  These figures show significantly less
isotropy and significantly more structure across the survey footprint than those
for rotSkyPos, likely due to the fact that, unlike rotSkyPos, the parallactic
angle is independent of the camera rotator position.  Hence, the parallactic
angle is more tightly constrained by geometry than rotSkyPos.  In fact, the only
mechanism by which the parallactic angle varies for a given field is through
variations in the hour angle at which that field is observed.

% \begin{figure}
% \centering\includegraphics[width=\linewidth]{figs/enigma1189RmsAnglerotSkyPosugrizybandallpropsOPSIComboHistogram.png}
% \caption{The relative angle of the detector plane with respect to the sky, RotSkyPos, as a histogram showing the number of fields vs. rms of the parameter.}
% \label{RotSkyPos}
% \end{figure}

% The distribution of rms values by filter is shown in
% \autoref{RotSkyPos} for the current candidate baseline simulation,
% enigma\_1189.  As shown, the rms values cluster around the value 1
% radian,  with typical values 1 +- 0.3 radian.  This compares to a
% completely uniform distribution over the half circle with an rms of
% 1.14.  As mentioned above, uniformity in cosine squared is the goal.
% Simulated observing of 100 visits to a field show this will produce
% a factor of 10 decrease in CCD-based shear systematics such as edge
% effects and the brighter-fatter x-y anisotropy.

\newcommand\plottwo[2]{{%
\typeout{Plottwo included the files #1 #2}
\centering
\leavevmode
\includegraphics*[width=0.45\columnwidth]{#1}%
\hfil
\includegraphics*[width=0.45\columnwidth]{#2}%
}}%

%  rotSkyPos metrics

\begin{figure}[tbh!]
\plottwo{figs/WL/minion_1016_AngularSpread_rotSkyPos_propID_54_and_i_HEAL_SkyMap.pdf}
        {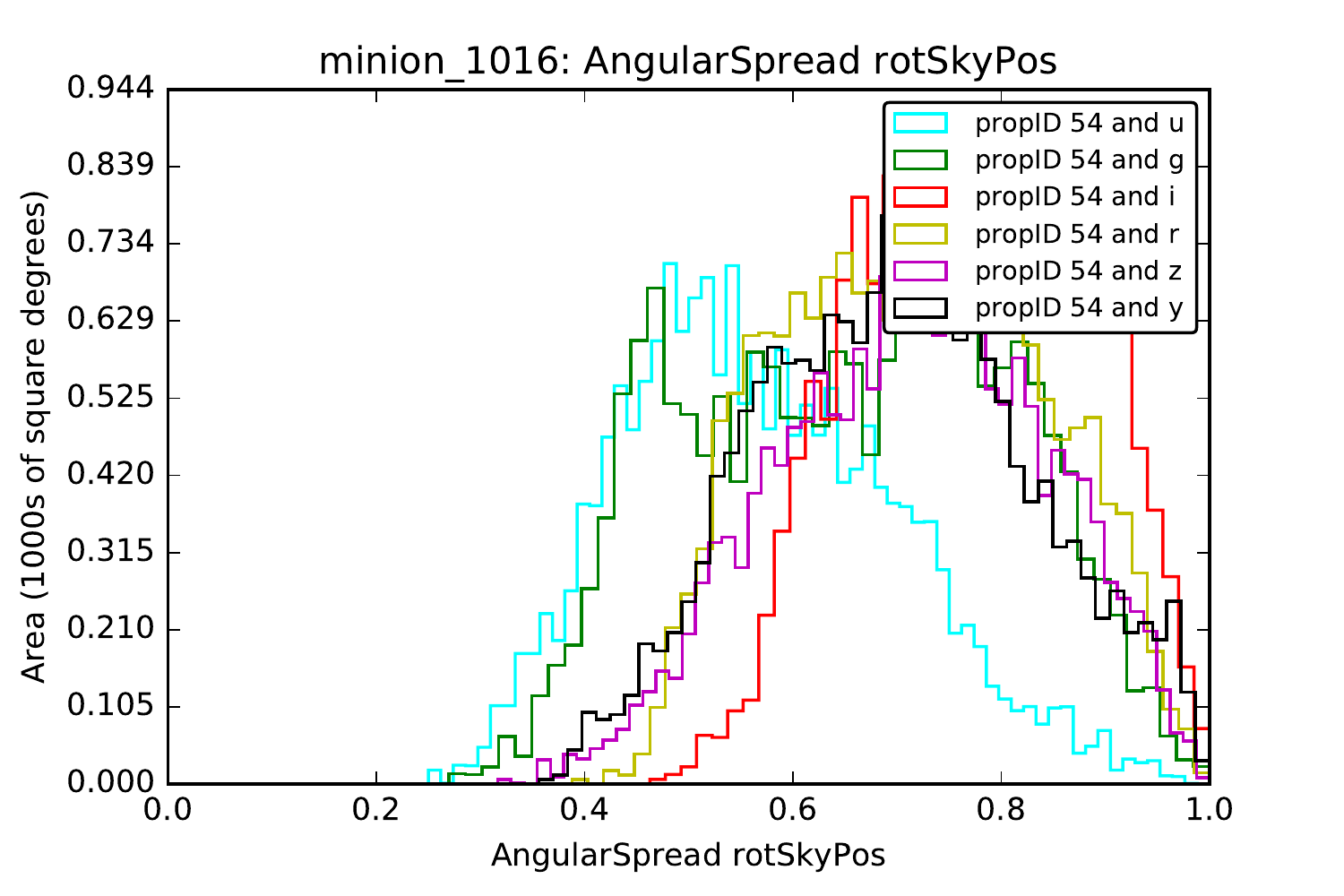}
\caption{\textbf{Left:} Sky map showing the distribution of the AngularSpread
    metric applied to the angle $2 \times$ rotSkyPos, where rotSkyPos is the
    angle between the $+y$ camera direction and North, and the factor of 2
    takes into account the degeneracy of angles separated by $\pi$ radians for
    spin-2 shear systematics.  An AngularSpread of 0 indicates a maximally
    anisotropic distribution (all visits have the same angle), while an
    AngularSpread of 1 indicates that visits are maximally balanced (the mean of
    $\cos \theta$ and $\sin \theta$ are both 0.) For the complete definition of
    the AngularSpread metric, please see the text.  To leading order, shear
    systematics permanently imprinted on the camera cancel when AngularSpread =
    1.  \textbf{Right:} Distribution of the AngularSpread metric applied to
    $2 \times$ rotSkyPos for all six LSST filters.}
\label{fig:WL_AngularSpread_rotSkyPos}
\end{figure}

\begin{figure}[tbh!]
\plottwo{figs/WL/minion_1016_Kuiper_rotSkyPos_propID_54_and_i_HEAL_SkyMap.pdf}
        {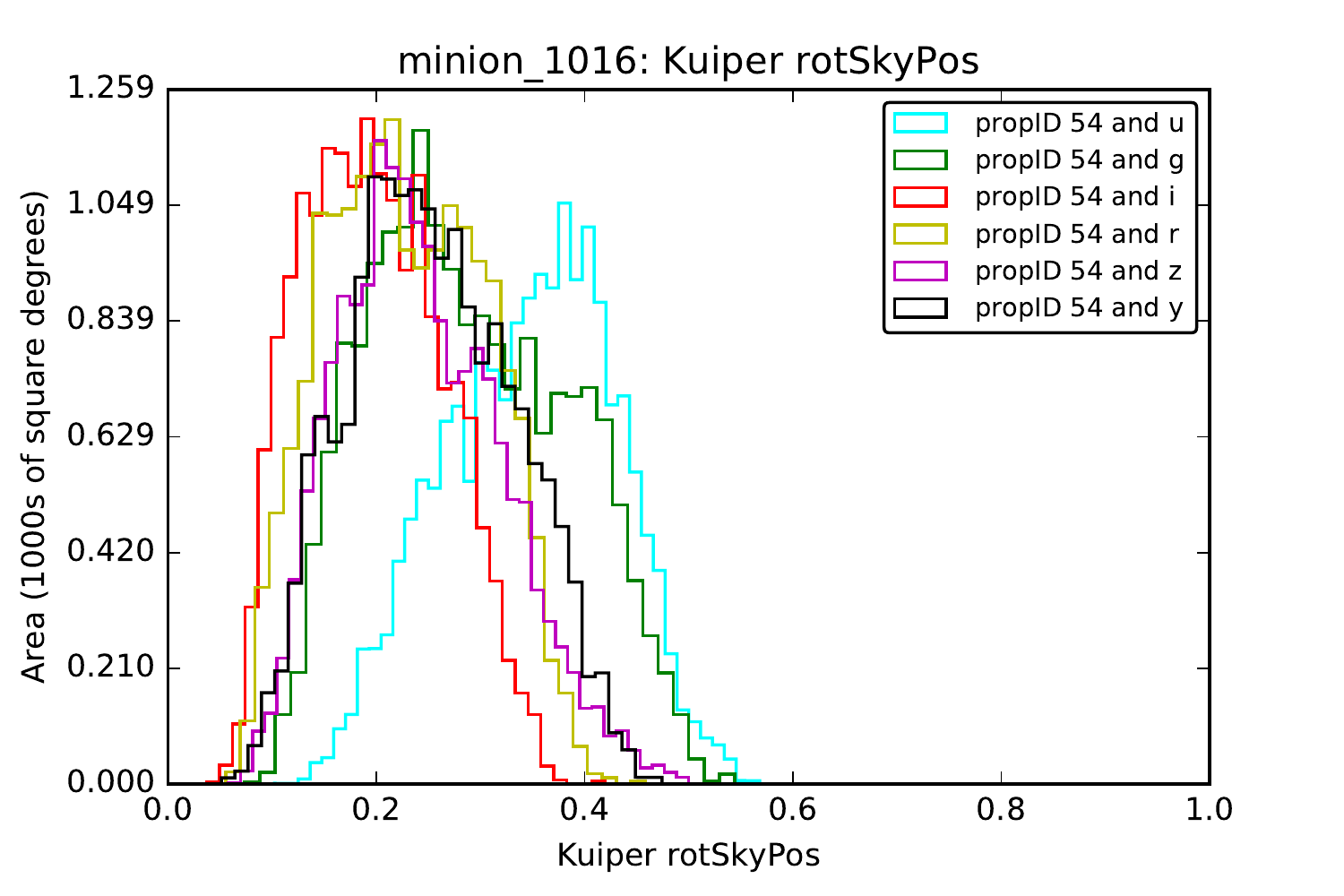}
\caption{\textbf{Left:} Sky map showing the distribution of the Kuiper metric
    (see text for definition) applied to the angle rotSkyPos mod $\pi$.  A
    Kuiper value of 0 indicates an isotropic distribution of angles (mod $\pi$),
    while a Kuiper value of 1 indicates a maximally anisotropic distribution.
    \textbf{Right:} Distribution of the Kuiper metric applied to (rotSkyPos mod
    $\pi$) for all six LSST filters.}
\label{fig:WL_Kuiper_rotSkyPos}
\end{figure}

%  ParallacticAngle metrics

\begin{figure}[tbh!]
\plottwo{figs/WL/minion_1016_AngularSpread_ParallacticAngle_propID_54_and_i_HEAL_SkyMap.pdf}
        {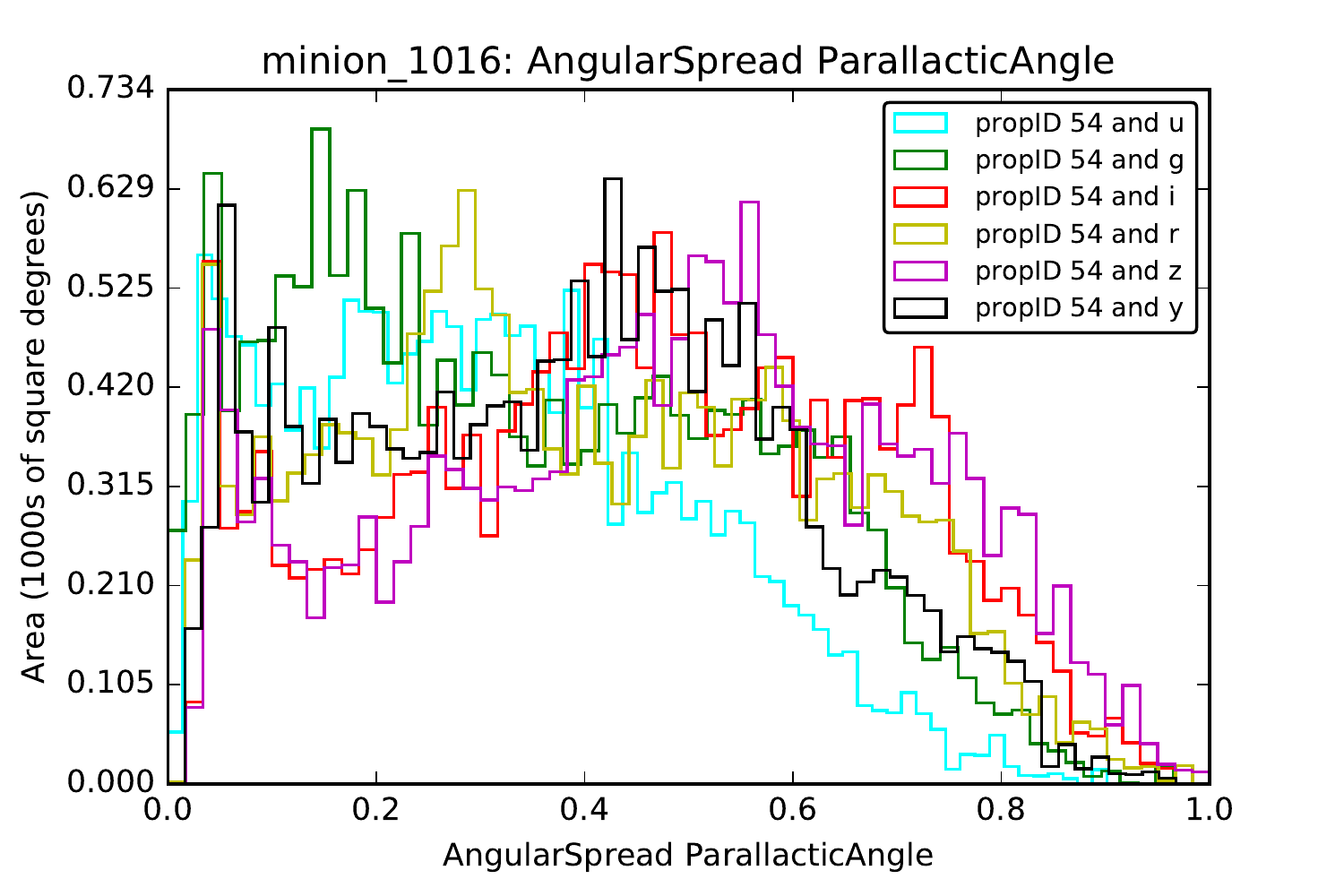}
\caption{Same as Fig. \ref{fig:WL_AngularSpread_rotSkyPos}, but for the
    parallactic angle (the angle between North and zenith) instead of rotSkyPos.
    The isotropy of the parallactic angle affects the impact of shear
    systematics due to telescope aberrations and differential chromatic
    refraction.}
\label{fig:WL_AngularSpread_ParallacticAngle}
\end{figure}

\begin{figure}[tbh!]
\plottwo{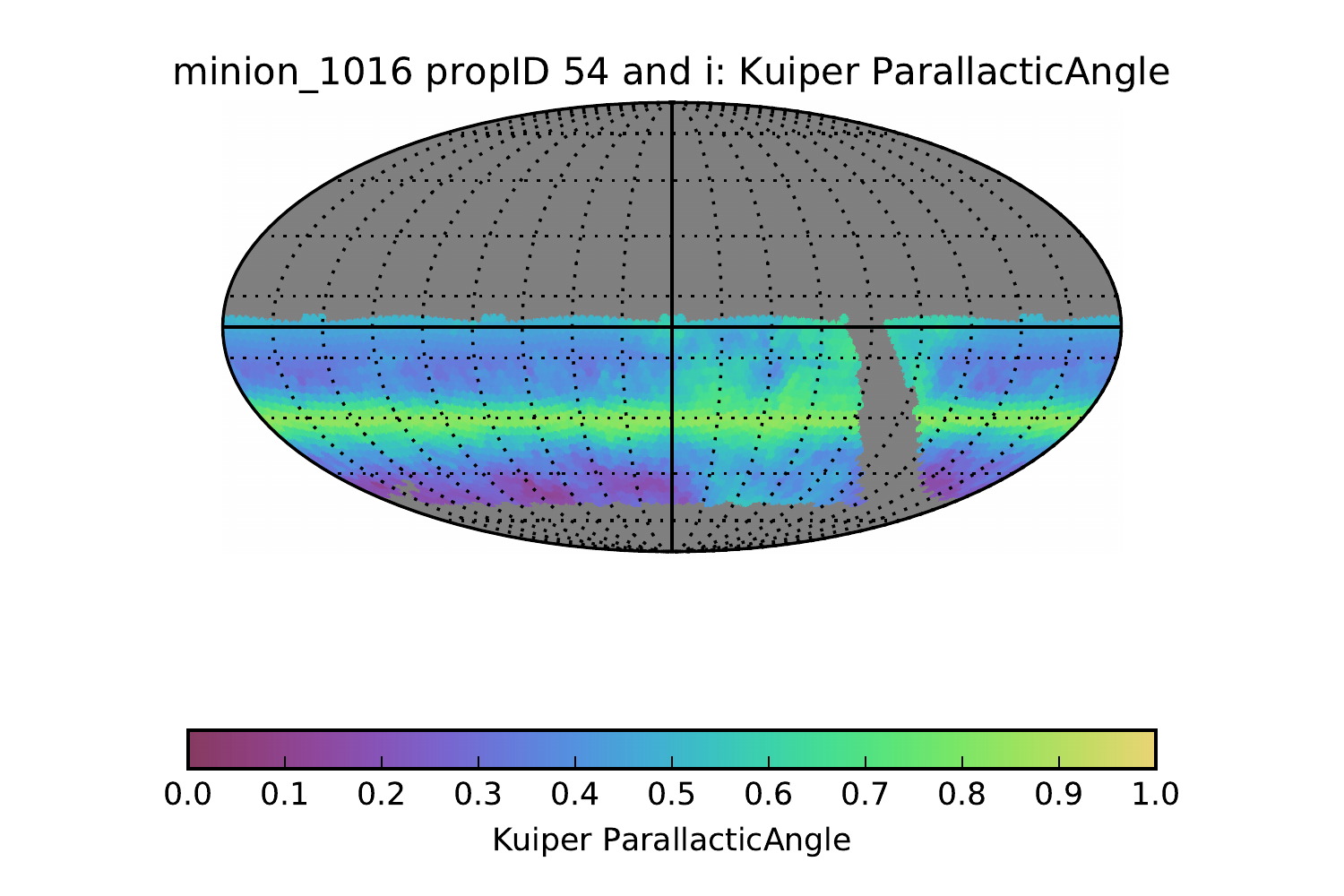}
        {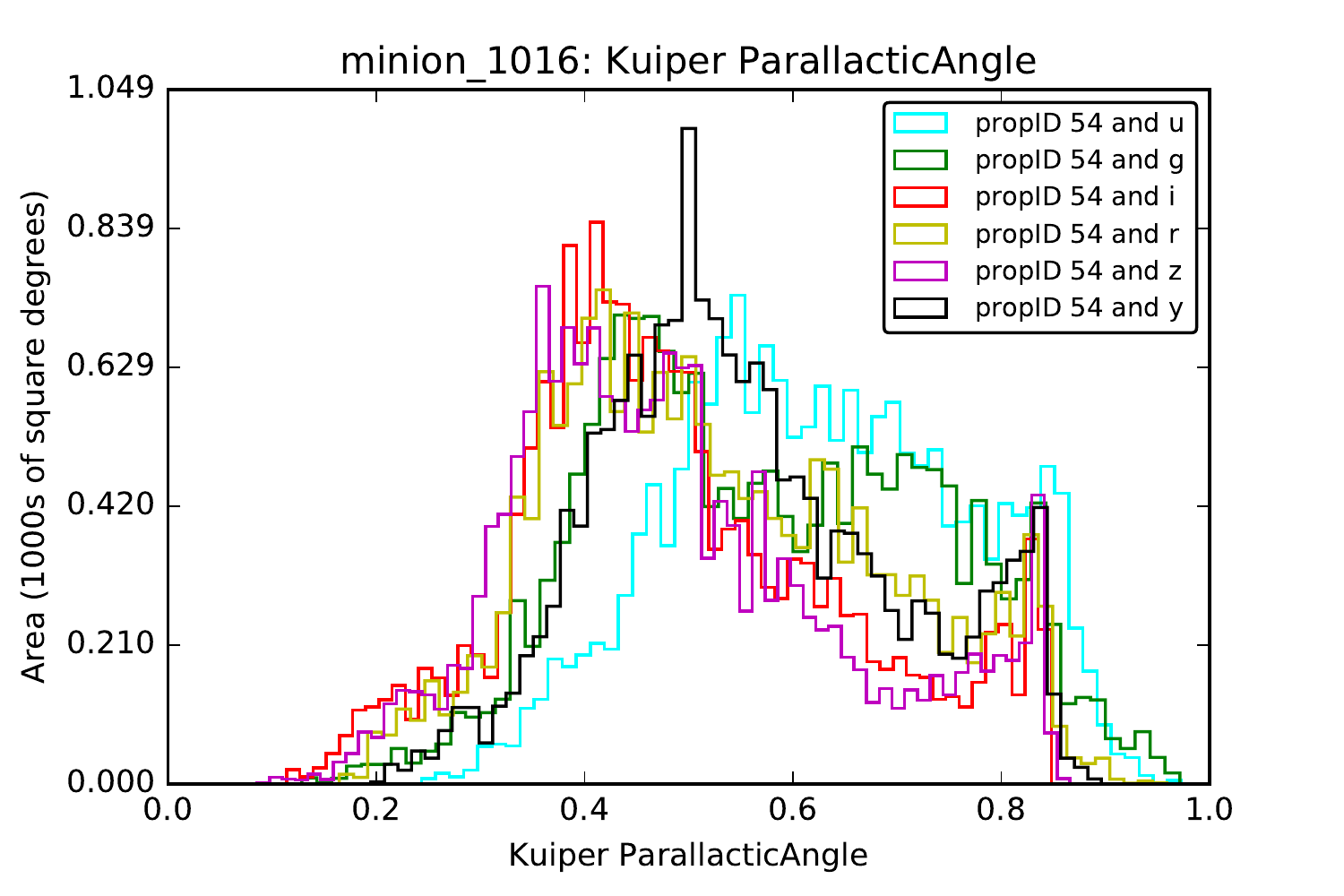}
\caption{Same as Fig. \ref{fig:WL_Kuiper_rotSkyPos}, but for the parallactic
    angle instead of rotSkyPos.}
\label{fig:WL_Kuiper_ParallacticAngle}
\end{figure}

\subsection{Ancillary data}

Optimal observing strategy relies on the use of all relevant data, with current
and historical coverage per field.  Ancillary data, and auxiliary uses of the
science exposures, play an important role informing next-field strategy.  We can
use large scale patterns of distortions of the PSF over the 20,000 stars per
exposure for PSF regularization in the per-CCD PSF fitting.  In the per CCD
fits, there is a benefit to setting aside some stars for validation tests of PSF
extrapolation.  These data may be used as a metric for image quality and thus
the ranked value of an exposure for shear systematics removal.  Fields with poor
image quality rise to the top of the priority list for re-observing at that
camera angle.  In addition to using all the stars in a given visit, there is
useful information in the wavefront sensors and the guide CCDs that may be used
to regularize the PSF reconstruction in a visit.  We might read out guider CCDs
in different ways to better monitor the atmosphere.

\subsection{Deep vs Wide}

% LSST Review by Nelson Padilla: This section points out the different advantages that come from deep vs. wide observations. Are there quantitative metrics that show the best balance between the two? A plot showing this would give much more weight to the important requirement of a deep 1000 sq. degree survey during commissioning.

As outlined above, many revisits to each field spanning many RotSkyPos angles
aids the suppression of additive shear systematics.  For a given per-visit
exposure time this leads to a deep survey, due to the multitude of visits.
There are several advantages to a deep survey over a shallow-wide survey for
weak lensing science, especially for dark energy where a range of lens redshifts
is required to sample the growth of dark matter structure from redshift beyond 1
to low redshift.  A strategy question for LSST is whether to go wide first and
then deep, or the reverse.  There are actually several drivers for depth over
area, given fixed observing time and camera+telescope etendue.  Provided that
sufficient area is covered to overcome sample variance at lower redshifts and to
adequately cover the important angular scales, a deep observing strategy
maximizes the cosmological signal-to-noise ratio both by maximizing the signal
and minimizing the noise (see, for example, Figure 26 from \citep{Jee2013}).

First, a deep survey strategy boosts the amplitude of the cosmic shear signals
due to the increased amplitude of the lensing kernel.  Given the same lens mass,
the amplitude of lensing signals is a simple function of the two distances:
observer-to-lens versus lens-to-source.  For example, when a lens is at $z =
0.5$, the shear for a source at $z = 1.1$ is nearly twice that at $z = 0.7$,
leading to a factor of 4 ratio in correlation function amplitudes.  Thus, a deep
survey can take advantage of the geometric effect of gravitational lensing more
efficiently.

Second, a deep survey strategy enables a longer redshift baseline to break
parameter degeneracies.  For example, shallower surveys with wider area are not
efficient in shrinking the length of the ``banana'' in the $\Omega_M$-$\sigma_8$
plane because of the $\sigma_8 \Omega_M^\alpha = \mathrm{const}$ degeneracy.
The deep strategy compensates for the loss of volume due to a reduced area by
extending the volume along the line of sight.

% LSST Review by Nelson Padilla: is there a way to be more quantitative about the advantage of depth over area at fixed volume? The advantages are explained qualitatively in the following paragraphs.

Third, deep surveys provide more useful galaxies per area.  This merit does not
entirely overlap with the second point.  In addition to new sources at higher
redshift, the fainter limiting magnitude enables detection of fainter galaxies
at a given redshift and better signal-to-noise ratio.  This argument requires
that photometric redshifts of this population of galaxies is as well-calibrated
as those at lower redshift.  This may be done via angular cross-correlation with
the brighter photometric sample galaxies with known spectra, in test patches.
Needless to say, the increased source density reduces the impact of shape noise
caused by intrinsic ellipticity dispersion.

Fourth, it mitigates the effects of intrinsic alignments (IA), an important
theoretical systematic in precision cosmic shear.  Current studies
\citep{Heymans2013} indicate that IA effects are dominated by luminous red
galaxies (LRGs).  Because a deep survey can access fainter galaxies, the net IA
systematic decreases because the fraction of the LRGs decreases at fainter
limiting magnitude.  Using the approach of \citet{Joachimi2011}, at $z = 0.5$
the amplitude of the IA power spectrum is reduced by a factor of two for an
increase in the limiting $r$ magnitude from 24 to 27 mag.

For cosmic shear the volume at $\sim 5$ arcminute to several degree scales and
over a wide range of redshift is most important.  Once a fair sampling of
“cosmic variance” is achieved the depth matters most, because of the z-width of
the lensing kernel.  With fixed survey duration on a camera+telescope of fixed
etendue, it is better to prioritize depth in order to maximize the weak lensing
cosmological signal-to-noise ratio.  In particular, for the LSST survey it would
be helpful to have several widely spaced deep drilling fields done early and
deeper than the main survey, in order to explore detailed observing strategies --
particularly the angle dither scheme.  Deep drilling data is useful to a variety
of LSST science drivers, including Bayesian analyses.  By the same reasoning, it
would be important to cover a significant area [perhaps 2000 sq. deg] to full
depth during the first year of the survey.  This would allow full assessment of
systematics, and could be chosen to overlap the WFIRST footprint.  Such an
overlap may enable additional systematics checks of photometric redshift accuracy
(exploiting WFIRST infrared photometry) and the impact of blends (exploiting the
smaller WFIRST PSF).

% LSST Review by Nelson Padilla: a brief list of advantages of overlap with WFIRST would be a definite plus here.

Metrics for reaching the SRD goals for WL related science are needed. Long
before LSST is on the sky such metrics can be applied to realistic simulations
of LSST observing, in order to inform the scheduler proposal ranking algorithm.
One metric is the sample variance in cosmic shear in 10 sq. deg DD fields versus
that, for say, 1000 sq. deg.  A related systematics metric would be the
suppression of additive PSF systematic residuals in 1000 sq. deg worth of
pointings over the sky, normalized by that in one field visited 100 times. These
important simulations will be enabled by the deep-wide full cosmological
simulation deliverables in DESC Data Challenges 2 and 3, together with \OpSim
runs.

\subsection{Discussion}

The RotSkyPos metrics show that the majority of fields have good randomization
of detector angles projected on the sky.  The randomization of parallactic
angles is less successful, though this is to be expected due to fewer knobs
available to adjust the parallactic angle of observations of a given field.  In
both cases, however, a significant fraction of fields show metric values lower
than expected for a uniform distribution.  Regardless of the \emph{per field}
criterion adopted, it is desirable to avoid the incidence of individual
discrepant fields.  The recommended criterion for randomization of RotSkyPos and
parallactic angle is not the behavior of the majority of the fields, but of the
minority with the least random behavior -- the number of non-random fields
should be minimized.

It is certain that actively controlling the statistics of RotSkyPos will require
additional slewing of the camera rotator.  At present, the operations plan is to
only slew (beyond that required to track the sky during exposures) when
necessary to prepare for a filter change -- that could be estimated at the
equivalent of $\simeq 3$ complete rotations per night.  To engage the rotator by
up to $\simeq 30$ degrees per visit would require $\simeq 300$ complete
rotations per night - an upper limit to the additional rotations, but one which might be approached by a requirement for
rigorous distribution over relatively short timescales.  On the other hand, a very significant improvement
in distribution could be achieved by simply introducing a constrained randomized rotator
offset after each filter change, with little or possibly no loss in observing efficiency.
Guidance is needed from simulations for the randomization requirement(s), and from science strategy
for the time scale over which it must be achieved.

Increasing the isotropy of the parallactic angle is trickier, since the
parallactic angle is only affected by the hour angle of observations (for a
given field).  It may be possible, however, to adjust the scheduler cost
function to better favor parallactic angle isotropy.

In summary, image quality weighted randomization of RotSkyPos is required at
10-20\% scatter in image quality over the sample of visits (2015 simulation
mentioned above).  Randomization of parallactic angle is also desired, but a
requirement is yet to be determined.  Interestingly, there is an excellent
opportunity to combine these two randomizations in an efficient observing
strategy which actually minimizes the number of camera rotations.  Re-visits to
a field over a range of hour angles and with camera rotations assures full 180
degree range coverage for CCD coordinates relative to sky.  A metric for this
can be written and run with \OpSim to explore optimization, including minimizing
camera rotations.

\subsection{Conclusions}

Randomization of camera angle covering 180 deg can effectively suppress
residual shear systematics remaining after DM pipeline first-order
correction.   More simulations are required, using intelligent
dithering, in order to assess the efficiency and the remaining WL shear
systematics.

We can now provide answers to the ten questions posed in %
\autoref{sec:intro:evaluation:caseConclusions}:

\begin{description}

\item[Q1:] {\it Does the science case place any constraints on the
tradeoff between the sky coverage and coadded depth? For example, should
the sky coverage be maximized (to $\sim$30,000 deg$^2$, as e.g., in
Pan-STARRS) or the number of detected galaxies (the current baseline
of 18,000 deg$^2$)?}

\item[A1:] On the scale of thousands of sq.deg, the WL signal and
control of systematics is enhanced by deeper (larger cosmological
volume) rather than wider surveying. These considerations led to the
18,000 sq.deg per ten years baseline.

\item[Q2:] {\it Does the science case place any constraints on the
tradeoff between uniformity of sampling and frequency of  sampling? For
example, a rolling cadence can provide enhanced sample rates over a part
of the survey or the entire survey for a designated time at the cost of
reduced sample rate the rest of the time (while maintaining the nominal
total visit counts).}

\item[A2:] WL is generally agnostic on this issue. Total good IQ x depth
is most important, and this might benefit from rolling cadences if the
mean airmass is lower, resulting in a higher IQ.

\item[Q3:] {\it Does the science case place any constraints on the
tradeoff between the single-visit depth and the number of visits
(especially in the $u$-band where longer exposures would minimize the
impact of the readout noise)?}

\item[A3:] 30-sec or longer u band exposures will help S/N.

\item[Q4:] {\it Does the science case place any constraints on the
Galactic plane coverage (spatial coverage, temporal sampling, visits per
band)?}

\item[A4:] WL ugrizy imaging must avoid low Galactic latitudes due to
stellar crowding, bright stars, and dust. Data at low latitudes will not
be used for WL.

\item[Q5:] {\it Does the science case place any constraints on the
fraction of observing time allocated to each band?}

\item[A5:] This depends on the relative system throughput vs wavelength.
For the case of high u band QE CCDs, approximately u10\%, g10\%, r22\%,
i22\%, z18\%, y18\%. The comes from 80, 80, 184, 184, 160, 160 ugrizy
30-sec visits per 9.6 sq.deg fields over 18,000 sq.deg per ten years --
from the SRD, driven by required low surface brightness shear
measurement and the needed S/N for photo-z for the gold sample of
galaxies. The longer integrations at longer wavelengths is due to the
red color of the sky background.

\item[Q6:] {\it Does the science case place any constraints on the
cadence for deep drilling fields?}

\item[A6:] It will be helpful if the individual exposure times in all
bands are similar to the main survey. A long series of exposures in each
filter would be good. Dithering in x-y-theta between exposures using
“intelligent dithering” is needed in order to detect low level
systematics and to help calibrate the main survey galaxy blending.
Emphasis on excellent seeing is important. Spreading the DD observing
over many half nights, maximizing the HA coverage, will be needed.

\item[Q7:] {\it Assuming two visits per night, would the science case
benefit if they are obtained in the same band or not?}

\item[A7:] Not necessarily. In fact, occasional back-to back gr
exposures help calibrate systematics from chromatic refraction.

\item[Q8:] {\it Will the case science benefit from a special cadence
prescription during commissioning or early in the survey, such as:
acquiring a full 10-year count of visits for a small area (either in all
the bands or in a  selected set); a greatly enhanced cadence for a small
area?}

\item[A8:] Yes, ugrizy full depth over at least 1000 sq.deg. Suppression
of WL systematics will depend on early detection of issues. This, and
the development and validation of MultiFit pipline, shear measurement
algorithms, cosmological analysis algorithms, and associated covariances
will rely on early deep surveying of a few thousand square degrees. This
also benefits co-analysis with WFIRST or Euclid data.

\item[Q9:] {\it Does the science case place any constraints on the
sampling of observing conditions (e.g., seeing, dark sky, airmass),
possibly as a function of band, etc.?}

\item[A9:] WL benefits significantly from good uniform seeing, IQ, and
depth in the r and I bands. For example, imaging in the r and i bands in
seeing worse than 1 arcsec is not productive, and revisits to that field
should await better seeing. The associated photo-z relies on uniform
depth spatially. This means low airmass.

\item[Q10:] {\it Does the case have science drivers that would require
real-time exposure time optimization to obtain nearly constant
single-visit limiting depth?}

\item[A10:] Yes. Uniform residual shear is even more important than
depth. Dithering in x-y-theta between exposures using “intelligent
dithering” is needed in order to detect low level systematics and to
help calibrate the main survey galaxy blending. In intelligent dithering
the scheduler is aware of the past history of the IQ in that field in
the two bands used for WL shear (nominally r and i). The algorithm tries
to uniformly sample camera angles relative to north with uniformly good
IQ. Rotation angles of previous poor IQ visits need to be repeated.

\end{description}

\navigationbar

\clearpage
\section{Photometric Redshifts}
\def\secname{photoz}\label{sec:\secname}

\credit{MelissaGraham},
\credit{SamSchmidt},
\credit{connolly},
\credit{ivezic}

\subsection{Introduction}

Photometric redshifts are an essential part of
every cosmology probe within LSST.  The principal concern for LSST
photo-$z$ performance is to meet the stringent requirements on redshift
uncertainty, bias, and catastrophic outlier rate as laid out in the
Science Requirements document. Photo-$z$'s are dependent on precise
measurements of galaxy colors, thus cadence and depth variations must be
examined as a function of all six LSST filter bandpasses.  Overall image
depth and signal-to-noise is our primary concern. For studies of Large
Scale Structure, Weak Lensing, Clusters, and Supernova host galaxies,
survey uniformity is desired for the full depth survey, while the
temporal details of how we reach full depth are not as important as
uniformity both as a function of sky position and observing conditions.
However, as we desire science-grade photometric redshifts after one year
of operations, two years, and so forth, the cadence must meet some basic
requirements for the six-band system at least on the timescales of the
yearly data releases.

\textbf{Specifications.} The Science Requirements Document (SRD; \citealt{LPM-17}) defines
the minimum statistical specifications for photometric redshifts for an
$i<25$, magnitude-limited sample of $4\times10^9$ galaxies from
$0.3<z<3.0$ as: (1) the root-mean-square of the error in photo-$z$
must be $\sigma < 0.02$; (2) the fraction of outliers must be $<10\%$;
and (3) the average bias must be $<0.003$.
The details for how these statistics are calculated are discussed in the Metrics section below.
With this in mind, we are developing software to show that our photo-$z$ algorithms
can meet specifications for LSST baseline parameters and to simulate the
impact of deviations from the 10-year baseline plan on photo-$z$
statistics.

\textbf{Planned Experiments.} This software is designed to allow the
user to modify LSST baseline parameters, simulate a set of test galaxy
observations (i.e., magnitudes with errors appropriate for the given
LSST parameters) from a training catalog with ``true" magnitudes and
redshifts and a realistic intrinsic dispersion in color, magnitude, and
redshift, run a photometric redshift algorithm on the test galaxies
(i.e., matching in color-space to the training catalog), and output
statistics for analysis. Modifiable LSST input parameters will include:
the limiting magnitude applied to the galaxy catalogs (e.g., $i<25$);
the number of visits per filter; the number of years of LSST
observations that have passed (this can be a fraction of a year);
systematic offsets to the magnitudes in each filter (default $=0$); and
coefficients for the magnitude uncertainties in each filter (default
$=1$).  Output for user analysis will include catalogs of $z_{\rm phot}$
and the metrics for photo-$z$
in any desired redshift range. For example, we will be able to vary the
total number of $u$-band visits and examine how this affects the
fraction of outliers at 1, 5, and 10-years of the survey. In this
software, parameters of the photo-$z$ algorithm itself will also be
modifiable, allowing us to test options in the algorithms against
various LSST observing strategies.

\textbf{Currently implemented photo-$z$ algorithm.} We draw $N_{\rm
test}$ ``test" and $N_{\rm train}$ ``training" galaxies from a catalog of
simulated galaxies, ensuring no overlap. We determine their
magnitude uncertainties as appropriate for the LSST parameters, randomly
scatter their magnitudes to induce an observational error, and calculate
the associated colors and color errors. 
We calculate the Mahalanobis distance \citep{Mahalanobis1936} in color space
between each test galaxy and the 10\% of training set galaxies closest matched
in apparent $i$-band magnitude (i.e., thereby applying a crude prior on magnitude),
and then use it to identify a color-matched subset of
training galaxies using a threshold defined by the $\chi^2$ percentage point function at 68\%.
We draw a random training galaxy from this subset and use its redshift as the
photo-$z$ for that test galaxy. We then calculate our statistical
metrics on the photometric redshifts for the test sample, using each
test galaxy's original catalog redshift as the ``true'' redshift.
Our adopted method is not necessarily the ``best" way to calculate photometric redshifts,
but due to its relatively direct relation between the photometric errors
and the uncertainty in the estimated photo-$z$, it is especially appropriate analyzing the impact of changes
to the LSST observing strategy.
Also, this process is open to substituting alternate photometric redshift
algorithms, a variety of galaxy catalogs, and/or adding different priors.
Note that for this experiment, we always assume that the
training set galaxies have photometric errors equivalent to that of the
full-depth (i.e., 10-year) LSST catalog. 

\subsection{Metrics}

The primary metrics we will use to evaluate LSST
observing strategies with respect to the SRD photo-z specifications are the
robust standard deviation, the robus bias, and the fraction of outliers. For all test
galaxies we calculate $\Delta z = (z_{\rm true} - z_{\rm phot}) /
(1+z_{\rm phot})$, and identify galaxies in the interquartile range (IQR) of $\Delta z$.
The robust standard deviation is calculated by dividing the full-width of the IQR by 1.349 (i.e., assuming a Gaussian distribution, which we find to be appropriate).
The robust bias is the mean value of $\Delta z$ for IQR galaxies.
Outlier galaxies are identified as those with $\Delta z$ exceeding the larger of 0.06 or 3$\sigma$, 
where $\sigma$ is the robust standard deviation for all galaxies with $0.3 \leq z_{\rm phot} \leq 3.0$.
The IQR is used to exclude outliers from influencing these statistics; in other words, the robust
standard deviation and bias are for a subset of ``good" photo-$z$'s.

\subsection{Initial Results}

To demonstrate this software with a
preliminary analysis, we apply the currently implemented photo-$z$
algorithm to a galaxy catalog based on the Millennium simulation \citep{2005Natur.435..629S},
which uses the galaxy formation models of \citep{2014MNRAS.439..264G}
and was constructed using the lightcone techniques described by \cite{2013MNRAS.429..556M}.
We use $N_{\rm train}=1000000$ and $N_{\rm test}=50000$ galaxies, and apply the 
condition of $i<25$ after scattering based on their magnitude errors in order
to generate a sample for which the LSST SRD specifications apply.
For this demonstration we show how the photo-$z$ metrics evolve with respect to
two of the basic LSST parameters: the year of the survey, and the number
of $u$-band visits. When we simulate results in a given year of LSST, we
assume uniform progression in all filters (i.e., the total number of
visits per filter, \texttt{[56, 80, 184, 184, 160, 160]} in
\texttt{[u,g,r,i,z,y]}, is distributed evenly over all years). When we
simulate the LSST 10-year results for a given number of $u$-band visits,
the visits removed/added to $u$-band are added/subtracted evenly to/from
the other five filters. The results of these tests are presented in
Figure~\ref{fig:redshifts} and~\ref{fig:metrics}. For example, in this
demonstration we can see how the photo-$z$ results will improve
over time, and how the fraction of outliers at low $z_{\rm phot}$ actually increases
for the first half of the survey as the standard deviation improves.
We can also see that a lack of $u$-band data induces more scatter 
in the photo-$z$ results at $z<0.5$ and $z>2.0$, but that allotting 
more than 56 visits to $u$-band does not necessarily improve the results,
since this comes at the expense of visits in the other filters.

\begin{figure}[h]
\begin{center}
\includegraphics[width=5cm]{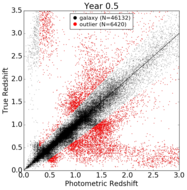}
\includegraphics[width=5cm]{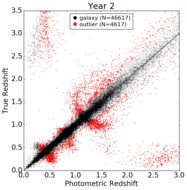}
\includegraphics[width=5cm]{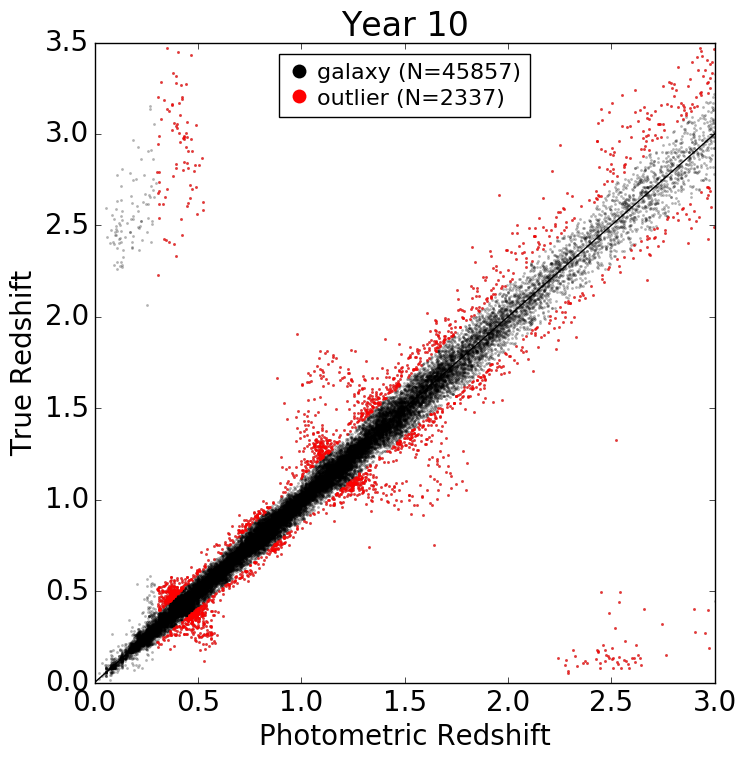}
\includegraphics[width=5cm]{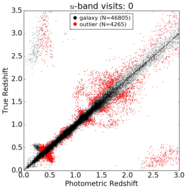}
\includegraphics[width=5cm]{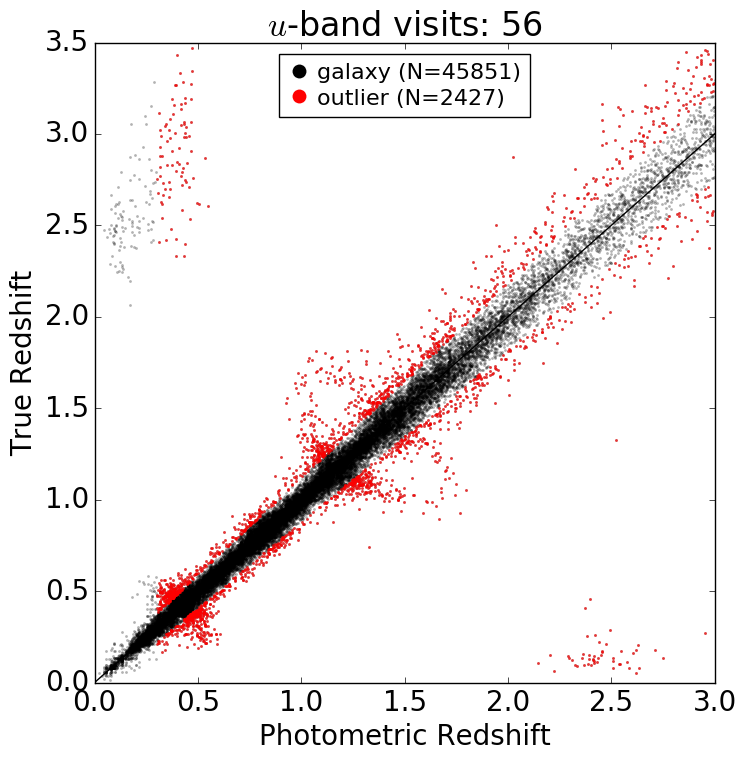}
\includegraphics[width=5cm]{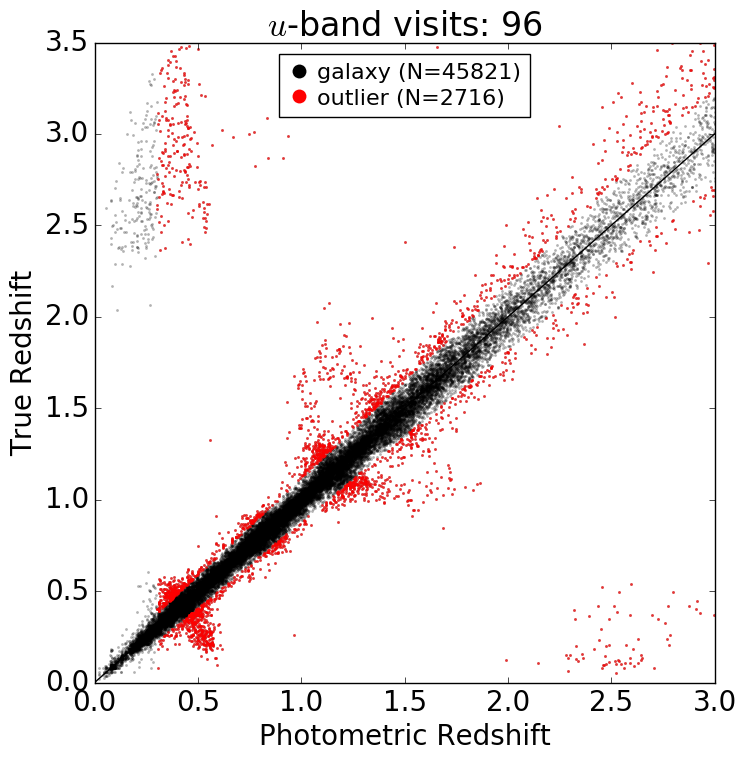}
\caption{Photometric vs. true (i.e., catalog) redshifts. Across the top row we show results from
0.5, 2.0 and 10.0 years of the LSST survey, and across the bottom row we show results for $0$,
$56$ (baseline), and $96$ $u$-band visits. 
\label{fig:redshifts}}
\end{center}
\end{figure}

\begin{figure}[h]
\begin{center}
\includegraphics[width=5cm]{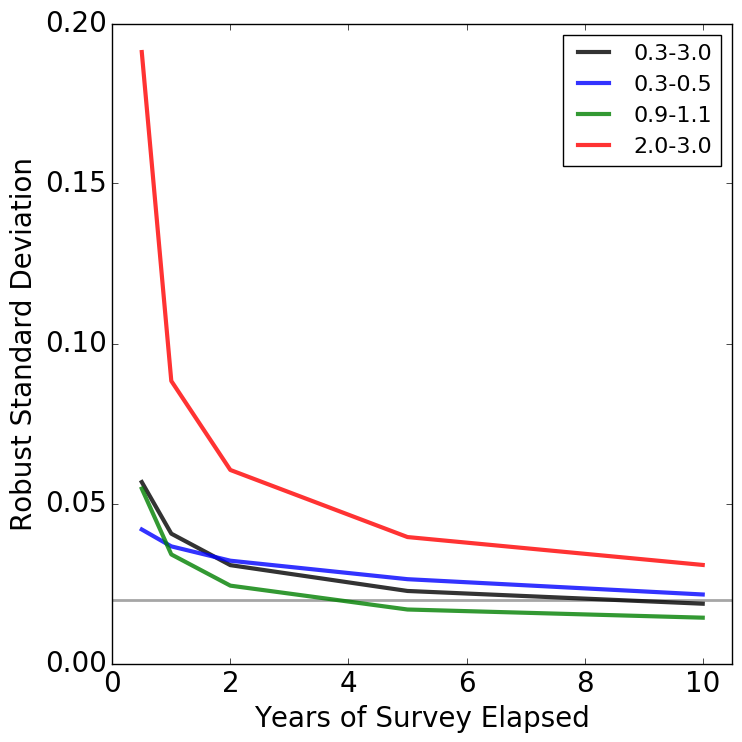}
\includegraphics[width=5cm]{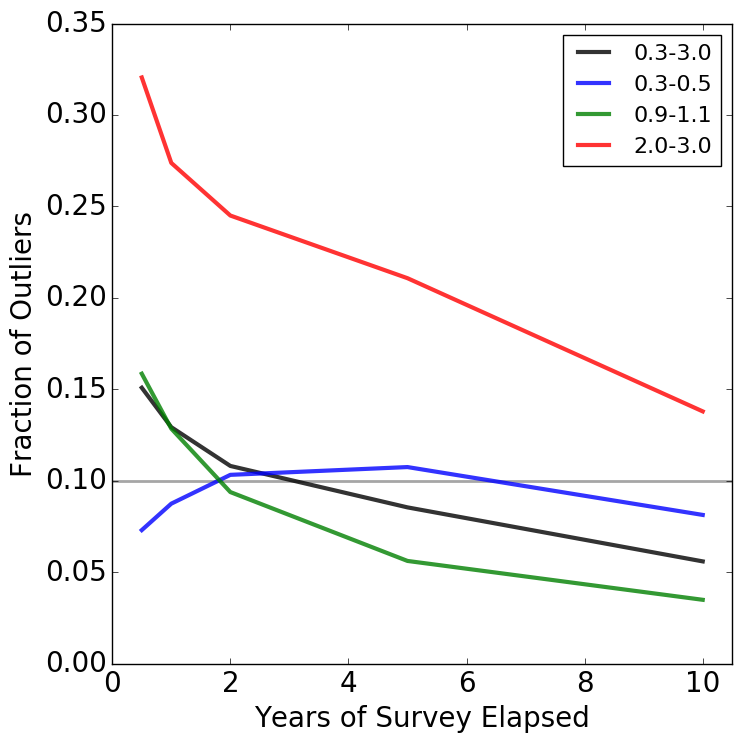}
\includegraphics[width=5cm]{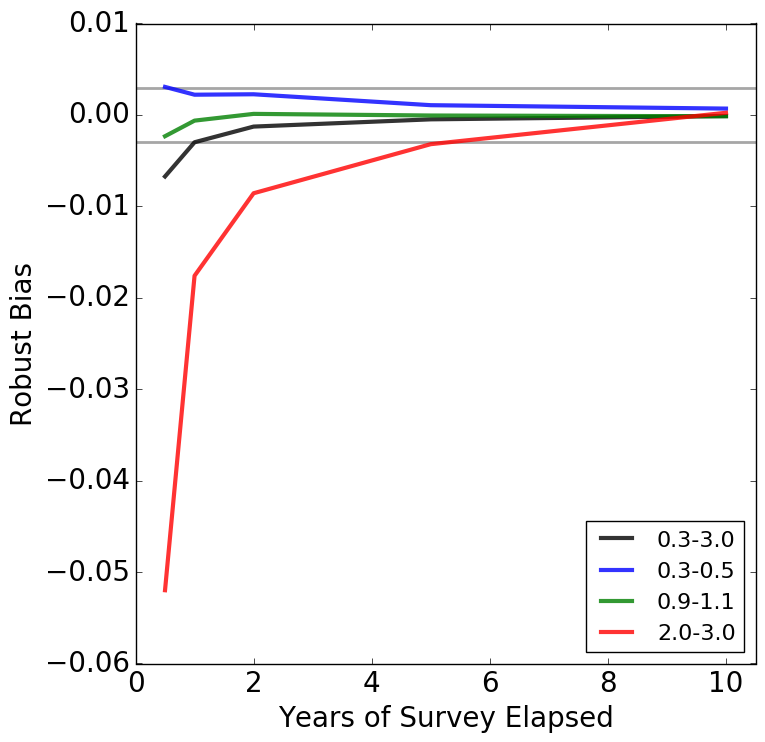}
\includegraphics[width=5cm]{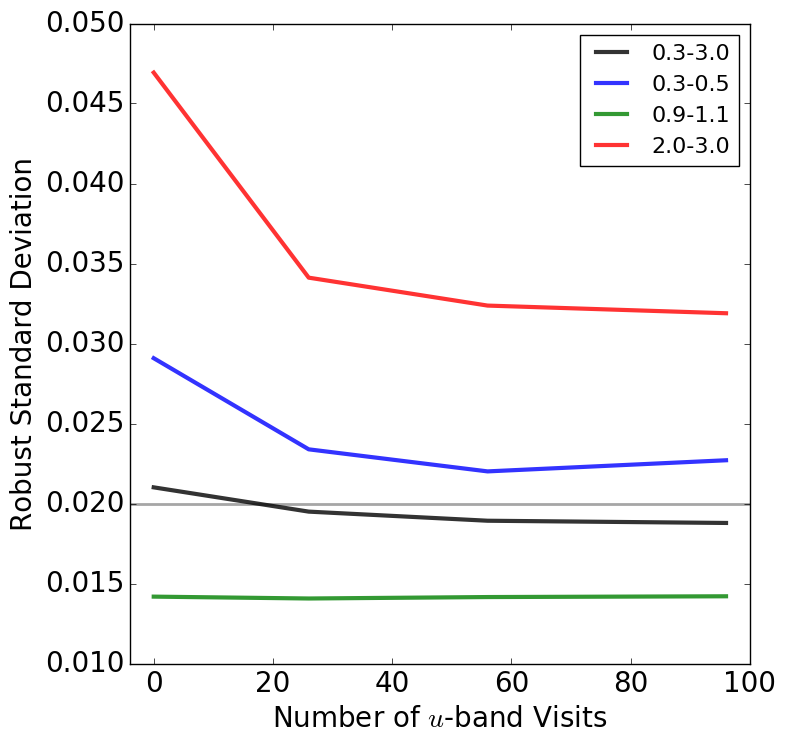}
\includegraphics[width=5cm]{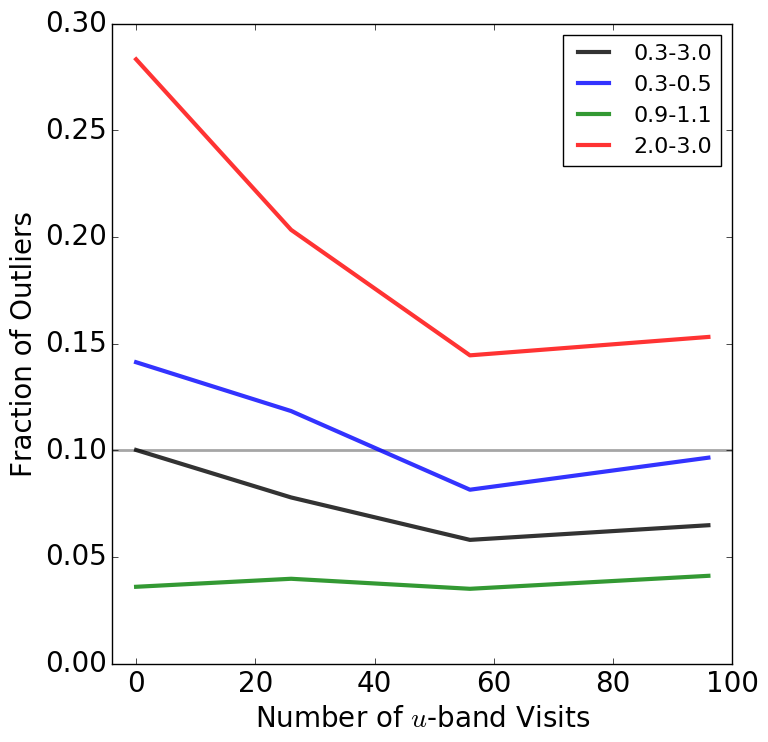}
\includegraphics[width=5cm]{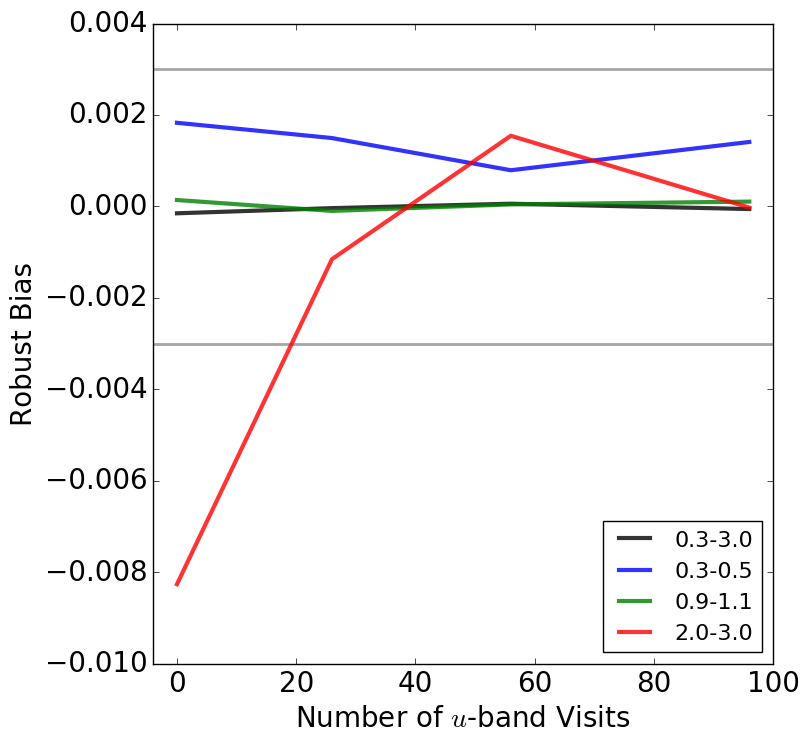}
\caption{Three photo-$z$ metrics as a function of LSST parameters. From
left to right, the y-axis is the robust standard deviation, the fraction of
outliers, and the robus bias. The top row shows these statistics as a function
of the number of years of LSST survey, and the bottom row shows them as
a function of the number of $u$-band visits. Colors show these relations
for four bins in redshift: 0.3--3.0 (black), 0.3--0.5 (blue), 0.9--1.1
(green), and 2.0--3.0 (red). Grey lines mark the SRD specification for
each statistical measure.
\label{fig:metrics}}
\end{center}
\end{figure}

\subsection{Discussion}

\textbf{Additional considerations for observing strategy.} As mentioned
above, overall image depth and signal-to-noise is our primary concern,
so we are not testing changes in e.g., the inter-night gap time or the
exposure time of individual visits.  Our software is instead focused on
modifying other LSST parameters such as systematic offsets to the
magnitudes in each filter and/or coefficients for the magnitude
uncertainties in each filter in order to simulate improvements or
degradations the system throughput, sky background brightness, and other
such factors. We also aim to test airmass distributions (i.e., changes
to the effective filter functions), different progression rates for
filters (e.g., a scenario in which we complete all $u$-band by year 2),
scenarios in which some areas of sky have better/worse coverage at any
given time, and so forth. In all respects we are open to suggestions
from the community, and direct interested readers towards Graham et al. (2017; in prep.).
% Should be submitted by the end of May 2017, and then the reference can be updated.

\textbf{Considerations for building the real training catalog.} All
photometric redshift algorithms require training set data consisting of
objects with secure spectroscopic redshifts.  For LSST, many of these
will be contained in a small number of training/calibration fields (e.g.
COSMOS, VVDS).  Imaging these fields to full depth in all six bands
early in the survey (but under the range of observing conditions
expected for the ten year survey) will be key to characterizing
performance.  Inclusion of these patches of full-depth imaging must be
included in any cadence design. Future simulations of photo-$z$ results
can include varying the quality of the training catalog obtained by
LSST.

\textbf{Integration with MAF.} One way to extend our program to be able
to evaluate observing strategies simulated with \OpSim could be to use
the MAF to enable us to simulate representative samples of galaxies
across the mock LSST sky, and compute the metrics we have defined.
It may be possible to avoid such a large computation by first defining
some intermediate diagnostic metrics, such as the $u$-band coverage, and
working out how our higher level metrics depend on them, using some
approximate interpolation formulae.

\textbf{Connecting to the Dark Energy Figure of Merit.} The metrics we
have defined here should be able to be related to the DETF Figure of
Merit, but because photo-zs affect all of the LSS, WL and CL
cosmological probes, this step may need to wait until a joint
Figure of Merit MAF metric is developed.

% --------------------------------------------------------------------
%
 \subsection{Conclusions}

 Here we answer the ten questions posed in
 \autoref{sec:intro:evaluation:caseConclusions}:

 \begin{description}

 \item[Q1:] {\it Does the science case place any constraints on the
 tradeoff between the sky coverage and coadded depth? For example, should
 the sky coverage be maximized (to $\sim$30,000 deg$^2$, as e.g., in
 Pan-STARRS) or the number of detected galaxies (the current baseline 
 of 18,000 deg$^2$)?}

 \item[A1:] Since increasing the areal coverage comes at the expense of depth per pointing, this will adversely affect the photometric redshift quality. We estimate that only ~80\% of the galaxies would then meet the signal-to-noise threshold for photo-z analysis, but that this would be more than offset by the extra area and the final catalog would be ~108\% the size. Although this appears to be a net gain, it would change the redshift range for analysis.

 \item[Q2:] {\it Does the science case place any constraints on the
 tradeoff between uniformity of sampling and frequency of  sampling? For
 example, a rolling cadence can provide enhanced sample rates over a part
 of the survey or the entire survey for a designated time at the cost of
 reduced sample rate the rest of the time (while maintaining the nominal
 total visit counts).}

 \item[A2:] The sampling frequency is not important to photo-z, but building a uniform depth as a function of time would enable early science that relies on photo-z, so long as this depth is distributed across all filters. This distribution probably does not need to be exactly even, but tolerances have yet to be studied (e.g. rolling cadence may lead to seasonal variations that induce large-scale patchiness in the depth). However, sampling to full depth in all of the bands except g would have a negative impact, for example.

 \item[Q3:] {\it Does the science case place any constraints on the
 tradeoff between the single-visit depth and the number of visits
 (especially in the $u$-band where longer exposures would minimize the
 impact of the readout noise)?}

 \item[A3:] Photo-z would be improved if the u-band magnitude uncertainties could be further minimized, so long as this does not take away visits from other filters. If this could be done by longer u-band exposures during single-visits but less overall u-band visits, that would be good for photo-z.

 \item[Q4:] {\it Does the science case place any constraints on the
 Galactic plane coverage (spatial coverage, temporal sampling, visits per
 band)?}

 \item[A4:] Galactic plane coverage is not applicable to photo-z.

 \item[Q5:] {\it Does the science case place any constraints on the
 fraction of observing time allocated to each band?}

 \item[A5:] We find that at least ~20 u-band visits are necessary for the photo-z statistics to meet specifications.

 \item[Q6:] {\it Does the science case place any constraints on the
 cadence for deep drilling fields?}

 \item[A6:] Cadence of the DDF is not applicable to photo-z.

 \item[Q7:] {\it Assuming two visits per night, would the science case
 benefit if they are obtained in the same band or not?}

 \item[A7:] Intra-night filter changes do not affect photo-z.

 \item[Q8:] {\it Will the case science benefit from a special cadence
 prescription during commissioning or early in the survey, such as:
 acquiring a full 10-year count of visits for a small area (either in all
 the bands or in a  selected set); a greatly enhanced cadence for a small
 area?}

 \item[A8:] Photometric redshift analysis would be greatly assisted by acquiring a full 10-yr count of visits in all six bands for a small area during commissioning or early in the survey, especially if these assets are in regions covered by existing spectroscopic surveys.

 \item[Q9:] {\it Does the science case place any constraints on the
 sampling of observing conditions (e.g., seeing, dark sky, airmass),
 possibly as a function of band, etc.?}

 \item[A9:] Preliminary analysis shows that photometric redshifts may benefit by sampling in airmass, especially in the u-band, but a full assessment of the tradeoffs is pending. In general, a more uniform distribution of conditions is a guard against systematics.

 \item[Q10:] {\it Does the case have science drivers that would require
 real-time exposure time optimization to obtain nearly constant
 single-visit limiting depth?}

 \item[A10:] No

 \end{description}

\navigationbar

% ====================================================================

% --------------------------------------------------------------------

%+
% SECTION:
%    supernovacosmology.tex
%
% CHAPTER:
%    cosmology.tex
%
% ELEVATOR PITCH:
%    SNIa cosmology, approach to evaluating dependence of science on cadence
%
%-
% ====================================================================
\clearpage
\section{Supernova Cosmology and Physics}
\def\secname{supernovae}\label{sec:\secname}

\newcommand{\ml}[1]{\textcolor{red}{[{\bf ML}: #1]}}

Lead authors:
\credit{jhrlsst},
\credit{rbiswas4},
\credit{MichelleLochner}

Contributing authors:
\credit{sethdigel},
\credit{RobFirth},
\credit{astrofoley},
\credit{lgalbany},
\credit{pgris},
\credit{ReneeHlozek},
\credit{ivezic},
\credit{saurabhwjha},
\credit{RickKessler},
\credit{AlexGKim},
\credit{aimalz},
\credit{jasonmcewen},
\credit{janewman-pitt-edu},
\credit{hiranyapeiris},
\credit{kponder},
\credit{rlschuhmann},
\credit{astrostubbs},
\credit{msullivan318},
\credit{wmwv}

\subsection{Introduction}
The acceleration of the rate of expansion of the Universe at late times is one of the
most exciting and fundamental discoveries\citep{Riess1998,Perlmutter1999} in recent times.
This discovery was made using Type Ia supernovae (SNIa) as standardizable candles, and
implies that 76\% of the energy density of the Universe is composed of dark energy
\citep{Frieman2008}. Type Ia supernovae are believed to be explosions of white
dwarfs that have approached the Chandrasekhar mass and are disrupted by
thermonuclear fusion of carbon and oxygen.

This section is concerned with the detection and characterization of
supernovae (SNe) over time with LSST and their various scientific
applications. A crucial application is the use of SNIA %and potentially some core-collapse SNe
(like type IIP) to trace
to trace the recent expansion history of the universe, and confront models of the
physics driving the late time accelerated expansion of the universe.

LSST will improve on past surveys by observing a substantial number of well-characterized
supernovae, at high redshift. This large sample is not necessarily tied to the large area of LSST
and can rather be obtained from a relatively small spatial region, such as the Deep Drilling Fields
(DDF) with larger numbers of well-measured light curves due to the long time interval and high volume
at high redshifts.

On the other hand, the Wide-Fast-Deep (WFD) aspect will make the LSST survey
the first to scan a very large area of the sky for SNe. SNe that are detected
and well characterized by the WFD will provide
\begin{itemize}
    \item a large, well-calibrated low redshift sample ($z \lesssim 0.1$) to replace/supplement the current  set of low redshift supernovae from a mixture of surveys. Such a large, clean low redshift sample is crucial in {\emph{providing a longer lever arm for the determination of cosmological parameters from supernovae.}}
    \item  a low and medium redshift ($z \lesssim 0.8$ and peaking at $z \sim 0.4$ ) sample spanning large areas of
the sky and therefore with the ability of {\emph{tracing large scale structure}} in a novel way, particularly due to
the inclusion of estimates radial distances. This will be possible by combining redshift estimates from supernova light
curves in conjunction with photometric redshifts from host galaxies.  Such a sample could also be used to probe the
{\emph{isotropy of the late time universe.}}
    \item This large sample of SNe will also enable further sharpening of our understanding of the properties of the SN population of both Type Ia and core-collapse SNe (see \autoref{sec:transients:SNtransients}). Aside from the science described in \autoref{sec:transients:SNtransients}, this understanding will also be extremely important to the goal of SN cosmology from LSST. When selecting supernovae satisfying specific criteria from observations in magnitude limited surveys, a lack of understanding of the population properties leads to selection biases in SNIa cosmology as well as the steps in photometric classification~\cite{2017ApJ...836...56K,2016ApJ...822L..35S}.
The WFD SN Ia
sample will dramatically increase the size of the sample available to
train such an empirical model, as well as understand the probability of
deviations and scatter from this model. Aside from issues like
calibration which need to be addressed separately, a larger sample of
such well measured SNe is probably the only way to address `systematics'
due to deviations from the empirical model.
\end{itemize}
%The anticipated WFD sample can
%be thought of as consisting of two components:  the low-redshift sample
%which is more likely to be complete, and the higher-redshift sample that
%will be able to constrain evolution.

% --------------------------------------------------------------------

\subsection{Target measurements and discoveries}
\label{sec:\secname:targets}

SNe of different types are visible over observer time scales of about a few
weeks (e.g., type Ia) to nearly a year (type IIP).  During the full
ten-year survey, LSST will scan the entire southern sky repeatedly with
a WFD cadence, and certain specific locations of the sky called the Deep
Drilling Fields (DDF) with special enhanced cadence.

This spatio-temporal window should contain millions
of SNe. However, the actual sequence of
observations by LSST, defined by the series of field pointing as a
function of time in filter bands (along with weather conditions), and
conditions used for detection will determine the extent to which each
SN can be detected and characterized well.  Characterization of the SNe
is at the core of a number of science
programs that use them as bright, abundant objects with empirically
determined intrinsic brightnesses. While type IIP supernovae may prove to be useful standard
candles \citep{Sanders2014}, we will focus this work on the more well-established type Ia SNe.

Ultimately, the study of dark energy using LSST supernovae will be performed by an analysis inferring the parameters of a cosmological model using a sample of supernovae constructed from the LSST observations, and astrophysical models of supernovae that allow one to relate the peak intrinsic brightness to observed properties of the supernova. We should emphasize that while such analyses have been performed on previous SN surveys, the analyses that would be performed on a really large dataset would be different and is currently under study. Estimates of the potential of such surveys have often been measured using figures of merit such as the Dark Energy Task Force figure of merit~\citep{Albrecht2006}. The main goal of such a metric is to estimate the potential of a survey in elucidating the understanding of dark energy. However, our primary aim here is to study the relative impact that different LSST Observing strategies would have on such dark energy analysis, rather than absolute impact, and to communicate the characteristics of the survey that make a strategy better than others. This necessitates studying the quality of observations corresponding to typical individual supernovae or groups of supernovae rather than producing a single output for a survey. Therefore we identify some of the key steps in the supernova analysis which are directly related to observational characteristics of survey, and define metrics in terms of such quantities. In fact, these metrics should be  thought of as a total of scores, where these scores characterize the sequences of observation on small patches of the sky in small time windows. The key steps we have chosen are
\begin{enumerate}
\renewcommand{\theenumi}{\alph{enumi}}
\item The detection of SNe \label{it:detection}
\item Estimating the intrinsic peak brightness of a supernovae and its redshift
\item Photometric classification \label{it:typing}
\end{enumerate}
The efficacy of photometric typing, redshifts and estimation of intrinsic brightnesses
are all dependent on the amount of information available in the observed
light curves of SNe. While these steps are not necessarily independent, it
is useful to think of the requirements on some of these steps separately;
it is not unlikely  that combinations of some of the steps would still be
affected by similar requirements. Further, in the absence of a complete analysis, we opt for certain ad-hoc criteria in determining the threshold for such observations. This would be necessary even if we were to perform an analysis similar to past surveys.

{\emph{Supernova detection}}\\
Supernova light curves consisting of flux measurements at different times are built through photometry
at specific locations on each of the observed images. A finite list of such specific locations is
constructed through a transient detection pipeline studying difference images. In brief, this process
consists of studying subtractions between a  `template' image (coadded over time so that a supernova
flux averages to a small value) and single visit images called `science images' at different times,
after correcting for differences in resolution, observing conditions and pixel registration. In such
difference images, one expects to obtain non zero pixel values at locations of transients including
supernovae, and pixel values at other locations (including locations of static astrophysical
sources) to be
consistent with zero aside from noise. The efficiency of detecting a supernova in a single
exposure
depends on a number of factors, the most significant of which is the signal (brightness of the supernova
in the science image compared to the template) to noise (SNR) in the relevant image.

%Thus, the probability
%of a supernova being detected in at least a fixed number of the images can be calculated from
%transients such as supernovae, with the probability of inferring the presence
%of an object with flux changing in these exposures being related to the change in
%brightness of the object and noise in the image. Combining a set of difference images
%along a supernova light curve results in higher probability of detecting a transient.

{\emph{Supernova classification}}\\
Because LSST will discover significantly more SNe than can be spectroscopically confirmed,
photometric classification of supernova type from multi-band light curves is crucial. While cosmology
with a photometric SNe sample with contamination from core collapse SNe is possible (see for
example
\citet{Kunz2007,Newling2011,Hlozek2012,Knights2013,Bernstein2012,Gjergo2013,Campbell2013,Rubin2015,Jones2016})
,
these methods still benefit from accurate class probabilities from classification algorithms. To
investigate the effect of observing strategy on SNe classification, we use the multi-faceted machine
learning pipeline developed in \citet{Lochner2016}.

{\emph{Estimating intrinsic supernova brightness at peak}}\\
The ultimate goal of using SNe (type SN Ia or
SN IIP) for cosmology requires estimating the intrinsic brightness of each SN at peak. The first (and
sometimes only, depending on the light curve
model) step is fitting the calibrated fluxes to a light curve model with
a set of parameters. According to the ansatz used in SN cosmology, the
intrinsic brightness of SNe is largely determined by the parameters of
the light curve model; hence the uncertainties on the inferred
parameters largely determine the uncertainties on the inferred peak
intrinsic brightness or distance moduli of the SNe. This means the error on the fitted distance
modulus parameter is a useful proxy for the quality of the light curve and the accuracy of
the resulting cosmological inference.

% --------------------------------------------------------------------

\subsection{Metrics}
\label{sec:\secname:metrics}

Since the steps described above are all necessary for the determination of
SN intrinsic brightnesses, a metric for supernova cosmology must
quantify the ability to perform these steps on each supernova of the
sample. To connect this to the output of OpSim, we propose the
following strategy:
\begin{itemize}
    \item Study the sequences of observations in small spatial regions
    of the sky so that the sequences of observations relate to positions
    of astrophysical objects like supernovae. This capability is already
    built into \texttt{MAF} with multiple slicers like the
    \texttt{OpSimFieldSlicer} or the \texttt{Healpixslicer}. For example, in
    \autoref{fig:SN_sampling}, we show such a sequence for a WFD and
    DDF field for a single year.
    \item On each such spatial region, we look at sliding time windows,
    each time window of size about 70 days (corresponding roughly to a
    supernova Type Ia  lifetime starting 20 days before peak and
    extending to about 50 days after peak). As an example, we choose a
    time window around the night=570, which has an MJD value of 49923
    for both the fields (fieldID: 744 and 309) shown in
    \autoref{fig:SN_sampling} and show the time window in
    \autoref{fig:TimeWindow}.
    \item  We assign a metric value that we call \textbf{perSNMetric}
    $PM$ to each of these time windows to estimate the quality of
    observations for a supernova whose rough lifetime matches that time
    window. The prescription for assigning these values to each
    time window defines our metric and should quantify the success of
    the steps mentioned above. We would expect this value to be a
    function of the properties of the sequence of observations and the
    properties of the transients (SN) being studied. $$ PM =
    PM(\rm{observation Sequence, SN properties})$$
    \item We add up the \textbf{perSNMetric} for the time windows to
    estimate the metric values $M$ for the spatial region of the sky
    surveyed. $$M = \sum_i PM_i. $$ This gives us our final metric $M$.
\end{itemize}

\begin{figure}
 \centering
 \includegraphics[width=\textwidth]{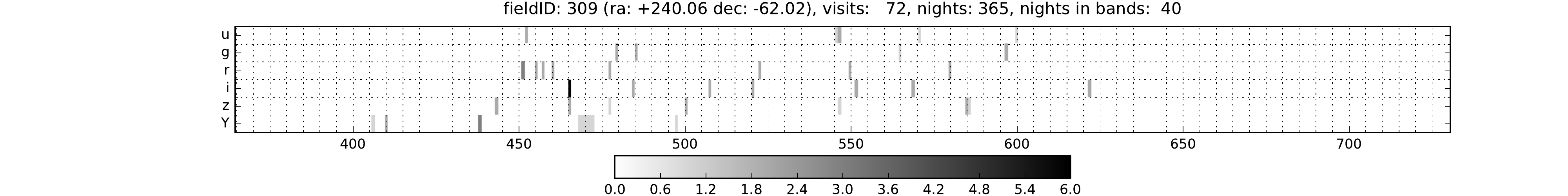}
 \includegraphics[width=\textwidth]{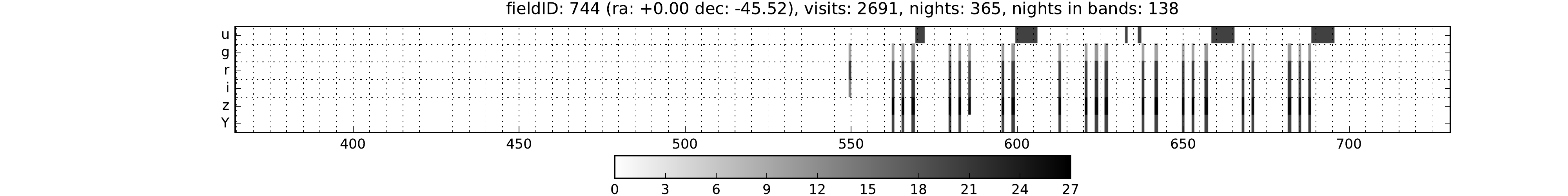}
 \includegraphics[height=0.2\textheight]{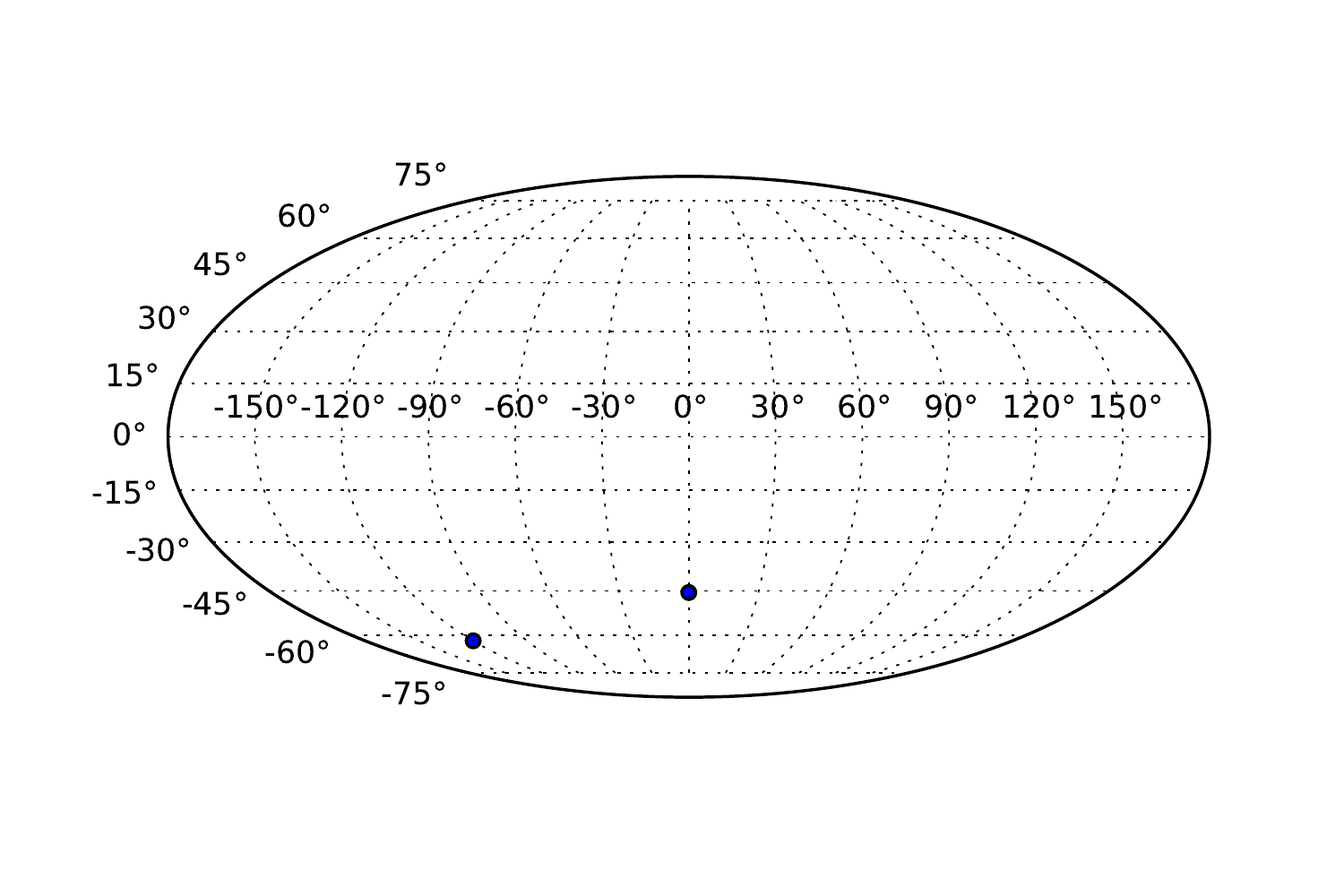}
 \caption{Example of the cadence in the 2nd season in a WFD Field
 (fieldID 309) (top-panel) and a Deep Drilling Field (fieldID 744)
 (middle panel) and the spatial location of these two fields shown on a
 map. The cadence plots show a heatmap of the number of observations per
 night during the second season in each filter u, g, r, i, z, y.
 The header shows the fieldID and location of
 the field, the total number of visits during that period, the number of
 distinct nights on which observations are taken, and the number of
 distinct observations (where observations are considered indistinct
 if they are on the same night and use the same band).}
  \label{fig:SN_sampling}
\end{figure}

\begin{figure}
\centering
 \includegraphics[width=\textwidth]{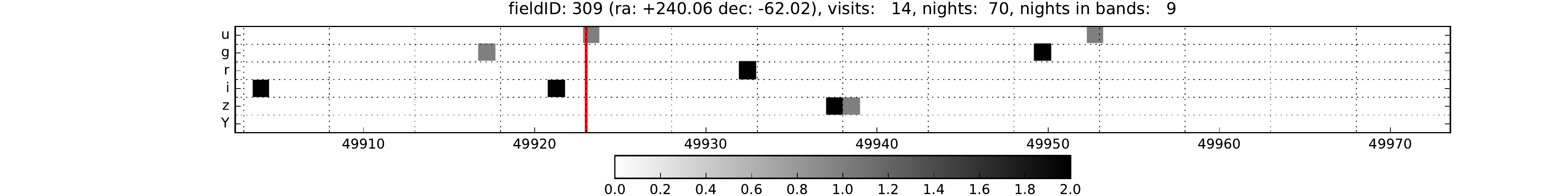}
 \includegraphics[width=\textwidth]{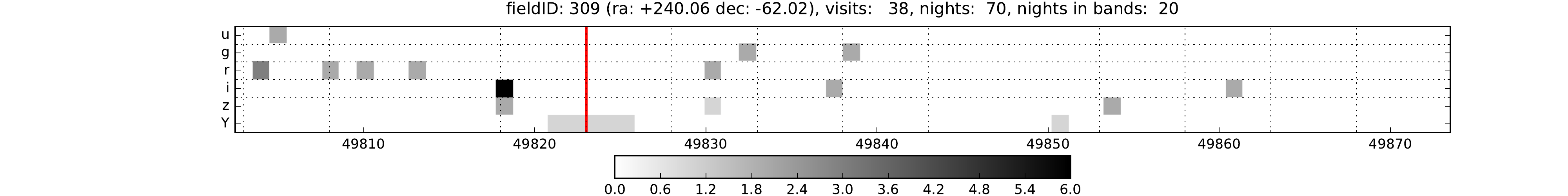}
 \includegraphics[width=\textwidth]{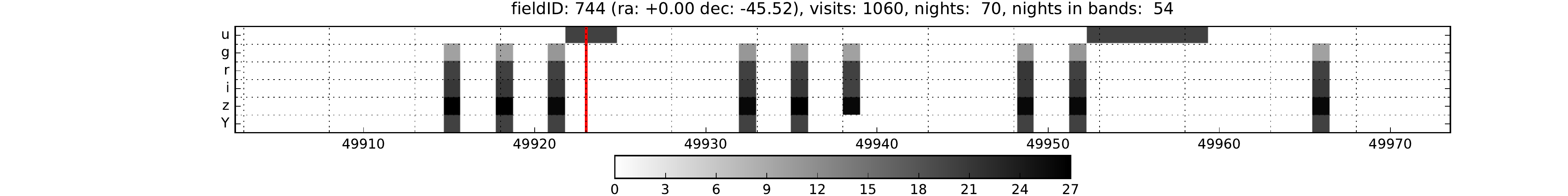}
 \caption{Example of a time window in a WFD Field (fieldID 309)
 (top-panel) and a second time window (middle panel) on the same field,
 and a Deep Drilling Field (fieldID 744) (bottom panel) all extending
 -20 days before and 50 days after a chosen night or MJD. For the Deep
 Drilling Field in the bottom panel, and the WFD field in the top
 panel, the chosen date around which the time window is constructed is
 the MJD of 49923, which is also 570 nights into the survey and marked
 by a red vertical line (which can be used to compare the location to
 \autoref{fig:SN_sampling}. The middle panel shows a window in Field
 309 centered around an MJD of 49823 or a night of 470 which may also be
 compared to \autoref{fig:SN_sampling}. The plots again show the
 heatmap of observations in each filter in each night as in the cadence
 plots of \autoref{fig:SN_sampling}.}
  \label{fig:TimeWindow}
\end{figure}

To define the metric $M,$ we need to define the perSNMetric. Two
different approaches to defining the perSNMetric for a given OpSim run are possible: a) Use
a simulated supernovae Type Ia with specific parameters, observed with
the sequence of observations in the above time window, and evaluate the
success of each step. b) Study heuristics of the observation sequences by using large simulations
with randomized parameter values. Here we will discuss the simpler approach (a).

%The SN metric in a spatial region
%reflects the contribution of the sample of SNe observed in that spatial
%region towards inferring the cosmological parameters. Let us
%consider a case where each SN observed with conditions better than a
%certain threshold contributes equally to the inference. Then the relevant
%metric would be a function of the number of SN in the sample passing
%such selection criteria. More generally, when the quality of all the
%supernovae are not similar, the metric should be thought of as
%the weighted sum of supernovae, with the weights being related to the
%inverse of the effective variance of the distance modulus:
%begin{equation}
%M\sim \sum_i w_i , \qquad  w_i \sim 1.0 /\sigma^2_\mu.
%\end{equation}
%By comparing with the form of the perSNMetric, we see that the
%perSNMetric should be a proxy for $1.0/\sigma^2_\mu,$ where
%$\sigma^2_\mu$ is the effective variance on the distance
%modulus of the supernova, as determined by fitting an empirical model to the supernova light curve.

%\subsubsection{ Steps in the PerSNMetric}
\subsubsection{Steps in the PerSNMetric}
\label{sec:persnmetric}
As described before, the measurement of the distance modulus is the
result of several steps. Therefore, we expect the perSNMetric to be a
product of metrics in each of the steps:

\begin{equation}
PM_i = \prod_{\rm{steps}} PM_i^{\rm{steps}}
\end{equation}

These components of perSNMetric constructed in different steps are
described in \autoref{tab:stepsAndMetrics}.
\begin{center}
 \begin{table}
\begin{tabular}{| p{5cm} |p{10cm}| }
\hline Metric & Description \\
\hline
I. SN discovery  (SNDM) &  Given the observations in a time window corresponding to the lifetime of
a supernova, evaluate the  probability of detecting a
transient. \\
II. SN classification (SNCM) & Given the observations in a time window corresponding to the
lifetime of a
supernova, evaluate the probability of accurately classifying the transient as a type Ia.\\
III. SN light curve characterization quality (SNQM) & Given the observations in a time window
corresponding to the lifetime of a supernova, evaluate the quality of characterization.\\
\hline
\end{tabular}
\caption{Components of the perSNMetric}
\label{tab:stepsAndMetrics}
\end{table}
\end{center}

\emph{I. Discovery Metric}

This metric is designed to gauge the performance of detection of SNe
discussed in \autoref{sec:\secname:targets}.
This metric is a proxy for the potential for a supernova to be detected
 during its lifetime by the set of images taken in different bands by LSST. The actual
 detection of SN during LSST is likely to use more stringent criteria, leading to
 smaller numbers of supernovae in order to deal with possibly large numbers of false positives in detection. A larger
 number of images taken at a time when the supernova is bright enough increases the
 probability of detection. Technically, assuming that a single detection in any of the images containing
 the supernova is sufficient to trigger photometry at the location, one can find the
 probability of detection from an SNR vs. efficiency of detection curve. The signal-to-noise ratio
(SNR) can be determined given properties of a supernova (redshift, intrinsic brightness etc.)
 and the five sigma depth provided in OpSim. While such a SNR-efficiency curve does not
 yet exist for the LSST pipeline, one can use such a curve from previous surveys, in particular a
 SNR-efficiency curve constructed during a stage of the Dark Energy Survey for g, r, i, z bands of DES
 \citep{Kessler2015}. This is shown in \autoref{fig:SNR_detection}.
\begin{figure}
 \centering
 \includegraphics[width=\textwidth]{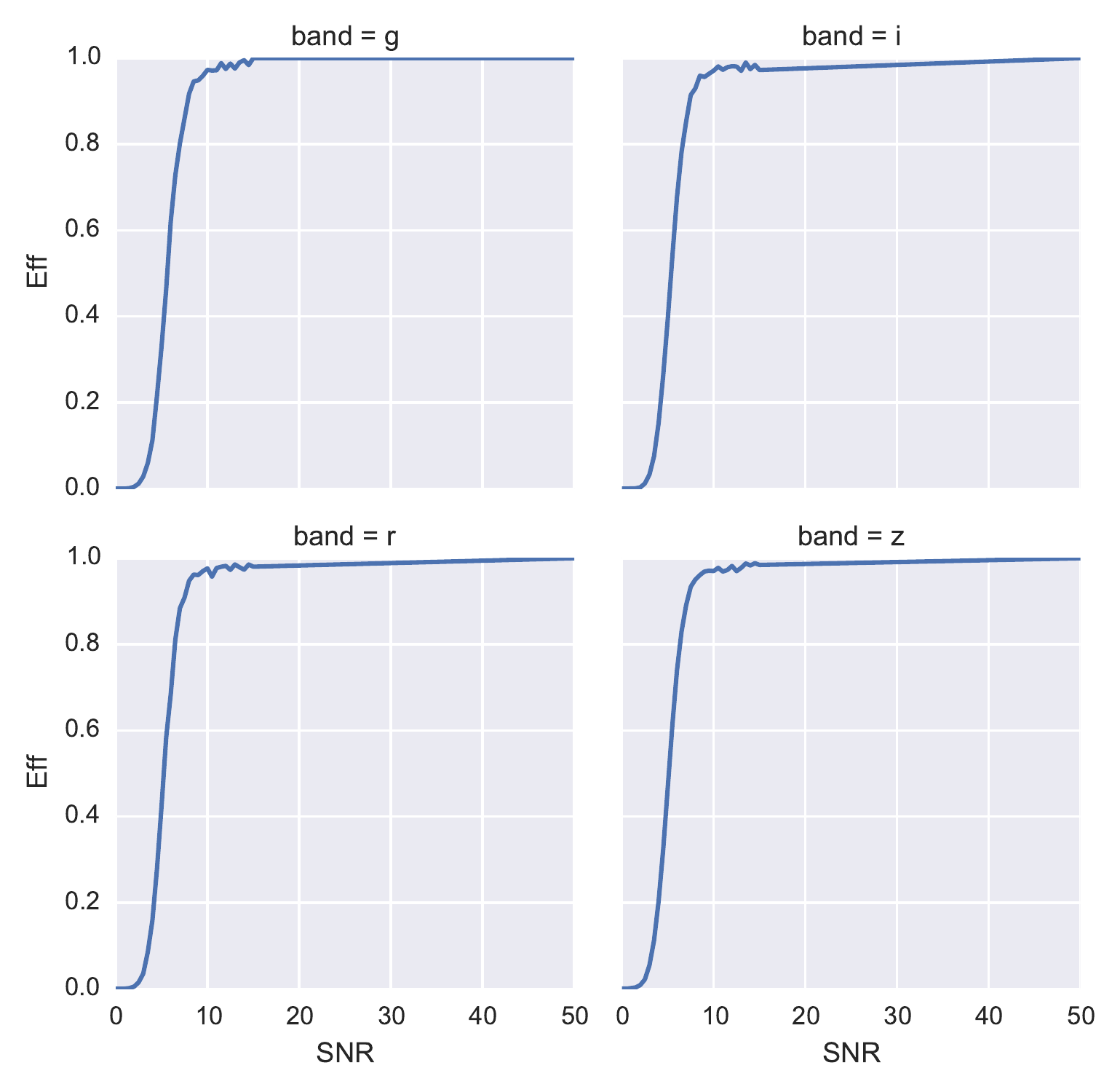}
 \caption{Probability of detecting a transient from a single difference image in different bands as
a
 function of the signal-to-noise as obtained from Dark Energy Survey \citep{Kessler2015}. It shows,
that
for high SNR greater than $\sim 10,$ a single exposure used in difference imaging may be sufficient to detect the SN, while for
lower SNR, several such image differences may be necessary to have a high probability of
detection.}
 \label{fig:SNR_detection}
\end{figure}

Using this information, one can compute the probability that a SN with given properties
will be discovered if from a set of images where the SN have a known signal to noise ratio. We use this as the
value for the supernova discovery metric (SNDM). While very high redshift (and thus faint)
supernovae will not be discovered, it is
potentially possible to discover many supernovae whose light curves will in turn not be well
characterized or hard to classify.

\emph{II. Classification Metric}

Separating supernovae from other detected transients is being considered in
\autoref{chp:transients}. Here we concern ourselves with problem of classifying subclasses of
SNe. Multiple techniques have been proposed to solve this problem \citep{Frieman2008,sako2008, kessler2010b,
ishida2012, sako2014} and it is not yet clear how the
relative success of these techniques are affected by observing strategy. Work is ongoing to use the
multifaceted, machine learning pipeline developed in \citet{Lochner2016} to compare alternative
observing strategies. As this pipeline employs a variety of different feature extraction and machine learning
techniques, it is ideal to investigate the effect of observing strategy on the supernova classification. The exact
metric used to determine the efficacy of the classification depends
on the exact problem at hand. For producing a general purpose, well-classified set of all types of
supernovae (for example, to study supernova population statistics), one could use the AUC metric
used in \citet{Lochner2016}, which is a good balance between purity and completeness. Ideally,
one would like a metric to be evaluated per object and included in the PerSNMetric. This is,
however, challenging as the success of classification will depend strongly on the nature of the
available training set, which will in turn depend on the spectroscopic follow-up program of LSST
and the availability of additional training data. Once a training set is determined and the
classification algorithm trained, a useful per-object metric is the classification probability of an
object being a Ia, which will be higher if the light curve is well-measured. This probability is
confounded by other factors (for example, other types appearing similar to Ia's), making
classification a very difficult metric to include on the per-object level. This is therefore left for future work.

\emph{III. Quality Metric}

We construct the quality metric for the perSNMetric by obtaining the
light curve of the SN in the time window described above. We fit the
light curve, using the SALT2 model \citep{Guy2007,2014A&A...568A..22B}, and approximately estimate the uncertainty in
distance from
the light curve fit alone. Of course, as is well known, luminosity
distance estimates of supernova Type Ia also show an intrinsic scatter
of around $0.1$ in previous surveys, which may be expected to decrease
with better training samples and understanding of underlying
correlations of SN Ia properties and their environments. We compute a
quality metric for each SN Ia as the ratio of the square of the
intrinsic dispersion to variance of the distance indicator from the supernova.
$$ QM = 0.05^2/\sigma^2_{\mu}.$$ If our sample had a perfect discovery rate, and good classification (for example if we had spectroscopic classification), the uncertainty on
cosmological parameters would be entirely due to this quality metric and would be expected to scale with the quality metric as
$$\frac{1}{\sigma} \sim \sqrt{\sum{\frac{1.0}{1.0 + 1.0 / QM}}}$$
where the sum is over the SNe in the sample.

%Move this to OpSim analysis
% The quality metric evaluated on the example SN plotted is $1.0$ if observed in the deep field
% (\autoref{fig:SNIaLCopsimdeep}), and $0.002$ in the WFD field (\autoref{fig:SNIaLCopsimmain}).

% --------------------------------------------------------------------

\subsection{OpSim Analysis}
\label{sec:\secname:analysis}
The scientific goal of characterizing SNe is, to a large extent, dependent
on how well the light curves of individual SNe are sampled in time and
filters. To study this, we re-index the OpSim output on spatial
locations rather than use the temporal index. Here, we first illustrate
in terms the cadence in two example LSST fields.

% {\bf Analysis, Results and Discussion}

% \begin{figure}[tbh!]
% %\vskip -1.3in
% \includegraphics[angle=0,width=0.99\hsize:,clip]{figs/SN_309_lcavg.pdf}
% %\vskip -1.3in
% \caption{Time-interval averaged light curve of
% \autoref{fig:SNIaLCopsimmain}. The light curve shows only a small number
% of the data points, which is insufficient to classify this object as a
% Type Ia and may be also difficult to classify this object as a SN. }
% \label{fig:SNIaLCopsimmain2}
% \end{figure}

% \begin{figure*}[!hb]
%     \begin{minipage}[b]{\linewidth}
%         \includegraphics[width=\textwidth]{figs/supernova/fig_firstSeason_0}
%         \includegraphics[width=\textwidth]{figs/supernova/fig_firstSeason_1}
%         \includegraphics[width=\textwidth]{figs/supernova/fig_firstSeason_2}
%         \includegraphics[width=\textwidth]{figs/supernova/fig_firstSeason_3}
%         \includegraphics[width=\textwidth]{figs/supernova/fig_firstSeason_4}
%     \end{minipage}
% \label{fig:opsimSummary}
% \caption{Cadence of Observations in the timewindow of a year towards a few sample of
% positions. Grey-scale indicates the number of visits. {\it add details}
% }
% \end{figure*}

We analyzed the OpSim output of the Baseline Observing Strategy,
enigma$\_$1189$\_$sqlite.db{\footnote
{\url{http://ops2.tuc.noao.edu/runs/}}} which includes Deep Drilling
Fields (DDF) and the main survey (WFD). While this is no longer the current baseline observing
strategy, we do not anticipate our conclusions would change with the use of
\texttt{minion\_1016}. Future work will include repeating our analyses with new OpSim runs, such as
\texttt{kraken\_1043}, which does not enforce visit pairs, and the new experimental rolling
cadences.

Using the OpSim output \texttt{enigma\_1189}, we generated
light curves for a type Ia supernova with a redshift of z=0.5 at a few different
locations. A date in MJD is chosen where the LSST simulated data are
reasonably well populated. We used the SALT2-extended model
with $x_0$, $x_1$ and $c$ set so that the SN Ia would have a specific
magnitude of $-19.3$ in the rest frame BessellB band. This was performed
using a version of \texttt{SNCosmo} to interpolate the SALT2 surfaces, and the
LSST catalog simulation package to calculate the flux for LSST
bandpasses.

\autoref{fig:SNIaLCopsimdeep} shows the light curve in
different filters in a deep drilling field. The
number of visits for 50 days (which is the period of the simulated SN Ia light curve in
rest-frame, which translates to 75 days at z=0.5) is 53 per filter. For this light
curve, the supernova quality metric (SNQM) and the
discovery metric (SNDM), are both equal to 1. SNDM=1 indicates that this object is a transient that
will be definitely discovered, and SNQM=1 indicates that the light
curves will be of high quality enough to contribute extremely well to
the inference of cosmological parameters. The light curves and
quantified metric demonstrate that data from Deep Drilling Fields would
generate high quality light curves, allowing a high rate of supernova
discovery.

In contrast, \autoref{fig:SNIaLCopsimmain} shows a light curve from the WFD survey. This light
curve is
generated in Field 290 and has an average number
of data points in the light curve of 2 per filter. Using these light
curves, the probability (SNDM) of detecting this supernova is less
than 0.1. \autoref{fig:perSNCadence} directly compares the light curves and cadences of the two
fields considered, from the DDF and WFD.

\begin{figure}
%\vskip -1.3in
%*\includegraphics[angle=0,width=0.99\hsize:,clip]{figs/SN_290_lc.pdf}
\centering
\includegraphics[angle=0,width=14truecm]{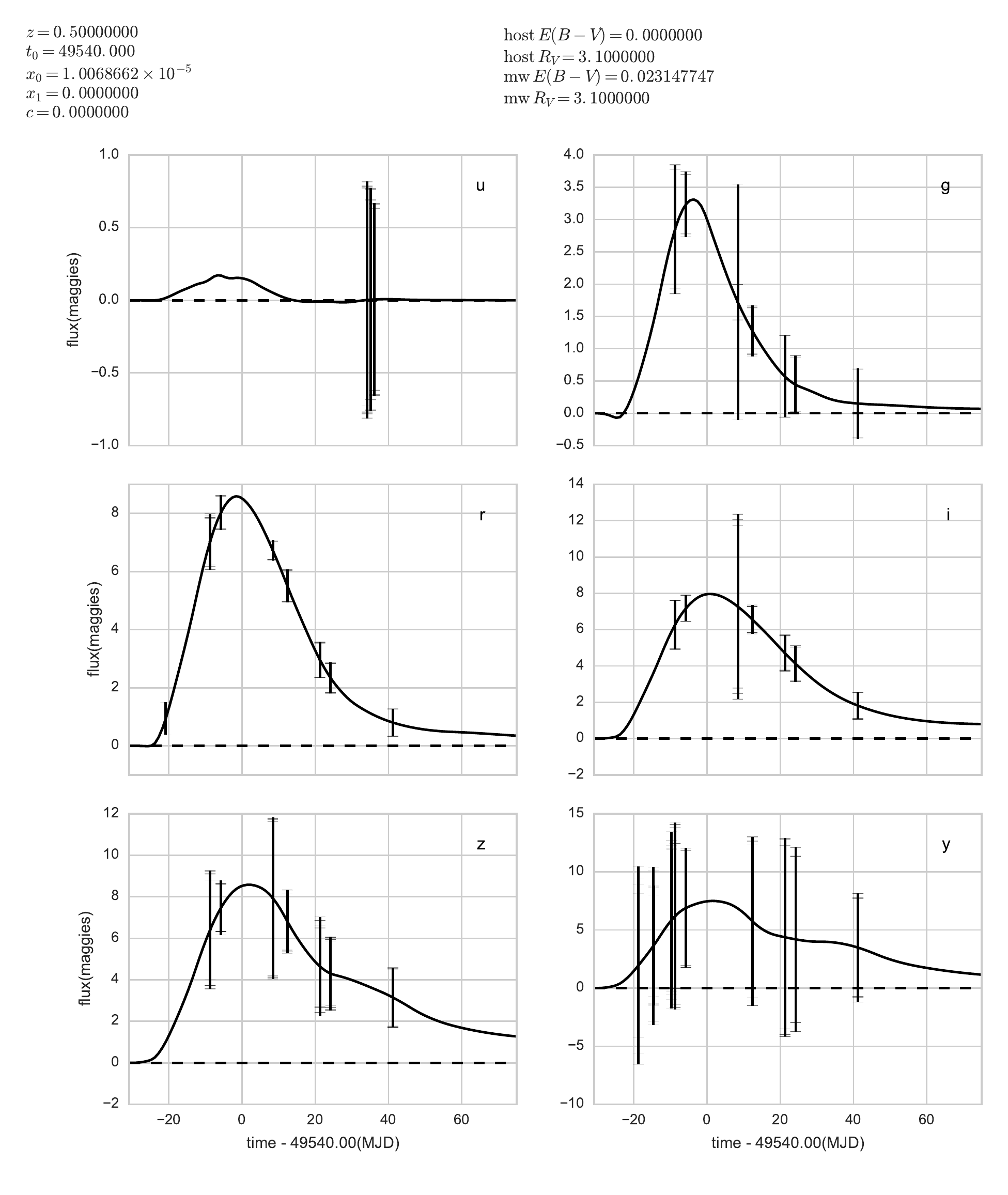}
%\vskip -1.3in
\caption{An example of a light curve, in six filter bands, of a SN Ia from the DDF field with fieldID 290 in
\texttt{enigma\_1189}.
}
\label{fig:SNIaLCopsimdeep}
\end{figure}

\begin{figure}
%\vskip -1.3in
\centering
\includegraphics[angle=0,width=0.99\hsize:,clip]{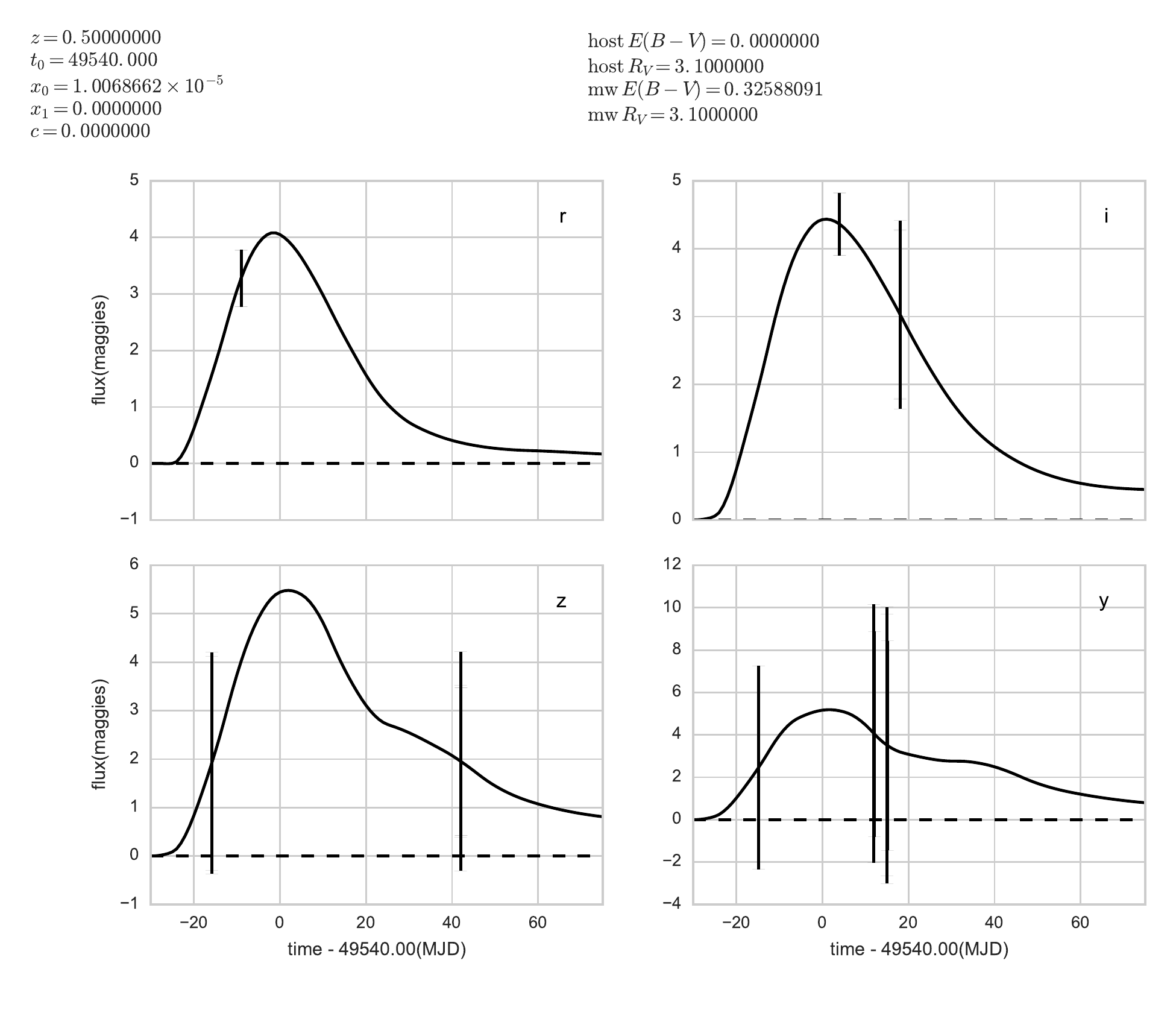}
%\vskip -1.3in
\caption{An example of a light curve, where only four filter bands are available, of a SN Ia from
the WFD survey in
\texttt{enigma\_1189}.
}
\label{fig:SNIaLCopsimmain}
\end{figure}

\begin{figure}
\centering
\includegraphics[angle=0,width=\textwidth,clip]{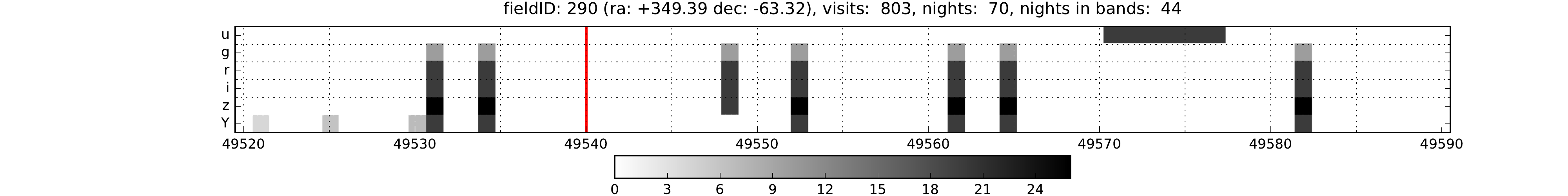}
\includegraphics[angle=0,width=\textwidth,clip]{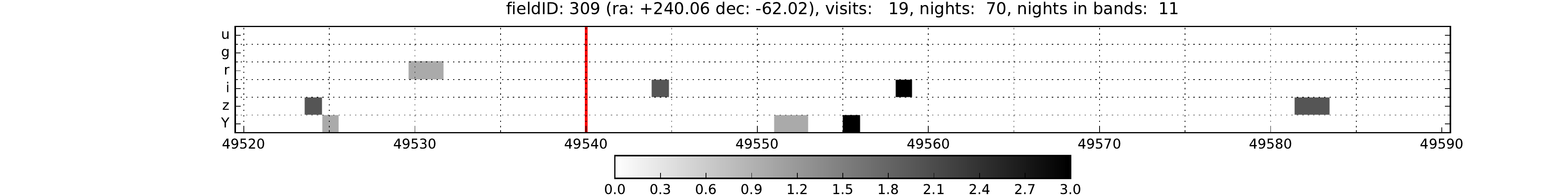}
\caption{Cadence of observations in the time window of a representative
supernova at redshift of $z=0.5$ in a DDF (top) field (fieldID: 290) and
a WFD (bottom) field (fieldID: 309). The red lines show the date of
peak, and the shades show the number of observations in a night in
a distinct filter.}
\label{fig:perSNCadence}
\end{figure}

{\bf Analysis Results:}
To further study the quality of light curves across the survey, we simulate the same type Ia
supernova in 16 different fields, including DDF 290 already studied, from the \texttt{enigma\_1189}
OpSim run and record the average number of visits per 50 day time window
(\autoref{tab:lcpositions}). A well-known rule of thumb (R. Foley, priv. communication) for good
quality SN light curves is to
demand 7-10 epochs per light curve spread over 50 days or so for more than one filter.
\autoref{tab:lcpositions} list characteristics of light curves towards 15 fields in the OpSim
output.

We summarize results of light curve analysis using OpSim output as follows:
\begin{itemize}
\item A high portion of the light curves, in both the WFD and DDF, can be identified as transients
using  the SN
discovery metric (SNDM$>$0.5 in \autoref{tab:lcpositions}).
This metric is currently using SNR of detections.
\item Light curves from WFD show poor qualities (SNQM$<$0.3) whereas the light curves from DDF show
high quality ($\geq$1). Note that we examined only a few cases of SNe at redshift of 0.5 and the
metric will be applied to a much larger range of SNe in future simulations.
\item We have explored ways to improve SN classification \citep{Lochner2016} and the classification
metric is described in Section
\ref{sec:\secname:metrics}.
In this analysis, we have not yet considered classification based on individual light curves.
This is because, unlike the discovery and quality metric, it is difficult to discuss
classification of a single object without having a well-defined training set, which is yet to be
determined for LSST.
\item In conclusion, based on the SNQM, we have showed that WFD data alone using the Baseline
Observing Strategy are not useful for SN studies.
\end{itemize}

%Although
%the averages in \autoref{tab:lcpositions} are only approximations because the cadence is
%non-uniform, they give a clear indication that with the \texttt{enigma\_1189} observing strategy,
%the WFD will not be useful for SNe studies.

{\bf Need for LSST observing strategy optimization for SN science:}

The results from our OpSim analysis motivate our proposal of a rolling cadence
in order to improve the sampling of SNe over a much larger area than the DDFs.
The DDF are useful for SNe at redshifts between 0.6 and 1.2, and the WFD will help us to increase
the number of SNe at low redshift
as shown in Figure 11.1 of LSST science book \citep{2009arXiv0912.0201L}. For example, a higher
number of SNe at z $=0.4$ is expected from WFD than DDF.
Consequently, DDF will have only a few SNe at low redshift,
since the DDF sample is limited to small patches of sky.
A significant scatter in distance modulus of low redshift SNe is found \citep[e.g.][]{Suzuki2012} which hinders deriving accurate
cosmological parameters.
Since the cosmology measurement requires low as well as high redshift distance moduli, a large
sample
of low redshift SNe will improve the cosmological parameter inference. The WFD would also provide a
useful sample of nearby supernovae to better constrain variations in SNe populations (for both
type Ia and core-collapse SNe).
Light curves from WFD
with reasonable quality can be achieved using our new proposed
observing strategy.

A larger sample of well-characterized light curves from the WFD can address two key science
goals:
\begin{itemize}
\item Tightly constraining cosmological parameters ($\Omega_M$ and $\Omega_\Lambda$)
by improving the measurement of distance modulus of low redshifted SNe. Without implementing
the new observing strategy that we recommend below, LSST will not be able to
make a significant contribution to SN cosmology below z$<$0.5.
\item The WFD offers a unique opportunity for isotropy studies (complementary to large-scale galaxy
surveys) and dynamical dark energy studies. Apparent luminosities and number counts can be
estimated for a large range of angular scales and
redshifts with WFD. However, this science case will only be possible with a large sample of
well-characterized SN Ia in a large field of view, providing further support for our
recommendation of a rolling cadence observing strategy.
\end{itemize}

% [ML] I've removed this (fairly arbitrary) chategorization, it's more confusing than helpful
% and it's clear how bad WFD is from the numbers alone

% In column 5, we categorize the positions into 5 categories
% base on the data points (the value in column 3) per filter for 50 days.
% When the number of LSST data points (the value in column 3) per filter
% for 50 days is $>$9 as the category A, 5-9 as B, 1.8 -- 5 for C, 1 --
% 1.8 for D, and $<$1 for E. An example of light curve of Category A has
% been shown in Figure \ref{fig:SNIaLCopsimdeep}, Examples light curves of
% Category B have been shown in Figures \ref{fig:SNIaLCopsimmain} and
% \ref{fig:SNIaLCopsimmain2}, respectively. An example of a light curve of
% Category C (ex. Dec.=-40 ;No. 7 in Table \ref{tab:lcpositions}) is shown
% in \autoref{fig:SNIaLCminus40}.

% \new{An example of Category B toward -66$^{\circ}$ is shown
% \autoref{fig:SNIaLCminus66} (No. 3 in \autoref{tab:lcpositions}). The
% light curves in {\it u, g} and {\it r} bands have 3 -- 5 data points,
% while {\it i} band has 8 or more than data points. The band z and y show
% relatively large error bars. In order to increase probability to
% recognize the light curve as a SN and Type Ia, simulated LSST data
% points could be increased by a factor of 2-3. Our goal is to improve the
% light curves which belong to Category B and C so that we can distinguish
% them as a SN and particularly as a Type Ia SN. This will improve values
% in SNDM and SNQM.}

\begin{center}
\begin{table}
%\tabletypesize{\scriptsize}
\centering
\begin{tabular}{|p{1.3cm} |p{3.3cm}|p{4cm}|p{1.9cm}|p{1.7cm}|p{1.7cm}|}
%\begin{tabular}{|p{0.9cm} |p{3.3cm}|p{4cm}|p{1.9cm}|}
\hline
 FieldID & (R.A., Decl.) & Number of LSST visits per year (u,g,r,i,z,y)       & Number of Avg. data
per SN period per filter & SNDM  & SNQM\\
\hline
290 DDF  & (349.386,-63.321)  & 2363 (398,229,402,414, 522,396) & 53 & 1.0  & 1.6 \\
 17      & (190,-83) &    239 (38,41,41,44,33,42) & 5.3 & .. & ..\\
% 290  &(20,-83) & 252 (52,56,40,21,37,44)  &5.7 &&\\
 22      &(20,-83) & 252 (52,56,40,21,37,44)  &5.7 &..&..\\
%old 3      &(120.012,-71.879) &  220(36,38,37,32,44,33) & 5.0 & B &\\
  217      &(116,-66) &  220 (36,38,37,32,44,33) & 5.0  & .. & ..\\
% 290(+DDF)  &(116,-66) &  220 (36,38,37,32,44,33) & 5.0  && 1.0 & 8.28\\
 309      &(240.05,-62.02) &101 (2,5,11,19,19,45) & 2.2 & 1.00 & 0.10  \\
 645      &(120,-50)  &80 (4,7,9,18,24,18)        & 1.8 &0.54 & 0.01\\
 949      & (80,-40)  &      96 (5,8,15,17,27,24) &  2.2 & 0.99 & 0.09\\
 948      & (280,-40) &      86 (4,2,6,4,24,18)   &  2.0 & 0.99 & $<$0.002\\
 1401      & (58,-27)   & 98 (3,3,9,26,31,26) &   2.2 & 0.77 & 0.002\\
 1754      & (30,-20)  &      86 (3,4,10,21,27,21) &  2.0 & 1.00 & 0.05\\
 1720      & (100,-20) &      58 (4,2,6,4,24,18)   &  1.3 & 0.99 & 0.03 \\
% 000  & (358.41,+0.18)  &                      &     & C({\it 0.1}) & \ref{fig:SNIaLCDecp18}\\
 2556  & (6.097, -1.105) & 80 (4,7,9,18,24,18)  & 1.8 & 1.00 & 0.05\\
% 290  & (6.097, -1.105) &                      &     & C({\it 0.1}) & \ref{fig:SNIaLCDecp18}\\
 2718     & (50,+1.5) &      72 (3,6,10,12,22,19) &  1.6 & 1.00 & 0.09\\
 2751     & (320, +5) &      7 (0,0,2,0,4,0)      &  0.2 & 0.09 & 0.02\\
% 2910     & (60,+5)   &      66 (0,7,11,20,28,0)  &  1.5 &.. & ..&\\
 3606     & (60,+20)  &      72 (0,8,13,22,29,0)  &  1.6 &..& ..\\
% 3959     & (60,+30)  &      44 (0,5,6,15,18,0)   &  1.0 &&\\
\hline
\end{tabular}
\caption{Table of 15 fields in the OpSim run \texttt{enigma\_1189}. The first column is simply an
index, with the special example fields of the DDF field 290 indicated. The
position of the fields is shown in column 2. The third column contains the total and per filter
band number of visits per year and this is averaged per filter per 50 day time window in column 4.
It can be seen that with this observing strategy, only the deep drilling fields are suitable for
supernova cosmology, where 7-9 data points per filter band is considered adequate quality. SNDM and
SNQM are SN discovery and quality metric, respectively, in the last two columns. Note that the
metric measurements are missing for some entries due to a spatial mismatch in fields that will be
rectified in future.}
\label{tab:lcpositions}
\end{table}
\end{center}

\subsection{Discussion}
For \texttt{enigma\_1189} (and also, most likely, for the current baseline strategy), the DDF
fields will produce an exquisite sample of
well-characterized SNe for cosmology and astrophysics studies. Further analysis is required to
determine exactly how many (useful) supernovae will be detected and what the resulting cosmological
constraints will be, but in this section we have discussed and motivated several important
intermediate metrics.

\subsubsection{Scientific motivation for rolling cadence}
It is clear from the above analysis, that the WFD component of the LSST survey will not be useful
for supernova cosmology. However, with some changes to the observing strategy, it is likely that a
large part of the WFD can be leveraged by implementing a rolling cadence strategy. The idea is to
sample a particular field with much higher cadence, at the expense of other fields, for a period of
time (such as 50-100 days) and change fields throughout the survey to preserve uniformity by the end
of the 10 year period.
%Ideally, we would propose changing the filter every day, and trebling the
%average WFD cadence in these smaller fields.
Our aim is to achieve 7-9 points per light
curve over a 50 day period, which would produce light curves of reasonably high quality, that
can be well-classified and well-characterized for cosmological parameter estimation.

\subsubsection{Proposed observing strategy optimized for SN cosmology}

We propose a new observing strategy that provides a dense sampling in time, to improve the
observing strategy of the WFD survey in order to produce SN light curves which are
more useful for SN studies.
We suggest to increase the sampling rate by a factor of 2 or more, and we choose the best option for the sampling rate
to be enhanced by a factor of 3 ($\times$3 hereafter).
%We use an increase of the sampling rate by a factor of 3 for our proposed Cadence, hereafter.
An example of light curve generated with current Universal cadence
and with $\times$3
enhanced light curve is shown in Fig. \ref{fig:LCrandom}.
The enhanced light curve is generated by drawing randomly from a uniform distribution over the
length of the light curve and predicting the flux from the underlying Ia model.

\begin{figure}
\centering
\includegraphics[width=13truecm]{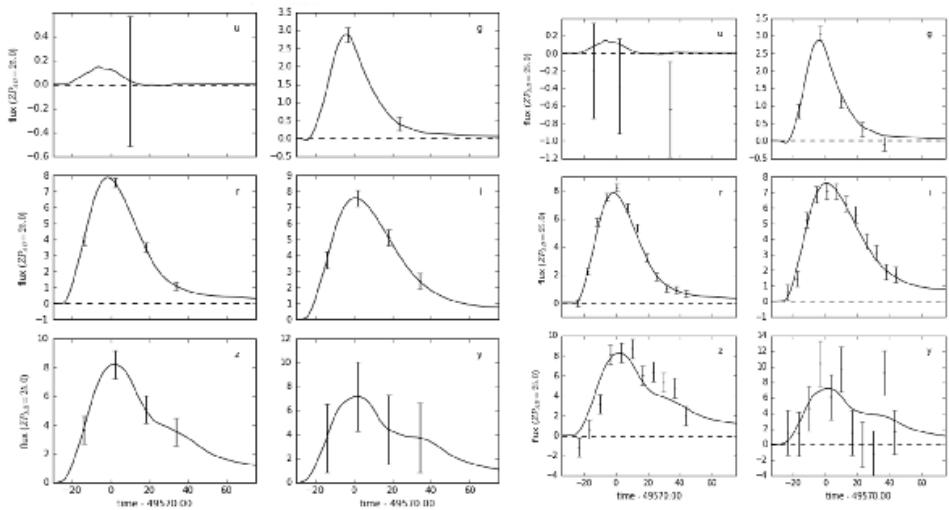}
\caption{An example of light curve of SN Type Ia generated using OpSim output
at RA. of 58$^{\circ}$ and Decl. of -27$^{\circ}$ (left) and the same light curve
after the sampling rate is enhanced by a factor of 3 (right).}
  \label{fig:LCrandom}
\end{figure}

\subsubsection{Details of rolling cadence of WFD optimized for SN cosmology }
%Ideally, we would propose changing the filter every day, and trebling the
%average WFD cadence in these smaller fields. Our primary goal is to  achieve our goal of 7-9 points per light
There are likely many possible LSST observing strategies one could propose to achieve our
goal of increasing the sampling of light curves by a factor of three. We hope to investigate
multiple OpSim runs in the future with a variety of implementations of rolling cadence.
In this section, we show a plausible observing strategy
that can achieve a factor of 3 increase in sampling rate in order to obtain reasonable-quality SN light curves.

The observing strategy of WDF is defined (see Section 1.6.2 in the LSST science book) as follows.

\noindent $\bullet$ A revisit time of three days on average per 10,000 deg$^2$ of sky (i.e., the area visible at any given time of the year), with two visits per night (particularly useful for establishing proper motion vectors for fast moving asteroids).

We define the 10,000 deg$^2$ survey area (the visible sky for a given time of the
year) simply as the ``visible sky".
In order to increase the sampling rate by a factor of 3, we propose to make the revisit time of $\sim$1 day (1/3 of three days above)
on average visible
sky. This means one third of the visible sky ($\sim$3300 deg$^2$) can be chosen. Preference
may be given to the part of
sky with low air-mass, but at the same time
a uniform coverage of LSST needs to be considered.
Ideally we would propose a filter change every day, and observe approximately the same
patch of visible sky for
$\sim$ 50 days before
going to another part.
As a result,
 the light curves will have  a sampling rate ($\times$3) in time.
A total of 1.4$\times$10$^5$ SNe Ia (Section 11.2.2 in \citet{2009arXiv0912.0201L}) expected from
WDF would become a factor of 2-3 lower
(details are to be investigated in future), but
the SNe that LSST WFD discovered will have  meaningful, reasonable-quality light curves which can be used to
classify the types of SNe and to improve cosmological parameters.

\subsection{Conclusion}

Here we answer the ten questions posed in Sub-section 1.2.1:

\begin{description}
\item[Q1:] {\it Does the science case place any constraints on the
tradeoff between the sky coverage and coadded depth? For example, should
the sky coverage be maximized (to $\sim$30,000 deg$^2$, as e.g., in
Pan-STARRS) or the number of detected galaxies (the current baseline
of 18,000 deg$^2$)?}

\item[A1:] For supernova observations, the most important aspect is cadence: we need a
good sampling of well-measured light curves of supernovae. If the sky
coverage can be increased without sacrificing the number of observations
per supernova (i.e., over one season), such as using a rolling cadence
strategy, this will greatly improve supernova science. Inasmuch as
requiring increased sky coverage could negatively affect cadence over one
season, our preference is not to increase to a larger sky coverage.

\item[Q2:] {\it Does the science case place any constraints on the
tradeoff between uniformity of sampling and frequency of  sampling? For
example, a rolling cadence can provide enhanced sample rates over a part
of the survey or the entire survey for a designated time at the cost of
reduced sample rate the rest of the time (while maintaining the nominal
total visit counts).}

\item[A2:] Frequency of sampling is much more important than uniformity over long time
scales for all forms of supernova science. For SN Ia cosmology, light-curve
sampling should be about three times as frequent as for the current
baseline WFD cadence. While further investigation is needed, a reasonable
length of a season with enhanced rates is around 120-150 days. A rolling
cadence will allow this improved sampling while still keeping the sky
coverage and co-added depth, by concentrating on different fields in
different seasons (see Section 9.5 for details).

\item[Q3:] {\it Does the science case place any constraints on the
tradeoff between the single-visit depth and the number of visits
(especially in the $u$-band where longer exposures would minimize the
impact of the readout noise)?}

\item[A3:] In general for supernova science, the number of visits is more important to
the single-visit depth, though we would not advocate any decrease in
single-visit exposure time (below 2x15 sec). The U band  is useful for the
low-z (wide-area) sample and and their calibration of the Hubble diagram.
It is not obvious if longer exposures are helpful at this time for
supernova science.

\item[Q4:] {\it Does the science case place any constraints on the
Galactic plane coverage (spatial coverage, temporal sampling, visits per
band)?}

\item[A4:] Our supernova science is extragalactic and will use high Galactic-latitude
fields almost exclusively (to minimize Milky Way extinction systematic
uncertainties). Survey time spent on Galactic fields takes away from survey
time for our science; we would optimize for less time in the Galactic
plane.

\item[Q5:] {\it Does the science case place any constraints on the
fraction of observing time allocated to each band?}

\item[A5:] For supernova science, we have no strong preference for any band and would
benefit from roughly uniform depth per band. While redder bands are more
useful for higher redshift supernovae, the bluer bands are helpful
especially for lower redshift objects. Visits in each band should be
uniformly spread in time. A higher number of visits to redder bands is
preferred for DDF.

\item[Q6:] {\it Does the science case place any constraints on the
cadence for deep drilling fields?}

\item[A6:] Our optimal cadence would for the DDFs would be nightly or every other
night; every three nights would be acceptable. Beyond this we run the risk
of large gaps more than $\sim$7 days, which would be detrimental to supernova
science. Ideally, the DDFs would be observed in all available filters,
though not all filters are needed every night (so for instance some filters
could be observed on night 1, the others on night 2, and repeat). The
single-night depth should be 1-2 mag deeper than a WFD field, but not much
more; any extra time is better spent on a different night.

We would like a suitable number of extragalactic DDFs in total so that at
least a few ($\sim$3 to 5) are being observed at any time in the survey: we
would limit an increase in the number of DDFs if it came at the expense of
our preferred high cadence. Each field should be observed for a $``$season" of
length at least $\sim$120-150 days, staggering the fields so new fields cycle in
as old ones cycle out. In addition, for DDFs, the u-band exposure time
could be increased to minimize readout noise.

\item[Q7:] {\it Assuming two visits per night, would the science case
benefit if they are obtained in the same band or not?}

\item[A7:]Supernova science benefits enormously from having color information,
particularly in discovery and early epochs to help classification. Two
visits in the same band will be less useful than visits in different bands.

\item[Q8:] {\it Will the case science benefit from a special cadence
prescription during commissioning or early in the survey, such as:
acquiring a full 10-year count of visits for a small area (either in all
the bands or in a  selected set); a greatly enhanced cadence for a small
area?}

\item[A8:] During commissioning, we would like as many DDFs as possible observed in
all filters (with few times the per-night exposure time) so that templates
can be created for each of the DD fields we consider. This will allow us to
find useful supernovae in these DDFs as soon as the survey starts, and
provide us with useful time to deal with any issues in the image
subtraction etc. before the survey begins.

Early in the survey, we must build templates for all fields (more broadly
than just the DDFs). Our favored method for doing this is to devote the
bulk of year 1 to covering the full sky area, with a rolling cadence survey
commencing thereafter. As above, the templates should be at least a few
times the exposure time of a single visit (so that the template noise does
not dominate the subtraction). This is particularly important in the first
3-5 weeks of the survey, to ensure that we get useful data from the rest of
the critical first year.

\item[Q9:] {\it Does the science case place any constraints on the
sampling of observing conditions (e.g., seeing, dark sky, airmass),
possibly as a function of band, etc.?}

\item[A9:] While supernova observations benefit from good seeing, it is not as strong
a requirement as it is for other science cases. Moreover we would like to limit
any image coaddition issues for very good seeing (under the assumption that
these remain): we can make good use of observing conditions with slightly
poorer seeing.

\item[Q10:] {\it Does the case have science drivers that would require
real-time exposure time optimization to obtain nearly constant
single-visit limiting depth?}

\item[A10:]A sufficiently long (or nominal), fixed exposure time should be adequate
for supernova science. No real-time optimization is necessary, and
single-visit science is not the limiting factor for SNe science.

\end{description}

\navigationbar

%\end{document}

% --------------------------------------------------------------------

% ====================================================================
%+
% SECTION NAME:
%    lenstimedelays.tex
%
% CHAPTER:
%    cosmology.tex
%
% ELEVATOR PITCH:
%    Lensed quasars and supernovae provide distance measurements for
%    cosmology. They are a few days to a few weeks in length. To
%    measure them well we need long campaigns (>~3 years) with high
%    night-to-night cadence (better than the standard 5 days if
%    possible, especially as combining all filters might be difficult.)
%
% AUTHORS:
%   Phil Marshall (@drphilmarshall)
%-
% ====================================================================

\section{ Strong Gravitational Lens Time Delays }
\def\secname{lenstimedelays}\label{sec:\secname}

\credit{drphilmarshall},
\credit{rhiannonlynne},
\credit{tanguita}

The multiple images of strongly lensed quasars and supernovae have
delayed arrival times: variability in the first image will be observed
in the second image some time later, as the photons take different
paths around the deflector galaxy, and through different depths of
gravitational potential. If the lens mass distribution can be modeled
independently, using a combination of high resolution imaging of the
distorted quasar/SN host galaxy and stellar dynamics in the lens
galaxy, the measured time delays can be used to infer the ``time delay
distance'' in the system. This distance enables a direct physical measurement
of the Hubble constant, independent of the distance ladder.  \citet{TM16} provide a recent review of this cosmological probe, including its the main systematic errors and observational follow-up demands. High resolution follow-up imaging and spectroscopy is needed to constrain the lens galaxy mass distribution, and this is expected to dominate the systematic error budget in systems with good measured time delays. In tis section we investigate the cadence needed for LSST to provide these good time delays, in lensed quasar systems. We leave the assessment of lensed supernova time delays to future work.

% --------------------------------------------------------------------

\subsection{Target measurements and discoveries}
\label{sec:\secname:targets}

For this cosmological probe to be competitive with LSST's others, the
time delays of several hundred systems (which will be distributed
uniformly over the extragalactic sky) will need to be measured with
bias below the sub-percent level, while the precision required is a
few percent per lens.  In galaxy-scale lenses, the kind that are most
accurately modeled, these time delays are typically between several
days and several weeks long, and so are measurable in monitoring
campaigns having night-to-night cadence of between one and a few days,
and seasons lasting several months or more.

This size of sample seems plausible: \citet{OM10} predicted that several thousand lensed quasar systems should be detectable at LSST single visit depth and resolution, and \citet{LiaoEtal2015} found that around 400 of these should yield time delay measurements of high enough quality for cosmography.
To obtain accurate as well as precise lensed quasar time delays, several
monitoring seasons are required. Lensed supernova time delays have not
yet been measured, but their transient nature means that their time
delay measurements may be more sensitive to cadence than season or
campaign length.

% --------------------------------------------------------------------

\subsection{Metrics}
\label{sec:\secname:metrics}

Anticipating that the time delay accuracy would depend on night-to-night
cadence, season length, and campaign length, we carried out a large
scale simulation and measurement program that coarsely sampled these
schedule properties. In \citet{LiaoEtal2015}, we simulated 5 different
light curve datasets, each containing 1000 lenses, and presented them to
the strong lensing community in a ``Time Delay Challenge.'' These 5
challenge ``rungs'' differed by their schedule properties, in the ways
shown in \autoref{tab:tdcrungs}. Entries to the challenge consisted of samples of measured time delays, the quality of which the challenge team then measured via three primary diagnostic metrics: time delay accuracy, time delay
precision, and useable sample fraction (\ie the number of lenses that could be measured well, divided by the number of simulated lenses in the dataset). The accuracy of a sample was defined to be the mean fractional offset between the estimated and true time delays within the sample. The precision of a sample was defined to be the mean reported fractional uncertainty within the sample.

Focusing on the best challenge
submissions made by the community, we derived a simple power law model
for the variation of each of the time delay accuracy, time delay
precision, and useable sample fraction, with the schedule properties
cadence, season length and campaign length. These models are shown in
\autoref{fig:tdcresults}, reproduced from \citet{LiaoEtal2015}, and are
given by the following equations:
\begin{align}
|A|_{\rm model} &\approx 0.06\% \left(\frac{\rm cad} {\rm 3 days}  \right)^{0.0}
                          \left(\frac{\rm sea}  {\rm 4 months}\right)^{-1.0}
                          \left(\frac{\rm camp}{\rm 5 years} \right)^{-1.1} \notag \\
  P_{\rm model} &\approx 4.0\% \left(\frac{\rm cad} {\rm 3 days}  \right)^{ 0.7}
                         \left(\frac{\rm sea}  {\rm 4 months}\right)^{-0.3}
                         \left(\frac{\rm camp}{\rm 5 years} \right)^{-0.6} \notag \\
  f_{\rm model} &\approx 30\% \left(\frac{\rm cad} {\rm 3 days}  \right)^{-0.4}
                        \left(\frac{\rm sea}  {\rm 4 months}\right)^{ 0.8}
                        \left(\frac{\rm camp}{\rm 5 years} \right)^{-0.2} \notag
\end{align}

%%%%%%%%%%%%%%%%%%%%%%%%%%%%%%%%%%%%
\begin{table*}
\begin{center}
\capstart
\begin{tabular}{cccccc} \hline\hline
  Rung &  Mean Cadence & Cadence Dispersion & Season   & Campaign & Length   \\
       &  (days)       & (days)             & (months) & (years)  & (epochs) \\ \hline
  0    &    3.0        &   1.0              &   8.0    &    5     & 400      \\
  1    &    3.0        &   1.0              &   4.0    &    10    & 400      \\
  2    &    3.0        &   0.0              &   4.0    &    5     & 200      \\
  3    &    3.0        &   1.0              &   4.0    &    5     & 200      \\
  4    &    6.0        &   1.0              &   4.0    &    10    & 200      \\
\hline\hline
\end{tabular}
\end{center}
\caption{The observing parameters for the five rungs of the Time Delay
Challenge. Reproduced from \citet{LiaoEtal2015}.\label{tab:tdcrungs}}
\end{table*}
%%%%%%%%%%%%%%%%%%%%%%%%%%%%%%%%%%%%

%%%%%%%%%%%%%%%%%%%%%%%%%%%%%%%%%%%
\begin{figure*}[!ht]
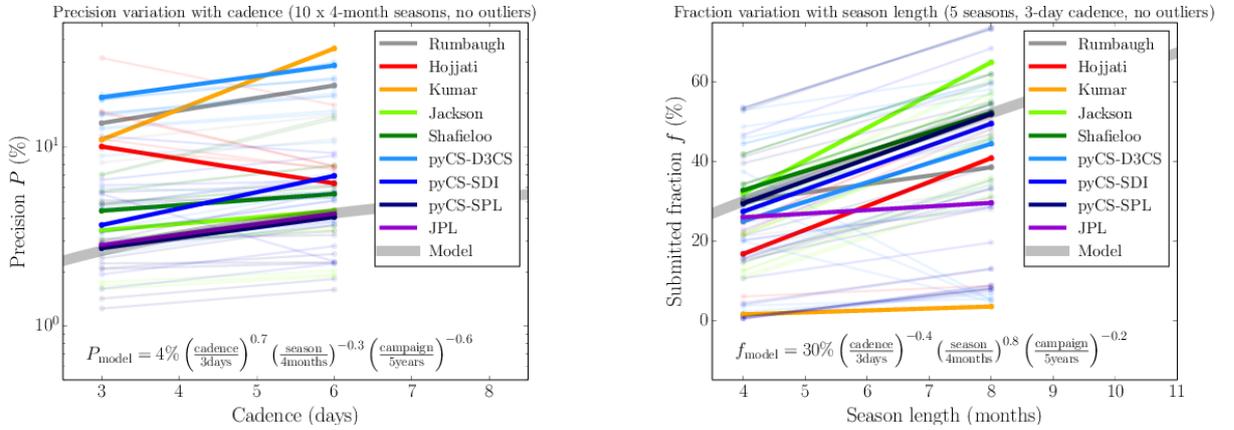

  \capstart
  \begin{minipage}[b]{\linewidth}
    \begin{minipage}[b]{0.48\linewidth}
      \centering\includegraphics[width=\linewidth]{figs/Precision_cadence_nca.pdf}
    \end{minipage} \hfill
    \begin{minipage}[b]{0.48\linewidth}
      \centering\includegraphics[width=\linewidth]{figs/Fraction_season_nca.pdf}
    \end{minipage}
  \end{minipage}
\caption{Examples of changes in precision $P$
(left) and success fraction $f$ (right) with schedule properties, as
seen in the different TDC submissions. The gray approximate power law
model was derived by visual inspection of the pyCS-SPL results; the
signs of the indices were pre-determined according to our expectations.
Reproduced from \citet{LiaoEtal2015}.}
\label{fig:tdcresults}
\end{figure*}
%%%%%%%%%%%%%%%%%%%%%%%%%%%%%%%%%%%

All three of these diagnostic metrics would, in an ideal world, be
optimized: this could be achieved by decreasing the night-to-night
cadence (to better sample the light curves), extending the observing
season length (to maximize the chances of capturing a strong variation
and its echo), and extending the campaign length (to increase the number
of effective time delay measurements).

The quantity of greatest scientific interest is the {\it accuracy in
cosmological parameters}: this could be computed as follows. Setting a
required accuracy threshold  defines the available number of lenses,
which in turns gives us the mean precision per lens there. Combining the
whole sample, we would get the error on the weighted mean time delay, as
used by \citet{Coe+Moustakas2009}. This uncertainty, which scales as one
over the square root of the number of available lenses,  can be roughly
equated to the statistical uncertainty on the Hubble constant \citep{Coe+Moustakas2009,TM16}. The
Figure of Merit would be the final percentage precision on $H_0$, as a
way to sum up the sample size and time delay measurability (at fixed
accuracy requirement).

% --------------------------------------------------------------------

\subsection{\OpSim Analysis}
\label{sec:\secname:analysis}

% \OpSim analysis: how good would the default observing strategy be, at
% the time of writing for this science project?

In this section we present the results of our ongoing \OpSim / MAF
analysis, as we start to try to
answer the question ``how good would the proposed observing
strategies be, for time delay lens cosmography?''

\autoref{fig:lenstimedelays:accuracymaps} shows maps of TDC2 time delay
measurement accuracy from our MAF analysis of two \OpSim databases, the
baseline cadence \opsimdbref{db:baseCadence}, and a ``No Visit Pairs''
strategy, \opsimdbref{db:NoVisitPairs}. We use the \metric{TdcMetric} to compare three different
analysis scenarios, differing by a) whether or not we can combine all 6
filters' light curves such that they behave like the TDC2 single-filter
simulations (as was assumed by \citeauthor{LiaoEtal2015}), and b)
whether we wait 5 or 10 years before making the time delay measurement.\footnote{Here, ``years'' means ``seasons:'' we used the
\metric{SeasonStacker} to work
with seasons, rather than calendar years.}

These sky maps saturate at a threshold of 0.04\%, chosen conservatively
to be 5 times stricter than that used by
\citeauthor{Hojjati+Linder2014}. We can see that the area of sky
providing lenses measurable at this accuracy or better increases
markedly as we move from $ri$ to $ugrizy$. These maps predict that while
the accuracy should increase between DR5 and DR10, the sky area yielding
accuracy better than 0.04\% should already be close to the full WFD area
(18000 square degrees) by DR5: this bodes well for our ability to make
use of shorter campaign sky areas observed at higher frequency, as would
emerge from a rolling cadence strategy.

To summarize the diagnostic metric results, we first compute the area of
this ``high accuracy'' ($A < 0.04\%$) sky. We can then compute the cadence, season, and
campaign length just in these areas; these values are reported in
\autoref{tab:lenstimedelays:results}. The high accuracy area can be used
to define a ``Gold Sample'' of lenses, whose mean precision per lens we
can compute. The TDC2 useable fraction averaged over this area gives us
the approximate size of this sample: we simply re-scale the 400 lenses
predicted by \citet{LiaoEtal2015} by this fraction over the 30\% found
in TDC2. While these numbers are approximate, the ratios between
different observing and analysis strategies provide a useful indication of relative merit. In this table, the impacts of higher night-to-night sampling rate and survey length can be seen.\footnote{The small decrease in the number of useable lenses with $ugrizy$ going from 5 years to 10 years is due to the slightly lower precision in 5 years, and is an artifact of how the metrics are calculated (as global means rather than running totals).}

As described above, we follow \citet{Coe+Moustakas2009} and compute a
very simple time delay distance Figure of Merit ``\texttt{DPrecision}''
as follows. We first combine the fractional time delay precision in
quadrature with an assumed 4\% ``modeling uncertainty,'' and then divide
this by the square root of the number of Gold Sample lenses. This
estimated ensemble distance precision can be straightforwardly related
to cosmological parameter precision, as \citet{Coe+Moustakas2009} show
(it's very roughly the precision on the Hubble constant).  This distance
precision Figure of Merit is given in the final column of
\autoref{tab:lenstimedelays:results}. We see that in DR5, being able to
combine all filters instead of just $ri$ should give a FoM of 0.29\%
instead of 1.25\%; between DR5 and DR10 this then should improve to
0.24. This suggests that time delay measurement is primarily {\it
analysis-limited}. In the ``No Visit Pairs'' strategy, we seet that the
DR5 all-band FoM is 0.26\%, just a 10\% improvement. This might be
because the ``No Visit Pairs'' does not {\it insist} that the visits in
the visit pairs are split over different nights, only that they don't
{\it have} to be taken on the same night. We might expect a more
aggressive approach to splitting visit pairs to make a bigger difference --
but it seems unlikely that it would be a factor of two improvement.

%%%%%%%%%%%%%%%%%%%%%%%%%%%%%%%%%%%
\begin{figure*}[!ht]
  \capstart
  \begin{minipage}[b]{\linewidth}
    \begin{minipage}[b]{0.48\linewidth}
       \centering\includegraphics[width=\linewidth]{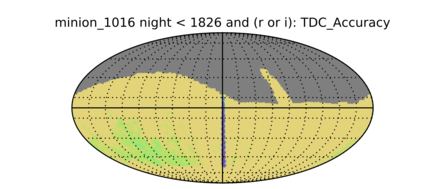}
    \end{minipage} \hfill
    \begin{minipage}[b]{0.48\linewidth}
       \centering\includegraphics[width=\linewidth]{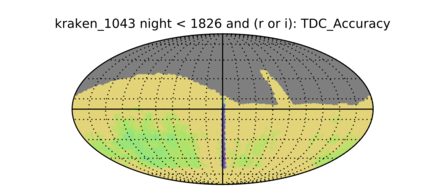}
    \end{minipage}
  \end{minipage}
  \begin{minipage}[b]{\linewidth}
    \begin{minipage}[b]{0.48\linewidth}
       \centering\includegraphics[width=\linewidth]{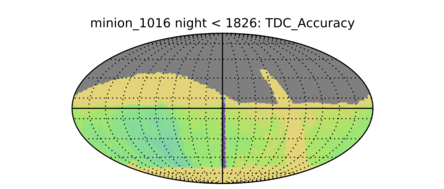}
    \end{minipage} \hfill
    \begin{minipage}[b]{0.48\linewidth}
       \centering\includegraphics[width=\linewidth]{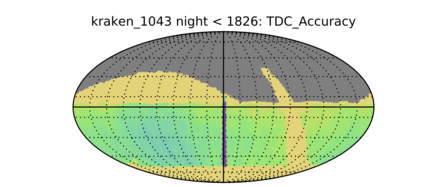}
    \end{minipage}
  \end{minipage}
  \begin{minipage}[b]{\linewidth}
    \begin{minipage}[b]{0.48\linewidth}
       \centering\includegraphics[width=\linewidth]{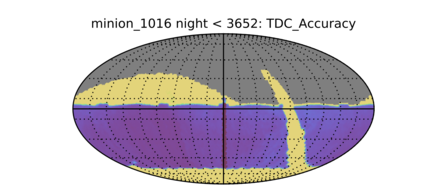}
    \end{minipage} \hfill
    \begin{minipage}[b]{0.48\linewidth}
       \centering\includegraphics[width=\linewidth]{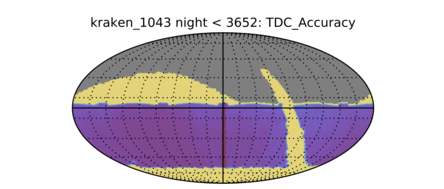}
    \end{minipage}
  \end{minipage}
\caption{Sky maps of the TDC2 time delay measurement accuracy metric $A$ for the baseline cadence \opsimdbref{db:baseCadence} (left) and the ``No Visit Pairs'' strategy, \opsimdbref{db:NoVisitPairs} (right). The rows show the build up of data quality with time and analysis capability, from 5 years of $r$ and $i$-band light curve data only (top row), to 5 years of hypothetically-combined $ugrizy$ light curve data (middle row), to 10 years of hypothetically-combined $ugrizy$ light curve data (bottom row). The maps saturate at the threshold accuracy of 0.04, such that any regions that are {\it not yellow} should yield high accuracy lens time delays, while the yellow regions show where high accuracy time delay measurement is not possible.}
\label{fig:lenstimedelays:accuracymaps}
\end{figure*}
%%%%%%%%%%%%%%%%%%%%%%%%%%%%%%%%%%%

%%%%%%%%%%%%%%%%%%%%%%%%%%%%%%%%%%%%%%

\begin{table*}
\begin{center}
\caption{Lens Time Delay Metric Analysis Results.}
\label{tab:lenstimedelays:results}
\footnotesize
\begin{tabularx}{\linewidth}{ccccccccc}
  \hline
  \OpSim run                       % runName -> db
   & Filters                       % filters
    & Years                        % Nyears
     & \texttt{cadence}            % high_accuracy_cadence
      & \texttt{season}            % high_accuracy_season
       & \texttt{Area}             % high_accuracy_area
        & \texttt{dtPrecision}     % precision_per_lens
         & \texttt{Nlenses}        % N_lenses
          & \texttt{DPrecision} \\ % distance_precision
  \hline\hline
  \opsimdbref{db:baseCadence}
   & $ugrizy$
    & $10$
     & $4.5$
      & $6.9$
       & $19004$
        & $5.09$
         & $468$
          & $0.24$ \\
  \opsimdbref{db:baseCadence}
   & $ugrizy$
    & $5$
     & $5.1$
      & $6.6$
       & $17926$
        & $6.29$
         & $472$
          & $0.29$ \\
  \opsimdbref{db:baseCadence}
   & $ri$
    & $10$
     & $10.4$
      & $5.7$
       & $18566$
        & $7.07$
         & $285$
          & $0.42$ \\
  \opsimdbref{db:baseCadence}
   & $ri$
    & $5$
     & $14.2$
      & $5.5$
       & $4841$
        & $10.81$
         & $75$
          & $1.25$ \\
  \opsimdbref{db:NoVisitPairs}
   & $ugrizy$
    & $10$
     & $3.9$
      & $7.1$
       & $18907$
        & $4.88$
         & $502$
          & $0.22$ \\
  \opsimdbref{db:NoVisitPairs}
   & $ugrizy$
    & $5$
     & $4.5$
      & $6.7$
       & $18093$
        & $5.93$
         & $504$
          & $0.26$ \\
  \opsimdbref{db:NoVisitPairs}
   & $ri$
    & $10$
     & $8.5$
      & $6.1$
       & $18617$
        & $6.34$
         & $329$
          & $0.35$ \\
  \opsimdbref{db:NoVisitPairs}
   & $ri$
    & $5$
     & $10.7$
      & $6.1$
       & $9358$
        & $9.35$
         & $171$
          & $0.71$ \\
   \hline

\multicolumn{9}{p{\linewidth}}{\scriptsize Notes: see the text for
the definitions of each metric.}
\end{tabularx}
\normalsize
\medskip\\
\end{center}
\end{table*}
%%%%%%%%%%%%%%%%%%%%%%%%%%%%%%%%%%%%%%

% --------------------------------------------------------------------

\subsection{Discussion}
\label{sec:\secname:discussion}

The main risk involved with this science case is that the
``multi-filter'' light curve analysis presented here, which extrapolates
from the results of the single-filter TDC1, does not represent
accurately the real-life combination of all 6 filters together. The
second time delay challenge (TDC2) will help answer this question.  For
now, just using 2 filters gives an upper limit on the overall precision
we should expect.

Naively, we would expect the relaxation of the visit pairs requirement
to increase the  night-to-night cadence by a factor of two, if the
visits are redistributed randomly in time. However, it seems \OpSim is
not as liberal as this, such that we do not see much improvement over
the baseline cadence. Efficiency maximization could be preventing visits
being fully split. We are interested in any changes to the WFD survey
time sampling that reduce the inter-night gaps: these  would include
rolling cadence schemes.

% ====================================================================

\subsection{Conclusions}

Based on the above results, we now answer the ten questions posed in
\autoref{sec:intro:evaluation:caseConclusions}:

\begin{description}

\item[Q1:] {\it Does the science case place any constraints on the
tradeoff between the sky coverage and coadded depth? For example, should
the sky coverage be maximized (to $\sim$30,000 deg$^2$, as e.g., in
Pan-STARRS) or the number of detected galaxies (the current baseline
of 18,000 deg$^2$)?}

\item[A1:] Yes: it's probably better to have smaller area and better
data, especially if the multi-filter light curve analysis turns out to
be difficult. This conclusion is partly informed by the apparently small
difference between 5 years and 10 years campaign length, although we
need to be careful: COSMOGRAIL studies show that the longer monitoring
campaigns yield significantly higher accuracy results.

\item[Q2:] {\it Does the science case place any constraints on the
tradeoff between uniformity of sampling and frequency of sampling? For
example, a rolling cadence can provide enhanced sample rates over a part
of the survey or the entire survey for a designated time at the cost of
reduced sample rate the rest of the time (while maintaining the nominal
total visit counts).}

\item[A2:] Yes: higher frequency is better, up to the point where the
visit separation becomes less than one night. We would like to see the
additional visits gained through rolling cadence used to fill in the
nights in the campaign rather than going deeper by taking more visits
per night.

\item[Q3:] {\it Does the science case place any constraints on the
tradeoff between the single-visit depth and the number of visits
(especially in the $u$-band where longer exposures would minimize the
impact of the readout noise)?}

\item[A3:] Yes: we have not investigated this in detail, but it's very
likely that we'd prefer more shallow visits than fewer deeper visits.

\item[Q4:] {\it Does the science case place any constraints on the
Galactic plane coverage (spatial coverage, temporal sampling, visits per
band)?}

\item[A4:] No: only inasmuch as that time spent in the plane could have
been used to improve the night-to-night sampling frequency. The effect
is probably not large.

\item[Q5:] {\it Does the science case place any constraints on the
fraction of observing time allocated to each band?}

\item[A5:] Yes, but: in principle an all $i$-band survey would make time
delay estimation easier. However, since we need the other filters to
assist with lens finding and lens environment characterization, we'd be
reluctant to advocate any move away from the universal $ugrizy$
coverage.

\item[Q6:] {\it Does the science case place any constraints on the
cadence for deep drilling fields?}

\item[A6:] To first order, no.

\item[Q7:] {\it Assuming two visits per night, would the science case
benefit if they are obtained in the same band or not?}

\item[A7:] Unclear: different bands are probably going to be
better, but this has not been tested. It would clearly provide good
information on the AGN color variability model.

\item[Q8:] {\it Will the case science benefit from a special cadence
prescription during commissioning or early in the survey, such as:
acquiring a full 10-year count of visits for a small area (either in all
the bands or in a  selected set); a greatly enhanced cadence for a small
area?}

\item[A8:] Not really. We rely on the difference imaging working well,
so making a good template across as much sky as possible would be good.
Including a known lens system (from DES, perhaps) in any deep field
would be useful too: we would be happy if the cadence to be similar to
WDF to be able to test our software.

\item[Q9:] {\it Does the science case place any constraints on the
sampling of observing conditions (e.g., seeing, dark sky, airmass),
possibly as a function of band, etc.?}

\item[A9:] No.

\item[Q10:] {\it Does the case have science drivers that would require
real-time exposure time optimization to obtain nearly constant
single-visit limiting depth?}

\item[A10:] No.

\end{description}

\navigationbar

% ====================================================================

% --------------------------------------------------------------------

% --------------------------------------------------------------------

\chapter[Special Surveys]{Special Surveys}
\def\chpname{specialsurveys}\label{chp:\chpname}

Chapter editors:
\credit{dnidever},
\credit{knutago}.

% Confirmed leads for LMC/SMC: Knut Olsen, David Nidever

% Confirmed leads for special surveys:

% \section*{Summary}
% \addcontentsline{toc}{section}{~~~~~~~~~Summary}
%
% Executive summary goes here, highlighting the primary conclusions from
% the chapter's science cases. This should be abstract length, no more:
% say, 200 words.

% --------------------------------------------------------------------

\section{Introduction}
\label{sec:specials:intro}

The four main LSST science themes, as defined by the Science Book,
drive the design of LSST's main Wide-Fast-Deep survey.  However, it
has always been recognized that many important scientific projects,
including some that are highly relevant to LSST's main science themes,
are not well served by the areal coverage and/or cadence constraints
placed on the WFD survey.  To this end, the LSST Project set aside a
nominal 10\% of the observing time to serve what are collectively
called ``special surveys.''

Projects that
will certainly make use of this $\approx10\%$ time (that is not dedicated to the WFD
survey) include the Deep Drilling fields and the Galactic Plane surveys,
as well as any survey wishing to
observe at declinations below $-60^\circ$, such as the Magellanic
Clouds.  These special programs have the potential to
heavily oversubscribe the nominal 10\%
of time assigned to them.  It is of thus critical importance for these
programs to define compelling science cases, clearly justify their
observing requirements, and derive metrics to quantify the performance
of a given schedule for the program. This chapter provides a venue for
such investigations.

A minimal set of 4 ``extragalactic'' Deep Drilling Fields have been
included in many of the \OpSim runs to date, including the baseline cadence simulation, \opsimdbref{db:baseCadence} (\autoref{chp:cadexp}), and have
been evaluated in various science sections throughout this paper.
As described in \autoref{sec:intro:timeline}, there will be a further call for Deep Drilling Field proposals, and so we defer discussion of such observing programs to that activity. In this chapter we collect together various ideas for additional special surveys or otherwise unusual observing programs, and the discussion of their metric evaluation.

In \autoref{sec:solar_system_specials} a number of different concepts for special surveys that would support solar system science is described. This is followed by an extensive discussion of short exposure observing at twighlight (\autoref{sec:shortexp}). If short exposure times will indeed be a possibility, it won't just be solar system science that benefits: an example of a small program that could make good use of such a capability is the open star cluster special survey proposed in \autoref{sec:M67_special}. We expect the ideas in this chapter to be developed (or discarded) over the time before the survey starts: this white paper aims to provide a (potentially temporary) home for them during that period.

% Add sections below, one science investigation per section, one
% section per file.

% --------------------------------------------------------------------

% PJM: commented out for now, for lack of content:
% \input{SpecialSurveys/deepdrilling.tex}

% --------------------------------------------------------------------

% PJM: This is currently in the Magellanic Clouds Chapter, but
% could be moved back here soon...
% \input{SpecialSurveys/magcloudsurvey.tex}

% --------------------------------------------------------------------

% ====================================================================
%+
% NAME:
%    section-name.tex
%
% ELEVATOR PITCH:
%    Explain in a few sentences what the relevant discovery or
%    measurement is going to be discussed, and what will be important
%    about it. This is for the browsing reader to get a quick feel
%    for what this section is about.
%
% COMMENTS:
%
%
% BUGS:
%
%
% AUTHORS:
%    David Nidever (@dnidever)
%    Knut Olsen (@knutago)
%-
% ====================================================================

\section{Solar System Special Surveys}
\def\secname{solar_system_specials}\label{sec:\secname}

\credit{davidtrilling},
\credit{rhiannonlynne}.

% This individual section will need to describe the particular
% discoveries and measurements that are being targeted in this section's
% science case. It will be helpful to think of a ``science case" as a
% ``science project" that the authors {\it actually plan to do}. Then,
% the sections can follow the tried and tested format of an observing
% proposal: a brief description of the investigation, with references,
% followed by a technical feasibility piece. This latter part will need
% to be quantified using the MAF framework, via a set of metrics that
% need to be computed for any given observing strategy to quantify its
% impact on the described science case. Ideally, these metrics would be
% combined in a well-motivated figure of merit. The section can conclude
% with a discussion of any risks that have been identified, and how
% these could be mitigated.

%A short preamble goes here. What's the context for this science
%project? Where does it fit in the big picture?

There are several populations of Near Earth Objects (Solar System bodies
whose orbits bring them close to the Earth's orbit) that, because of
their orbital properties, would not be easily detected in the
wide-fast-deep survey. These populations are very interesting for both
scientific and sociological purposes, though, due to their close
proximity to the Earth, and in fact their potential for impacting the
Earth. LSST will have the capability to carry out surveys for these
populations by using a small amount of special survey time. Two of
these special surveys have pointings that fall within the nominal
wide-fast-deep plan, and simply require a modification of the cadence.
The third program is a twilight program, with a special cadence (though
all twilight programs are likely to  have special cadences). These three
programs are listed here and described below. The three special surveys are
the following:

\begin{itemize}
\item A special survey to look for mini-moons, which are temporarily captured
satellites of the Earth;
\item A special survey to find meter-sized impactors up to two weeks prior to impact.
This would allow telescopic characterization of these impactors, which could
be compared to laboratory measurements of the meteorites derived from
the impactor. Advanced warning of an impactor also allows detailed
study of impact physics by being on location when the impact
occurs;
\item A special survey to observe the ``sweetspot'' in twilight fields
to look for NEOs in very Earth-like orbits that would otherwise not
be found in opposition fields.
\end{itemize}

% Need the figure and caption
These surveys will support three important scientific investigations:
\begin{enumerate}
\item What are the properties of the population of objects that is
nearest to the Earth?
\item What is the impact risk from NEOs in populations that
have not yet been well characterized (mini-moons, sweetspot objects)?
\item How do the telescopic properties of an impactor relate to the
laboratory-measured properties of the ensuing meteorites?
\end{enumerate}

%Many different types of objects and measurements with their own cadence
%``requirements'' will fall into these two broad categories (with some
%overlap).  These will be outlined in the next section.
Some details of the special cadence requirements for these
science investigations are described in the following section.

% --------------------------------------------------------------------

\subsection{Target measurements and discoveries}
\label{sec:\secname:targets}

\subsubsection{Special cadences}

Each of the three Solar System special surveys requires a special
cadence. These cadences are described here.

\begin{itemize}

\item{{\bf Mini-moons}}
Mini-moons are objects that are temporary satellites of the Earth
\citep{2014Icar..241..280B, 2017Icar..285...83F}
% bolin et al, icarus, 241, 280 http://adsabs.harvard.edu/abs/2014Icar..241..280B
% fedorets et al. icarus 285 83 http://adsabs.harvard.edu/abs/2017Icar..285...83F
Therefore, they have orbital motions similar to the Earth's moon,
and much faster than other Solar System populations. Therefore,
a special cadence is required to detect these objects enough
times to link objects, create tracklets, and determine orbits.
A suggested cadence for a mini-moon survey is a series
of 3~second exposures, with each pointing visited at least
twice per night. Such a survey would cover essentially
all of the opposition sky each night. The opposition sky should
be re-observed several nights in a row in order to
link objects from night to night and determine their orbits.
While the details of this special cadence are not yet
fully refined, this special survey would likely have little impact
on the overall LSST program since this small-scale
program, which extends over a small number of nights,
is effectively a compressed rolling cadence in which
the aggregate field coverage is unchanged.

\item{{\bf Impactors}}
The Earth is struck by meter-sized impactors about
once a month \citep[\eg][Trilling \etal 2017 submitted]{Boslough2015, 2017Icar..284..416T}.
% boslough et al. 2015 in Aerospace Conference, 2015 IEEE, 1-12
% tricarico 2017 icarus 284 416
% trilling et al 2017 AAS journals submitted
On two occasions, impacting asteroids have
been discovered some hours before impact, but
there are no existing surveys that are dedicated to finding
impactors.
% xxx ATLAS xxx.
Impactors generally have small apparent motions
on the sky (because their orbits are not too different
than the Earth's). The single exposure depth of LSST
images suggests that a meter-sized NEO could be
discovered perhaps a week before impact, given
the typical Earth-relative velocity of such a body
\citep[\eg][]{2017arXiv170506209C}.
% chesley & veres 2017 https://arxiv.org/abs/1705.06209
A suggested cadence for an impactor survey would be
to survey the opposition patch four times per night.
This is more visits than in the nominal cadence, and
would allow high fidelity linking of observations to
find orbits. The nominal wide-fast-deep cadence
(twice per night, three times during a lunation) has
a latency of orbit determination of up to two weeks,
which is not acceptable for the impactor survey, as an
impact would occur on a timescale of just a few days
from discovery.
The cadence of four observations/night should be repeated
roughly every three days, so that an object on an
inbound trajectory could be observed at least once,
and possibly twice, before impact.
Note that this cadence is compatible with
the wide-fast-deep survey, in that the fields and
exposure times are nominal; the only difference is that
each field is visited four times in a night, and that
the fields are revisited every few nights. The overall
impact of this special survey on the wide-fast-deep
survey is likely to be small, and possibly negligible.
Given the importance of this small but
significant investigation, it is critical that the
survey simulators be capable of including such
a special survey in planning for LSST operations.

\item{{\bf Twilight/sweetspot survey}}

NEOs on very Earth-like orbits are relatively
unlikely to come to opposition, and therefore
are relatively unlikely to appear in data
obtained in the wide-fast-deep survey.
These objects are particularly interesting
since, having very Earth-like orbits, they
are the most likely objects to be Earth
impactors.
These objects are most likely to be detected
in a twilight survey that looks at the ``sweetspot'' ---
a location at around 60~degrees Solar
elongation that is only visible at twilight.
Because these sweetspot fields are only visible
for 30--60~twilight minutes each night,
a special
cadence is required to find and link these objects
to determine their orbits.
These observations would be best carried out
in the $z$ filter (because the observations are
made in twilight, when the sky is still relatively
bright). Fields should be revisited at 15~minute
intervals, and each field should be revisited
every other night during this experiment, so that
observations can be linked.
(A long interval
between observations prohibits linking.)
The total experiment
should last roughly one week, so that each
object would have a tracklet on four nights
(nights 1,3,5,7).
During twilight, some 25~pointings could be visited
after the sky is sufficiently dark but
before the fields have set.
Because these observations are made during twilight,
there may be no significant impact on the
nominal wide-fast-deep survey. See \autoref{sec:shortexp} for a more extensive discussion of short exposure and twilight observing in special survey mode.
\end{itemize}

\subsubsection{Measurements}

For each of these three programs, the most important measurement
to be made is the position of any object as a function of time.
In other words, the usual measurements of moving
objects from LSST images is also the requirement for
the source detections for these special surveys. As usual
for Solar System surveys, there is a trade-off of
sensitivity (Solar System objects are most easily
detected in $r$ band) against characterization (observing
a given object in multiple filters yields an estimate
of composition). For these three cases, discovery and
good orbit determination is probably more important than
immediate characterization from LSST measurements,
so the nominal expectation is that
the nighttime special surveys would be carried out in
$r$ band and the twilight program in $z$ band.

% --------------------------------------------------------------------

\subsection{Metrics}
\label{sec:\secname:metrics}

The metrics to be used to determine the efficacy of LSST
at scientific success of these special surveys are identical
to those employed in \autoref{chp:solarsystem}.
The most important of these metrics
include the completeness as a function of size; the
number of detections over a given length of time (for instance,
the one week approach timescale of impactors); and
the quality of the derived orbit. These metrics are defined
in more detail in \autoref{chp:solarsystem}. The important question is:
how much value do the special surveys add?

%
% % --------------------------------------------------------------------
%
% \subsection{OpSim Analysis}
% \label{sec:\secname:analysis}

The current default observing strategy does not include
any of these special surveys. Therefore, the scientific yield,
at this default, is zero. Both the mini-moons and impactor
surveys are relatively small experiments, on the scale of
the LSST project, at something like 10--20~hours total
per instance of the experiment. (The impactor experiment,
for example, might be carried out one or several times a year,
both to build up statistics and to identify further potential
impactors.) Furthermore, the impactors survey cadence
is different from the nominal wide-fast-deep survey,
but could be a simple modification of the nominal wide-fast-deep survey
cadence.

The twilight/sweetspot survey is also not included in
the current baseline \OpSim strategy, and nor are any twilight observations (\autoref{sec:shortexp}).
It is critical to ensure that \OpSim can handle the kind
of dedicated cadences described above in order to assess
the global impact of these small-scale but highly
important Solar System special surveys.

\navigationbar

% --------------------------------------------------------------------

% ====================================================================
%+
% NAME:
%    short_exposures.tex
%
% CHAPTER:
%    specialsurveys.tex
%
% ELEVATOR PITCH:
%
% AUTHORS:
%    Chris Stubbs (@astrostubbs))
%-
% ====================================================================

\section{Short Exposure Surveying}
\def\secname{shortexp}\label{sec:\secname}

\credit{astrostubbs}

The current LSST requirements stipulate a minimum exposure time of 5
seconds, with an expected default exposure time of 15 seconds. This
document advocates for decreasing the minimum exposure time requirement
from 5 to 0.1 seconds. This would increase the dynamic range for bright
sources (compared to the default 15 sec time) by about 5 magnitudes, to
a total of 13 astronomical magnitudes (where dynamic range is the
difference between the brightest unsaturated source and the faintest
point source detectable at 5 sigma). This is a large factor, and would
enable a wide range of science goals, outlined below. One interesting
aspect of this is that it would allow us to operate the LSST system
during twilight times that would otherwise saturate the array due to
background sky brightness. This would allow a number of the goals
described below to be carried out without impacting the primary survey
by conducting observations during twilight sky conditions that would
saturate the array at longer exposure times.

% ----------------------------------------------------------------------

\subsection{Introduction}
\label{sec:\secname:intro}

Since the twilight sky brightness is an important factor discussed
below, we provide here a very brief outline of the temporal evolution of
the background sky brightness.

\citet{1993AJ....105.1206T}
provide a simple framework that serves our purposes well. They provide
observational data as well as a simple model for the evolution of
twilight sky brightness. Figure~1 from that paper is included below, as \autoref{fig:Tyson}.
They show that a good model for the sky brightness evolution is given by
an exponential with
$\log_{10}(S)=(k/\tau)t+C$,
where S is the sky brightness in electrons per pixel per second, C is
the dark sky background, k = (10.6 minutes)$^{-1}$  is a universal
(band-independent) timescale during which the sky's surface brightness
changes by a factor of ten (at latitude $-$30 degrees), and $\tau$ is a
season-dependent factor that ranges from 1.0 at the equinox to 1.07 in
austral winter and 1.20 in austral summer. So the rule of thumb is that
we should expect it to take 4.25 minutes for the sky background to
change by one magnitude per square arc sec. (In what follows we'll
ignore the increased twilight time in summer and winter.)

For current generation typical astronomical camera systems that take
over a minute to read out, this 4.2 minute time scale means that only a
handful of images can be obtained during twilight time. But for the LSST
camera with a 2 second readout time, we can obtain hundreds of short
exposures during twilight. Even if we are limited to a 15 second cadence
due to thermal stability or data transfer limitations there is a large
amount of time opened up that we can use.

What do we stand to gain in operational time with shorter exposures? If
the standard survey terminates taking 15 second exposures due to some
sky brightness criterion, by shifting to 0.1 sec images at that point we
will have changed the sky flux per pixel by 2.5 $\log_{10}(150)$ = 5.4
magnitudes. This brings us back into a high dynamic range regime, as
described below.

\begin{figure}[htbp]
\begin{center}
\includegraphics[trim = 0 7cm 0 1mm, clip, width=\textwidth]{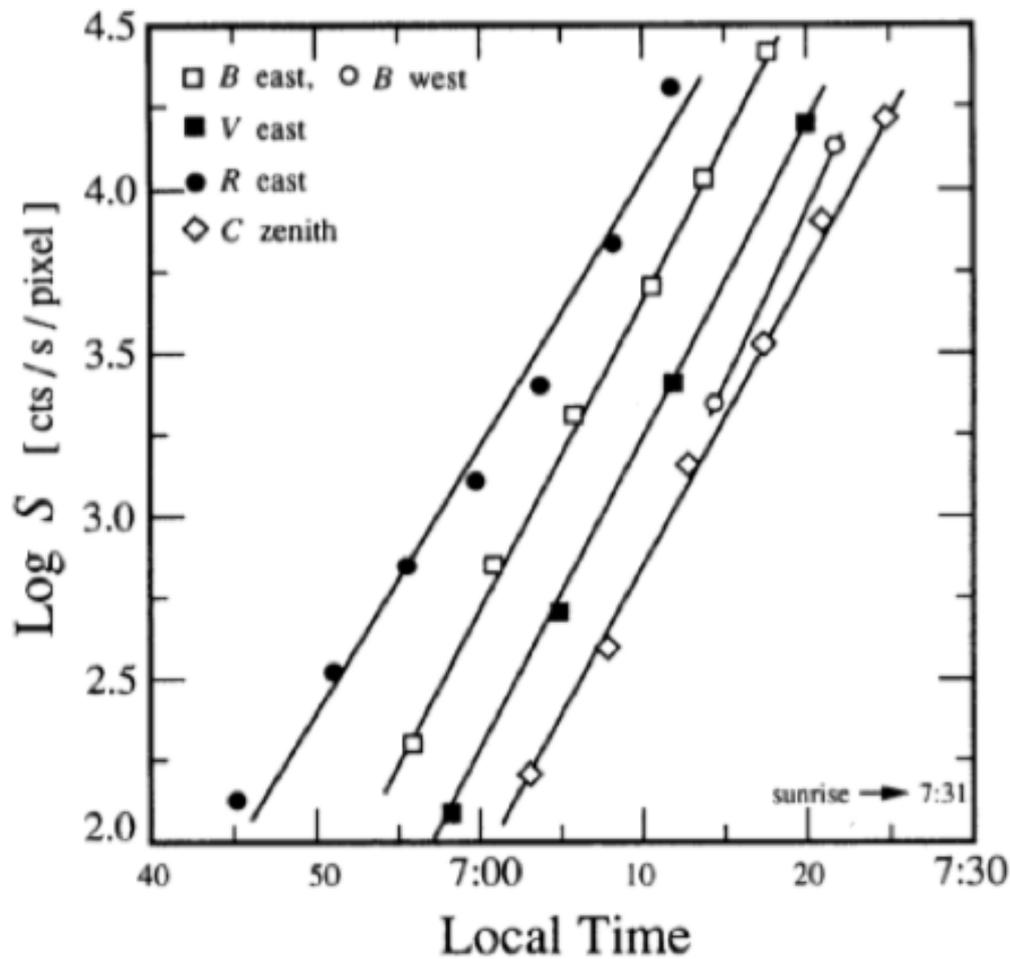}
\caption{(reproduced from Tyson et al, 1993). This plot shows the
  twilight sky surface brightness as a function of local time for four
  broadband filters (C, B, V and R) and different pointing directions.
  The surface brightness changes by one magnitude in a 4.2 minute interval,
essentially independently of the passband and pointing.}
\label{fig:Tyson}
\end{center}
\end{figure}

\autoref{fig:twilight} illustrates the principles that underpin this proposal. LSST is
a unique combination of hardware and software, that will deliver
reliable catalogs of both the static and the dynamic sky. By pushing
towards shorter integration times we can greatly expand the scientific
reach of the system.

The dynamic range in magnitudes that we can achieve for a given
integration time depends on the sky background, the read noise, and the
full well depth per pixel. We will adopt a typical value of 100Ke for
the full well depth, but the arguments presented below are essentially
independent of this value. The dynamic range in magnitudes is limited on
the bright end by the point source whose PSF peak exceeds full well, and
on the faint end by the 5$\sigma$ point source sensitivity, which
depends on sky brightness per pixel. So we are squeezed between the two
parameters of full well depth and sky background.

\begin{figure}[htbp]
\begin{center}
\includegraphics[width=6in]{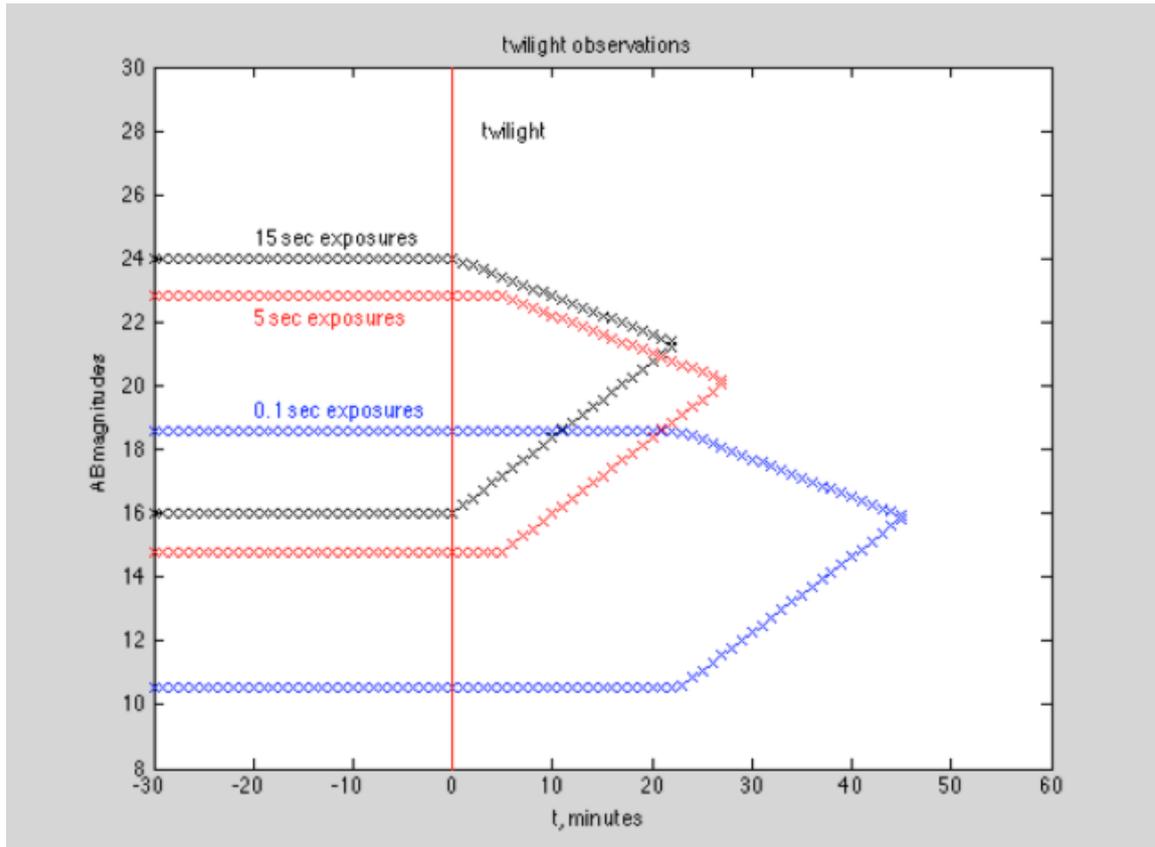}
\caption{Twilight dynamic range. As we enter morning twilight time, the increasing sky brightness requires brighter sources for 5 sigma detection, and also limits unsaturated objects to increasingly fainter sources. Eventually the gap between these goes to zero. But operating at shorter exposure times allows us to push useful survey operations into brighter twilight time, and also to increase the dynamic range of the LSST survey products. The black lines correspond to 15 second integrations (nominally in the r band), the red lines to 5 second exposures, and the blue curves to 0.1 second exposures. The upper lines in each case represent the 5 sigma point source detection threshold while the lower line corresponds to the source brightness that produces saturation in the peak pixel of the PSF. Adding shorter exposure times increases our dynamic range in flux, and adds valuable observing time.}
\label{fig:twilight}
\end{center}
\end{figure}

The 5-sigma limiting flux scales as the square root of the sky
brightness, while the saturation flux decreases linearly as sky
brightness increases. So the two curves in \autoref{fig:twilight} have
slopes that differ by a factor of two. Operating during bright-sky time
with short exposures adds about 20 minutes of observing per twilight, or
40 minutes per night. This is a non-trivial resource!

\autoref{fig:twilight} shows one reason why it is not advantageous to go
below 0.1 second exposures- we would lose the overlap between a twilight
survey and the standard LSST object catalog.

% ----------------------------------------------------------------------

\subsection{Science Drivers for Shorter Exposures}
\label{sec:\secname:drivers}

Having set the stage for the opportunity to operate at shorter exposure
times either during dark sky time, or during twilight, or both, we now
describe some of the scientific motivations for doing so.

\subsubsection{Discovery space at short time scales.}

LSST is a time domain discovery machine. It is hard to anticipate the
importance of being able to detect astronomical variability on short
time scales. By extending the time domain sensitivity to phenomena with
a characteristic time of less than 5 seconds, we will have added 1.5
orders of magnitude in time domain sensitivity.

Taking short exposures does not necessarily imply a requirement on fast
image cadence. Periodic variability can be readily detected and
characterized with a succession of short images that do not satisfy the
Nyquist criterion, as long as we know the time associated with each data
point to adequate accuracy. But it does seem appropriate to investigate
the maximum possible rapid-fire imaging rate for LSST, presumably
limited by either data transfer bottlenecks or by thermal issues within
the camera.

\subsubsection{Distances to Nearby SN Ia- an essential ingredient in using supernovae to probe dark energy.}

The determination of the equation of state parameter of the Dark Energy
using type Ia supernovae entails measuring the redshift dependence of
the luminosity distances to objects over a range of redshifts. The low
end of this redshift range is limited by peculiar velocities to
considering supernovae at redshifts z$>$0.01. At this distance (distance
modulus of $\mu$ =33) the peak brightness of a type Ia supernova is r=15
and exceeds the expected LSST point source saturation limit.

Moreover, the rate on the sky of these bright nearby supernovae is so
low that in the standard cadence we don't expect to obtain well-sampled
multiband light curves for them. But we will discover many of them on
the rise. Using twilight time with short exposures to obtain appropriate
temporal and passband coverage will allow us to extend the LSST SN
Hubble diagram across the entire redshift range of 0.01 to 1.

It is vitally important that we obtain these nearby-SN light curves on
the same photometric system, reduced with the same data reduction
pipeline, as the distant sample. This means we really must use the LSST
instrument and software in order to avoid systematic errors arising from
differences in photometric systems or algorithmic issues.

We stress that this twilight SN followup campaign can be accomplished
without impacting the main survey, during the roughly 20 minutes per
night of twilight that would otherwise unusable at the default exposure
time. We would use the brighter twilight time to obtain pointed
observations on nearby supernovae, motivated by the importance of
photometric uniformity described above.

\subsubsection{A Bright Star Survey for Galactic Science.}

We could also use the added twilight time to conduct a bright star
survey, and the precise astrometry and photometry from LSST can then be
used in conjunction with archived data ranging from 11th to 27th AB
magnitudes. This short-exposure domain would extend the LSST dynamic
range in fluxes by two orders of magnitude, towards the bright end.
Moreover, obtaining precise positions, fluxes and variability at these
brighter magnitudes would greatly increase the overlap with the
historical archive of astronomical information, including from digitized
plate data. We would be able to obtain astrometric and color information
to high precision, as well time series for variability studies.

An example of an application to Milky Way structure studies comes from
RR Lyrae variable stars. With a saturation magnitude of around 16th in
the standard LSST survey, RR Lyrae closer than 20 kpc will be saturated
in the standard LSST images. So we will lose nearly all Galactic RR
Lyrae. Extending the survey's bright limit to 11th magnitude will allow
us to collect light curves for RR Lyrae beyond $\sim$ 100 parsecs,
collecting essentially all Southern hemisphere Galactic RR Lyrae.

Another application for stellar population studies is measuring the
fraction of binary stars as a function of stellar type, metallicity, age
and environment. By conducting a variability survey in the 11-18
magnitude range we can capitalize on temperature and metallicity data
already in hand for many of these objects.

Another application of a bright star survey would be to search for
planetary transits in the magnitude range appropriate for radial
velocity followup observations using 30 meter class telescopes. For high
dispersion spectrographs at the 4m aperture class, most targets are
currently around 8th magnitude, so we should expect 30m telescopes to
attain similar radial velocity precisions for sources of magnitude  8 +
5log(30/4) = 12. By going to shorter exposures we obtain almost an
hour's additional observing time per night when these sources don't
saturate, whereas they are far beyond saturation in the default 15
second LSST survey images.

A typical (r$-$K) color between SDSS and 2MASS is r$-$K=3. The 2MASS
catalog is complete down to K$\sim$14 which corresponds to r$\sim$17. So
most 2MASS stars will be saturated in the standard LSST 15 second
observations. A bright star survey will allow a multiband match to the
2MASS data, as well as an astrometric comparison between the two
catalogs.

Finally, as pointed out in \autoref{sec:solar_system_specials}, the apparent magnitude of solar system objects depends on their
distance from us and from the sun, as well as illumination and
observation geometry. Extending the bright limit will allow us to track
asteroid positions as they approach opposition -- see \autoref{sec:solar_system_specials} for more details.

% ----------------------------------------------------------------------

\subsection{Counterarguments}
\label{sec:\secname:counter}

\subsubsection{What About Scintillation Effects?}

Short exposure times suffer from scintillation effects. An estimate for
uncertainty due to scintillation is provided by
\url{http://astro.corlan.net/gcx/scint.txt}. For a 0.1 second
integration we expect a fractional flux uncertainty of  0.15 at 2
airmasses and 0.043 at 1 airmass, for a 10 cm aperture. Scaling this up
to the 8.5m aperture of LSST by a factor D$^{2/3}$ predicts fractional
flux variations of below one percent, even at two airmasses, for a 0.1
second exposure. So scintillation should not impact our ability to make
precision measurements of flux and position.

\subsubsection{What about just doing this with smaller telescopes?}

A possible counter-argument to the proposal of allowing for shorter
exposure times is that much of this can be done with smaller telescopes.
But it's important to bear in mind that LSST is a system, and the data
reduction and dissemination tools are as important as the hardware. We
intend to deliver accessible, high-quality, well-calibrated photometry
on a common photometric system and correspondingly good positions. If we
do so from a co-added point source depth of 27th to the short-exposure
bright limit of 11th magnitude we will span over six decades in flux on
a well-calibrated flux scale. We would also have the ability to study
astrophysical variability on time scales from 0.1 second to 10 years,
which is nine decades in the time domain. This combination of temporal
and flux dynamic range would be a truly remarkable  achievement, and
would yield science benefits far beyond the illustrative examples
provided above. Much of this discovery space is enabled by going to
shorter exposures.

\subsection{Proposed Implementation and Impacts}

The implementation of this would simply entail taking short-exposure
images during twilight time that would otherwise go unused. The data
rate would go up, and the number of shutter cycles per night would also
increase, and so both the telescope and data management teams would need to advise on the cost of the program.

\navigationbar

% --------------------------------------------------------------------

% ====================================================================
%+
% NAME:
%    M67_special.tex
%
% CHAPTER:
%    specialsurveys.tex
%
% ELEVATOR PITCH:
%    As coeval, equidistant, and chemically homogeneous collections of stars,
%    open star clusters are ideal for studying the dependence of astrophysical
%    phenomena on the most fundamental stellar parameters - age and mass.
%
% AUTHORS:
%    Suzanne Hawley
%    Ruth Angus
%    Derek Buzasi
%    Jim Davenport
%    Mark Giampapa
%    Vinay Kashyap
%    Soren Meibom
%-
% ====================================================================

\section{A Mini-Survey of the Old Open Cluster M67}
\def\secname{M67_special}\label{sec:\secname}

\credit{suzannehawley},
\credit{ruthangus},
\credit{derekbuzasi},
\credit{jimdavenport},
\credit{markgiampapa},
\credit{vinaykashyap},
\credit{sorenmeibom}.

% LSST Review by Jason Kalirai:
% The kinds of effects that are being measured are things that will vary with age and mass, so you really want to survey a range of clusters with different ages. There are other clusters that satisfy the selection criteria (low extinction, well populated, nearby) that could also be added to the science case. I think building this up to a more comprehensive program offers some advantages. Otherwise, it seems to be the type of thing that could be done today with Subaru/HSC as a stand alone proposal.
% PJM: added footnote in the Science Case section.

\subsection{Introduction}

As coeval, equidistant, and chemically homogeneous collections of stars, open
star clusters are ideal for studying the dependence of astrophysical phenomena
on the most fundamental stellar parameters - age and mass.
Indeed, there are few fields in astronomy that do not rely on results from
cluster studies, and clusters play a central role in establishing how stellar
rotation and magnetic activity can be used to constrain the ages of stars and
stellar populations.
From an observational perspective, because of their angular extent they are
accessible to efficient surveys in both imaging and multi-object spectroscopy.
A selection of clusters representing a sequence in age can be used to
establish critical empirical relationships such as the dependence of activity
on rotation, the relationships between activity, rotation and stellar age, the
evolution of activity cycles, and the nature and evolution of flare
activity\textemdash{}an urgent area of investigation in view of the potential
impacts on the structure and evolution of exoplanet atmospheres in systems
with late-type host stars.

Unfortunately for observers, open clusters dissipate on timescales which are
generally comparable to stellar evolution timescales on the lower main
sequence, so older clusters are relatively rare.
In addition, most clusters lie close to the galactic plane, where determining
membership is significantly complicated by the large numbers of foreground and
background stars.

In this section, we suggest an LSST survey of M67, an open cluster whose
relative compactness, age, and location above the galactic plane combine to
make it the ideal cluster for a closer look. This proposed program could be adapted to include more clusters; we leave that more ambitious investigation to future work.

\subsection{Science Case }

The evolution of the rotation rate and magnetic activity in solar-type
stars are intimately connected. Stellar rotation drives a magnetic
dynamo, producing a surface magnetic field and magnetic activity which
manifests as starspots, chromospheric (Ca II HK, H$\alpha$) and coronal
(X-ray) emission, and flares. The magnetic field also drives a stellar
wind causing angular momentum loss (\textquotedblleft{}magnetic braking'')
which in turn slows the rotation rate over time, leading to decreased
magnetic activity. More magnetically active stars (larger spots, stronger
Ca II HK, H$\alpha$ and X-ray emission, more flares) therefore tend
to be younger and to rotate faster. The rotation-age relationship
is known as gyrochronology, and the correlation between rotation,
age and magnetic activity for solar-type stars was first codified
by Skumanich (1972). However, the decrease in rotation rate and magnetic
field strength over long time-scales is poorly understood and, in
some cases, hotly contested (Angus et al. 2015, Van Saders et al.
2016). Recent asteroseismic data from the Kepler spacecraft have revealed
that magnetic braking may cease at around solar Rossby number, implying
that gyrochronology is not applicable to older stars (Van Saders 2016).

In addition, the rotational behavior of lower mass stars is largely
unknown due to the faintness of mid-late type M dwarfs. There is reason
to believe that M dwarfs cooler than spectral type $~\mathrm{M}4$
may behave differently from the G, K and early M stars, since that
spectral type marks the boundary where the star becomes fully convective,
and a solar-type shell dynamo (which requires an interface region
between the convective envelope and radiative core of the star) can
no longer operate. Using chromospheric H$\alpha$ emission as a proxy,
West et al. (2008) studied a large sample of M dwarfs from SDSS and
showed that magnetic activity in mid-late M dwarfs lasts much longer
than in the earlier type stars.

The difficulties inherent in understanding the evolution of stellar
rotation and activity on the lower main sequence are further increased
by our inability to obtain accurate ages for field stars with ages
comparable to that of the Sun, which appears to be just the range
of ages for which our understanding of the phenomena are most suspect.
While asteroseismology can address this situation with exquisite precision,
it can only do so for the brighter stars accessible to space missions
such as Kepler. Making use of older open clusters is a way to fill
this gap.

The solar-age and solar-metallicity open cluster, M67, is a benchmark
cluster for understanding stellar evolution and the nature of late-type
stars at solar age. M67 is unique due to its solar chemical composition,
the fact that it is relatively nearby ($\sim900$ pc), and its relatively
low extinction due to its location above the galactic plane. Extensive
proper motion, radial velocity and photometric surveys have been carried
out (e.g., Girard et al. 1989, Montgomery et al. 1993, Yadav et al.
2008, Geller et al. 2015), while Giampapa et al. (2006) conducted
a survey of chromospheric activity in the solar-type members of M67
which yielded interesting insights on the range of magnetic activity
on sun-like stars in comparison with the range exhibited by the Sun
during the sunspot cycle. Nehag et al. (2011) find that solar twins
in M67 have photospheric spectra that are virtually indistinguishable
from the Sun\textquoteright{}s at echelle resolutions.

Located in the sky at approximately $\mathrm{RA}=9\mathrm{h}$ and
$\mathrm{Dec}=+12^{\circ}$, M67 is an exceptionally meritorious and
accessible candidate for an LSST special survey, which would also enable
productive follow-up observations by an array of OIR facilities.
% PJM: Note following Jason Kalirai's LSST Review:
(While we would like to study multiple clusters with LSST, we focus here on M67 in the special survey section because it is very slightly outside of the nominal footprint of the survey. Extending this science case to multiple clusters is a topic for future work.) LSST
observations of M67 would yield data on the rotation periods and variability
of its members at high precisions, particularly for dwarfs later than
about K0 ($V>16$). Little is known about the nature of variability
on short and long time scales for low-mass dwarfs at solar age. For
example, the frequency of \textquoteleft{}superflaring\textquoteright{}
at solar age could be investigated for the first time. Furthermore,
the combination of LSST observations and OIR synoptic datasets for
M67 would enable the characterization of the conditions of the habitable
zones in late- type stars at solar age.

In addition to sun-like stars, M67 includes an array of interesting
objects such as blue stragglers (Shetrone \& Sandquist 2000), an AM
Her star (Gilliland et al. 1991, Pasquini et al. 1994), a red straggler,
two subgiants (Mathieu et al. 2003), and detected X-ray sources due
to stellar coronal emission (e.g., Pasquini \& Belloni 1998). Davenport
\& Sandquist (2010) found a minimum binary fraction of 45\% in the
cluster. Other investigations include studies of the white dwarf cooling
sequence (Richer et al. 1998), angular momentum evolution near the
turnoff (Melo et al. 2001), and the behavior of key light elements
such as lithium and beryllium (e.g., Randich et al. 2007).

\subsection{Observing Plans }

Performing the special survey of M67 which we advocate would require
two modifications to the baseline LSST operations mode. LSST does
not plan to observe as far north as $\mathrm{Dec}=+12^{\circ}$ in
its main survey, but the M67 field should certainly be accessible
for a special survey as a single pointing. Since imaging the entire cluster
would require less than a single LSST field, we view this additional
pointing as being of minimal inconvenience relative to the expected
scientific gain. As we anticipate rotation periods ranging from $\sim\mathrm{days}$
up to several months, we would require sampling over all of these
timescales, though it need not be continuous.

A second potential complication is that the cluster is relatively
bright. While dwarfs below about spectral K0 in M67 are fainter than
the LSST bright limit of $\sim16$, the cluster G dwarfs will saturate
the LSST detectors in a 15-second integration. We suggest two alternative
approaches to address this issue. First, if the short exposure surveying mode
suggested in \autoref{sec:shortexp} is adopted,
then the new LSST minimum exposure time of 0.1 seconds would easily
accommodate the entire M67 main sequence. Alternatively, or if the
short exposure mode is not adopted, we note that work with the Kepler
mission (e.g., Haas et al. 2011) has shown success using custom pixel
masks to accurately perform photometry on stars as much as 6 magnitudes
brighter than the saturation level. Similar techniques applied to
the LSST fields should enable photometry for the G dwarfs, particularly
those in less-crowded portions of the field.

\navigationbar

% --------------------------------------------------------------------

% --------------------------------------------------------------------

\chapter[Synergy with WFIRST]{Synergy with WFIRST}
\def\chpname{wfirst}\label{chp:\chpname}

Chapter editor:
\credit{jasondrhodes}.

Contributing authors:
\credit{rubind},
\credit{davidpbennett},
\credit{mtpenny},
{\it Rachel Street}.

\section*{Summary}
\addcontentsline{toc}{section}{~~~~~~~~~Summary}

WFIRST will launch in $\sim$ 2025 for a 5 year mission to explore dark energy, find and characterize exoplanets, and take wide, deep infrared surveys of the galactic and extragalactic sky.  WFIRST was recognized by the Astro2010 Decadal Survey as an excellent NIR complement to LSST's optical capabilities. More recent work has recognized the strong synergy these two projects will have in helping to address cosmological questions in the 2020s \citep{2015arXiv150107897J}.
  Together, the two observatories can accomplish significantly more (and better) science than either can alone in the areas of weak lensing, large-scale structure studies, strong lensing, supernova studies, exoplanet investigations,  and photometric redshift determination. Accomplishing this will require coordinated observations.  We have identified three areas of proposed coordination: 1. Early coverage of the $>2000$ square degree WFIRST High Latitude Survey to the full optical depth for enhanced photometric redshifts for both LSST and WFIRST; 2. Coordinated LSST optical observations in the WFIRST supernova discovery fields; 3. Precursor, simultaneous, and follow-up observations of the WFIRST microlensing fields near the galactic bulge.

  We note here that the focus of the WFIRST  input to this paper is on suggesting changes or enhancements to the LSST observing strategy (cadence or survey overlap) that will provide mutual benefit for WFIRST and LSST.  The observing strategy suggestions here are not exhaustive of the possibilities for synergy between WFIRST and LSST; rather the suggestions here are based on the planned primary WFIRST surveys that are driving the WFIRST mission requirements.  As the plans and possibilities for WFIRST Guest Observer surveys evolve, and the ideas for additional science investigations that will make use of planned WFIRST survey data mature, the community should continue to look for areas of survey synergy.  Both WFIRST and LSST should consider survey modifications that would increase the global combined science output of the two surveys.

% --------------------------------------------------------------------

\section{Introduction}
\label{sec:wfirst:intro}

% Introduce, with a very broad brush, this chapter's science projects,
% and why it makes sense for them to be considered together.

The Wide Field Infrared Survey Telescope (WFIRST) is a NASA mission that
entered Phase A in February 2016.  WFIRST was the highest recommendation
for large space missions in the 2010 New Worlds New Horizons Decadal
Survey.  That recommendation envisioned a wide-field observatory with
near infrared (NIR) capabilities to complement LSST's optical
capabilities; the Decadal Survey recognized the obvious synergy between
WFIRST and LSST.  WFIRST's design has evolved since 2010 and the design
being pursued for a mid-2020s launch uses an existing $2.4$m telescope
donated to NASA, giving WFIRST capabilities not envisioned by the
Decadal Survey.  WFIRST has 3 primary science objectives:

\begin{itemize}
\item Determine the nature of the dark energy that is driving the
current accelerating expansion of the universe using a combination of
weak lensing, galaxy clustering (including Baryon Acoustic Oscillations
and Redshift Space Distortions), and supernovae type Ia (SN).
\item Study exoplanets through a statistical microlensing survey and via
direct imaging and spectroscopy with a coronagraph.
\item Perform NIR surveys of the galactic and extragalactic sky via a
Guest Observer program.
\end{itemize}

WFIRST will be at L2 to enable the thermal stability needed for the
precise astrometric, photometric, and morphological measurements
required for these science goals. The baseline WFIRST mission
architecture is described in detail in the final report of the WFIRST
Science Definition Team \citep{2015arXiv150303757S}
%(arxiv/1503.03757).
The WFIRST Wide Field
Instrument(WFI) has a NIR focal plane with a $\sim0.28$ square degree
field of view made up 18 4k$\times$4k Teledyne H4RG NIR detectors and will
have imaging capabilities from $0.5-2$ microns and grism spectroscopy
capabilities from $1-2$ microns with $R\sim461\lambda$.  The WFI
also contains an Integral Field Channel (IFC) spectrometer with $R\sim100$
resoluton over the range $0.6-2$ microns for SN follow up. The exoplanet
coronagraph will have imaging ($0.43-0.97$ microns) and spectroscopic
($0.6-0.97 $ microns) capabilities with a contrast ratio of 1 part in a
billion.

WFIRST's  5 year primary mission is envisioned to have $\sim2$years dedicated to a
$\sim2200$ square degree High Latitude Survey (HLS) for weak lensing and
galaxy clustering,  $\sim1$ year of microlensing observations divided into 6
seasons, $\sim0.5$ years of SN search and follow-up, $\sim0.8$ years dedicated to
the coronagraph and the remaining time dedicated to competitively selected Guest
Observer observations. WFIRST has no expendables that would prevent an
extended mission of 10 years or longer, and an extended mission will likely be
given over entirely to Guest Observer observations.

The synergy with LSST is very promising indeed. In this chapter we aim
to  lay out three  specific projects in the three main WFIRST science
areas, and test the simulated LSST Observing
Strategies for their performance in each case. Then, we use these
results to design a suite of modified LSST Observing Strategies, which
we propose as new \OpSim simulation runs.

\navigationbar

% --------------------------------------------------------------------

% ====================================================================
%+
% SECTION:
%    WFIRST_weaklensing.tex
%
% CHAPTER:
%    wfirst.tex
%
% ELEVATOR PITCH:
%
%
% AUTHORS:
%    Jason Rhodes @jasondrhodes
%-
% ====================================================================

\section{Cosmology with the WFIRST HLS and LSST}
\def\secname{\chpname:weaklensing}\label{sec:\secname}

\credit{jasondrhodes}

WFIRST's High Latitude Survey (HLS) will cover
2200 square degrees in 4 NIR photometric filters
(3 of which will be sufficiently sampled for weak lensing shape
measurements) and NIR grism spectroscopy.  The benefits of overlapping
spectroscopic and photometric surveys for dark energy constraints and
systematics mitigation are strong.  The primary scientific driver of the
photometric portion of the WFIRST HLS is weak gravitational lensing,
but there is a wide range of ancillary science that will be possible
with the publicly available WFIRST HLS data (see for instance, the SDT
report mentioned above).  However, the requirements on the HLS are
largely set by constraints from weak lensing measurements.  Each galaxy
in the WFIRST weak lensing survey needs to have an accurate photometric
redshift.  This requires optical photometry that reaches the depth of
the NIR photometry WFIRST will acquire ($J~27AB$).  \emph{Thus, the
WFIRST weak lensing survey will require the full  10-year LSST depth in
4 optical bands for optimal photometric redsfhift determination}.

There is strong benefit not just to WFIRST, but to LSST, in coordinating
observations of the WFIRST HLS survey field. The combination of
full-depth LSST data and WFIRST HLS NIR data will provide the gold
standard in photo-zs.  Furthermore, WFIRST grism observations over the
same area will provide many millions of high quality slitless spectra
and WFIRST's IFC can be run in parallel with WFI observations to provide
many more very accurate spectroscopic redshifts in the survey area.
Thus, the WFIRST photometric data will help to provide better LSST
photo-zs and  WFIRST will also provide many of the spectra needed as a
training set to calibrate the photo-zs for both missions.  A further
benefit to LSST might be the reduced need for LSST observations at the
reddest end of the LSST wavelength range (the $z$ and $y$ filters), where
both the atmosphere and the physics of CCDs make ground-based
observations less efficient than what WFIRST can achieve. Further work is needed to quantify this benefit, especially as the WFIRST proposed filter set is evolving.Finally, the
joint processing of LSST and WFIRST data will provide better object
deblending parameters than LSST can achieve alone; WFIRST will be able
to provide a morphological prior for the deblending of LSST images.

% --------------------------------------------------------------------

\subsection{Measuring Dark Energy Parameters}
\label{sec:\secname:targets}

The goal in this science project would be to measure Dark Energy parameters from various weak lensing
probes, capitalizing on the improved photometric redshifts that a joint
analysis of LSST and WFIRST photometry would provide. A useful
Figure of Merit is the usual Dark Energy Task Force figure of merit,
quantifying the available precision on the equation of state
parameters $w_0$ and $w_a$. For example, we are interested in
improvements in the weak lensing DE FoM as the LSST photometric redshifts
and galaxy shapes are improved over the whole LSST survey area
via joint analysis with WFIRST. We are also interested in increasing the
WFIRST DE FoM as quickly as possible.
Indeed, the basic problem we face is one of timing: being able to combine
the WFIRST data with the LSST sooner will accelerate the production of
cosmological results.

Therefore, we propose an acceleration of the LSST survey over about $10\%$ of the
LSST survey area (the $\sim2200$ WFIRST HLS) such that the full LSST ten
years survey depth is reached on a timescale that maximizes the joint
usefulness of LSST and WFIRST data on that area.  Assuming the two year
WFIRST HLS is taken in the first four years of a WFIRST mission that
launches in 2024, this would require reaching full LSST depth over that
area in $\sim2028$ rather than $\sim2032$. Since the HLS area is roughly
$1/8$ as large as the LSST ``Main Survey"'' region, this could be
achieved by devoting 1.25 years of LSST observations to the HLS area,
assuming that it covers a wide enough range of Right Ascension.  More
practically, it could be achieved by devoting 25\% of LSST observing
time to this area during each of the first 5 years of the LSST survey,
which doubles the time it would naturally be observed during those years
at a modest reduction in coverage of the rest of the Main Survey area
during that time period.   Given existing plans to speed up the LSST
cadence over small sub-areas of the LSST survey, this may only require
coordination of the locations of the accelerated LSST area and the
WFIRST HLS. As LSST and WFIRST progress, there is a mutual benefit in
continuing discussions about the optimal joint observation schedule.

A simple, first order diagnostic metric would be the amount of LSST/WFIRST
overlapping survey area that reaches the full LSST depth when the WFIRST
HLS is completed.  Such a metric is straightforward, but not
able to be meanigfully encoded until the 2020s, when the WFIRST launch date and survey
plan is more definite.  Strawman survey plans could, however, be
evaluated to help with LSST schedule planning.
A slightly more complicated metric could include
the pace at which the overlapping LSST/WFIRST survey areas are both
taken to full depth, since this would make each data set maximally
useful to the US community (or anyone with immediate access to both
WFIRST and LSST data).  WFIRST data is unlikely to have any proprietary
period.  Current plans call for the WFIRST HLS to be conducted in
multiple passes, but the exact survey pattern is still undecided, so this
metric is also not quantifiable yet.

There may be some reduced need for the the LSST reddest bands in the
WFIRST HLS overlap area, which should also be folded into the metric.
We note that
the default survey strategy would only achieve the full LSST photometric
depth over the WFIRST HLS after 10 years of survey ($\sim2032$).

% % --------------------------------------------------------------------
%
% \subsection{OpSim Analysis}
% \label{sec:\secname:analysis}
%
%
% % --------------------------------------------------------------------

\subsection{Discussion}
\label{sec:\secname:discussion}

Increasing the cadence of the LSST survey over $\sim10\%$ of the LSST
survey has science benefits that go far beyond the LSST/WFIRST synergy
described here: an observing strategy that met the joint WFIRST/LSST cosmology goals could
also provide the kind of ``rolling cadence'' favored by other science teams.
There are benefits to certain aspects of time-domain
science.  Every effort should be made to coordinate all discussions of
increased survey cadence (resulting in full LSST depth well before 10
years) over sub-areas of the LSST survey footprint.  Specific attention
should be paid to whether the accelerated portions of the LSST survey
can completely overlap the WFIRST HLS, and whether the position of the
WFIRST HLS can be determined, in part, by other science drivers within
LSST.  This will require close LSST and WFIRST coordination at the
Project levels.

% % --------------------------------------------------------------------

\subsection{Conclusions}
\label{sec:\secname:conclusions}

Even before implementing the metrics decsribed above, we can offer
tentative answers to the ten questions posed in
\autoref{sec:intro:evaluation:caseConclusions}:

\begin{description}

\item[Q1:] {\it Does the science case place any constraints on the
tradeoff between the sky coverage and coadded depth?}

\item[A1:] WFIRST requires the full 10 year LSST depth over the $\sim$ 2000 square degree WFIRST High Latitude Survey.

\item[Q2:] {\it Does the science case place any constraints on the
tradeoff between uniformity of sampling and frequency of  sampling? For
example, a rolling cadence can provide enhanced sample rates over a part
of the survey or the entire survey for a designated time at the cost of
reduced sample rate the rest of the time (while maintaining the nominal
total visit counts).}

\item[A2:] The WFIRST HLS synergy does not place constraints on the uniformity of time sampling.

\item[Q3:] {\it Does the science case place any constraints on the
tradeoff between the single-visit depth and the number of visits?}

\item[A3:] The WFIRST HLS only places requirements on the total depth in
the 5 LSST photometric filters.

\item[Q4:] {\it Does the science case place any constraints on the
Galactic plane coverage (spatial coverage, temporal sampling, visits per
band)?}

\item[A4:] No: the WFIRST HLS will not be in the Galactic plane.

\item[Q5:] {\it Does the science case place any constraints on the
fraction of observing time allocated to each band?}

\item[A5:] There are multiple solutions to the allocation of depth to each of the 5 LSST and 3 WFIRST bands that will be used for optimzed photometric redshifts. This is a area of ongoing study.

\item[Q6:] {\it Does the science case place any constraints on the
cadence for deep drilling fields?}

\item[A6:] No.

\item[Q7:] {\it Assuming two visits per night, would the science case
benefit if they are obtained in the same band or not?}

\item[A7:] The WFIRST HLS is largely agnostic about the timing of the different filters.

\item[Q8:] {\it Will the case science benefit from a special cadence
prescription during commissioning or early in the survey, such as:
acquiring a full 10-year count of visits for a small area (either in all
the bands or in a  selected set); a greatly enhanced cadence for a small
area?}

\item[A8:] The WFIRST HLS synergy would be globally maximized (for both
LSST and WFIRST) if the full LSST depth in all 5 bands is reached in the
first 5 years of LSST operations.

\item[Q9:] {\it Does the science case place any constraints on the
sampling of observing conditions (e.g., seeing, dark sky, airmass),
possibly as a function of band, etc.?}

\item[A9:] The benefit to LSST of having high precision space based
galaxy shape measurements would be maximized if the observing conditions
allowed for the best possible LSST shapes for cross-calibration of shear
measurements.

\item[Q10:] {\it Does the case have science drivers that would require
real-time exposure time optimization to obtain nearly constant
single-visit limiting depth?}

\item[A10:] No.

\end{description}

% ====================================================================

\navigationbar

% ====================================================================
%+
% SECTION:
%    WFIRST_supernovae.tex
%
% CHAPTER:
%    wfirst.tex
%
% ELEVATOR PITCH:
%-
% ====================================================================

\section{Supernova Cosmology with WFIRST and LSST}
\def\secname{\chpname:supernovae}\label{sec:\secname}

\credit{rubind}

% This individual section will need to describe the particular
% discoveries and measurements that are being targeted in this section's
% science case. It will be helpful to think of a ``science case" as a
% ``science project" that the authors {\it actually plan to do}. Then,
% the sections can follow the tried and tested format of an observing
% proposal: a brief description of the investigation, with references,
% followed by a technical feasibility piece. This latter part will need
% to be quantified using the MAF framework, via a set of metrics that
% need to be computed for any given observing strategy to quantify its
% impact on the described science case. Ideally, these metrics would be
% combined in a well-motivated figure of merit. The section can conclude
% with a discussion of any risks that have been identified, and how
% these could be mitigated.
%
% A short preamble goes here. What's the context for this science
% project? Where does it fit in the big picture?

The WFIRST SN survey seeks to measure thousands of SNe Ia with excellent systematics control over a two-year period. The Science Definition Team (SDT) outlined a three-tiered cadenced imaging survey: wide to $z=0.4$ (27.44 square degrees), medium to $z=0.8$ (8.96 square degrees), deep to $z=1.7$ (5.04 square degrees). SNe discovered in the imaging would be followed with IFU spectrophotometry, helping to monitor changes in SN physical parameters and the extinction distribution with redshift.
% However, due to the slew time and high read noise in short exposures, the wide survey would be very inefficient, spending a bit more than half of its time on slews, while the medium survey would spend a significant fraction of its time slewing.
However, the LSST DDFs offer a path to high signal-to-noise, well calibrated, multi-band optical imaging over an even larger area than WFIRST can survey. If the wide and medium tiers are replaced with LSST DDF discoveries, then WFIRST can offer spectrophotometry (with good host-galaxy subtraction) for $\sim$ 2,000 LSST SNe, with screening spectra for $\sim$ 1-2,000 more. As the WFI and IFU operate in parallel, this survey could provide sparsely sampled NIR imaging for $\sim$ 5,000 SNe up to $z = 1$ at the same time as the spectroscopy. The joint survey would thus provide systematics control (almost certainly better than either survey alone), as well as a cross-check of LSST photometric typing and host-galaxy-only redshift assignment.

% --------------------------------------------------------------------

\subsection{Target measurements and discoveries}
\label{sec:\secname:targets}

% Describe the discoveries and measurements you want to make.
%
% Now, describe their response to the observing strategy. Qualitatively,
% how will the science project be affected by the observing schedule and
% conditions? In broad terms, how would we expect the observing strategy
% to be optimized for this science?

The targets of the measurements are related to those enumerated in
\autoref{sec:supernovae:targets}. The SNe must be detected $\sim$ 10
observer-frame days before maximum light, so that there is time for a
shallow screening spectrum before deeper spectrophotmetry around
maximum. There should be enough visits per filter so that some
photometric screening can be done before WFIRST triggers any
spectroscopy. There should be an identification of the host galaxy (if
seen), so that joint WFIRST/LSST photometric redshifts can be used to
provide a distance-limited sample (minimizing selection effects).
Finally, the light curve should continue after the SN has been sent to
WFIRST, so that important light-curve parameters (date of maximum, rise
time and decline time, etc.) can be measured.

%All these goals can be likely be met with $\sim 3$ day rest-frame cadence ($\sim 5$ observer-frame days). LSST would measure NUV to rest-frame $V$-band (with WFIRST providing redder wavelength coverage), or observer-frame $grizY$. For a plausible SN Ia (based on the rising light curve), a series of typing/sub-typing spectra would be triggered, with increasing depth, as the confidence grew that the transient was a SN Ia. DR: in my simulations, I've assumed a depth for each filter of 26th magnitude (probably not realistic for $Y$-band, but very feasible for the other filters); is this too shallow? LSST would contribute 4 transients per day to the pool of objects observed by WFIRST. In practice, the LSST DDFs will contain more SNe Ia than this, so a random sample (perhaps sculpted in redshift) should be sent for observations.

% --------------------------------------------------------------------

\subsection{Metrics}
\label{sec:\secname:metrics}

% Quantifying the response via MAF metrics: definition of the metrics,
% and any derived overall figure of merit.

The primary metrics are based on constraining cosmological parameters;
the DETF FoM is standard. For the joint observations proposed here, this
FoM increases about 20\%, from $\sim 300$ for WFIRST alone (with a Stage
IV CMB constraint) to $\sim 370$. However, the number of SNe at $z \sim
0.5$ increases by $\sim$ 50\% over a WFIRST-only survey, improving some
$w(z)$-derived FoM values by 40\%.

The cosmological metric will essentially depend on the number of SNe
meeting the above targets. It will degrade if core-collapse SNe are
mistakenly sent to WFIRST for followup, if SNe Ia are sent to WFIRST but
the LSST light curve is lost due to weather gaps, or if the cadence and
depth simply do not allow the measurement of light curve parameters.
These metrics will be strongly related to those in
\autoref{sec:supernovae:metrics}, but with more emphasis on the rising
portion of the light curve.

% --------------------------------------------------------------------

%\subsection{OpSim Analysis}
%\label{sec:\secname:analysis}

% OpSim analysis: how good would the default observing strategy be, at
% the time of writing for this science project?

% --------------------------------------------------------------------

%\subsection{Discussion}
%\label{sec:\secname:discussion}

% Discussion: what risks have been identified? What suggestions could be
% made to improve this science project's figure of merit, and mitigate
% the identified risks?

% ====================================================================
%
\subsection{Conclusions}

Here we answer the ten questions posed in
\autoref{sec:intro:evaluation:caseConclusions}. As WFIRST will only cadence a few 10's of square degrees, we assume that the overlap region will be contained in the DDFs, not in the main survey. We thus place no constraints on the main survey.

\begin{description}

\item[Q1:] {\it Does the science case place any constraints on the
tradeoff between the sky coverage and coadded depth?}

\item[A1:] In terms of direct overlap with WFIRST, there is no constraint. There are secondary considerations (such as constraining  Milky Way extinction and probing isotropy) that may prefer a larger number of square degrees, but this has not been investigated.

\item[Q2:] {\it Does the science case place any constraints on the
tradeoff between uniformity of sampling and frequency of sampling?}

\item[A2:] No constraints.

\item[Q3:] {\it Does the science case place any constraints on the
tradeoff between the single-visit depth and the number of visits
(especially in the $u$-band where longer exposures would minimize the
impact of the readout noise)?}

\item[A3:] In the DDFs, the SN Ia science would not be harmed by having longer exposures and less-frequent visits, at least if there was one observation every few days in each filter. No constraint on the main survey.

\item[Q4:] {\it Does the science case place any constraints on the
Galactic plane coverage (spatial coverage, temporal sampling, visits per
band)?}

\item[A4:] No constraints.

\item[Q5:] {\it Does the science case place any constraints on the
fraction of observing time allocated to each band?}

\item[A5:] This has not been quantitatively investigated. Based on experience with other programs, the reddest filters (z and Y) should receive the most time.

\item[Q6:] {\it Does the science case place any constraints on the
cadence for deep drilling fields?}

\item[A6:] To obtain good light curves, we would want at least one observation every few days per filter. This is also a requirement for photometric typing to obtain reasonable purity if WFIRST spectroscopy is to be triggered. Note: of all the questions, this one is the most important to the program.

\item[Q7:] {\it Assuming two visits per night, would the science case
benefit if they are obtained in the same band or not?}

\item[A7:] No benefit.

\item[Q8:] {\it Will the case science benefit from a special cadence
prescription during commissioning or early in the survey, such as:
acquiring a full 10-year count of visits for a small area (either in all
the bands or in a  selected set); a greatly enhanced cadence for a small
area?}

\item[A8:] For the DDFs, such observations might help provide deep SN-free (template) images or host-galaxy photometric redshifts. This has not been quantitatively investigated. No constraints on the main survey.

\item[Q9:] {\it Does the science case place any constraints on the
sampling of observing conditions (e.g., seeing, dark sky, airmass),
possibly as a function of band, etc.?}

\item[A9:] As long as the DDF observations reach the targeted depths, there is no benefit to the SN Ia science from having specific observing conditions.

\item[Q10:] {\it Does the case have science drivers that would require
real-time exposure time optimization to obtain nearly constant
single-visit limiting depth?}

\item[A10:] We do not currently know of any benefit to the SN Ia science from doing this.

\end{description}

% ====================================================================

\navigationbar

% ====================================================================
%+
% SECTION:
%    WFIRST_microlensing.tex
%
% CHAPTER:
%    wfirst.tex
%
% ELEVATOR PITCH:
%
%
% AUTHORS:
%    David Bennett(@davidpbennett)
%-
% ====================================================================

\section{Exoplanetary Microlensing with WFIRST and LSST}
\def\secname{\chpname:microlensing}\label{sec:\secname}

\credit{davidpbennett},
\credit{mtpenny},
\credit{poleski},
\credit{rachelstreet}.

One of the most exciting prospects of future micolensing surveys is the
characterization of the population of ``rogue'' or free-floating planets.
While recent estimates of the rogue planet abundance (Mroz et al. 2017 in prep)
 have cast doubt on the large population of rogue planets inferred by
\citet{2011Natur.473..349S}, they have clearly demonstrated that microlensing surveys
have the sensitivity necessary to detect them. Firm detections of the rogue
planet population that might be predicted by planet formation simulations
awaits the accumulation of sufficient survey duration for giant rogue planets,
and the WFIRST microlensing survey for smaller planets~\citep{2015arXiv150303757S,2017arXiv170408749B}.
% of discovery to come out of gravitational
%microlensing surveys is the discovery of a large population of ``rogue''
%planets by the MOA Collaboration \citep{2011Natur.473..349S}, though more
%recent analysis by the OGLE Collaboration (Mroz et al. 2017 in prep) does not confirm this result. Nevertheless, WFIRST will detect many rogue planets between Mars and Jupiter mass even if their abundance is more than an order of magnitude smaller than inferred by MOA or a factor of a few smaller than expected from theoretical studies~\citep{2015arXiv150303757S, 2017arXiv170408749B}.

Microlensing's rogue planets
are isolated in the sense that no host star can be detected
by microlensing. Depending on the peak magnification and light curve
coverage, this can imply that a host must be $> 10\,$AU or $> 100\,$AU away,
and \citet{2012ApJ...757..119B} have argued that the median separation
of possible host stars is likely to be $> 30\,$AU. The presence and properties
of a host star can be constrained either with WFIRST's diffraction limited survey images separated by the mission duration, or adaptive optics observations on large telescopes ($\ge 8$~m) taken $4$--$8$ years after the rogue planet event~\citep{2016JKAS...49..123G,2016AJ....152...96H}.
%Further observations by both the MOA and OGLE collaborations provide
%a qualitative confirmation of this result, as dozens of additional
%short timescale events have been discovered by the MOA and OGLE
%alert systems, but full details of the implied rogue planet
%populations  will only come from detailed analyses of both the MOA
%and OGLE samples.

A major weakness with the microlensing survey data sets that can be used to detect rogue planets is that, thus far, the properties
of the population can only been inferred by their Einstein radius
crossing time, $t_{\rm E}$, distribution. But, the Einstein radius crossing
time depends not only on the lens mass, but also on its distance and
transverse velocity. As a result, we cannot directly infer the mass or distance
distribution of the rogue planet sample.

Our understanding of the rogue planet distribution can be greatly improved
by measuring the microlensing parallax effect \citep{1992ApJ...392..442G,1995ApJ...454L.125A}.
Combining $t_{\rm E}$ with the microlens parallax $\pi_{\rm E}$ yields a mass-distance relationship, a narrower range of possible lens masses and a projection of the lens-source relative velocity, which can in certain cases be used to identify the lens as belonging to the Galactic disk or bulge~\citep{2015ApJ...802...76Y}. The mass and distance of the lens can be uniquely solved for if, in addition to parallax, the angular diameter of the Einstein ring can be measured via finite source effects \citep{1994ApJ...424L..21N}. Finite source effects will be measured in a substantial fraction of events involving bound and free-floating planets.
%The microlensing parallax effect can be described
%by the transverse relative lens-source velocity, ${\bf v}_{\rm \perp}$, projected
%to the position of the observer,
%\begin{equation}
%\tilde{\bf v} = {\bf v}_{\rm \perp} D_S/(D_S-D_L) \ , \label{eq-vp}
%\end{equation}
%where $D_L$ and $D_S$ are the lens and source distances, respectively.

Typically, microlensing parallax
is measured using the orbital motion of the Earth, but it can also be
measured using light curve observations from telescopes at different locations
in the Solar System \citep{2007ApJ...664..862D,2015ApJ...804...20C} or
even different locations on Earth \citep{2009ApJ...698L.147G}. In the
case of microlensing by planetary mass objects, the event durations are
too short to allow a significant light curve change due to the Earth's
orbital motion, but near simultaneous observations from Earth and a
satellite orbiting at the Earth-Sun L2 point (where WFIRST will orbit) allows the measurement of microlensing parallax signals for planetary mass
lenses \citep{2003ApJ...591L..53G}. The principles of the technique are identical to those employed by the {\it K2} Campaign 9 microlensing parallax survey~\citep{2016PASP..128l4401H,2016AJ....152...96H,2017AJ....153..161P}.
\autoref{fig-lc} shows an example of the light curves for one of the
rogue planets with a mass determined by simultaneous WFIRST and LSST
observations.

\begin{figure}[t]
\centering\includegraphics[width=0.5\linewidth]{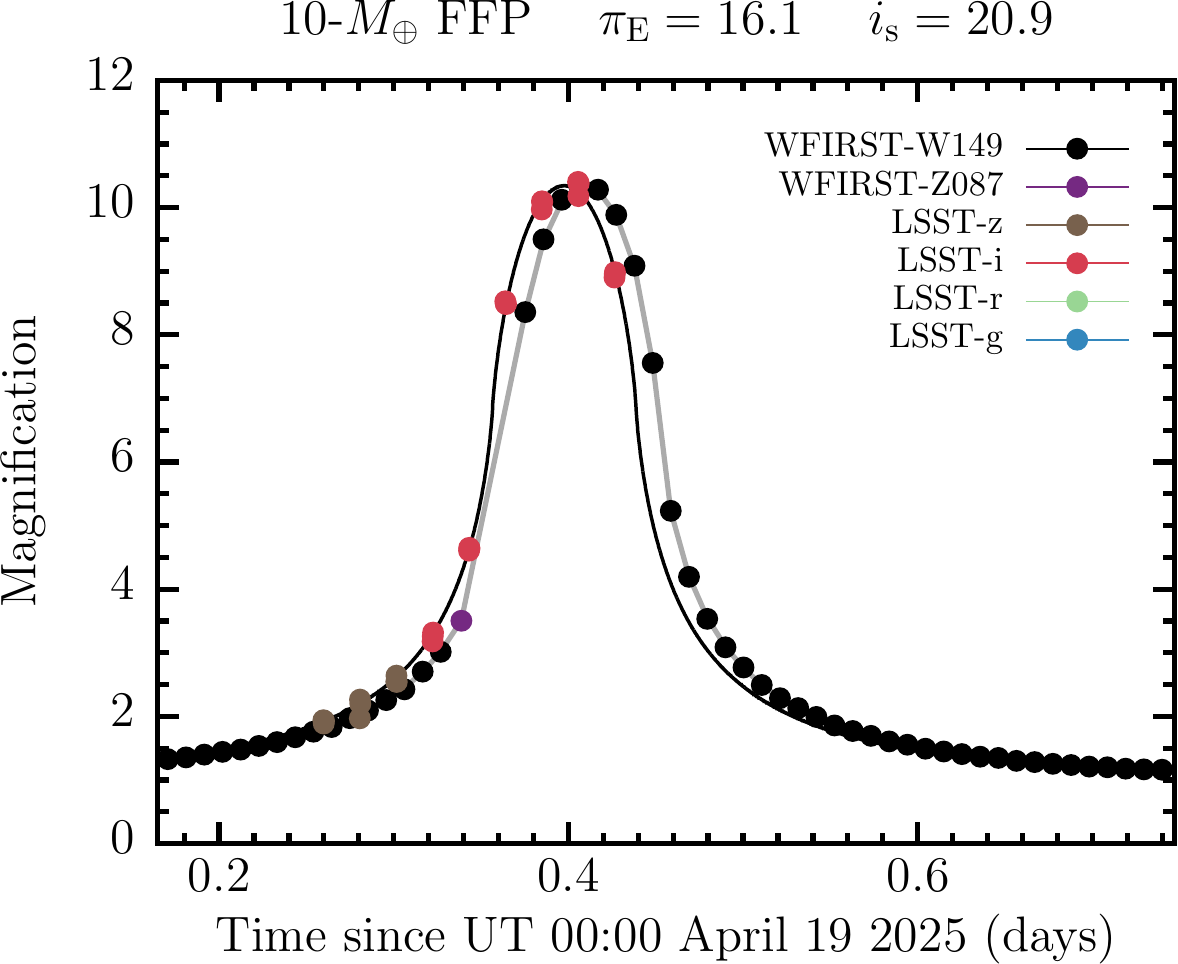}
\caption{A simulated light curve of a $10 M_{\oplus}$ planet with a microlensing parallax
mass measurement from simultaneous WFIRST and LSST observations. The short duration of WFIRST planetary microlensing events and the high extinction in WFIRST's fields requires high-cadence (ideally every $15$-$30$ min) observations in the redder bands ($i$ or $z$, preferably). During WFIRST observations around the equinoxes, the LSST-WFIRST microlensing field will only be visible for a few hours per night.}
\label{fig-lc}
\end{figure}

WFIRST planetary microlensing events will greatly improve our
understanding of not only rogue planets population but also of planets in wide orbits similar to those of
Uranus or Neptune and with a wide range of masses~\citep{2015arXiv150303757S}. Such cold planets
cannot be studied using techniques other than microlensing~\citep[e.g.,][]{2012ARAA..50..411G} and
WFIRST gives us a unique opportunity to detect a statistically significant
number of wide orbit planets.
%Even though the wide orbit planets will be detected,
%their properties may not be well constrained by WFIRST or it may be hard to distinguish
%between a wide orbit planet and a rogue planet, particularly for planets on the widest
%orbits or when the source trajectory does not closely approach the host star.
%The two main parameters that can be derived for bound planets are
%the mass ratio, $q$, and the projected separation, $s$, in units of Einstein radius.
A wide-orbit planet adds an additional peak to the otherwise
standard microlensing light curve produced by the host star that occurs a few Einstein crossing timescales before or after the host event's peak~\citep[e.g.,][]{2014ApJ...795...42P}. Typical host event timescales are ${\sim}25$~days, so it will be common for the host peak of a wide-orbit planet discovered by WFIRST to lie outside of WFIRST's $72$~day observing window, but inside the time when LSST can observe it. Without ground-based observations outside the WFIRST window, the properties of many wide-orbit planet detections by WFIRST will be poorly constrained, particularly the mass ratio of the planet to its host, and its projected separation from the host. However, deep, low-cadence LSST
monitoring of the WFIRST microlensing field over whole bulge observing season
will reveal the peaks of planet host microlensing events that are unobservable
to the WFIRST and will enable measurement of the projected separation and mass ratio.
%help to constrain projected separation and mass ratio.

%The time separation of the two peaks, $\Delta t$, is directly related
%to the planet projected separation:
%\begin{equation}
% \Delta t =  t_{\rm E}(s - {1/s}) \ .
%\end{equation}
%sTypical value of $t_{\rm E}$ is $25~{\rm days}$, hence, for a wide
%separation planet ($s$ on the order of a few) $\Delta t$ will be
%comparable to the length of a single WFIRST session i.e., $72~{\rm days}$.
%Even though WFIRST will be able to observe the planetary anomaly,
%it may not observe the host event and hence poorly constrain planet properties.

% --------------------------------------------------------------------

\subsection{A Proposed Observing Strategy}
\label{sec:\secname:proposal}

Based on the desiderata above, we
propose simultaneous high cadence observations of the WFIRST microlensing fields
by LSST during each
of the six 72-day WFIRST exoplanet microlensing survey sessions. These
will allow microlensing parallax measurements to determine the distances
and masses of a representative sub-sample of the rogue planets found by
the WFIRST microlensing survey. These measurements will be crucial for
the interpretation of WFIRST's rogue planet discoveries, and they
cannot be obtained by another method.

We also propose continuous monitoring of the WFIRST microlensing fields
at a cadence of one observation per day
starting a year before and ending a year after the WFIRST microlensing survey.
This will allow us to constrain the presence of a host star for rogue planet candidates and,
if the host is detected, measure the planet-host mass ratio and projected separation.

A single LSST pointing, centered on the WFIRST microlensing
fields, would cover all 10 WFIRST microlensing fields.

For our preliminary estimates of the high-cadence observing, we assume
that the bulge is observed every 30 minutes when the bulge is at
an airmass of $< 2.5$ for 76-day observing runs (each 72-day WFIRST observing
season plus 2 days on either side). Each visit consists of 3 exposures,
one 2 sec exposure followed by two 15 sec exposures. With a 2 sec readout
and 1 sec for the shutter to open and close, this comes to 39 sec on target
per visit (since the final readout can be done while slewing).
If we assume a 30 deg slew in Azimuth before and after each microlensing pointing, the slews
to and from the target should take 22 sec, which is 12 sec above the average. So,
each visit will take 61 sec out of the regular observing sequence.
The number of observations per night, assuming a 30 minute cadence, for
a Spring 2025 observing session are given in \autoref{tab:wfirst_ml_survey}. We will require
that these observations be taken in the $riz$ or $y$ filters with at
least 3 (or 0) observations in each filter per night. The total number
of observations with this observing plan is 649 or 11.0 hour per
76-day observing session or 3894 observations and 66.0 hours for
all the high cadence observations that we propose.

\begin{table}
\begin{tabular}{ c c }
  {\bf Dates} & {\bf Number of observations per night}\\
\hline
Feb 10-16     &  3 \\
Feb 17-23     &  4 \\
Feb 24-Mar 1  &  5 \\
Mar 2-8       &  6 \\
Mar 9-14      &  7 \\
Mar 15-21     &  8 \\
Mar 22-28     &  9 \\
Mar 29-Apr 4  & 10 \\
Apr 5-10      & 11 \\
Apr 11-17     & 12 \\
Apr 18-24     & 13 \\
Apr 25-28     & 14 \\
\end{tabular}
\caption{Observations per night at 30 minute cadence for a Spring
WFIRST microlensing survey. This is also approximately the number of minutes required per night for exoposure time and overheads.}
\label{tab:wfirst_ml_survey}
\end{table}

These observing plans can be altered by changing the cadence of the high
cadence observations from once every 30 minutes to once every 15, 60,
or 120 minutes, or we could change the number of WFIRST microlensing
observing seasons that were covered. We have not yet simulated the
different observing cadences, however.

The low-cadence (1 observation per day) observations taken when WFIRST
is not observing, would consist of 1270 visits if we assume that
the observations are not taken during the time when the $u$ filter is
on the telescope (this is assumed to be 1/6 of the time). The low-cadence
off-season observations then total 21.5 hours. Low- and high-cadence
observations will take a total of 87.5 hours over 8 years.

% --------------------------------------------------------------------

\subsection{Metrics}
\label{sec:\secname:metrics}

\subsubsection{High-cadence observations}

\autoref{tab:wfirst_ml_results}
shows the results of our simulations of the combined proposed WFIRST-LSST
observing program. We assume that there is 1 planet per main sequence
star at each of $1\,M_\oplus$, $10\,M_\oplus$, and $100\,M_\oplus$.
%This is the 1-$\sigma$ lower limit found by \citet{2011Natur.473..349S} at
%$M_{\rm L} \approx 300\,M_\oplus$, and the rogue planet mass function is
%thought to increase toward lower masses, so this is a conservative
%assumption.
The first row gives the number of events that will be
observed by WFIRST. The second row gives the number of these events
with source SDSS-$i \leq 23$, which were the only events included in the
LSST simulations. The third and fourth rows give the number of these events
with LSST-WFIRST microlensing parallax measurements and the number
with full mass measurements. It is this row that indicates the
value of the LSST observations.

\begin{table}
\begin{tabular}{lcccc}
Category & $100\,M_\oplus$ & $10\,M_\oplus$ & $1\,M_\oplus$ & Total \\
\hline
WFIRST-events    &   417   &         127    &         33    &  577  \\
$i \leq 23$      &    88   &          30    &         13    &  131  \\
$\pi_{\rm E}$ measured &    22   &           8.2  &          2.7  &   32.9 \\
$M_{\rm L}$ measured   &    5.9  &           3.4  &          1.5  &   10.8 \\
\end{tabular}
\caption{Number of rogue planets of the given mass detected, assuming
one such planet per main sequence star, and our proposed LSST observation
program.}
\label{tab:wfirst_ml_results}
\end{table}

One can see from the final column that the LSST observations should
yield more than 30 rogue planet microlensing parallax measurements and
more than 10 rogue planet mass measurements.
These are measurements that cannot be made by other methods, as there is
currently no other known way to measure the mass distribution and abundance
of dark, isolated objects.
% \textbf{THE PREVIOUS SENTENCE NEEDS MORE JUSTIFICATION.}
%
In addition, this program would
also yield masses for a somewhat larger number of bound planets
\citep{2003ApJ...591L..53G}, although many of these will have their
masses determined by other means as well.

For a proxy for a Figure of Merit, we select the product of the numbers of
rogue planet microlensing parallax measurements $\pi_{\rm E}$
and rogue planet mass measurements $M_{\rm L}$.
% Added by PJM:
We expect this quantity to be simply related to the ultimate precision
on any rogue planet population hyperparameter we may try to infer from
the LSST-WFIRST sample.
Some additional
work is still needed to test and develop this metric so that it can be easily incorporated into the \MAF based on \OpSim runs.
The value of this product is 355 for our straw man program.

\subsubsection{Low-cadence observations}

LSST observations will improve models for WFIRST planetary events
in a number of ways but here we focus on detection of the planet
host peak if it was missed and poorly constrained by WFIRST. We simulate
a population of planets with masses of $0.1\,M_\oplus$, $1\,M_\oplus$,
and $10\,M_\oplus$ that follow logarithmic mass function with
normalization extrapolated from \citep{2012Natur.481..167C}. We narrow the sample of
planets detected by WFIRST in a number of steps. First, we reject the events
with host peak during the LSST high-cadence observations, i.e., during one of six
76-day long campaigns. Second, we exclude the events for which flux increase
during the WFIRST observing season is at least $0.75$ of the event maximum
flux increase, because in these cases WFIRST
data will strongly constrain the host peak even though it is not fully observed.
Third, we require at least a single LSST observation within $\pm0.05t_{\rm E}$ of
the host peak and the magnified source flux to be brighter than SDSS-$z$ of $23~{\rm mag}$.
The expected number of $0.1\,M_\oplus$, $1\,M_\oplus$, and $10\,M_\oplus$
planets for which host peak is detected are 2.1, 8.7, and 36.9, respectively.
We chose a sum of those as a proxy for a Figure of Merit. For the observing strategy
presented above the value of this Figure of Merit is 47.7. If a maximum airmass is chosen to be
2.0 instead of 2.5, then the requested time decreases by $6\%$ and Figure of Merit decreases by $12\%$.
The metric is based solely on the distribution of peak times and source magnitudes
for a given WFIRST simulation, so once this is set, the metric can
be easily computed for any \OpSim run.

% Quantifying the response via MAF metrics: definition of the metrics,
% and any derived overall figure of merit.

% % --------------------------------------------------------------------
%
% \subsection{OpSim Analysis}
% \label{sec:\secname:analysis}
%
% OpSim analysis: how good would the default observing strategy be, at
% the time of writing for this science project?
%
%
% % -------------------------------------------------------------------- % %

\subsection{Discussion}
\label{sec:\secname:discussion}

The rough estimates of our metrics and Figure of Merit given above were
carried out using a very simple model for the LSST observations we
have proposed. We would hope to refine this analysis using the output
of an \OpSim run where the WFIRST+LSST microlensing program was included
as an additional special survey, and our Figure of Merit coded in the
MAF. We do not expect the impact of this special survey on other science cases
to be high, as it would only need $12$--$24$ hours per year in total. The risk
seems similar, but smaller than that of a Deep Drilling Field.

WFIRST microlesning field will have the highest cadence among LSST bulge fields,
hence, should be significantly affect Figures of Merit for Milky Way,
galactic transients, and stellar variability.

% ====================================================================

\subsection{Conclusions}

Here we answer the ten questions posed in
\autoref{sec:intro:evaluation:caseConclusions}:

% Answered by DB 08/09/16,
% minor changes by RP 08/10/16, added by MP 08/15/16

\begin{description}

\item[Q1:] {\it Does the science case place any constraints on the
tradeoff between the sky coverage and coadded depth?}

\item[A1:] No: this is irrelevant to the microlensing survey. We have
specific cadence requirements and much less interest in coadded depth.

\item[Q2:] {\it Does the science case place any constraints on the
tradeoff between uniformity of sampling and frequency of sampling?}

\item[A2:] During the microlensing survey, we must maintain the
requested 30 or 15 minute cadence to within 3 minutes. Outside of the
microlensing survey, we must maintain our daily cadence to within 4
hours.

\item[Q3:] {\it Does the science case place any constraints on the
tradeoff between the single-visit depth and the number of visits
(especially in the $u$-band where longer exposures would minimize the
impact of the readout noise)?}

\item[A3:] There is no trade-off: we cannot reduce our cadence.

\item[Q4:] {\it Does the science case place any constraints on the
Galactic plane coverage (spatial coverage, temporal sampling, visits per
band)?}

\item[A4:] All of the observations must take place in a field that we
will specify near the Galactic Center, at Galactic coordinates of
roughly $l = 1$~deg, $b = -1.5$~deg.

\item[Q5:] {\it Does the science case place any constraints on the
fraction of observing time allocated to each band?}

\item[A5:] We  require observations to be in a passband at least as red
as the $r$-band, i.e. that 0\% of the observing time be in the $u$ or
$g$-bands.

\item[Q6:] {\it Does the science case place any constraints on the
cadence for deep drilling fields?}

\item[A6:] Our proposal may amount to a new deep drilling field in the
central Galactic bulge. In this case, the constraints would be 15 or 30
min cadence during 6 WFIRST microlensing seasons, and 1 day cadence
otherwise.

\item[Q7:] {\it Assuming two visits per night, would the science case
benefit if they are obtained in the same band or not?}

\item[A7:] We propose between 3 and 14 visits per night during the
WFIRST microlensing survey, and 1 visit per night otherwise. We have no
restrictions other than the restriction to no visits in the $u$ and
$g$-bands.

\item[Q8:] {\it Will the case science benefit from a special cadence
prescription during commissioning or early in the survey, such as:
acquiring a full 10-year count of visits for a small area (either in all
the bands or in a  selected set); a greatly enhanced cadence for a small
area?}

\item[A8:] We require many visits during the WFIRST exoplanet
microlensing seasons, starting in 2025. Before that, we require only one
observation per night.

\item[Q9:] {\it Does the science case place any constraints on the
sampling of observing conditions (e.g., seeing, dark sky, airmass),
possibly as a function of band, etc.?}

\item[A9:] There is value in relaxing a default airmass constraint,
where it would mean no observations were taken. We only have strict
cadence requirements.

\item[Q10:] {\it Does the case have science drivers that would require
real-time exposure time optimization to obtain nearly constant
single-visit limiting depth?}

\item[A10:] No.

\end{description}

% ====================================================================

\navigationbar

% --------------------------------------------------------------------

% ====================================================================
%+
% SECTION:
%    WFIRST_proposals.tex
%
% CHAPTER:
%    wfirst.tex
%
% ELEVATOR PITCH:
%    Maximizing the overlap between LSST and WFIRST is likely to be a fruitful
%    approach to modifying the LSST observing strategy. Let's pull together the
%    findings from the three WFIRST science cases and propose some OpSim
%    experiments.
%
%-
% ====================================================================
% 
% \section{Maximizing the Synergy between WFIRST and LSST}
% \def\secname{\chpname:proposals}\label{sec:\secname}
%
% \credit{jasondrhodes}
%
% In the previous sections, we introduced figures of merit for each WFIRST
% science project, and tested the existing LSST observing strategies for
% their performance. In the process we learned some of the shortcomings of the
% baseline LSST strategy, and suggested some alternative cadence
% options. In this section, we will pull those suggestions together to propose a
% suite of new \OpSim experiments.
%
% % Make table here.
%
% % ====================================================================
%
% \navigationbar

% --------------------------------------------------------------------

% --------------------------------------------------------------------

% ====================================================================
\chapter[Conclusions, Tensions and Trade-offs]{Conclusions, Tensions and Trade-offs}
\def\chpname{tradeoffs}\label{chp:\chpname}

Chapter editors:
\credit{ivezic},
\credit{StephenRidgway},
\credit{drphilmarshall}

In this chapter we summarize the findings of the science cases in this
white paper, and propose some actionable conclusions for the LSST Project.
We then discuss the tensions and tradeoffs apparent in these findings.

% ---------------------------------------------------------------------

\section{Summary of Cadence Constraints}

\credit{ivezic}

The authors of the preceding chapters' science cases provided guidelines
for improving the baseline LSST cadence, via their 10-question
conclusions sections (see \autoref{sec:intro:evaluation:caseConclusions}).  We summarize this input below, extracting a number
of recommendations for the Project to consider. The most important suggested action
items for the Project Team include i) implementation, analysis and
optimization of the ``rolling cadence'' idea, and ii) execution of a
{\it systematic} effort to further optimize the ultimate LSST cadence
strategy.

\begin{description}

\item[Q1:] {\it Increased sky coverage is possible, at the price of fewer visits per field and shallower
coadded depth.  Does the science case place any constraints on the
tradeoff between the sky coverage and depth? For example, should
the sky coverage be maximized (to $\sim$30,000 deg$^2$, as e.g., in
Pan-STARRS) or the number of detected galaxies (the current baseline of 18,000 deg$^2$)?}

A general conclusion from many science cases is that the sky coverage
can be increased only if the baseline single-visit and coadded depths
are not significantly degraded. For example, the number of galaxies above
some threshold photo-z quality would increase by about 8\% with the larger
sky area, but the redshift range would be diminished a bit. Milky Way mapping and large-scale
structure studies (\eg galaxy clustering, BAO) are
area-limited more than depth-limited and thus a larger sky coverage is
preferred for them. In time-domain cases where the LSST baseline performance is already
more than adequate, a larger sky coverage is preferred, too (AGNs,
extra-galactic cepheids); when cadence performance is barely sufficient
or insufficient, smaller sky area with better temporal sampling is suggested
(supernovae, strong lensing).

The Project would enable finer optimization by providing several more
simulations with the sky coverage in between the baseline cadence sky
area of 18,000 deg$^2$) and the so-called ``Pan-STARRS cadence'' with
$\sim$30,000 deg$^2$. MAF analysis should be extended to compute the
effective number of galaxies good for weak lensing, as well as to track
the performance of star-galaxy separation.

\item[Q2:] {\it Does the science case place any constraints on the
tradeoff between uniformity of sampling and frequency of  sampling? For
example, a rolling cadence can provide enhanced sample rates over a part
of the survey or the entire survey for a designated time at the cost of
reduced sample rate the rest of the time (while maintaining the nominal
total visit counts).}

Supernovae provide one of the strongest drivers for the implementation of
rolling cadence. A sampling rate about three times higher than the
uniform sampling implemented in the baseline cadence (revisit time scale of
about one day), and lasting 3-4 months, is suggested. It is likely that
such a rolling cadence would also be beneficial for constraining
asteroid orbits and for studying short time scale variables (e.g.
cataclysmic variables). Extreme cases of a rolling cadence, with roughly
week-long campaigns, for special sky regions would be beneficial for
studies of Young Stellar Objects, for example.

Other types of transients also require a rolling cadence, with necessary compromises.  A faster cadence in
bluer filters in a shorter rolling window, and slower cadence in redder filters, may help address the
trade-offs.

The production and analysis of several families of rolling cadence
simulations should be a high priority, because this
baseline cadence modification might provide more significant science
benefits than any other proposed modification.

\item[Q3:] {\it Does the science case place any constraints on the
tradeoff between the single-visit depth and the number of visits
(especially in the $u$-band where longer exposures would minimize the
impact of the readout noise)?}

A large number of science cases would benefit from deeper $u$ band data
(both single-visit and co-added depth), as long as the total number of visits
is not decreased (of course, increasing the number of $u$ band visits
would be beneficial for $u$ band survey science, too).

Simulated cadence \opsimdbref{db:DoubleUbandExptimeSameVisits}, which
doubled the $u$ band exposures from 30 sec to 60 sec while retaining the Baseline number of visits, demonstrated that an
improvement of $u$ band single-visit depth of 0.5 mag can be achieved
with only a minor loss of coadded depth and the number of visits in other bands.
We advocate taking $u$ band exposures that are long enough to be sky noise dominated.
The Project should investigate this option further and potentially
modify the baseline per-band exposure time allocation.

\item[Q4:] {\it Does the science case place any constraints on the
Galactic plane coverage (spatial coverage, temporal sampling, visits per
band)?}

A general comment made by science cases that depend on the Galactic
plane coverage is that Figures of Merit (FoM) exist, but that ``many
OpSim runs with a variety of temporal distributions of exposures within
the ten-year survey lifetime [are needed] to make these FOMs useful.''

The Project could consider producing a family of simulations that explore
different cadence strategies for the Galactic disk coverage.

\item[Q5:] {\it Does the science case place any constraints on the
fraction of observing time allocated to each band?}

Generally, the baseline time allocation is satisfactory. Potential
improvements depend on studied population (e.g. young stellar objects)
and the required imaging depth.
The case for deeper $u$ band data (both single-visit and coadded depth)
has the strongest drivers. The fraction of visits in red bands should be
higher for deep drilling fields than for the main survey.

When producing a new generation of deep drilling simulations, the
Project should revisit the per-band exposure time allocation for deep
drilling fields.

\item[Q6:] {\it Does the science case place any constraints on the
cadence for deep drilling fields?}

The strongest constraints for the deep drilling field (DDF) cadence come
from supernovae. DDFs should be preferrably visited nightly,
1-2 mag deeper than for the main survey visits. At least a few
extragalactic fields should be observed at any month, with
each field observed for a ``season'' of length at least 120-150 days,
staggering the fields so new fields cycle in as old ones cycle out, and
avoiding the moon. For
potential deep drilling fields within the inner Galactic plane, short
temporal baselines would be very useful. In the case of Young Stellar
Populations, for example, a single week of denser monitoring for specific regions is
suggested, with the goal of having several observations per night,
separated in time by at least an hour from one another. A more extensive study of the value of a Galactic deep drilling field is required.

\item[Q7:] {\it Assuming two visits per night, would the science case
benefit if they are obtained in the same band or not?}

Most science cases prefer the visits in a pair to be taken in two different bands.
Exceptions are GRBs and asteroids; in the case of GRBs the second visit is
needed to promptly establish variability and in the case of asteroids
different bands result in decreased sample completeness.

The Project could investigate whether it is possible to increase the
fraction of visit pairs with different bandpasses.

\item[Q8:] {\it Will the case science benefit from a special cadence
prescription during commissioning or early in the survey, such as:
acquiring a full 10-year count of visits for a small area (either in all
the bands or in a  selected set); a greatly enhanced cadence for a small
area?}

A lot of excellent suggestions were made by multiple science cases.
For example, photo-z analysis would be greatly assisted by acquiring
a full 10-yr count of visits for a small area during commissioning or
early in the survey, especially if these regions would be also covered by
existing spectroscopic surveys.
The Comissioning Science Verification Team should incorporate these
suggestions, when possible, into the Commissioning plan, as well
as to socialize the existing plan for ``mini surveys'' with Science
Collaborations.

\item[Q9:] {\it Does the science case place any constraints on the
sampling of observing conditions (e.g., seeing, dark sky, airmass),
possibly as a function of band, etc.?}

The strongest drivers are formulated for obtaining good seeing data (the
weak lensing case for good seeing in the $r$ and $i$ bands is already
implemented in the simulator) and for minimizing various correlations
(e.g. parallax factor vs. hour angle).

The Project could analyze the impact of secular effects on randomizing
various correlations vs. an explicit algorithmic driver to do so
implemented in the Scheduler.

\item[Q10:] {\it Does the case have science drivers that would require
real-time exposure time optimization to obtain nearly constant
single-visit limiting depth?}

No strong driver for real-time exposure time optimization was
identified.

The Project could monitor the fate of a similar proposal considered by
the upcoming Zwicky Transient Facility.

\end{description}

\navigationbar

% --------------------------------------------------------------------

\section{Tensions and Tradeoffs}

\credit{StephenRidgway}

The LSST survey will be carried out with physical and operational
constraints that will impact all science objectives.  These include
design limitations, such as the aperture of the telescope, the detector
noise and readout time, and the limited number of filter changes that
can be supported during the lifetime of the survey.  They include
natural constraints, such as the quantity of useful observing time
in 10 years, and system optimization constraints, such as
exposure time and sky coverage.

The LSST observing schedule can be designed, to some extent, to minimize
the impact that these limitations have on any one or few science
objectives. As the science objectives become more numerous and more
complex, the optimization becomes more difficult and the chances
increase that significant compromises may be required.

In the science chapters of this report, science objectives are
described, and for each, diagnostic metrics, and in some cases figures
of merit, are designed to represent quantitatively the interests of that
topic in a schedule optimization.  Not all of these metric sets are
fully worked out, and in most cases they are provisional pending further
analysis and community review and input.  However, they do already
suffice to bring attention to many special requirements.

The design of the LSST scheduler, and of the algorithms that will select
the visit sequences, has a considerable distance to go before hard
cadence questions must be confronted and resolved.  However, it is
already possible to survey the reach of science needs, and to identify
areas of competition which may become candidates for careful trades and
decisions in the years before the survey begins.

In the rest of this chapter, we review the possible tensions that are
now evident, and where tradeoffs may become necessary. Most potential
tensions within the main survey concern {\it temporal sampling for
variable targets}.  Tensions among static science objectives, and
between static science and variable science, may be less likely and
mild. The strongest points of tension {\it may} turn out to be between
mini-surveys and the main survey; similarly, rolling cadences (\autoref{sec:rolling}) are also likely to introduce new tensions.

% --------------------------------------------------------------------

\subsection{Discussion: Variable Targets -- Where's The Tension?}

Strictly periodic targets are relatively neutral to cadence speed, since
successive periods can be combined to improve phase coverage steadily
through the survey.  The only odd cases are ultra-short ($<~ $1 minute)
and ultra long ($> $10 years) periods, and periods very, very close to
one sidereal day.  However, with precision measurements over a long
term, some of the very interesting results for periodic variables will
be in period drifts or slight deviations from periodicity. Furthermore,
even periodic targets benefit from an early interval of higher frequency
sampling, at least in some sky regions, as this can accelerate the ramp
up of the science.

However, by far the majority of variables and transients, stellar and galactic,
are not periodic. The analysis of their aperiodic light curves will be greatly simplified (or may
absolutely require) sufficient sampling within a characteristic interval that depends
on the target type. One would like to satisfy the sampling theorem, with
visits at twice the frequency of the highest frequency content, but this
is only a conceptual guide: knowledge of, and experience with, the
targets and the science objectives can provide practical criteria.

A truly uniform cadence provides a revisit rate of one visit pair every
16 days (in the $r$ or $i$ bands), or every 3.7 days (in any filter) --
assuming a 5 month observing season.  This is a sparse sample rate for
many variable types.  Achieving higher sample rates requires (possibly
very strong) deviations from complete uniformity.  Thus the obvious
conclusion: rapid cadences cannot apply everywhere all the time. Rapid
cadences must be designed, executed selectively, bounded by the number
of visits available, and coordinated with all other such cadences, as
well as more general survey requirements.

\subsubsection{Examples}

Several examples will illustrate the diversity of cadences that are
represented in the science programs described in this white paper.

QSO variability tends to be stronger in low temporal frequencies   A uniform distribution of the
LSST visits, with minimal seasonal gaps, provides fairly good support
for identifying QSOs from their variability pattern.

For a SN, sufficient sampling must be acquired during the life of the
event. A good cadence in at least one filter is required to support
classification, and multiple colors to support photometric redshift
determination.  A uniform LSST cadence, even with large seasonal gaps,
does not provide a sufficient sampling rate for SN science - an
enhancement of a factor of 2 or greater is strongly requested.

To determine the rotational period of stars with spots, sampling must
resolve light variations sufficiently to constrain periodicity {\it
within} the spot lifetime, which is typically several weeks. This
cadence is much more rapid than provided by a uniformly distributed WFD
visit pattern.

Flaring stars and interacting binaries, and new fast transients in general, may show dramatic flux changes
in minutes to hours, and correct identification of such events may
require several data points, and possibly more than one filter, on a
similar time scale.

The solar system small body case is particularly complex.  The science
is one of the main LSST drivers.  Detection of PHAs has a non-scientific
and even political component. Asteroid confusion can interfere with
transient discovery. The density of targets is a strong function of
position on the sky.  Characterization of solar system objects, by
determination of orbits, requires visit patterns on short timescale
($\approx$ hour return) and intermediate time scale ($\approx$2 weeks) -
long timescale confirmation occurs naturally later in the survey.  The
number and pattern of rapid revisits required for positive
identification depends strongly on the false positive rate.

\subsubsection{How to provide a range of cadence speeds}

The problem of sampling diverse events was of course recognized very
early in survey planning. Previous cadence development has explored the
following special cadence options:

\begin{description}

\item{Rapid revisits} -- this feature was introduced for study of solar
system bodies and most schedule simulations give high priority to
acquiring visits in pairs with $~$30 minute separation.  Experiments
have been done with triples, in case that should prove necessary for
asteroid characterization.  The possibility of using a different revisit
pattern in different parts of the sky (e.g. less frequent away from the
ecliptic) has been mentioned but has not yet been implemented in \OpSim.  Different patterns for different filters
(e.g. not using pairs for visits in the $u$ and $y$ bands) have been
suggested and investigated. The use of visit pairs is clearly a very
high impact decision, for practical purposes reducing by a factor of 2
the effective revisit rate for other targets. However, rapid pairs are
very effective for measuring brightness gradients for rapidly varying
objects, and thus particularly valuable for the difficult problem of
characterizing blank sky, short timescale, transients.

\item{Mini-surveys} --  these can include special cadences. The deep
drilling concept utilizes rapid visit sequences to achieve greater depth
without saturation of detector wells, giving sky-limited true time
series with $\approx$30 second sampling steps.  The possibilities for
mini-surveys are limitless, but of course they are bounded by the amount
of time available outside the main survey. The trade between
mini-surveys and the main survey is discussed below.

\item{Rolling cadence} -- this would allow for the possibility of
re-deploying visits within the main survey so as to respond to special
cadence demands without compromising main survey goals (or, perhaps in
principle, trading against main survey goals in a measured and optimum
way). Rolling cadences were introduced in \autoref{sec:rolling}.  As an example, the
average 9 visit pairs per year in the $r$ filter, which would be
distributed over a season in a uniform survey, could be distributed over
2 months, 1 month, 1 week, or 1 day, in a rolling cadence (leaving no
visits in $r$ for the rest of the season).  Or, more conservatively for
the main survey, half (4-5 visit pairs) could be spent in an enhanced
visit rate, reserving the other half to maintain visit pairs in the rest
of the season.  Also, a rolling cadence can concern any number of
filters; for example, one filter could be used to provide short bursts
of rapid sampling, while other filters could maintain a uniform
distribution.  In particular, for discovering transients a simulation with faster cadence in bluer filters in a shorter rolling window and slower cadence in redder filters in a longer rolling window may help address many of the trade-offs involved. Different rolling cadences can be used in different parts
of the sky, or at different times during the survey.  There is an
immense range of possibilities for rolling cadence, and the surface has
barely been scratched: \OpSim 4 should allow this approach to be explored in detail.

\item{Commissioning survey} -- the highest priority of the commissioning
schedule is, of course, commissioning the hardware and software of the LSST system.  However, a secondary objective
is to demonstrate system operation in the planned survey mode -
presumably including main survey, deep drilling, rolling cadences, etc.
There have been other suggestions, going beyond these basics, such as
integrating some fields to the full survey depth, or acquiring some
special cadences.  However, there is no assurance that all of these will
be possible, and they will be prioritized following formal commissioning
objectives.

\end{description}

\subsubsection{Other options for special cadences}

\begin{description}

\item{Pre-survey options} -- there are a number of survey instruments
(\eg CTIO, CFHT, Subaru) that can easily reach the single visit depth of
LSST. These resources could be used to explore limited sky regions
($\approx$1\% of the LSST sky) with cadences that are planned for LSST
(or cadences that are not planned for LSST), providing touchstone
datasets especially for the more common target types that will dominate
the survey.

\item{Twilight survey} -- \autoref{sec:shortexp} describes a concept for
twilight data acquisition, using short exposures to tolerate bright sky and extend the dynamic range of the survey overall.
This time is not required for current LSST science, and in principle
could be allocated to $z$, $y$ filters in short bursts (20 minutes) of
fast cadence ($<~$15 seconds) imaging, within the sensitivity limits of
the twilight sky.

\item{Follow-up} -- LSST is, in large part, a discovery machine. It is
not realistic to expect LSST to provide its own follow-up for {\it all}
possible target types and characteristics. Fortunately, many of the most
useful discoveries will be bright enough to follow-up with smaller, more
accessible apertures.  Follow-up can be far more customized to the
science needs than a general purpose LSST cadence.  Faint targets of
sufficient value may likewise merit followup with exceptional  ground
and space-based resources  \citep[see][for analysis and discussion]{NajitaEtal2016}.

\end{description}

\subsubsection{Frequency of filter changes}

Multi-color visits are a very special case for LSST.  Changing the
huge optical filters will be a substantial mechanical task.
While filter change time is not fully characterized, it will be slow
enough (about 2 minutes) that filter change frequency will compete directly
with efficiency. Also, the mechanisms have a finite lifetime: ``rapid''
multi-color sequences will be the exception rather than the rule.

Many science objectives for variable targets would be best served by
simultaneous (or nearly so) color information.   It is important to identify
when and if rapid filter changes are of value or essential, and to trade this
against the efficiency and limited lifetime of the filter change mechanism.

\subsubsection{Tension between rapid and slow cadences}

In summary, we can readily identify competing demands for very different
cadences, including fast cadences in multiple ranges. For characteristic
times ranging from  $\lesssim$1 minute to $\sim$1 month, a uniform visit
distribution cannot be fully satisfactory, and in some cases it may turn
out to be totally unsatisfactory.  A number of concepts for alternate
cadences are available.  None can provide rapid cadence all the time
over all the sky. It may be possible to provide cadences matched to most
requirements over part of the sky all of the time, and over all of the
sky at some time. For transient targets, a complex survey cadence may
obtain limited duration but ``appropriate'' sampling of a fraction of
the actual events, with the exact fraction still to be decided but
inevitably significantly less than one.

The tension in scheduling is between science objectives and limited
scheduling flexibility. The confrontation between science requirements
and schedule performance leans on the metrics and merit functions that
are the major goal of this white paper.  It should be clear from the
foregoing chapters that the difficult goal of metrics analysis is not in
describing sampling for the science, which is ``easy.''  The more
difficult part is in determining the number of science targets for which
adequate sampling can be provided by a simulation, and perhaps the
greatest challenge is determining how many such targets with the
specified sampling are required for a science objective.  It is only
when this step has been accomplished for a large part of the science
that competition between the science objectives can become a
quantitative process.

% --------------------------------------------------------------------

\subsubsection{Discussion: Static Target Science -- Is There Any Tension?}

The needs of static target science appear to have fewer points of
potential tension among them than variable targets.  The major
cadence-related concerns are:

\begin{description}

\item{Photometry} -- the best photometric performance will be achieved
after the calibration has been closed around the sky, with superior image
quality and superior photometric quality visits to every field.  The
sooner that it is achieved during the survey, the sooner high quality
photometry will be available.  This could be a target of active schedule
control, with corresponding decreasing flexibility in some other
schedule variables.

\item{Astrometry} -- both proper motions and parallaxes are served by
any schedule that spreads visits well over the duration of the survey.
Parallaxes benefit from observing at a range of hour angles (for control of HA-dependent biases),
which is slightly in competition with the preference to observe at small airmass
for best image quality, but typical simulated schedules show good
astrometric performance. A rolling cadence that moved a significant
fraction of visits from one time period of the survey to another could
impact the astrometric performance (either for better or for worse),
though as long as the fraction of visits concerned was small, the effect
would correspondingly be small.

\item{Homogeneity} -- an  example of required homogeneity is image
quality. Just as the atmosphere offers a range of image quality during a
night and from night to night, each point on the sky will be observed
with a range of image quality.  To enable understanding of selection
effects, and to compare sky regions on an even playing field, it is
desirable that for each filter, all parts of the sky should be observed
with a similar distribution of image quality, and in particular with
similar best image quality. Achieving homogeneity of conditions actively
could be quite challenging, but simulations show that with a large
number of independent visits, it occurs naturally to good approximation.
(In \autoref{chp:cadexp} we saw that some higher order declination and seasonal effects persist in the current Baseline Cadence: these are expected to be reduced further with \OpSim~4.)
Any cadence that relied on concentrated bursts of visits in a short
interval would tend to reduce the spread of conditions observed.
However, such an extreme has not been proposed or studied.

\item{Randomization} -- closely related to homogeneity, randomization is
means of achieving homogeneity in some observing parameters.  Examples
are the projected angles of camera and telescope optics on the sky.
These are less random than sky conditions, as they depend on instrument
setup and schedule history.  Simulations show that optics angles are
well randomized passively (i.e. without scheduler optimization) for most
points on the sky, but not for all.  Randomization could be improved,
for example by actively running the camera rotator when advancing from
one sky position to the next, in order to populate under represented
camera angles. The rotation takes time, and could reduce the overall
efficiency of the survey.  Only simulations can explore the impact of
these additional mechanical motions.

\item{Dithering} -- dithering of visits is a powerful method of
improving homogeneity of sky coverage passively. Few compromises have
been identified with dithering thus far.  Dithering for small regions
has a price. Imagine the loss in depth due to large dithers with a
single FOV, e.g. a deep drilling field (this has not been proposed).
Due to this effect, certain rolling cadences can have a potential small
loss of efficiency or efficacity when implemented with dithering.

\end{description}

The foregoing shows that within the static science domain, there are few
and mild points of tension and potential competition.

% --------------------------------------------------------------------

\subsection{Discussion: Tension Between Static And Variable Science}

For the most part, the tensions between static science and variable
science are modest and easily understood.  A variable-driven cadence
that requires special timing of visits may result in loss of efficiency
due to increased slew times, or observing under less optimum conditions
(larger airmass).  Special cadences are likely to reduce randomization
and homogenization to a small degree. However, except for very
aggressive cadence implementations, these are second-order effects.
Furthermore, they are readily measurable with simple metrics: the impact of
variable science schedule considerations on static science should be
small, although this situation may well change under a rolling cadence.

% --------------------------------------------------------------------

\subsection{Discussion: Mini-surveys and the Main Survey -- Tension for Sure}

The LSST proposal and current plan allow a fraction of the total survey
time for mini-surveys. These may cover either special sky regions,
special cadences, or both. The fraction 10\% has been carried for
mini-surveys, but this is not a sacred number. In
\autoref{chp:cadexp}, current scheduling experience has shown that the
main survey program, to design depth, can be accomplished in $\approx$
85\% of the available time. However, improvements in simulations could
move this estimate up or down. Adequacy of the design depths could be
reconsidered.  And of course the execution of the survey could encounter
unprecedented circumstances.

Proposals for mini-surveys include deep drilling fields, the northern
ecliptic, and the Magellanic clouds. Notional suggestions for deep
drilling fields alone would exceed 10\%.   Most schedule simulations
have allocated a limited number of visits to the Galactic plane (based
on the expectation that crowding would limit the useful stacked depth).
However, as detailed in \autoref{chp:galaxy}, many areas of Galactic
science could benefit from a more aggressive visit plan (as they are not driven by stacked image depth), perhaps similar
to the main survey.

At present there is no evidence that the trade between main and
mini-surveys will require difficult compromises.  But it is a natural
area of tension, and since is likely to be exacerbated by weather disruptions (not to
mention the evolution of the science), it is likely to be with us through
the life of the survey.

% --------------------------------------------------------------------

\subsection{Final Thoughts}

The likely points of technical and scientific tension in scheduling are
apparent from the schedule simulation experience (\autoref{chp:cadexp})
and the science objectives and metrics
(\autoref{chp:solarsystem}--\autoref{chp:specialsurveys}).  Static
science has relatively few and mild points of concern.  Variable science
has little and moderate tension with static science, although this situation may change with rolling cadences.  Variable science
has many points of tension between different variable science
objectives, owing to the wide range of time scales. These lead to
contrasting technical demands, and may or may not prove to be areas of
scientific competition.

% The essential information needed to clarify tensions is the
% determination, for each science objective, of the following three things:
% \begin{itemize}
% 	\item The number of such targets required for the science;
% 	\item For a simulated schedule, the number of instances of targets satisfactorily observed;
% 	\item The cadence requirements.
% \end{itemize}
% This information is the key to calibrating the metrics in terms of
% absolute and relative  sufficiency.

The science teams are urged to continue to define and refine metrics, and in particular their all-important Figures of Merit, and the \MAF developers to provide
tools for the large-scale comparison of simulated schedules by whole suites of metrics. \OpSim~4 will allow  exploration of the kinds of rolling cadences preferred by various science teams, and it will be important to be able to evaluate these more flexible simulations of the LSST observing strategy across the whole range of science projects, so that the prioritization decisions that need to be made are as well-informed as possible.

% ====================================================================

\navigationbar

% --------------------------------------------------------------------

\section*{Acknowledgements}

The authors of this paper are a subset of the membership of the LSST
Science Collaborations: we thank our colleagues for many useful
discussions on the topic of LSST observing strategy, and the LSST
Corporation for providing financial support for our various
collaboration meetings during which so many of these conversations take
place. We are very grateful to Tim Jenness (LSST Publication Board
Chair), Michael Reuter, Chris Stubbs, and Sandrine Thomas for arranging
and carrying out the Project review of the introductory, \OpSim, and
conclusions chapters of this paper, and to the members of the Science
Advisory Council for reviewing the science case sections: Timo Anguita,
Niel Brandt, Jason Kalirai, Mansi Kasliwal, David Kirkby, Charles Liu,
Renu Malhotra, Nelson Padilla, An\^{z}e Slosar, and Michael Strauss. We
acknowledge the support of the LSST Project in the form of three
``cadence workshops'' (in Phoenix in August 2015, Bremerton in August
2016, and Tucson in November 2016), and for the creation and maintenance of
the \OpSim and \MAF tools. This paper was, and is still being,
researched and written collaboratively using the GitHub software
development platform.

% --------------------------------------------------------------------

\appendix

\setcounter{chapter}{0}
\chapter*{Appendix: \MAF Metrics}
\def\chpname{metrics}\label{chp:\chpname}
\addcontentsline{toc}{section}{Appendix: \MAF Metrics}
\markboth{}{}

This appendix contains two tables, listing the titles and one-line summaries of each metric provided with the \MAF package (\autoref{tab:maf_metrics}) and contributed by the community (\autoref{tab:contrib_metrics}). The Python code for both the \MAF metrics amd the contributed metrics can be found on GitHub.\footnote{\MAF: \url{https://github.com/lsst/sims_maf/tree/master/python/lsst/sims/maf}
\newline\noindent\code{sims\_maf\_contrib}: \url{https://github.com/LSST-nonproject/sims\_maf\_contrib/tree/master/mafContrib}
\newline\noindent Documentation at \url{ https://sims-maf.lsst.io/}}

\begin{table}[!h]
\scriptsize
\setcounter{table}{0}
\makeatletter
\renewcommand{\thetable}{A.\arabic{table}}
\caption{Metrics provided by the
\href{https://github.com/lsst/sims_maf/tree/master/python/lsst/sims/maf}{\MAF package}.
\label{tab:maf_metrics}}
\medskip
\begin{tabular}{ll}
\hline
                Metric Name &                                                                       Description \\
\hline\hline
 AccumulateCountMetric &  Calculate the number of visits over time. \\
 AccumulateM5Metric &  The 5-sigma depth accumulated over time. \\
 AccumulateMetric &  Calculate the accumulated stat. \\
 AccumulateUniformityMetric &  Make a 2D version of UniformityMetric. \\
 ActivityOverPeriodMetric &  Count the fraction of the orbit that receive observations, such that activity is \\
  &  detectable. \\
 ActivityOverTimeMetric &  Count the time periods where we would have a chance to detect activity on a \\
  &  moving object. \\
 AveGapMetric &  Calculate the gap between consecutive observations, in hours. \\
 AveSlewFracMetric &  Average time for slew activity, multiplied by percent of total slews. \\
 BaseMetric &  Base class for the metrics. \\
 BaseMoMetric &  Base class for the moving object metrics. \\
 BinaryMetric &  Return 1 if there is data. \\
 ChipVendorMetric &  Examine coverage with a mixed chip vendor focal plane. \\
 Coaddm5Metric &  Calculate the coadded m5 value at this gridpoint. \\
 ColorDeterminationMetric &  Identify SS objects which could have observations suitable to determine colors. \\
 CompletenessMetric &  Compute the completeness and joint completeness of requested observations. \\
 CountMetric &  Count the length of a simData column slice. \\
 CountRatioMetric &  Count the length of a simData column slice, then divide by a normalization \\
  &  value. \\
 CountSubsetMetric &  Count the length of a simData column slice which matches ``subset". \\
 CountUniqueMetric &  Return the number of unique values \\
 CrowdingMagUncertMetric &  Given a stellar magnitude, calculate the mean uncertainty on the magnitude from \\
  &  crowding. \\
 CrowdingMetric &  Calculate whether the coadded depth in r has exceeded the confusion limit. \\
\end{tabular}
\end{table}

\begin{table}[!t]
\scriptsize
\setcounter{table}{0}
\makeatletter
\renewcommand{\thetable}{A.\arabic{table}}
\caption{Metrics provided by the
\href{https://github.com/lsst/sims_maf/tree/master/python/lsst/sims/maf}{\MAF package}, continued...}
\medskip
\begin{tabular}{ll}
\hline
            Metric Name &                                                                       Description \\
\hline\hline
  &  observations. \\
 DiscoveryChancesMetric &  Count the number of discovery opportunities  for an SS object. \\
 DiscoveryMetric &  Identify the discovery opportunities for an SS object. \\
 Discovery\_EcLonLatMetric &  Returns the ecliptic lon/lat and solar elongation (deg) of the i-th SS object \\
  &  discovery opportunity. \\
 Discovery\_N\_ChancesMetric &  Count the number of discovery opportunities for SS object in a window between \\
  &  nightStart/nightEnd. \\
 Discovery\_N\_ObsMetric &  Calculates the number of observations of SS object in the i-th discovery track. \\
 Discovery\_RADecMetric &  Returns the RA/Dec of the i-th SS object discovery opportunity. \\
 Discovery\_TimeMetric &  Returns the time of the i-th SS object discovery opportunity. \\
 Discovery\_VelocityMetric &  Returns the sky velocity of the i-th SS object discovery opportunity. \\
 FftMetric &  Calculate a truncated FFT of the exposure times. \\
 FilterColorsMetric &  Calculate an RGBA value that accounts for the filters used up to time t0. \\
 FracAboveMetric &  Find the fraction above a certain value. \\
 FracBelowMetric &  Find the fraction below a certain value. \\
 FullRangeAngleMetric &  Calculate the full range of an angular (radians) simData column slice. \\
 FullRangeMetric &  Calculate the range of a simData column slice. \\
 HighVelocityMetric &  Count the number of observations with high velocities. \\
 HighVelocityNightsMetric &  Count the number of nights with nObsPerNight number of high velocity detections. \\
 HistogramM5Metric &  Calculate the coadded depth for each bin (e.g., per night). \\
 HistogramMetric &  A wrapper to stats.binned\_statistic. \\
 HourglassMetric &  Plot the filters used as a function of time. \\
 IdentityMetric &  Return the input value itself. \\
 InterNightGapsMetric &  Calculate the gap between consecutive observations between nights, in days. \\
 IntraNightGapsMetric &  Calculate the gap between consecutive observations within a night, in hours. \\
 KnownObjectsMetric &  Identify SS objects which could be classified as ``previously known" based on \\
  &  their peak V magnitude. \\
 LightcurveInversionMetric &  Identify SS objects which would have observations suitable to do lightcurve \\
  &  inversion. \\
 LongGapAGNMetric &  Compute the max delta-t and average of the top-10 longest observation gaps. \\
 MagicDiscoveryMetric &  Count the number of SS object discovery opportunities with very good software. \\
 MaxMetric &  Calculate the maximum of a simData column slice. \\
 MaxPercentMetric &  Return the percent of the data which has the maximum value. \\
 MaxStateChangesWithinMetric &  Compute the maximum number of changes of state that occur within a given \\
  &  timespan. \\
 MeanAngleMetric &  Calculate the mean of an angular (radians) simData column slice. \\
 MeanMetric &  Calculate the mean of a simData column slice. \\
 MeanValueAtHMetric &  Return the mean value of a metric at a given H. \\
 MedianAbsMetric &  Calculate the median of the absolute value of a simData column slice. \\
 MedianMetric &  Calculate the median of a simData column slice. \\
 MinMetric &  Calculate the minimum of a simData column slice. \\
 MinTimeBetweenStatesMetric &  Compute the minimum time between changes of state in a column value. \\
\end{tabular}
\end{table}

\begin{table}[!t]
\scriptsize
\setcounter{table}{0}
\makeatletter
\renewcommand{\thetable}{A.\arabic{table}}
\caption{Metrics provided by the
\href{https://github.com/lsst/sims_maf/tree/master/python/lsst/sims/maf}{\MAF package}, continued...}
\medskip
\begin{tabular}{ll}
\hline
                   Metric Name &                                                                       Description \\
\hline\hline
MoCompletenessAtTimeMetric &  Calculate the completeness (relative to the entire population) <= a given H as a \\
  &  function of time, \\
 MoCompletenessMetric &  Calculate the completeness (relative to the entire population), given the counts \\
  &  of discovery chances. \\
 NChangesMetric &  Compute the number of times a column value changes. \\
 NNightsMetric &  Count the number of distinct nights an SS object is observed. \\
 NObsMetric &  Count the total number of observations where an SS object was ``visible". \\
 NObsNoSinglesMetric &  Count the number of observations for an SS object, but not if it was a single \\
  &  observation on a night. \\
 NRevisitsMetric &  Calculate the number of (consecutive) visits with time differences less than dT. \\
 NStateChangesFasterThanMetric &  Compute the number of changes of state that happen faster than ``cutoff". \\
 NightPointingMetric &  Gather relevant information for a single night to plot. \\
 NormalizeMetric &  Return a metric values divided by ``normVal". \\
 NoutliersNsigmaMetric &  Calculate the number of visits outside the given sigma threshold. \\
 ObsArcMetric &  Calculate the time difference between the first and last observation of an SS \\
  &  object. \\
 OpenShutterFractionMetric &  Compute the fraction of time the shutter is open compared to the total time the \\
  &  dome is open. \\
 OptimalM5Metric &  Compare the co-added depth of the survey to one where all the observations were \\
  &  taken on the meridian. \\
 PairMetric &  Count the number of pairs that could be used for Solar System object detection. \\
 ParallaxCoverageMetric &  Check how well the parallax factor is distributed. \\
 ParallaxDcrDegenMetric &  Compute parallax and DCR displacement vectors to find if they are degenerate. \\
 ParallaxMetric &  Calculate the uncertainty in a parallax measures given a serries of \\
 PassMetric &  Just pass the entire array. \\
 PeakVMagMetric &  Pull out the peak V magnitude of all observations of an SS object. \\
 PercentileMetric &  Find the value of a column at a given percentile. \\
 PhaseGapMetric &  Measure the maximum gap in phase coverage for observations of periodic \\
  &  variables. \\
 ProperMotionMetric &  Calculate the uncertainty in the fitted proper motion assuming Gaussian errors. \\
 RadiusObsMetric &  Find the radius a point falls in the focal plane. \\
 RapidRevisitMetric &  Calculate uniformity of time between consecutive visits on short timescales (for \\
  &  RAV1). \\
 RmsAngleMetric &  Calculate the standard deviation of an angular (radians) simData column slice. \\
 RmsMetric &  Calculate the standard deviation of a simData column slice. \\
 RobustRmsMetric &  Use the inter-quartile range of the data to estimate the RMS. \\
 SlewContributionMetric &  Average time a slew activity is in the critical path. \\
 StarDensityMetric &  Interpolate the stellar luminosity function to return the number of \\
 SumMetric &  Calculate the sum of a simData column slice. \\
 TableFractionMetric &  Compute a table for the completeness of requested observations. \\
 TeffMetric &  Effective exposure time for a given set of visits based on fiducial 5-sigma \\
  &  depth expectations. \\
 TemplateExistsMetric &  Calculate the fraction of images with a previous template image of desired \\
  &  quality. \\
\end{tabular}
\end{table}

\begin{table}[!ht]
\scriptsize
\setcounter{table}{0}
\makeatletter
\renewcommand{\thetable}{A.\arabic{table}}
\caption{Metrics provided by the
\href{https://github.com/lsst/sims_maf/tree/master/python/lsst/sims/maf}{\MAF package}, continued...}
\medskip
\begin{tabular}{ll}
\hline
                   Metric Name &                                                                       Description \\
\hline\hline
 TgapsMetric &  Histogram up all the time gaps. \\
 TotalPowerMetric &  Calculate the total power in the angular power spectrum between lmin/lmax. \\
 TransientMetric &  Calculate what fraction of the transients would be detected. \\
 UniformityMetric &  Calculate how uniformly observations are spaced in time. \\
 UniqueRatioMetric &  Return the number of unique values divided by the total \\
 ValueAtHMetric &  Return the metric value at a given H value. \\
 VisitGroupsMetric &  Count the number of visits per night within deltaTmin and deltaTmax. \\
 ZeropointMetric &  Return a metric values with the addition of zeropoint. \\
 fOAreaMetric &  Metric to calculate the FO Area. \\
\hline
\end{tabular}
\end{table}

\begin{table}[!t]
\scriptsize
\setcounter{table}{1}
\makeatletter
\renewcommand{\thetable}{A.\arabic{table}}
\caption{Community-contributed MAF metrics in the \href{https://github.com/LSST-nonproject/sims\_maf\_contrib/tree/master/mafContrib}{\code{sims\_maf\_contrib}  package}.\label{tab:contrib_metrics}}
\medskip
\begin{tabular}{ll}
\hline
                 Metric Name &                                                                       Description \\
\hline\hline
 AngularSpreadMetric &  Compute the angular spread statistic which measures uniformity of a distribution \\
  &  angles accounting for 2pi periodicity. \\
 CampaignLengthMetric &  The campaign length, in seasons. \\
 GRBTransientMetric &  Detections for on-axis GRB afterglows decaying as \\
 GalaxyCountsMetric &  Estimate the number of galaxies expected at a particular coadded depth. \\
 GalaxyCountsMetric\_extended &  Estimate galaxy counts per HEALpix pixel. Accomodates for dust extinction, \\
  &  magnitude cuts, \\
 MeanNightSeparationMetric &  The mean separation between nights within a season, and then the mean over the \\
  &  campaign. \\
 NumObsMetric &  Calculate the number of observations per data slice. \\
 PeriodDeviationMetric &  Measure the percentage deviation of recovered periods for pure sine wave \\
  &  variability (in magnitude). \\
 PeriodicMetric &  From a set of observation times, uses code provided by Robert Siverd (LCOGT) to \\
  &  calculate the spectral window function. \\
 PeriodicStarMetric &  At each slicePoint, run a Monte Carlo simulation to see how well a periodic \\
  &  source can be fit. \\
 RelRmsMetric &  Relative scatter metric (RMS over median). \\
 SEDSNMetric &  Computes the S/Ns for a given SED. \\
 SNMetric &  Calculate the signal to noise metric in a given filter for an object of a given \\
  &  magnitude. \\
 SeasonLengthMetric &  The mean season length, in months. \\
 StarCountMassMetric &  Find the number of stars in a given field in the mass range fainter than \\
  &  magnitude 16 and bright enough to have noise less than 0.03 in a given band. M1 \\
  &  and M2 are the upper and lower limits of the mass range. ``band" is the band to \\
  &  be observed. \\
 StarCountMetric &  Find the number of stars in a given field between D1 and D2 in parsecs. \\
 TdcMetric &  Combine campaign length, season length, and mean night speartion into a single \\
  &  metric. \\
 ThreshSEDSNMetric &  Computes the metric whether the S/N is bigger than the threshold in all the \\
  &  bands for a given SED \\
 TransientAsciiMetric &  Based on the transientMetric, but uses an ascii input file and provides option \\
  &  to write out lightcurve. \\
 TripletBandMetric &  Find the number of ``triplets" of three images taken in the same band, based on \\
  &  user-selected minimum and maximum intervals (in hours), \\
 TripletMetric &  Find the number of ``triplets" of three images taken in any band, based on user- \\
  &  selected minimum and maximum intervals (in hours), \\
\hline
\end{tabular}
\end{table}

% --------------------------------------------------------------------

\bibliographystyle{yahapj}
\bibliography{references}

% --------------------------------------------------------------------

\end{document}

% ====================================================================